\documentclass[12pt]{article}
\pdfoutput=1
\usepackage{geometry,comment,bm,url,hyperref,makeidx}
\usepackage{cite,graphicx,amsmath,amssymb,bbm}
\usepackage{graphicx,graphbox}
\usepackage{multicol,slashed}
\usepackage[usenames,dvipsnames]{xcolor}

\newcommand{\be}{\begin{equation}}
\newcommand{\ee}{\end{equation}}
\newcommand{\bea}{\begin{eqnarray}}
\newcommand{\eea}{\end{eqnarray}}

\newcommand{\ol}{\overline}

\numberwithin{equation}{section}

\begin{document}

\thispagestyle{empty}

\vspace*{1cm}

\begin{center}

{\LARGE \bf Lectures on Naturalness,\\[.3cm]
String Landscape and Multiverse}

\vspace{1.5cm}
{\large Arthur Hebecker}\\

\vspace{0.5cm}

{\it Institute for Theoretical Physics, Heidelberg University, Philosophenweg 19,\\[.1cm]
 D-69120 Heidelberg, Germany}

\vspace*{0.5cm}

\large{24 August, 2020}

\vspace{1.5cm}

\begin{abstract}

\noindent The cosmological constant and electroweak hierarchy problem have been a great inspiration for research. Nevertheless, the resolution of these two naturalness problems remains mysterious from the perspective of a low-energy effective field theorist.  The string theory landscape and a possible string-based multiverse offer partial answers, but they are also controversial for both technical and conceptual reasons. The present lecture notes, suitable for a one-semester course or for self-study, attempt to provide a technical introduction to these subjects. They are aimed at graduate students and researchers with a solid background in quantum field theory and general relativity who would like to understand the string landscape and its relation to hierarchy problems and naturalness at a reasonably technical level. Necessary basics of string theory are introduced as part of the course. This text will also benefit graduate students who are in the process of studying string theory at a deeper level. In this case, the present notes may serve as additional reading beyond a formal string theory course.
\end{abstract}
\end{center}

\newpage

\section*{Preface}
This course intends to give a concise but technical introduction to `Physics Beyond the Standard Model' and early cosmology as seen from the perspective of string theory. Basics of string theory will be taught as part of the course. As a central physics theme, the two hierarchy problems (of the cosmological constant and of the electroweak scale) will be discussed in view of ideas like supersymmetry, string theory landscape, eternal inflation and multiverse. The presentation will include critical points of view and alternative ideas and explanations. Problems with solutions are also provided to facilitate the use of these notes in classroom and for self-study.

Basic knowledge of quantum field theory (QFT), general relativity and cosmology will be assumed. Supersymmetry, elements of supergravity and fundamentals of string theory will be taught together with a number of geometrical concepts needed to study string compactifications. However, given the limited scope of a one-semester lecture series, this can clearly not replace a full string theory course or the detailed study of string geometry.

The author has taught this course at Heidelberg University with the intention to prepare students who have taken a two-semester QFT and a one-semester relativity course for Master thesis research in string phenomenology. Another goal was to allow students who intend to do research in particle phenomenology, cosmology or formal (mathematical) string theory to develop some basic understanding of the possible relation of string theory to `real world' physics and its most fundamental problems.

For students who had the privilege of enjoying a complete graduate-level education (with full lecture courses on strings and on supersymmetry/supergravity) before embarking on research, most of the material in the first part of this course will be familiar. Still, depending on the focus of their string and cosmology courses, they may find useful additional information about landscape, multiverse, eternal inflation and alternative perspectives in the second half of the course.

The detailed plan of the lecture notes is as follows: We will start in Sect.~\ref{smhp} with a brief tour of the Standard Model, emphasising the perspective of a low-energy effective field theory, the coupling to gravity, and the electroweak hierarchy and cosmological constant problems. Section~\ref{sands} introduces supersymmetry and supergravity which, however, offer only a partial resolution of the fine-tuning problems of the Standard Model discussed above. It becomes apparent that the highest relevant energy scales, at the cutoff of the effective field theory, have to be involved. This motivates the study of string theory as the best-explored candidate quantum gravity theory in Sects.~\ref{stbs} and~\ref{stss}. We will see that the bosonic string, which suffers from the absence of fermions and from an unstable vacuum, has to be promoted to the superstring. The latter provides all desired ingredients for a consistent theory of gravity and particle physics, albeit in ten spacetime dimensions and with far too much supersymmetry. Compactifications to four dimensions are the subject of Sects.~\ref{comp} and~\ref{tfl}: First, we consider pure Calabi-Yau geometries, leading to highly supersymmetric and unrealistic 4d Minkowski-space models. Then, the inclusion of non-perturbative objects and fluxes leads to the proper string landscape with supersymmetric and non-supersymmetric models with non-zero cosmological constants of either sign. The key insight is the enormous number of such solutions (a recent analysis arriving at $\sim 10^{272,000}$), each corresponding to a different 4d effective theory. We now see that, assuming the non-trivial constructions with broken supersymmetry and positive cosmological constant stand up to further scrutiny, the `fine-tuned' or `unnatural' parameters of the Standard Model may indeed be accommodated by the string landscape. Section~\ref{eimp} deals with the important but complicated and speculative question of how the landscape gets populated during eternal inflation and whether statistical predictions for future observations can be derived. A number of alternative perspectives on the hierarchy problems and on quantum gravity are discussed in Sect.~\ref{alt} before, in Sect.~\ref{summ}, we end by summarising the overall picture and the challenges that should have crystallised during the study of this course.

While useful references for background material and deeper exploration will be provided as we go along, it may not hurt to give some essential literature right away: Good sources for the background knowledge in QFT and relativity are \cite{Peskin:1995ev} and \cite{Wald:1984rg} respectively. For more details on Standard Model and particle-physics related topics, Refs. \cite{Cheng:1985bj, Donoghue:1992dd} represent useful sources. For supersymmetry and supergravity see \cite{Wess:1992cp, Freedman:2012zz}. Two of the most complete modern string theory textbooks are \cite{Polchinski:1998rq, Blumenhagen:2013fgp}. Concerning string phenomenology, \cite{Ibanez:2012zz} represents a very comprehensive monograph which, in particular, covers the important subjects of how specific gauge and matter sectors are realised in string compactifications - a topic which we treat very superficially in this course. A very useful set of notes emphasising the geometric side of how the landscape arises from string theory is \cite{Denef:2008wq}. For a detailed review of string landscape physics see \cite{Schellekens:2013bpa}.

\tableofcontents

\section{The Standard Model and its Hierarchy Problem(s)}
\label{smhp}

As already stated in the Preface, we assume some familiarity with quantum field theory (QFT), including basics of regularisation and renormalisation. There exists a large number of excellent textbooks on this subject, such as~\cite{Peskin:1995ev, Weinberg:1995mt, Itzykson:1980rh, Srednicki:2007qs, Schwartz:2013pla}. The reader familiar with this topic will most probably also have some basic understanding of the Standard Model of Particle Physics, although this will not be strictly necessary since we will introduce this so-called Standard Model momentarily. It is also treated at different levels of detail in most QFT texts, most notably in~\cite{Weinberg:1995mt, Schwartz:2013pla}. Books devoted specifically to theoretical particle physics and the Standard Model include~\cite{Cheng:1985bj, Donoghue:1992dd, Nachtmann:1990ta}. A set of lecture notes covering the Standard Model and going beyond it is \cite{sche}. We will to refer to some more specialised texts as we go along.

\subsection{Standard Model - the basic structure}
A possible definition of the Standard Model \index{Standard Model} is as follows: It is the most general renormalisable field theory with gauge group
\be
G_{SM}= SU(3)\times SU(2)\times U(1)\,,\label{gro}
\ee
three generations of fermions, and a scalar. These fields transform in the representations 
\be
({\bf 3},{\bf
2})_{1/6}\,+\,({\bf {\bar 3}},{\bf 1})_{-2/3}\,+\, ({\bf {\bar
3}},{\bf 1})_{1/3}\,+\,({\bf 1},{\bf 2})_{-1/2}\,+\, ({\bf 1},{\bf
1})_{1}\qquad\quad\mbox{and}\qquad\quad ({\bf 1},{\bf 2})_{1/2}\label{mat}
\ee
respectively. Here the boldface numbers specify the representations of $SU(3)$ and $SU(2)$ via its dimension (in our case only singlet, fundamental or anti-fundamental occur, the latter denoted by an overline) and the index gives the $U(1)$ charge $Y$, also known as \index{hypercharge} hypercharge. The overall normalisation of the latter is clearly convention dependent.\footnote{
By this we mean the freedom to rescale the gauge potential $A_\mu$ by a constant, such that the values of $Y$ and of the gauge coupling change correspondingly. In contrast to the non-abelian case (see below), there is no preferred choice intrinsic to the abelian gauge lagrangian $F_{\mu\nu}F^{\mu\nu}$ since the latter is homogeneous in $A_\mu$. In our conventions the electric charge is given by $Q=T_3+Y$, with $T_3$ the third $SU(2)$ generator.
}

If one adds gravity in its simplest and essentially unique theoretical formulation (Einstein's general relativity), then this data offers an almost complete fundamental description of the material world. This structural simplicity and the resulting small number of fundamental parameters (to be specified in a moment) is very remarkable. What is even more remarkable is the enormous underlying unification: So many very different macroscopic and microscopic phenomena which we observe in everyday life and in many natural sciences follow from such a (relatively) simple underlying theory.

Clearly, important caveats have already been noted above: The description is {\it almost} complete, the theory is {\it relatively} simple (not as simple as one would wish) and, maybe most importantly, it is only fundamental {\it to the extent that we can test it at the moment}. Quite possibly, more fundamental building blocks can be identified in the future. The rest of this course is about exactly these caveats and whether, based on those, theoretical progress is possible. 

But first let us be more precise and explicit and turn the defining statements (\ref{gro}) and (\ref{mat}) into a field-theoretic lagrangian. Given the theoretically well-understood and experimentally tested rules of quantum field theory (QFT), this can be done unambiguously. The structure of the lagrangian is 
\be
{\cal L}_{SM}={\cal L}_{gauge}+{\cal L}_{matter}+{\cal L}_{Higgs}+{\cal L}_{Yukawa}\,.
\ee
The gauge part is completely standard,
\be
{\cal L}_{gauge}\,=\,-\frac{1}{4g_1^2}F_{\mu\nu}^{(1)}F^{(1)\,\mu\nu}
\,-\,\frac{1}{2g_2^2}\,\mbox{tr}\,F_{\mu\nu}^{(2)}F^{(2)\,\mu\nu}
\,-\,\frac{1}{2g_3^2}\,\mbox{tr}\,F_{\mu\nu}^{(3)}F^{(3)\,\mu\nu}\,,
\ee
with the upper index $(i)$ running over $U(1)$, $SU(2)$, $SU(3)$, in this order. The field strengths are defined as $F_{\mu\nu}=i[D_\mu,D_\nu]$ with $D_\mu=\partial_\mu-iA_\mu$ and, in the non-abelian cases, $A_\mu=A_\mu^A T_A$. One should also remember the standard normalisation tr$(T_AT_B)=\delta_{AB}/2$ of the $SU(N)$ generators in the fundamental representation. 

The matter \index{Standard Model!matter} or, more precisely, the {\it fermionic} matter contribution reads
\be
{\cal L}_{matter}=\sum_j \ol{\psi}_j i\slashed{D}_j\psi_j\qquad\mbox{with}\qquad
(D_j)_\mu=\partial_\mu-iR_j(A_\mu)
\ee
with $j$ running over left-handed quark doublets, right-handed up- and 
down-type quarks, lepton-doublet and right-handed leptons (each coming in three generations or families):
\be
\psi_j\in\{\,\{q_L^a,(u_R^a)^c,(d_R^a)^c,l_L^a,(e_R^a)^c\},\,a=1,2,3\,\}\,.
\label{fie}
\ee
The five types of fermions from $q_L$ to $e_R^c$ correspond precisely to the five terms in the direct sum in (\ref{mat}). Furthermore, $R_j(A_\mu)$ denotes the representation of $A_\mu\in Lie(G_{SM})$ appropriate for the fermion of type $j$. To make our conventions unambiguous, we have to specify in detail how we describe the spinor fields. One convenient choice (the one implicitly used above) is to always work with left-handed 4-component or Dirac spinors. \index{Dirac spinor} In other words, we do {\it not} use general Dirac spinors built from Weyl spinors \index{Weyl!spinor} according to 
\be
\psi_D=\left(\!\begin{array}{c}\psi_\alpha \\ \ol{\chi}^{\dot{\alpha}}\end{array}\!\right)\qquad\mbox{with}\qquad \alpha,\dot{\alpha}=1,2\,.
\ee
Instead, all our 4-spinors are left-handed:
\be
\psi=\left(\!\begin{array}{c}\psi_\alpha \\ 0\end{array}\!\right)\,.
\ee
In particular, this explains why we use the charge-conjugate of right-handed quarks and leptons as our fundamental fields, cf.
\be
q_L=\left(\!\begin{array}{c}(q_L)_\alpha \\ 0\end{array}\!\right) \qquad\mbox{vs.}\qquad
u_R^c=\left(\!\begin{array}{c}(u_R)_\alpha \\ 0\end{array}\!\right)=
\left(\!\begin{array}{c}0 \\ (\ol{u}_R)^{\dot{\alpha}}\end{array}\!\right)^c\,.
\ee
We see that the quantum numbers given in (\ref{mat}) can be viewed as referring to either these left-handed fields or to the corresponding 2-component Weyl spinors. The latter will in any case be the most useful way to describe fermions when talking about supersymmetry below.

The scalar or Higgs lagrangian is \index{Higgs potential}
\be
{\cal L}_{Higgs}=-(D_\mu\Phi)^\dagger(D^\mu\Phi)-V(\Phi)\qquad\mbox{with}\qquad
V(\Phi)=-m_H^2\Phi^\dagger\Phi+\lambda_H(\Phi^\dagger\Phi)^2\,,
\ee
where $\Phi$ is an $SU(2)$ doublet with charge $1/2$ under $U(1)$ (the hypercharge-$U(1)$ or $U(1)_Y$). Finally, there are the Yukawa terms \index{Yukawa coupling}
\be
{\cal L}_{Yukawa} = -\sum_{jk}\lambda_{jk}\ol{\psi}_j\psi^c_k\,\Phi \,\,+\,\,\mbox{h.c.}\,,
\label{gy}
\ee
where the sum runs over all combinations of fields for which the relevant product of representations contains a gauge singlet. We left the group indices and their corresponding contraction implicit.

Note that, since all our fields are l.h. 4-component spinors, we have to write $\ol{\psi}\psi^c$ rather than simply $\ol{\psi}\psi$. The latter would be identically zero. Note also that the 4-spinor expression $\ol{\psi}\psi^c$ corresponds to $\ol{\psi}_{\dot{\alpha}}\ol{\psi}^{\dot{\alpha}}$ in terms of the Weyl spinor $\psi_\alpha$ contained in the 4-spinor $\psi$. 

Crucially, the Higgs potential has a minimum with $S^3$ topology at $|\Phi|=v\simeq 174$~GeV, leading to spontaneous \index{spontaneous symmetry breaking} gauge symmetry breaking. One can choose the VEV to be real and aligned with the lower component of $\Phi$, leading to the parameterisation
\be
\Phi=\left(\! \begin{array}{l} 0 \\ v+h/\sqrt{2} \end{array}
\!\right)\,.
\ee
It is easy to see that the symmetry breaking pattern is $SU(2)\times U(1)_Y\,\to\, U(1)_{em}$ (see problems). Three would-be Goldstone-bosons along the $S^3$ directions are `eaten' by three of the four vector bosons of $SU(2)\times U(1)_Y$. This leads to the $W^\pm$ and $Z$ bosons with masses $m_{W^\pm}\simeq 80$~GeV and $m_Z\simeq 90$~GeV. The surviving real Higgs scalar $h$ is governed by
\be
{\cal L}\supset -\frac{1}{2}(\partial h)^2-\frac{1}{2}m_h^2 h^2\,.
\ee
Our notation `$\supset$' means that we are displaying only the subset of terms in ${\cal L}$ which is most important in the present context. One can easily relate the parameters after symmetry breaking to those of the original lagrangian:
\be
v^2=m_H^2/(2\lambda_H)\qquad,\qquad m_h^2=4\lambda_H v^2\,.
\ee
We have $m_h\simeq 125$~GeV, $m_H=m_h/\sqrt{2}=88$~GeV, and $\lambda_H\simeq 0.13$.\footnote{We note that a slightly different convention, $v\to v'$,  with $\Phi_2=(v'+h)/\sqrt{2}$ and hence $v'=\sqrt{2}v\simeq 246$~GeV, is also widely used. Ours has the advantage that $m_{top}\equiv m_t\simeq v$.}

The surviving massless gauge boson is, of course, the familiar photon. 
Finally, one can check that the allowed Yukawa terms suffice to give all charged fermions a mass proportional to $v$. The three lightest quark masses are not directly visible to experiment since the confinement dynamics of the $SU(3)$ gauge theory (QCD) hides their effect. The three upper components of the lepton-doublets -- the neutrinos -- have $Q=0$ and remain massless. 

The reader should check explicitly which (three) Yukawa coupling terms are allowed and that {\it no} further renormalisable operators (i.e. operators with mass dimension $\leq 4$) consistent with the gauge symmetry of the Standard Model exist.

\subsection{Standard Model - parameter count}
\index{Standard Model!parameters}
The most obvious parameters are the three gauge couplings $g_i$. Then there is of course the Higgs quartic coupling $\lambda_H$ and the Higgs mass parameter $m_H$ (defining the negative quadratic term $-m_H^2|\Phi|^2$ in the potential). It is not so easy to count the independent Yukawa couplings contained in the three terms \index{Yukawa coupling}
\be
\sum_{a,b=1}^3\left(\lambda^u_{ab}\,\ol{q}_L^{\,a}\Phi^* u_R^b+
\lambda^d_{ab}\,\ol{q}_L^{\,a} \Phi d_R^b+ \lambda^e_{ab}\,\ol{l}_L^{\,a} \Phi e_R^b \right)\,\,+\,\,\mbox{h.c.}\label{sy}
\ee
The reader should check that these and only these terms are $G_{SM}$-invariant, given the general Yukawa-term structure displayed in (\ref{gy}). One frequently sees the notation (suppressing generation indices)
\be
\ol{q}_L\tilde{\Phi}u_R \qquad\mbox{with}\qquad \tilde{\Phi}_\alpha=\epsilon_{\alpha\beta}(\Phi^\beta)^* \label{pst}
\ee
for the first term above. Introducing this 2-component vector $\tilde{\Phi}$ is necessary if one wants to read (\ref{sy}) in terms of $SU(2)$ matrix notation. If one simply says that `group indices are left implicit' (as we do), writing $\Phi^*$ is sufficient. Of course, we could also have avoided the explicit appearance of $\Phi^*$ in (\ref{sy}) altogether by exchanging it with its complex conjugate, implicit in `h.c.' This is a matter of convention and the form given in (\ref{sy}), (\ref{pst}) is close what most authors use.

Maybe the easiest way to count the parameters in (\ref{sy}) is to think in terms of the low-energy theory with $\Phi$ replaced by its VEV. Then the above expression contains three $3\!\times\!3$ complex mass matrices. Furthermore, these mass matrices relate six independent sets of fermions (since the first term only contains $u_L$ and the second only $d_L$). Thus, the matrices can be diagonalised using bi-unitary transformations - i.e. a basis change of the fermion fields. We are then left with $3\!\times\! 3=9$ mass parameters for three sets of up- and down-type quarks and three leptons. 

However, due to $SU(2)$ gauge interactions, one of terms in
\be
\sum_{a=1}^3\ol{q}_L^a\slashed D q_L^a
\ee
contains both $u_L$ and $d_L$. It originates in the off-diagonal terms of $\sigma^{1,2}$ which are contained in $\slashed{D}$. In this $u_L/d_L$ term, the unitary transformation used above does not cancel and a physical $3\!\times\! 3$ matrix describing `flavour\index{flavour} changing charged currents'\index{flavour-changing charged current} (the CKM matrix)\index{CKM matrix} is left.\footnote{
These flavour-changing currents correspond to vertices with a (charged) $W$ boson and two left-handed fermions with different flavour (one up and one down, either both from the same or from different generations). 
}
Let us write the relevant term symbolically as 
\be
\sum_{a,b=1}^3\ol{u}_L^a\gamma^\mu U^{ab}d_L^b\,.
\ee
The matrix $U$ arises as the product of two unitary matrices from the bi-unitary transformations above. Hence it is unitary.

It will be useful to pause and think more generally about parameterising a unitary \mbox{$n\times n$} matrix $U$. 
First, a general complex matrix $U$ has $2n^2$ real parameters. The matrix $UU^\dagger$ is always hermitian, 
so imposing the hermiticity requirement $UU^\dagger=\mathbbm{1}$ imposes $n^2$ real constraints. Then $n^2$ parameters are left. Next, recall that orthogonal matrices have $n(n-1)/2$ real parameters or rotation angles. Thus, since unitary matrices are a superset of orthogonal matrices, we may think of characterising unitary matrices by $n(n-1)/2$ angles and $n^2-n(n-1)/2$ phases.\footnote{This is not a proof. One needs to show that such a  parameterisation in terms of angles and phases exists. We will not touch the interesting subject of parameterisations of unitary matrices.} Now, in our concrete case, we are free to transform our unitary matrix (in an $n$-generation
Standard Model) according to
\be
U\to D_u U \ol{D}_d\,,\label{reph}
\ee
where $D_{u,d}$ are diagonal matrices made purely of phases. This is clear since we may freely rephase the fields $u_L^a$ and $d_L^a$ (together with their mass partners  $u_R^a$ and $d_R^a$ -- to keep the masses real). The rephasing freedom of (\ref{reph}) can be used to remove $2n-1$ phases from $U$. The `$-1$' arises since one overall common phase of $D_u$ and $D_d$ cancels and hence does not affect $U$. So we are left with $n^2-n(n-1)/2-(2n-1)=(n-1)(n-2)/2$ physically significant phases.

Now we return to $U$ as part of our Standard Model lagrangian with real, diagonal fermion mass matrix. Here $n=3$ and, according to the above, the CKM matrix has 3 real ``mixing angles'' and one complex phase (characterising $CP$ \index{$CP$ violation} violation in the weak sector of the Standard Model). For more details, see e.g.~\cite{Cheng:1985bj}, Chapter 11.3.\footnote{
For a broader discussion of C, P, CP and its violation see e.g.~\cite{Blanke:2017ohr, Grinstein:2017pvg, kt, Bigi:2000yz, Fleischer:2006fx, Branco:1999fs} and references therein.
}

This brings our total parameter count to $3+2+9+4=18$. However, we are not yet done since we completely omitted a whole general type of term in gauge theories, the so-called topological or $\theta$-term \index{$\theta$-term}
\be
{\cal L}\supset \theta\,\mbox{tr}\,F\wedge F\sim \theta\, \epsilon^{\mu\nu\rho\sigma}F_{\mu\nu}^a F_{\rho\sigma}^a \sim \theta\,\mbox{tr}\,F\tilde{F}\,.
\ee
Most naively, this adds 3 new parameters, one for each factor group. However, these terms are total derivatives if expressed in terms of $A_\mu$. Thus, they are invisible in perturbation theory and do not contribute to the Feynman rules. In the non-abelian case, there exist gauge field configurations localised in space and time (called instantons) for which 
\be
\int \mbox{tr}  F\wedge F
\ee
is non-zero. This does not clash with the total-derivative feature since no globally defined $A_\mu$ exists for such instanton configuations. We will return to instantons in more detail later. For the $U(1)$, such configurations do not exist, which severely limits the potential observability of the $\theta$-term in $U(1)$ gauge theories. 

Furthermore, and maybe most importantly, the $\theta$-term is precisely of the type that the non-invariance of the fermionic path integral measure induces if chiral fermion fields are re-phased. Thus, in the presence of charged fermions without mass terms (or analogous Yukawa-type couplings preventing a re-phasing) such $\theta$ parameters are unphysical. The upshot of a more detailed analysis in the Standard Model case (where some but not all conceivable fermionic mass terms are present) is that the $SU(2)$ and $U(1)$ $\theta$-terms are unobservable (see e.g.\cite{Anselm:1993uj}) but the QCD $\theta$-term is physical (for some non-trivial issues in this context see \cite{Cao:2017ocv} and refs.~therein). If one goes beyond the Standard Model by adding more fields or even just higher-dimension operators, the electroweak $\theta$-terms may become physical. 

A non-zero value of $\theta_{QCD}$ breaks CP. This is directly visible from the $\epsilon$-tensor in the definition of the $\theta$-term as well as from its equivalence (through re-phasing) to complex fermion mass parameters.\footnote{
Recall that, at the lagrangian level, charge conjugation is related to complex conjugation. In particular, it is broken by complex lagrangian parameters which can not be removed by field redefinitions.
}
Now, let us assume that $\theta_{QCD}={\cal O}(1)$ and that, as a result, CP is broken at the ${\cal O}(1)$ level by the theory of strong interactions. If light-quark masses were $\sim$~GeV, one would then expect the electric dipole moment of the neutron to be ${\cal O}(1)$ (in GeV units). Allowing for the suppression by the tiny light-quark masses $\sim 10^{-3}$ GeV, the dipole moment should still be large enough to be detected if $\theta_{QCD}$ where ${\cal O}(1)$. However, corresponding search experiments have so far only produced an {\it extremely} small upper bound. The detailed analysis of this bound implies roughly $\theta_{QCD}<10^{-10}$. 

In any case, we now arrived at our final result of 19 parameters. However, the status of these parameters is very different. Most notably, 18 of them correspond to dimension-4 (or marginal) operators, while one -- the Higgs mass term -- is dimension-2 and hence relevant. (We recall that the term `relevant' refers to `relevant in the IR'.) \index{marginal operator} \index{relevant operator}

Let us try to make the same point from a more intuitive and physical perspective: Since the theory is renormalisable, one can imagine studying it at a very high energy scale, $E\gg v\sim m_H$. At this scale the Higgs mass is entirely unimportant and we are dealing with a theory of massless fields characterised by 18 dimensionless coupling constants. Classically, this structure is scale invariant since only dimensionless couplings are present. At the quantum level, even without the Higgs mass term, this scale invariance is badly broken by the non-zero beta-functions, most notably of the gauge couplings. Indeed the gauge couplings run quite significantly and, even in the absence of the Higgs $|\Phi|^2$ term, QCD would still confine at about 1~GeV and break the approximate scale invariance completely in the IR.

However, this `high-scale' Standard Model described above is very peculiar in the following sense: One perfectly acceptable operator, $-m_H^2\Phi^2$, is missing entirely. More precisely, if we characterise the theory at a scale $\mu$ by dimensionless couplings, e.g. $g_i^2(\mu)$, $\lambda_H(\mu)$ etc., then we should include a parameter $m_H^2(\mu)/\mu^2$. If we start at some very high scale (e.g. the Planck scale $M_P\sim 10^{18}$~GeV -- more on this point later), then this parameter has to be chosen extremely small,
\be
m_H^2(\mu)/\mu^2 \sim 10^{-32}\qquad\mbox{at}\qquad \mu\sim M_P\,,
\ee 
to describe our world. Indeed, running down from that scale it keeps growing as $1/\mu^2$ until, at about $\mu\sim 100$~GeV, it starts dominating the theory and completely changes its structure. This is our first encounter with the hierarchy problem, which we will discuss in much more detail below.

\subsection{Effective field theories - cutoff perspective}\label{eft} \index{effective field theory} \index{EFT}

In this course, we assume familiarity with basic QFT. The language of (low-energy) effective field theory can be viewed as an important part of QFT
and hence many readers will be familiar with it. Nevertheless, since this subject is of such an outstanding importance for what follows, we devote some space to recalling the most fundamental ideas of effective field theory (EFT). In addition to chapters in the various QFT books already mentioned, the reader will be able to find many sets of lecture notes devoted specifically to the subject of EFTs, e.g.~\cite{Georgi:1994qn, Manohar:1996cq, Pich:1998xt, Luty:2005sn, Kaplan:2005es, Cohen:2019wxr}. For a wider perspective on effective theories (not restricted to QFT), see e.g.~\cite{Wells:2012rla}.

To begin, let us assume that our QFT is defined with some UV cutoff $\Lambda_{UV}$ (and, if one wants, in finite spatial volume $\sim 1/\Lambda_{IR}$), such there can be no doubt that we are dealing with a conventional quantum mechanical system. Of course, the larger the ratio $\Lambda_{UV}/\Lambda_{IR}$, the more degrees of freedom this system has. The possible IR cutoff will not be relevant for us and we will not discuss it further. The best example of a UV cutoff (though not very practical in perturbative calculations) is presumably the lattice cutoff. It is e.g. well established that this leads to a good description of gauge theories, including all perturbative as well as non-perturbative effects. Next, it is also well-known and tested in many cases that the lattice regularisation can be set up in such a way that Poincare-symmetry is recovered in the IR. Of course, we could use Poincare-invariant cutoffs (e.g. dimensional regularisation, Pauli-Villars or even string theory) from the beginning, but the lattice is conceptually simpler and more intuitive. Thus, we will be slightly cavalier concerning this point and assume that we can disregard Poincare-breaking effects in the IR of our system.

As a result (and here we clearly assume a large amount of non-trivial QFT intuition to be developed by reading standard texts) our low-energy physics can be characterised by an action of the symbolic structure 
\be
S=\int d^4\,x\,\left(-\frac{1}{2g^2}\mbox{tr}F^2+\frac{\theta}{8\pi^2}\, \mbox{tr}F\tilde{F} +\frac{c_1}{\Lambda^4}\mbox{tr}F^4 +\frac{c_2}{\Lambda^4}(\mbox{tr}F^2)^2+\cdots\right)\,.\label{geft}
\ee
Here we focussed on the gauge theory case and wrote $\Lambda\equiv \Lambda_{UV}$ for our cutoff scale. In other words, we expect that generically all terms allowed by the symmetries are present and that, on dimensional grounds, whenever a dimensionful parameter is needed, it is supplied by the cutoff scale $\Lambda$. At low energies, only terms not suppressed by powers of $\Lambda$ will be important, hence we will always encounter renormalisable theories in the IR. The relevance of terms in the IR decreases as their mass dimension \index{mass dimension} grows. This is obvious if one thinks, e.g., in terms of the contribution a given operator makes to a 4-gluon amplitude: The first term in (\ref{geft}) will contribute $\sim g^2$; the third will contribute $\sim g^4k^4/\Lambda^4$. Clearly, at small typical momentum $k$, only the first term is important.\footnote{The 
second term is a total derivative and hence does not contribute in perturbation theory.
} 
To see this explicitly one needs to split off the propagator from the first term and to rescale $A_\mu\to gA_\mu$. The lagrangian will then contain terms of the type
\be
A\,\partial^2 A\,+\,g\, A^2\,\partial\,A\,+\,g^2\,A^4\,+\,g^4\,(c_1/\Lambda^4)\,(\partial A)^4+\cdots\,,
\ee
confirming our claim about the 4-gluon amplitude.

The numerical coefficients in (\ref{geft}) depend on the details of the regularisation (e.g. the lattice model) or, in more physical terms, on the UV definition of our theory at the scale $\Lambda$. Indeed, recall our assumption that, what we perceive as a QFT at low energies, is defined as a finite quantum mechanical system at the scale $\Lambda$. There are in general many discrete and continuous choices hidden in this definition. They will be reflected in the values of $g,\theta$ and the $c_i$. Some of these terms can hence be unusually large or small and this can to a certain extent overthrow the ordering by dimension advertised above. However, in the mathematical limit $k/\Lambda\to 0$, the power of $k/\Lambda$ wins over numerical prefactors. An exception arises if one coefficient is exactly zero. This  important possibility will be discussed below.

Let us add to our gauge theory example given above the apparently much simpler example of a real scalar field, symmetric under $\phi\to -\phi$:
\be
S=\int d^4\,x\,\left(c_0\Lambda^2\phi^2-\frac{1}{2}(\partial\phi)^2- \lambda\phi^4+\frac{c_1}{\Lambda^2}\phi^6+\cdots\right)\,.
\ee
The key novelty is that we have a term proportional to a positive power of $\Lambda$ (a relevant operator). In the gauge theory case, the most important operators were merely marginal. Moreover, this term is a mass term and for $c_0={\cal O}(1)$ the EFT below the scale $\Lambda$ is simply empty. Thus, we must assume that a very particular UV completion exists which allows for either $c_0=0$ (for some qualitative reason) or at least for the possibility to {\it tune} this coefficient to a very small value, $c_0\ll 1$. We now see that this has some similarity to the Standard Model, where (assuming that the Standard Model continues to be the right theory above the TeV-scale), a similar tuning might be needed to keep $m_H^2$ small. \index{fine tuning}

Arguing that there is a `tuning' or `fine-tuning problem' based only on the above is not very convincing. One of the reasons is that we were vague about the UV completion at the scale $\Lambda$. It appears possible that the right UV completion will effortlessly allow for $c_0\ll 1$ or maybe even predict such a small value. Indeed, we have to admit right away that we will not be able to rule this out during this whole course. But we will try to explain why many researchers have remained pessimistic concerning this option.

\subsection{Effective field theories - QFT$_{\bf UV}$ vs. QFT$_{\bf IR}$} \index{effective field theory} \index{EFT}

To do so, we will now modify the use of the word effective field theory: In the above, we assumed some finite (non-QFT) UV completion and called EFT what remains of it in the IR. Now, we want to start with some QFT in the UV (to be itself regularised or UV-completed at even higher scales) and consider how it transits to another QFT (which we will call EFT) in the IR. The simplest way in which this can happen is as follows: Let our QFT$_{UV}$ contain a particle with mass $M$ and focus on the physics at $k\ll M$. In other words, we `integrate out' the heavy (from the IR perspective) particle and arrive at a theory we might want to call QFT$_{IR}$ -- our low-energy EFT. 

Let us start with a particularly simple example, borrowed from \cite{Luty:2005sn}:
\be
{\cal L}=\ol{\psi}i\slashed{\partial}\psi-m\ol{\psi}\psi-\frac{1}{2}(\partial \phi)^2 -\frac{1}{2}M^2\phi^2+y\phi\ol{\psi}\psi-\frac{\lambda}{4!}\phi^4\,.
\label{tmp}
\ee
We assume $m\ll M\ll \Lambda$ and we have already ignored all terms suppressed by $\Lambda$. The above lagrangian is then renormalisable, such that we may indeed view (\ref{tmp}) as {\it defining} our theory through some parameter choice at a very high\footnote{
We 
do not insist on being able to take the mathematical limit $\Lambda\to \infty$ or $\mu_1\to \infty$ since we do not want to deal with issues like a possible Landau pole or a sign change of $\lambda$ in the far UV.
} 
scale $\mu_1\gg M$. We are interested in the EFT at $\mu_2$ with $m\ll \mu_2\ll M$.

The correct procedure (`running and matching') \index{running and matching} would be as follows: One writes down the most general lagrangian ${\cal L}_{EFT}$ for $\psi$ at the scale $\mu_2$ and calculates (at some desired loop order) a sufficiently large set of observables (e.g. mass, 4-point-amplitude etc.). Then one calculates the same observables using the full theory defined by (\ref{tmp}). This includes tree level diagrams and loops involving $\phi$ as well as the renormalisation group (RG) evolution. Finally, one determines the parameters of ${\cal L}_{EFT}$ such that the two results agree.

\begin{figure}[ht]
\begin{center} 
\includegraphics[width=4cm]{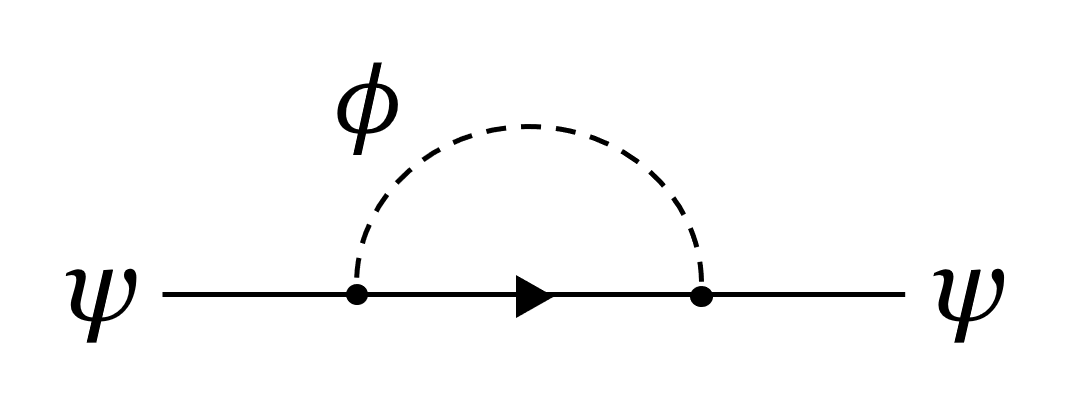}
\caption{One-loop fermion self energy in the Yukawa theory.}
\label{sep} 
\end{center}
\end{figure}

Our course is not primarily about EFTs and we will take a shortcut. First, we set $\lambda=0$ since it will not be essential in what we have to say. Second, we integrate out $\phi$ classically: We ignore the $(\partial\phi)^2$ term since we are at low energies and we extremise the relevant part of ${\cal L}$ with respect to $\phi$:
\be
\frac{\delta}{\delta\phi}\left(-\frac{1}{2}M^2\phi^2+y\phi\ol{\psi}\psi
\right)=0\qquad\Rightarrow\qquad \phi=\frac{y}{M^2}\,\ol{\psi}\psi\,.
\ee
Inserting this back into our lagrangian we obtain
\be
{\cal L}_{EFT}=\ol{\psi}i\slashed{\partial}\psi-m\ol{\psi}\psi+\frac{y^2}{2M^2} (\ol{\psi}\psi)^2+\cdots\,.\label{left}
\ee
Finally, we calculate loop corrections involving the heavy field $\phi$ to all operators that potentially appear in ${\cal L}_{EFT}$. In this last step, the correction which is most critical for us is the mass (or more generally the self energy correction) for $\psi$, cf.~Fig.~\ref{sep}.

Dropping all numerical prefactors, this gives (for details see e.g.~\cite{Srednicki:2007qs})
\be
\Sigma(\slashed{p})\sim y^2 \int d^4\,k\,\frac{-\slashed{k}+m}{(k^2+m^2)[(k+p)^2+M^2 ]}\,.\label{sig}
\ee
After summing, in the standard way, all such self-energy corrections to the propagator, one obtains \index{self-energy}
\be
\frac{i}{\slashed{p}-m-\Sigma(\slashed{p})}\,.
\ee
This resummed propagator can be viewed as a function of the matrix-valued argument $\slashed{p}$. Its pole then determines the corrected mass $m_c=m+\delta m$. This can be made explicit by Taylor expanding $\Sigma(\slashed{p})$ around $\slashed{p}=m_c$:
\be
\Sigma(\slashed{p})=\Sigma(m_c)+\Sigma'(m_c)(\slashed{p}-m_c)+\frac{1}{2}
\Sigma''(m)(\slashed{p}-m_c)^2+\cdots\,.
\ee
Now the propagator takes the form
\be
\frac{i}{(\slashed{p}-m_c)-\Sigma'(m_c)(\slashed{p}-m_c)+\cdots}\qquad
\mbox{with}\qquad m_c=m+\Sigma(m_c)\,.
\ee
As usual in perturbation theory we estimate
\be
\delta m = \Sigma(m)\,.
\ee

Thus, we need to evaluate~(\ref{sig}) for $\slashed{p}=m$. Introducing a cutoff $\Lambda$, we have in total three scales: $m$, $M$ and $\Lambda$. We see right away that the (naively possible) linear divergence arising from the term $\sim\slashed{k}$ will vanish on symmetry grounds as long as our cutoff respects Lorentz symmetry. Morover, the contribution from the term $\sim\slashed{k}$ in the regime $k\sim m\ll M$ is suppressed by $1/M^2$. Thus, we may disregard the term $\sim \slashed{k}$ altogether. 

We may then focus on the term $\sim m$. It gets a small contribution from the momentum region $k\lesssim M$ and is log-divergent for $k\gg M$. We can finally conclude that the leading result for the mass correction extracted from (\ref{sig}) must be proportional to $m$. Any enhancement beyond this can at best be logarithmic, but still proportional to $m$. For the moment this is all we need: We learn that (\ref{left}) is the right lagrangian after the replacement
\be
m\,\,\to\,\,m_{EFT}\equiv m_c \equiv m+\delta m\equiv m(1+y^2\times {\cal O}(1))\,.
\ee
Here `${\cal O}(1)$' may include a logarithmic cutoff dependence, like in particular $\ln(\Lambda/M)$. Moreover, as noted earlier, we may define our theory at a finite scale $\mu_1\gg M$. Then the log-divergence is traded for 
$\ln(\mu_1/M)$.

We could have argued the same without even drawing any Feynman diagram: Indeed, writing our model in terms of left- and right-handed spinors,
\be
\ol{\psi}\psi=\ol{\psi}_L\psi_R+\ol{\psi}_R\psi_L\,,
\ee
one sees immediately that for $m=0$ it possesses the $\mathbb{Z}_2$ symmetry 
\be
\psi_L\to\psi_L\,,\qquad \psi_R\to-\psi_R\,,\qquad \phi\to -\phi\,.
\ee
The mass term $\sim m$ breaks this symmetry. Thus, we expect that both the UV theory and the EFT regain this symmetry in the limit $m\to 0$. The loop correction $\delta m$ of
\be
m_{EFT}=m+\delta m
\ee
must hence itself be proprtional to $m$. The punchline is that integrating out $\phi$ does not clash with the lightness (or masslessness) of $\psi$. 

It is interesting and important to develop this language further by considering the low-energy EFT of the Standard Model below the scale of Higgs, $W$ and $Z$-bosons or the pion EFT below the confinement scale $\Lambda_{QCD}$. We leave it to the reader to explore this using the vast literature.

\subsection{The Standard Model as an effective field theory}

Let us now apply the above language to the Standard Model. We first assume that a finite cutoff $\Lambda\gg$~TeV is present and that the Standard Model is the effective theory valid below this cutoff. At the moment, we allow this cutoff to either be the scale at which the framework of QFT becomes insufficient (string or some other fundamental cutoff scale) or, alternatively, the scale at which the Standard Model is replaced by a different, more fundamental, ultraviolet QFT. It is natural to view $\Lambda$ as our main dimensionful parameter and organise the langrangian as
\bea
{\cal L} &=& {\cal L}_2+{\cal L}_4+ {\cal L}_5 + {\cal L}_6 +\cdots
\\
&=&c_0\Lambda^2|\Phi|^2-|D\Phi|^2-\lambda_H|\Phi|^4+ {\cal L}_4'+{\cal L}_5 + {\cal L}_6 +\cdots
\eea
Here, in the first line, we have organised the lagrangian in groups of dimension-2, dimension-4 (and so on) operators.
In the second line, we have displayed ${\cal L}_2$, which is just the Higgs mass term, and the rest of the Higgs lagrangian explicitly. Thus, ${\cal L}_4'$ is our familiar renormalizable Standard Model lagrangian without the Higgs part. We also have $m_H^2=c_0\Lambda^2$ and we note that $|c_0|\ll 1$ is necessary for the Higgs to be a dynamical field below the cutoff scale. But our discussion in the previous section has not lead to an unambiguous conclusion about whether this should be viewed as a problem. 

Since we now think of the Standard Model as of an EFT, we included terms of mass dimension 5, mass dimension 6, and so on. It turns out that, at mass dimension 5, the allowed operator is essentially unique (up to the flavour stucture). It is known as the {\bf Weinberg operator}. \index{Weinberg operator} We write it down for the case of a single family and using a two-component (Weyl) spinor notation (cf.~Problem~\ref{wsp}). The l.h. 4-component lepton-spinor $l_L$ then takes the form
\be
l_L=\left(\begin{array}{c}l_\alpha\\ 0 \end{array}\right)\,,
\qquad \mbox{with}\qquad \alpha=1,2\,.
\ee
The Weinberg-operator reads
\be
{\cal L}_5=\frac{c}{\Lambda}(l\cdot\Phi)^2+\mbox{h.c.}=\frac{c}{\Lambda}\,l_i^\alpha l_{\alpha j}\epsilon^{ik}\epsilon^{jl}\Phi_k\Phi_l+\mbox{h.c.}\label{wop}
\ee
Here we used the fact that two Weyl spinors can form a Lorentz invariant as 
\be
\psi^\alpha\psi_\alpha=\epsilon^{\alpha\beta}\psi_\beta\psi_\alpha\,,
\ee
where the $\epsilon$ tensor appears in its role as an invariant tensor of the Lorentz group $SL(2,\mathbb{C})$. By contrast, the $\epsilon$-tensors in (\ref{wop}) appear in their role as invariant tensors of the $SU(2)$ factor in $G_{SM}$ and allow us to combine two doublets (Higgs and leptons) into a singlet.

Now, since 
\be
\langle \Phi\rangle =\left(\begin{array}{c}0\\ v \end{array}\right)
\qquad\mbox{and}\qquad 
l_\alpha = \left(\begin{array}{c}\nu_\alpha\\ e_\alpha \end{array}\right)\,,
\ee
the low energy effect of the above operator is to give mass to the upper component of the lepton doublet, i.e. to the neutrino:
\be
{\cal L}_5=\frac{cv^2}{\Lambda}\nu^\alpha\nu_\alpha+\mbox{h.c.}
\ee
Writing the neutrino as a Majorana rather than a Weyl fermion, this becomes the familiar Majorana mass term. Introducing three families, the constant $c$ is promoted to a $3\times 3$ matrix $c_{ab}$. \index{neutrino mass} \index{Majorana mass}

Given our knowledge that neutrino masses are non-zero and (without going into the non-trivial details of the experimental situation) are of the order $m_\nu\sim 0.1$~eV, an effective field theorist can interpret the situation as follows: The neutrino mass measurements represent the detection of the first higher-dimension operator of the Standard Model as an EFT. As such, they determine the scale $\Lambda$ via the relation (assuming $c={\cal O}(1)$) 
\be
m_\nu\sim v^2/\Lambda\qquad\Rightarrow \qquad \Lambda\sim 3\times 10^{14}\,\,\mbox{GeV}\,.
\ee
On the one hand, this is discouragingly high. On the other hand, it is significantly below the (reduced) Planck scale of $M_P\simeq 2.4\times 10^{18}$~GeV. It is also relatively close to, though still significantly below, the supersymmetric Grand Unification scale $M_{GUT}\sim 10^{16}$~GeV to which will return later. Let us note that, without supersymmetry, the GUT scale is less precisely defined and one may argue that the UV scale derived from the Weinberg operator above is actually intriguingly close to such a more general GUT scale. 

It is very remarkable that the Standard Model with the Weinberg operator allows for a simple UV completion at the scale $\Lambda$. This so-called {\bf seesaw mechanism} \cite{Minkowski:1977sc, Yanagida:1979as, GellMann:1980vs} \index{seesaw mechanism} involves (we discuss the one-generation case for simplicity) the addition of just a single massive fermion, uncharged under $G_{SM}$. The relevant part of the high-scale lagrangian is (in Weyl notation for spinors)
\be
{\cal L}\supset \beta l\Phi\nu_R-\frac{1}{2}M\nu_R\nu_R+\mbox{h.c.}\label{ses}
\ee
Integrating out the extra fermion (often referred to as the right-handed neutrino $\nu_R$), one obtains precisely the previously given Weinberg operator with 
\be
c\sim \beta^2\qquad \mbox{and}\qquad \Lambda\sim M\,.
\ee
In other words, the observed neutrino masses behave as 
\be
m_\nu\sim \beta^2v^2/M\,.
\ee
As a result, we can make $M$ (and thus $\Lambda$) smaller, bringing it closer to experimental tests, at the expense of also lowering $\beta$. Of course, one has to be lucky to actually discover $\nu_R$ at colliders, given that then $\beta$ would have to take the rather extreme value of $\sqrt{100\,\,\mbox{GeV}/ 10^{14}\,\,\mbox{GeV}}\sim 10^{-6}$.

An even more extreme option, which however has its own structural appeal, is to set $M$ to zero. This can be justified, e.g., by declaring lepton number to be a good, global symmetry of the Standard Model (extended by r.h. neutrinos). By this we mean the $U(1)$ symmetry $l\to e^{i\chi}l$, $\nu_R\to e^{-i\chi}\nu_R$. 
Now the Standard Model has an extra field, the fermionic singlet $\nu_R$ (more precisely three copies of it). The first term in (\ref{ses}) is just another Yukawa coupling (given here in Weyl notation, but otherwise completely analogous to the e.g. the electron Yukawa term). The second term is missing. This version of the Standard Model, extended by r.h. neutrinos, is again a renormalisable theory and it can account for the observed neutrino masses. The latter do not arise from the seesaw mechanism sketched above, but correspond simply to a tiny new Yukawa coupling. In this case $\beta\sim m_\nu/v \sim 10^{-12}$, which may be perceived as uncomfortably small. The second smallest coupling would be that of the electron, $\beta_e\sim 0.5$~MeV$\,/v\sim 10^{-5}$.

At mass dimension 6, there are many further terms that can be added to ${\cal L}_{SM}$. For example, any term of ${\cal L}_4$ can simply be multiplied by $|\Phi|^2$. The arguably most interesting terms are the 4-fermion-operators. They include terms like (now again in Dirac notation)
\be
{\cal L}_6\supset \sum_{ijkl}c_{ijkl}(\ol{\psi}_i\psi_j)(\ol{\psi}_k\psi_l)
\label{4fo}
\ee
as well as similar operators involving gamma matrices. Even with the restriction by gauge invariance, there are many such terms and we will not discuss them in any detail. Crucially, many of them are very strongly constrained experimentally. First, if one does not impose the global symmetries of lepton and baryon number, some of these operators induce proton decay. (We recall that with baryon number we refer to a $U(1)$ symmetry acting on quarks, with a prefactor 1/3 in the exponent.) The extraordinary stability of the proton would then push $\Lambda$ up beyond $10^{16}$~GeV. But even imposing baryon and lepton number as additional selection rules\footnote{
To 
be precise, the two corresponding $U(1)$ symmetries, known as $U(1)_B$ and $U(1)_L$ are so-called accidental symmetries of the Standard Model. This means that, given just gauge symmetry and particle content, and writing down allowed renormalisable operators, these symmetries are automatically preserved at the classical level. It is hence not unreasonable to assume that they hold also in certain UV completions and may constrain 4-fermion operators.
} 
for (\ref{4fo}), strong constraints remain. These are mostly due to so-called flavour-changing neutral currents\index{flavour-changing neutral current}\index{FCNC} or FCNCs\footnote{
The name characterises processes which change flavour and have a structure that could arise from integrating our a neutral gauge boson, like the $\gamma$ or $Z$. It is an important fact that, in the Standard Model, such processes or the corresponding 4-fermion-operators are extremely suppressed. They are hence an important signal of new physics.
} 
(the analogues of the flavour-changing charged currents mentioned earlier) and to lepton flavour violation (e.g. the decay $\mu^-\to e^++2e^-$). Such constraints push $\Lambda$ to roughy $10^3$~TeV. Of course, the new-physics scale can be much lower if the relevant new physics has the right `flavour properties' not to clash with data.

\subsection{The electroweak hierarchy problem}
\index{electroweak hierarchy problem} \index{hierarchy problem}
Now we come in more detail to what is widely considered the main problem of the Standard Model as an effective theory: the smallness of the Higgs mass term. So far, we have only pointed out that, in the EFT approach with cutoff $\Lambda$, it is natural to write
\be
m_H^2\sim c_0\Lambda^2\,.
\ee
We have many reasons to think that $\Lambda$ is large compared to the weak scale, implying $|c_0|\ll 1$. The main question hence appears to be whether we can invent a more fundamental theory at scale $\Lambda$ in which $|c_0|\ll 1$ can be understood.

\begin{figure}[ht]
\begin{center} 
\includegraphics[width=8cm]{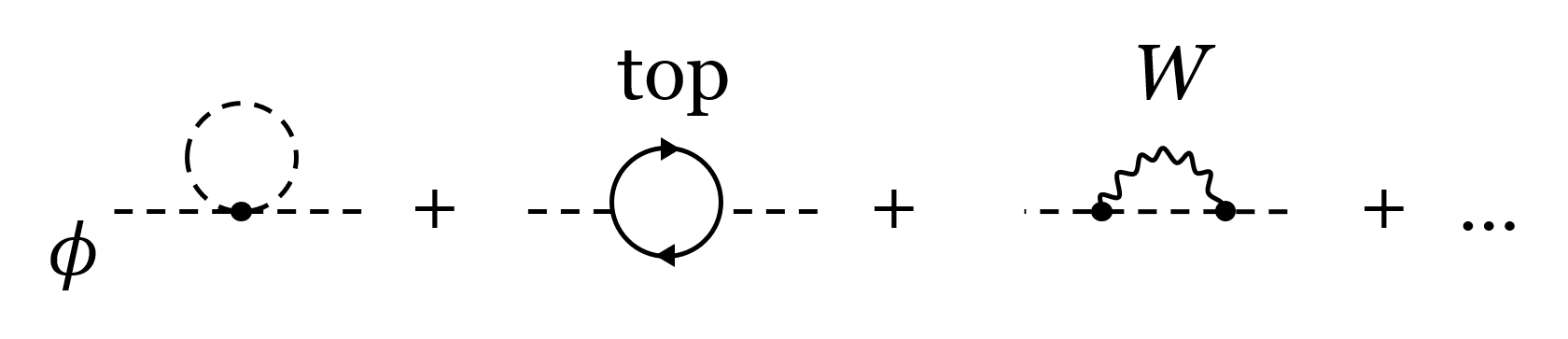}
\caption{Contributions to the Higgs self energy.}
\label{hse} 
\end{center}
\end{figure}

Let us first give a very simple argument (though possibly not very strong) why this is not easy. Namely, consider the theory as given by a classical lagrangian at $\Lambda$ and ask for low-energy observables. The most obvious is maybe a gauge coupling,
\be
\alpha_i^{-1}(\mu)\simeq \alpha_i^{-1}(\Lambda)+\frac{b_i}{2\pi}\ln\left(\frac{\Lambda}{\mu}\right)+{\cal O}(1)\,,
\ee
where we restricted attention to the one-loop level. The relevant diagrams are just the self-energy diagrams of the corresponding gauge boson with scalars, fermions and (in the non-abelian case) gauge bosons running in the loop. We see that, for $\Lambda\gg\mu$, the correction becomes large, but it grows only logarithmically. This goes together with the logarithmic divergence of the relevant diagrams, which is in turn related to the vanishing mass dimension of the coupling or operator coefficient that we are correcting. By contrast, for the Higgs mass we find
\cite{Gildener:1976ai, Veltman:1980mj}.
\be
m_H^2(\mu)=m_H^2(\Lambda)+\frac{c_H}{16\pi^2}\Lambda^2+{\cal O}(\Lambda^0)\,,
\label{hmf}
\ee
with $c_H$ a coupling-dependent dimensionless parameter to be extracted from diagrams like those in Fig.~\ref{hse}. We see that, suppressing ${\cal O}(1)$ coefficients and disregarding the logarithmic running of the dimensionless couplings between $\mu$ and $\Lambda$, we have $c_H=\lambda_H+\lambda_t^2+g_2^2+\cdots$.

Thus, $c_H$ is an ${\cal O}(1)$ number and if we set $\Lambda=1$~TeV, only an ${\cal O}(1)$ cancellation between the two terms on the r.h. side of (\ref{hmf}) is required to get the right Higgs mass parameter of the order of $(100\,\,\mbox{GeV})^2$. Things are actually a bit worse since there is a color factor of 3 coming with the top and other numerical factors. But, much more importantly, we can {\it not} simply declare 1~TeV to be the scale where our weakly coupled QFT breaks down and some totally unknown new physics (discrete space time, string theory etc.) sets in. One but not the only reason is the issue of flavour-changing neutral currents mentioned above. If we take the (still rather optimistic) value $\Lambda\sim 10$~TeV, we already require a compensation at the level of 1$\%$ or less between the two leading terms on the r.h. side of (\ref{hmf}). This starts to deserve the name fine-tuning or hierarchy problem.

A cautionary remark concerning expressions like $m_H^2(\mu)$ or $m_H^2(\Lambda)$ is in order. Such dimensionful parameters sometimes (not always) have power-divergent loop corrections. The momentum integral implicit in the loop correction is then dominated in the UV and changes by an ${\cal O}(1)$ factor if the regularization procedure changes. This is in contrast to e.g. $\alpha^{-1}(\mu)$ which is, at leading order, independent of how precisely the scale $\mu$ is defined. One can see that most easily by noting that $\ln(\Lambda/\mu)$ does not change significantly in the regime $\Lambda/\mu\gg 1$ if $\Lambda$ or $\mu$ are multiplied by, say, a factor of 2. Thus, a possibly less misleading way to write (\ref{hmf}) is
\be
m_H^2=m_{H,\,0}^2+\frac{c_H}{16\pi^2}\Lambda^2+{\cal O}(\Lambda^0)\,,
\label{hmf2}
\ee
Here $m_H^2$ is, by definition, the value of this operator in the IR and $m_{H,\,0}^2$ is the bare or classical value in the UV lagrangian. 

Still, the fine-tuning argument is not very convincing since, in (\ref{hmf2}), the two crucial terms between which a cancellation is required both depend on the cutoff or regularization used. For example, in dimensional regularization with minimal subtraction, the second term is simply zero and no cancellation appears necessary. Now this is clearly unphysical, but one may entertain the hope that some physical cutoff with similar features will eventually be established, defining a UV theory with a `naturally' small $m_H^2$ in spite of large $\Lambda$. 

But a much more technical and stronger argument making the fine-tuning explicit can be given. We make it using a toy model, but the relevance to the Standard Model will be apparent. The toy model is in essence something like `the inverse' of the Yukawa model of (\ref{tmp}). There, we considered the mass correction a fermion obtains when a heavy scalar is integrated out. We found that no large correction to the small fermion mass arises. Now consider (again following \cite{Luty:2005sn}), 
\be
{\cal L}=-\frac{1}{2}(\partial\phi)^2-\frac{1}{2}m^2\phi^2-\frac{\lambda}{4!}\phi^4+\ol{\psi}(i\slashed{\partial}-M)\psi+y\phi\ol{\psi}\psi\,.
\ee
We literally simply renamed $m\leftrightarrow M$, having of course in mind that now $m\ll M$. As before, we will not go through a careful procedure of `running and matching' to derive the low-energy EFT, but take the shortcut of integrating out the heavy field classically and adding loop corrections to the low-energy lagrangian terms. 

Since the fermion appears only quadratically in the action, its equations of motion are solved by $\psi=0$ for any field configuration $\phi(x)$. Hence, the first step consists in just dropping all terms with $\psi$. When considering loops, we focus only on corrections to the scalar mass proportional to $y^2$, finding
\be
{\cal L}_{EFT}=-\frac{1}{2}(\partial\phi)^2-\frac{1}{2}m_{EFT}^2\phi^2-
\frac{\lambda}{4!}\phi^4+\cdots\,,
\ee
with
\be
m^2_{EFT}\simeq m^2+\frac{y^2}{16\pi^2}\int_0^{\Lambda^2}k^2\,d(k^2)\,\frac{\mbox{tr}(\slashed{k}-M)^2}{(k^2+M^2)^2}\,.
\ee
This integral corresponds to the second diagrams of Fig.~\ref{hse}. It is immediately clear that both terms proportional to $\Lambda^2$ as well as to $M^2$ will arise:
\be
m^2_{EFT}\simeq m^2+\frac{y^2}{16\pi^2}\left(c_1\Lambda^2+c_2M^2\ln(\Lambda^2/M^2)+c_3M^2+\cdots\right)\,.\label{meft}
\ee
See e.g.~\cite{Srednicki:2007qs} for a corresponding analysis in dimensional regularisation. (Note that, while a quadratic divergence in 4 dimensions does not show up as a pole at $d=4$, it corresponds to a logarithmic divergence in 2 dimensions and hence shows up as a pole at $d=2$.)

Crucially, we now see that if, by some `UV miracle', the $m^2$ and $\Lambda^2$ terms always cancel to make $m_{EFT}^2$ very small, the tuning issue still remains: Even a very tiny relative change of $M^2$ (assuming that $M^2\gg m_{EFT}^2)$, would upset this cancellation. Of course, we can not rule out a UV model where {\it everything}, including masses of particles at intermediate scale (like our $M$ with $m_{EFT}\ll M\ll \Lambda$) are automatically correctly adjusted to ensure the necessary cancellation in (\ref{meft}). But now it becomes more apparent how tricky any mechanism accomplishing that would have to be. 

Concretely in the Standard Model with a seesaw mechanism for neutrino masses, the scale $M$ might be that of the heavy r.h. neutrino and one has, given the above, a strong argument for fine-tuning. Alternatively, one can of course avoid any such heavy particles (also giving up on Grand Unifcation - see below) and imagine that the Standard Model directly runs into a new theory in the UV where, at some scale $\Lambda$, a massless scalar is explained without tuning. I am not aware of any sufficiently concrete and convincing scenario of this sort. Nevertheless, we will return to a more detailed discussion of this and related logical possibilities in Sect.~\ref{alt}.

For now, let us accept that, from an EFT perspective, the Standard Model with UV scale $\Lambda$ is fine tuned and try to quantify the problem.

\subsection{Fine tuning}\label{fts}
\index{fine tuning}

Let us first emphasise that, having a small (dimensionful or dimensionless) parameter in an EFT is not in itself problematic or related to tuning. Indeed, the electron mass is small, but it comes from a dimensionless Yukawa coupling which only runs logarithmically. Thus, once small in the UV, it will stay small in the IR `naturally'.

Moreover, the relevant coupling of type
\be
\lambda_e\ol{l}_L\Phi e_R
\ee
is forbidden by chiral symmetry transformations, e.g. $e_R\to e^{i\alpha}e_R$. One can view $\lambda_e$ as a small effect in the UV lagrangian breaking this symmetry. Hence, the above operator will only receive loop corrections proportional to this symmetry breaking effect, i.e. to $\lambda_e$ itself. 

The same argument can be made for fermion masses even when they are viewed as dimensionful parameters. We have seen one example in Sect.~\ref{eft}. 
Another example is the Standard Model below the electroweak symmetry breaking scale, where the electron mass term
\be
m_e\ol{e}e=m_e\ol{e}_Le_R+\mbox{h.c.}
\ee
can be forbidden by the global $U(1)$ symmetry $e_R\to e^{i\alpha}e_R$, as above. Hence, there will be no loop corrections driving $m_e$ up to the electroweak scale, given that the tree-level value is small.

Small parameters with this feature are called `technically natural', a notion due to 't~Hooft~\cite{tHooft:1980xss}. More precisely, a small parameter is {\bf technically natural}\index{technically natural} if, by setting it to zero, the symmetry of the system is enhanced. The crucial point for us is that finding such a symmetry for the Higgs mass term turns out to be difficult if not impossible: One obvious candidate is a shift symmetry\index{shift symmetry}, $\Phi\to \Phi+\alpha$, with $\alpha=\,\,$const. But this forbids all non-derivative couplings and hence clashes with the main role the Higgs plays in the Standard Model, most notably with the top-Yukawa coupling, which is ${\cal O}(1)$. Nevertheless, attempts to at least alleviate the hierarchy problem using this idea have been made and we will discuss them in Sect.~\ref{lhm}. Another option is scale-invariance but, once again, the Standard Model as a quantum theory is not scale invariant - couplings run very significantly. Moreover, in the UV, most ideas for how the unification with gravity will work break scale invariance completely. Again, attempts along these lines nevertheless exist and will be mentioned. However, at our present `leading order' level of discussion it is fair to say that the smallness of $m_H^2$ is probably not technically natural. 

Somewhat more vaguely, one may say that the Higgs mass term is unnaturally small. To make this statement more precise, the notion of tuning or {\bf fine tuning} has been introduced. Roughly speaking, a theory is tuned if the  parameters in the UV theory (at the scale $\Lambda$) have to be adjusted very finely to realise the observed low-energy EFT.

It is not immediately obvious how to implement this in terms of formulae since, as just explained, e.g. the electron mass is known with high accuracy and even a tiny change of the UV-scale Yukawa coupling will lead to drastic disagreement with experiment. The main point one wants to make is that, as we have seen, the smallness of the Higgs mass apparently arises from the {\bf compensation} between two terms,
\be
m_H^2=m_{H,\,0}^2+\frac{c_H}{16\pi^2}\Lambda^2+\cdots\,.\label{hmfr}
\ee
Clearly, in such a situation, a small {\it relative} change, e.g., $m_{H,\,0}^2$ induces a much larger {\it relative} change of $m_H^2$. 

A widely used formula implementing this is known as the {\bf Barbieri-Giudice measure for fine-tuning}~\cite{Barbieri:1987fn} (see also \cite{Ellis:1986yg} and, for modern perspectives and further references, \cite{Wells:2018sus, Azhar:2018lzd}):
\index{Barbieri-Giudice measure}
\be
FT=\left|\frac{x}{\cal O}\,\frac{\partial {\cal O}}{\partial x}\right|=
\left|
\frac{\partial \ln({\cal O})}{\partial \ln(x)}\right|\,.
\ee
Here $x$ is the theory parameter and ${\cal O}$ the relevant observable. In our case, $x=m_{H,\,0}^2$ and ${\cal O}=m_H^2$ is given by (\ref{hmfr}), such that
\be
FT=m_{H,\,0}^2\,\frac{\partial \ln(m_{H,\,0}^2+c_H\Lambda^2/ (16\pi^2))}{\partial m_{H,\,0}^2}=\frac{m_{H,\,0}^2}{m_{H,\,0}^2+c_H\Lambda^2/ (16\pi^2))}\sim \frac{\Lambda^2/(16\pi^2)}{m_H^2}\,.
\ee
Here, in the last step, we assumed that $m_H^2\ll c_H\Lambda^2/(16\pi^2)$, such that $m_{H,\,0}^2\sim c_H\Lambda^2/(16\pi^2)$. Moreover, we have used that $c_H={\cal O}(1)$. As already noted earlier, this just formalises what we said at the intuitive level earlier: The fine tuning is roughly $\Lambda^2/(1~\mbox{TeV})^2$. 

For completeness, we record the natural multi-particle generalisation of the  Barbieri-Giudice measure. In this more general context, one may call it a `fine tuning functional', defined as a functional on the space of theories $T$ (following~\cite{Wells:2018sus}):
\be
FT[T]=\sum_{ij} \left|\frac{x_i}{\cal O}_j\,\frac{\partial {\cal O}_j}{\partial x_i}\right|\,.
\ee
We also note that our discussion was somewhat oversimplified and less concrete than in \cite{Barbieri:1987fn}. There, the observable was $m_Z^2$ (this is clearly tied to $m_H^2$, which is however not directly observable). Furthermore, the UV theory was not some rather vague cutoff-QFT, but it was a concrete model: The supersymmetric, in fact even supergravity-extended version of the Standard Model. We will get at least a glimpse of this below, after we have introduced supersymmetry.

Unfortunately, the above definition of fine-tuning has many problems. First, it is clearly not reparametrisation independent. In other words, it crucially depends on our ad hoc choice of $x_i$ as operator coefficients in a perturbative QFT and of the ${\cal O}_i$ as, roughly speaking, particle masses. Thus, one is justified in looking for other, possibly related, definitions. One such alternative definition is probabilistic: Choose a (probability) {\bf measure on the set of UV theories} and ask how likely it is to find a particular low-energy observable to lie in a certain range. For example, we might consider $m_{H,\,0}^2$ to have a flat distribution between zero and $2\Lambda^2/(16\pi^2)$ (where we also set $c_H=-1$). Then we obtain a small Higgs mass only if $m_{H,\,0}^2$ happens, by chance, to fall very close to the center point of its allowed range. 

To make this quantitative, we need to distinguish notationally between the Higgs mass parameter $m_H^2$ in our statistical set of theories and the concrete Standard Model value of this quantity. Let us call the latter $m_{H,\,obs.}^2$. Then, we may ask for the probability to find $m_H^2$ in the interval $[-m_{H,\,obs.}^2,m_{H,\,obs.}^2]$, or equivalently $|m_H^2|\lesssim m_{H,\,obs.}^2$. We obtain
\be
p(m_{H,\,obs.}^2)\simeq \frac{m_{H,\,obs.}^2}{\Lambda^2/(16\pi^2)}\,.
\ee
This is just the inverse of the Barbieri-Giudice fine tuning value, confirming at least at some intuitive level that the above definitions make sense. However, it becomes even more apparent that some ad hoc assumptions have come in. In particular, we required a measure on the space of UV theories or UV parameters.\index{UV parameters}

Finally, another ambiguity of the probabilistic view on fine-tuning is related to the choice of the allowed interval of the EFT observable. In the above, things were rather clear since our task was to quantify the problematic {\it smallness} of the Higgs mass relative to the cutoff. It was then natural to define all theories with $|m_H^2|\lesssim m_{H,\,obs.}^2$ as `successful'. However, the Higgs mass is by now known rather precisely, $m_h=125.18\pm 0.16$~GeV~\cite{Tanabashi:2018oca}, which translates into a similarly small allowed interval for $m_{H,\,obs.}^2$. If we had defined successful theories as those with $m^2_{H,\,obs.}$ falling into that interval, a much larger fine-tuning would result. Even worse, one could consider the very precisely known electron mass in the same way and would find a huge fine tuning of the UV-scale coupling $\lambda_e$, in spite of the logarithmic running and the technical naturalness.\footnote{This issue can be overcome by using {\it Bayesian inference} (see e.g.~\cite{Kass:1995loi, Trotta:2008qt}). Here, one  ascribes certain prior probabilities  to different models, each in turn predicting a certain distribution of Higgs masses. One may then use Bayes' theorem to derive the posterior (i.e.~after the Higgs mass has been measured) probability of each model. The models in question may for example be the Standard Model, with fine-tuned Higgs mass as above, and a supersymmetric model (see below) in which the Higgs mass is naturally small. The result will of course be a much higher probability for the `natural' model, see e.g.~\cite{Allanach:2007qk, Cabrera:2008tj, Fichet:2012sn, Fowlie:2014xha}. Crucially, the ratio of probabilities will automatically be insensitive to the precision with which the Higgs mass has been measured.
}\index{Bayesian inference}

Thus, one has to be careful with both definitions and it may well be that the final word about this has not yet been spoken. A suggestion for sharpening the probabilistic perspective is as follows: Consider the manifold of UV couplings (with some measure) and the map to the manifold of observables. On the latter, let ${\cal O}_0$ be some {\bf qualitatively} distinguished point, in our case the point of vanishing Higgs mass term. This point is distinguished since it specifies the boundary between two qualitatively different regimes -- that with spontaneously broken and unbroken $SU(2)$ gauge symmetry.  Let us assume that for any other point ${\cal O}$, we can in some way measure the distance to this special point, $|{\cal O}-{\cal O}_0|$. Now one may say that an {\it observed} EFT, corresponding to a point ${\cal O}_{obs.}$ on the manifold of observables, is fine tuned to the extent that the probability for all theories with 
\be
|{\cal O}-{\cal O}_0|<|{\cal O}_{obs.}-{\cal O}_0|
\ee
is small. In other words, we measure how unlikely it is that a randomly chosen theory falls more closely to the special point ${\cal O}_0$ than our observed EFT. 

Before closing this section, let us introduce some terminology that might be useful to sharpen our understanding: Simply the fact that the weak scale $m_{ew}$ is so much smaller than the Planck scale $M_P$ may be perceived as requiring explanation. It is common to characterise this as the {\bf large hierarchy problem}\index{large hierarchy problem}. This problem of very different energy scales in the same theory becomes an actual fine tuning problem if the Standard Model (or some minimal extension not affecting the quadratic Higgs mass divergence) is valid up to $M_P$. However, we only have strong experimental evidence that the cutoff of the Standard Model as an EFT is above $\sim 10\,$TeV. It could even be lower, but this leads to various phenomenological problems with flavour violation and precision data. Taking such a low cutoff $\Lambda$, one is only faced with a mild fine tuning $\sim m_H^2/(\Lambda^2/16\pi^2) \sim 10^{-2}$ for $\Lambda\sim 10\,$TeV. This is known as the {\bf little hierarchy problem}\index{little hierarchy problem}, which is on the one hand much less severe but on the other hand much more concrete and data-driven.

We close by recommending to the reader the lecture notes \cite{Wells:2009kq}, which discuss Higgs boson physics in much more detail, emphasising in particular the electroweak hierarchy problem. See also \cite{Giudice:2013yca} for brief, less technical discussions of the concept of {\bf Naturalness}\index{Naturalness} and \cite{Craig:2022uua} for a broad overview of recent work on Naturalness and further refs.

\subsection{Gravity and the cosmological constant problem}
\index{cosmological constant}\index{cosmological constant!problem}

Including gravity\index{gravity} in a minimalist approach amounts to the substitution
\be
{\cal L}_{SM}[\psi,\eta_{\mu\nu}]\qquad\to\qquad 
\sqrt{g}\,{\cal L}_{SM}[\psi,g_{\mu\nu}]+\frac{1}{2}M_P^2\sqrt{g}\,{\cal R}[g_{\mu\nu}]-\sqrt{g}\,\lambda\,.
\ee
As a result, two essential modifications of the discussion above arise: First, we learn that the Higgs mass problem is just one of two instances of very similar hierarchy problems - the other being the cosmological constant problem. Second, gravity sets an upper bound on the cutoff $\Lambda$, in a way that sharpens the Higgs mass hierarchy problem. 

In more detail, let us start by recalling what we need to know about gravity~\cite{Wald:1984rg, wei, mtw, car, strau}. On the one hand, gravity changes the picture very deeply: The arena of our Standard-Model QFT changes from $\mathbb{R}^4$ (with flat Lorentzian metric) to a Lorentzian manifold with dynamical metric, horizons, singularities in the cosmic past or future, or possibly even with topology change. The causal structure, which is so crucial for the definition of a QFT, becomes dynamical together with the metric. In particular, if one takes the metric itself to be a dynamical quantum field, the quantisation of this field depends on the causal structure, which follows from the (then a priori unknown) dynamics of this field itself. Diffeomorphism\index{diffeomorphism} invariance makes it very hard to define what a local observable in the usual QFT sense is supposed to be. Finally, to just mention one more issue, QFTs are most easily defined in euclidean metric. But this is extremely problematic in gravity since, even for a topologically trivial 4d euclidean manifold, the local value of ${\cal R}[g_{\mu\nu}]$ can take either sign. Thus, fluctuations around a flat euclidean background do not necessarily suppress the weight factor $\exp(-S_E)$ in the path integral. As a result, the whole euclidean approach\index{euclidean approach} may become problematic \cite{Gibbons:1978ac}.

But, on the other hand, one may also ignore most of the deep conceptual problems above and pretend that one has added to the Standard Model QFT just another gauge theory (see e.g.~\cite{vel}). We can not develop this approach here in any detail but only sketch the results: One expands the metric around flat space,
\be
g_{\mu\nu}=\eta_{\mu\nu}+h_{\mu\nu}\,.
\ee
and tries to think of $h_{\mu\nu}$ as of a gauge potential, analogous to $A_\mu$. The reader will recall the covariant derivative acting on a vector field,
\be
D_\mu v_\nu=\partial_\mu v_\nu-\Gamma_{\mu\nu}{}^\rho v_\rho\qquad \mbox{with}\qquad \Gamma_{\mu\nu}{}^\rho=\frac{1}{2}g^{\rho\sigma}(\partial_\mu
g_{\nu\sigma}+  \partial_\nu
g_{\mu\sigma} -  \partial_\sigma
g_{\mu\nu} )\,
\ee
and the curvature tensor written as a commutator, in analogy to $F_{\mu\nu}$:
\be
R_{\mu\nu\rho}{}^\sigma\,v_\sigma = [D_\mu,D_\mu]\,v_\rho\,,\qquad \mbox{or symbolically:}\qquad R \sim [\partial-\Gamma,\partial-\Gamma]\,.
\ee
From this, it is clear that the gravitational lagrangian takes the symbolic form (ignoring index structure and numerical factors)
\be
M_P^2\,[\,h\partial^2 h+h(\partial h)^2+h^2(\partial h)^2+\cdots]\,.
\ee
Defining $\kappa\equiv 1/M_P$ and rescaling $h\to \kappa h$, this becomes
\be
h\partial^2h+\kappa h(\partial h)^2+\kappa^2h^2(\partial h)^2+\cdots\,.
\ee
This is already quite analogous to the gauge theory structure (we are thinking of the non-abelian case, but are suppressing the group and Lorentz indices for brevity)
\be
A \partial^2 A+ gA^2\partial A+ g^2 A^4\,.
\ee
The crucial differences are that $g$ is dimensionless and the series of higher terms terminates at the quartic vertex. By contrast, in gravity the coupling has mass dimension $-1$ and the series goes on to all orders (both from $R$ as well as from the $R^2$, $R^3$ terms etc. which have to be added to the lagrangian to absorb all divergences arising at loop level). We will not discuss the technicalities of this -- suffice it to say that the Fadeev-Popov procedure and the introduction of ghosts work, at least in principle, as in gauge theories. 

We also recall that, for any observable that we can calculate in perturbation theory, the expansion reads
\be
c_0+c_1\kappa\Lambda+c_2\kappa^2\Lambda^2+\cdots
\ee
on dimensional grounds. From this we see that we have to expect power divergences and that higher loops are more and more divergent, consistent with the well-known fact that quantum gravity is perturbatively non-renormalisable. 

Finally, coming closer to our main point, we remember that $g_{\mu\nu}$ or, in our approach, $h_{\mu\nu}$ appears also in ${\cal L}_{SM}[\psi,g_{\mu\nu}]$. Since, as we know, the energy-momentum tensor\index{energy-momentum tensor} is\index{stress-energy tensor (see energy-momentum tensor)} defined as the variation of $S_{SM}$ with respect to $g_{\mu\nu}$ at the point $g_{\mu\nu}=\eta_{\mu\nu}$, it is clear that the leading order coupling of $h$ with matter is given by
\be
{\cal L}\supset \kappa h_{\mu\nu}T^{\mu\nu}\,.
\ee
This is, once again, completely analogous to the gauge theory coupling to matter via
\be
{\cal L}\supset g A_\mu^a j^\mu_a \qquad\mbox{with, e.g.}\qquad j_a^\mu=\ol{\psi}\gamma^\mu T_a\psi\,.
\ee

\begin{figure}[ht]
\begin{center} 
\includegraphics[width=7cm]{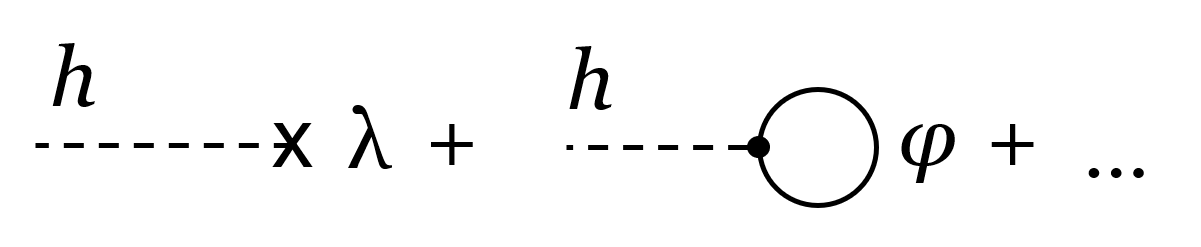}
\caption{Tree level and loop effect of the cosmological constant term on the metric field $h_{\mu\nu}$.}
\label{tad} 
\end{center}
\end{figure}

What is essential for us is that the cosmological constant term gives rise to an energy-momentum tensor
\be
T^{\mu\nu}=- \eta^{\mu\nu}\lambda\,.
\ee
If $\lambda$ is non-zero, then this corresponds to a non-zero source or `tadpole term'\index{tadpole} for the metric (gauge) field $h_{\mu\nu}$:
\be
{\cal L}\supset -\kappa h_{\mu\nu}\eta^{\mu\nu}\lambda\,.
\ee
The meaning of the word tadpole in this context becomes obvious if one considers the above as a tree-level diagrammatic effect and adds the first loop correction (due e.g.~to a scalar particle minimally coupled to gravity). This is illustrated in Fig.~\ref{tad}. 

One may think of the loop diagram in Fig.~\ref{tad} as a correction to $\lambda$, in direct analogy to the loop corrections to the Higgs mass from integrating out heavy particles which we discussed before. Thus, in analogy to e.g.~(\ref{hmf2}) and renaming our original cosmological term in the tree-level action to $\lambda_0$, we have
\be
\lambda=\lambda_0+\frac{c_\lambda}{16\pi^2}\Lambda^4\,.\label{cc}
\ee
The coefficient $c_\lambda$ does not include a small coupling constant but is merely proportional to the sum of bosonic and fermionic degrees of freedom (one may interpret this sign difference either as being due to the usual `minus' for each fermion loop or as the negative sign of the vacuum energy of the fermionic harmonic oscillator). Related to this, one can of course interpret the divergence as a sum over the vacuum energies of the oscillators corresponding to free field momentum modes. 

Famously, if one compares the observed value of the vacuum energy,
\be
\lambda\simeq (2.2\,\mbox{meV})^4\,,
\ee
with the expectation from (\ref{cc}) based on $\Lambda=M_P\simeq 2.4\times 10^{18}$~GeV, one finds a mismatch (i.e. a required fine tuning) of $10^{120}$. As in the Higgs mass case, there are caveats to this argument: Indeed, the value of the loop correction depends completely on the UV regularisation and one may imagine schemes where it would simply be zero. Also, as in Higgs case, there are counterarguments to this suggestion. Indeed, any massive particle contributes to the loop correction in a way that depends on its mass. Thus, the observed value changes dramatically if, e.g., the mass of the heavy r.h. neutrino needed in the seesaw mechanism changes.

To see this more explicitly, it is useful to give an explicit covariant formula for the one-loop correction to $\lambda$ (see problems for a derivation). For a single real scalar and in euclidean signature, one has
\be
\delta\lambda=\frac{1}{2}\int \frac{d^4k}{(2\pi)^4}\,\ln(k^2+m^2)= c_0\Lambda^4+c_1\Lambda^2m^2+\cdots\,.
\ee
We see that even the sub-leading term proportional to the mass is still also proportional to $\Lambda^2$ and hence huge. In fact, this is true even for the light particles of the Standard Model. Furthermore, there are effects due to the Higgs potential, the non-perturbative gluon-condensate of QCD and from all couplings (which enter at the two and higher-loop level). Thus, the case for an actual fine-tuning appears to be very strong indeed. Clearly, the amount of fine tuning may be significantly reduced compared to what we just estimated: We could add to the Standard Model heavy bosons and fermions, such that above a certain mass scale $M$ the number of fermions and bosons is equal and at least the leading $\Lambda^4$ term disappears. 

This last idea turns out to work much better than expected. It is realised in a systematic way in supersymmetry (SUSY) or supergravity (SUGRA). It still does not solve the cosmological constant problem, even in principle. The reason is that the scale of supersymmetry breaking is much too high. It does, however, solve the Higgs mass or electroweak hierarchy problem in principle. The fact that this solution does not work (at least not very well) in practice is due to fairly recent data, especially from the LHC. Nevertheless, it will be important for us to study SUSY in general and to a certain extent the SUSY version of the Standard Model. The reasons are twofold. First, as noted, SUSY is an excellent example for how things {\it could} work out nicely at the cutoff scale $\Lambda$. The precise understanding of how our apparent fine tunings could disappear or at least be mitigated will help us to evaluate their technical content and physical meaning. Second, if one wants quantum gravity divergences to also be tamed at the cutoff scale, SUSY is not enough and string theory is required. But the relation of the latter to real-world physics relies (at least in the best understood cases) on SUSY, which we hence have to understand at least at an introductory level. 

For reviews of the cosmological constant problem, see e.g.~\cite{Weinberg:1988cp, Weinberg:2000yb, Padmanabhan:2002ji, Padmanabhan:2006ag}. In particular, the arguments of  \cite{Weinberg:1988cp} 
against adjustment mechanisms\index{adjustment mechanism} for a zero cosmological constant(sometimes referred to as Weinberg's no-go theorem)\index{Weinberg's no-go theorem} are noteworthy.

\subsection{Problems}

\subsubsection{Electroweak symmetry breaking}
\index{electroweak symmetry breaking}

{\bf Task:} Calculate $W$ and $Z$ boson masses as well as the electromagnetic coupling $e$ in terms of $v$ and $g_{1,2}$. Derive the formula for the electric charge $Q=T_3+Y$, where $Y$ is the $U(1)$ hypercharge and $T_3$ is the so-called `isospin generator'\index{isospin} of the $SU(2)$ gauge group.

\noindent
{\bf Hints:} Apply the covariant derivative (for uncoloured fields)
\be
D_\mu=\partial_\mu - ig_2 A_\mu^a R(T^a) - ig_1 R(Y) B_\mu
\ee
to the Higgs VEV to derive the mass terms for $W^{\pm}$ and $Z$. Identify the massless field (the linear combination orthogonal to the massive vectors) as the photon and express the covariant derivative in terms of these fields.

\noindent
{\bf Solution:} The Higgs transforms in the fundamental representation of $SU(2)$, hence $R(T^a)=\sigma^a/2$. It has hypercharge $1/2$, hence $R(Y)=1/2$. 
It is convenient to work with $W^\pm=(A^1\mp iA^2)/\sqrt{2}$. Then one has
\be
A^1\sigma^1+A^2\sigma^2\,=\,
(A^1+iA^2)(\sigma^1-i\sigma^2)/2\,+\,(A^1-iA^2)(\sigma^1+i\sigma^2)/2
\,=\,\sqrt{2}W^-\sigma^- + \sqrt{2}W^+\sigma^+\,,
\ee
where
\be
\sigma^+=(\sigma^1+i\sigma^2)/2=\left(\begin{array}{cc}0&1\\0&0\end{array}
\right)\,,
\qquad
\sigma^-=(\sigma^1-i\sigma^2)/2=\left(\begin{array}{cc}0&0\\1&0\end{array}
\right)\,.
\ee
In the symmetry-broken vacuum, one then finds:
\be
D_\mu H=D_\mu\left(\begin{array}{c} 0 \\ v \end{array}\right)
=-
\frac{i}{2}\left(g_2\sqrt{2}W^+_\mu\left(\begin{array}{c} v \\ 0 \end{array}\right)
+\left[
-g_2A^3_\mu+g_1B_\mu
\right] \left(\begin{array}{c} 0 \\ v \end{array}\right)
\right)\,.
\ee
This gives rise to the mass term
\be
{\cal L}\supset -|D_\mu H|^2 = -\frac{v^2}{4}\left(2g_2^2|W^+_\mu|^2 + (g_1^2+g_2^2)(Z_\mu)^2 \right)\,.
\ee
We have to recall that $W^-=(W^+)^*$ and the complex $W$ boson\index{$W$ boson}\index{$Z$ boson} is normalised like a complex scalar field, i.e. without a factor $1/2$ in kinetic and mass term. Moreover, we introduced the canonically normalised massive vector
\be
Z_\mu=\frac{1}{\sqrt{g_1^2+g_2^2}}(g_2A^3_\mu-g_1 B_\mu)\,.
\ee
Thus, the mass term is
\be
{\cal L}\supset -m_W^2|W^+_\mu|^2-\frac{1}{2}m_Z^2(B_\mu)^2\,,
\ee
from which we read off
\be
m_W=g_2v/\sqrt{2}\qquad\mbox{and}\qquad m_Z=\sqrt{g_1^2+g_2^2}\cdot v/\sqrt{2}\,.
\ee

Next, we note that the linear combination of $A^3$ and $B$ orthogonal to $Z$ is
\be
A_\mu=\frac{1}{\sqrt{g_1^2+g_2^2}}(g_1A^3_\mu+g_2B_\mu)\,.
\ee
It is then immediate to express $A^3$ and $B$ through $Z$ and $A$:
\be
A^3_\mu=\frac{1}{\sqrt{g_1^2+g_2^2}}(g_1A_\mu+g_2Z_\mu)
\qquad \mbox{and}\qquad
B_\mu=\frac{1}{\sqrt{g_1^2+g_2^2}}(g_2A_\mu-g_1Z_\mu)\,.
\ee

Now the covariant derivative for a general field takes the form
\bea
D_\mu&=&\partial_\mu-ig_2\sqrt{2}(W^+_\mu R(T^+)+W^-_\mu R(T^-))
\\
&&-\frac{i}{\sqrt{g_1^2+g_2^2}}Z_\mu(g_2^2R(T^3)-g_1^2R(Y))
-\frac{ig_1g_2}{\sqrt{g_1^2+g_2^2}}A_\mu(R(T^3)+R(Y))\,.
\eea
It is clear that the transition between $A^3,B$ and $Z,A$ may be interpreted as an SO(2) rotation. The corresponding angle is known as the weak mixing angle or Weinberg angle $\theta_W$.  It is defined by
\be
\sin\theta_W=\frac{g_1}{\sqrt{g_1^2+g_2^2}}\,.
\ee
In terms of this angle, the electromagnetic charge (i.e. the prefactor of $A_\mu$ in the covariant derivative) is given by $e=g_2\sin\theta_W$. The group-theoretic coefficient is $Q=R(T^3)+R(Y)$. One often keeps the necessary use of the appropriate representation implicit, writing simply $Q=T_3+Y$.

\subsubsection{The Standard Model is anomaly free}
\index{anomaly}

{\bf Task:} Confirm this statement.

\begin{figure}[ht]
\begin{center} 
\includegraphics[width=6cm]{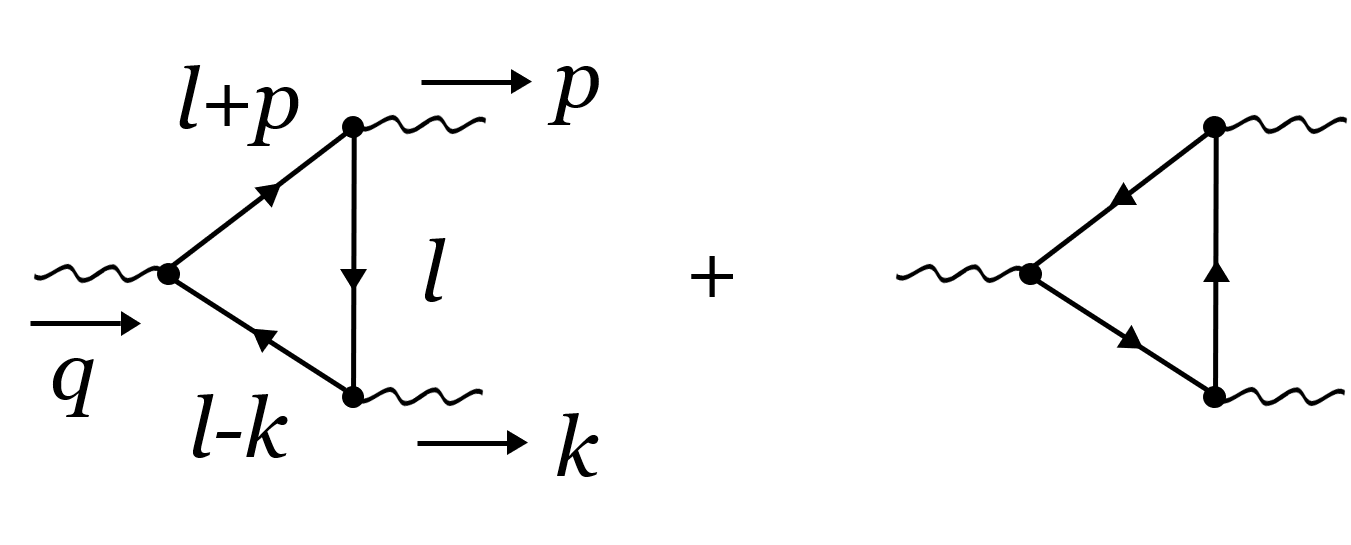}
\caption{Scattering amplitude interpretation of the expectation value of the axial current\index{axial current}. The momentum $q$ is related by Fourier transformation to the argument $x$ of $j^\mu(x)$. In (\ref{abjf}), $x$ has been set to zero.}
\label{abj} 
\end{center}
\end{figure}

\noindent
{\bf Hints:} Famously, in a theory with a single l.h. fermion $\psi$ (or, equivalently, a single Weyl fermion), the anomalous current non-conservation for 
\be
j_\mu\equiv \ol{\psi}\gamma_\mu\psi
\ee
reads
\be
\partial_\mu j^\mu=- \frac{1}{32\pi^2}F\tilde{F}\,.
\ee
A classical way to derive this is to consider the corresponding amplitude relation
\be
\langle p,k|\partial_\mu j^\mu(0)|0\rangle = -\frac{1}{32\pi^2}\,\langle p,k|\epsilon^{\alpha\nu\beta\rho} F_{\alpha\nu} F_{\beta\rho}(0)|0\rangle\,.
\label{abjf}
\ee
Here $\langle p,k|$ stands for a final state with two outgoing gauge bosons with momenta $p$ and $k$. The l.h. side of this equality is evaluated according to the diagrams in Fig.~\ref{abj}, and the r.h. side simply by expanding the fields in terms of creation and annihilation operators.

Given this diagrammatic understanding, it is very easy to see what the right generalisation to the non-abelian case is: At each vertex, the abelian gauge group generator `$1$' has to be replaced by the corresponding non-abelian generator $(T_a)_{ij}$. As a result, one has
\be
\partial_\mu j^\mu_a=-\frac{1}{32\pi^2}\,D_{abc}\epsilon^{\mu\nu\rho\sigma}
F_{\mu\nu}^b F_{\rho\sigma}^c\quad\mbox{with}\quad D_{abc}\equiv\frac{1}{2}\mbox{tr} [T_a\{T_b,T_c\}]\,.
\label{naa}
\ee

It should now be clear how to proceed: Consider the Standard Model fermions as one l.h. fermion field $\psi$ transforming the appropriate (very large) representation of $G_{SM}$. Let $T_a$ run over all 12 generators of this group. With this interpretation of $T_a$ and of the trace in (\ref{naa}), one only needs to check that $D_{abc}=0$. A lot of this is repetitive and can be simplified. For instance, the threefold repetition due to the three generations can be dropped -- even a single generation is anomaly free. Furthermore, rather than thinking about a complicated block-diagonal $T_a$, one can just sum over the different corresponding fermions in the loop. Finally, we clearly only need to show that $D_{abc}=0$ for all the 3$\,\times\, 3=9$ possible different assignments of $a$, $b$ and $c$ to the factor groups $SU(3)$, $SU(2)$ and $U(1)$. Which particular generator of e.g. $SU(3)$ one then chooses is immaterial. As a result, the amount of work is actually rather limited.

\noindent
{\bf Solution:} As explained above, we need to go through all possible ways of assigning the three generators corresponding to the three vertices of the triangle to the factors of $G_{SM}$. Thus, symbolically, we have to consider
\be
U(1)^3\,,\qquad U(1)^2\,SU(2)\,,\qquad U(1)\,SU(2)^2\,,\qquad U(1)\,SU(2)\,SU(3)\,,\cdots
\ee
and so on. But the generators of $SU(N)$ groups are all traceless, such that e.g. in the $U(1)^2\,SU(2)$ case we have (for each fermion species or, equivalently, each block)
\be
\mbox{tr}[T_{U(1)}^2 T^A_{SU(2)}]=\mbox{tr}[T_{U(1)}^2]\,\mbox{tr} [T^A_{SU(2)}]=0\,.
\ee
Thus, we only need to consider combinations where all three generators come from the same factor or where two come from the same factor and the third from the $U(1)$:
\be
U(1)^3\,,\qquad U(1)\,SU(2)^2\,,\qquad U(1)\,SU(3)^2\,,\qquad SU(2)^3\,,\qquad SU(3)^3\,.
\ee

Now let us go through this case by case. In the first case, we simply have to sum the cubes of the charges of all fermions. The anti-commutator is, of course, irrelevant. Using the list at the beginning of Sect.~1.1 of the notes, this gives
\bea
&3\times 2\times \left(\frac{1}{6}\right)^3
+3\times \left(-\frac{2}{3}\right)^3
+3\times \left(\frac{1}{3}\right)^3
+2\times \left(-\frac{1}{2}\right)^3
+\left(1\right)^3&\nonumber
\\
&=\frac{1}{36}-\frac{8}{9}+\frac{1}{9}-\frac{1}{4}+1=0\,.&
\eea
Note that the $SU(3)$ and $SU(2)$ representations are only relevant to determine the multiplicities corresponding to each set of fermions. 

In the second case, the anti-commutator is again irrelevant. Indeed,
\be
\mbox{tr}[T_{U(1)}\{T^a_{SU(2)},T^b_{SU(2)}\}]=2\, \mbox{tr}[T_{U(1)}]\,\mbox{tr}[
T^a_{SU(2)}T^b_{SU(2)}]\,,
\ee
and
\be
\mbox{tr}[T^a_{SU(2)}\{T_{U(1)},T^b_{SU(2)}\}]= 2\, \mbox{tr}[T_{U(1)}]\,\mbox{tr}[
T^a_{SU(2)}T^b_{SU(2)}]\,.
\ee
Since the $SU(2)$-trace always gives $\delta^{ab}/2$, we just need to sum the $U(1)$ charges of all $SU(2)$ doublets:
\be
3\times\frac{1}{6}+0+0-\frac{1}{2}+0=0\,.
\ee

The third case is analogous: We have to sum over the $U(1)$ charges of all $SU(3)$ triplets. (It does not matter whether it is a triplet or anti-triplet since tr$[T^a_{SU(3)}T^b_{SU(3)}]=\delta^{ab}/2$ holds for both). This gives
\be
2\times\frac{1}{6}-\frac{2}{3}+\frac{1}{3}+0+0=0\,.
\ee

In the fourth case we have $T^a_{SU(2)}=\sigma^a/2$ and hence
\be
\mbox{tr}[T^a_{SU(2)}\{T^b_{SU(2)},T^c_{SU(2)}\}]=\frac{1}{8}
\mbox{tr}[\sigma^a\{\sigma^b,\sigma^c\}]=\frac{1}{8}\mbox{tr}[\sigma^a]\,2\delta^{bc}=0\,.
\ee
Thus, we see that any theory with only fundamental representations (the antifundamental is equivalent to the fundamental) of $SU(2)$ is {\it trivially} free of the triangle anomaly. In fact, this extends to {\it all} representations of $SU(2)$ due to the reality-properties of its representations. 

Finally, the fifth and last case is the only one where we need to take into consideration that different representations of the same non-abelian group appear. We write
\be
T^a_{SU(3),\,fund.}=T^a_3\qquad \mbox{and}\qquad T^b_{SU(3),\,anti-fund.}=T^a_{\ol{3}}\,.
\ee
Now, since for a fundamental field $\Phi$ we have
\be
\Phi\,\to\, \exp(i\epsilon T)\,\Phi\qquad\mbox{and}\qquad \Phi^*\,\to\, \exp(-i\epsilon T^*)\,\Phi^*\,=\,\exp(-i\epsilon T^T)\,\Phi^*\,,
\ee
we can conclude that
\be
T^a_{\ol{3}}=-(T^a_3)^T\,.
\ee
As a result, we find
\be
\mbox{tr}[T^a_{\ol{3}}\{T^b_{\ol{3}},T^c_{\ol{3}}\}]=
\mbox{tr}[(-T^a_3)^T\{(-T^b_3)^T,(-T^c_3)^T\}]= -\mbox{tr}[T^a_3\{T^b_3,T^c_3 \}] \,.
\ee
Thus, we have to add the $SU(3)$-triplets and subtract the anti-triplets, each with its multiplicity:
\be
2-1-1=0\,.
\ee

We finally note that triangle anomalies (as considered above) involving different gauge group factors are called `mixed'. Without going into details, we also record the fact that a so-called mixed $U(1)$-gravitational anomaly exists. It comes from a triangle diagram\index{triangle diagram} involving one gauge-boson and two gravitons. To allow for a consistent coupling of the Standard Model to gravity, this anomaly also has to vanish. The calculation is similar to the $U(1)SU(2)^2$ and the $U(1)SU(3)^2$ case. Since all fermions couple to gravity in the same way, we simply have to add all $U(1)$ charges:
\be
6\times\frac{1}{6}-3\times\frac{2}{3}+3\times\frac{1}{3}-2\times\frac{1}{2}+1=0\,.
\ee

\subsubsection{The Standard Model and $SU(5)$}
\label{smsu5}

{\bf Task:} Embed $G_{SM}$ in a natural way in $SU(5)$\index{$SU(5)$} and show that the matter content of one generation (with all its gauge charges) follows from the ${\bf 10}+\ol{\bf 5}$ of $SU(5)$, where ${\bf 10}$ stands for the antisymmetric second rank tensor and $\ol{\bf 5}$ for the antifundamental representation. Consider a situation where the Standard Model follows from such an $SU(5)$ gauge theory (a {\bf Grand Unified Theory}\index{Grand Unified Theory} or {\bf GUT})\index{GUT} realised at some higher energy scale. Derive the tree-level prediction for the relative strength of the three Standard Model gauge couplings.

\noindent
{\bf Hints:} The `natural embedding' corresponds, of course, to identifying the upper-left $3\times 3$ block of $5\times 5$ $SU(5)$ matrices with $SU(3)$ and the lower-right $2\times 2$ block with $SU(2)$. The inverse would be equivalent - this is merely a convention. Hence, when viewed as generators of $SU(5)$, the $SU(3)$, $SU(2)$ and $U(1)$ generators are
\be
\left(\begin{array}{cc}
\left(T_{SU(3)}^a\right)_{3\times 3} & 0_{3\times 2}\\
0_{2\times 3} & 0_{2\times 2}
\end{array}\right)\,\,,\quad
\left(\begin{array}{cc}
0_{3\times 3} & 0_{3\times 2}\\
0_{2\times 3} & \left(T_{SU(2)}^a\right)_{2\times 2} 
\end{array}\right)\,\,,\quad
\frac{1}{\sqrt{60}}
\left(\begin{array}{ccccc}
-2&&&&\\
&-2&&&\\
&&-2&&\\
&&&3&\\
&&&&3
\end{array}\right).
\ee
The prefactor of the $U(1)$ generator ensures the standard non-abelian normalisation tr$(T^aT^b)=\delta^{ab}/2$. With this, it is immediate to write down the {\bf branching rule}\index{branching rule}
\be
{\bf 5}=({\bf 3},{\bf 1})_{-2}+({\bf 1},{\bf 2})_3\qquad\mbox{under}\qquad SU(5)\to SU(3)\times SU(2)\times U(1)\,.
\ee
Here we have rescaled the $U(1)$ generator in an obvious way for notational convenience. All one now needs to do is to infer the branching rules for the $\ol{\bf 5}$ and ${\bf 10}$ and to determine the gauge couplings $g_i$ of the Standard Model in the normalisation given in the lecture. (We note that, as is probably well-known, this unification scheme can not work without significant loop corrections -- cf.~Sect.~\ref{quni})

A classical introduction to group theory for physicists is \cite{Georgi:1999wka}. For an extensive collection of group and representation theory data see \cite{Slansky:1981yr}. The physics of Grand Unification, which builds on the technical observation discussed in this problem, is reviewed e.g.~in \cite{Ross:1985ai, Nath:2006ut, Raby:2017ucc, hh, Croon:2019kpe}.

\noindent
{\bf Solution:} The branching rule for $\ol{\bf 5}$ follows trivially from complex conjugation of the above:
\be
\ol{\bf 5}=(\ol{\bf 3},{\bf 1})_2+({\bf 1},{\bf 2})_{-3}\,.
\ee
Here we have used the fact that $\ol{\bf 2}={\bf 2}$ for $SU(2)$. This is obvious since $Lie(SU(2))=Lie(SO(3))$ and since, as derived in quantum mechanics, $SO(3)$ has a unique 2-dimensional representation. It can also be demonstrated explicitly by showing that, if 
\be
\psi_i\to U_{ij}\psi_j\,,\qquad\mbox{and}\qquad \psi_i^*\to \to U^*_{ij} \psi_i^*\,,
\ee
then the field $\chi_i\equiv \epsilon_{ij}\psi^*_j$ transforms exactly as $\psi_i$. We leave that to the reader. 

Formally speaking, we are claiming that the two representations ${\bf 2}$ and its complex conjugate, $\ol{\bf 2}$, are equivalent. This implies an isomorphism between the two vector spaces which commutes with the group action. In our case, the isomorphism is the multiplication with $\epsilon$. We will see a less trivial example of this below, which we will work out and after which it will be even more clear how to finish the $SU(2)$ discussion.

At this point, just looking at the pure $SU(2)$ doublet (there is only one such field in the Standard Model!), we can already identify the $U(1)$ charges with those of the Standard Model. We have the covariant derivative as it follows from the GUT:
\be
D_\mu=\partial_\mu-igT^a_{SU(2)}(A_2)_\mu^a-igY_{GUT}(A_1)_\mu\,.
\ee
According to the above,
\be
Y_{GUT}=\frac{-3}{\sqrt{60}}\,.\label{ygut}
\ee
On the Standard Model side, we have
\be
D_\mu=\partial_\mu-ig_2T^a_{SU(2)}(A_2)_\mu^a-ig_Y\,Y\,(A_1)_\mu
\ee
with
\be
Y=-1/2\label{ysm}
\ee
for the pure doublet (the lepton doublet).  Thus, we learn that 
\be
g\,Y_{GUT}=g_Y\,Y \qquad \mbox{or}\qquad \frac{g_Y^2}{g^2}=\frac{3}{5}\,.
\ee
This is the famous normalisation change between the Standard Model hypercharge $U(1)$ and the $SU(5)$-normalised $U(1)$. Note that we call the Standard Model gauge couplings $g_Y$, $g_2$ and $g_3$ at this point since, very frequently, the name $g_1$ is reserved for the hypercharge coupling in GUT normalisation, i.e. $g_1=\sqrt{5/3}\,g_Y$. 

We also see that the down-type r.h. quarks have the correct charge to be the $SU(3)$ anti-triplet coming with this $SU(2)$ doublet. (Their hypercharge differs by a factor -2/3, as it follows from $SU(5)$.)

As for the numerical outcome, we have the GUT prediction that $g_1=g_2=g_3$ at the GUT scale. This has to be compared to the observed values of roughly
\be
\alpha_1^{-1}\simeq 60\,,\qquad \alpha_2^{-1}\simeq 30\,,\qquad \alpha_3^{-1}\simeq 8
\ee
at the scale $m_Z$. Here the first two values follow from $\alpha_{em}^{-1}\simeq 127$, $e=g_2\sin\theta_W$, $\sin^2\theta_W=g_Y^2/(g_Y^2+g_2^2)$ and $\sin^2\theta_W\simeq 0.23$ together with $g_1=\sqrt{5/3}g_Y$ as explained above. Thus, as already noted, significant loop corrections (most plausibly from running over a large energy range) are needed for this unification scheme to work. 

Finally, three Standard Model fields are missing and we hope to get them from the ${\bf 10}$. To check this, let us first write down the tensor product 
\be
{\bf 5}\times {\bf 5}=[({\bf 3},{\bf 1})_{-2}+({\bf 1},{\bf 2})_3]\times [({\bf 3},{\bf 1})_{-2}+({\bf 1},{\bf 2})_3]
\ee
and anti-symmetrise:
\be
({\bf 5}\times {\bf 5})_A=(({\bf 3}\times {\bf 3})_A,{\bf 1})_{-4}+({\bf 1},({\bf 2}\times {\bf 2})_A)_6+({\bf 3},{\bf 2})_1\,.
\ee
Here the last representation only appears once since the other, equivalent term belongs to the symmetric part of the rank-2 tensor. In giving the $U(1)$ charges we have, as before, suppressed the factor $1/\sqrt{60}$. Comparing Eqs.(\ref{ygut}) and (\ref{ysm}), we see that the Standard Model $U(1)_Y$ charges in this normalisation follow by multiplication with a factor $(-1/2)/(-3)=1/6$. Given that $({\bf 2}\times {\bf 2})_A$ is clearly a singlet, we recognise the middle term as the r.h. electron. The last term is clearly the l.h. quark doublet. The first term should then be the r.h. up-type quark. 

\begin{figure}[ht]
\begin{center} 
\includegraphics[width=5cm]{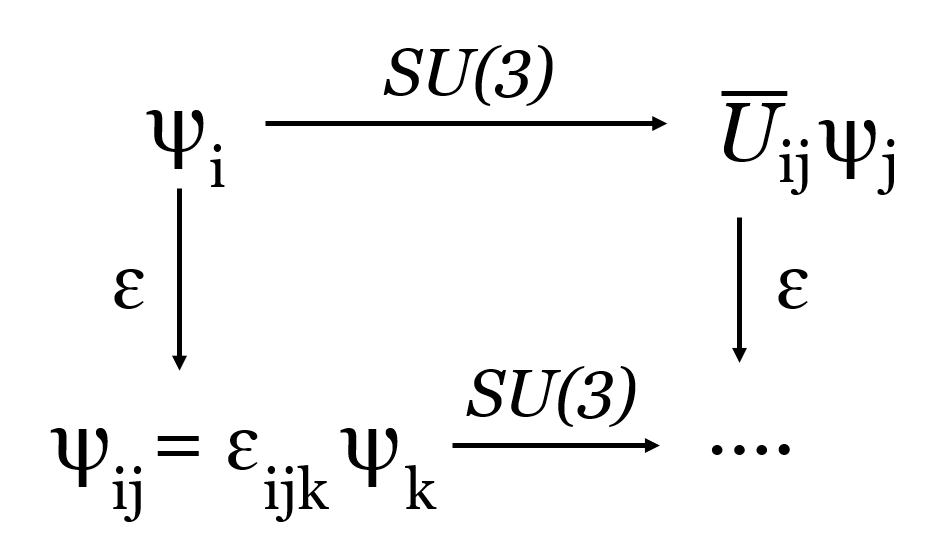}
\caption{Commuting diagram demonstrating the equivalence of two representations.}
\label{cd} 
\end{center}
\end{figure}

All we need to establish is that
\be
({\bf 3}\times {\bf 3})_A=\ol{\bf 3}\,.
\ee
To do so, we first identify the vector spaces of antisymmetric $SU(3)$ tensor and (anti-)vector by
\be
\psi_{ij}=\epsilon_{ijk}\psi_k\,.
\ee
Then we just need to show that they transform consistently, i.e., that the diagram in Fig.~\ref{cd} commutes. This implies
\be
U_{ik}U_{jl}\epsilon_{klm}\psi_m=\epsilon_{ijk}U^*_{km}\psi_m\,.
\ee
To verify this equality, we remove $\psi_m$ and multiply by $(U^T)_{mn}$:
\be
U_{ik}U_{jl}U_{nm}\epsilon_{klm}=\epsilon_{ijk}U^*_{km}(U^T)_{mn}\,.
\ee
On the r.h. side the two mutually inverse matrices cancel; the l.h. side is just the epsilon tensor multiplied by the determinant of $U$, the latter being unity. Thus, we are done.

\subsubsection{Weyl spinors}\label{wsp}
\index{Weyl!spinor}

{\bf Tasks:} 

\noindent
(1) Define the canonical map $SL(2,\mathbb{C})\to SO(1,3)$ using the vector of four sigma matrices $\sigma_\mu=(\mathbbm{1},\sigma_1,\sigma_2, \sigma_3)$. Then go on to show that $(\sigma_\mu)_{\alpha\dot{\alpha}}$ is an invariant tensor of the Lorentz group. Build the Dirac spinor and gamma-matrices from Weyl spinors and sigma matrices and express the transformation of a Dirac spinor\index{Dirac spinor} under a Lorentz rotation in terms of a given $SL(2,\mathbb{C})$ matrix $M$. 

\noindent
(2) Rewrite the Dirac spinor invariants 
\be
\ol{\psi}^{(1)}_D\psi^{(2)}_D\equiv \psi^{(1)\dagger}_D\gamma_0\psi^{(2)}_D\,\,, 
\qquad
\ol{\psi}^{(1)}_D\gamma_5\psi^{(2)}_D\,\,, 
\qquad
\ol{\psi}^{(1)}_D\gamma_\mu\psi^{(2)}_D\,\,, 
\qquad
\ol{\psi}^{(1)}_D\gamma_\mu\gamma_5\psi^{(2)}_D
\ee 
in terms of Weyl spinors. Use the upper/lower and lower/upper index summation convention for undotted and dotted Weyl indices respectively:
\be
\psi\chi\equiv \psi^\alpha\chi_\alpha\,\,,\qquad \ol{\psi}\ol{\chi}\equiv \ol{\psi}_{\dot{\alpha}}\ol{\chi}^{\dot{\alpha}}\,.
\ee

\noindent
(3) Check the crucial identity 
\be
\ol{\sigma}_\mu\sigma_\nu+\ol{\sigma}_\nu\sigma_\mu=-2\eta_{\mu\nu}\mathbbm{1} \label{smao}
\ee
and derive the Clifford algebra\index{Clifford algebra} relation for the $\gamma$ matrices from it.
\noindent

\noindent
{\bf Hints:} 

\noindent
(1) The first part is a direct generalisation of the construction of the map $SU(2)\to SO(3)$ which should be familiar from quantum mechanics. The second step is a straightforward calculation using only the fact that the indices $\alpha$ and $\dot{\alpha}$ transform with $SL(2,\mathbb{C})$ matrices and with complex conjugate $SL(2,\mathbb{C})$ matrices respectively. In the last step you need to use the convention that the upper / lower two components of a Dirac spinor are given by a Weyl spinor with lower undotted / upper dotted index. 

A convenient set of conventions is that of the Appendix of the book by Wess and Bagger \cite{Wess:1992cp}, in particular
\be
\epsilon_{\alpha\beta}=\left(\begin{array}{cc} 0 & -1 \\ 1 & 0 \end{array}
\right)\,\,,\qquad 
\epsilon^{\alpha\beta}=\left(\begin{array}{cc} 0 & 1 \\ -1 & 0 \end{array}
\right)\,\,,\qquad \mbox{such that} \qquad 
\epsilon_{\alpha\beta}\epsilon^{\beta\gamma}=\delta_\alpha{}^\gamma\,.
\ee
This allows us to raise and lower Weyl indices with the $\epsilon$ tensor. Of course one needs to use the fact -- please check if not obvious -- that $\epsilon$ is an invariant tensor of $SL(2,\mathbb{C}$.)

\noindent
(2) This is completely straightforward. Deviating from the Wess-Bagger conventions, it may be convenient to define $\gamma^5\sim \gamma^0\gamma^1\gamma^2\gamma^3$ with a prefactor which ensures that the l.h. projector (i.e. the projector on the undotted Weyl spinor) is $P_L=(\mathbbm{1}-\gamma^5)/2$. 

\noindent
(3) Use that $\epsilon^{\alpha\beta}=-i(\sigma_2)^{\alpha\beta}$ together with the familiar commutation relations of the Pauli matrices. 

\noindent
{\bf Solution:} 

\noindent
(1) Given a 4-vector $v$, define $\hat{v}\equiv v^\mu\sigma_\mu$. The matrix $\hat{v}$ is hermitian, as is the matrix
\be
\hat{v}'=M\hat{v}M^\dagger\,,\label{str}
\ee
where $M\in SL(2,\mathbb{C})$. Since $\{\sigma_\mu\}$ is a basis of hermitian $2\times 2$ matrices, there exists a unique decomposition 
\be
\hat{v}'=v'^\mu\sigma_\mu\,,
\ee
which defines the $SL(2,\mathbb{C})$-transformed vector $v'$. To see that this an $SO(1,3)$ transformation, it suffices to check that $v^2$ is preserved. This follows immediately from 
\be
\det\hat{v}=\left(\begin{array}{cc}v^0+v^3 & v^1-iv^2 \\ v^1+iv_2 & v^0-v^3
\end{array}\right)=(v^0)^2-\ol{v}^2=-v^2
\ee
together with the obvious fact that the $SL(2,\mathbb{C})$ transformation (\ref{str}) preserves the determinant. 

With this, we are ready to check that $(\sigma_\mu)_{\alpha\dot{\alpha}}$\index{dotted Weyl index} is an invariant tensor. To do so, let $M\in SL(2,\mathbb{C})$ and let $\Lambda\in SO(1,3)$ be its image under the map defined above. We have
\be
\Lambda_\mu{}^\nu M_\alpha{}^\beta \ol{M}_{\dot{\alpha}}{}^{\dot{\beta}} (\sigma_\nu)_{\beta\dot{\beta}}=\Lambda_\mu{}^\nu (M\sigma_\nu M^\dagger)_{\alpha\dot{\alpha}}\,.\label{ite}
\ee
Note that undotted/dotted Weyl indices by definition transform with $M/\ol{M}$. We also know that
\be
M\sigma_\mu M^\dagger v^\mu = \sigma_\mu v'^\mu=\sigma_\mu\Lambda^\mu{}_\nu v^\nu
\ee
for any $v$ and hence
\be
M\sigma_\nu M^\dagger=\sigma_\mu\Lambda^\mu{}_\nu\,.
\ee
With this, we return to (\ref{ite}) and continue the calculation according to
\be
\Lambda_\mu{}^\nu M_\alpha{}^\beta \ol{M}_{\dot{\alpha}}{}^{\dot{\beta}} (\sigma_\nu)_{\beta\dot{\beta}} = \Lambda_\mu{}^\nu (\sigma_\rho)_{\alpha\dot{\alpha}} \Lambda^\rho{}_\nu=\eta_{\mu\sigma} \Lambda^\sigma{}_\tau \eta^{\tau\nu}\, (\sigma_\rho)_{\alpha\dot{\alpha}} \Lambda^\rho{}_\nu =\eta_{\mu\sigma}\eta^{\sigma\rho} (\sigma_\rho)_{\alpha\dot{\alpha}} = (\sigma_\mu)_{\alpha\dot{\alpha}}\,.
\ee
Thus, we are indeed dealing with an invariant tensor. 

Finally, we take
\be
\psi_D=\left(\begin{array}{c}\psi_\alpha\\ \ol{\chi}^{\dot{\alpha}} \end{array}\right)
\ee
as a definition of a Dirac spinor. For covariance reasons (and up to possible convention-dependent prefactors), we have
\be
\gamma_\mu=\left(\begin{array}{cc} 0 & (\sigma_\mu)_{\alpha\dot{\alpha}} \\ (\ol{\sigma}_\mu)^{\dot{\beta}\beta} & 0 \end{array}\right)\,.
\ee
The Lorentz transformation matrix is
\be
D(M)=\left(\begin{array}{cc} M_\alpha{}^\beta & 0 \\ 0 & \ol{M}^{\dot{\alpha}}{}_{\dot{\beta}} \end{array}\right)\,,
\ee
where the $\ol{M}^{\dot{\alpha}}{}_{\dot{\beta}}$ is obtained from $M_\alpha{}^\beta$ by complex conjugation and raising/lowering of the indices.

\noindent 
(2) Using our suggestion to define $\gamma^5=\mbox{diag}(-\mathbbm{1},\mathbbm{1})$, the result follows from the definitions:
\bea
\ol{\psi}^{(1)}_D\psi^{(2)}_D = \chi^{(1)}\psi^{(2)}+\ol{\psi}^{(1)}\ol{\chi}^{(2)}\,\,,\quad&\quad 
\ol{\psi}^{(1)}_D\gamma_5\psi^{(2)}_D = -\chi^{(1)}\psi^{(2)}+\ol{\psi}^{(1)}\ol{\chi}^{(2)}
\\
\ol{\psi}^{(1)}_D\gamma_\mu\psi^{(2)}_D = \ol{\psi}^{(1)}\ol{\sigma}_\mu\psi^{(2)}+\chi^{(1)}\sigma_\mu \ol{\chi}^{(2)}
\,\,,\quad&\quad
\ol{\psi}^{(1)}_D\gamma_\mu\gamma_5\psi^{(2)}_D =
- \ol{\psi}^{(1)}\ol{\sigma}_\mu\psi^{(2)}+\chi^{(1)}\sigma_\mu \ol{\chi}^{(2)}\,.
\eea

\noindent
(3) Write
\bea
&&(\ol{\sigma}_\mu)^{\dot{\alpha}\alpha}(\sigma_\nu)_{\alpha\dot{\beta}}+\{\mu\leftrightarrow\nu\}\,=\,\epsilon^{\dot{\alpha}\dot{\gamma}}\epsilon^{\alpha\beta}
(\ol{\sigma}_\mu)_{\dot{\gamma}\beta}(\sigma_\nu)_{\alpha\dot{\beta}}+\{\mu\leftrightarrow\nu\}
\\
&=&[(-i\sigma_2)\ol{\sigma}_\mu(-i\sigma_2)^T(\sigma_\nu)]^{\dot{\alpha}}{}_{\dot{\beta}}+\{\mu\leftrightarrow\nu\}\,=\,
[(\sigma_2)\ol{\sigma}_\mu(\sigma_2)(\sigma_\nu)]^{\dot{\alpha}}{}_{\dot{\beta}}+\{\mu\leftrightarrow\nu\}\,,
\eea
where in the last two expressions $\ol{\sigma}_\mu$ and $\sigma_\nu$ are assumed to have lower indices. Now use that
\be
\sigma_2\ol{\sigma}_0\sigma_2=\sigma_0\qquad \mbox{and}\qquad
\sigma_2\ol{\sigma}_i\sigma_2=-\sigma_i\,.
\ee
Using this minus sign, it becomes clear that the expressions with $\{\mu\nu\}=\{0i\}$ and $\{\mu\nu\}=\{i0\}$ vanish after symmetrization. The case $\{\mu\nu\}=\{00\}$ obviously gives the right answer. For  $\{\mu\nu\}=\{ij\}$ one needs to use $\sigma_i\sigma_j+\sigma_j\sigma_i=2\delta_{ij}$ to find the result. The Clifford algebra for $\gamma$ matrices is a direct consequence.

\subsubsection{Covariant expression for the 1-loop vacuum energy}
\index{vacuum energy}

{\bf Task:} Derive the covariant expression ($\sim \int d^4k\,\ln(k^2+m^2)$) for the vacuum energy given in the lecture.

\noindent
{\bf Hints:} Write down the path integral for gravity and a real scalar and integrate out the scalar, including in particular its vacuum fluctuations. Focus only on the dependence on rescalings of the metric, i.e.~on metrics of the form $g_{\mu\nu}=\alpha\eta_{\mu\nu}$. This can in turn be interpreted as a dependence on the 4-volume.

\noindent
{\bf Solution:} The complete partition function (we suppress any source terms for simplicity) reads
\be
Z=\int Dg\,D\phi\,\exp\left[i\int d^4x\,\sqrt{g}\left(\frac{1}{2}M_P^2{\cal R}- (\partial\phi)^2-m^2\phi^2\right)\right]\,.
\ee
Here $Dg$ stands for the integration over all metrics and $\sqrt{g}$ is the square root of the modulus of the determinant of $g_{\mu\nu}$. The $\phi$-part of the action can be rewritten as 
\be
-i\int d^4x\,\sqrt{g}\,\phi M \phi\qquad\mbox{with}\qquad M\equiv -\partial^2+m^2\,.
\ee
For our purposes, it will be sufficient to understand how the $\phi$ part of the partition function changes with $\alpha$ if $g_{\mu\nu}=\alpha\eta_{\mu\nu}$. Under this restriction, we can reparameterise our spacetime such that $g_{\mu\nu}= \eta_{\mu\nu}$ and only keep track of the dependence on the total 4-volume $V$. 

After Wick rotation\index{Wick rotation} (deformation of the $x^0$ integration contour from real to imaginary axis by clockwise rotation and subsequent renaming $x^0=-ix^4$), we have
\be
-\int_V d^4x\,\phi M_E \phi\qquad\mbox{with}\qquad M_E\equiv -\partial^2 +m^2 \qquad \mbox{and} \qquad \partial^2=\delta^{\mu\nu} \partial_\mu\partial_\nu\,.
\ee
Now we are dealing with a Gaussian integral with a matrix in the exponent, giving us
\be
\int D\phi\,\exp\left[-\int_V d^4x\,\phi M_E \phi\right]= \frac{1}{\sqrt{\det(M_E)}}=\exp\left(-\frac{1}{2}\mbox{tr}\ln M_E \right)\,.
\ee
Here we have absorbed an infinite constant factor in the definition of $D\phi$ in the first step und applied the identity $\ln\det=\mbox{tr}\ln$ in the second step. 

Now we note that in Fourier space
\be
M_E(k,p)=\delta^4(k-p)\,(k^2+m^2)\,,
\ee
and hence
\be
\mbox{tr}\,\ln\,M_E=\int d^4k\,\delta^4(k-k)\ln(k^2+m^2)=V\int\frac{d^4k}{(2\pi)^4}\,\ln(k^2+m^2)\,.
\ee
Here, in the first step, the $\delta$ function has remained outside the log since it only signals that the matrix in question is diagonal. In the second step, we used
\be
(2\pi)^4\delta^4(k-p)=\int d^4x\,e^{ix(k-p)}=V\qquad\mbox{for}\qquad p=k\,.
\ee
Undoing the Wick rotation and reinstating $\int d^4x\,\sqrt{g}$ instead of $V$, we find 
\be
Z=\int Dg\,\exp\left[i\int d^4x\,\sqrt{g}\left(\frac{1}{2}M_P^2{\cal R}- \lambda\right)\right]\quad\mbox{with}\quad \lambda=\frac{1}{2}\int \frac{d^4k}{(2\pi)^4}\,\ln(k^2+m^2)\,.
\ee
Note that the intermediate transition to euclidean space could have been avoided by regularising the oscillating Gaussian (with an $i$ in the exponent) in some other way.

\section{Supersymmetry and Supergravity}\label{sands}
\index{supersymmetry (see SUSY)}

There are many motivations to learn about SUSY\index{SUSY}. Let us give a few: SUSY is the only known symmetry relating fermions and bosons and may as such be a logical next step in the historical road towards unification in fundamental physics. String theory is the best-understood model of quantum gravity (or indeed the true theory underlying quantum gravity) and its stable versions all rely on SUSY (in 2d and in 10d). The only presently controlled roads from 10d strings to the 4d Standard Model involve 4d SUSY theories as an intermediate step. SUSY can, at least in principle, resolve the hierarchy problem at the scale where it becomes manifest. In other words: If it had been discovered at the electroweak scale, we could have found ourselves in a world without the hierarchy problem. Even if that happens at, say, 10 TeV, the tuning would be much less severe than without SUSY. Finally, SUSY is a central tool in formal field theory research since SUSY theories usually involve many cancellations at the loop-level making them much better controlled. For example the, best-understood example of the famous AdS/CFT correspondence (to be explained later) involves an ${\cal N}\!=\!4\,$ 4d super-Yang-Mills (SYM) theory. Here ${\cal N}\!=\!4$ stands for four times the minimal amount of supersymmetry in the given dimension.

The structure and notation of what follows will be strongly influenced by the classic text \cite{Wess:1992cp}, but there are many other useful books \cite{west, weinberg3, turning, shifman}.

\subsection{SUSY algebra and superspace}
\index{superspace}
Recall the Poincare algebra 
\bea
\left[ P_\mu,P_\nu \right]&=&0
\label{poi1}\\
\left[ M_{\mu\nu} , P_\rho \right] & = & i\eta_{\mu\rho}P_\nu-i\eta_{\nu\rho}P_\mu
\label{poi2}\\
\left[ M_{\mu\nu},M_{\rho\sigma}\right] &=&i\eta_{\mu\rho}M_{\nu\sigma}  
- i\eta_{\nu\rho}M_{\mu\sigma} 
- i\eta_{\mu\sigma}M_{\nu\rho}  
+ i\eta_{\nu\sigma}M_{\mu\rho} 
\label{poi3}
\eea
as the symmetry algebra of $\mathbb{R}^{1,3}$. This algebra can be represented by differential operators acting on functions on $\mathbb{R}^{1,3}$, e.g.
\be
P_\mu=-i\partial_\mu\qquad\qquad \left(\partial_\mu=\frac{\partial}{\partial x^\mu}\right)\,.
\ee
Indeed, these operators generate translations according to
\be
\exp[i\epsilon^\mu P_\mu]\,f(x)=f(x)+\epsilon^\mu \partial_\mu f(x)+\cdots = f(x+\epsilon)\,.
\ee
Finite rotations in $\mathbb{R}^{1,3}$ are analogously generated by $M_{\mu\nu}$. 

Any relativistic QFT has the above symmetry, but it may have additional (`internal') symmetries acting on the fields. Examples are a shift symmetry $\phi\to \phi+\epsilon$ or rotations in field space $\Phi\to \exp(i\epsilon_a T_a)\Phi$ with $\Phi\in \mathbb{C}^N$ and $T_a$ the $SU(N)$ generators. Here `additional' means that the full symmetry algebra is the direct sum of Poincare and internal Lie algebra. The {\bf Coleman-Mandula theorem}~\cite{Coleman:1967ad}\index{Coleman-Mandula theorem} claims that such a direct sum structure is the only possibility for how the Poincare-Algebra can be extended to a larger symmetry of a QFT (more precisely, of the S-matrix).

This theorem can be avoided if one generalises the definition of a symmetry by a Lie algebra: One replaces the latter by a so-called Lie superalgebra. Moreover, the resulting extension of the Poincare algebra is (essentially) unique and is called the supersymmetry algebra. This uniqueness is the statement of the {\bf Haag-Lopuszanski-Sohnius theorem}~\cite{Haag:1974qh}\index{Haag-Lopuszanski-Sohnius theorem}.

We will not demonstrate uniqueness but only present the result of the analysis: The new generators to be added are (Weyl) spinors $Q_\alpha$ and the crucial new algebra relations are
\be
\{Q_\alpha,\ol{Q}_{\dot{\alpha}}\}=2(\sigma^\mu)_{\alpha\dot{\alpha}}P_\mu\,, \qquad \{Q_\alpha,Q_\beta\}=0\,,\qquad \{\ol{Q}_{\dot{\alpha}} ,\ol{Q}_{\dot{\beta}}\}=0\,.\label{qc}
\ee
The main novelty is that for these generators one does not provide commutators but anti-commutators, hence we are now dealing with a Lie superalgebra.\index{Lie superalgebra}

The object $(\sigma^\mu)_{\alpha\dot{\alpha}}$ is defined as
\be
\sigma^\mu=(-\mathbbm{1},\sigma^1,\sigma^2,\sigma^3)
\ee
and is an invariant tensor of $SL(2,\mathbb{C})$ just like $(\gamma^\mu)_{ab}$ is an invariant tensor of $SO(1,3)$. In fact, these two statements are of course related since the Lie algebras are the same and, roughly speaking, $\gamma$ consists of two blocks of $\sigma$'s (cf.~Problem~\ref{wsp}). One can avoid $\sigma$'s and Weyl spinors and formulate everything using left-handed 4-spinors, but Weyl spinors are very convenient in this context. 

Relations between two bosonic generators remain commutators and relations between the new fermionic and the old bosonic commutators are also formulated in terms of commutators:
\be
[P_\mu,Q_\alpha]=0\,,\qquad [M_{\mu\nu},Q_\alpha]=i(\sigma_{\mu\nu})_\alpha{}^\beta Q_\beta\,,\label{mqc}
\ee
where
\be
\sigma_{\mu\nu}\equiv -\frac{1}{4}(\sigma_\mu\ol{\sigma}_\nu -\sigma_\nu\ol{\sigma}_\mu)\qquad \mbox{and}\qquad (\ol{\sigma}_\mu)^{\dot{\alpha}\alpha} \equiv \epsilon^{\dot{\alpha}\dot{\beta}}\epsilon^{\alpha\beta} (\sigma_\mu^*)_{\dot{\beta}\beta}\,.\label{sba}
\ee
We will often use an overline instead of the star (or dagger) for complex conjugation (or the adjoint operator). The overline on $\ol{\sigma}$ does not specify whether upper or lower indices are assumed. Indices can be raised or lowered using the $\epsilon$ tensor. Given that we need the lower-upper index version of $(\sigma_{\mu\nu})_\alpha{}^\beta$ in (\ref{mqc}), the expression on the l.h.~side of (\ref{sba}) should be read as defining precisely this version. Hence, it involves a upper-upper index version of $\ol{\sigma}$, which is provided on the r.h.~side of (\ref{sba}).

The full SUSY algebra is defined by (\ref{poi1})-(\ref{poi3}) together with (\ref{qc}) and (\ref{mqc}). Thus, we see that it consists of the Poincare algebra, the $Q$ anticommutators, and the claim that the $Q$'s transform under the Poincare algebra as space-time-independent spinors. It may at this point also be useful to say more formally what a {\bf Lie superalgebra} is: It is a vector space with a $\mathbb{Z}_2$ grading (it splits in an even and odd part) and with a binary operation that obeys the rules even$\times$even$\,\to\,$even, even$\times$odd$\,\to\,$odd and odd$\times$odd$\,\to\,$even. Furthermore, there are rules concerning the symmetries of these operations and Jacobi-like identities. These are, however, automatically fulfilled if the operations are explicitly realised through (anti)commutators, as in our case.

Next we want to represent this algebra on a larger space, called superspace. Its coordinates are
\be
x^\mu\,\,(\mu=0\cdots 3)\qquad \mbox{and} \qquad \theta^\alpha\,\,(\alpha=1,2)\,,
\ee
the latter being fermionic (Grassmann variables)\index{Grassmann variable} which form a Weyl spinor. The key relations for our purposes are
\be
(\theta^\alpha)^*=\ol{\theta}^{\dot{\alpha}}\,,\qquad \{\theta^\alpha,\theta^\beta\}=0\quad\mbox{and}\quad\mbox{h.c.}\,,\qquad 
\{\theta^\alpha,\ol{\theta}^{\dot{\alpha}}\}=0
\ee
or, more explicitly,
\be
(\theta^1)^2=(\theta^2)^2=0\,\,,\qquad \theta^1\theta^2=-\theta^2\theta^1\,, \qquad\mbox{etc.}
\ee
One also defines partial derivatives
\be
\partial_\alpha=\frac{\partial}{\partial\theta^\alpha}\,\,\qquad \ol{\partial}_{\dot{\alpha}}=\frac{\partial}{\partial \ol{\theta}^{\dot{\alpha}}}
\ee
together with the obvious rules
\be
\partial_\alpha\theta^\beta=\delta_\alpha{}^\beta\,\,,\qquad
\ol{\partial}_{\dot{\alpha}}\ol{\theta}^{\dot{\beta}}= \delta_{\dot{\alpha}}{}^{\dot{\beta}}\,\,,\qquad
\partial_\alpha\ol{\theta}^{\dot{\beta}}=0\,\,\qquad
\ol{\partial}_{\dot{\alpha}}\theta^\beta=0\,.
\ee
The reader should check that, as a result of the anticommutation relations for the $\theta$'s, the $\partial$'s also anticommute.

The space parameterised by the $x^\mu$ and $\theta^\alpha$ is called {\bf superspace}, in this case $\mathbb{R}^{4|4}$, with 4 bosonic and 4 real fermionic (or two complex fermionic) dimensions. Intuitively, one may want to think of $\mathbb{R}^4$ not as a set of points but, equivalently, as the algebra of functions on $\mathbb{R}^4$: $\{1,x^\mu,x^\mu x^\nu,\cdots\}$.\footnote{According to the Gelfand-Naimark theorem, one may in fact (under very general circumstances) always think of the algebra of functions on a space rather than of the space itself. These objects encode the same information.} The generalisation to superspace is then obvious: One simply thinks of the algebra of functions including $\theta$'s, i.e.~$\{1,x^\mu,\theta^\alpha,x^\mu x^\nu,x^\mu\theta^\alpha, \cdots\}$. This is not a commutative algebra anymore and hence it is not really the space of functions on some set of points. However, in QFT we anyway mostly work with the space of functions on our spacetime. So the formal generalisation from $\mathbb{R}^4$ to $\mathbb{R}^{4|4}$ on the basis of the respective function spaces should not prevent us from doing all relevant manipulations.

Next, we naturally expect that a symmetry of this enlarged space will involve some analogue of the familiar generators of translations, i.e. $Q_\alpha \sim
\partial_\alpha+\cdots$. The ellipsis stands for extra terms which must come in to ensure that $Q$'s anticommute to give the $P$'s. The correct formulae turn out to be
\be
Q_\alpha=\partial_\alpha-i(\sigma^\mu)_{\alpha\dot{\alpha}} \ol{\theta}^{\dot{\alpha}}\partial_\mu\,,\qquad 
\ol{Q}_{\dot{\alpha}}=-\ol{\partial}_{\dot{\alpha}}+i\theta^\alpha (\sigma^\mu)_{\alpha\dot{\alpha}} \partial_\mu\,.
\ee
It is a straightforward but very important exercise to derive the essential part of the SUSY algebra from this:
\be
\{Q_\alpha,\ol{Q}_{\dot{\alpha}}\}=2i(\sigma^\mu)_{\alpha\dot{\alpha}} \partial_\mu\,, \qquad \{Q_\alpha,Q_\beta\}=0\,,\qquad \{\ol{Q}_{\dot{\alpha}} ,\ol{Q}_{\dot{\alpha}}\}=0\,.\label{algd}
\ee

Let us pause for a small, technical comment: The reader will have noticed that, with the standard identification $P_\mu=-i\partial_\mu$ (recall that we are using a mostly-plus metric), the algebras of (\ref{qc}) and (\ref{algd}) differ by a sign. This is nothing deep but merely a result of two different ways of defining the operators $Q$ and $P$. On the one hand, one may think of them as acting on functions. On the other hand, as acting on coordinates. To make this clear, one may consider the relation 
\be
(A_\partial B_\partial f)(x)=(B_\partial f)(Ax)=f(BAx)
\ee
between operators $A$, $B$ acting on coordinates $x$ and the corresponding differential operators $A_\partial$, $B_\partial$ acting on functions of $x$. It is immediately clear from the above that the Lie algebras characterising the action on functions and on the space itself differ by a sign. In our context, (\ref{algd}) corresponds to the action on functions and (\ref{qc}) to that on $\mathbb{R}^{4|4}$. The latter is also relevant for the action on the Hilbert space, where states are e.g. of the type $\Phi(x)|0\rangle$, with $\Phi$ a QFT field operator. The reader may also want to recall that $\hat{A}\hat{B}\Phi(x)\hat{B}^{-1}\hat{A}^{-1}=\Phi(ABx)$, with $\hat{A}$,$\hat{B}$ operators acting on the Hilbert space. Thus, there is no further sign change when comparing the action on $\mathbb{R}^{4|4}$ to that on the Hilbert space.

\subsection{Superfields}\index{superfield}
Now one builds a field theory on this enlarged space. A (complex) general superfield\index{general superfield} is a function
\bea
F(x,\theta,\ol{\theta})& =&f(x)+\theta\phi(x)+\ol{\theta}\ol{\chi}(x)+ \theta^2m(x) +\ol{\theta}^2n(x)+\theta\sigma^\mu\ol{\theta}v_\mu(x)
\nonumber \\
&&+\theta^2\ol{\theta}\ol{\lambda}(x)+\ol{\theta}^2\theta\psi(x)+\theta^2 \ol{\theta}^2d(x)\,.\label{gsf}
\eea
Here the r.h. side is a Taylor expansion of the l.h. side where, however, all higher terms vanish. The coefficient functions $\phi$, $\chi$, $\lambda$ and $\psi$ are Weyl spinors, anticommuting among each other and with the $\theta$'s. 

We have started to use a very convenient shorthand notation for the product of Weyl spinors, for example
\be
\theta\phi\equiv \theta^\alpha\phi_\alpha=\epsilon^{\alpha\beta}\theta_\alpha \phi_\beta\,\,,\quad\mbox{and analogously}\quad \theta^2=\theta^\alpha\theta_\alpha\,.
\ee
It is an essential part of this convention that suppressed undotted indices are always summed from upper-left to lower-right. For dotted indices, the rule is inverse:
\be
\ol{\theta}\ol{\chi}\equiv \ol{\theta}_{\dot{\alpha}}\ol{\chi}^{\dot{\alpha}}\,.
\ee
This convention goes together with certain $\epsilon$ tensor conventions:
\be
\epsilon_{\alpha\beta}=\left(\begin{array}{cc} 0 & -1 \\ 1 & 0 \end{array}
\right)\,\,,\qquad 
\epsilon^{\alpha\beta}=\left(\begin{array}{cc} 0 & 1 \\ -1 & 0 \end{array}
\right)\,\,,\qquad \mbox{with} \qquad 
\epsilon_{\alpha\beta}\epsilon^{\beta\gamma}=\delta_\alpha{}^\gamma\,.
\ee
With this contraction one has, in spite of the anticommutation relations,
\be
\psi\chi=\chi\psi\,,
\ee
as the reader should check. 

It goes without saying that the Poincare algebra acts on superfields in the usual way, e.g.
\be
\delta_\epsilon F=i\epsilon^\mu P_\mu F = \epsilon^\mu\partial_\mu F\,.
\ee
By analogy, we define the {\bf SUSY transformation}\index{SUSY!transformation}
\be
\delta_\xi F=(\xi Q+\ol{\xi}\ol{Q})F=[\,(\xi\partial-i\xi \sigma^\mu\ol{\theta} \partial_\mu)\,+\,\mbox{h.c.}\,]F\,.
\ee

Here by `h.c.' we mean the application of a formal $*$-operation on the algebra of functions and differential operators. In essence, this is just complex conjugation and its obvious extension to the $\theta$'s. A crucial exception is the rule
\be
(\partial_\alpha)^*=-\ol{\partial}_{\dot{\alpha}}\,,
\ee
which is required by consistency. The reader should check this by carefully thinking about the possible ways to evaluate $(\partial_\alpha\theta^\beta)^*$. 

Returning to our SUSY transformations, we note that the abstract concept of the superfield $F$ mainly serves the purpose of defining SUSY transformations on the set of `component' fields\index{component fields} $f$, $\phi_\alpha$ etc. The latter are conventional quantum fields. Concretely, after calculating $\delta_\xi F$, we expand it in a Taylor series and define $\delta_\xi f$, $\delta_\xi\phi$, etc. as the coefficients of the various terms with growing powers of $\theta$:
\be
\delta_\xi F=\delta_\xi f+\theta^\alpha(\delta_\xi\phi)_\alpha+\ol{\theta}_{\dot{\alpha}}(\delta_\xi\ol{\chi})^{\dot{\alpha}}\cdots\,.
\ee
This defines the SUSY transformation of the component fields.

\subsection{Chiral superfields}\index{chiral superfield}
The general superfield is too large to be practically useful and it does indeed correspond to a reducible representation of the SUSY algebra. Simpler superfields exist and are sufficient to write down the most general SUSY lagrangian. 

To define the chiral superfield, it is useful to first introduce SUSY-covariant derivatives (in a way very similar to the $Q$'s):
\be
D_\alpha=\partial_\alpha+i(\sigma^\mu)_{\alpha\dot{\alpha}} \ol{\theta}^{\dot{\alpha}}\partial_\mu\,,\qquad 
\ol{D}_{\dot{\alpha}}=-\ol{\partial}_{\dot{\alpha}}-i\theta^\alpha (\sigma^\mu)_{\alpha\dot{\alpha}} \partial_\mu\,.
\ee
They obey 
\be
\{D_\alpha,\ol{D}_{\dot{\alpha}}\}=-2i(\sigma^\mu)_{\alpha\dot{\alpha}} \partial_\mu\,, \qquad \{D_\alpha,D_\beta\}=0\,,\qquad \{\ol{D}_{\dot{\alpha}} ,\ol{D}_{\dot{\alpha}}\}=0
\ee
and, crucially, any $D$ or $\ol{D}$ anticommutes with any $Q$ or $\ol{Q}$,
\be
\{D_\alpha,\ol{Q}_{\dot{\alpha}}\}=0\qquad \mbox{etc.}
\ee
This last feature implies that
\be
\ol{D}_{\dot{\alpha}}F=0\qquad\Rightarrow\qquad \ol{D}_{\dot{\alpha}} \delta_\xi F=0\,.
\ee
In other words, superfields fulfilling the condition $\ol{D}_{\dot{\alpha}}F=0$ form a subrepresentation of the Lie superalgebra representation provided by general superfields. They are called {\bf chiral superfields}.

One may show that chiral superfields can always be written as 
\be
\Phi=\Phi(y,\theta)\qquad \mbox{with} \qquad y^\mu=x^\mu+i\theta\sigma^\mu\ol{\theta}\,,
\ee
where $\Phi$ is an unconstrained function. It can be expanded according to
\be
\Phi=A(y)+\sqrt{2}\theta\psi(y)+\theta^2 F(y)\,.\label{csfe}
\ee
As explained above for the general superfield, one obtains the SUSY transformations of the component fields by expanding $\delta_\xi\Phi$ in the same way as $\Phi$. The result reads
\bea
\delta_\xi A&=&\sqrt{2}\psi\xi
\nonumber \\
\delta_\xi\psi&=&i\sqrt{2}\sigma^\mu\ol{\xi}\partial_\mu A+\sqrt{2}\xi F
\\
\delta_\xi F&=&i\sqrt{2}\ol{\xi}\ol{\sigma}^\mu\partial_\mu \psi\,. \nonumber
\eea
We note that one can analogously define antichiral superfields, $D_\alpha\ol{\Phi}=0$, and that the conjugate of a chiral superfield is antichiral.

\subsection{SUSY-invariant lagrangians}
We state without proof that the most general such lagrangian, at the 2-derivative-level and built from chiral superfields $\{\Phi^1,\cdots,\Phi^n\}$ only, reads
\be
{\cal L}=K(\Phi^i,\ol{\Phi}^{\ol{\imath}})\Big|_{\theta^2\ol{\theta}^2}+
W(\Phi^i)\Big|_{\theta^2}+\mbox{h.c.}\,.\label{sil}
\ee
Here $K$ is a real function of a set of complex variables $\Phi^i$. With $\Phi^i$ being chiral superfields, $K$ becomes a general superfield. It is not chiral since both $\Phi^i$ and $\ol{\Phi}^{\ol{\imath}}$ are involved. The first term in ${\cal L}$ is the projection of the general superfield $K$ on its highest component, i.e., it is the analogue of the function $d(x)$ appearing in the Taylor expansion (\ref{gsf}). 

The function $K$ is called the Kahler potential\index{Kahler!potential} (for those who know this term from complex geometry: the relevance in the present context will become clear momentarily). The expression $K\Big|_{\theta^2\ol{\theta}^2}$, viewed as part of the lagrangian, is called the $D$ term.\index{$D$ term} This name comes simply from the traditional use of the variable $d(x)$ for the highest component. The key point in this non-trivial way of writing a lagrangian is, of course, its required invariance under SUSY transformations. For this, we need to recall that the commutator of $Q$ and $\ol{Q}$ is $P$. Hence the mass dimension of $Q$ is $1/2$. Since $Q$ involves $\partial/\partial\theta$, the mass dimension of $\theta$ is $-1/2$ (one may think of it very vaguely as the square root of $x$). Thus, in the Taylor expansion of superfields in powers of $\theta$ the mass dimensions of components grow. As a result, due also to the linear nature of SUSY transformations, the highest component can not transform into any other component -- there simply is no component with a suitably high mass dimension. The only way it can transform is into a {\it derivative} of another component. Thus, the first term of the above lagrangian is invariant up to total derivatives, as one would have hoped. 

Similarly, $W$ is called the superpotential\index{superpotential} and it is an analytic (or  holomorphic) function of the $\Phi^i$. This makes $W$ a chiral superfield. In its Taylor expansion in $\theta$, with the coefficients being functions of $y$, the highest component is traditionally called $F$ (cf.~(\ref{csfe})). Hence the corresponding two terms in (\ref{sil}) are sometimes called $F$ terms.\index{$F$ term} Concretely, to get these terms one expands the chiral superfield $W(\Phi^i)$ in $\theta$ (with the coefficients being functions of $y$), extracts the coefficient of $\theta^2$, and replaces $y$ by $x$. The result, together with its hermitian conjugate, is the $F$-term lagrangian. It is SUSY invariant up to a total derivative for the same reason as explained in the case of the $D$ term.

An equivalent way of writing this lagrangian is as
\be
{\cal L}=\int d^2\theta\,d^2\ol{\theta}\, K(\Phi^i,\ol{\Phi}^{\ol{\imath}})
+\int d^2\theta\, W(\Phi^i)+\mbox{h.c.}
\ee
Using standard integration rules for Grassmann variables,
\be
\int d \theta_1\,\theta_1=1\qquad \mbox{and}\qquad \int d\theta_1\,1=0\,,
\ee
and the analogous formulae for $\theta_2$, one can easily check that the integral formulation is equivalent to the projection formulation of ${\cal L}$. The SUSY invariance is particularly easily seen in the integral formulation: The SUSY generator $Q$ is a linear combination of $x$ derivatives and $\theta$ derivatives. The $x$ derivative of any lagrangian is, by definition, a total derivative and thus leaves the action invariant. The $\theta$ derivative of any expression in $\theta$ integrates to zero,
\be
\int d\theta_1\,\frac{\partial}{\partial\theta_1}\Big(\cdots\Big)=0\,,
\ee
as one can easily convince oneself. Thus, any action which is an integral over the full superspace is invariant. Similarly, any action built as the integral of an expression in $\theta$ (not $\ol{\theta}$) and integrated over half the superspace is invariant. (Here it is important to note that we can replace $y$ with $x$ by appealing to the Taylor expansion and the irrelevance of total derivatives in $x$.)

\subsection{Wess-Zumino-type models}
The possibly simplest interesting SUSY model is the\index{Wess-Zumino!model} Wess-Zumino model \cite{Wess:1974tw}.\footnote{For a very interesting earlier model, with supersymmetry realised non-linearly on fermionic fields only, see \cite{Volkov:1973ix}.}
It is defined by 
\be
K=\ol{\Phi}\Phi\,\,,\qquad W=\frac{m}{2}\Phi^2+\frac{\lambda}{3}\Phi^3\,.
\ee
A straightforward explicit calculation according to the rules above gives the following component form of the lagrangian:
\be
{\cal L}=-|\partial A|^2-i\ol{\psi}\ol{\sigma}^\mu\partial_\mu\psi-
\left(\frac{m}{2}\psi^2+\lambda\psi^2A\right)+\mbox{h.c.}+(mA+\lambda A^2)F +\mbox{h.c.}+|F|^2\,.
\ee
Since $F$ has no kinetic term (and thus does not propagate) we can integrate it out by purely algebraic operations and without any approximation. Such fields are called auxiliary fields.\index{auxiliary field} The equation of motion for $F$ is
\be
F=-m\ol{A}-\lambda\ol{A}^2\,,
\ee
and inserting this into the original lagrangian gives
\be
{\cal L}=-|\partial A|^2-i\ol{\psi}\ol{\sigma}^\mu\partial_\mu\psi-
\left(\frac{m}{2}\psi^2+\lambda\psi^2A\right)+\mbox{h.c.}-V(A,\ol{A})\,,
\ee
with the scalar potential (or $F$-term potential)
\be
V(A,\ol{A})=|F|^2=|mA+\lambda A^2|^2\,.
\ee

This is easily generalised to (non-renormalizable and multi-field) models of the type
\be
K=K(\Phi^i,\ol{\Phi}^{\ol{\imath}})\,\,,\qquad W=W(\Phi^i)\,.
\ee
We only display the purely bosonic part of the resulting component lagrangian. 
More details are given in the problems. With the auxiliary fields already integrated out, one has:
\be
{\cal L}=K_{i\ol{\jmath}}(A,\ol{A})\,(\partial A^i)(\partial \ol{A}^{\ol{\jmath}}) +K^{i\ol{\jmath}}(A,\ol{A})\,(\partial_i W(A))(\ol{\partial}_{\ol{\jmath}} \ol{W}(\ol{A}))+\cdots\,.
\ee
Here
\be
K_{i\ol{\jmath}}=\partial_i \ol{\partial}_{\ol{\jmath}} K\qquad\mbox{and} \qquad K_{i\ol{\jmath}}K^{k\ol{\jmath}}=\delta_i{}^k\,,
\ee
in other words, indices denote partial derivatives and the upper-index matrix is defined as the inverse. To simplify notation, we have have suppressed field indices in the arguments such that, when writing e.g.~$W(A)$, we mean $W(A^1,\cdots,A^n)$.

We note that the scalar components $A^i$ parametrise a complex manifold (as in so-called sigma-models) and, in supersymmetry, the metric on this field space is the Kahler metric\index{Kahler!metric} $K_{i\ol{\jmath}}$, defined with the help of the Kahler potential $K$. The superpotential $W$ is locally a holomorphic function on this manifold; globally it is a section in an appropriate complex line bundle.

\subsection{Real Superfields}\index{real superfield}
We have to discuss real superfields, another subrepresentation contained in that of the general superfield, since they are needed to describe gauge theories. But we will be very brief since, conceptually, the procedure is similar to that used in the chiral superfield case.

A real superfield $V=V(x,\theta,\ol{\theta})$ is defined by the condition $V=\ol{V}$. It can be Taylor expanded in $\theta$ and $\ol{\theta}$. We will build lagrangians which are invariant under the SUSY gauge transformation\index{SUSY!gauge transformation}
\be
2V\to 2V+\Lambda+\ol{\Lambda}\,,
\ee
with $\Lambda$ a chiral superfield. Using this transformation, $V$ can be brought to a form where certain components vanish (the so-called Wess-Zumino gauge):\index{Wess-Zumino!gauge}
\be
V=-\theta\sigma^\mu\ol{\theta}A_\mu+i\theta^2\,\ol{\theta}\ol{\lambda}-i \ol{\theta}^2\,\theta\lambda+\frac{1}{2}\theta^2\ol{\theta}^2 D\,.
\ee
Moreover, applying SUSY-covariant derivatives to $V$, one constucts the so-called field-strength superfield
\be
W_\alpha=-\frac{1}{4}\ol{D}^2D_\alpha V\,,
\ee
which can be shown to be chiral and gauge invariant. Its name is justified since it does indeed contain the field strength $F_{\mu\nu}=\partial_\mu A_\nu- \partial_\nu A_\mu$ in one of its components:
\be
W=i\lambda(y)+\left[D(y)+i\sigma^{\mu\nu}F_{\mu\nu}(y)\right]\cdot\theta+ \theta^2\sigma^\mu\partial_\mu\ol{\lambda}(y)\,,
\ee
where $4\sigma^{\mu\nu}=\sigma^\mu\ol{\sigma}^\nu-\sigma^\nu\ol{\sigma}^\mu$. One can show that SUSY gauge transformations contain standard gauge transformations as a subset. Hence, it is natural to look for SUSY-invariant and SUSY-gauge-invariant lagrangians. At the 2-derivative level, the unique option is
\be
{\cal L}=\frac{1}{4g^2}\left(W^\alpha W_\alpha\Big|_{\theta^2} +\ol{W}_{\dot{\alpha}} \ol{W}^{\dot{\alpha}}\Big|_{\ol{\theta}^2}\right)
=\frac{1}{g^2}\left\{ -\frac{1}{4}F_{\mu\nu}F^{\mu\nu}-i\ol{\lambda} \ol{\sigma}^\mu\partial_\mu\lambda +\frac{1}{2}D^2\right\}\,,
\ee
where $\lambda$ is the gaugino and $D$ the real auxiliary field that has already appeared above.

It is straightforward to extend this to the non-abelian case, where $V$ and $W$ are matrix-valued superfields, taking values in the Lie Algebra of the gauge group. Let us write the corresponding lagrangian including also a charged matter superfield $\Phi$, to be thought of as a column-vector in some appropriate representation. We have
\be
{\cal L}=\frac{1}{2g^2}\,\mbox{tr}\left( W^2\Big|_{\theta^2}+\mbox{h.c.}\right)
+\Phi^\dagger e^{2V}\Phi\Big|_{\theta^2\ol{\theta}^2}+{\cal W}(\Phi)\Big|_{\theta^2}+\mbox{h.c.}\,.
\ee
Here $e^{2V}$ has to be taken in the representation of $\Phi$, and $\Phi^\dagger$ has to be interpreted as a row vector. The superpotential ${\cal W}$ is a holomorphic, gauge invariant function of $\Phi$. We denote it by a caligraphic letter ${\cal W}$ to avoid confusion with the field strength superfield $W$. This lagrangian is invariant under the non-abelian super gauge transformations
\be
e^{2V}\to e^{\Lambda^\dagger} e^{2V}e^{\Lambda}\,\,,\qquad \Phi\to e^{-\Lambda}\Phi\,.
\ee

One frequently uses the naming conventions for components
\be
\Phi=\{\Phi,\psi,F\}\,\,,\qquad V=\{A_\mu,\lambda,D\}\,.
\ee
It is a slight abuse of notation to denote the scalar matter component by the same name as the superfield, but this convention is widespread and it is usually clear from the context which meaning is intended. With these conventions, the component form of the lagrangian reads
\bea
{\cal L}&=&\frac{1}{g^2}\,\mbox{tr}\left\{ -\frac{1}{2}F_{\mu\nu}F^{\mu\nu} -2i\ol{\lambda}\ol{\sigma}^\mu D_\mu\lambda+D^2 \right\}
\\
\nonumber
\\
&&-|D_\mu\Phi|^2-i\ol{\psi}\ol{\sigma}^\mu D_\mu\psi+|F|^2+i\sqrt{2}\left(
\Phi^\dagger\lambda\psi-\ol{\psi}\ol{\lambda}\Phi\right) + \Phi^\dagger D \Phi\,, \nonumber
\eea
where we have set ${\cal W}=0$ for simplicity.
This lagrangian is called off-shell since it is SUSY invariant without using the equations of motion. Integrating out the auxiliary field, one arrives at the on-shell lagrangian. Concerning $F$, this step is trivial in the present simple example: $F$ is just set to zero. By contrast, integrating out $D=D_a T_a$ induces a quartic term in the scalar fields, the so-called $D$-term potential.\index{$D$-term potential}

\subsection{SUSY breaking}\label{susybr}\index{SUSY!breaking}
We have so far defined the spinor $Q$ as a differential operator on superspace. Hence, it is an operator on the space of superfields, hence an operator transforming different component fields into each other. After quantisation, we will thus be able to define a corresponding operator $Q$ on the Hilbert space. This operator will mix bosons and fermions and, since
\be
[Q_\alpha,P_\mu P^\mu]=0\,,
\ee
this implies that the mass of fermions and bosons (in the same superfield or multiplet) is the same. Thus, to be relevant for the real world, supersymmetry must be spontaneously broken. In other words, while the action should be supersymmetric, the vacuum should not be invariant under supersymmetry transformations.

At the perturbative level, this simply means that the lowest-energy field configuration should not be invariant under SUSY. In the context of chiral superfields, the r.h. side of 
\bea
\delta_\xi A&=&\sqrt{2}\psi\xi
\nonumber \\
\delta_\xi\psi&=&i\sqrt{2}\sigma^\mu\ol{\xi}\partial_\mu A+\sqrt{2}\xi F
\\
\delta_\xi F&=&i\sqrt{2}\ol{\xi}\ol{\sigma}^\mu\partial_\mu \psi \nonumber
\eea
should hence be non-zero. Maintaining Lorentz-invariance, this can only be achieved if $F\neq 0$ in the vacuum. This is called $F$-term breaking and the simplest lagrangian with this feature is 
\be
{\cal L}=\ol{\Phi}\Phi\Big|_{\theta^2\ol{\theta}^2}+c\Phi\Big|_{\theta^2}
+ \mbox{h.c.}\label{lmo}
\ee
The relevant terms in component form are 
\be
{\cal L}=\ol{F}F+cF+\mbox{h.c.}+\cdots\,,
\ee
which implies $\ol{F}=-c$ in the vacuum. However, while SUSY is formally broken, the theory is free, thus $F$ does not couple to other fields and hence the spectrum remains supersymmetric. 

This is easily remedied adding a higher-dimension operator, 
\be
{\cal L}=\left[\ol{\Phi}\Phi-\frac{1}{M^2}(\ol{\Phi}\Phi)^2\right] 
\Big|_{\theta^2\ol{\theta}^2}+c\Phi\Big|_{\theta^2} + \mbox{h.c.}
\ee
Now, ignoring fermions and derivative terms, the component lagrangian reads 
\be
{\cal L}=\ol{F}F-4\ol{F}F\ol{\Phi}\Phi/M^2+cF+\mbox{h.c.}+\cdots\,.
\ee
The vacuum is again at $\Phi=0$ and $\ol{F}=-c$, but now this non-zero $F$ introduces scalar masses and supersymmetry is broken in the spectrum of the theory. 

We note that apparently simpler models which extend (\ref{lmo}) by adding terms $\sim \Phi^2$ or $\sim \Phi^3$ to the superpotential do not work in our context. They reinstate a SUSY-preserving vacuum, which is obvious since in such models the linear term can be absorbed in a shift of $\Phi$. In fact, the simplest renormalisable model with chiral superfields and spontaneous SUSY breaking is the {\bf O'Raifeartaigh model}\index{O'Raifeartaigh model} with lagrangian \cite{ORaifeartaigh:1975nky}
\be
{\cal L}=\sum_{i=1}^3 \ol{\Phi}^i\Phi^i \Big|_{\theta^2\ol{\theta}^2}+
\left[\Phi^1(m^2+\lambda(\Phi^3)^2)+\mu\Phi^2\Phi^3\right]
\Big|_{\theta^2} + \mbox{h.c.}\,.
\ee
It is easy to write down the $F$-term potential and minimise it to find spontaneous SUSY breaking. Sometimes the name  O'Raifeartaigh model is used more generally for any model with $F$-term breaking.  

A completely analogous story can be developed for real superfields, i.e. (abelian) gauge theories, where SUSY breaking is signalled by a non-zero VEV of the $D$ term. The simplest model realising this is 
\be
{\cal L}=\frac{1}{2g^2}W^2\Big|_{\theta^2}+2\kappa V\Big|_{\theta^2\ol{\theta^2}}\,,
\ee
where the new term linear in $V$ is known as {\bf Fayet-Iliopoulos}\index{Fayet-Iliopoulos term} or FI term\index{FI term} \cite{Fayet:1974jb}. At the component level one finds
\be
{\cal L}=\frac{1}{2g^2}D^2+\kappa D\qquad\Rightarrow\qquad D=-\kappa g^2\neq 0\,.
\ee
As before, the model needs to be enriched to see this formally present SUSY breaking in the spectrum. This can be achieved e.g. by adding two chiral superfields (to avoid anomalies) with charge $\pm 1$ and mass $m$. One finds that the fermions remain massless while the boson masses split according to $m_{1,2}^2=m^2\pm \kappa g^2$. See e.g.~\cite{Wess:1992cp} for details.

\subsection{Supersymmetrizing the Standard Model}
The Minimal Supersymmetric Standard Model\index{Minimal Supersymmetric Standard Model} or MSSM\index{MSSM} is obtained basically by promoting all fermions and scalars of the Standard Model to chiral superfields and all vectors to real superfields. The additional components introduced in this way are made heavy by an appropriate mechanism of SUSY breaking, to be discussed shortly. The only exception to this rule arises in the Higgs sector, where one now needs two different Higgs doublets and hence two corresponding superfields: $H_u$ and $H_d$. The reason will become clear immediately. Of the many reviews of this wider subject we refer in particular to~\cite{Luty:2005sn, Martin:1997ns, Giudice:1998bp}.

After these preliminaries, we give the set of chiral superfields:
\be
\Phi^a=\{Q,U,D,L,E,H_u,H_d\}\,.
\ee
The gauge representations are as in (\ref{mat}), up to $H_u$, which has opposite $U(1)$ charge. Our naming conventions follow (\ref{fie}) and we have suppressed the generation index on the matter superfields $Q,\cdots\!,E$ for brevity. The lagrangian can be organised in three pieces. First,
\be
{\cal L}_{gauge}=\sum_{i=1}^3\mbox{tr}\left(\frac{1}{2g_i^2}(W_i)^\alpha (W_i)_\alpha\Big|_{\theta^2}+\mbox{h.c.}
\right)\,,
\ee
with `tr' to be replaced by `1/2' in the $U(1)$ case. Second,
\be
{\cal L}_K=\sum_{a=1}^7\Phi_a^\dagger e^{2V}\Phi_a\Big|_{\theta^2\ol{\theta^2}}
\,,
\ee
where $K$ stands for kinetic or Kahler potential term and where the superfield $V=V_1+V_2+V_3$ contains the three real superfields corresponding to three factors of $G_{SM}$. In each term, one must use the representation appropriate for $\Phi^a$. Third, we have the superpotential term
\be 
{\cal L}_W=(W_\mu+W_Y+W_e)\Big|_{\theta^2}+\mbox{h.c.}
\ee
with
\be
W_\mu=\mu H_u H_d\,,\qquad W_Y=\lambda_uQH_uU+\lambda_d QH_dD+\lambda_eLH_dE\,,
\ee
and
\be
W_e=aLH_u+bQLD+cUUD+dLLE\,.
\ee
The structure of ${\cal L}_{gauge}$ and ${\cal L}_K$ requires no further comments: They simply provide the necessary kinetic terms and gauge interactions. The Standard Model Yukawa couplings come from $W_Y$, together with new interactions that are not present in the Standard Model. To give masses to all fermions, we were forced to introduce two Higgs fields. Indeed, holomorphicity forbids the appearance of the $\tilde{\Phi}$ variable used in the up-type Yukawa term of the non-supersymmetric Standard Model. Hence, a new Higgs multiplet $H_u$ with opposite $U(1)$ charge has to be introduced. An independent reason for this second doublet is the need to cancel the $U(1)$ anomaly introduced by the fermion (the `Higgsino')\index{Higgsino} contained in $H_d$. 

Finally, there are {\it extra} terms without a Standard Model analogue, allowed due to the enlarged field content. We have collected these terms in $W_e$ but, since some of them induce proton decay and lepton number violation, we basically want to forbid them. We also note that we have limited ourselves to the renormalisable level -- hence $W$ is truncated at cubic order. To see that cubic terms correspond to marginal operators, recall that $\theta^2$ has mass dimension $-1$. Hence, projection on the $\theta^2$ component corresponds to raising the mass dimension by one unit. Thus, mass dimension 3 in $W$ corresponds to mass dimension 4 in ${\cal L}$. 

To forbid $W_e$, the concept of an {\bf $\bm R$-symmetry}\index{$R$-symmetry} (which is crucial in SUSY independently of phenomenology) is useful. To explain this concept, we define standard (global) $U(1)$ and $U(1)_R$ transformations of chiral superfields as follows:
\be
U(1):\,\,\Phi(y,\theta)\to e^{im\epsilon}\Phi(y,\theta)\,\,,\qquad \quad
U(1)_R:\,\,\Phi(y,\theta)\to e^{in\epsilon}\Phi(y,e^{-i\epsilon}\theta)\,.
\ee
Here $m$ and $n\equiv R(\Phi)$ are the $U(1)$ and $U(1)_R$ charges of $\Phi$ respectively. It follows immediately that (and this is the crucial feature of an $R$-symmetry!) the components transform differently, depending on their mass dimension:
\be
A\to e^{in\epsilon}A\,,\qquad \psi\to e^{i(n-1)\epsilon}\psi\,,\qquad
F\to e^{i(n-2)\epsilon}F\,.
\ee

Invariance of the lagrangian requires $R(K)=0$ and $R(W)=2$. The former is clear since the projection on the $\theta^2\ol{\theta}^2$ component does not change the $R$-charge. By contrast, projection on the $\theta^2$ component lowers the $R$-charge by 2. 

For our purposes, the interesting assignment is:
\be
\mbox{For}\,\,\,\,Q,U,D,L,E:\,\,\,R=1\qquad\qquad\mbox{and for}\,\,\,\,H_u,H_d:\,\,\,R=0\,.
\ee
This restricts $W$ to the Yukawa terms. However, this is too strong since it also forbids the so-called $\mu$ term $\mu H_u H_d$. But the latter is needed since even after SUSY breaking (see below) it is the only source for Higgsino masses. (Higgsinos are the - so far unobserved and hence heavy - fermionic partners of the Higgs.) 

A possible resolution is the breaking of $U(1)_R$ to its $\mathbb{Z}_2$ subgroup. By this we mean restricting $U(1)\equiv\{\,e^{i\epsilon}\,\}$ to the two elements with $\epsilon=0$ and $\epsilon=\pi$. After this breaking to $\mathbb{Z}_2$, $R$-charges are identified modulo 2. Indeed, superfields with $R$-charge $m$ and $m+2$ now transform identically. In particular the selection rule $R(W)=2$ for superpotential terms is modified to $\mbox{$R(W)=2\!\mod{2}\,$}$. In other words, one now only demands $R(W)\in 2\mathbb{N}$. As a result, the $\mu$ term is allowed while all terms in $W_e$ are still forbidden. Moreover, the transformation rules of the Standard Model fields and their superpartners under this so-called {\bf $\mathbb{Z}_2$ $\bm R$-parity}\index{$R$-parity} are\index{sfermion}\index{gaugino}
\bea
\mbox{\bf Even:}&&\mbox{Higgs scalars, fermions, gauge bosons}
\\
\mbox{\bf Odd:}&&\mbox{Higgsinos, sfermions, gauginos}\,.
\eea
Here by sfermion one refers to the superpartner of a Standard Model fermion. A sfermion is hence a boson. For example, a selectron is one of the many sleptons. The top-squark is the SUSY partner of the top-quark etc.

The above $R$-parity assignments imply that any of the so-called superpartners can not decay into a combination of Standard Model particles. Hence the lightest superpartner (the {\bf lightest supersymmetric particle or LSP})\index{lightest supersymmetric particle}\index{LSP} is absolutely stable and provides a natural dark matter candidate. Unfortunately, with growing LHC-bounds on its mass the abundance predicted from its so-called freeze-out in early cosmology tends to become too high, calling for extensions of the simplest settings. For more details see e.g.~\cite{Plehn:2017fdg, bertone, Hooper:2009zm, Olive:2003iq}.

\subsection{Supersymmetric and SUSY breaking masses and non-re\-nor\-ma\-lisation}\index{non-renormalisation}
The simplest way to make the above construction realistic is to add mass terms to the supersymmetric Standard Model such that all the superpartners of Standard-Model particles become sufficiently heavy. (Recall that the Higgsino can be made heavy by a sufficiently large $\mu$ term.) While technically correct, such an approach of explicit SUSY breaking is not very satisfying or illuminating concerning the resolution of the hierarchy problem.

Hence, we will introduce somewhat more structure and try to arrive at the MSSM using spontaneuous SUSY breaking. Specifically, we introduce a {\bf hidden sector}\index{hidden sector} in which SUSY is broken spontaneously.\footnote{
In 
principle, one may imagine situations where SUSY is broken spontaneously in the supersymmetrised Standard Model, without introducing any additional fields. However, it turns out that this does not work in practice, taking into account experimental constraints on masses and the phenomenologically required gauge symmetry breaking.
} 
It will then be communicated to the Standard Model by higher-dimension operators. To illustrate this structure, we start with the toy-model lagrangian
\be
{\cal L}=\left(\ol{S}S-c_1(\ol{S}S)^2\right)\Big|_{\theta^2\ol{\theta}^2}+c_2 S
\Big|_{\theta^2}+\mbox{h.c.}+\ol{\Phi}\Phi\Big|_{\theta^2\ol{\theta}^2}+ \frac{1}{M^2}\ol{\Phi}\Phi\ol{S}S\Big|_{\theta^2\ol{\theta}^2}\,.\label{tml}
\ee
We recognise a model with a chiral superfield $S$ and with spontaneous SUSY breaking ($F_S\neq 0$). In addition, we have a free and massless chiral superfield $\Phi$. The latter represents the Standard Model or, more specifically, its Higgs sector. Finally, the last term is a higher-dimension operator, suppressed by a large mass scale $M$, coupling the two sectors. All we need to know about the hidden sector is that $S=0$ and $F_S\neq 0$ in the vacuum. Inserting this in our lagrangian and focussing on the $\Phi$-sector only, we have
\be
{\cal L}\supset \ol{\Phi}\Phi\Big|_{\theta^2\ol{\theta}^2}+\frac{1}{M^2}\ol{A}_\Phi A_\Phi \ol{F}_SF_S\,,
\ee
where we also ignored a quartic fermionic term arising from the superfield higher-dimension operator. We see that the result is equivalent to just having added a (`soft') SUSY-breaking scalar mass term to the $\Phi$ sector\index{soft terms}
\be
{\cal L}\supset m_{soft}^2|A_\Phi|^2\qquad\mbox{with}\qquad m_{soft}^2=|F_S|^2/M^2\,.
\ee
Crucially, in our approach we see right away that this term is radiatively stable - it is secretly a higher-dimension operator and does as such not receive power-divergent loop corrections.\footnote{
The intuitive reason is very simple and can be thought of as the opposite of the more familiar fact that operators with mass-dimension less than four do receive power-like loop corrections. Indeed, to correct a higher-dimension operator in an otherwise renormalisable model the loop must produce a coefficient of negative mass dimension. This could be the result of the tree-level coefficient multiplied by dimensionless couplings and cutoff-dependent logarithms. Any positive power of the cutoff would have to come with a mass term in the denominator. The latter would imply an infrared divergence, which however does not arise in 4d and in the present context.
} 
This explains the name `soft'. In fact, the two sectors decouple completely as $M\to \infty$, making it clear that the coupling operator can only renormalise proportionally to itself. (We see here another possibility, in addition to symmetries, why a certain coefficient in the lagrangian may be zero in a natural way: In its absence, the model becomes the sum of two completely independent theories.)

Our point about the mass term not being quadratically divergent may appear trivial - after all the $\Phi$ sector itself is a free theory, so of course nothing renormalises. However, it is immediate to enrich our model by e.g. $W(\Phi)\sim \Phi^3$, leading to quartic self interactions. Alternatively, $\Phi$ may be charged under some gauge group, like the Higgs in the Standard Model is. Nothing in our argument changes: The operator $\sim 1/M^2$ inducing the mass can not have power-divergences. 

However, one could clearly add a term $W(\Phi)\sim m_{SUSY}\Phi^2$ to our action, in other words, a supersymmetric mass term. We have to be sure that interactions in the $\Phi$ sector will not, if such a term is absent in the beginning, induce it through loop corrections. This, as it turns out, is in fact the main point where SUSY saves us: {\bf The superpotential does not renormalise}. This so called non-renormalisation theorem is, at least at a superficial level and in our simple model, easy to understand~\cite{Seiberg:1993vc}: 

Indeed, consider the Wess-Zumino model with tree level superpotential 
\be
W=\frac{m}{2}\Phi^2+\frac{\lambda}{3}\Phi^3\,.\label{mlf}
\ee
Introduce a $U(1)$ and $U(1)_R$-symmetry under which $\Phi$ has charges $(1,1)$. Clearly, this is respected by our canonical Kahler potential $K$, but the superpotential breaks both symmetries. One can interpret this breaking as being due to non-zero VEVs of superfields $m$ and $\lambda$, the scalar components of which have acquired non-zero VEVs. For this interpretation to work, one needs to assign to $m$ the charges $(-2,0)$ and to $\lambda$ the charges $(-3,-1)$. Now, assuming that perturbative loop corrections break neither these $U(1)$ symmetries nor SUSY, we expect that the effective superpotential (relevant for the Wilsonian effective action) will still respect the two $U(1)$ symmetries. Using holomorphicity and the fact that each term in $W$ must have charges $(0,2)$, we have
\be
W_{\rm eff}=\sum_{ijk}c_{ijk}m^i\lambda^j\Phi^k=m\Phi^2\,f(\lambda\Phi/m)\,.\label{ssp}
\ee
In the second step, we used the fact that, under the symmetry constraints, the triple sum collapses to a single sum, which can then be viewed as a power series in $(\lambda\Phi/m)$. This last combination of fields can appear to any power, since both its $U(1)$ and $U(1)_R$ charge vanish.

Now, the constant and linear term in $f$ correspond to the terms already present at tree level - their values are $1/2$ and $1/3$ by assumption. We see that higher terms in $\lambda$, which may in principle arise from loop corrections, always come with higher powers of $\Phi$ and hence do not affect mass and trilinear coupling. Moreover, it is easy to convince oneself that such higher terms in $\lambda$, as derived from (\ref{ssp}), correspond precisely to terms following from tree-level exchange of $\Phi$. But such tree-level effects should not be included in $W_{\rm eff}$. They appear in the calculation of observables if one uses only the tree-level expression for $W$ together with the standard Feynman rules. Including them in $W_{\rm eff}$ would lead to double counting. Now, compared to tree-level effects, loop effects always have a higher power of $\lambda$ (given a certain number of external legs). Hence such loop effects are not described by the higher-$\lambda$ terms in $f$. As a result, we learn that $W_{\rm eff}=W$ and no loop corrections arise. 

In summary, we have learned that the structure of (\ref{tml}), supplemented by a superpotential of type (\ref{mlf}), is radiatively stable. In particular, the supersymmetric and supersymmetry-breaking mass terms can both be chosen small compared to the cutoff scale and are not subject to power-like divergences.

\subsection{The Minimal Supersymmetric Standard Model (MSSM)}
\index{Minimal Supersymmetric Standard Model}\index{MSSM}

With this, it is straightforward to introduce SUSY breaking by a \index{spurion}spurion\footnote{
The field is spurious in that we only introduced it to parameterise a certain symmetry breaking effect. Its actual dynamics is not important for us.
}
 superfield $S$ into the SUSY Standard Model. Without aiming at completeness, we give four types of higher-dimension terms which are sufficient to generate all essential SUSY breaking effects:
\bea
{\cal L}_1 &=& \frac{1}{M^2}\ol{Q}Q\ol{S}S\Big|_{\theta^2\ol{\theta}^2}\qquad 
\qquad,\qquad\qquad {\cal L}_2 \,=\, \frac{1}{M^2}Q^2\ol{S}S\Big|_{\theta^2\ol{\theta}^2}
\nonumber \\
\\
{\cal L}_3 &=& \frac{1}{M} Q^3 S\Big|_{\theta^2}+\mbox{h.c.}\qquad\qquad,\qquad \qquad {\cal L}_4 = \frac{1}{M} W^\alpha W_\alpha S\Big|_{\theta^2}+\mbox{h.c.}
\nonumber
\eea
Here $Q$ stands for generic Standard Model chiral superfields. The different factors of $Q$ in one term may also be replaced by different Standard Model fields, e.g. $Q^3\to QH_uU$. 

The effects of these different terms are easy to read off. For example,
\be
{\cal L}_1\supset \frac{|F_S|^2}{M^2}|A_Q|^2\equiv M_0^2|A_Q|^2\,,
\ee
where we refer to $M_0$ as the soft mass which $A_Q$ acquires. Similarly, ${\cal L}_2$ induces a holomorphic soft mass, which due to symmetry constraints arises only in the Higgs sector, with $Q^2\to H_u H_d$. Furthermore, ${\cal L}_3$ induces soft trilinear\index{trilinear term} or `$A$-terms':\index{$A$-terms}
\be
{\cal L}_3\supset \frac{F_S}{M}A_Q^3\equiv A\cdot A_Q^3\,. 
\ee
Finally, the last term induces gaugino masses $M_{1/2}$,
\be
{\cal L}_4\supset \frac{F_S}{M}\lambda^\alpha \lambda_\alpha \equiv M_{1/2}\lambda^\alpha\lambda_\alpha\,.
\ee

A standard scenario, known as `Gravity Mediation'\index{gravity mediation} has $M\sim M_P\sim 10^{18}$~GeV, a value which corresponds the scale at which one may expect quantum gravity to induce all allowed higher-dimension operators. Then one would need the SUSY breaking scale in the hidden sector to be $\sqrt{|F_S|}\sim 3\times 10^{10}$~GeV (which is sometimes referred to as an `intermediate scale')\index{intermediate scale} to obtain 
\be
M_0\sim A\sim M_{1/2} \sim 1\,\mbox{TeV}\,.
\ee
Of course, many new parameters are introduced in this way. In particular, there are as many $A$-terms as there are entries in the Yukawa coupling matrices, and the soft masses come as $3\times 3$ matrices in generation space. If the scale of the soft terms (sometimes referred to as the SUSY breaking scale) is low - e.g. in the TeV range, then generic values for the soft terms are ruled out by flavour-changing neutral currents\index{flavour-changing neutral current} (FCNCs)\index{FCNC} and other experimental signatures. Some symmetry-based model building is needed to make this scenario realistic. 

It is crucial that no renormalisable couplings between hidden and visible sector are present. In particular, a superpotential term $SQ^2$ (or concretely $SH_uH_d$) would induce a Higgs mass $\sim \sqrt{|F_S|}$, destabilising the hierarchy. Furthermore, we need a non-zero $\mu$ term for the Higgs, but it should not be too large, again to avoid a hierarchy destablisation.  

Thus, the task is to induce a supersymmetric $\mu$ term of the same size as the (otherwise very similar) SUSY-breaking holomorphic mass term $\sim H_uH_d$ (where $H_u$, $H_d$ are the scalar components, not the superfields). There is a very elegant solution to this problem known as the {\bf Giudice-Masiero mechanism}\index{Giudice-Masiero mechanism} \cite{Giudice:1988yz}. It is based on the higher-dimension couplings
\be
{\cal L}\supset \left(\frac{1}{M}\ol{S}H_uH_d+\frac{1}{M^2}\ol{S}S H_u H_d\right)\Big|_{\theta^2\ol{\theta}^2}\,.
\ee
They induce terms
\be
{\cal L}\supset \frac{\ol{F}_S}{M}H_uH_d\Big|_{\theta^2}+\frac{|F_S|^2}{M^2}H_uH_d\,. \label{babm}
\ee
Here, in the first term, $H_u$, $H_d$ represent superfields and, in the second term, the same symbols are used for their scalar components. Clearly, the first term in (\ref{babm}) is the previously discussed $\mu$ term\index{$\mu$ term}, but with a coefficient governed by the SUSY-breaking scale, $\mu\sim \ol{F}_S/M$. The second term is the so-called $B_\mu$ term\index{$B_\mu$ term}, the previously mentioned holomorphic mass term for the Higgs:
\be
{\cal L}\supset B_\mu H_uH_d\qquad \mbox{with}\qquad B_\mu\sim |F_S|^2/M^2\,.
\ee

Upon integrating out the $F$-terms of the Higgs superfields, the $\mu$ term also contributes to the quadratic Higgs scalar potential, which in total reads
\bea
V_2 &=& (|\mu|^2+m_{H_u}^2)|H_u|^2+(|\mu|^2+m_{H_d}^2)|H_d|^2
+ B_\mu H_u H_d+\mbox{h.c.}\nonumber \\ \label{hmm}
\\ \nonumber
&=& \left( \begin{array}{c}H_u \\ \epsilon\ol{H}_d \end{array} \right)^\dagger
\left( \begin{array}{cc} |\mu|^2+m_{H_u}^2 & B_\mu \\ B_\mu &  |\mu|^2+m_{H_d}^2\end{array} \right)
\left( \begin{array}{c}H_u \\ \epsilon\ol{H}_d \end{array} \right)\,.
\eea
The second line makes it apparent that we are dealing simply with a $4\times 4$ complex mass matrix, giving mass to the 4 scalars contained in $(H_u,\epsilon \ol{H}_d)^T$. Due to $SU(2)$ symmetry, this matrix has a $2\times 2$ block structure and hence only two distinct eigenvalues. Electroweak symmetry breaking requires one of the eigenvalues to be negative.

An independent quartic Higgs interaction is not present in the SUSY Standard Model since no cubic Higgs superpotential is allowed. However, the $D$ term of the $SU(2)\times U(1)$ SUSY gauge theory does the important job of creating such a coupling:
\be
V_4= \frac{1}{8}(g_1^2+g_2^2)\,(|H_u|^2+|H_d|^2)^2+\frac{1}{2}g_2^2|H_u \ol{H}_d|^2\,.
\ee
Assuming soft terms are close to the weak scale, the scalar potential $V_2+V_4$ and its symmetry breaking structure has been analysed in great detail, but we will not discuss this. Suffice it to say that electroweak symmetry can be broken as required, both Higgs doublets generically develop VEVs (the ratio being parameterised by $\tan\beta\equiv v_u/v_d$)\index{$\tan\beta$}, and the Higgs mass is predicted in terms of this mixing angle and the gauge couplings. This is a great success, given in particular that all parameters of this model are now protected from power-divergences, the SUSY breaking and weak scale are naturally small, and the model is renormalisable and can, in principle, be valid all the way to the Planck scale. In addition, extrapolating the Standard Model gauge couplings to high energy scales \cite{Dimopoulos:1981yj, Dimopoulos:1981zb, Ibanez:1981yh, Sakai:1981gr}, one finds that they meet rather precisely at the GUT scale $M_{GUT}\simeq 10^{16}$~GeV (see Problem~\ref{quni}). This has been known since about 1990 and has given a lot of credibility to the model \cite{Amaldi:1991cn}.

However, the predicted Higgs mass is bounded by the $Z$-boson mass at tree level, which is clearly incompatible with observations. The correction needed to bring the Higgs mass up to its observed value of 125~GeV can be provided by loops, but this requires a large mass of the stop quark (also `top squark' or simply `stop') or large trilinear terms. This drives (again through loops) the Higgs VEV to higher values and partially spoils the success of the hierarchy problem resolution. In addition, the non-discovery of superpartners at the LHC has raised the lower limits for soft terms, also limiting the success of the supersymmetric resolution of the hierarchy problem. Thus, the phenomenological status of this model has deteriorated. From a modern perspective, it may be appropriate to view the MSSM not as a weak-scale model but rather as a model at a significantly higher scale, $m_{soft}\gg m_{ew}$.

This perspective implies that one integrates out all SUSY partners and the second Higgs at $m_{soft}$ and is left with just the Standard Model below that scale. More precisely, this requires that the lowest eigenvalue of the Higgs mass matrix in (\ref{hmm}) is smaller than the typical entries (which are $\sim m_{soft}^2$). This is a fine tuning of the order $m_{ew}^2/m_{soft}^2$ which one may have to accept. This fine tuning ensures that $m_H^2$ of the Standard Model Higgs, which sets the weak scale, is somewhat below the SUSY breaking scale. One may refer to this as a `high-scale' or `split' MSSM \cite{ArkaniHamed:2004fb, Giudice:2004tc}, and it is not implausible that such a model (or some variant thereof) arises in string theory (see e.g.~\cite{Denef:2004cf, Hebecker:2012qp}).

We may here return to the terminology introduced at the end of Sect.~\ref{fts}: We have learned that low-scale SUSY can solve the large hierarchy problem. (Here `low' refers to the TeV range, including say 10 TeV or even higher.) SUSY does, however, suffer from a little-hierarchy problem. This is related to the detailed interplay between SUSY breaking and electroweak symmetry breaking, which force $m_{soft}$ to go up to 10 TeV or above. The severity of this little hierarchy problem of low-scale SUSY depends on the details of the model and is still under debate.

\subsection{Supergravity - superspace approach}
\index{supergravity}

All that was said above must, of course, be consistently embedded in a generally relativistic framework. The resulting structure, known as supergravity, is equally elegant and unique, though technically much more complicated than flat-space SUSY. We can only give a brief summary of results. Since we described flat SUSY using the superspace approach, let us start by noting that a similar (curved) superspace approach can also be used to derive supergravity \cite{Wess:1992cp, bk}. For a brief discussion of this see also \cite{Quevedo:2010ui}.

One starts, as before, with coordinates 
\be
z^M=(x^\mu,\theta^\tau,\ol{\theta}_{\dot{\tau}})
\ee
with the above indices being `Einstein indices', as in conventional general relativity. Then one introduces a vielbein, $E_A{}^M(z)$, i.e. a basis of vectors, labelled by the `Lorentz indices'
\be
A=(a,\alpha,\dot{\alpha})\,.
\ee
As in general relativity, one defines a connection, introduces constraints (such as the vanishing torsion constraint), and removes gauge redundancies. This is very cumbersome in the present case, but it eventually leads to a supergravity superspace action
\be
S=\int d^8z \,E\, \Omega(\Phi,\ol{\Phi})+\int d^6 z\, \varphi^3\, W(\Phi)+ \mbox{h.c.}
\ee
Here $E$ is the determinant of the vielbein $E_A{}^M$. The latter contains a real vector superfield and an (auxiliary) chiral superfield \cite{bk}
\bea
{\cal H}^\mu&=&\theta\sigma^a\ol{\theta}e_a{}^\mu+ i\ol{\theta}^2\theta\psi^\mu + \mbox{h.c.} + \theta^2\ol{\theta}^2 A^\mu
\\
\varphi&=& e^{-1}\left(1-2i\theta\sigma_\mu\ol{\psi}^\mu+F_\varphi\theta^2\right)
\eea
with $e=\mbox{det}(e_a{}^\mu)$ and $\sigma_\mu = \sigma_a e^a{}_\mu$. We thus have the component fields
\be
e_a{}^\mu(x)\,,\qquad \psi_\alpha{}^\mu(x)\,,\qquad A^\mu(x)\,,\qquad  F_\varphi(x)\,.
\ee
Here the first is the familar vielbein of Einstein's theory, and the last two are auxiliaries (some authors use the notation $B(x)\equiv F_\varphi(x)$). The crucial new feature is a physical, propagating spin-(3/2) field $\psi_\alpha{}^\mu$, called the {\bf gravitino}\index{gravitino}, which is the superpartner of the vielbein (or equivalently of the metric or graviton). The $z$ integrations are over the full or half of the Grassmann part of superspace, as in the flat case. The argument $\Phi$ stands for as many chiral superfields, containing matter degrees of freedom, as one wants. The function $\Omega$ is real. 

It goes far beyond the scope of these notes to derive the component action. However, to get a glimpse of what is going on, we can consider the flat-space limit:
\be
e_a{}^\mu=\delta_a{}^\mu\,,\quad \psi_\alpha{}^\mu=0\,,\quad A^\mu=0\,,\quad
\varphi=1+\theta^2F_\varphi\,.
\ee
Then the action takes the form
\be
S=\int d^8z\,\varphi\ol{\varphi}\, \Omega(\Phi,\ol{\Phi})+\int d^6 z\, \varphi^3\, W(\Phi)+ \mbox{h.c.}\label{ssa}
\ee
From this, integrating out $F_\Phi$ and $F_\varphi$, one can straightforwardly obtain the supergravity scalar potential. To be specific, one finds the potential in the Brans-Dicke frame\index{Brans-Dicke frame}. This is so because, in the curved case, one would have also have found
\be
S \supset \int d^4x\,\sqrt{g}\,\frac{1}{2}M_P^2\,{\cal R}\cdot\frac{\Omega(\Phi,\ol{\Phi})}{3}\,,
\ee
i.e., the Einstein-Hilbert term in the Brans-Dicke frame. Rescaling the metric to absorb the factor $\Omega/3$, one arrives at an Einstein-frame curvature term together with the supergravity scalar potential
\be
V=e^{K/M_P^2}\Big(K^{i\ol{\jmath}}(D_iW)(\ol{D}_{\ol{\jmath}}\ol{W})-3|W|^2 /M_P^2 \Big)\label{sgp}
\ee
where
\be
D_iW=\partial_iW+K_iW
\ee
and
\be
K=-3M_P^2\ln[-\Omega/(3M_P^2)]\qquad \mbox{or}\qquad \Omega=-3M_P^2\exp[-K/(3M_P^2)]\,.
\ee
This goes together with conventional kinetic terms for the fields $\Phi^i$, based on the supergravity Kahler metric $K_{i\ol{\jmath}}$. We have given all of the above keeping $M_P$ explicit to make it easy to see that the flat space limit, $M_P\to \infty$, takes us back to previous formulae. In particular, one can see that the first term in (\ref{sgp}) corresponds to the familiar $F$-term scalar potential while the second term is supergravity-specific. It is non-zero even if $W$ is just a number and thus allows for the introduction of a cosmological constant, albeit only a negative one. This is consistent with the fact that the Poincare superalgebra can be generalised to Anti-de-Sitter but not to de-Sitter space.

In practice, one mostly works with the above formulae in units in which $M_P=1$. This is much more economical and we will do so from now on.

Let us note that, among many other terms, one has
\be
{\cal L}\supset -e^{K/2}\ol{W}\psi_\mu\sigma^{\mu\nu}\psi_\nu+\mbox{h.c.}\,,
\ee
which implies a gravitino mass
\be
m_{3/2}=e^{K_0/2}W_0\,,
\ee
where $W_0$ and $K_0$ are the vacuum values of $W$ and $K$. We will suppress the indices `0' from now on since it will be clear from the context whether the vacuum value or some other dynamical value is meant. Supersymmetry breaking is, as before, governed by non-zero VEVs of (some of) the $F$-terms,
\be
F^i=e^{K/2}D^iW \equiv e^{K/2}K^{i\ol{\jmath}}\ol{D}_{\ol{\jmath}}\ol{W}\,.
\ee
Realistically, we have $\lambda=V_0\simeq 0$ (the non-zero meV-scale cosmological-constant value is negligible compared to particle-physics scales). Hence, the positive-definite $F$-term piece and the negative $|W|^2$ piece must compensate with high precision in the formula for $V$. We thus have
\be
|F|\sim e^{K/2}|W| \qquad \mbox{and hence}\qquad m_{3/2}\sim |F|\,.
\ee
Here $|F|$ is the length of the vector $F^i$, calculated using the Kahler metric $K_{i\ol{\jmath}}$. We note, however, that this is in Planck units and, reinstating $M_P$, one has $m_{3/2}\sim |F|/M_P$. Thus, if one takes the hidden-sector $F$ very low, near the weak scale (as is in principle consistent with our SUSY-breaking discussion), the gravitino can still be very light. This, however, requires that it couples to Standard Model fields only very weakly. 

We note that the SUSY solution to the weak-scale hierarchy problem works as before: All that we said remains valid since we are working at an EFT scale $\mu\ll M_P$ and the rigid limit (supplemented by the gravitino, if it is sufficiently light) can be used. The non-renormalisation theorem extends to the $W$ of supergravity. What is more, the presence of higher-dimension operators which was central in the communication of SUSY breaking from hidden to visible sector can be argued to be generic in the supergravity context: After all, the theory is non-renormalisable, so all in principle allowed operators are expected to be present with $M_P$-suppression. Also, the non-linear structure of $\Omega$ expressed in terms of $K$ suggests such operators. In other words, even if $K=\ol{\Phi}\Phi$, the presence of factors like $\exp(K)$ introduces many higher-dimension operators. The corresponding, very generic way of SUSY breaking mediation (through Planck suppressed higher-dimension operators) is called {\bf gravity mediation}.\index{gravity mediation}

\subsection{Supergravity - component approach}
\index{supergravity}

Before closing this Section, we should note that we only discussed the superfield approach to supergravity since it fits the previous analysis of rigid supersymmetry best. It is not the most economical or widely used approach, which is instead based on the component form of SUSY multiplets and (superconformal) tensor calculus~\cite{Freedman:2012zz}.

Very briefly, the story can be told as follows: In general relativity, Lorentz symmetry becomes local. Since the SUSY parameter $\xi$, being a spinor, transforms non-trivially under the Lorentz group, it would be inconsistent to consider it a global object. Instead, it must be promoted to a space-time dependent quantity,
\be
\xi\quad\to \xi(x)\,,
\ee
such that supersymmetry becomes a gauge symmetry. But now we are clearly missing a gauge field defining the connection associated with our gauge symmetry. By analogy to
\be
A_\mu(x)\quad\to\quad A_\mu(x)+\partial_\mu\epsilon(x)\,,
\ee
one writes
\be
\psi_\mu(x)\quad\to\quad \psi_\mu(x)+\partial_\mu\xi(x)\,.
\ee
The new field $\psi_\mu$ is a vector-spinor, also known as gravitino. We here interpret both $\xi$ and $\psi_\mu$ as 4-component spinors, specifically Majorana spinors. 

The presence of the gravitino can also be motivated in a different way: Indeed, we are clearly missing a superpartner for the graviton. As it turns out, the right object is $\psi_\mu$. To understand this better, we take a step back, forget about superfields, and recall the SUSY algebra with its generators $Q$ and $\ol{Q}$ (that come on top of the Poincare generators). They have spin and hence raise or lower the spin of objects on which they act. Indeed, developing the representation theory of the SUSY Poincare algebra one finds multiplets including particles with different spin or, in the massless case, helicity. We aready know the multiplets
\be
(0,1/2)\qquad\mbox{and}\qquad (1/2,1)
\ee
corresponding to the chiral and real superfield (or the scalar and vector multiplet). Naturally, one expects and indeed finds the multiplet
\be
(3/2,2)
\ee
containing gravitino and graviton. For this to be consistent, one needs the gravitino to contain 2 degrees of freedom on shell, to match those of the graviton. Indeed, the general expressions for numbers of degrees of freedom of a vector spinor, initially and after taking into account gauge redundancy, constraints and the on-shell condition, are
\be
d\cdot 2^{[d/2]}\qquad \to\qquad \frac{1}{2}(d-3)\cdot 2^{[d/2]}\,.
\ee
Here the exponent $[d/2]$ (the integer fraction of $d/2$) characterises the dimension of a general spinor, the reduction from $d$ to $d-3$ is associated with gauge freedom and constraints, and the prefactor $1/2$ is the usual reduction from off-shell to on-shell degrees of freedom affecting any spinor (due to the equation of motion being first order). 

We record for completeness the underlying action and equation of motion (the Rarita-Schwinger equation),\index{Rarita-Schwinger equation}
\be
S=-\int d^d x\,\ol{\psi}_\mu \gamma^{\mu\nu\rho}\partial_\nu\psi_\rho \qquad
\mbox{and} \qquad \gamma^{\mu\nu\rho}\partial_\nu\psi_\rho=0\,,
\ee
although we will not have time to discuss the derivation of the physical degrees of freedom from this dynamical description. Furthermore, we should note that the modern way of deriving actions in this context is the so-called {\bf tensor calculus}\index{tensor calculus}. By this one means rules for multiplying (combining) multiplets to obtain new multiplets. We saw an example of this when we noted that $\Phi_1(y,\theta)\Phi_2(y,\theta)$, with $\Phi_1$ and $\Phi_2$ chiral, defines a new chiral superfield. This can be formulated without superspace, just on the basis of the components. With this method, the full action of supergravity, including supergravity coupled to chiral and vector multiplets, can be derived. 

More specifically, the method of choice is `superconformal tensor calculus', which first extends the theory to a conformal supergravity, then breaks scale invariance by a VEV and removes the extra degrees of freedom by constraints. (The non-SUSY version of this would be to replace the Planck scale by a field and then recover usual gravity by giving this field a VEV.) In fact, this superconformal method is also used in the superspace approach and we saw a trace of the field whose VEV eventually breaks scaling symmetry in the chiral compensator\index{chiral compensator} $\varphi(y,\theta)$ of (\ref{ssa}).

Let us end with part of the general 4d supergravity action (the full action being given e.g. in~\cite{Wess:1992cp, Freedman:2012zz}). The input are three functions, the (real) Kahler potential $K$, the holomorphic superpotential $W$ and the (also holomorphic) gauge-kinetic function $f_{ab}$. Returning also to the Weyl description of spinors, one has:
\bea
\frac{1}{\sqrt{g}}{\cal L} &=& \frac{1}{2}R
+\epsilon^{\mu\nu\rho\sigma}\ol{\psi}_\mu\ol{\sigma}_\nu D_\rho\psi_\sigma
+K_{i\ol{\jmath}}\left[(D_\mu \phi^i)(D^\mu\ol{\phi}^{\ol{\jmath}})-i\ol{\chi}^{\ol{\jmath}}\ol{\sigma}^\mu D_\mu \chi^i\right]
\nonumber \\
\nonumber \\
&& +(\mbox{Re}f_{ab})\left[-\frac{1}{4}F^a{}_{\mu\nu}F^{b\,\mu\nu} -\ol{\lambda}^a\ol{\sigma}^\mu D_\mu\lambda^a\right]
+\frac{1}{4}(\mbox{Im}f_{ab})F^a{}_{\mu\nu}\tilde{F}^{b\,\mu\nu}
\nonumber \\
\nonumber \\
&& 
-e^{K/2}\left[\left(\ol{W}\psi_\mu\sigma^{\mu\nu}\psi_\nu
+\frac{i}{\sqrt{2}}(D_iW)\chi^i\sigma^\mu\ol{\psi}_\mu
+\frac{1}{2}(D_i D_jW)\chi^i\chi^j\right)+\mbox{h.c.}
\right]
\nonumber \\
\nonumber \\
&& -V_F-V_D+\{\,\mbox{further fermionic terms}\,\}\,.
\eea
Here 
\bea
D_iW&=&W_i+K_iW
\\
D_iD_jW&=&W_{ij}+K_{ij}W+K_iD_j W+K_jD_iW-K_iK_j W-\Gamma_{ij}{}^kD_k W\,.
\nonumber
\eea
We already know the $F$-term potential
\be
V_F=e^K\Big(K^{i\ol{\jmath}}(D_iW)(\ol{D}_{\ol{\jmath}}\ol{W})-3|W|^2\Big)\,.
\ee
The $D$-term potential\index{$D$-term potential} has until now only been given implicitly and in a special case. More generally, it reads (cf.~\cite{Villadoro:2005yq} for a very compact discussion)
\be
V_D=\frac{1}{2}[(\mbox{Re}\,f)^{-1}]^{ab}D_a D_b\,.
\ee
To define the $D$ terms, we recall that the scalars parameterise a Kahler manifold which, to be gauged, must have some so-called (holomorphic) Killing vector\index{Killing vector} fields
\be
X_a=X_a^i(\phi)\frac{\partial}{\partial \phi^i}\,.
\ee
They define the direction in which the manifold can be mapped to itself by the gauge transformation corresponding to the index $a$. They also appear in the general formula for the covariant derivatives:
\be
(D_\mu\phi)^i=\partial_\mu \phi^i-A_\mu^a X_a^i(\phi)\,.
\ee
Now, the $D$ terms are defined as real solutions of the differential equations (the Killing equations)
\be
X_a{}^i=-iK^{i\ol{\jmath}}\,\frac{\partial D_a(\phi,\ol{\phi})}{\partial \ol{\phi}^{\ol{\jmath}}}\,.
\ee
Mathematically, they are the Killing potentials. They can be given explicitly as 
\be
D_a=iK_i X_a{}^i+\xi_a\,,
\ee
where the $\xi_a$ are so-called supergravity FI terms. The latter are only allowed for abelian generators and they are believed to cause problems for a quantum-gravity UV completion.\footnote{
Such `constant' FI terms require the supergravity to be `gauged', i.e.~the gravitino to be charged under the $U(1)$ responsible for the FI term. Concretely, there is a mixing between this gauged $U(1)$ and a global $U(1)$ $R$-symmetry such that, in the end, a certain global $U(1)$ survives. This is problematic as global symmetries are expected to be inconsistent with quantum gravity, see e.g.~\cite{Komargodski:2009pc, Dienes:2009td}.}

The terms we omitted when writing the action involve kinetic mixings between matter fermions, gauginos, and gravitino (which become relevant in the presence of gauge-symmetry or SUSY breaking) as well as 4-fermion-terms and couplings between fermions and the gauge field strength.

\subsection{Problems}

\subsubsection{Simple manipulations within the superspace approach}

{\bf Tasks:}

\noindent
(1) Check that, with our upper-left/lower-right convention for contracting Weyl indices, $\psi\chi=\chi\psi$. Check that consistency requires $\partial_\alpha\partial_\beta=-\partial_\beta\partial_\alpha$. Check that, again for consistency, one must have $(\partial_\alpha)^*=-\ol{\partial}_{\dot{\alpha}}$.

\noindent
(2) Check as many of the anticommutation relations between $Q$, $\ol{Q}$, $D$ and $\ol{D}$ as you need to feel confident.

\noindent
(3) Derive the transformation rules for the components of the chiral superfield.

\noindent
{\bf Hints:} Mostly straightforward manipulations - no hints needed. Recall that $(AB)^*=B^*A^*$ for an abstract algebra with a $*$-operation. When solving (3), it is useful to first work out $\delta_\xi\theta$, $\delta_\xi\theta^2$, $\delta_\xi y^\mu$ and $\delta_\xi f(y)$ for a generic function $y$. 

\noindent
{\bf Solution:}

\noindent
(1) One immediately finds
\be
\psi\chi=\psi^\alpha\chi_\alpha=\psi^\alpha \epsilon_{\alpha\beta}\chi^\beta=
\chi^\beta\epsilon_{\beta\alpha}\psi^\alpha=\chi^\beta\psi_\beta=\chi\psi\,.
\ee
Furthermore,
\be
\partial_1\partial_2\theta^2\theta^1=1\qquad \mbox{\qquad} \partial_2\partial_1\theta^2\theta^1=-\partial_2\partial_1\theta^1\theta^2=-1\,.
\ee
Next, one obviously has $(\partial_\alpha\theta^\beta)^*=\delta_\alpha{}^\beta$. By contrast, one may also evaluate this by first using the rules of an abstract algebra with a `$*$' and differentiating only after that. In other words, consider 
\be
(\partial_\alpha\theta^\beta)^*=\overleftarrow{\ol{\theta}^{\dot{\beta}} (-\ol{\partial}_{\dot{\alpha}})}\,,
\ee
where the arrow indicates that the derivative still acts on the variable. Also, we have to impose $\alpha=\dot{\alpha}$ and $\beta=\dot{\beta}$. Now, since Grassmann objects always anticommute, we also have
\be
\overleftarrow{\ol{\theta}^{\dot{\beta}} (-\ol{\partial}_{\dot{\alpha}})}
=\ol{\partial}_{\dot{\alpha}}\ol{\theta}^{\dot{\beta}}=\delta_{\dot{\alpha}}{}^{\dot{\beta}}=\delta_\alpha{}^\beta\,,
\ee
as desired. Clearly, the minus sign in the action of the `$*$' on derivatives was needed to get this consistent result. 

\noindent
(2) Using the definitions in the lecture, we have
\be
\{Q_\alpha,\ol{Q}_{\dot{\alpha}}\}=\{\partial_\alpha-i(\sigma^\mu)_{\alpha\dot{\beta}}\ol{\theta}^{\dot{\beta}}\partial_\mu\,,\,-\ol{\partial}_{\dot{\alpha}}+i\theta^\beta (\sigma^\nu)_{\beta\dot{\alpha}}\partial_\nu\} = i(\sigma^\nu)_{\alpha\dot{\alpha}}\partial_\nu+(-1)(-i)(\sigma^\mu)_{\alpha\dot{\alpha}}\partial_\mu\,.\label{cca}
\ee
Here we used the fact that non-zero contributions only arise from the first term of $Q$ acting on the second term of $\ol{Q}$ and vice versa. The resulting contributions add up giving the overall factor of 2 in the commutator given in the lecture. It is clear that, for two $Q$'s, the result will be zero since each term vanishes separately. Also, for $Q$ and $\ol{D}$ the result is zero on account of the sign flip in the definition of $\ol{D}$: the analogues of the final two terms in (\ref{cca}) cancel in this case.

\noindent
(3) We need to calculate
\be
\delta_\xi\Phi(y,\theta)=[(\xi\partial-i\xi\sigma^\mu\ol{\theta}\partial_\mu) + \mbox{h.c.}]\,(A(y)+\sqrt{2}\theta\psi(y)+\theta^2 F(y))\,.
\ee
We first note that
\be
(\delta_\xi\theta)^\alpha=\xi^\beta\partial_\beta\theta^\alpha=\xi^\beta\qquad
\mbox{or}\qquad \delta_\xi\theta=\xi\,.
\ee
Similarly,
\be
\delta_\xi\theta^2=\xi^\alpha\partial_\alpha\theta^\beta\theta_\beta=
\xi\theta+\theta^\beta\xi^\alpha\partial_\alpha\theta_\beta = \xi\theta-
\theta_\beta\xi^\alpha\partial_\alpha\theta^\beta=2\xi\theta\,.
\ee
Furthermore, for a generic function $f(y)$, we have
\be
\delta_\xi f(y)=(\partial_\mu f(y))\,\delta_\xi y^\mu
\ee
and
\bea
\delta_\xi y^\mu&=& [(\xi\partial-i\xi\sigma^\nu\ol{\theta}\partial_\nu)
+(-\ol{\xi}\ol{\partial}+i\theta\sigma^\nu\ol{\xi}\partial_\nu)]\,(x^\mu+i\theta \sigma^\mu\ol{\theta})
\nonumber \\
&=&i\xi\sigma^\mu\ol{\theta}-i\xi\sigma^\mu\ol{\theta}+i\theta\sigma^\mu \ol{\xi} + i\theta\sigma^\mu \ol{\xi} = 2 i\theta\sigma^\mu \ol{\xi}\,.
\eea
Note that, to get the sign of the third term in the second line right, one needs to take into account that
\be
\ol{\xi}\ol{\partial}\ol{\theta}^{\dot{\beta}}=\ol{\xi}_{\dot{\alpha}} \ol{\partial}^{\dot{\alpha}}\ol{\theta}^{\dot{\beta}}=-\ol{\xi}^{\dot{\alpha}} \ol{\partial}_{\dot{\alpha}}\ol{\theta}^{\dot{\beta}}=-\ol{\xi}^{\dot{\beta}} \,.
\ee

After these preliminaries, one immediately finds
\bea
\delta_\xi \Phi &=& (\partial_\mu A)\,(2i\theta\sigma^\mu\ol{\xi}) +\sqrt{2}\xi\psi + \sqrt{2}(\theta\partial_\mu\psi)\,(2i\theta \sigma^\mu\ol{\xi}) +2(\xi\theta)F
\label{cfd} \\
&=& 1\cdot(\sqrt{2}\xi\psi)\,+\,\sqrt{2}\theta\,(\sqrt{2}i\sigma^\mu\ol{\xi} \partial_\mu A\,+\,\sqrt{2}\xi F)\,+\,\theta^2(-\sqrt{2}i(\partial_\mu\psi) \sigma^\mu\ol{\xi})\,.\nonumber
\eea
Here, to derive the last term, we used
\be
\theta^\alpha\theta^\beta = -\frac{1}{2}\epsilon^{\alpha\beta}\theta^2\,.
\label{tde}
\ee
The second line of (\ref{cfd}) is already in a form which allows one to directly read off the quantities $\delta_\xi A$, $\delta_\xi \psi$ and $\delta_\xi F$ as the coefficients of 1, $\sqrt{2}\theta$, and $\theta^2$. To match this with the formula given in the lecture, one also needs to use the relation 
\be
\psi\sigma^\mu\ol{\xi}=-\ol{\xi}\ol{\sigma}^\mu\psi. 
\ee
This relation is easily derived using the definition of $\ol{\sigma}$ in the lectures through complex conjugation. One also needs the hermiticity of Pauli matrices.

\subsubsection{Deriving component actions}
\index{component actions}

{\bf Task:} Consider a generic chiral superfield model defined by a Kahler potential $K(\Phi^i,\ol{\Phi}^{\ol{\jmath}})$ and a superpotential $W(\Phi^i)$. The full component lagrangian reads
\bea
{\cal L} &=& -g_{i\ol{\jmath}}(\partial_\mu A^i)(\partial^\mu \ol{A}^{\ol{\jmath}})- ig_{i\ol{\jmath}} \ol{\psi}^{\ol{\jmath}}\ol{\sigma}^\mu D_\mu\psi^i +\frac{1}{4} R_{i\ol{\jmath}k\ol{l}}\psi^i\psi^k \ol{\psi}^{\ol{\jmath}} \ol{\psi}^{\ol{l}}
\nonumber \\
&& -\frac{1}{2}(D_iD_j W)\psi^i \psi^j+\mbox{h.c.}\,\,-\,\,g^{i\ol{\jmath}} (D_i W)
(D_{\ol{\jmath}} \ol{W})\,. \label{cmf}
\eea
Here 
\bea
&&\hspace*{-.5cm}\partial_i=\frac{\partial}{\partial \Phi^i}\,,\qquad 
\ol{\partial}_{\ol{\imath}}=\frac{\partial}{\partial \ol{\Phi}^{\ol{\imath}}}\,,\qquad
g_{i\ol{\jmath}}=\partial_i\ol{\partial}_{\ol{\jmath}} K=K_{i\ol{\jmath}}\,,\qquad
\Gamma_{ij}{}^k=g^{k\ol{l}}\partial_i g_{j\ol{l}}\,,\qquad
R_{i\ol{\jmath}k\ol{l}} = g_{m\ol{l}} \partial_{\ol{\jmath}} \Gamma_{ik}{}^m\,,
\nonumber\\ 
&& \hspace*{-0.8cm} D_i W=\partial_i W\,,\qquad
D_i D_j W=\partial_i(D_j W)-\Gamma_{ij}{}^k(D_k W)\,,\qquad
D_\mu \psi^i=\partial_\mu \psi^i+\Gamma_{jk}{}^i(\partial_\mu A^j)\psi^k\,.
\eea
Note that $\Gamma$ and $R$ are exactly the same Christoffel symbols and Riemann tensor that are familiar from general relativity. The formulae only look slightly different since we parameterise the manifold using complex coordinates and they are slightly simpler than usual because the metric is not generic but a Kahler metric. The covariant derivative $D_\mu$ has nothing to do with spacetime being curved (it is not) but rather related to the fact that $\psi$ lives in a bundle over the scalar manifold parameterised by $A$. Thus, comparing $\psi$ at two different points in $x$ requires knowledge of the values of $A$ at these points.

Derive the first two and the last term in (\ref{cmf}). If you wish, try also the others. 

\noindent
{\bf Hints:} You can save work by shifting $x$ under the integral:
\be
x^\mu+i\theta\sigma^\mu\ol{\theta}\,\,,\,\,x^\mu-i\theta\sigma^\mu\ol{\theta} \qquad \longrightarrow \qquad x^\mu\,\,,\,\, x^\mu-2i\theta\sigma^\mu \ol{\theta}\,.
\ee
Independently, prove and use the formula
\be
(\theta \sigma^\mu\ol{\theta}) (\theta \sigma^\nu\ol{\theta})
=-\frac{1}{2} \theta^2\ol{\theta}^2\eta^{\mu\nu}\,.\label{tef}
\ee

\noindent
{\bf Solution:} We start with the last formula. It is clear that the l.h. side must be proportional to $\theta^1\theta^2\ol{\theta}^1\ol{\theta}^2$ and hence to $\theta^2\ol{\theta}^2$. The latter is a scalar, so it must be multiplied by an invariant tensor with indices $\mu$ and $\nu$, where $\eta^{\mu\nu}$ is the only choice. Thus, one only needs to check normalisation. This is done most easily by focussing on $\mu=\nu=0$:
\be
(\theta\sigma^0\ol{\theta})^2=\left(\theta^1\ol{\theta}^1+\theta^2\ol{\theta}^2 \right)^2 = -2\theta^1\theta^2\ol{\theta}^1\ol{\theta}^2\,. 
\ee
We also have
\be
\theta^2=\theta^\alpha\epsilon_{\alpha\beta}\theta^\beta=2\theta^1\theta^2
\ee
and hence
\be
\theta^2\ol{\theta}^2=-4\theta^1\theta^2\ol{\theta}^1\ol{\theta}^2\,.
\ee
Recalling that we use the mostly-plus metric, the result follows. 

Now we proceed to evaluate the $D$ term of the Kahler potential. Since we are only interested in the kinetic term of $A$, we can set $\psi$ and $F$ to zero. Thus, with the shift of variables explained above, we have to evaluate
\be
K\Big(A^i(x),\ol{A}^{\ol{\imath}}(x-2i\theta\sigma\ol{\theta})\Big)\Big|_{\theta^2 \ol{\theta}^2}\,.
\ee
This is done by first Taylor expanding $\ol{A}$,
\be
K\Big(A^i,\ol{A}^{\ol{\imath}}-2i\theta\sigma^\mu\ol{\theta}\partial_\mu \ol{A}^{\ol{\imath}}+\theta^2\ol{\theta}^2\partial^2\ol{A}^{\ol{\imath}}\Big)\,,
\ee
where we used (\ref{tef}) to simplify the quadratic term in the expansion. Next we Taylor expand $K$, keeping only what will contribute to the $D$ term:
\be
K\Big|_{\theta^2\ol{\theta}^2}=\left(K_{\ol{\imath}}(A,\ol{A})\,\theta^2\ol{\theta}^2 \partial^2 \ol{A}^{\ol{\imath}} +\frac{1}{2}K_{\ol{\imath}\ol{\jmath}} (A,\ol{A})\,(2i\theta \sigma^\mu \ol{\theta}\partial_\mu \ol{A}^{\ol{\imath}})\, (2i\theta\sigma^\nu \ol{\theta}\partial_\nu \ol{A}^{\ol{\jmath}})\right) \Big|_{\theta^2\ol{\theta}^2}\,.
\ee
The second term can again be simplified using (\ref{tef}), which gives
\bea
K\Big|_{\theta^2\ol{\theta}^2}&=&K_{\ol{\imath}}(A,\ol{A})\,\partial^2 \ol{A}^{\ol{\imath}} + K_{\ol{\imath}\ol{\jmath}}(A,\ol{A})\, (\partial_\mu \ol{A}^{\ol{\imath}})\, (\partial^\mu \ol{A}^{\ol{\jmath}})
\\
&=& -\partial_\mu(K_{\ol{\imath}}(A,\ol{A})\,\partial^\mu \ol{A}^{\ol{\imath}} + K_{\ol{\imath}\ol{\jmath}}(A,\ol{A})\,(\partial_\mu \ol{A}^{\ol{\imath}})\, (\partial^\mu \ol{A}^{\ol{\jmath}})\,\,+\,\,\mbox{total derivative}
\nonumber \\
&=& -K_{j\ol{\imath}}(\partial_\mu A^j)(\partial^\mu \ol{A}^{\ol{\imath}}) 
\,\,+\,\,\mbox{total derivative}\,.\nonumber
\eea
This is our desired result.

To derive the last term in (\ref{cmf}), we only need to consider the terms involving $F$. It is clear that the Taylor expansion in $\theta^2$ and $\ol{\theta}^2$ gives
\be
K\Big|_{\theta^2\ol{\theta}^2}\supset K_{i\ol{\jmath}}F^i\ol{F}^{\ol{\jmath}}
\qquad\mbox{and}\qquad
W\Big|_{\theta^2}\supset W_i F^i +\mbox{h.c.}\label{trt}
\ee
Varying w.r.t. $\ol{F}^{\ol{\jmath}}$ one finds 
\be
\ol{W}_{\ol{\jmath}}+K_{i\ol{\jmath}}F^i=0\qquad\mbox{and hence}\qquad
F^i=-g^{i\ol{\jmath}}\ol{W}_{\ol{\jmath}}\,.
\ee
Inserting this in the three terms of (\ref{trt}), the result
\be
{\cal L}\supset -g^{i\ol{\jmath}} W_i \ol{W}_{\ol{\jmath}}
\ee
eventually follows.

Let us finally consider the fermion kinetic term. It will be convenient to shift the variable such that we have to deal with
\be
K(\Phi(x+2i\theta\sigma\ol{\theta}),\ol{\Phi}(x))\Big|_{\theta^2\ol{\theta}^2}\,.
\ee
Now, suppressing the spacetime arguments and the projection on the highest component for brevity, we expand the chiral superfields in the fermionic components:
\be
2K_{i\ol{\jmath}}\,(\ol{\theta}\ol{\psi})\,(\theta\psi)\,.
\ee
Then we expand $\psi$ to linear order in the quantity $2i\theta\sigma^\mu\ol{\theta}$:
\be
4K_{i\ol{\jmath}}\,(\ol{\theta}\ol{\psi}^{\ol{\jmath}})\,
(\theta\partial_\mu\psi^i)\,(i\theta \sigma^\mu\ol{\theta})
\ee
At this point we have to employ (\ref{tde}) and the hermitian conjugate relation 
\be
\ol{\theta}^{\dot{\beta}}\ol{\theta}^{\dot{\alpha}}=-\frac{1}{2}\epsilon^{\alpha\beta}\ol{\theta}^2\,.
\ee
Thus, we have
\be
(\ol{\theta}_{\dot{\alpha}}\ol{\psi}^{\dot{\alpha}})\,(\theta^\alpha
\partial_\mu\psi_\alpha)\,(i\theta^\beta\sigma^\mu_{\beta\dot{\beta}}
\ol{\theta}^{\dot{\beta}})=-\frac{1}{2}\,i (\ol{\theta}_{\dot{\alpha}}\ol{\psi}^{\dot{\alpha}})\,(\partial_\mu\psi^\beta
\sigma^\mu_{\beta\dot{\beta}}
\ol{\theta}^{\dot{\beta}})\,\theta^2=\frac{1}{4}\,i\partial_\mu\psi
\sigma^\mu\ol{\psi}\,\theta^2\ol{\theta}^2\,.
\ee
Now we use the relation
\be
\psi\sigma^\mu\ol{\chi}=-\ol{\chi}\ol{\sigma}^\mu\psi\,,
\ee
which follows from the hermiticity of $\sigma$ matrices and the anticommutation of spinors. Moreover, we implement the $\theta^2\ol{\theta}^2$ projection. This gives
\be
-i\,K_{i\ol{j}}\,\ol{\psi}^{\ol{\jmath}}\ol{\sigma}^\mu\partial_\mu\psi^i\,.
\ee
With the renaming $K_{i\ol{\jmath}}\to g_{i\ol{\jmath}}$ this is the partial-derivative part of our kinetic term.

We still have to find the term responsible for its covariantisation. For this, we note that we obtained the term $2i\theta\sigma^\mu\ol{\theta}$ from expanding $\psi$. But we could equally well have expanded $A$ in $K_{i\ol{\jmath}}$ to obtain this term. The calculation proceeds precisely as above, but in the final formula $\partial_\mu$ acting on $\psi$ is dropped. Instead, one has to replace $K_{i\ol{\jmath}}$ by
\be
\partial_k K_{i\ol{\jmath}} \partial_\mu A^k\,.
\ee
Thus, we finally have the term
\be
-i\partial_k g_{i\ol{\jmath}}\,\partial_\mu A^k \ol{\psi}^{\ol{\jmath}}\ol{\sigma}^\mu \psi^i\,.
\ee
To see that this is what we want, we work backward from (\ref{cmf}) and rewrite the corresponding term:
\be
-ig_{i\ol{\jmath}}\ol{\psi}^{\ol{\jmath}}\ol{\sigma}^\mu \Gamma_{jk}{}^i (\partial_\mu A^j)\psi^k=
-ig_{i\ol{\jmath}}\ol{\psi}^{\ol{\jmath}}\ol{\sigma}^\mu g^{i\ol{l}}\partial_j g_{k\ol{l}} (\partial_\mu A^j)\psi^k
\,.
\ee
Now the agreement is apparent.

\subsubsection{Fierz identities for Weyl spinors}
\index{Fierz identities}

{\bf Task:} Derive the covariant orthonormality condition for $\sigma$ matrices
\be
(\sigma_\mu)_{\alpha\dot{\alpha}}(\ol{\sigma}_\nu)^{\dot{\alpha}\alpha}=-2\eta_{\mu\nu}\,.
\label{sse}
\ee
Use it to simplify expressions like $(\sigma_\mu)_{\alpha\dot{\alpha}}(\sigma^\mu)^{\beta\dot{\beta}}$ and $(\sigma_\mu)_{\alpha\dot{\alpha}}(\ol{\sigma}^\mu)^{\dot{\beta}\beta}$. From this, Fierz identities like
\be
(\phi\sigma^\mu\ol{\chi})(\psi\sigma_\mu\ol{\eta})=-2(\psi\phi)(\ol{\chi}\ol{\eta})
\qquad\mbox{and}\qquad
(\phi\sigma^\mu\ol{\chi})(\ol{\eta}\ol{\sigma}_\mu\psi)=2(\psi\phi)(\ol{\chi}\ol{\eta})\label{2fi}
\ee
immediately follow. One can use those to replace bi-spinors within some longer expressions according to
\be
(\cdots\ol{\chi}\psi\cdots)=\frac{1}{2}(\cdots \ol{\sigma}^\mu\cdots)(\psi\sigma_\mu\ol{\chi})
\qquad\mbox{and}\qquad
(\cdots\psi\ol{\chi}\cdots)=\frac{1}{2}(\cdots \sigma^\mu\cdots)(\ol{\chi}\ol{\sigma}_\mu\psi)\,.\label{fiis}
\ee

\noindent
{\bf Hints and background:} Fierz identities are probably familiar in the context of Dirac spinors, where they are also used to rewrite expressions with four spinors in such a way that the pairs connected by index contraction (possibly through $\gamma$ matrices) change. The basic underlying idea making this possible is the completeness of $\{\mathbbm{1},\gamma_\mu,\gamma_5,\gamma_\mu\gamma_5,[\gamma_\mu,\gamma_\nu]\}$ in the space of 4$\times 4$ matrices. In our context, things are much simpler since the 4 $\sigma$-matrices already provide a basis of the space of $2\times 2$ matrices.

\noindent
{\bf Solution:} Let us start by rewriting the second matrix on the l.h.~side of (\ref{sse}) according to
\be
(\ol{\sigma}_\nu)^{\dot{\alpha}\alpha}=
(\sigma_\nu)^{\alpha\dot{\alpha}}=\epsilon^{\alpha\beta}
\epsilon^{\dot{\alpha}\dot{\beta}}(\sigma_\nu)_{\beta\dot{\beta}}= [(i\sigma_2)\sigma_\nu(-i\sigma_2)]^{\alpha\dot{\alpha}}
= [\{\sigma_0,-\sigma_1,\sigma_2,-\sigma_3\}]^{\alpha\dot{\alpha}}= [\{\sigma_0,-\sigma_i\}]^{\dot{\alpha}\alpha}\,.
\ee
With this and the usual orthonormality relations between the Pauli matrices and the unit matrix, the r.h.~side of (\ref{sse}) immediately follows.

Now we recall that the $\sigma$ matrices form a basis of 2$\times 2$ hermitian matrices. In fact, over the complex numbers they are a basis of all 2$\times 2$ matrices. Hence we have
\be
M_{\alpha\dot{\alpha}}=M^\mu(\sigma_\mu)_{\alpha\dot{\alpha}} \label{mmr}
\ee
for generic $M_{\alpha\dot{\alpha}}$ and appropriate coefficients $M^\mu$. Multiplying by $(\ol{\sigma}_\nu)^{\dot{\alpha}\alpha}$ and using (\ref{sse}), one finds
\be
M_{\alpha\dot{\alpha}}(\ol{\sigma}_\nu)^{\dot{\alpha}\alpha}=-2M^\mu\eta_{\mu\nu}\,.
\ee
Solving this for $M^\mu$ and inserting in (\ref{mmr}) gives
\be
-\frac{1}{2}M_{\beta\dot{\beta}}(\ol{\sigma}^\mu)^{\dot{\beta}\beta}(\sigma_\mu)_{\alpha\dot{\alpha}}= M_{\alpha\dot{\alpha}} = M_{\beta\dot{\beta}} \delta_\alpha{}^\beta \delta_{\dot{\alpha}}{}^{\dot{\beta}}
\ee
or, since $M$ was generic,
\be
(\sigma_\mu)_{\alpha\dot{\alpha}}
(\ol{\sigma}^\mu)^{\dot{\beta}\beta} = -2 
\delta_\alpha{}^\beta \delta_{\dot{\alpha}}{}^{\dot{\beta}} \,.
\ee
Using the hermiticity of $\sigma$ matrices and lowering the indices, one then also has
\be
(\sigma_\mu)_{\alpha\dot{\alpha}}
(\sigma^\mu)_{\beta\dot{\beta}} = -2 
\epsilon_{\alpha\beta} \epsilon_{\dot{\alpha}\dot{\beta}} \,.
\ee
From this, the identities in (\ref{2fi}) follow straightforwardly by multiplication and contraction with four spinors, where one has of course to be very careful with the spinor ordering and signs. Finally, (\ref{fiis}) provides two different ways for reinterpreting (\ref{2fi}) as a method for replacing two spinors within a longer string of Weyl spinor expressions.

\subsubsection{SUSY in components}

{\bf Task:} Demonstrate that the SUSY algebra is represented on the scalar (or chiral) multiplet, without using superspace. Realise SUSY without the auxiliary field (just on $A$ and $\psi$) by allowing yourself to use the equations of motion (i.e. working on-shell).

\noindent
{\bf Hints:} As explained, while SUSY is very conveniently derived in superspace, it can also be discussed entirely at the level of component fields. This is important since in many cases (in higher dimensions, in many supergravity theories, or in situations with more than the minimal set of $Q$'s, also known as ${\cal N}=2$ or ${\cal N}=4$ SUSY), no superspace description exists or is not efficient. To discuss this component description, one focuses on the bosonic generators 
\be
\delta_\xi=\xi Q+\ol{\xi}\ol{Q}\,.
\ee
Their algebra, defined with commutators, is equivalent ot the SUSY algebra. Start by calculating $[\delta_\xi,\delta_\eta]$ using the known algebra of the $Q$'s. Then check that the algebra is represented on the components by using the explicit expressions for $\delta_\xi A$, $\delta_\xi \psi$ and $\delta_\xi F$ that were given in the lecture and that have already been derived in a previous exercise. Show also that the algebra still `closes' (a common synonym for being represented on a certain set of fields) if $\delta_\xi F$ is dropped and, in the other expressions, $F$ is replaced using the equations of motion. (For simplicity, we consider the free case and hence free equations of motion.) Note that in this latter case one has to use equations of motion `to close the algebra'. One also says that the algebra is only realised `on-shell'.

Use the Fierz identities and try not to get lost in the many spinors and indices, especially when evaluating the algebra on $\psi$. 

\noindent
{\bf Solution:} First, one has
\be
[\xi Q,\ol{\eta}\ol{Q}]=\xi^\alpha Q_\alpha\ol{Q}_{\dot{\alpha}} \ol{\eta}^{\dot{\alpha}} - \ol{\eta}_{\dot{\alpha}} \ol{Q}^{\dot{\alpha}} Q^\alpha \xi_\alpha
= \xi^\alpha \{Q_\alpha,\ol{Q}_{\dot{\alpha}}\} \ol{\eta}^{\dot{\alpha}} = 2 \xi\sigma^\mu \ol{\eta}\,P_\mu = -2i\, \xi\sigma^\mu\ol{\eta}\, \partial_\mu
\ee
and hence
\be
[\delta_\xi,\delta_\eta]=[\xi Q,\ol{\eta}\ol{Q}]+[\ol{\xi}\ol{Q}, \eta Q] = [\xi Q,\ol{\eta}\ol{Q}] - (\xi\leftrightarrow \eta) = -2i (\xi\sigma^\mu\ol{\eta} - \eta\sigma^\mu \ol{\xi}) \,\partial_\mu\,.
\ee
To see that this explicitly holds for the scalar multiplet, we start with the scalar component that gives this multiplet its name:
\bea
[\delta_\xi,\delta_\eta]A &=& \delta_\xi \delta_\eta A- (\xi\leftrightarrow \eta) = \delta_\xi \sqrt{2} \eta\psi - (\xi\leftrightarrow \eta) = \sqrt{2}\eta (i\sqrt{2}\eta \sigma^\mu\ol{\xi}\partial_\mu A +\sqrt{2}\xi F) - (\xi\leftrightarrow \eta)
\nonumber \\
&=& 2i\eta \sigma^\mu\ol{\xi}\,\partial_\mu A 
- (\xi\leftrightarrow \eta) = -2i\xi \sigma^\mu\ol{\eta}\,\partial_\mu A - (\xi\leftrightarrow \eta)
\,. \eea
This is the desired result.

The analogous calculation for the fermion is slightly more involved:
\bea
[\delta_\xi,\delta_\eta] \psi &=& \delta_\xi ( i\sqrt{2}\sigma^\mu\ol{\eta}\partial_\mu A +\sqrt{2} \eta F) - (\xi\leftrightarrow \eta)
\nonumber
\\
&=& i\sqrt{2} \sigma^\mu \ol{\eta}\,\partial_\mu(\sqrt{2}\xi \psi) + \sqrt{2}\eta\,i\sqrt{2}\,\ol{\xi}\ol{\sigma}^\mu\partial_\mu \psi - (\xi\leftrightarrow \eta)
\label{fcal}
\\
&=& 2i (\sigma^\mu\ol{\eta})(\xi \partial_\mu \psi) + 2i\eta (\ol{\xi}\ol{\sigma}^\mu\partial_\mu \psi) - (\xi\leftrightarrow \eta)\,.
\nonumber
\eea
Here in the last line we have introduced (formally superfluous) brackets to emphasise where the consecutive contraction of Weyl indices is interrupted.
Now, using the two Fierz-type identities in (\ref{fiis}), we 
rewrite the terms in such a way that $\xi$ and $\ol{\eta}$ (or $\eta$ and $\ol{\xi}$) are contracted with each other through one $\sigma$ matrix:
\be
[\delta_\xi,\delta_\eta] \psi = i(\xi\sigma_\nu\ol{\eta})(\sigma^\mu \ol{\sigma}^\nu\partial_\mu\psi)+i(\ol{\xi}\ol{\sigma}_\nu\eta)(\sigma^\nu\ol{\sigma}^\mu \partial_\mu\psi) - (\xi\leftrightarrow \eta)\,.
\ee
At this point, it is convenient to make the two explicitly written terms more similar by exchanging $\xi$ and $\eta$ in the second term (together with a sign change):
\be
[\delta_\xi,\delta_\eta] \psi = i(\xi\sigma_\nu\ol{\eta})(\sigma^\mu \ol{\sigma}^\nu\partial_\mu\psi)-i(\ol{\eta}\ol{\sigma}_\nu\xi)(\sigma^\nu\ol{\sigma}^\mu\partial_\mu\psi) - (\xi\leftrightarrow \eta)\,.
\ee
Next, employing the hermiticity of $\sigma$ matrices, we may replace $\ol{\sigma}$ by $\sigma$ in the second term. The re-ordering of spinors which is then also necessary introduces a further sign change:
\be
[\delta_\xi,\delta_\eta] \psi = i(\xi\sigma_\nu\ol{\eta})(\sigma^\mu \ol{\sigma}^\nu\partial_\mu\psi) + i(\xi \sigma_\nu\ol{\eta})(\sigma^\nu\ol{\sigma}^\mu\partial_\mu\psi) - (\xi\leftrightarrow \eta)\,.
\ee
Finally, using the Clifford-algebra-type relation $\sigma_\mu\ol{\sigma}_\nu + \sigma_\nu\ol{\sigma}_\mu = -2\eta_{\mu\nu}\mathbbm{1}$
(analogous to (\ref{smao})), the desired result follows.

The calculation for the auxiliary field is again simpler:
\bea
\hspace*{-1cm}[\delta_\xi,\delta_\eta] F = \delta_\xi i\sqrt{2} \ol{\eta}\ol{\sigma}^\mu \partial_\mu \psi - (\xi\leftrightarrow \eta)
&=& i\sqrt{2}\ol{\eta}\ol{\sigma}^\mu \partial_\mu \,[ i\sqrt{2} \sigma^\nu\ol{\xi}\partial_\nu A+\sqrt{2}\xi F ] - (\xi\leftrightarrow \eta) \label{ftc}
\\ \nonumber
&=& -2\ol{\eta}\ol{\sigma}^\mu \sigma^\nu \ol{\xi}\,\partial_\mu \partial_\nu A + 2 i \ol{\eta}\ol{\sigma}^\mu \xi\,\partial_\mu F - (\xi\leftrightarrow \eta)\,. 
\eea
Here, the first term in the second line simplifies if one uses the symmetry of $\partial_\mu\partial_\nu$ to replace the product of $\sigma$ matrices by $-\eta_{\mu\nu}\mathbbm{1}$. After this, the expression is proportional to $\ol{\xi}\ol{\eta}$ and vanishes upon $\xi$-$\eta$-antisymmetrisation. The second term in the last line of (\ref{ftc}) provides, after rewriting in terms of $\sigma^\mu$, our desired result.

Finally, we want to repeat the calculations for $[\delta_\xi,\delta_\eta]$ on $A$ and on $\psi$ with the auxiliary replaced according to the equations of motion. Specifically for the free theory, that means
\be
F=-m\ol{A}\,,
\ee
such that we now work with the SUSY transformation rules
\bea
\delta_\xi A&=&\sqrt{2}\xi\psi
\\
\delta_\xi \psi &=& i\sqrt{2}\sigma^\mu\ol{\xi}\partial_\mu A - m\sqrt{2}\xi \ol{A}\,.
\eea
In the analysis of $[\delta_\xi,\delta_\eta]A$ we do not even need the term with $m$ which formerly involved $F$. As we can see be revisiting our calculation above, this term simply drops out under $\xi$-$\eta$-antisymmetrisation. By contrast, in the fermion case the last line of (\ref{fcal}) is replaced by
\be
[\delta_\xi,\delta_\eta] \psi = 2i (\sigma^\mu\ol{\eta})(\xi \partial_\mu \psi) - \delta_\xi \,m\sqrt{2}\eta \ol{A} - (\xi\leftrightarrow \eta)\,. \label{onshr}
\ee
We can employ the equation of motion 
$i\ol{\sigma}^\mu\partial_\mu\psi + m\ol{\psi}=0$ to perform the rewriting
\be
m\,\delta_\xi \ol{A} = \sqrt{2}m(\ol{\xi}\ol{\psi}) = -i\sqrt{2}(\ol{\xi}\ol{\sigma}^\mu\partial_\mu\psi)\,,
\ee
After that, (\ref{onshr}) takes precisely the form of the last line of (\ref{fcal}). But from there, we already know how to arrive at the desired result, so we are done. In summary, if one is prepared to use the equations of motion, one can indeed live without the auxiliary field (on-shell SUSY). 

The reader may want to continue this exercise independently by also checking the invariance of the free lagrangian, off-shell and on-shell.

\subsubsection{Gauge coupling unification}\label{quni}
\index{gauge coupling unification}

{\bf Task:} Demonstrate that precision gauge coupling unification in the $SU(5)$ scheme does not work well in the Standard Model but, by contrast, works extremely well with low-scale supersymmetry.

\noindent
{\bf Hints:} Recall that the beta function\index{beta function} of a gauge theory with coupling $g$ is commonly defined as
\be
\beta(g)=\frac{dg}{d\,\ln\mu}=\frac{b\,g^3}{16\pi^2}+\cdots\,.
\ee
Here in the last expression we gave the leading-order result with the widely used `beta-function-coefficient' $b$ encoding the numerical prefactor. For a $U(1)$ gauge theory one explicitly finds
\be
b=\frac{q^2}{6}\,c\quad\mbox{with}\quad c=2\,/\,4\,/-22\quad\mbox{for}
\quad\mbox{a complex scalar$\,$/$\,$Weyl fermion$\,$/$\,$real vector}
\ee
with charge $q$ running in the loop. Here the last option is somewhat formal: Indeed, while a charged scalar or Weyl fermion is easy to add to an abelian gauge theory, adding a charged complex vector is somewhat artificial. More naturally, one would view such a vector as the combination of two real vectors, each of which corresponds to an extra $U(1)$ gauge theory. This type of charged matter does, in turn, appear naturally if our original $U(1)$ is viewed as a subgroup of a non-abelian gauge group. For this reason it is in fact useful to know the above numerical value of `$-22$'. The derivation of these three values of $c$ needs only the calculation of the log-divergence in the familiar vacuum-polarisation or self-energy diagram and can be found in many QFT textbooks,~e.g.~\cite{Peskin:1995ev}.

Obviously, the non-abelian case requires the substitution
\be
q^2\,\,\,\to\,\,\,\mbox{tr}(T_R^a T_R^b)\equiv T_R\,\delta^{ab}
\ee
in the relevant self-energy diagram, where $R$ stands for the representation in which the matter in the loop transforms. Here $T_R$ is the so-called Dynkin-index\index{Dynkin-index} of the representation $R$. The corresponding substitution in the beta function coefficient hence reads $q^2\to T_R$. Concretely, one has $T_F=1/2$ and $T_A=N$ for the fundamental and adjoint of $SU(N)$. One sometimes also refers to $T_A=T(A)=C_2(A)$ as the quadratic Casmir\index{Casmir operator} of the adjoint representation. 

It is now straightforward to obtain the values of $b_{1,2,3}$ and $b'_{1,2,3}$ for the running of the couplings of $U(1)$, $SU(2)$ and $SU(3)$ in the Standard Model and the MSSM. It is convenient to work with quantities like $\alpha_i^{-1}$ since solving the renormalisation group equation for these inverse squared couplings is particularly easy. Moreover, it is useful to work with $\Delta\alpha_{12}\equiv \alpha_1^{-1}-\alpha_2^{-1}$ etc. Also, please use $SU(5)$-normalisation\index{$SU(5)$} for the $U(1)$ gauge coupling. Calculate the values of the mass scales $M_{12}$, $M_{23}$ and $M_{13}$ at which the various gauge couplings meet in the Standard Model and the MSSM (with initial values for $\alpha_i$ and SUSY breaking at $m_Z$, to keep things simple). Finally, turn the logic around and derive the predicted value of $\alpha_3$ at $m_Z$ as it follows from the GUT hypothesis and the values of $\alpha_{1,2}$ at $m_Z$. 

\noindent
{\bf Solution:} Let us start with the Standard Model and with $b_3$. We have contributions from the triplets (or equivalently anti-triplets) corresponding to l.h. and r.h. up and down-type quarks as well as from the gluons:
\be
b_3=\frac{1}{6}\left(4\cdot 2\cdot 2 \cdot N_f\cdot\frac{1}{2}-22\cdot 3\right)=\frac{4}{3}N_f-11=-7\,.
\ee
Here, in the first term, the 4 comes from the Weyl fermion nature of our matter, the 2$\cdot$2 from l.h./r.h. and up/down, the $N_f=3$ from the three families, and the $1/2$ from $T_F=1/2$. In the second term we have the $-22$ from the vector nature of the gluons and the $3$ from $T_A=N=3$. 

Next, we consider $SU(2)$:
\be
b_2=\frac{1}{6}\left(4\cdot (3+1)\cdot N_f \cdot\frac{1}{2} + 2\cdot \frac{1}{2} - 22\cdot 2\right)=
\frac{4}{3}N_f-\frac{43}{6}=-\frac{19}{6}\,.
\ee
Here, in the first term we have again a 4 from the Weyl fermion nature, a $(3+1)$ from the 3 colours of the quark doublet and the $1$ lepton doublet, as well as $N_f/2$ as above. In the second term we have a 2 from the scalar nature of the Higgs as well as $T_F=1/2$. The third term is self explanatory, with $T_A=N=2$. 

Finally, for $U(1)$ we have:
\bea
b_1&=&\frac{1}{6}\left( 4\left[6\cdot\left(\frac{1}{6}\right)^2+ 3\left(\frac{2}{3}\right)^2+3\left(\frac{1}{3}\right)^2+ 2\left(\frac{1}{2}\right)^2+1^2 \right]N_f+2\cdot 2\left(\frac{1}{2}\right)^2\right)\,\frac{3}{5}
\nonumber
\\
&=&\frac{4}{3}N_f+\frac{1}{10}=\frac{41}{10}\,.
\eea
Here the 5 terms inside the square bracket correspond to the contributions from quark doublet, up and down-quark, lepton doublet and r.h.~electron. The additional contribution outside the square bracket comes from the Higgs, with a factor 2 because it is a complex scalar and another 2 because it is a doublet. Finally, the charges are given in Standard Model hypercharge normalisation, which is corrected by the explicit factor of $3/5$ to bring us to the right normalisation for the beta function coefficient of the $U(1)$ as a subgroup of $SU(5)$ (cf. the solution to Problem~\ref{smsu5}). 

The reader will not be surprised to note that the matter contribution to all $b_i$ is the same since, as we already know, matter comes in complete $SU(5)$ multiplets. 

To get the SUSY version of the above, one needs to add the effects of gauginos, extra Higgs and Higgsino fields, and sfermions. The gauginos give
\be
\Delta b_1^g=0\,,\qquad \Delta b_2^g=\frac{1}{6}\cdot 4\cdot 2 = \frac{4}{3}\,,\qquad \Delta b_3^g=\frac{1}{6}\cdot 4\cdot 3 = 2\,.
\ee
Here the 4 comes from the gauginos being Weyl fermions and the $(0,2,3)$ are the relevant values of $T_A$. 

The Higgs contribution receives a factor of two compared to the Standard Model because we now have two Higgs doublets. In addition, we have to replace $2\to 2+4$, since instead of a complex scalar we now have a complex scalar and a Weyl fermion. This amounts to a total factor of 6 or, equivalently, an additional term worth 5 times the Standard Model Higgs effect. Using the Higgs part of the previous analysis, this gives
\be
\Delta b_1^h=\frac{1}{2}\,,\qquad \Delta b_2^h=\frac{5}{6}\,,\qquad \Delta b_3^h=0\,.
\ee

Finally, the matter part suffers the substitution $4\to 4+2$, i.e., an additional term worth one half of the previous value. This means
\be
\Delta b_1^m=\Delta b_2^m=\Delta b_3^m=2\,.
\ee

Adding everything up and also displaying the Standard Model coefficients again for easier reference, we now finally have
\be
b_i=\left(\frac{41}{10},-\frac{19}{6},-7\right)\qquad \mbox{and}\qquad 
b_i'=\left(\frac{33}{5},1,-3\right)
\ee
for the Standard Model and the MSSM respectively.

For the rest of the exercise, our basic numerical input is
\be
\frac{2\pi}{\alpha_1}=370.7\,,\qquad 
\frac{2\pi}{\alpha_2}=185.8\,,\qquad 
\frac{2\pi}{\alpha_3}=53.2\,.
\ee
The first two values are in fact known with much more precision than displayed above. The last corresponds to $\alpha_3=0.118$ at $m_Z$ -- by now also a very well measured quantity. We have already discussed these numbers very roughly in Problem~\ref{smsu5}, but here we wanted to be a bit more precise. The standard source for such data is the Review of Particle Properties of the Particle Data Group (PDG) \cite{Tanabashi:2018oca}.\index{Particle Data Group}\index{PDG}

\begin{figure}[ht]
\begin{center} 
\includegraphics[width=8cm]{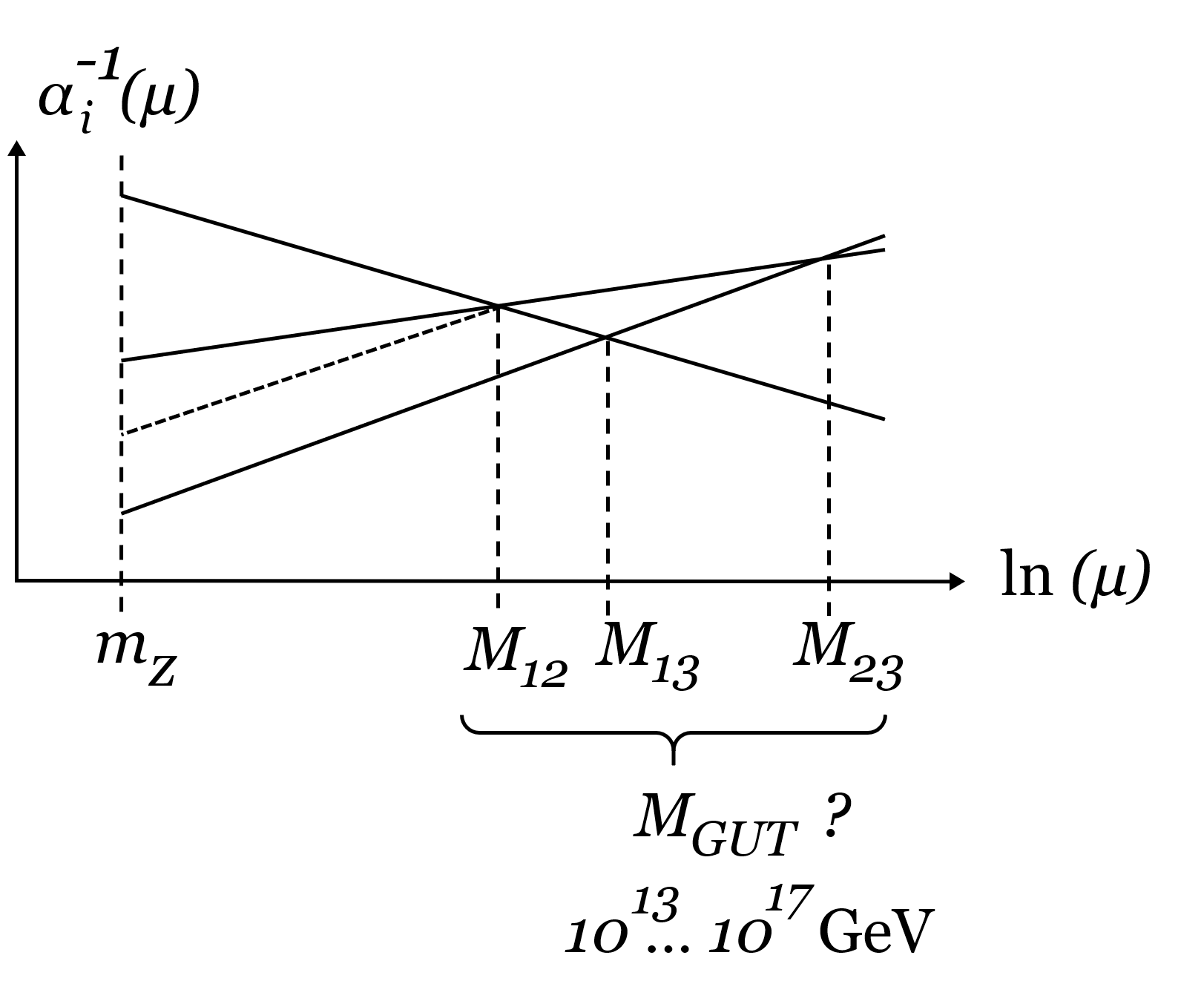}
\caption{One-loop running of inverse gauge couplings in the Standard Model.}
\label{smun} 
\end{center}
\end{figure}

On the analytic side, our main input are the three equations
\be
\alpha_i^{-1}(\mu)=-\frac{b_i}{2\pi}\,\ln(\mu)+(\mbox{const.})_i\,\,.
\ee
Starting at some high scale $M$ and running down to $m_Z$ this gives
\be
\alpha_i^{-1}(m_Z)=\alpha_i^{-1}(M)+\frac{b_i}{2\pi}\, \ln\left(\frac{M}{m_Z}\right)\,.
\ee
Specifically, if we assume that the two couplings $\alpha_1$ and $\alpha_2$ become equal at the scale $M=M_{12}$, then we deduce 
\be
\Delta\alpha_{12}(m_Z)=\frac{\Delta b_{12}}{2\pi} \ln\left(\frac{M_{12}}{m_Z}\right)\,,\label{a12}
\ee
where $\Delta\alpha_{12}\equiv\alpha_1^{-1}-\alpha_2^{-1}$ and $\Delta b_{12}\equiv b_1-b_2$. We find
\be
M_{12}=m_Z\,\exp\left(\frac{2\pi\Delta\alpha_{12}(m_Z)}{\Delta b_{12}}\right)
=90\,\mbox{GeV}\,\exp\left(\frac{370.7-185.8}{41/10-(-19/6))}\right)
\ee
and, using analogous formulae for the other `unification scales', we find
\be
M_{12}=1.0\times 10^{13}\,\mbox{GeV}\,,\qquad
M_{23}=9.5\times 10^{16}\,\mbox{GeV}\,,\qquad
M_{13}=2.4\times 10^{14}\,\mbox{GeV}
\ee
in the Standard Model. The running of inverse gauge couplings that corresponds to these results is sketched in Fig.~\ref{smun}. We see that gauge couplings do not really unify and the so-called grand unification scale $M_G$ remains somewhat vague, with a value in the range of $10^{13}\cdots 10^{17}\,$GeV. Alternatively, one may define $M_G$ by the unification of $\alpha_1$ and $\alpha_2$, and attempt to predict $\alpha_3$ at the weak scale by running it backwards from that point using $b_3$. This is illustrated in the figure by the dashed line, and it is apparent that this prediction will not be very good.

By contrast, as one now immediately verifies using the formulae above, the same analysis in the MSSM with SUSY breaking at $m_Z$ gives
\be
M'_{12}=2.0\times 10^{16}\,\mbox{GeV}\,,\qquad
M'_{23}=2.2\times 10^{16}\,\mbox{GeV}\,,\qquad
M'_{13}=2.1\times 10^{16}\,\mbox{GeV}\,.
\ee
This has been celebrated as a great success of the SUSY-GUT\index{SUSY!-GUT} idea, the scale of which is hence quantitatively fixed: $M_G\simeq 2\times 10^{16}\,$GeV. However, to a certain extent this perfection is accidental, as we will explain after turning the argument around to predict $\alpha_3(m_Z)$. 

To derive this prediction, one combines (\ref{a12}) with its analogue for $\alpha_{13}$, under the assumption that $M_{12}=M_{13}=M_G$. Eliminating $M_G$, one finds
\be
\Delta\alpha_{12}(m_Z)/b_{12}=\Delta\alpha_{13}(m_Z)/b_{13}
\ee
or
\be
\alpha_3^{-1}(m_Z)=\alpha_1^{-1}(m_Z)-\frac{b_{13}}{b_{12}}\, \Delta\alpha_{12}(m_Z)\,,
\ee
implying the predicted value $\alpha^{pred.}_3(m_Z)\simeq 0.117\,$. The corresponding non-SUSY prediction would be 0.071, i.e. completely off.

But one should not overstate the perfection of the result above: There are 2-loop corrections to the running, which are very well understood and lift the prediction to $\alpha^{pred.}_3(m_Z)\simeq 0.129$, which is about 10\% too large. This becomes slightly better but still not perfect if one takes into account that SUSY is broken not at $m_Z$ but at least at about a TeV. Finally, there are threshold corrections both at the SUSY breaking and the GUT scale, which also affect unification. By this we mean effects arising because not all SUSY partners and not all new GUT scale particles are degenerate at the respective scales $m_{soft}$ and $M_G$. Thus, SUSY unification works well but not as perfectly as the naive 1-loop analysis suggests. It does in fact become even slightly better if the SUSY breaking scale is raised above 1~TeV. However, one has to be honest and admit that, once one gives up on the SUSY solution of the hierarchy  problem, the SUSY breaking scale could be anywhere and one can not really claim any more that one predicts $\alpha_3(m_Z)$. A few more details and references to many much more detailed analyses can be found in the PDG review section on Grand Unification.

\subsubsection{Graviton spin (helicity)}
\index{graviton}\index{helicity}

{\bf Task:} Show that, under transverse rotations by an angle $\phi$, a linear superposition of the two physical photon states can be represented by a complex number rotating with a phase $\exp(i\phi)$. Show that, analogously, the general physical graviton state rotates twice as fast (i.e., that `the graviton has spin 2'). 

\noindent
{\bf Hints:} Let the photon momentum be $k\sim (1,1,0,0)^T$. Then transversality $\epsilon\cdot k=0$ together with the gauge choice $\epsilon^0=0$ leaves the two basis polarisations
\be
\epsilon_{(1)}=\left(\begin{array}{c}0 \\ 0 \\ 1\\ 0 \end{array}\right)\,\,,
\qquad
\epsilon_{(2)}=\left(\begin{array}{c}0 \\ 0 \\ 0\\ 1 \end{array}\right)\,.
\ee
The general state can be characterised by $\alpha\epsilon_{(1)} +\beta\epsilon_{(2)}$ or, equivalently, by
\be
\left(\begin{array}{c}\alpha \\ \beta\end{array}\right)\in \mathbb{R}^2\qquad 
\mbox{or}\qquad \alpha+i\beta\in \mathbb{C}\,.\label{rcr}
\ee
Similarly, under the constraints of transversality and tracelessness ($\epsilon^{\mu\nu}\eta_{\mu\nu}=0$), the graviton polarisation basis for $k\sim (1,1,0,0)^T$ is
\be
\epsilon_{(1)}=\left(\begin{array}{cccr}0&0&0&0 \\ 0&0&0&0 \\ 0&0&1&0 \\ 0&0&0&-1 \end{array}\right)\,\,,
\qquad
\epsilon_{(2)}=\left(\begin{array}{cccc}0&0&0&0 \\ 0&0&0&0 \\ 0&0&0&1 \\ 0&0&1&0 \end{array}\right)\,.
\ee
Again, the general state is $\alpha\epsilon_{(1)} +\beta\epsilon_{(2)}$ and the real or complex representation is provided by (\ref{rcr}). 

\noindent 
{\bf Solution:} The relevant Lorentz transformation reads
\be
\epsilon^\mu\to \Lambda^\mu{}_\nu\epsilon^\nu\qquad\mbox{with}\qquad
\Lambda=\left(\begin{array}{cccr}1&0&0&0 \\ 0&1&0&0 \\ 0&0&c&-s \\ 0&0&s&c \end{array}\right)\qquad\mbox{and}\qquad
\begin{array}{c}c=\cos\phi \\ s = \sin\phi\end{array}\,.
\ee
The vector $(\alpha,\beta)^T$ transforms by a $\phi$-rotation by definition. Elementary complex algebra then implies that 
\be
\alpha+i\beta\,\,\to\,\,\alpha'+i\beta'=e^{i\phi}(\alpha+i\beta)\,.
\ee
For the graviton, the general state can be represented by
\be
\left(\begin{array}{rr}\alpha & \beta \\ \beta & -\alpha\end{array}\right)
\ee
and the transformed state is
\bea
\left(\begin{array}{rr}\alpha' & \beta' \\ \beta' & -\alpha'\end{array}\right)
&=&
\left(\begin{array}{rr}c & -s \\ s & c\end{array}\right)
\left(\begin{array}{rr}\alpha & \beta \\ \beta & -\alpha\end{array}\right)
\left(\begin{array}{rr}c & s \\ -s & c\end{array}\right)
\,=\,
\left(\begin{array}{rr}c\alpha-s\beta & c\beta+s\alpha \\ s\alpha+c\beta & s\beta-c\alpha\end{array}\right)
\left(\begin{array}{rr}c & s \\ -s & c\end{array}\right)
\nonumber \\
\nonumber \\
&=&\left(\begin{array}{rr}c^2\alpha-sc\beta-cs\beta-s^2\alpha & sc\alpha-s^2\beta+c^2\beta+sc\alpha \\ cs\alpha+c^2\beta-s^2\beta+cs\alpha & s^2\alpha+sc\beta+sc\beta-c^2\alpha\end{array}\right)
\nonumber\\
\nonumber\\
&=& \left(\begin{array}{rr}c'\alpha-s'\beta & s'\alpha+c'\beta \\ s'\alpha+c'\beta & -(c'\alpha-s'\beta)\end{array}\right)
\eea
with $c'=c^2-s^2=\cos 2\phi$ and $s'=2sc=\sin 2\phi$. Hence
\be
\left(\begin{array}{r}\alpha' \\ \beta'\end{array}\right)=
\left(\begin{array}{rr}c' & -s' \\ s' & c'\end{array}\right)
\left(\begin{array}{r}\alpha \\ \beta\end{array}\right)
\ee
and, in the complex plane,
\be
\alpha+i\beta\,\,\to\,\,\alpha'+i\beta'=e^{2i\phi}(\alpha+i\beta)\,.
\ee

\section{String Theory: Bosonic String}\label{stbs}
\index{bosonic string}

\subsection{Strings -- basic ideas}

What we have achieved so far is not entirely satisfactory: Supersymmetry (more precisely, the broader framework of supergravity) offers a partial solution to the weak-scale hierarchy problem. Partial refers to the fact that SUSY partners have not been discovered (yet?) and hence some fine-tuning is probably needed after all. Supergravity is needed to combine this with general relativity, but it does not help with the cosmological constant problem, which unavoidably shows up in this context. Technically, the cosmological constant can be anything in supergravity: It can be negative due to the $-3|W|^2$ term, or positive due to a dominant $|DW|^2$ term (with SUSY spontaneously broken). It is also affected by UV divergences since (in spite of the non-renormalisation theorems for $W$), the Kahler potential $K$ is loop corrected. Moreover, the UV problems of gravity (all operators being generated at the scale $M_P$ -- i.e. formal `non-renormalisability') are not resolved by the prefix `super'. 

\begin{figure}[ht]
\begin{center} 
\includegraphics[width=12cm]{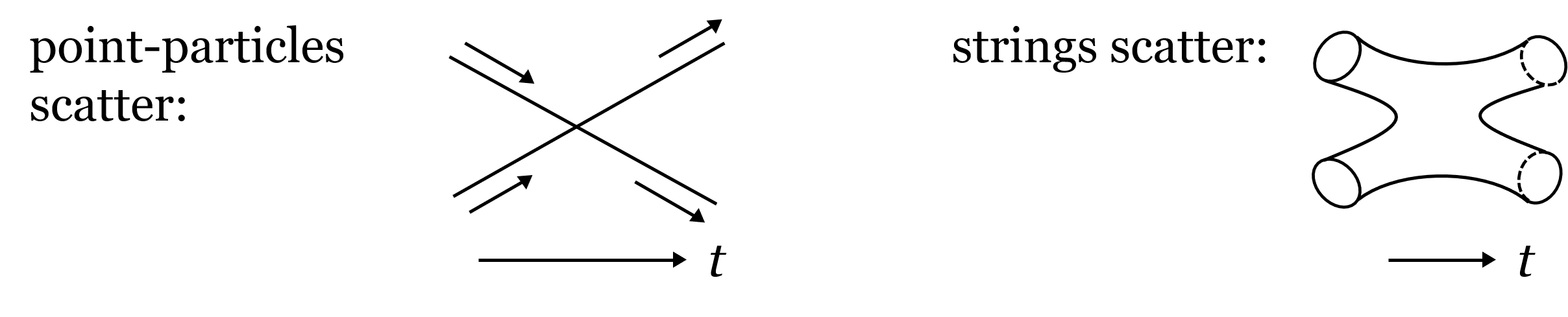}
\caption{Point particle scattering vs. string scattering.}
\label{ssc} 
\end{center}
\end{figure}

The string idea is illustrated in Fig.~\ref{ssc} and states simply that point-particles should be replaced by little loops of {\it fundamental string}. This might help with UV divergences (especially in gravity) since the interaction point is gone. Hence, when calculating a loop, there is no way in which this loop can go to zero size by the (e.g. two) interaction points 
becoming infinitely close. Some of the many standard textbooks are \cite{Green:1987sp, Polchinski:1998rq, Blumenhagen:2013fgp, bbs, kir}. 

\begin{figure}[ht]
\begin{center} 
\includegraphics[width=6cm]{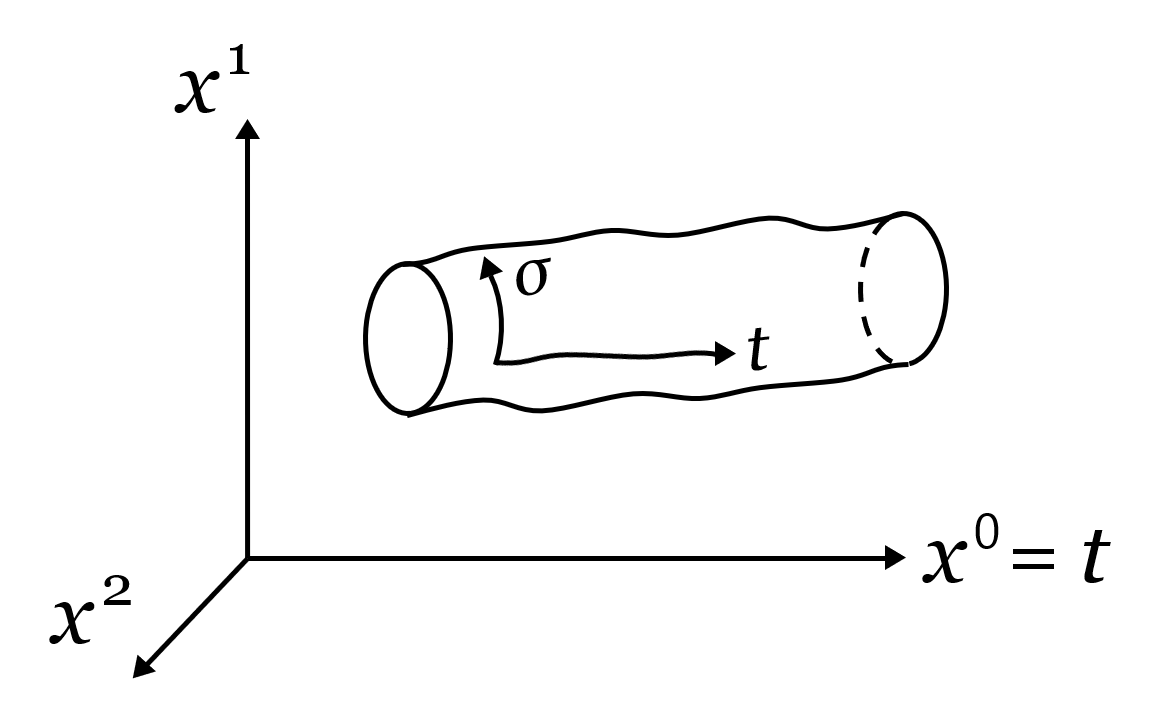}
\caption{String moving through target space.}
\label{its} 
\end{center}
\end{figure}

\begin{figure}[ht]
\begin{center} 
\includegraphics[width=12cm]{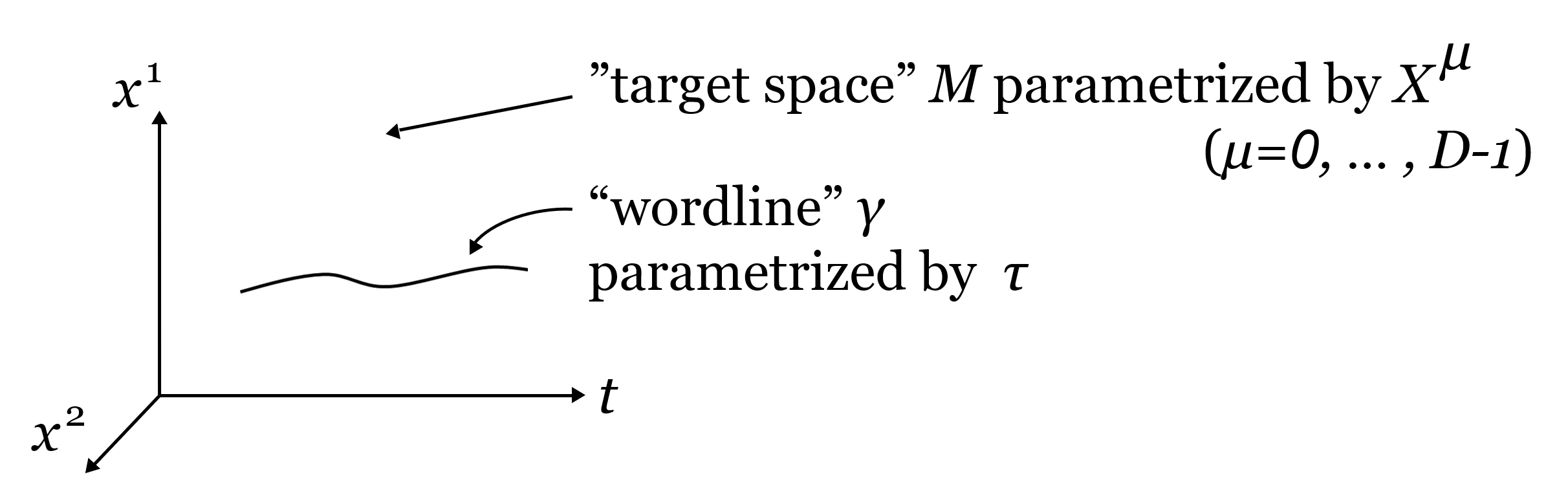}
\caption{Point particle through target space.}
\label{ppp} 
\end{center}
\end{figure}

But before discussing scattering, we will of course have to understand how a single string loop moves through space (in other words, how its {\bf worldsheet}\index{worldsheet} is embedded in {\bf target-space}\index{target-space}, more precisely, in target spacetime), see Fig.~\ref{its}. Before doing so, let us consider the more familiar case of a point particle, cf.~Fig.~\ref{ppp}. The embedding of the worldline $\gamma$ in target space is specified by the set of functions $X^\mu(\tau)$ and the natural action is 
\be
S=-m\int_\gamma ds\qquad\mbox{with}\qquad ds^2=-\eta_{\mu\nu}dX^\mu dX^\nu\qquad \mbox{and}\qquad dX^\mu=\dot{X}^\mu\,d\tau\,.
\ee
More explicitly, this action can be written as
\be
S=-m\int d\tau\,\sqrt{-\eta_{\mu\nu}\dot{X}^\mu\dot{X}^\nu}\,.
\ee
One can easily check that this is reparameterisation invariant under $\tau\to \tau'=\tau'(\tau)$ and that the non-relativistic limit is 
\be
S=\int dt\,\left(\frac{m}{2}\ol{v}^2-m\right)\,.
\ee
Much more could be said about this simple and familiar system (see e.g.~\cite{zw}), but for now this will suffice to motivate the Nambu-Goto action\index{Nambu-Goto action} for the string.

In complete analogy to the point particle, the {\bf Nambu-Goto action} for the bosonic string measures the surface area of the worldsheet embedded in target space:
\be
S_{NG}=-T\int_\Sigma df\,.
\ee
To write this more explicitly, one parametrises the worldsheet by (cf.~Fig.~\ref{its})
\be
\xi \equiv (\xi^0,\xi^1) \equiv (\tau,\sigma)\,.
\ee
The surface area is nothing but the volume of the 2d manifold, parameterised by $\xi$, measured with the induced metric $G_{ab}$. The latter is defined by
\be
ds^2=\eta_{\mu\nu}dX^\mu dX^\nu=\eta_{\mu\nu}\,\partial_aX^\mu\,\partial_b\,X^\nu d\xi^a\,d\xi^b  
\equiv G_{ab}\,d\xi^a d\xi^b\,.
\ee
Hence
\be
S_{NG}=-T\int_\Sigma d^2\xi\,\sqrt{-G}\qquad\mbox{with}\qquad G\equiv\mbox{det}(G_{ab})\,. 
\ee
The prefactor $T$ specifies the string tension.\index{string!tension} 

Due to the square root, the system is hard to quantise on the basis of this action. Instead, one uses the classically equivalent {\bf Polyakov action}\index{Polyakov action} \cite{Deser:1976rb, Brink:1976sc, Polyakov:1981rd}
\be
S_P=-\frac{T}{2}\int_\Sigma d^2 \xi\,\sqrt{-h}\,h^{ab}\partial_aX^\mu\,\partial_b X^\nu\,\eta_{\mu\nu}\,.
\ee
Here we introduced a new degree of freedom -- the worldsheet metric $h_{ab}$. To see the equivalence, one integrates out $h_{ab}$ by solving its equations of motion
\be
0=\delta_h\big[\sqrt{-h}h^{ab}G_{ab}\big]=-\frac{\delta h}{2\sqrt{-h}}h^{ab}G_{ab}+\sqrt{-h}\,\delta h^{ab}\,G_{ab}\,.
\ee
One proceeds by observing that, for a generic matrix $A$,
\be
\delta(\mbox{det}A)/(\mbox{det}A)=\delta\ln(\mbox{det}A)=\delta\,\mbox{tr}\,
\ln(A)
\ee
and hence
\be
\delta\,(\mbox{det}A)=(\mbox{det}A)\,\mbox{tr}(A^{-1}\,\delta A)= -(\mbox{det}A)\,\mbox{tr}(A\,\delta A^{-1})\,.
\ee
Applying this to $\delta h$, the equation of motion for $h_{ab}$ becomes
\be
0=\delta h^{ab}\,\left[\frac{h}{2\sqrt{-h}}h_{ab}\,h^{cd}G_{cd}+\sqrt{-h}G_{ab}
\right]
\ee
or, using the identity $h/\sqrt{-h}=-\sqrt{-h}$, 
\be
\frac{1}{2}h_{ab}\,h^{cd}G_{cd}=G_{ab}\,.
\ee
It is solved by $h_{ab}=\alpha G_{ab}$ for any $\alpha$. Inserting this in the Polyakov action,
\be
S_P=-\frac{T}{2}\int d^2\xi\,\sqrt{-h}\,h^{cd}G_{cd}=-\frac{T}{2}\int d^2\xi\,\sqrt{-\alpha^2G}\,2\alpha^{-1}=S_{NG}\,,
\ee
one obtains the Nambu-Goto action.

At this point, jumping somewhat ahead, we can sketch what will follow: The Polyakov action describes simply a 2d field theory of $D$ free scalars, living on a cylinder ($S^1\times\,\mbox{[Time]}$). This is a quantum mechanical system and its states have the interpretation of particles living in the  $D$-dimensional target spacetime. Consistency will require $D=26$, and the spectrum will contain a massless graviton and other massless (as well as many heavy) fields. However, it will also contain a particle with negative mass squared, a tachyon. Thus, the vacuum of the 26d gravitational field theory which this bosonic string describes is unstable. This instability problem will be cured if we move on to the superstring (based on a 2d supersymmetric worldsheet theory). The target spacetime will then have to be 10d and contact with the real world will be based on compactifying this 10d supergravity to 4d. The last step means considering geometries $M_6\times \mathbb{R}^4$, with $M_6$ a compact 6d manifold.

\subsection{Symmetries, equations of motion, gauge choice}

It is convenient to view the worldsheet theory as a 2d QFT with metric $h_{ab}$ and $D$ free scalars $X^\mu$:
\be
S_P=-\frac{T}{2}\int d^2\xi\,\sqrt{-h}\,(\partial X)^2\,\,,\qquad
(\partial X)^2=h^{ab}(\partial_a X^\mu)(\partial_b X^\nu)\eta_{\mu\nu}\,.
\ee
The three key symmetries of this theory are

\noindent
{\bf (1) Diffeomorphism:}$\,\,\xi^a\,\to\,\xi'^a(\xi^0,\xi^1).$\index{diffeomorphism}

\noindent
{\bf (2) Poincare symmetry:}$\,\,X^\mu\,\to\,X'^\mu=\Lambda^\mu{}_\nu X^\nu+V^\nu\,\,$with$\,\,\Lambda\in SO(1,D-1)$.

\noindent
{\bf (3) Weyl rescalings:}$\,\,h_{ab}(\xi)\,\,\to\,\,h'_{ab}(\xi)=h_{ab}(\xi) \exp[2\omega(\xi)]$, with $\omega$ an arbitrary real function.\index{Weyl!rescaling}

The first and second are obvious and follow immediately from the structure of our worldsheet action. It is noteworthy that target-space Poincare symmetry is an internal symmetry from the worldsheet perspective. The third is a specialty of the string. In other words, for a similar theory of moving $p$-branes, parameterised by $\xi^0,\xi^1,\cdots,\xi^p$, this symmetry does not exist unless $p=1$. 

To move on, it is convenient to use the\index{stress-energy tensor (see energy-momentum tensor)} energy-momentum tensor,\index{energy-momentum tensor}
\be
T^{MN}=\frac{2}{\sqrt{-g}}\cdot\frac{\delta S}{\delta g_{MN}}\qquad\mbox{or, equivalently}\qquad T_{MN}=\frac{-2}{\sqrt{-g}}\cdot\frac{\delta S}{\delta g^{MN}}\,,
\ee
which takes the form $T_{MN}=\mbox{diag}(\rho,p,\cdots,p)$ for an isotropic fluid. On the string worldsheet, a slightly different normalisation is common:
\be
T^{ab}=\frac{-4\pi}{\sqrt{-h}}\cdot\frac{\delta S_P}{\delta h_{ab}}\,.
\ee
One easily checks that 
\be
T^{ab}=-\frac{1}{\alpha'}\left(G^{ab}-\frac{1}{2}h^{ab}(h^{cd}G_{cd})\right)\,,
\ee
where we also introduced the {\bf Regge slope}\index{Regge slope}
\be
\alpha'\equiv \frac{1}{2\pi T}\,.
\ee
The latter is a different way to parameterise the string tension.\index{string!tension} It goes back to the early days of string theory, when the focus was on string theory as a model of hadronic physics. This is nicely explained in the first chapter of \cite{Green:1987sp}. 

It follows both from our discussion in the last section as well as from the general definition of $T^{ab}$ that the equation of motion of $h_{ab}$ is
\be
T^{ab}=0\,.
\ee
Moreover, tracelessness holds as an identity, i.e. independently of whether the field configuration obeys the equations of motion:
\be
T^a{}_a=0\qquad\mbox{for any}\qquad h_{ab}\,.
\ee
The reader should convince herself that this generally follows from symmetry (3). Finally, the equations of motion of $X$ are
\be
\Box X^\mu=0\qquad\mbox{with}\qquad \Box=D^a\partial_a\,.
\ee

It is crucial for what follows that {\bf diffeomorphisms} and {\bf Weyl rescalings} are (by definition) not just symmetries but {\bf gauge redundancies}. This allows one to work in the {\bf flat gauge},\index{flat gauge}
\be
h_{ab}=\mbox{diag}(-1,1)\,.\label{flme}
\ee
Indeed, very superficially one can argue as follows: A 2d metric contains 3 real functions. Diffeomorphisms and Weyl rescalings also contain $2+1=3$ real functions. Hence, it should be possible to bring $h_{ab}$ to any desired form. 

In somewhat more detail, one can explicitly check that
\be
\sqrt{-h'}{\cal R}[h']=\sqrt{-h}\,({\cal R}[h]-2\,\Box\,\omega)\qquad \mbox{for}\qquad h_{ab}'=e^{2\omega}h_{ab}\,.
\ee
Now, starting from any metric $h$, one may try to solve the equation $2\,\Box\,\omega={\cal R}$. This can always be achieved (in non-compact space with localised source ${\cal R}$) since it only requires the inversion of the Klein-Gordon operator. Without proof, we simply state that this holds also on the cylinder, which is our case of interest. For more details, see e.g.~\cite{Blumenhagen:2013fgp}. 

Once $2\,\Box\,\omega={\cal R}$ is solved, one can Weyl rescale $h$ using the solution $\omega$. The resulting metric will have vanishing Ricci scalar and, since in $d=2$
\be
{\cal R}_{abcd}=\frac{1}{2}(h_{ac}h_{bd}-h_{ad}h_{bc}){\cal R}\,,
\ee
it will be flat. More precisely, the worldsheet is a flat metric manifold and hence there exist coordinates in which the metric is manifestly flat in the sense of (\ref{flme}). 

\begin{figure}[ht]
\begin{center} 
\includegraphics[width=3.5cm]{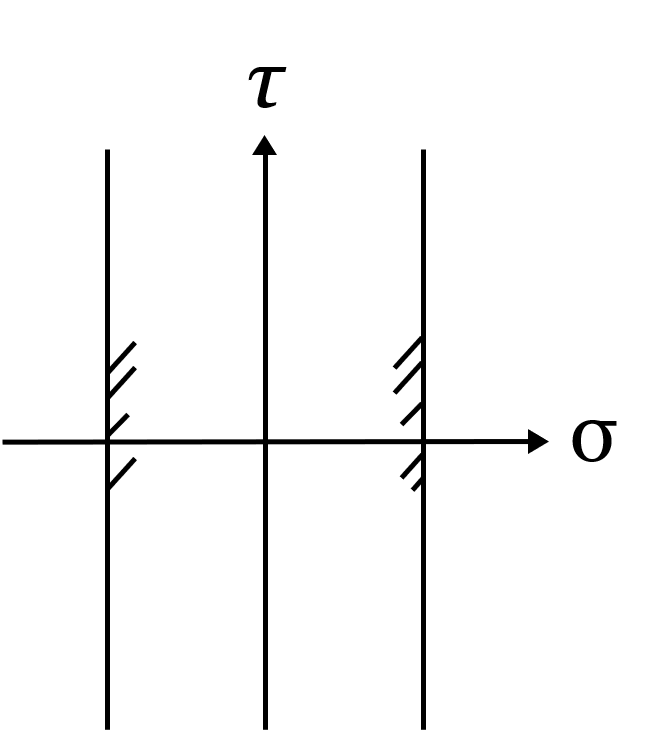}
\caption{The cylinder\index{cylinder}, on which the $X^\mu$ field theory lives, viewed as a strip with periodic boundary conditions.  In this picture, $X^\mu_L$ and $X^\mu_R$ correspond to left and right-moving waves.}
\label{strip} 
\end{center}
\end{figure}

Let us now focus on a flat worldsheet and on the corresponding equations of motion
\be
(\partial_\tau^2-\partial_\sigma^2)X^\mu=0\,.
\ee
It is convenient to use light-cone coordinates\index{light-cone coordinates} $\sigma^{\pm}=\tau\pm\sigma$, such that
\be
ds^2=-d\tau^2+d\sigma^2=-d\sigma^+d\sigma^-\qquad\mbox{and}\qquad
h_{++}=h_{--}=0\,\,,\qquad h_{+-}=h_{-+}=-\frac{1}{2}
\ee
and
\be
\Box=-4\partial_+\partial_-\qquad\mbox{with}\qquad \partial_\pm= \frac{\partial}{\partial \sigma^{\pm}}\,.
\ee
The equations of motion take the form
\be
\partial_-\partial_+X^\mu=0
\ee
and have the general solution
\be
X^\mu(\sigma^+,\sigma^-)=X^\mu_L(\sigma^+)+X_R^\mu(\sigma^-)\,,
\ee
being further constrained by $X^\mu(\tau,\sigma)=X^\mu(\tau,\sigma+\pi)$, cf.~Fig.~\ref{strip}. Here we have used the reparameterisation freedom to set the circumference of the cylinder to $\pi$. This is a convention used in many string theory texts, in particular in \cite{Green:1987sp} which we mostly follow.

Periodicity of $X^\mu$ implies periodicity of $\partial_+ X^\mu=\partial_+ X^\mu_L$ and of $\partial_- X^\mu=\partial_- X^\mu_R$. The latter depend only on $\sigma^+$ and $\sigma^-$ respectively and can therefore be represented as Fourier series in these two variables:
\be
\partial_+X_L^\mu\,\,\sim\,\, \mbox{const.}_L+\sum_{n\neq 0} f_{L,\,n} e^{-2in\sigma^+}\quad,\qquad
\partial_-X_R^\mu\,\,\sim\,\, \mbox{const.}_R+\sum_{n\neq 0} f_{R,\,n} e^{-2in\sigma^-}\,.
\ee
Returning to $X^\mu_L$ and $X^\mu_R$ by integration, the exponentials remain exponentials and the constants translate into linear terms. Moreover two integration constants appear. Hence, with a certain choice of prefactors, one finds the general solution or {\bf mode decomposition}\index{mode decomposition}
\bea
X_L^\mu&=&\frac{1}{2}x^\mu+\frac{l^2}{2}p^\mu\sigma^+ +\frac{il}{2}\sum_{n\neq 0}\frac{1}{n}\,\tilde{\alpha}_n^\mu\,e^{-2in\sigma^+}
\\
X_R^\mu&=&\frac{1}{2}x^\mu+\frac{l^2}{2}p^\mu\sigma^- +\frac{il}{2}\sum_{n\neq 0}\frac{1}{n}\,\alpha_n^\mu\,e^{-2in\sigma^-}\,.
\eea
Here we introduced $l=\sqrt{2\alpha'}$, the so-called string length\index{string!length}. One should be aware that the precise definition (the numerical prefactor) may vary from author to author and from context to context. 

The constants $x^\mu/2$ in the mode decomposition are chosen to be equal by convention. It is only their sum that has physical meaning, characterising the position of the center of mass of the string at worldsheet time $\tau=0$. Note that the coefficients of the two terms linear in $\sigma^+$ and $\sigma^-$ are forced to be equal by the periodicity of $X^\mu$. They describe how the position of the centre of mass changes as a function of $\tau$. It is hence natural to identify these coefficients, up to the proportionality factor $l^2/2$, with the target-space momentum $p^\mu$. One could easily convince oneself at the present, classical level of analysis that the proportionality factor has been chosen correctly for $p^\mu$ to be the standard momentum variable. But this will become clear anyway in a moment. Reality of $X^\mu$ implies that $x^\mu$ and $p^\mu$ are real, consistently with their physical meaning which we pointed out above. The oscillator modes\index{oscillator modes} have to satisfy
\be
(\alpha_n^\mu)^*=\alpha_{-n}^\mu\,.
\ee

\subsection{Open string}
\index{open string}

\begin{figure}[ht]
\begin{center} 
\includegraphics[width=6cm]{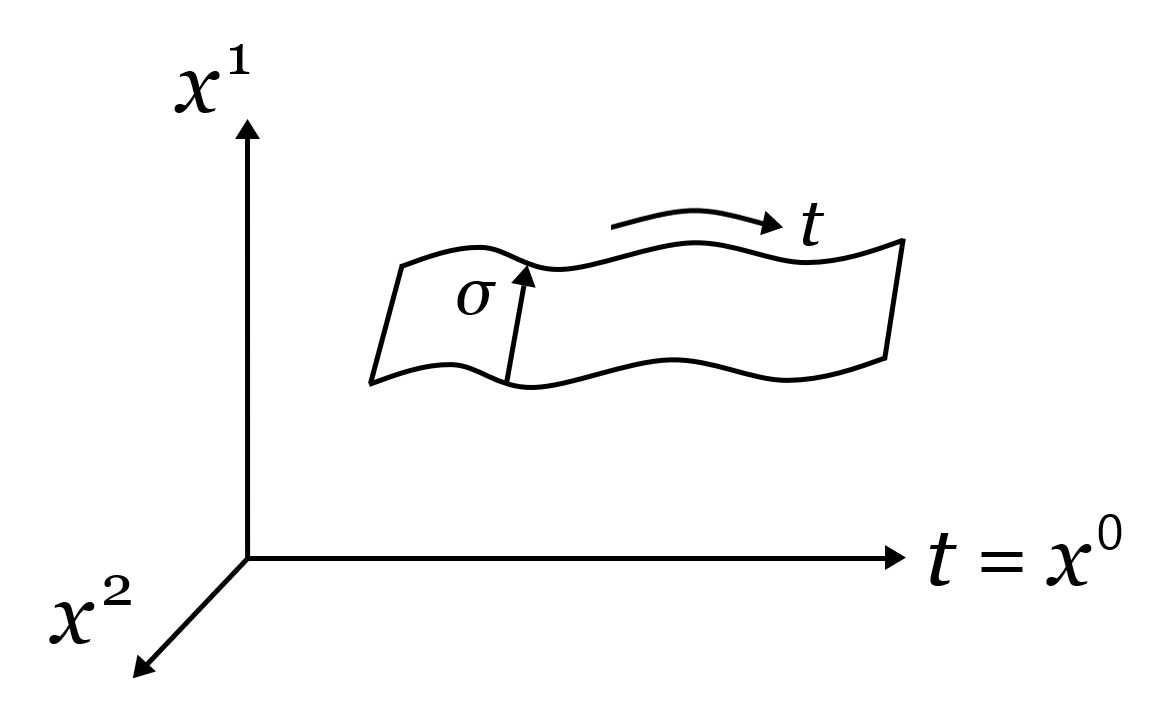}
\caption{Open string.}
\label{ops} 
\end{center}
\end{figure}

It will later on be crucial to also consider open strings. We introduce them already now since they are in fact a simpler version of the closed string\index{closed string} -- they basically carry half of the degrees of freedom. Instead of a cylinder, one now has to think of a strip (parameterised transversely by $\sigma\in (0,\pi)$) embedded in target space, cf.~Fig.~\ref{ops}. 

\begin{figure}[ht]
\begin{center} 
\includegraphics[width=8cm]{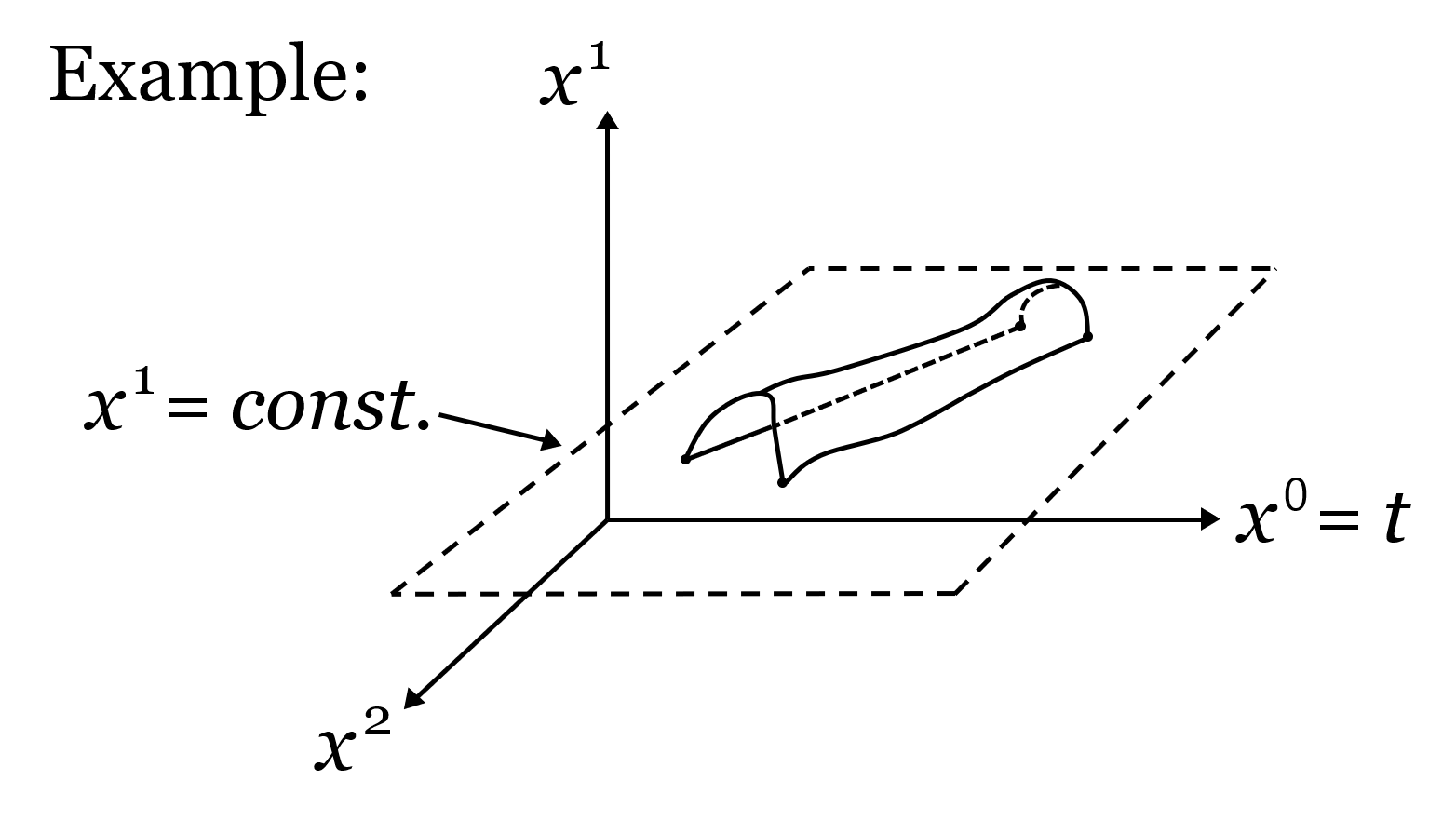}
\caption{Open string living on a D1-brane filling out the $X^2$ direction.}
\label{exo} 
\end{center}
\end{figure}

The variation of the action,
\be
\delta S=\frac{1}{2\pi\alpha'}\int d^2\sigma\,(\partial^2 X)\cdot\delta X-\frac{1}{2\pi\alpha'}\int d\tau\int_0^\pi d\sigma\,\partial_\sigma(\partial_\sigma X\cdot\delta X)\,,
\ee
now includes boundary terms. Indeed, while the first term vanishes if the equations of motion are obeyed, the second gives
\be
-\frac{1}{2\pi\alpha'}\int d\tau\,(\partial_\sigma X^\mu)\cdot \delta X_\mu \Big|^{\sigma=\pi}_{\sigma=0}\,.
\ee
To avoid introducing new degrees of freedom living at the boundary, we need that expression to vanish as well. This can be achieved by two different types of boundary conditions,\index{boundary conditions}
\be
\partial_\sigma X^\mu=0\quad (\mbox{Neumann})\,\,,\qquad\qquad \delta X_\mu=0\quad (\mbox{Dirichlet})\,.
\label{ndbc}
\ee
In the first case the string end moves freely (no momentum is lost at the end of the string), in the second it is confined to lie in a fixed hyperplane. For example (cf.~Fig.~\ref{exo}), one can enforce Neumann boundary conditions\index{Neumann boundary conditions} for $X^0$, $X^2$ and Dirichlet boundary conditions\index{Dirichlet boundary conditions} for $X^1$. One is then dealing with an open string living on a {\bf D-brane}\index{D-brane} (where D stands for Dirichlet) filling out the $X^0$ and $X^2$ directions of target spacetime. More generally, if a brane fills out $p$ spatial dimensions, i.e. if it is a $p$-dimensional object in the usual, spatial sense, one calls it a {\bf D$p$-brane}
(see \cite{Johnson:2000ch} for a dedicated textbook and \cite{Dai:1989ua, Leigh:1989jq} for two foundational original papers). For target space to be stationary, branes always have to fill out the time or $X^0$ direction. This, of course, does not contribute to their dimensionality as a spatial object. However, in spacetime a D$p$-brane is a $(p+1)$-dimensional object. Quite generally, an appropriate combination of Neumann and Dirichlet boundary conditions as given in (\ref{ndbc}) characterises open strings ending on such different D$p$-branes.

\begin{figure}[ht]
\begin{center} 
\includegraphics[width=9cm]{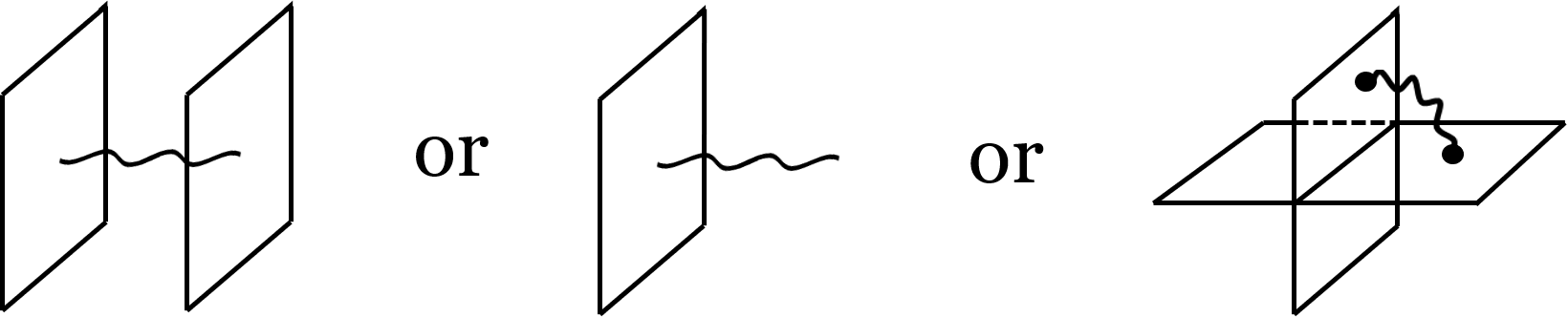}
\caption{Various brane configurations with strings attached.}
\label{brc} 
\end{center}
\end{figure}

We furthermore note that configurations with various, also intersecting branes are permitted, cf.~Fig.~\ref{brc}. Jumping ahead, we record that, analogously to the closed string states containing the target-space graviton, the open string states contain a massless vector particle: a $U(1)$ gauge boson. Thus, on every D$p$ brane one has a localised $(p+1)$-dimensional gauge theory. Moreover, one may have stacks of branes, for example $N$ D-branes filling out exactly the same hyperplane, i.e. lying on top of each other. On such a stack, there are $N^2$ distinct string states since each string can begin or end on any one of these $N$ coincident branes. This gives rise to a $U(N)$ gauge theory. If branes or brane stacks intersect, then the string living at the intersection (as in the last picture in Fig.~\ref{brc}) gives rise to states (target-space particles or fields) which are charged under the two gauge groups corresponding to the two branes. These states are confined to the intersection locus of the two branes or brane stacks. This is how Standard Model matter fields arise in some of the simplest phenomenologically interesting string models -- the so-called {\bf intersecting brane models}\index{intersecting brane models} (see \cite{Ibanez:2012zz} for a textbook, \cite{Blumenhagen:2005mu, Blumenhagen:2006ci} for reviews and \cite{Ibanez:2001nd, Blumenhagen:2001te, Cvetic:2001nr} for some of the original papers).

What is interesting for us at the moment is that the mode decomposition of the open string is simpler than that of the closed string. Indeed, while one needs sines and cosines (or equivalently exponentials) to Fourier decompose a periodic function, on an interval one can do with just sines or just cosines. Technically, one may say (and it is easy to demonstrate this explicitly) that, for the open string, the left and right-moving modes are identified: One arises from the other by reflection on the boundary. Explicitly, for the case of Neumann boundary conditions, one has the mode decomposition
\be
X^\mu=x^\mu+l^2p^\mu\tau+il\sum_{n\neq 0}\frac{1}{n}\alpha_n^\mu e^{-in\tau}\cos(n\sigma)\,.
\ee
Thus, it is often simpler to discuss the open string and then `double' the result to go over to the closed case. 

We also note that the case of Neumann boundary conditions for {\it all} $X^\mu$ should actually be viewed as a situation with spacetime filling branes. Thus, open strings {\it generally} end on D-branes.

\subsection{Quantisation}

We will only present the old covariant approach\index{old covariant approach}\index{old covariant quantisation}, briefly commenting on light-cone and modern covariant approach\index{modern covariant approach}\index{modern covariant quantisation} (also known as path integral or BRST quantisation\index{BRST quantisation}) at the end. The starting point is the flat-gauge Polyakov action which, breaking 2d covariance, can be written as
\be
S=\frac{1}{4\pi\alpha'}\int d^2\sigma (\dot{X}^2-X'^2)\qquad \mbox{with}\qquad 
d^2\sigma =d\tau d\sigma\,.
\ee
Here we have left the index $\mu$ and its contraction implicit. Nevertheless, the above describes $D$ free bosons and we have to keep in mind that one of them ($X^0$) has a wrong-sign kinetic term.

The canonical variables are 
\be
X^\mu \qquad \mbox{and} \qquad \Pi_\mu=\frac{\partial {\cal L}}{\partial \dot{X}^\mu}=\frac{1}{2\pi \alpha'}\,\dot{X}_\mu\,,
\ee
with equal-time commutation relations
\be
[\hat{\Pi}_\mu(\tau,\sigma),\,\hat{X}^\nu(\tau,\sigma')] = -i\delta(\sigma-\sigma')\delta_\mu{}^\nu\,\,,\qquad [\hat{X}^\mu,\hat{X}^\nu]=[\hat{\Pi}_\mu,\hat{\Pi}_\nu]=0\,.
\ee
Promoting our previous mode decomposition of $X^\mu$ (and a corresponding decomposition of $\Pi^\mu$) to the operator level, one finds
\be
[\hat{p}^\mu,\hat{x}^\nu]=-i\eta^{\mu\nu}\,\,,\qquad
[\hat{\alpha}_m^\mu,\hat{\alpha}_n^\nu]=m\,\delta_{m+n}\eta^{\mu\nu}\,\,,\qquad
[\hat{\tilde{\alpha}}_m^\mu,\hat{\tilde{\alpha}}_n^\nu]=m\,\delta_{m+n} \eta^{\mu\nu}\,,
\ee
where
\be
\delta_{m+n}\equiv \delta_{m+n\,,\,0}\,.
\ee
We will drop the hats from now on, assuming that it will always be clear from the context whether the operator or the classical variable is meant. The above commutators make it apparent that $p^\mu$ was correctly normalised to be the target space momentum of the string.

As usual in quantum mechanics, we now need a Hilbert space representation of our operator algebra. Given the non-trivial commutation relations of $p$ and $x$, we can only choose one of them to be diagonal. Since we are interested in a particle interpretation of string states, it is natural to choose $p$ and write
\be
{\cal H}=\sum_p{\cal H}(p)\,,
\ee
where ${\cal H}(p)$ is the eigenspace of the operators $\{\hat{p}^0,\cdots,\hat{p}^{D-1}\}$ with eigenvalues $\{p^0,\cdots,p^{D-1}\}\equiv p$.

We now focus on the subspace corresponding to one particular value of $p$ and rewrite the mode-algebra acting on it:
\be
[\alpha_m^\mu,\alpha_n^\nu]=m\,\delta_{m+n}\eta^{\mu\nu}\qquad\qquad
\to\qquad \qquad
[\alpha_m^\mu,\alpha_n^{\nu\,\dagger}]=|m|\delta_{m,n}\eta^{\mu\nu}\,.
\ee
We see that we are dealing simply with a very large set of oscillators, labelled by $\mu$ and $m>0$. We define a vacuum state and find the Fock space:\index{Fock space}
\be
{\cal H}(p)=\mbox{Span}\Big\{\,\alpha_m^\mu\alpha_n^\nu\cdots |0,p\rangle\,\,\Big| \,\, \mbox{any number of $\alpha$'s; any $\mu,\nu,\cdots$; any $m,n,\cdots <0$}\Big\}\,.
\ee

The situation we arrive at is very similar to the initial step of Gupta-Bleuler quantisation\index{Gupta-Bleuler quantisation} of electrodynamics: There, on account of the vector index of $A_\mu$ and the non-positive-definite metric $\eta_{\mu\nu}$, the Fock space includes negative norm states. They are removed by a physical state condition or constraint, related to the gauge invariance of the theory. Here, the same issue arises due to the vector index of $\alpha^\mu_m$. As will become clear momentarily, the resolution is similar to case of quantum electrodynamics (QED).

We fixed part of the gauge freedom by eliminating $h_{ab}$. The corresponding equation of motion was $T_{ab}=0$, which now has to be implemented as a constraint. It is convenient to do this in light-cone coordinates. One sees immediately that the vanishing trace condition takes a particularly simple form:
\be
T_a{}^a=0\qquad\Leftrightarrow \qquad T_{+-}=0\,.
\ee
Now, since in our theory the trace vanishes identically, one only needs to enforce the constraints
\be
T_{++}=T_{--}=0\,.
\ee
It is straightforward to check that
\be
T_{++}=(\partial_+ X_L)\cdot (\partial_+ X_L)\qquad \mbox{and}\qquad T_{--}=(\partial_- X_R)\cdot (\partial_- X_R)
\ee
and that, using the mode decomposition, the Fourier modes of the these quantities read
\bea
L_m &\equiv& \frac{1}{4\pi\alpha'}\int_0^\pi d\sigma\,T_{--} \, e^{-2im\sigma} = \frac{1}{2} \sum_{n=-\infty}^\infty \alpha_{m-n}\cdot\alpha_n\label{odl}
\\
\tilde{L}_m &\equiv & \frac{1}{4\pi\alpha'}\int_0^\pi d\sigma\,T_{++} \, e^{2im\sigma} = \frac{1}{2} \sum_{n=-\infty}^\infty \tilde{\alpha}_{m-n}\cdot\tilde{\alpha}_n\,.
\eea
Here we also used the simplifying notation
\be
\alpha_0^\mu=\tilde{\alpha}_0^\mu=\frac{l}{2}\,p^\mu\,.
\ee
For the open string, one defines 
\be
\tilde{L}_m \equiv  \frac{1}{2\pi\alpha'}\int_0^\pi d\sigma\,T_{++} \, e^{2im\sigma} = \frac{1}{2} \sum_{n=-\infty}^\infty \tilde{\alpha}_{m-n}\cdot\tilde{\alpha}_n\,,
\ee
with $T_{--}$ being a dependent quantity. It is easy to see that
\be
H=L_0+\tilde{L}_0\quad\mbox{(closed string)}\,\,\,,\qquad \qquad
H=L_0\quad\mbox{(open string)}\,.
\ee
Note that $\alpha_0^\mu=lp^\mu$ for the open string.

One can check that the operators $L_m$ (with or without tilde) satisfy the {\bf Virasoro algebra}\index{Virasoro algebra}
\be
[L_m,L_n]=(m-n)L_{m+n}+A(m)\delta_{m+n}\,\,,\qquad\mbox{with}\qquad 
A(m)=(m^3-m)D/12\,.\label{val}
\ee
Here the term proportional to $D$ is called the anomaly term and $D$ is the central charge. Note that this term depends on a possible additive redefinition of $L_0$, which is related to the ordering ambiguity present in all the terms of type $\alpha_{-k}\alpha_k$ in $L_0$. The form given above assumes normal ordering, i.e. $\langle 0,0|L_0|0,0\rangle=0$. 

\begin{figure}[ht]
\begin{center} 
\includegraphics[width=7cm]{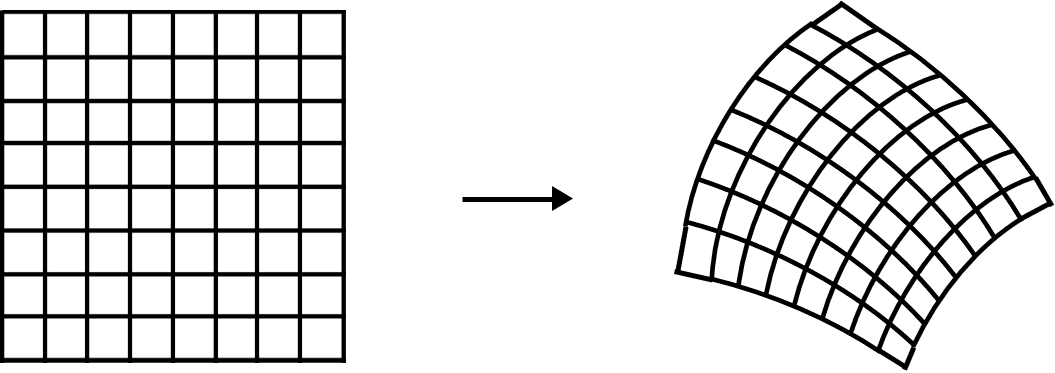}
\caption{Illustration of a conformal mapping of a given field configuration to a new one.}
\label{conf} 
\end{center}
\end{figure}

The classical part of this algebra, i.e. \eqref{val} without the anomaly term, is called {\bf Witt algebra}\index{Witt algebra}. It is satisfied by the differential operators 
\be 
D_m=ie^{im\theta}\frac{d}{d\theta}\,,
\ee
which generate diffeomorphisms on an $S^1$ parameterised by $\theta\in (0,2\pi)$. These remarks are the beginning of a long and important chapter of a proper string theory course -- 2d conformal field theory. However, we are not going to discuss this, such that a few comments will have to suffice:

When we fixed the gauge (diffeomorphisms and Weyl scalings), a residual gauge freedom was left. It consists of diffeomorphisms under which the metric changes only by Weyl scaling. Now it is useful to insist on the point of view that, after going the flat gauge, we are in a fixed-background QFT and coordinate reparameterisations are forbidden. From this perspective, the residual gauge freedom noted above corresponds to space-time dependent translations of the field configuration which preserve angles, i.e., conformal transformations (Fig.~\ref{conf}). Our theory is invariant under those and hence is a conformal field theory\index{conformal field theory} or CFT\index{CFT}~\cite{Belavin:1984vu,fms, Schottenloher:2008zz , Blumenhagen:2009zz, Ginsparg:1988ui, sche1}. The Virasoro algebra is the corresponding symmetry algebra. It is clear that conformal transformations can be generated as spacetime dependent translations. Given that $T_{ab}$ generates translations, we are not surprised to find that the Fourier modes of $T_{ab}$ are the desired symmetry generators. It is also natural that the Witt algebra, as introduced above, is the classical counterpart. 

The conformal symmetry just introduced is a central tool in developing string theory and, in particular, in deriving scattering amplitudes, loop corrections etc. We will have no time for this. But it may be useful to note that, when studying CFTs in their own right, the anomaly term or, equivalently, a non-zero central charge\index{central charge} do not represent a problem. However, in string theory the conformal symmetry is part of an underlying gauge symmetry and this term must vanish. It indeed does, in the so-called critical dimensions, but to see this one needs to do the gauge fixing more carefully, introducing Fadeev-Popov ghosts\index{Fadeev-Popov ghosts}. They cancel the central charge coming from the scalars.

Returning to our main line of development, we now want to be more explicit about the physical state condition. As in QED, it is sufficient to demand that the `annihilator part' of the constraint vanishes on physical states, i.e. $L_m|\,phys\,\rangle=0$ for $m\ge 0$. But it turns out that, at this point, a divergence present in the definition of $L_0$ has to be resolved. This has to do with operator ordering.

Indeed, our definition so far was 
\be
(L_0)_{tot} = \frac{1}{2}\sum_{n=-\infty}^\infty \alpha_{-n}\alpha_n=
\frac{1}{2}\alpha_0^2+\frac{1}{2}\sum_{n \neq 0} \alpha_{-n}\alpha_n\,.
\ee
We gave this quantity an index for `total' since we are going to separate the normal ordered part from it in a moment. We also note that the ordering of the creation and annihilation operators used above comes directly from the original definition 
\be
(L_0)_{tot}=H_{tot}=\frac{1}{4\pi\alpha'}\int_0^\pi d\sigma\,(\dot{X}^2+X'^2)\,.
\ee
Here, for simplicity, we think of the open string or, equivalently, just the right-moving part of the closed string.

To evaluate a constraint like $(L_0)_{tot}|\,phys\,\rangle=0$, we want to work instead with a normal-ordered operator. Hence, we define
\be
L_0\equiv \frac{1}{2}\sum_{n=-\infty}^\infty:\alpha_{-n}\alpha_n:=\frac{1}{2}\alpha_0^2 +\sum_{n>0}\alpha_{-n}\alpha_n\,.
\ee
Note that this {\bf supersedes} our previous definition in (\ref{odl}). The two definitions differ by a divergent normal ordering constant,
\be
(L_{0})_{tot}=L_0-a\qquad \mbox{with} \qquad a=-\frac{1}{2}(D-2) \sum_{n=1}^\infty n\,,\label{l0def}
\ee
following simply from $(1/2)(\alpha_n\alpha_n^\dagger+\alpha_n^\dagger \alpha_n)=\alpha_n^\dagger \alpha_n+n/2$\,. The prefactor $(D-2)$ counts the number of oscillators that contribute. The direct calculation gives, of course, $D$, but we have corrected this to $(D-2)$ on account of the wrong-sign scalar $X^0$. This is necessary since this wrong sign-scalar is associated with negative norm states, which are connected with the still unfixed (residual) gauge freedom. The latter corresponds to conformal transformations or reparametrisations of type $\sigma_+\to \sigma_+'=\sigma_+'(\sigma)$ and $\sigma_-\to \sigma_-'=\sigma_-'(\sigma)$, which preserve the flat gauge. One can fix this further gauge freedom by working in the so-called light-cone gauge. The corresponding quantisation procedure, called light-cone quantisation, manifestly gets rid of all oscillators except the $(D-2)$ transverse ones. We skip this important and useful chapter and ask the reader to trust that (\ref{l0def}) with the prefactor $(D-2)$ is the correct definition of $(L_0)_{tot}$.

Alternatively, one can use the Fadeev-Popov method and introduce ghosts, which will precisely cancel the two modes which we removed by hand. This is known as modern covariant quantisation, another method which we will not describe for reasons of time but which can be found in dedicated string theory textbooks.

The reader familiar with QED will immediately see that the above substitution of the naive prefactor $D$ with $(D-2)$ is analogous to the photon case: Of the formally four degrees of freedom associated with the vector $A_\mu$, only two transverse modes contribute to physical quantities like Casimir effect\index{Casimir effect} or vacuum free energy. This happens for exactly the same reason as here and to see it explicitly in a covariant QED calculation one also needs ghosts.

The simplest way to explicitly calculate the normal ordering constant $a$ is through $\zeta$ function regularisation:
\be
\left\{\sum_{n=1}^\infty n \right\}_{reg.}=\lim_{s\to-1}\left(\sum_{n=1}^\infty
n^{-s}\right)=\lim_{s\to -1}\zeta(s)=\zeta(-1)=-\frac{1}{12}\,.
\ee
This is of course quite formal and not very satisfying. Since the result is important, we want to spend some time to explain why the normal-ordering constant does in fact have a physical and a-priori finite definition. To see this, we give the infinite strip on which our 2d field theory lives a proper, physical width: $\pi\to \pi R$. Then we have 
\be
H_{tot}=\frac{1}{2R}\sum_{n=-\infty}^\infty\alpha_{-n}\alpha_n+\pi R\lambda\,.
\ee
Crucially, we have here also introduced a cosmological constant counterterm.

We are now dealing with a standard QFT problem -- the calculation of the total energy of a 2d theory on a strip of width $\pi R$. The sum over zero modes has a UV divergence, to be regularised by introducing a cutoff scale $\Lambda$. A cutoff dependence must also be assigned to the counterterm, $\lambda\to \lambda(\Lambda)$. Its form is determined by the requirement that the divergence for $\Lambda\to \infty$ cancels. Moreover, no finite ambiguity will arise since we know from Weyl invariance that the renormalised cosmological constant must vanish. Thus, we have
\be
H_{tot}=\frac{1}{R}\left( \frac{1}{2}\alpha_0^2+\sum_{n>0}\alpha_{-n}\alpha_n  \right)\,\,+\,\,\lim_{\Lambda\to\infty}\left[ \frac{D-2}{2R}\left\{\sum_{n-1}^\infty n \right\}_\Lambda +\pi R\lambda(\Lambda) \right]\,.
\ee
A very intuitive way of regularising this is to think in terms of physical modes with momenta $k_n=n/R$ and to multiply the contribution of each mode by $\exp(-k_n/\Lambda)$. It is then a straightforward exercise to do the summation, find the appropriate counterterm $\lambda(\Lambda)$, and to obtain the finite result
(cf.~Problem~\ref{noc})
\be
H_{tot}=\frac{1}{R}\left( \frac{1}{2}\alpha_0^2+\sum_{n>0}\alpha_{-n}\alpha_n  \right)\,\,-\,\,\frac{D-2}{24R}\,.
\ee
The physical interpretation of this finite correction is clear: This is a one-loop {\bf Casimir energy}\index{Casimir energy}, associated with the finite size of the space on which the QFT lives.\footnote{
Demonstrating 
that this result is in fact independent of the precise form of the regularising function, $\exp(-k/\Lambda)\to f(k/\Lambda)$, is not entirely trivial. See e.g.~\cite{Itzykson:1980rh} for a discussion of the corresponding 4d problem.
}

Returning to our stringy convention with $R=1$ and to the notation $L_0$ instead of $H$, we have
\be
(L_0)_{tot}=L_0-a\qquad \mbox{with}\qquad L_0=\frac{1}{2}\alpha_0^2+\sum_{n>0}\alpha_{-n}\alpha_n \qquad\mbox{and}\qquad
a=\frac{D-2}{24}\,.
\ee
The physical state condition hence reads 
\be
(L_m-a\delta_m)|\,phys\,\rangle=0\qquad\mbox{for}\qquad m\ge 0\,\,,
\ee
where it is crucial to remember that $L_0$ is, by definition, normal ordered.

\subsection{Explicit construction of physical states -- open string}
\index{physical states}

We start with the open-string worldsheet {\bf vacuum},
\be
|0,p\rangle\,\,,\qquad\qquad\mbox{defined by}\qquad\qquad
\hat{p}^\mu|0,p\rangle=p^\mu|0,p\rangle\,.
\ee
Our physical state conditions with $m>0$ are automatically satisfied for any $p$,
\be
L_m|0,p\rangle=\frac{1}{2}\sum_n\alpha_{m-n}\alpha_n|0,p\rangle=0\,,
\ee
since in each term of this sum either $n>0$ or $m-n>0$. Thus, there is always an annihilator involved, giving zero if applied to the vacuum.

By contrast, the $m=0$ condition is non-trivial, giving
\be
\left(\alpha' p^2+\sum_{n>0}\alpha_{-n}\alpha_n-a\right)|0,p\rangle=0\,,
\ee
where we used that $\alpha_0=lp=\sqrt{2\alpha'}\,p$. With $M^2=-p^2$, this translates into
\be
M^2=-a/\alpha'\,.
\ee
Thus, $p$ can not be an arbitrary vector. Rather, it must satisfy the above {\bf mass-shell condition}.\index{mass-shell condition}

Moving to the {\bf first excited level}\index{excited level}, we have to consider states
\be
\zeta_\mu\alpha^\mu_{-1}|0,p\rangle\,,
\ee
with a polarisation vector $\zeta$. The mass shell condition now reads
\be
0=(L_0-a)\zeta_\mu\alpha^\mu_{-1}|0,p\rangle = \left(\alpha' p^2 +\alpha_{-1}\cdot \alpha_1-a\right) \zeta_\mu\alpha^\mu_{-1}|0,p\rangle = \left(\alpha' p^2 +1-a\right)
\zeta_\mu\alpha^\mu_{-1}|0,p\rangle\,,
\ee
implying
\be
M^2=(1-a)/\alpha'\,.
\ee
Of the $L_m$ conditions with $m>0$, now the first also becomes non-trivial:
\be
0=L_1\,\zeta\cdot\alpha_{-1}|0,p\rangle=\left(\frac{1}{2}\sum_n\alpha_{1-n}\cdot\alpha_n\right)\,\zeta\cdot\alpha_{-1}|0,p\rangle\,.
\ee
Of the various terms in the sum, only those can contribute where $n\leq 1$ and $1-n\leq 1$. This occurs only for $n=0,1$, such that we find
\be
0=\frac{1}{2}(\alpha_1\cdot\alpha_0+\alpha_0\cdot\alpha_1)\,\zeta\cdot \alpha_{-1}|0,p\rangle=\zeta\cdot\alpha_0|0,p\rangle=\zeta\cdot p|0,p\rangle\,.
\ee
The implication is that the polarisation has to be transverse. We also need the norm of the state, which is
\be
\langle 0,p|(\zeta_\mu\alpha^\mu_{-1})^\dagger(\zeta_\nu\alpha^\nu_{-1})|0,p \rangle = \langle 0,p|0,p\rangle\zeta_\mu\zeta^\mu=\zeta^2\,.
\ee
Here we chose $\zeta_\mu$ real, using the freedom to redefine the $\alpha_{-1}^\mu$ if necessary.

At the so-called {\bf second excited level}, one has to analyse states of the form
\be
(\epsilon_{\mu\nu}\alpha_{-1}^\mu\alpha_{-1}^\nu+\epsilon_\mu\alpha_{-2}^\mu) |0,p\rangle\,,
\ee
but we will not do so. 

Instead, we summarise the results by focussing on the {\bf first excited level}:

\noindent
{\bf (A)} For $a>1$ we have $M^2<0$. This means that $p$ is spacelike and hence timelike vectors $\zeta$ with $\zeta\cdot p=0$ exist. Thus, there are allowed states with $\zeta^2<0$ and hence negative norm. This is excluded.

The deeper reason for the problem is that a Weyl anomaly arises, which can only be cured by considering a background which is not simply flat $D$-dimensional Minkowski space. This is known as {\bf supercritical string theory}\index{supercritical string} and may (in the supersymmetric case) nevertheless be relevant phenomenologically, although this is not established (see e.g.~\cite{Antoniadis:1988vi, Tseytlin:1991xk, Silverstein:2001xn, Hellerman:2006nx} and refs.~therein). 

\noindent
{\bf (B)} For $a=1$, we have $M^2=-p^2=0$. This implies that the $(D-1)$ independent $\zeta$'s which satisfy $\zeta\cdot p=0$ fall into two classes:

\noindent
First, there is one longitudinal polarisation, $\zeta||p$, corresponding to a zero-norm state.

\noindent
Second, there are $(D-2)$ transverse polarisations, which are spacelike and give rise to positive-norm states. One may think e.g. of $p=(1,1,0,\cdots,0)$ and $\zeta_i=(0,0,\cdots,0,1,0,\cdots,0)$, with unity at position $1+i$ and $i\in\{1,2,\cdots,(D-2)\}$. This is consistent with Gupta-Bleuler quantisation in QED. It gives the correct description of a gauge theory with a massless vector. This case is known as {\bf critical string theory}\index{critical string}. In the following, we will completely focus on this case and the corresponding {\bf critical dimension}\index{critical dimension} $D=26$.

{\bf (C)} For $a<1$ we have $M^2>0$ for the first excited (and all higher) levels. Thus except possibly for the vacuum state, this case is in practice not very interesting. It is not inconsistent at the present level of analysis (giving rise to a massive vector with $(D-1)$ positive norm states). Problems, possibly solvable, arise in the interacting theory. This is known as the {\bf subcritical string}\index{subcritical string}. The Weyl anomaly is also present, as in the supercritical case. Together, cases (A) and (C) are known as non-critical string theory. 

We close by mentioning that the overall picture in the critical case is just like in gauge theory quantisation: We have restricted our Fock space by imposing the physical state condition. The resulting space has no negative norm states, but so-called null states are still present. The actual positive-definite {\bf Hilbert} space is constructed as a quotient
\be
{\cal H}_0\equiv {\cal H}_{phys}/{\cal H}_{null}\,.
\ee
The mass-shell condition, originating from $(L_0-1)|\,phys\,\rangle=0$, can be written as 
\be
M^2=-p^2=(N-1)/\alpha'\qquad\mbox{with}\qquad N\equiv \sum_{n>0}\alpha_{-n}\alpha_n\,.
\ee
The operator $N$ or its expectation value is called the {\bf level}. We have found a {\bf tachyon}\index{tachyon} at level 0, a massless vector at level 1, and we could have found massive string excitations at level 2 and higher. The tachyon corresponds to the statement that our assumed 26d Minkowski vaccum is unstable since a scalar with negative mass squared is present. It will decay by tachyon condensation, which is an interesting subject of research. But we will not discuss this since we use the bosonic string only as a toy model to get ready for the superstring.

\subsection{Explicit construction of physical states -- closed string}

A repetition of the analysis of the previous section will again single out the case $a=1$ or $D=26$. We focus right away on this case, recalling however that the number of operators and constraints is now doubled. We rewrite
\be
(L_0-a)|\,phys\,\rangle=0\qquad,\qquad (\tilde{L}_0-a)|\,phys\,\rangle=0
\ee
as
\be
(L_0-\tilde{L}_0)|\,phys\,\rangle=0\qquad,\qquad (L_0+\tilde{L}_0-2a)|\,phys\,\rangle=0\,. 
\ee
We recall that
\be
L_0=\frac{\alpha_0^2}{2}+N=\frac{p^2}{4}\alpha'+N\,,
\ee
where we used that $\alpha_0=p\,l/2=p\sqrt{\alpha'/2}$ in the closed-string case. Analogous equations hold for the left-movers. With this, the physical state conditions become
\be
(N-\tilde{N})|\,phys\,\rangle=0\qquad\mbox{and}\qquad (p^2\alpha'/2+N+\tilde{N}-2)|\,phys\,\rangle=0 \,,
\ee
known as {\bf level matching}\index{level matching} and {\bf mass shell conditions} respectively. The latter is also frequently given as
\be
M^2=2(N+\tilde{N}-2)/\alpha'\,.
\ee

Now one proceeds systematically, level by level, as before. At the {\bf vacuum} level one again finds a tachyon,
\be
|0,p\rangle\,\,,\qquad M^2=-4/\alpha'\,.
\ee

At the first excited level, due to the level matching condition, both $\alpha_{-1}$ and $\tilde{\alpha}_{-1}$ have to be used:
\be
\xi_{\mu\nu}\alpha_{-1}^\mu\tilde{\alpha}_{-1}^\nu|0,p\rangle\,\,,\qquad M^2=0\,.
\ee
Note that, as before, one really has $M^2=2(1+1-2a)/\alpha'$, such that masslessness follows only for $a=1$, i.e. in the critical dimension. At the first excited level, the $L_1$ and $\tilde{L}_1$ constraints are non-trivial. They read
\be
\xi_{\mu\nu}p^\mu=0\qquad\mbox{and}\qquad \xi_{\mu\nu}p^\nu=0\,.
\ee
It is also easy to check that the norm of our states is
\be
\langle \,phys\,|\,phys\,\rangle\sim \xi_{\mu\nu}\xi^{\mu\nu}\,,
\ee
which is always non-negative if the physical state conditions are satisfied. 

To classify the states, it is helpful to think of the polarisation tensor literally as of an element in the tensor product of two copies of ${\mathbb R}^D$,
\be
\xi^{\mu\nu}=\sum_{ab}v_{(a)}^\mu\otimes v_{(b)}^\nu\,.
\label{xitp}
\ee
In analogy to the standard treatment of the photon, one chooses a basis $v_{(a)}$ with one element $v_{(0)}\sim p$, one lightlike element $v_{(1)}$ with non-zero product with $p$, and $D-2$ spacelike elements orthogonal to $v_{(0)}$ and $v_{(1)}$. Of these, only the spacelike vector and $v_{(0)}$ are allowed to appear in~(\ref{xitp}). Hence we have $(D-1)^2$ physical basis states. Furthermore, $(D-2)^2$ of them (those built from spacelike vectors only) have positive norm. The rest corresponds to gauge freedom. 

Choosing $p\sim v_{(0)}\sim (1,1,0,\cdots,0)$ and $v_{(1)}\sim (1,-1,0,\cdots,0)$, we see explicitly how products of the $(D-2)^2$ transverse vectors form a basis for the transverse polarisations $\xi_t$. They correspond to the lower-right corner of the matrix $\xi$:
\be
\xi=\left(
\begin{array}{cc}
\begin{array}{cc}0&\,\,0\\0&\,\,0\end{array}
&
\begin{array}{ccc}0&\cdots&0\\0&\cdots&0\end{array}
\\
\begin{array}{cc}0&\,\,0\\ \cdot&\,\,\cdot\\0&\,\,0\end{array}
&
\xi_t
\end{array}
\right)\,.
\ee
The transverse physical polarisations $\xi_t$ transform under $SO(D-2)$, the group of rotations in the spacelike hyperplane transverse to $p$. This is called `little group'\index{little group} -- the subgroup of $SO(1,D-1)$ leaving $p$ invariant. Our rank-2 tensor representation of physical polarisations is not irreducible but decomposes into symmetric, antisymmetric and trace part. These 3 representations correspond to 3 different fields of the $D$-dimensional field theory which the string describes from the target space perspective. They are:

\noindent
(1) The graviton\index{graviton} $G_{\mu\nu}$, with $(D-1)(D-2)/2-1$ d.o.f.s (note that for $D=4$ this correctly reproduces the known result of 2 d.o.f.s).

\noindent
(2) The Kalb-Ramond field\index{Kalb-Ramond field} or antisymmetric tensor $B_{\mu\nu}$, with $(D-2)(D-3)/2$ d.o.f.s.

\noindent
(3) The dilaton\index{dilaton} $\phi$, with 1 d.o.f.

We could go on to discuss excited states, but all we will need to know is that there they form a tower with increasing mass and that the number of states at each consecutive level grows extremely fast. The mass spacing is $\Delta M^2=4/\alpha'$.

\subsection{The 26d action}

We are only interested in the critical case, $D=26$, and we focus on the closed string (for more details see e.g.~\cite{Polchinski:1998rq}). It is immediate to write down a quadratic-level action for the above fields (to be supplemented by the tachyon which, as we know, has negative mass squared and makes the 26d Minkowski-space solution unstable). Assuming that one also knows how to compute scattering amplitudes, one can supplement this action by interaction vertices and write down the full, non-linear expression at the 2-derivative level. It reads (suppressing the tachyon):
\be
S=\frac{1}{\kappa^2}\int d^{26}x\sqrt{-G}e^{-2\phi}\left[{\cal R}[G]-\frac{1}{12}H_{\mu\nu\rho}H^{\mu\nu\rho}+4(\partial\phi)^2\right]\,.
\label{10da}
\ee
where
\be
H=dB\,,
\ee
in complete analogy with $F=dA$ in the 1-form case.

Many important comments have to be made. First, it is apparent that the value of $\kappa^2$ can be changed by a shift of $\phi$. Thus, we can for example define $\kappa^2=c\alpha'^{12}$, with some numerical constant $c$. Then the choice of the background value of $\phi$ determines the 26d Planck mass relative to the mass of the first excited string modes, which is $2/\sqrt{\alpha'}$. 
It also governs the perturbativity of the theory, i.e. the importance of string loops, as we will discuss further down.

Second, the apparently wrong sign of the dilaton-kinetic term is misleading. Indeed, the above is called the string-frame\index{string!frame} action (similar to what is known as the Brans-Dicke frame\index{Brans-Dicke frame} action in the non-stringy gravitational literature). One can go to the Einstein frame\index{Einstein frame} by the Weyl rescaling\index{Weyl!rescaling}
\be
G_{\mu\nu}=\tilde{G}_{\mu\nu}\,e^{-\phi/6}\,.
\ee
The result is
\be
S=\frac{1}{\kappa^2}\int d^{26}x\sqrt{-\tilde{G}}\left[{\cal R}[\tilde{G}]-\frac{1}{12}e^{-\phi/3}\,H_{\mu\nu\rho}H^{\mu\nu\rho}-
\frac{1}{6}(\partial\phi)^2\right]\,.
\ee
In this frame, the Planck mass is manifestly fixed and the mass of the excited states changes with varying dilaton background. 

Third, this is the first (but not the last) time we encounter a higher-form gauge theory\index{higher-form gauge theory}. So it may be useful to remind the reader of some of the relevant basic notions. (A standard summary of conventions can be found in the Appendix of Volume II of~\cite{Polchinski:1998rq}.) It is convenient not to think just of an antisymmetric field or gauge potential $A_{\mu_1\cdots\mu_p}$ but of a differential form,\index{differential form}
\be
A_p=\frac{1}{p!}A_{\mu_1\cdots\mu_p}dx^{\mu_1}\wedge \cdots\wedge dx^{\mu_p}\,.
\ee
Our present case $p=2$ with $B_{\mu_1\mu_2}=A_{\mu_1\mu_2}$ is part of the more general structure of such gauge theories.

One should think of the $dx^\mu$ as basis vectors of the dual tangent space (the cotangent space) of a manifold, such that
\be
dx^\mu\left(\frac{\partial}{\partial x^\nu}\right)=\delta^\mu_\nu\,.
\ee
Higher $p$-forms\index{$p$-forms} take their values in the $p$-fold exterior product (the antisymmetric part of the tensor product) of the cotangent space. This is symbolised by the wedge, e.g.
\be
dx^1\wedge dx^2=dx^1\otimes dx^2-dx^2\otimes dx^1\,.
\ee
It generalises to 
\be
dx^1\wedge \cdots \wedge dx^p=p!\, dx^{[1}\otimes\cdots\otimes dx^{p]}\,,
\ee
where $[\cdots]$ stands for antisymmetrisation. The implication is that, for example,
\be
A_p(\partial_1,\cdots,\partial_p)=A_{[1\cdots p]}=A_{1\cdots p}\,.
\ee
Consistently with the above, one formally defines the product of two forms
\be
(A_p\wedge B_q)_{\mu_1\cdots\mu_{p+q}}=\frac{(p+q)!}{p!q!}A_{[\mu_1\cdots \mu_p} B_{\mu_{p+1}\cdots \mu_{p+q}]}\,.
\ee
Crucially, the natural map from functions (0-forms) to 1-forms,
\be
d:\,f\,\,\mapsto\,\,df=\partial_\mu f\,dx^\mu\qquad \mbox{with}\qquad 
df(\partial_\mu)\equiv \partial_\mu f\,,
\ee
has a generalisation to higher forms:
\be
(dA_p)_{\mu_1\cdots\mu_{p+1}}=(p+1)\partial_{[\mu_1}A_{\mu_2\cdots\mu_{p+1}]} \,.
\ee
This so-called {\it exterior derivative}\index{exterior derivative} is a central mathematical concept. It is immediate to convince oneself that $d\circ d =d^2=0$.

By its very definition, a $p$-form provides, at every point, a totally antisymmetric map from the $p$-fold tensor product of the tangent space to the real numbers. Thus, it can be used to define the volume of an infinitesimal parallelepiped (with orientation, i.e. ordering of the vectors by which it is spanned) at any point of the manifold. This gives rise to the possibility of integrating a $p$-form over a finite $p$-dimensional submanifold:
\be
V(C_p)=\int_{C_p}A_p\,.
\ee

After these preliminaries, it is clear how to interpret $A_p$ as a physical gauge potential. First, one has the gauge transformation and the gauge-invariant field strength
\be
A_p\,\,\to\,\,A_p+d\chi_{p-1} \qquad \mbox{and} \qquad F_{p+1}=dA_p\,.
\ee
The natural lagrangian is $\sim\,|F_{p+1}|^2\equiv F_{\mu_1\cdots\mu_{p+1}} F^{\mu_1\cdots\mu_{p+1}}$ and the natural coupling to charged objects is 
\be
S_{matter}\sim \int_{\Sigma_p} A_p\,.
\ee
This is completely analogous to electrodynamics, where the matter coupling is the integral of $A_1$ along the worldline of the electron. Here, it is the integral of $A_p$ along the $p$-dimensional worldvolume $\Sigma_p$ of a $(p-1)$-brane.\index{brane} (Recall the convention that the variable $p$ in the term $Dp$-brane counts only the spatial dimensions.)

In the case at hand, a charged object suitable as a source for $B_2$ is already present in the theory we have so far developed: It is the fundamental string itself. Thus, the term
\be
\int_{\Sigma_2}B_2
\ee
has to be added both to our 10d action and to our worldsheet action for the string. If $B_2$ is non-zero, this changes our 2d theory and its quantisation. 

Similarly, we see that (\ref{10da}), with $G_{\mu\nu}=\eta_{\mu\nu}$, $\phi=0$ and $B=0$ describes the solution in the background of which our fundamental string, introduced earlier, propagates. This is the 2d theory one is easily able to quantise. But clearly other solutions for this 10d action exist and the string can be quantised in their background as well. The 2d theory is then much more complicated, e.g. through
\be
\int d^2\sigma \sqrt{-h}h^{ab}(\partial_a X^\mu)(\partial_b X^\nu)\eta_{\mu\nu} \qquad\to \qquad \int d^2\sigma \sqrt{-h}h^{ab} (\partial_a X^\mu)(\partial_b X^\nu)G_{\mu\nu}(X)\,.
\ee
We see that this theory now ceases to be free or quadratic in the fields. For example, if near $X=0$ we can write 
\be
G_{\mu\nu}=\eta_{\mu\nu}+c\cdot(X^1)^2\eta_{\mu\nu}+\cdots\,,
\ee
we encounter a quartic interaction vertex in the worldsheet theory. Similarly, a non-zero $B_2=B_2(X)$ adds new terms to the worldsheet lagrangian. In particular, the $X$ dependence of $B_2$ leads to new interaction terms in the theory of scalars $X^\mu$ living on the worldsheet.

Before closing this section, we should discuss the role of the dilaton\index{dilaton} $\phi$ from the worldsheet perspective. This field is related to the Einstein-Hilbert term,
\be
\int d^2\sigma \sqrt{-h}\,{\cal R}\,,
\ee
of the worldsheet action. At first sight, this term is clearly allowed. It  respects all symmetries of the worldsheet, including in particular Weyl invariance. A more careful analysis reveals, however, that it can be written as a total derivative and hence does not affect the equations of motion. Indeed, following the standard derivation of Einstein's equations from the Einstein-Hilbert action, one finds
\be
\delta_h \int d^2\sigma \sqrt{-h}\,{\cal R} = \int d^2\sigma\,\left({\cal R}_{ab}-\frac{1}{2}h_{ab}{\cal R}\right)\,\delta h^{ab}\,\,+\,\,\mbox{boundary terms}\,.
\ee
But in $d=2$ one has 
\be
{\cal R}_{ab}-\frac{1}{2}h_{ab}{\cal R}=0
\ee
as an identity. This follows from the symmetries of the Riemann tensor which, as already noted above, can be expressed in terms of the Ricci scalar. Thus, the worldsheet Einstein-Hilbert term does not change under continuous deformations of the worldsheet metric. Its integral can however be non-zero, measuring topological features of the worldsheet (see below). 

Now, comparing the dynamics described by the target-space action given above and the role of the Einstein-Hilbert term on the worldsheet (we will see more details of this further down), one can establish that $\phi$ has to be identified with the coefficient of the worldsheet Einstein-Hilbert term:
\be
S_P\supset \frac{1}{4\pi}\int d^2\sigma\,\sqrt{-h}\,{\cal R}\,\phi\,.
\ee
As before, if $\phi=\phi(X)$ is non-constant, new interactions are introduced into the worldsheet theory. 

As a final remark, we note that backgrounds solving the 26d equations of motion are precisely those in which the Weyl invariance on the propagating string worldsheet remains unbroken. In this sense, the 2d theory can be used to directly determine the 26d dynamics, without calculating scattering amplitudes and comparing them to 26d EFT vertices.

\subsection{Problems}

\subsubsection{Point particle action}
\index{point particle}

{\bf Task:} Guess the `Polyakov action' for the point particle and derive the `Nambu-Goto action' given in the lecture. Determine the parameters to achieve consistency which what you know from your course in special relativity.

\noindent
{\bf Hints:} Introduce a worldline metric $h_{\tau\tau} \equiv h$, such that $ds^2= h_{\tau\tau}d\tau^2$. Allow a worldline cosmological constant term (which is forbidden in the string case by Weyl invariance, but permitted for the point particle).

\noindent
{\bf Solution:} The natural guess is
\be
S_P=c\int d\tau \sqrt{h}(h^{-1}\dot{X}^2-\lambda)\,, 
\ee
from which one derives the equations of motion for $h$:
\be
0=\frac{1}{2\sqrt{h}}(h^{-1}\dot{X}^2-\lambda)+\sqrt{h}(-h^{-2}\dot{X}^2)
= -\frac{1}{2}(h^{-3/2}\dot{X}^2+h^{-1/2}\lambda)\,.
\ee
It follows that 
\be
h=-\dot{X}^2/\lambda
\ee
and hence
\be
S_P=c\int d\tau\sqrt{h}(-h^{-1}h\lambda-\lambda)=-2c\lambda\int d\tau\sqrt{h}
=-2c\sqrt{\lambda}\int d\tau\,\sqrt{-\dot{X}^2}\,.
\ee
This reproduces the standard (`Nambu-Goto-type') relativistic point particle action for, e.g., 
$c=m/2$ and $\lambda=1$. Thus, the final result is
\be
S_P=\frac{m}{2}\int d\tau\sqrt{h_{\tau\tau}}\,\left(h_{\tau\tau}^{-1} \dot{X}^\mu \dot{X}_\mu -1\right)\,.
\ee

\subsubsection{Commutation relations of oscillator modes}
\index{oscillator modes}

{\bf Task:} Demonstrate the consistency of the commutation relations of $p^\mu$, $x^\mu$, $\alpha_n^\mu$, $\tilde{\alpha}_n^\mu$ with those of the $X^\mu$s and $\Pi^\mu$s at equal time.

\noindent
{\bf Hint:} It is efficient to first calculate the commutator of $X^\mu$ with $\Pi^\mu$, of $X^\mu$ with itself etc.~using the mode expansion and then apply a Fourier transformation to both sides.

\noindent
{\bf Solution:} Collecting formulae from the lecture notes we have
\bea
X^\mu(\tau,\sigma) &=& x^\mu+l^2p^\mu\tau+\frac{il}{2}\sum_{n\neq 0}\frac{1}{n}\left[ \tilde{\alpha}_n^\mu e^{-2 in\sigma^+} +\alpha _n^\mu e^{-2 in\sigma^-}\right] \,,
\\
\Pi^\nu(\tau,\sigma') &=& \frac{1}{2\pi\alpha'}\left\{l^2p^\nu + l \sum_{n\neq 0} \left[ \tilde{\alpha}_n^\nu e^{-2 in\sigma'^+} +\alpha_n^\nu e^{-2 in\sigma'^-}\right]\right\}\,.
\eea
When writing the commutator, we may right away focus on those pairings of terms from the mode expansion which have a chance of being non-zero:
\bea
\hspace*{-.3cm}[\Pi^\nu(\tau,\sigma'),X^\mu(\tau,\sigma)] &=& -\frac{l^2}{2\pi\alpha'}i\eta^{\mu\nu}-\frac{il^2}{4\pi \alpha'}\sum_{n\neq 0}\eta^{\mu\nu}\left[e^{-2 i n(\sigma^+ - \sigma'^+)} + e^{-2 i n(\sigma^- - \sigma'^-)} \right]
\nonumber
\\
&=& -\frac{i}{\pi}\eta^{\mu\nu} - \frac{i}{2\pi}\eta^{\mu\nu}\sum_{n\neq 0}\left[ e^{-2 i n(\sigma - \sigma')} + e^{2 i n(\sigma - \sigma')}
\right]\,.
\nonumber
\\
&=& -\frac{i}{\pi}\eta^{\mu\nu} - \frac{i}{\pi}\eta^{\mu\nu}\sum_{n\neq 0} e^{-2 i n(\sigma - \sigma')} \,\,=\,\,-\frac{i}{\pi}\eta^{\mu\nu}\!\!\sum_{n=-\infty}^{\infty}e^{-2 i n(\sigma - \sigma')}\,.
\eea
To get the sign right, it is crucial to note that $[\alpha_m^\mu,\alpha_n^\nu]/n=
m\delta_{m+n}\eta^{\mu\nu}/n=-\delta_{m+n}\eta^{\mu\nu}$. One may finish here by recognising the $\delta$ function in $\sigma-\sigma'$ on the r.h.~side.

But let us be fully explicit by finally applying a Fourier transformation in $\sigma'$ and $\sigma$ to both sides of our result. Using also the canonical commutation relations, the l.h.~side gives
\be
\int\limits_{0}^{\pi}d\sigma'\, e^{2 im\sigma'}
\int\limits_{0}^{\pi}d\sigma\, e^{2 ik\sigma}\big( -i\eta^{\mu\nu}\delta(\sigma-\sigma')\big)= -i\eta^{\mu\nu} 
\int\limits_{0}^{\pi}d\sigma\, e^{2 i(m+k)\sigma}
=-i\pi\eta^{\mu\nu}\delta_{m+k}\,.
\ee
Analogously, on the r.h.~side one finds
\be
-\frac{i}{\pi}\eta^{\mu\nu}\sum_n (\pi\delta_{n-k})(\pi\delta_{m+n}) = 
-i\pi \eta^{\mu\nu}\delta_{m+k}\,.
\ee
Thus, both sides agree.

The commutators $[X^\mu(\tau,\sigma),X^\nu(\tau,\sigma')]$ (and similarly for $\Pi^\mu$) vanish since the relevant sums contain explicit factors of the mode indices $n$. For example, dropping all prefactors and the manifestly vanishing $x^\mu/p^\mu$ contribution, one encounters expressions like
\be
\sum_{n\neq 0} \frac{1}{n^2} n \left(e^{-2in(\sigma-\sigma')}
+ e^{2in(\sigma-\sigma')}\right)\,.
\ee
But this is zero by antisymmetry in $n$.

\subsubsection{Trace of the energy-momentum tensor}
\index{energy-momentum tensor}\index{stress-energy tensor (see energy-momentum tensor)}

{\bf Task:} Use a symmetry argument to show that the trace of the energy-momentum tensor of the string vanishes identically (no hint needed).

\noindent
{\bf Solution:} By Weyl invariance,
\be
0=S_P[h_{ab}+\epsilon h_{ab}]-S_P[h_{ab}]\simeq \epsilon h_{ab}\frac{\delta S_P}{\delta h_{ab}}=\epsilon h_{ab}\,\left(-\frac{\sqrt{-h}}{4\pi}T^{ab} \right)=\epsilon \left(-\frac{\sqrt{-h}}{4\pi}\right)\,T^a{}_a\,.
\ee
Hence, $T^a{}_a=0$.

\subsubsection{Virasoro algebra}
\index{Virasoro algebra}

{\bf Task:} Derive the classical part of the Virasoro algebra using the mode expansion of the generators and the canonical commutation relations (or equivalently Poisson brackets) of the oscillator modes. Then also derive the anomaly under the assumption that the operator-ordering ambiguity in $L_0$ is resolved by normal ordering, i.e. that $\langle 0,0|L_0|0,0\rangle=0$.

\noindent
{\bf Hints:} For the first part, use the derivation or Leibniz rule for commutators: $[A,BC]=[A,B]C+B[A,C]$. For the second part, argue that only expressions with $L_0$ on the r.h. side are affected by the ordering ambiguity. Thus, the anomaly must take the form $A(m)\delta_{m+n}$. Then evaluate the commutator $[L_1,[L_m,L_{-m-1}]]$ directly and with the Jacobi identity (in derivation form). Derive from this a recursive formula for the $A(m)$. Show that the expression $A(m)=am^3+bm$ satisfies this relation and hence determines $A(m)$ unambiguously up to the coefficients $a$ and $b$. Fix $a,b$ by evaluating $[L_m,L_n]$ with $(m,n)$ being $(1,-1)$ and $(2,-2)$ in the zero-momentum vacuum $|0,0\rangle$. 

A very similar derivation can be found in \cite{Green:1987sp}, but try to succeed on your own before consulting that book or our text below.

\noindent
{\bf Solution:} We focus on $D=1$ and find from the known mode expansion
\be
[L_m,L_n]=\frac{1}{4}\sum_{k,l}[\alpha_{m-k}\alpha_k\,,\,\alpha_{n-l}\alpha_l]\,.
\ee
Applying the derivation or Leibniz rule for commutators once gives
\be
[L_m,L_n]=\frac{1}{4}\sum_{k,l}\left\{
[\alpha_{m-k}\alpha_k\,,\,\alpha_{n-l}]\,\alpha_l\,+\,
\alpha_{n-l}\,[\alpha_{m-k}\alpha_k\,,\,\alpha_l]
\right\}\,.
\ee
The second application together with the standard commutation relations gives
\bea
[L_m,L_n]\!\!\!&=&\!\!\!\frac{1}{4}\sum_{k,l}\left\{
\alpha_{m-k}[\alpha_k,\alpha_{n-l}]\alpha_l
+
[\alpha_{m-k},\alpha_{n-l}]\alpha_k \alpha_l
+
\alpha_{n-l}\alpha_{m-k}[\alpha_k,\alpha_l]
+
\alpha_{n-l}[\alpha_{m-k},\alpha_l]\alpha_k
\right\}
\nonumber \\
&=&\!\!\frac{1}{4}\sum_k\{
k\alpha_{m-k}\alpha_{n+k}+(m-k)\alpha_k\alpha_{m+n-k}+k\alpha_{n+k}\alpha_{m-k}
+(m-k)\alpha_{m+n-k}\alpha_k
\}\,.
\eea
Now let us shift the summation index according to $k\,\to\,k-n$ in the first and third term. If we were in addition allowed to change the order of the $\alpha$'s in the second and third terms, we would obtain
\be
[L_m,L_n]=\frac{1}{2}\sum_k\{
(k-n)\alpha_{m+n-k}\alpha_k\,+\,(m-k)\alpha_{m+n-k}\alpha_k
\}
= \frac{m-n}{2} \sum_k\alpha_{m+n-k}\alpha_k = (m-n)L_{m+n}\,.
\ee
It is clear that this operation is illegal precisely in situations with an ordering ambiguity on the r.h.~side, i.e. for $m+n=0$. Hence, we have shown that
\be
[L_m,L_n]=(m-n)L_{m+n}+A(m)\delta_{m+n}
\ee
with some so far unknown function $A$. 

Now we evaluate the commutator given in the hints directly,
\be
[L_1,[L_m,L_{-m-1}]]=(2m+1)[L_1,L_{-1}]=(2m+1)(2L_0+A(1))\,,
\ee
and through the derivation rule,
\bea
[L_1,[L_m,L_{-m-1}]]&=&[L_m,[L_1,L_{-m-1}]+[[L_1,L_m],L_{-m-1}]
\nonumber \\
&=& (2+m)[L_m,L_{-m}]+(1-m)[L_{m+1},L_{-m-1}]
\nonumber \\
&=& (2+m)(2mL_0+A(m))+(1-m)((2m+2)L_0+A(m+1))\,.
\eea
Comparing both results gives the recursion relation
\be
(m-1)A(m+1)=(2+m)A(m)-(2m+1)A(1)\,.
\ee
Given also that $A(m)=-A(-m)$ by its definition, it is clear that $A(1)$ and $A(2)$ are sufficient to determine all $A(m)$ unambiguously. Moreover, it is easy to check that $A(m)=am^3+bm$ solves the recursion:
\be
(m-1)(a(m+1)^3+b(m+1))=(2+m)(am^3+bm)-(2m+1)(a+b)
\ee
for all $a,b$. Thus, if we can fix $a,b$, we have found the unique solution.

This is easy to achieve: Note first that each term in $L_{-1}$ (and even more so in $L_1$) contains either an annihilator or a $p$. Hence
\be
\langle 0,0|[L_1,L_{-1}]|0,0\rangle=0\,,
\ee
implying $A(1)=0$. By contrast, $L_{-2}$ contains a single term without annihilators, hence
\be
\langle 0,0|[L_2,L_{-2}]|0,0\rangle=\langle 0,0|L_2L_{-2}|0,0\rangle=
\frac{1}{4}\langle 0,0|\alpha_1\alpha_1\alpha_{-1}\alpha_{-1}|0,0\rangle=\frac{1}{2}\,.\label{l2c}
\ee
This implies $A(2)=1/2$. Thus, we have to solve
\be
a+b=0 \qquad \mbox{and}\qquad 8a+2b=\frac{1}{2}\,,
\ee
giving $a=-b=1/12$\,. Clearly, if we generalise from one to $D$ bosons, nothing changes except that, in the very last step, one gets a factor of $D/2$ on the r.h.~side of (\ref{l2c}). Thus, the result given in the lecture follows.

\subsubsection{Normal ordering constant as Casimir energy}\label{noc}
\index{Casimir energy}

{\bf Task:} Finish the calculation of the normal ordering constant $a$ of the open string as the Casimir energy of 2d field theory on a strip,
\be
-a=\lim_{\Lambda\to\infty}\left[\frac{D-2}{2}\left\{\sum_{n=1}^\infty n\right\}_\Lambda+\pi R^2\lambda(\Lambda)\right]\,.
\ee
For hints see lecture notes.

\noindent
{\bf Solution:} As explained in the lecture, we think of the sum as of a sum over modes with physical momenta $k_n=n/R$, suggesting a regularization by a suppression factor $\exp(-k_n/\Lambda)$. The sum $S$ then reads
\bea
S(\Lambda)&=&\sum_{n=1}^\infty n\,e^{-n/\Lambda R}=-\frac{d}{d\alpha} \sum_{n=1}^\infty e^{-\alpha n}\qquad (\,\mbox{with}\,\,\, \alpha=1/\Lambda R)
\\
\nonumber \\
&=& -\frac{d}{d\alpha}\left(\frac{1}{1-e^{-\alpha}}\right)=\frac{e^{-\alpha}}{ \left(1-e^{-\alpha}\right)^2}=\frac{1}{(1-e^{-\alpha})(e^\alpha-1)}
\nonumber\\
\nonumber \\
&=& \frac{1}{(\alpha-\alpha^2/2+\alpha^3/6)(\alpha+\alpha^2/2+\alpha^3/6)}+{\cal O}(\alpha)
\nonumber\\
\nonumber\\
&=& \frac{1}{\alpha^2(1-\alpha/2+\alpha^2/6)(1+\alpha/2+\alpha^2/6)} +{\cal O}(\alpha)
\nonumber\\
\nonumber\\
&=& \frac{1}{\alpha^2(1+\alpha^2/12)} +{\cal O}(\alpha)=\frac{1}{\alpha^2}
\left(1-\frac{1}{12}\alpha^2\right) + {\cal O}(\alpha)
\nonumber\\
\nonumber\\
&=& \Lambda^2R^2-\frac{1}{12}+{\cal O}(1/\Lambda)\,.\nonumber
\eea
This gives rise to
\be
-a=\lim_{\Lambda\to\infty}\left[\frac{D-2}{2}\left(\Lambda^2R^2-\frac{1}{12} +{\cal O}(1/\Lambda)\right)+\pi R^2\lambda(\Lambda)\right]\,.
\ee
The cosmological-constant counterterm is unambiguously determined to be\\ 
$\lambda(\Lambda)=\Lambda^2(D-2)/(2\pi)$, such that finally
\be
a=\frac{D-2}{24}\,.
\ee

\subsubsection{Kalb-Ramond field from the worldsheet perspective}
\index{Kalb-Ramond field}

{\bf Task:} Work out the formal expression
\be
\int_{\Sigma_2} B_2\,,
\ee
such that it becomes a standard Riemann double integral in $d\sigma^1d\sigma^2$ with an integrand depending on the functions $X^\mu(\sigma)$ and $B_2(X)$.

\noindent
{\bf Hints:} Interpret and apply the equality (in two dimensions) 
\be
\int dx\wedge dy=\int dx\,dy\,\,(dx\wedge dy)(\partial_x,\partial_y)
\ee
between a form integral and a Riemann double integral.

\noindent
{\bf Solution:} In analogy to the formula for translating a form integral in a Riemann integral given in the hint, we have (suppressing the index $\Sigma_2$ which is always associated with our integral)
\be
\int B_2=\int d\sigma^1\,d\sigma^2\,B_2(\partial_1,\partial_2)\,.
\ee
Since $B_2$ is originally defined in target space rather than on the worldsheet, we need to push-forward the vectors $\partial_a$ to the target space using the embedding map $X^\mu(\sigma)$ before we can explicitly insert them in $B_2$:
\be
\int B_2= \int d\sigma^1\,d\sigma^2\,B_2\left(\partial_1 X^\mu\, \frac{\partial}{\partial X^\mu},\partial_2 X^\mu\, \frac{\partial}{\partial X^\mu}\right)\,.
\ee
With 
\be
B_2=\frac{1}{2!}\,B_{\mu\nu}\,dX^\mu\wedge dX^\nu
\ee
one now finds
\be
\int B_2= \int d\sigma^1\,d\sigma^2\,B_{\mu\nu}(X(\sigma))\,(\partial_1 X^\mu(\sigma))\,(\partial_2 X^\nu(\sigma))\,,
\ee
where $\sigma$ stands for $\{\sigma^1,\sigma^2\}$. The factor $1/2!$ disappeared since we dropped a second term, where $\partial_1$ and $\partial_2$ would have been exchanged.

\section{String Theory: Interactions and Superstring}\label{stss}
\index{superstring}

Before we can see what the string-theoretic UV completion of gravity has to say about the real world, a few more formal developments are necessary. First, we want to understand at least in principle how scattering amplitudes and loop effects are calculated. Second, we need to introduce fermions and get rid of the tachyon. The main textbook sources continue to be \cite{Green:1987sp, Polchinski:1998rq, Blumenhagen:2013fgp, bbs, kir}.

\subsection{State-operator correspondence}
\index{state-operator correspondence}

\begin{figure}[ht]
\begin{center} 
\includegraphics[width=6cm]{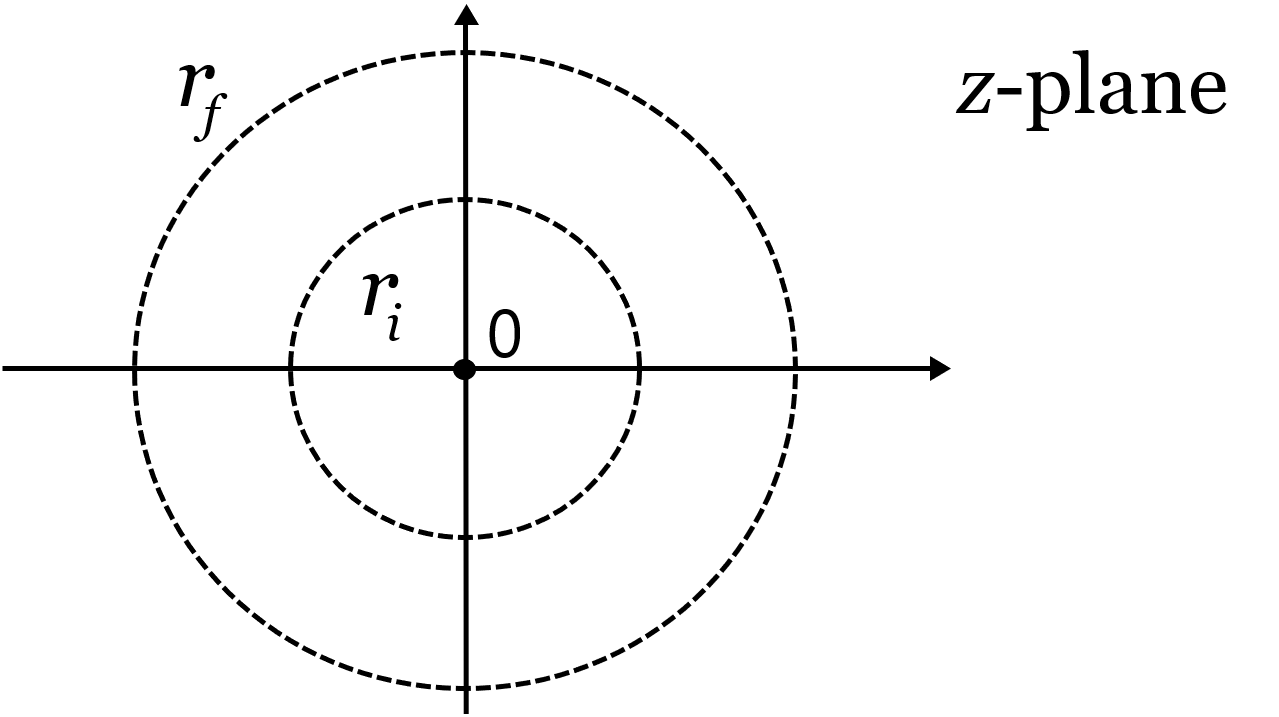}
\caption{String propagation mapped to the $z$-plane. The part of the cylinder\index{cylinder} between initial time $\tau_i$ and final time $\tau_f$ corresponds to the annulus (ring) between $r_i$ and $r_f$.}
\label{soc} 
\end{center}
\end{figure}

Before discussing scattering amplitudes and loops, a few more words about the worldsheet theory after gauge fixing are necessary. We learned that this is a CFT\index{CFT} and we will here work with the euclidean version of this theory. The symmetries of the CFT include angle-preserving deformations of the worldsheet. For example, we can map our fundamental cylinder corresponding to the propagation of the string to the $z$-plane, 
\be
z=r\,e^{i\varphi}\in\mathbb{C}\,.
\ee
Specifically, we want the map to be such that time runs radially and circles of constant $r$ correspond to constant-time cuts through our cylinder (cf.~Fig.~\ref{soc}). The reader is invited to consider the explicit map $z=\exp(iw)$ and identify a strip in the $w$-plane (with periodic boundary conditions, i.e.~a cylinder) that is mapped to the $z$-plane in the desired way.

Next, let us recall that a state in a 4d QFT may, analogously to the Schr\"odinger wave function of quantum mechanics, be described by a {\bf Schr\"odinger wave functional},\index{Schr\"odinger wave functional}
\be
\Psi:\,\phi\,\mapsto\, \Psi[\phi,t]\in\mathbbm{C}\,.
\ee
Here $\phi:\,\ol{x}\mapsto\,\phi(t,\ol{x})\in\mathbb{R}$ is a field configuration at fixed time. The evolution of such states is described by the QFT version of the Feynman path integral.

In our context, a string state at time $\tau_i$ is then represented, in the radial representation, by a wave functional
\be
\Psi_i[X_i,r_i]\,.
\ee
Here $X_i$ stands for any of the possible field configurations $X_i^\mu(r_i,\varphi)$. The wave functional obtained by Hamiltonian evolution at radial time $r_f$ reads
\be
\Psi_f[X_f,r_f]=\int DX_i \int_{X_i}^{X_f} DX\,e^{-S[X]}\,\Psi_i[X_i,r_i]\,.
\label{ram}
\ee
Here the labels $X_i$ and $X_f$ of the integral mean that we integrate over field configurations $X(r,\varphi)$ satisfying $X(r_i,\varphi)=X_i(\varphi)$ and $X(r_f,\varphi)=X_f(\varphi)$.

Now, consider the limit in which our evolution starts at $\tau_i=-\infty$, corresponding to $r_i=0$ or $z=0$. In this limit, we can write 
(\ref{ram}) as
\be
\Psi_f[X_f,r_f]=\int^{X_f} DX\,e^{-S[X]}\,\lim_{r_i\to 0}\,\Psi_i[X(r_i,\varphi),r_i]=\int^{X_f} DX\,e^{-S[X]}\,O(z=0)\,.\label{ram1}
\ee
Here, in the first step, we have absorbed the integral over $X_i$ in the integral over $X$, dropping the initial boundary condition. In other words, we now integrate  over functions $X$ which are also defined inside the inner circle of radius $r_i$ (possibly with a singularity at the origin). Nevertheless, they are weighted by the functional $\Psi_i$ according to their values on that circle. In the limit $r_i\to 0$, this becomes a weighting according to the local behaviour of the functions $X$ near $z=0$. Hence, in the second step, we have introduced the operator $O$ in the CFT, i.e., some functional of $X$ depending only on its local behaviour at the origin. In the simplest and most relevant cases, this is an expression involving $X(0)$ and its derivatives, as familiar from a conventional local QFT operator built from a field $X(z)$ (cf.~Problem~\ref{esom}).

By the above procedure, we have understood how a given state, in our case the state defined by $\Psi_i$, specifies an operator. The opposite direction is obvious: Clearly, Equation (\ref{ram1}) may be interpreted as describing the evolution of some state, defined by $O$, from $\tau=-\infty$ to $\tau_f$. 

Thus, we now know how to associate a CFT state with an operator and vice versa.

\subsection{Scattering amplitudes}
\begin{figure}[ht]
\begin{center} 
\includegraphics[width=12cm]{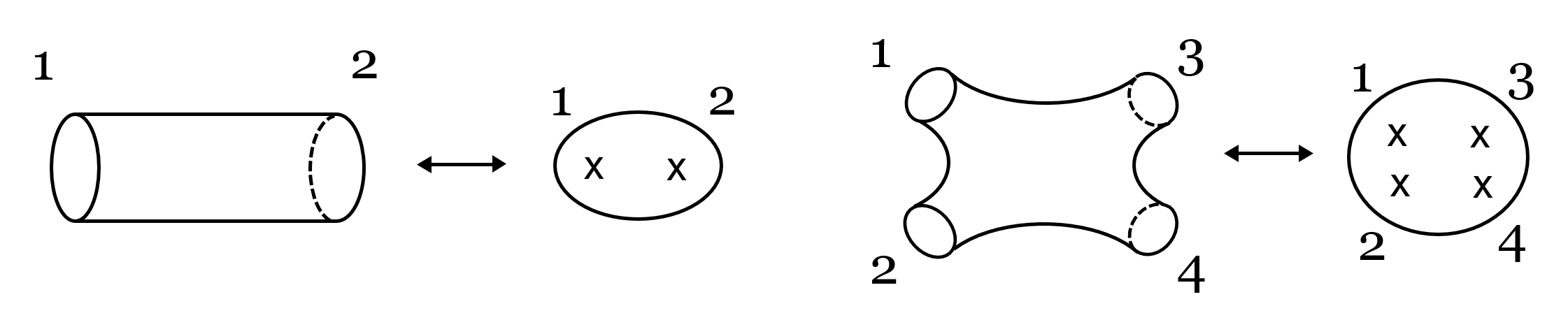}
\caption{Identification of non-compact worldsheets describing string propagation (left) or scattering (right) with appropriate compact worldsheets (in this case spheres) with operator insertions.}
\label{sca} 
\end{center}
\end{figure}

\index{scattering amplitudes}
After the discussion of the previous section, it should be at least intuitively clear that the integral over fields on a cylinder can be replaced by the integral over fields on the sphere, with appropriate operators inserted at the points which are mapped to $\tau=\pm \infty$. This is illustrated in Fig.~\ref{sca} together with an analogous map corresponding to the 2-to-2 scattering of string states.

\begin{figure}[ht]
\begin{center} 
\includegraphics[width=9cm]{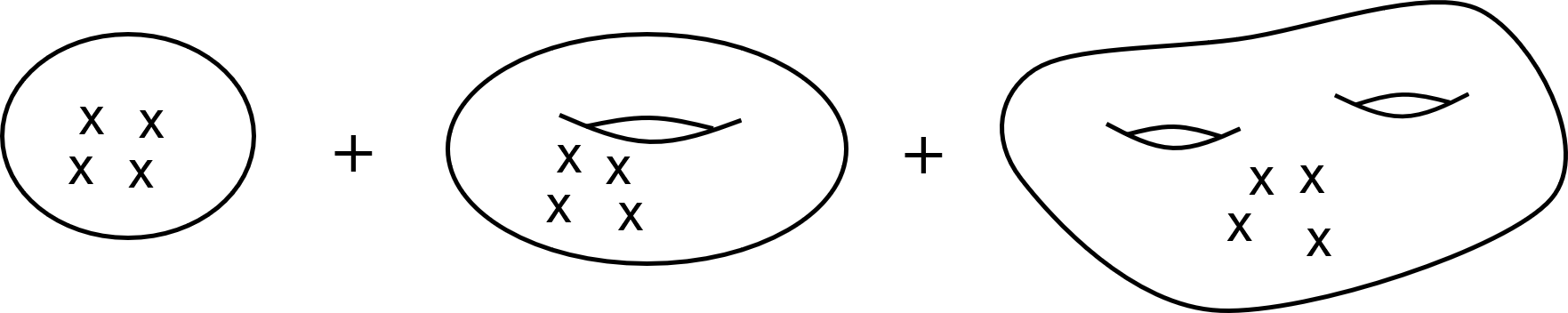}
\caption{
Contributions of worldsheets of genus\index{genus} zero, one and two to the four-point scattering amplitude.}
\label{rie} 
\end{center}
\end{figure}

This leads very naturally to the following fundamental formula for $n$-point scattering amplitudes in string theory, which one may view as the definition of the theory:
\be
{\cal A}_n=\sum_{g=0}^{\infty}\int\frac{Dh\,DX}{V\!ol_{Diff.\times Weyl}}\,e^{-S[X,h]}\int d^2 z_1\cdots d^2 z_n\,\,V_1(z_1,\ol{z}_1)\cdots 
V_n(z_n,\ol{z}_n)\,.
\ee
Here the sum is over all compact oriented 2d manifolds (Riemann surfaces)\index{Riemann surface}, as illustrated in Fig.~\ref{rie}. The terms are labelled by the genus $g$ of the worldsheet. The integration is not only over scalar field configurations $X$ but also over metrics $h$. This definition is more fundamental and the gauge-fixed integral just over $X$ (corresponding to the CFT language) must be carefully derived from it. The reason is that there is a non-trivial interplay between the topology of the manifold, the position of the vertex operators and the residual gauge freedom. In this process, one also has to divide out the infinite factor coming from gauge redundancies. This factor becomes manifest when one uses the Fadeev-Popov method to treat the functional integration. 

The action to be used in the above is
\be
S[X,h]=\frac{1}{4\pi\alpha'}\int_\Sigma d^2\sigma\,\sqrt{-h}(\partial X)^2+\frac{\phi}{4\pi}\int_\Sigma d^2\sigma\,\sqrt{-h}{\cal R}\,,\label{22a}
\ee
where we suppress boundary terms relevant in the open-string case. We have also assumed that the target space is flat and the dilaton constant. Restricting our attention to the oriented string, one has
\be
\frac{1}{4\pi}\int_\Sigma d^2\sigma\,\sqrt{-h}{\cal R}=\chi(\Sigma)=2-2g\,,
\ee
where $\chi$ is known as the Euler number. Thus, the second term on the r.h. side of (\ref{22a}) just supplies a factor
\be
g_s^{-\chi}=g_s^{-2+2g}\,\,\qquad\mbox{with}\qquad g_s\equiv e^{\phi}\,.
\ee
The quantity $g_s$ is known as the {\bf string coupling}.\index{string!coupling}

Finally, the so-called {\bf vertex operators}\index{vertex operators} $V_i$ have to be chosen appropriately to reflect the physical states in the scattering of which one is interested. They can be derived from our understanding of the physical states of the quantised string and the state-operator mapping. Here, we only provide as an example the vertex operator for the tachyon of momentum $k$. It is basically the simplest operator that has the desired transformation properties under translations:
\be
V(k,z,\ol{z})=g_s\,:e^{ik_\mu X^\mu(z,\ol{z})}:\,\,.
\ee
Indeed, a target-space translation $X^\mu\to X^\mu+\epsilon^\mu$ gives this operator (and hence the corresponding state) an extra factor $e^{ik_\mu \epsilon^\mu}$. So this is clearly a state of momentum $k^\mu$. The normalisation by $g_s$ ensures that free propagation is proportional to $g_s^0$, tree-level 4-point-scattering to $g_s^2$, one-loop 4-point-scattering to $g_s^4$, and so on. Apart from the exponential and $g_s$ prefactors, our operator is just the unit operator. It is shown in Problem~\ref{esom} that this corresponds to the vacuum state and hence the tachyon.

From here, it would be relatively straightforward to calculate some of the simplest amplitudes and loop diagrams (a.k.a. higher genus contributions) and convince oneself of the promised very soft UV behaviour and loop finiteness. But we have to move on.

\subsection{Worldsheet supersymmetry}
\index{worldsheet!supersymmetry}

We need to find a string theory which describes target space fermions and which has no tachyon. Both can be achieved by supersymmetrising the worldsheet. We will follow the presentation of \cite{Green:1987sp}, where the reader may also find a list of the most important original papers. Let us only mention the key contribution of Ramond, Neveu and Schwarz\index{Ramond-Neveu-Schwarz superstring} \cite{Ramond:1971gb, Neveu:1971rx}. Their approach, which also underlies the following discussion, is known as the  `RNS superstring'\index{RNS superstring}. A detailed presentation can also be found in \cite{Blumenhagen:2013fgp}.

As in 4d, we simply add fermionic worldsheet coordinates,
\be
\sigma^a\qquad \mbox{`+'} \qquad \theta_\alpha\,.
\ee
The 2d Lorentz transformations are completely analogous to the familiar 4d case, 
\be
\sigma^a\,\to\, \Lambda^a{}_b\sigma^b\,\,,\qquad \theta_\alpha\,\to\,
S_\alpha{}^\beta\theta_\beta\,,
\ee
with
\be
\Lambda=\exp(i\epsilon^{ab}J_{ab})\qquad\mbox{and}\qquad
S=\exp(i\epsilon^{ab}\left\{i[\gamma_a,\gamma_b]/4\right\})\,.
\ee
One may easily check that 
\be
\gamma^0=\left(\begin{array}{rr}0&-i\\i&0\end{array}\right)\,\,,\quad
\gamma^1=\left(\begin{array}{rr}0&i\\i&0\end{array}\right)
\ee
fulfil
\be
\{\gamma^a,\gamma^b\}=-2\eta^{ab}\,.
\ee
Of course, because of antisymmetry in $a,b$, there is only one independent generator of type $i[\gamma_a,\gamma_b]/4$ (spinor representation) or $J_{ab}$ (vector representation). The commutation relations are trivially the same: After all, $SO(1,1)$ is a one-parameter group. Nevertheless, the two representations are different, cf.~Problem~\ref{so2r}.

Furthermore, since $S$ is real, one may obviously demand that the spinor is real:
\be
\theta^*=\left(\begin{array}{c}\theta_-\\\theta_+\end{array}\right)^*
=\left(\begin{array}{c}\theta_-\\\theta_+\end{array}\right)=\theta\,.
\ee
This is a particularly simple version of the familiar Majorana condition
\be
\psi=\psi^c \qquad\qquad \qquad (\,\mbox{with}\,\,\psi^c\equiv C\,\ol{\psi}^T\,\,)\,.
\ee
The reader may want to consult the appendix of Volume 2 of \cite{Polchinski:1998rq} for a systematic discussion of spinors in various dimensions. 

Following very closely the familiar 4d procedure, one may promote the scalars to (general) superfields,
\be
X^\mu\,\,\to\,\,Y^\mu(\sigma,\theta)
\ee
with 
\be
Y^\mu(\sigma,\theta)=X^\mu(\sigma)+\ol{\theta}\psi^\mu(\sigma)+\frac{1}{2} \ol{\theta}\theta\,B^\mu(\sigma)\,.
\ee
Here $\ol{\theta}=\theta^\dagger\gamma^0$, as in 4d. We define SUSY generators
\be
Q_\alpha=\frac{\partial}{\partial\ol{\theta}^\alpha}+i(\gamma^a\theta)_\alpha\,\partial_a\,,
\ee
which are also Majorana spinors, and observe that they satisfy the SUSY algbera
relation
\be
\{Q_\alpha,\ol{Q}^\beta\}=-2i(\gamma^a)_\alpha{}^\beta\partial_a\,.
\ee

The SUSY transformation can be defined by
\be
\delta_\xi Y(\sigma,\theta)=(\ol{\xi}Q)\,Y(\sigma,\theta)\,,
\ee
leading to
\bea
\delta_\xi X^\mu &=& \ol{\xi}\psi^\mu
\\
\delta_\xi \psi^\mu &=& -i(\gamma^a \xi)\partial_a X^\mu+B^\mu\xi
\\
\delta_\xi B^\mu &=& -i\ol{\xi}\gamma^a\partial_a \psi^\mu\,.
\eea

To write down SUSY-invariant actions, it is sufficient to integrate any expression in the $Y^\mu$ over the full superspace. In our case this is simply a $d^2\sigma\,d^2\theta$ integral. However, we are looking for a specific action which would serve as a generalization of the (so far flat-space) bosonic Polyakov action. For this purpose, it is convenient to introduce the supercovariant derivative
\be
D_\alpha=\frac{\partial}{\partial \ol{\theta}^\alpha}-i(\gamma^a\theta)_\alpha
\partial_a\,.
\ee
The SUSY version of our bosonic action (with $l=1$) can then be given as
\be
S=\frac{i}{4\pi}\int d^2\sigma\, d^2\theta\, (\ol{D}^\alpha Y^\mu)(D_\alpha Y_\mu)=-\frac{1}{2\pi}\int d^2\sigma\,(\partial_a X^\mu\,\partial^a X_\mu-i\ol{\psi}^\mu\slashed{\partial}\psi_\mu-B^\mu B_\mu)\,.
\ee
The auxiliary field vanishes on-shell such that, in summary, we have simply added a free fermion $\psi^\mu$ for every scalar.

\subsection{Worldsheet supergravity}
\index{worldsheet!supergravity}

The next step is to introduce gravity (more precisely, to promote the metric to a field, since gravity in the sense of a dynamical theory does not really exist in $d=2$). This implies making SUSY local, as explained earlier. 

Since our theory contains spinors, we will need a vielbein, related to the metric by 
\be
h_{ab}=(e^m)_a (e^n)_b\,\eta_{mn}\,.
\ee
Here $a,b,\cdots$ are `curved' or `Einstein indices' as before and $m,n,\cdots$ are `frame' or `Lorentz indices'. Furthermore, since the Lorentz symmetry transforming the Lorentz indices is local, we require a spin connection, to define covariant derivatives of objects with frame indices:
\be
\nabla_a v^m=(\partial_a+\omega_a)v^m\qquad\mbox{with}\qquad \omega_a\in\,
Lie(SO(1,d-1))\,,
\ee
in our case with $d=2$. It is defined by demanding covariant constancy of the vielbein,
\be
0=\nabla_a e^m{}_b=\partial_a e^m{}_b+(\omega_a)^m{}_n e^n{}_b-\Gamma_{ab}{}^c e^m{}_c\,,
\ee
where $\Gamma$ stands for the usual Christoffel symbols. Clearly, the object on which $\nabla$ acts can transform in any representation of $SO(1,d-1)$, in which case $\omega_a$ has to be taken in that representation. 

With these preliminary remarks our action becomes, in the first step,
\be
S_2=-\frac{1}{2\pi}\int d^2\sigma\,e\,\left\{h^{ab}(\partial_a X^\mu) (\partial_b X_\mu) -i\ol{\psi}^\mu\gamma^a\nabla_a\psi_\mu\right\}\,.\label{sga}
\ee
The index $2$ stands for `quadratic order'. We want to make this invariant under a local version of the SUSY transformations above, i.e. with $\xi\to\xi(\sigma)$. In addition, we need to define SUSY transformations of our new field, the metric or, more appropriately, the vielbein. Working at leading order in perturbations around flat space, $e^m{}_a=\delta^m{}_a$, one postulates
\be
\delta_\xi e^m{}_a=-2i\ol{\xi}\gamma^m\chi_a\,.
\ee
Here $\chi_a$ is the gravitino. Its appearance on the r.h. side is natural since, as we argued earlier, it has to come to provide the superpartner for the metric. The rest of this relation is fixed (up to normalisation) by covariance. 

The action of (\ref{sga}) is not invariant under local SUSY but, since it was invariant under the global version, its non-invariance is controlled by the derivative of $\xi$. Thus, we have
\be
\delta_\xi S_2=\frac{2}{\pi} \int d^2\sigma\,\sqrt{-h}\,(\nabla_a\ol{\xi})\,J^a\,,
\ee
where $J^a$ is by definition the Noether current corresponding to the global version of the symmetry. Explicitly, one finds (at quadratic order in the fields) 
\be
J^a=\frac{1}{2}\gamma^b\gamma^a\psi^\mu\partial_b X_\mu\,,
\ee
known as the {\bf supercurrent}\index{supercurrent}. The non-invariance of $S_2$ can be compensated by adding a term
\be
S_3=-\frac{2}{\pi}\int d^2\sigma\,\sqrt{-h}\,\ol{\chi}_a\, J^a 
=-\frac{1}{\pi}\int d^2\sigma\,\sqrt{-h}\,\ol{\chi}_a\gamma^b\gamma^a\psi^\mu \partial_bX_\mu\,,
\ee
and introducing the transformation law
\be
\delta_\xi\chi_a=\nabla_a\xi\,.
\ee
For obvious reasons, this method of constructing supergravity actions is known as the Noether method.\index{Noether method}

Its implementation is not yet complete: Only after a modification of $\delta_\xi \psi$ by a term proportional to the gravitino and the addition of a quartic term,
\be
S_4=-\frac{1}{4\pi}\int d^2\sigma\,\sqrt{-h}\,(\ol{\psi}\psi)\,(\ol{\chi}_a\gamma^b\gamma^a \chi_b)\,,
\ee
does the theory become invariant under local SUSY. We recall that the Einstein-Hilbert term is a total derivative. This matches the fact that the gravitino kinetic term is identically zero in $d=2$ (since $\gamma^{[a}\gamma^b\gamma^{c]}=0$).

Finally, the theory is still Weyl invariant, with transformation laws
\be
\delta_\omega X=0\,\,,\qquad \delta_\omega e^m{}_a=\omega e^m{}_a \,\,,\qquad \delta_\omega \psi=\frac{1}{2}\omega\psi\,\,,\qquad \delta_\omega \chi_a=\frac{1}{2}\omega\chi_a\,.
\ee
Due to SUSY, this symmetry now has a fermionic counterpart, parameterised by the infinitesimal Majorana spinor $\eta$:
\be
\delta_\eta X=\delta_\eta e =\delta_\eta \psi=0\,\,,\qquad \delta_\eta\chi_a=i\gamma_a\eta\,.
\ee
This makes our theory {\bf super-Weyl-invariant}\index{super-Weyl-invariant} and, after gauge fixing, {\bf superconformal}.\index{superconformal}

Before closing this section, one important remark is in order: As already noted, the action described above and its quantisation to be discussed momentarily (the `RNS approach' to the superstring) are built on worldsheet supersymmetry. In this approach, the appearance of supersymmetry in the target-space theory (to be discussed later) remains somewhat miraculous. However, there exists another approach with a different superstring action, the so-called Green-Schwarz (GS) superstring\index{Green-Schwarz superstring}\index{GS-superstring} \cite{Green:1983wt, Schwarz:1982jn, Berkovits:2002zk}, which is built from the very beginning on the requirement of target-space SUSY. While it looks very different, it is equivalent to the RNS approach. For a textbook treatment see e.g.~\cite{bbs}.

\subsection{Quantisation of the superstring}

The large gauge symmetry make it possible to go to flat gauge: As before, diffeomorphism and Weyl invariance are sufficient to choose a flat metric and vielbein. The key new point is that the local transformations encoded in $\xi$ and $\eta$ allow one to set the gravitino to zero. Thus, we have to quantise the simple action
\be
S=-\frac{1}{2\pi}\int d^2\sigma\left[(\partial_a X^\mu)(\partial^a X_\mu)-i\ol{\psi}^\mu\gamma^a\partial_a\psi_\mu\right]\,.
\ee
As before, the equations of motion of the fields that have been eliminated by gauge fixing must be imposed as constraints. These are 
\be
T_{ab}=0\,,
\ee
as before, where we now have
\be
T_{ab}=(\partial_a X)\cdot(\partial_b X)+\frac{i}{2}\ol{\psi}\cdot \gamma_{\{a} \partial_{b\}}\psi-\frac{1}{2}h_{ab}\left(
(\partial X)^2+\frac{i}{2}\ol{\psi}\cdot\slashed{\partial} \psi\right)\,.
\ee
Here the curly brackets stand for symmetrisation. In addition, we have used local SUSY and super-Weyl invariance (see~\cite{Blumenhagen:2013fgp} for details) to set the gravitino $\chi^a$ to zero. But, as we have just seen in the last section, its equations of motion correspond, at leading order in $\chi$, to the vanishing of the supercurrent:
\be
(J_a)_\alpha=0\,.
\ee
This is the second, new constraint. 

The mode decomposition for the bosonic part is as before. To discuss the mode decomposition of the fermionic part, write
\be
S=S_B+S_F\qquad \mbox{with}\qquad S_F=\frac{i}{\pi}\int d^2\sigma\,(\psi_-\cdot \partial_+\psi_-+\psi_+\cdot\partial_-\psi_+)\,\,,\quad\mbox{where}\qquad
\gamma^\pm=\gamma^0\pm\gamma^1\,.
\ee
This explains our indexing convention in
\be
\psi=\left(\begin{array}{c}\psi_-\\ \psi_+\end{array}\right)
\ee
since we now see that $\psi_+$ and $\psi_-$ are left and right movers respectively. Due to the fermionic nature of $\psi_{\pm}$, a sign is not detectable (observables are always built from bilinears). Hence, the sign may not or may change when going once around the string. As a result, two different types of boundary conditions (known as {\bf Ramond}\index{Ramond boundary conditions} and {\bf Neveu-Schwarz})\index{Neveu-Schwarz boundary conditions} are possible. This leads to 4 sectors:
\bea
&&\psi_+(\sigma+\pi)=+\psi_+(\sigma)\quad;\quad \psi_-(\sigma+\pi)=+\psi_-(\sigma) \qquad \mbox{R-R}
\nonumber \\
&&\psi_+(\sigma+\pi)=+\psi_+(\sigma)\quad;\quad \psi_-(\sigma+\pi)=-\psi_-(\sigma) \qquad \mbox{R-NS}
\nonumber \\
&&\psi_+(\sigma+\pi)=-\psi_+(\sigma)\quad;\quad \psi_-(\sigma+\pi)=+\psi_-(\sigma) \qquad \mbox{NS-R}
\nonumber \\
&&\psi_+(\sigma+\pi)=-\psi_+(\sigma)\quad;\quad \psi_-(\sigma+\pi)=-\psi_-(\sigma) \qquad \mbox{NS-NS}\,.
\eea
Note that we could not have used an arbitrary phase $\exp(i\alpha)$ instead of the sign in the boundary conditions since our spinors are real. The mode decomposition in the R-NS sector reads
\be
\psi_+^\mu=\sum_{r\in\mathbb{Z}}\tilde{\psi}_r^\mu\,e^{-2ir(\tau+\sigma)} \quad,\quad \psi_-^\mu=\sum_{r\in\mathbb{Z}+\frac{1}{2}}\psi_r^\mu\,e^{-2ir(\tau -\sigma)}\,,
\ee
and analogously for the other 3 sectors. The reality constraint translates to the usual relation between modes with opposite frequency: $(\tilde{\psi}_r^\mu)^* = \tilde{\psi}_{-r}^\mu$ and $(\psi_r^\mu)^* = \psi_{-r}^\mu$.

On the open string, one only has R-R\index{R-R sector} and NS-NS sectors\index{NS-NS sector}. To understand this, one may think of the open string as coming from the closed string (to be viewed as a theory on $S^1$) by `modding out' a $\mathbb{Z}_2$ symmetry. In other words, one goes from $S^1$ to $S^1/\mathbb{Z}_2$. The $\mathbb{Z}_2$ acts by $\sigma\to -\sigma$ on the space, turning the spatial part of the worldsheet from a circle into an interval. Two boundaries are created at the so-called fixed points of the action, i.e. at $\sigma=0$ and $\sigma=\pi$ (if we start with a $2\pi$ circle). This $\mathbb{Z}_2$ action also exchanges left and right movers in terms of fields. But such an exchange would be inconsistent in a R-NS\index{R-NS sector} or NS-R\index{NS-R sector} sector.\footnote{Of course, this construction can be translated into the alternative picture where the open superstring is defined on an interval from the start. Then two consistent sets of boundary conditions at the two boundaries $\sigma=0$ and $\sigma=\pi$ have to be introduced. We leave that to the reader.}

Due to the above, one actually calls the two distinct open-string sectors simply R and NS.

Skipping the standard steps of canonical quantisation, we immediately display the commutation relations of the oscillator modes, promoted to operators:
\bea
[\alpha_m^\mu,\alpha_n^\nu] &=& m\,\delta_{m+n}\,\eta^{\mu\nu}
\\
\{\psi_r^\mu,\psi_s^\nu\} &=& \delta_{r+s}\,\eta^{\mu\nu} \quad\mbox{with}\quad
\left\{ \begin{array}{lr}
r,s\in\mathbb{Z} & \mbox{(R)}
\\
r,s\in\mathbb{Z}+\frac{1}{2} & \mbox{(NS)}
\end{array}\right.\,\,.
\nonumber
\eea
The different normalisation (manifest in the prefactor $m$ and the missing prefactor $r$) is conventional. As before, the operators responsible for the constraints are expanded in Fourier modes,
\be
L_m=\frac{1}{\pi}\int_{-\pi}^\pi d\sigma\,e^{im\sigma}\,T_{++}\qquad,\qquad
G_r=\frac{\sqrt{2}}{\pi}\int_{-\pi}^\pi d\sigma\,e^{ir\sigma}\,J_+\,,
\ee
with
\bea
L_m &=& \frac{1}{2}\,:\,\left\{ \sum_{n\in \mathbb{Z}}\alpha_{-n}\cdot \alpha_{m+n} + \sum_{r\in \mathbb{Z}+\nu}\left(r+\frac{m}{2}\right)
\psi_{-r}\cdot\psi_{m+r}\right\}\,:
\\
G_r &=& \sum_{n\in \mathbb{Z}}\alpha_{-n}\cdot\psi_{r+n}\qquad\qquad\mbox{where}\qquad
\nu\equiv\left\{\begin{array}{cc}0 & \mbox{(R)} \\ 1/2 & \mbox{(NS)} \end{array}\,.
\right.
\eea
These operators generate the {\bf super-Virasoro algebra}\index{super-Virasoro algebra}. More precisely, there are two different algebras, one for the Ramond case ($r,s$ even) and one for the Neveu-Schwarz case ($r,s$ odd):
\be
[L_m,L_n]= (m-n)L_{m+n}+A(m)\,\,,\qquad
\{G_r,G_s\}=2L_{r+s}+B(r)\delta_{r+s}\,\,,
\ee
\be
[L_m,G_r]=(m/2-r)\,G_{m+r}
\ee
with the anomaly terms
\be
A(m)=  D(m^3-m)/8 \qquad\mbox{and}\qquad B(r)=D(4r^2-1)/8\,.
\ee
As before, only the annihilator-part of the classical constraints is imposed quantum-mechanically:
\be
(L_m-a\delta_m)\,|phys\rangle=0\,\,\,\,\,\,(m\ge 0)\,\,\,,\qquad \qquad
G_r\,|phys\rangle=0\,\,\,\,\,\,(r\ge 0)\,,
\ee
where we note that there is no normal ordering ambiguity and hence no normal ordering constant associated with $G_0$. 

We do not repeat the derivation but simply quote the result for the normal ordering constant:
\be
a=(D-2)\left(\frac{1}{24}-\frac{1}{24}\right)=0\,\,\,\,\,\,\mbox{(R)}
\qquad, \qquad
a=(D-2)\left(\frac{1}{24}+\frac{1}{48}\right)=\frac{D-2}{16} \,\,\,\,\,\,\mbox{(NS)}
\,.
\ee
We see that, in the Ramond-case, the fermions precisely cancel the effect of the bosons. In the Neuveu-Schwarz case, this supersymmetric cancellation is upset by the non-trivial boundary conditions imposed on the fermions but not on the bosons.

Let us now turn concretely to the Fock space of the open-string {\bf NS sector}: We have
\be
\mbox{Vacuum:}\,\,|0,k\rangle\quad,\qquad\mbox{Creation operators:}\,\,\,
\alpha_{-m}^\mu\,\,;\,\,\psi_{-r}^\mu\,\,\,\,\,\,(m,r>0)\,.
\ee
The mass shell condition reads
\be
0=(L_0-a)\,|0,k\rangle = (\alpha'p^2+N^\alpha+N^\psi-a)\,|0,k\rangle\,,
\ee
where
\be
N^\alpha=\sum_{m=1,2,\cdots}\alpha_{-m}\alpha_m\qquad,\qquad 
N^\psi=\sum_{r=\frac{1}{2},\frac{3}{2},\cdots}r\,\psi_{-r}\psi_r\,.
\ee
This implies that there is a scalar at level zero,
\be
\alpha' M^2=-a\,,
\ee
and a (target-space!) vector corresponding to the physical $\psi_{-1/2}$ excitations at level $1/2$:
\be
\epsilon_\mu\psi^\mu_{-1/2}|0,k\rangle\qquad\mbox{with}\qquad
\alpha' M^2= \frac{1}{2}-a\,.
\ee
In analogy to the logic of the bosonic case, we expect that $D=10$ (with $a=1/2$) is the critical dimension, corresponding to the vector being massless (and the scalar a tachyon, as in the bosonic string).

Next, we turn to the open-string {\bf R-sector}, which superficially differs only very little in that
\be
N^\psi=\sum_{r=0,1,2,\cdots}r\,\psi_{-r}\cdot\psi_r=\sum_{r=1,2,\cdots} r\,\psi_{-r}\cdot\psi_r\,.
\ee
But this number operator leads to the very peculiar situation that the $\psi_0^\mu$ do not appear in $L_0$ and hence do not affect the energy (mass squared) of a state. They do, however, satisfy the non-trivial algebra ($D$-dimensional Clifford algebra) 
\be
\{\psi_0^\mu,\psi_0^\nu\}=\eta^{\mu\nu}\,.
\ee
Hence, every mass eigenspace must carry a representation of this algebra, i.e. it must be a target-space spinor:
\be
\mbox{Vacuum:}\,\,\,\,\,\,|\alpha,k\rangle\qquad \mbox{with} \qquad
\alpha= 1,2,3,\cdots, 2^{D/2}=32\,.
\ee 
Since $a=0$, this spinor is massless. To derive the critical dimension we would need to either consider heavier, excited states or involve ghosts and the vanishing central charge argument. We do not do this here and only assert that the critical dimension is still $D=10$.

\subsection{GSO or Gliozzi-Scherk-Olive projection}
\index{Gliozzi-Scherk-Olive projection}\index{GSO projection}

Before constructing the 10d superstring theories which may be relevant for the real world, we need a further technical ingredient. The underlying idea is that one may always use a projection operator (an operator $P$ with $P^2=P$) commuting with the Hamiltonian $H$ to reduce the Hilbert space ${\cal H}$ in a consistent manner. A familiar example is the projection on symmetric and antisymmetric subspaces of ${\cal H}\otimes {\cal H}$ to define bosons and fermions in 2-particle quantum mechanics. Another example (from this course) is the projection of functions on $S^1$ to functions on $S^1/\mathbb{Z}^2$, which corresponds to the projection to even and odd functions and hence to the projection from closed to open string (with Dirichlet or Neumann boundary conditions). The new Hilbert space after projection is, by definition, Image($P$). 

Here, we focus on the open superstring and consider 
\be
P=\frac{1}{2}(1+(-1)^F)\,\,,\qquad\mbox{where}\qquad F\,\equiv\,\mbox{Fermion number}\,.
\ee
This amounts to keeping only states with even $F$ (note that $F$ is only defined mod 2). Concretely, one defines
\bea
(-1)^F|0,k\rangle &=& -|0,k\rangle\quad\mbox{(NS)} \nonumber
\\
(-1)^F|\alpha,k\rangle &=& |\beta,k\rangle\,\Gamma_{\beta}{}^\alpha\quad\mbox{(R)} \qquad\mbox{where}\,\,\Gamma\equiv \Gamma^{11}\equiv \Gamma^0\Gamma^1\cdots\Gamma^9\,.
\label{ponv}
\eea
together with
\be
(-1)^FX^\mu=X^\mu(-1)^F\qquad\mbox{and}\qquad (-1)^F\psi^\mu=-\psi^\mu(-1)^F\,.
\ee
Here the minus sign in the first line of (\ref{ponv}) is a choice which eliminates the tachyon. In the second line, the non-trivial implementation of $(-1)^F$ through the matrix $\Gamma\equiv\Gamma^{11}$ (the 10d version of $\gamma^5$) is enforced by consistency. Indeed, $(-1)^F$ by its very definition anticommutes with all $\psi^\mu_r$s.
This includes $\psi^\mu_0$, which as we know act on the vacuum like the $\Gamma^\mu$s. Thus, $(-1)^F$ must be represented by a matrix anticommuting with all $\Gamma^\mu$s. But this, by definition, is $\Gamma\equiv\Gamma^{11}$.

After this projection, the tachyon is gone and the 32-component Majorana spinor has turned into a 16-component Majorana-Weyl spinor (in 10d both conditions may be imposed together). On shell and in terms of the appropriate representations of the little group SO(8), one has
\be
\mbox{massless vector}\,\,\,\,(8_v)\qquad+\qquad \mbox{Majorana-Weyl fermion}\,\,\,\,(8)\,.
\ee
Here the symbol $8_v$ stands for the (defining) vector representation of $SO(8)$, the symbol $8$ for the chiral Majorana spinor. We will later on also need the opposite-chirality Majorana spinor, which corresponds to an inequivalent representation. It is denoted by $8'$.

The $8_v+8$ found above fit a 10d supersymmetric gauge theory. But we will not develop this construction since it anyway must be coupled to a closed string sector. Our purpose was only to explain the idea of this particular projection on even (or similarly on odd) fermion number states.

\subsection{Consistent type II superstring theories}
\index{type II superstring}

We now turn to the closed string case. The name `type II' refers to the presence of two supersymmetries (equivalently two gravitinos) in 10d, as will become clear in a moment. We first recall the relevant mass-shell and level-matching conditions
\be
(L_0+\tilde{L}_0)\,|phys\rangle=0\qquad \mbox{and}\qquad 
(L_0-\tilde{L}_0)\,|phys\rangle=0
\ee
with
\be
L_0=\frac{\alpha'}{4}p^2+N-\nu\,\,\,,\qquad 
\tilde{L}_0=\frac{\alpha'}{4}p^2+\tilde{N}-\tilde{\nu}\qquad\mbox{and}\qquad
\nu/\tilde{\nu}=\left\{\begin{array}{cc}0&\mbox{(R)}\\ 1/2&\mbox{(NS)}
\end{array}\right.\,.
\ee
Note that the spacing between the different mass levels differs by a factor of 4 compared to the open string. The lowest levels in the four possible sectors are
\be
\begin{array}{lll}
\mbox{\bf Sector} & \mbox{\bf $SO(8)$ rep.} & \mbox{\bf mass}
\\
{\rm NS}\,- & 1 & {\rm tachyon}
\\
{\rm NS}\,+ & 8_v & {\rm massless}
\\
{\rm R}\,- & 8' & {\rm massless}
\\
{\rm R}\,+ & 8 & {\rm massless}\,,
\end{array}
\nonumber
\ee
where $\pm$ refers to the eigenvalue of $(-1)^F$ on which one can potentially project and $8/8'$ refer to the two inequivalent spinor representations of $SO(8)$. (Of course the `1' appearing in the row of the tachyon is only intended to say that this is a scalar with a single degree of freedom -- it is strictly speaking not appropriate to classify it using the little group of massless particles in 10d.) As a side remark, the existence of these in total three 8-dimensional, inequivalent representations of $SO(8)$ is related to the $\mathbb{Z}_3$ symmetry of its Dynkin diagram. 

When combining left and right-moving sectors, 
the level matching constraint allows the (NS$\,-$) sector to be paired only with itself. The other 3 sectors can be paired in any combination. This gives the unprojected spectrum
\be
\begin{array}{ll}
\mbox{\bf Sector} & \mbox{\bf $SO(8)$ rep.}
\\
({\rm NS}\,-,\,{\rm NS}\,-) & 1
\\
({\rm NS}\,+,\,{\rm NS}\,+) & 8_v\times 8_v
\\
({\rm NS}\,+,\,{\rm R}\,-) & 8_v\times 8'
\\
({\rm NS}\,+,\,{\rm R}\,+) & 8_v\times 8
\\
({\rm R}\,-,\,{\rm NS}\,+) & 8'\times 8_v
\\
\cdots & \cdots\,\,.
\end{array}
\nonumber
\ee
There are in total 10 sectors in this table and (independently of the specific fermion-number-projector), one might imagine building a consistent theory from any combination of them. Clearly, there are $2^{10}$ possibilities to select some subset of these sectors. But this selection can not be random: We want it 

\noindent
{\bf (1)} Not to contain a tachyon.

\begin{figure}[ht]
\begin{center} 
\includegraphics[width=4.5cm]{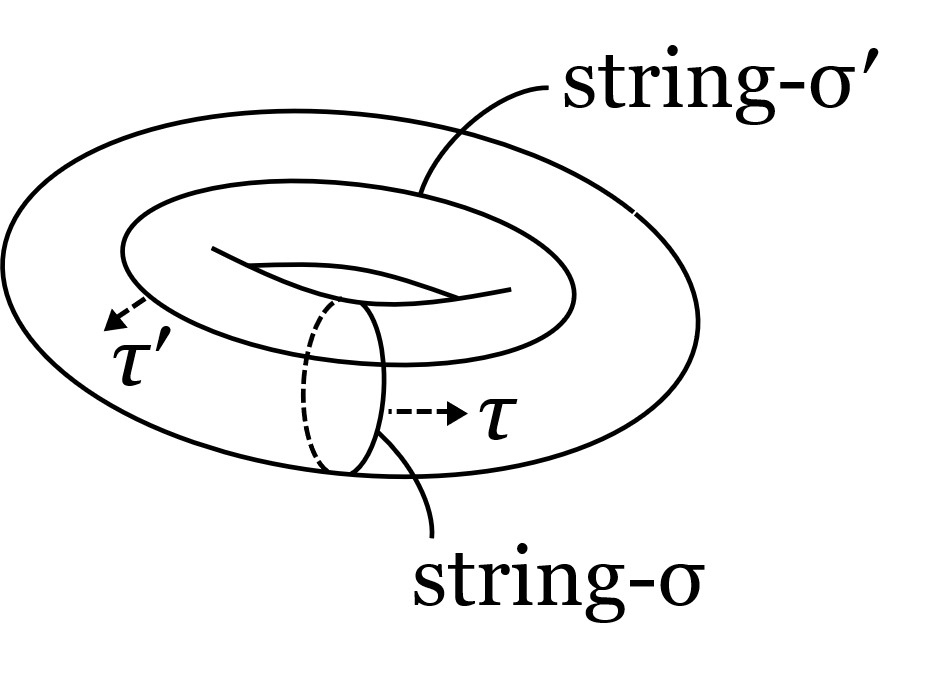}
\caption{Illustration of the intuitive meaning of modular invariance.\index{modular invariance}}
\label{mins} 
\end{center}
\end{figure}

\noindent
{\bf (2)} To be modular invariant (i.e. invariant under large diffeomorphisms\index{large diffeomorphisms} of a worldsheet torus, for example under exchange of $\tau$ and $\sigma$ and hence under reinterpretation of the direction of time flow, cf.~Fig.~\ref{mins})

\begin{figure}[ht]
\begin{center} 
\includegraphics[width=3.5cm]{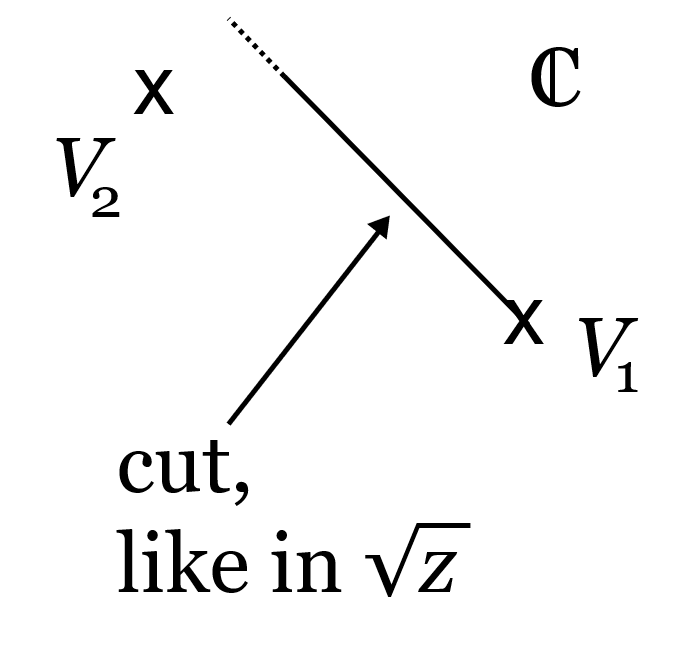}
\caption{In principle, branch cuts can arise in the correlation function between two vertex operators. This should, however, be forbidden since it makes the integration over all positions impossible.}
\label{bcu} 
\end{center}
\end{figure}

\noindent
{\bf (3)} To obey certain mutual consistency rules among the selected vertex operators on the worldsheet. (There should be no leftover phase or branch cut when one operator circles another, cf.~Fig.~\ref{bcu}. The operator product expansion should close or, in other words, it should not be possible to produce a state in scattering which we have excluded from our selection.)

With this, it can be shown that only two inequivalent possibilities of the $2^{10}$ are left~\cite{Polchinski:1998rq}. The corresponding selections are easily formulated using fermion number constraints or projections:

\noindent
{\bf Type IIA}: \hspace{1cm}left:\hspace{.5cm} $(-1)^F=1$ \hspace{1cm}right:\hspace{.5cm} $(-1)^{\tilde{F}}=1\,\,$(NS)$\,\,\,/\,\,\,
(-1)^{\tilde{F}}=-1\,\,$(R)\index{type IIA string theory}

\noindent
{\bf Type IIB}: \hspace{1cm}left:\hspace{.5cm} $(-1)^F=1$ \hspace{1cm}right:\hspace{2.7cm} $(-1)^{\tilde{F}}=1\,.$\index{type IIB string theory}

These general rules translate, specifically in {\bf type IIA} in:
\be
\begin{array}{ccclcl}
{\bf Sector} &&& SO(8)\,\,{\bf rep.} &&
\\
\\
{\rm (NS+,NS+)} & \qquad 8_v\times 8_v &=& 1+28+35 &=& [0]_\phi+[2]_{B_2}+(2)_G
\\
{\rm (NS+,R-)} & \qquad 8_v\times 8' &=& 8+56' &=& {\rm spinor}\,\,+\,\,{\rm vector\mbox{-}spinor}'
\\
{\rm (R+,NS+)} & \qquad 8\times 8_v &=&  8'+56 &=& {\rm spinor}'\,\,+\,\,{\rm vector\mbox{-}spinor}
\\
{\rm (R+,R-)} & \qquad 8\times 8' &=& 8_v+56_t &=& [1]_{C_1}+[3]_{C_3}
\end{array}\nonumber
\ee
To derive the last two columns of this table, one needs elementary representation theory (see e.g.~\cite{Slansky:1981yr}). We will only motivate and interpret the results. We note that $SO(8)$ has three inequivalent 56-dimensional representations: two vector-spinors and one antisymmetric rank-2 tensor. We used a square and round bracket for antisymmetric and traceless symmetric tensors of a given rank. Hence e.g. $[2]$ stands for the familiar Kalb-Ramond field and $(2)$ for the graviton. On the bosonic side, we have dilaton, $B_2$, $g_{\mu\nu}$ and two form-fields, $C_1$ and $C_3$. The latter are a new feature of the superstring and the correponding charged states are so-called D0 and D2 branes, which are non-perturbative objects (in the sense that they do not directly follow from the perturbative analysis of worldsheet degrees of freedom). They have to be introduced into the theory for consistency, have their own action and dynamics, and provide potential endpoints for open strings. 

We are finding a so-called ${\cal N}=2$ supersymmetric theory since we have two gravitinos which are both partners of the same, unique graviton. The other two spinors are known as dilatini. There are two independent SUSY generators and hence SUSY transformations relating the graviton to either one or the other gravitino. However, the overall structure is more involved and all degrees of freedom are needed to fully match fermions and bosons.

Analogously, one finds the field content of the {\bf type IIB} string:
\be
\begin{array}{ccclcl}
{\bf Sector} &&& SO(8)\,\,{\bf rep.} &&
\\
\\
{\rm (NS+,NS+)} & \qquad 8_v\times 8_v &=& 1+28+35 &=& [0]_\phi+[2]_{B_2}+(2)_G
\\
{\rm (NS+,R+)} & \qquad 8_v\times 8 &=& 8'+56 &=& {\rm spinor}'\,\,+\,\,{\rm vector\mbox{-}spinor}
\\
{\rm (R+,NS+)} & \qquad 8\times 8_v &=&  8'+56 &=& {\rm spinor}'\,\,+\,\,{\rm vector\mbox{-}spinor}
\\
{\rm (R+,R+)} & \qquad 8\times 8 &=& 1+28+35_+ &=& [0]_{C_0}+[2]_{C_2}+[4]_{+\,,\,C_4}
\end{array}\nonumber
\ee

The key differences are that this theory is chiral (a preference is given to one of the two different available chiralities of spinors and vector-spinors). Furthermore, the form-field and hence the brane content is different. It is easy to remember that type IIA and IIB theory contain odd and even $p$-form gauge potentials respectively and hence even and odd D$p$-branes. A further noteworthy specialty of the IIB theory is the fact that the $C_4$ theory is subject to a self-duality \index{self-duality} constraint, $F_5=* F_5$, which halves the number of degrees of freedom (cf.~the index `$+$' of $[4]_+$ and $35_+$).

\subsection{Other 10d theories}

The name type II refers to the two supersymmetries. There is also a minimally supersymmetric 10d superstring theory called {\bf type I}\index{type I string theory} with unoriented strings\index{unoriented strings}. It follows by modding out {\bf worldsheet parity}\index{worldsheet!parity}. By this one means introducing of an operator $\Omega$ which realises the classical transformation $\sigma\to -\sigma$ at the quantum level (hence $\Omega^2=1$) and projecting on its 1-eigenspace by 
\be
P=\frac{1}{2}(1+\Omega)\,.
\ee
A detailed analysis reveals that stability (`tadpole-cancellation') always requires the presence of 32 D9-branes, giving rise to gauge fields living in 10d. Due to the projection the group is not $U(32)$ but its `real subgroup', $SO(32)$. We will return to this when discussing such `orientifold projections' more generally in Sect.~\ref{asi}.

Furthermore, it is consistent (and allows for tachyon removal) to supersymmetrise only the left- or right-moving half of the worldsheet theory. For obvious reasons such theories are called heterotic and they come in two types, named after their non-abelian gauge group (which are present in both cases): {\bf heterotic $E_8\times E_8$} and {\bf heterotic $SO(32)$}\index{heterotic string theory}. The corresponding 10d supergravity theories are only ${\cal N}=1$ supersymmetric.

\begin{figure}[ht]
\begin{center} 
\includegraphics[width=10cm]{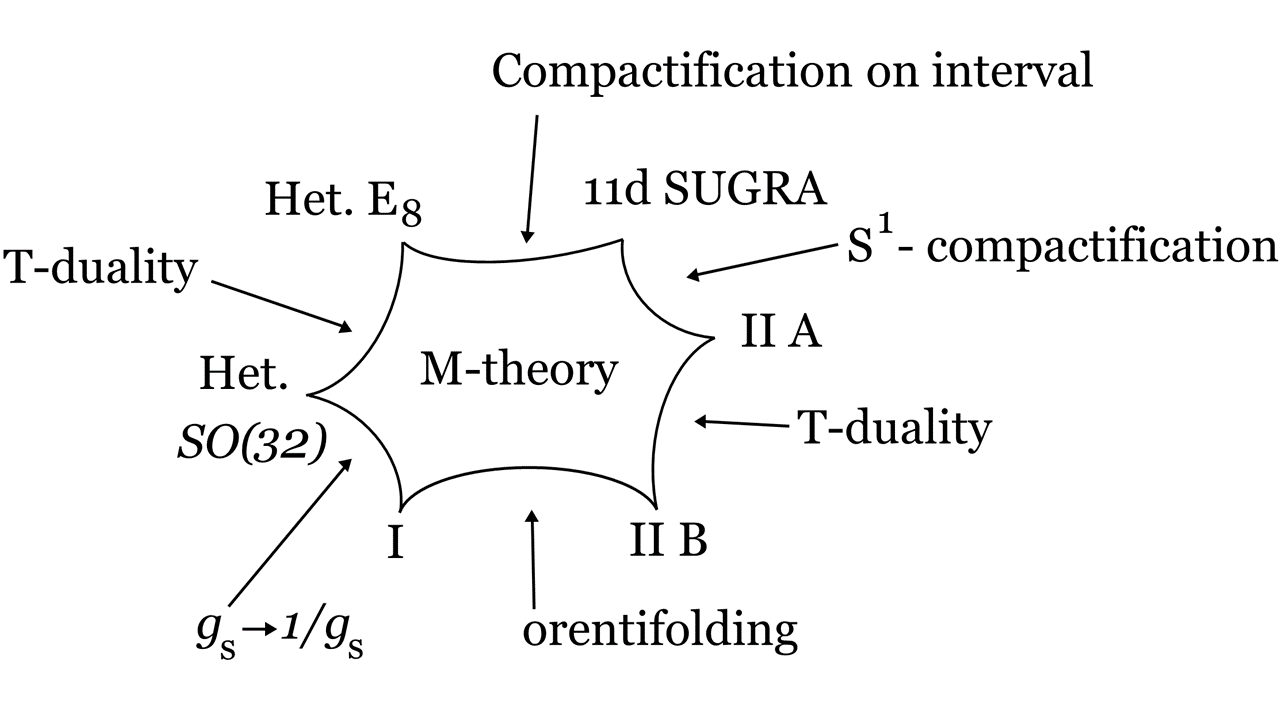}
\caption{Illustration of M-theory and its perturbative corners: the 5 superstring theories and 11d SUGRA.}
\label{mth} 
\end{center}
\end{figure}

Not surprisingly, the SO(32) heterotic theory is related to type I by a so-called {\bf duality}. In this particular case, it is a strong-weak duality, which\index{duality} means that type I at weak string coupling is identical to heterotic at strong coupling and vice versa. In fact, all of the five 10d theories above are related to each other by dualities, projections or compactifications (see~Fig.~\ref{mth}) and are sometimes referred to collectively as the (perturbative corners of) {\bf M-theory}.\index{M-theory}

One usually includes 11d supergravity in this set, although the fundamental objects there appear to be membranes (specifically {\bf M2-branes})\index{M2-branes} rather than strings and the theory is much less well understood in the ultraviolet. Occasionally, the name M-theory is also used to refer only to 11d-supergravity rather than to the whole set of theories. It is believed that these 6 theories are the calculable, perturbative corners of a more general and not yet fully understood structure: M-theory as `defined' by the inner region of the `amoeba' in Fig.~\ref{mth}. 

Two of the edges connecting neighbouring corners of the amoeba in Fig.~\ref{mth} have already been briefly mentioned in the main text. Two further edges are defined by different possibilities for compactifying from 11d to 10d, in one case with the introduction of two E$_8$ gauge theories localised at the boundaries of an interval. We now turn to the remaining two edges labelled {\bf T-duality}\index{T-duality}. This\index{duality} is a key concept in string theory deserving a whole section, but we have to limit ourselves to a short qualitative explanation for reasons of space. To be specific, focus on the connection between type IIA and IIB. In this case, the statement of T-duality amounts to the following:

Consider type IIA with coupling $g_s$ compactified on an $S^1$ of radius $R$ to 9d. At the same time, consider type IIB with coupling $g_s'$ compactified on an $S^1$ of radius $R'$. The two resulting 9d theories are exactly identical if 
\be
R'=\frac{\alpha'}{R} \qquad\mbox{and}\qquad g_s'=g_s\frac{\sqrt{\alpha'}}{R}\,.
\ee
In other words, one theory compactified on a very small (in string units) $S^1$ is equivalent to the other theory compactified on a large $S^1$. A simple first step in checking this statement is to identify the spectrum of states in the two models. This can be immediately realised with the tools already available to the reader. The key point is that the tower of so-called winding states\index{winding states} (the string wrapped around the $S^1$) on one side is mapped to the tower of Kaluza-Klein states (waves travelling around the $S^1$, cf.~Sect.~\ref{kkc}) on the other side. The winding number is mapped to the discrete momentum characterising the motion in a compact dimension. A more detailed (also 10d field-theoretic) set of rules for the identification, the Buscher rules\index{Buscher rules}, can be found in~\cite{Buscher:1987qj}. The procedure of T-dualisation can be iterated, such that e.g.~one theory on a small $T^6$ is identified with the {\it same} theory (because the number of dualisations is even) on a large $T^6$.  

A key conceptual point of T-duality is that it can be viewed as a manifestation of the fact that the stringy UV completion really abandons the concept of smooth spacetime at sub-stringy distances: Making some compact manifold small does not lead to a new sub-stringy geometry but rather to a familiar super-string-sized  geometry, possibly in a different theory.

\subsection{Problems}

\subsubsection{Explicit state-operator mapping in the free case}\label{esom}

{\bf Task:} Calculate explicitly the operators which, if inserted at $z=0$ in the radial description of the closed string, define the single-particle excited states created by $\alpha^\mu_{-m}$ with $m\ge 1$. 

\noindent
{\bf Hints:} Work with the euclidean (Wick-rotated) version of the theory, defining e.g.~$(\sigma^1,\sigma^2)=(\sigma^1,i\sigma^0)$. Write $w=\sigma^1+i\sigma^2$, such that the worldsheet cylinder corresponds to a vertical strip with width $\pi$ in the complex $w$ plane. Define $z=\exp(-2iw)$, such that constant-time cuts of the cylinder are mapped to circles in the $w$-plane. The origin of the $z$ plane now corresponds to the infinite past of the cylinder, $\sigma^0=-i\infty$. 

Express our mode expansion of $\partial_-X$ (we suppress the index $\mu$ for brevity) in terms of the variable $z$. Invert the result, expressing the oscillator modes in terms of integrals of $\partial X$ over a closed contour in the $z$ plane. 

Finally, use the expression obtained for a creation operator $\alpha_n$ under a path integral over fields on the $z$ plane. Assuming that the fields $X$ can be Taylor expanded in $z$ and $\ol{z}$ at the origin, obtain the desired expression for the vertex operators. Start by arguing why the vacuum state with momentum $p=0$ corresponds to the unit operator. 

\noindent
{\bf Solution:}
We start by rewriting our formula from the lecture (with $l=1$) as
\be
X_R=\frac{1}{2}x+\frac{1}{2}p\sigma_-+\frac{i}{2}\sum_{n\neq 0}\frac{1}{n}\,\alpha_n\,e^{-2in\sigma_-}
\ee
as
\be
\partial_- X=\sum_n\alpha_n\,e^{-2in\sigma_-}\,,
\ee
where we also set $p/2=\alpha_0$. Using $\sigma_-=\sigma^0-\sigma^1=-i\sigma^2-\sigma^1=-w$, this becomes
\be
-\partial_w X=\sum_n \alpha_n e^{2inw}\,.
\ee
Next, with $2w=i\ln z$, we have
\be
-2iz\partial_z X=\sum_n \alpha_n z^{-n}\qquad\mbox{or}\qquad
\partial_z X = \frac{i}{2}\sum_n\frac{\alpha_n}{z^{n+1}}\,.
\ee
The coefficient of $1/z^{n+1}$ is extracted, using the residue theorem, by performing a counter-clockwise contour integral with the measure $dz\,z^n/(2\pi i)$:
\be
\alpha_n=-2\oint\frac{dz}{2\pi}\,z^n\,\partial_z X\,.\label{alde}
\ee

In the above, $\alpha_n$ is an operator acting on a state, defined on one of the circles in the $z$ plane. Similarly, $X$ is a local field operator integrated over this circle. All of this is to be interpreted at the fixed radial time corresponding to this circle. Obviously, such an operator identity can be used under the path integral, with some operator inserted at $z=0$ to define the initial state and with the boundary conditions at $|z|\to\infty$ defining the final state. The latter will not be relevant for us and we will ignore them.

Start by inserting the unit operator at $z=0$ and calling the corresponding (so far unknown) state $|\Psi\rangle$:
\be
\lim_{t_i\to-\infty}e^{-(t_f-t)H}\,\alpha_n\,e^{-(t-t_i)H}\,|\Psi\rangle \quad\sim\quad  \int^{X_f(r_f)} DX\,e^{-S_P[X]}\oint\frac{dz}{2\pi}\,z^n\,\partial_z X\,.
\label{fifo}
\ee
Here the r.h.~side is to be interpreted as a functional depending on the boundary conditions $X=X_f$ at some largest circle of radius $r_f$. The l.h.~side is defined by evolving an initial state $|\Psi\rangle$ to the time $t$, corresponding to the radius $r$ of the contour integral on the r.h.~side, applying $\alpha_m$, and then evolving to the final time $t_f$ corresponding to $r_f$. The tilde means that we are not keeping track of normalisations.

It is immediately clear that, assuming that we integrate over well-behaved functions $X$, the r.h.~side vanishes for $n\ge 0$ since there are no appropriate poles inside the contour. But the state annihilated by all $\alpha_n$ with non-negative $n$ is, by definition, the vacuum: $|\Psi\rangle =|0,0\rangle$. 

Next, we consider creation operators, $\alpha_{-n}$ with $n>0$. We also use that the vacuum corresponds to the unit operator and repeat the step from (\ref{alde}) to (\ref{fifo}) for this case:
\be
\alpha_{-n}|0,0\rangle \quad\sim\quad  \int^{X_f(r_f)} DX\,e^{-S_P[X]}\,\oint\frac{dz}{2\pi \,z}\,\frac{1}{z^{n-1}}\,\partial_z X\,,
\ee
where now $n>0$. We have simplified the l.h.~side since, as noted, we do not keep track of the normalisation. Finally, we may Taylor expand $X(z,\ol{z})$ keeping only the term which will provide a non-zero contribution to the contour integral:
\be
\alpha_{-n}|0,0\rangle \quad\sim\quad  \int^{X_f(r_f)} DX\,e^{-S_P[X]}\,\frac{(\partial_z)^n X(0)}{(n-1)!}\,.
\ee
Thus, up to normalisation, $(\partial_z)^n X(0)/(n-1)!$ is our final result for the operator corresponding to the creation operator $\alpha_{-n}$.

\subsubsection{Euler number and genus of Riemann surfaces}
\index{Euler number}\index{Riemann surface}

{\bf Task:} Calculate explicitly the Ricci scalar ${\cal R}$ of a 2-sphere of radius $R$ and use this result to derive the formula
\be
\chi(\Sigma)=2-2g\label{chg}
\ee
for the Euler number 
\be
\chi(\Sigma)\equiv \frac{1}{4\pi}\int_\Sigma d^2\sigma\,\sqrt{\mbox{det}(g_{ab})}\,{\cal R}\,.\label{cint}
\ee
Here $g$ is the `number of holes' or `number of handles'' of the Riemann surface. 

\noindent
{\bf Hints:} Recall that the Riemann tensor in 2d is highly symmetric and that you hence do not need to calculate all components to obtain the Ricci scalar. In the second part of the problem, it will be sufficient if you give a `physicist's derivation', drawing lots of pictures and taking the existence of intuitively obvious limits for granted. 

\noindent
{\bf Solution:} We will use the standard parameterisation of the unit sphere by azimuthal and polar angle, such that
\be
ds^2=d\theta^2+\sin^2\theta\,d\phi^2\,.
\ee

Recalling our general 2d result
\be
{\cal R}_{abcd}=\frac{1}{2}(g_{ac}g_{bd}-g_{ad}g_{bc}){\cal R}\,,
\ee
from the discussion of the symmetries of the bosonic string, we have
\be
{\cal R}_{\theta\phi\theta\phi}=\frac{1}{2}\,g_{\theta\theta}g_{\phi\phi}\,{\cal R}=\frac{1}{2} \,\sin^2\theta\,{\cal R}\,.
\ee
Thus,
\be
{\cal R}=\frac{2}{\sin^2\theta}\,{\cal R}_{\theta\phi\theta\phi}=
\frac{2}{\sin^2\theta}\,{\cal R}_{\theta\phi\theta}{}^\phi\,g_{\phi\phi}
=2\,{\cal R}_{\theta\phi\theta}{}^\phi\,.
\ee

The required curvature coefficient can be obtained from the standard formula
\be
{\cal R}_{abc}{}^d=-\partial_a\Gamma_{bc}{}^d+\Gamma_{ac}{}^e\Gamma_{be}{}^d- \{\,a\,\leftrightarrow\,b\,\}\,.
\ee
It is explicitly given by
\be
{\cal R}_{\theta\phi\theta}{}^\phi=-\partial_\theta\Gamma_{\phi\theta}{}^\phi +\Gamma_{\theta\theta}{}^e\Gamma_{\phi e}{}^\phi+\partial_\phi \Gamma_{\theta\theta} {}^\phi-\Gamma_{\phi\theta}{}^e \Gamma_{\theta e}{}^\phi\,.
\ee
Using the standard formula
\be
\Gamma_{ab}{}^c=\frac{1}{2}g^{cd}(\partial_a g_{bd}+\partial_b g_{ad}-\partial_d g_{ab})
\ee
we calculate the Christoffel symbols
\bea
\Gamma_{\phi\theta}{}^\phi&=&\frac{1}{2}g^{\phi\phi}(\partial_\phi g_{\theta\phi}+\partial_\theta g_{\phi\phi}-\partial_\phi g_{\phi\theta}) = \frac{1}{2}g^{\phi\phi}\partial_\theta g_{\phi\phi}=\frac{\cos\theta}{\sin \theta}\,,
\\
\Gamma_{\theta\theta}{}^\phi&=&\frac{1}{2}g^{\phi\phi}(2\partial_\theta g_{\theta\phi} -\partial_\phi g_{\theta\theta}) =0\,,
\\
\Gamma_{\theta\theta}{}^\theta&=&0\,,
\\
\Gamma_{\phi\theta}{}^\theta&=&\frac{1}{2}g^{\theta\theta}(\partial_\phi g_{\theta\theta}+ \partial_\theta g_{\phi\theta} - \partial_\theta g_{\phi\theta}) = 0\,.
\eea
Here the zero result in the third line is obvious since the only non-zero derivative $\partial_\theta g_{\phi\phi}$ can not appear.

With this, we finally obtain
\be
{\cal R}_{\theta\phi\theta}{}^\phi=-\partial_\theta\left( \frac{\cos\theta}{\sin\theta}\right)-\frac{\cos^2\theta}{\sin^2\theta}=
1+\frac{\cos^2\theta}{\sin^2\theta}-\frac{\cos^2\theta}{\sin^2\theta}=1\,,
\ee
hence ${\cal R}=2$ for the unit sphere and
\be
{\cal R}=2/R^2
\ee
for a sphere of radius $R$.

Since the surface is $4\pi R^2$, we obtain the Euler number
\be
\chi(S^2)=2\,,
\ee
consistent with (\ref{chg}) and the absence of handles on a sphere.

\begin{figure}[ht]
\begin{center} 
\includegraphics[width=6cm]{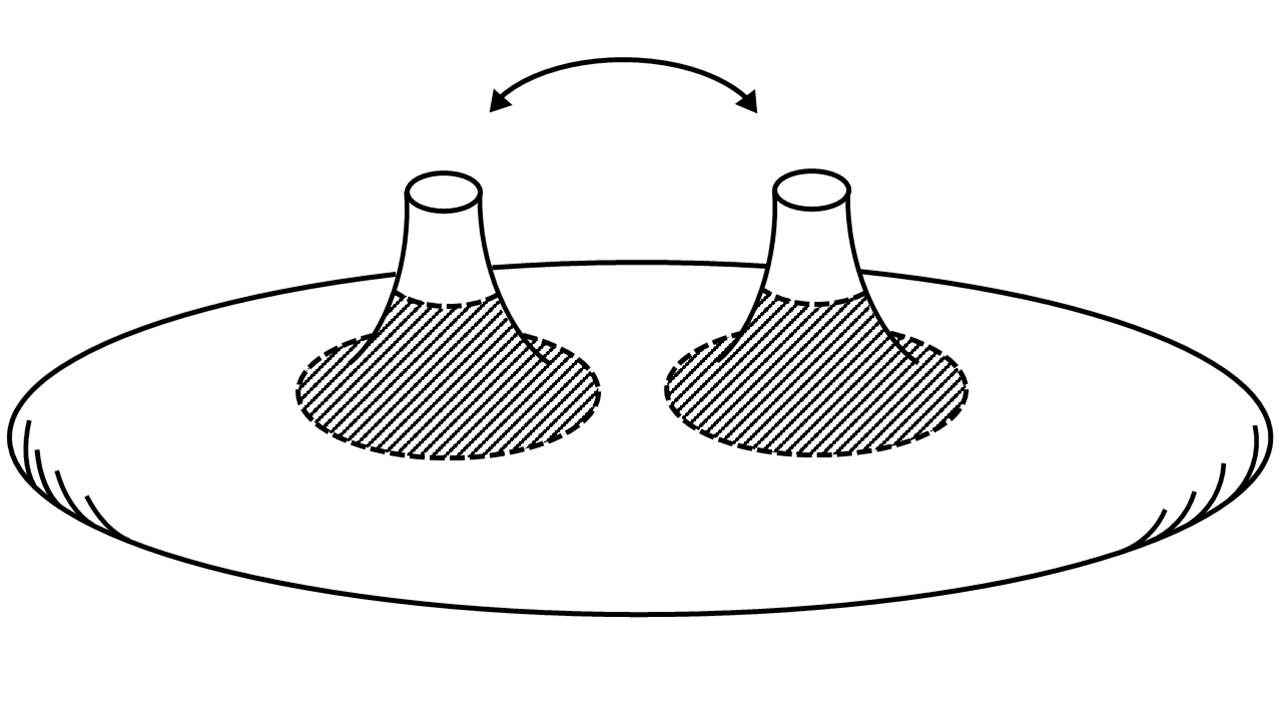}
\caption{
A torus deformed to a `pancake' with a handle attached. The handle is realised by two `smokestacks', to be identified at their edges. The only regions with non-zero curvature are at the edge of the pancake and in the hatched areas where the handle is attached. 
}
\label{pancake} 
\end{center}
\end{figure}

Now let us move on to the case of a torus, i.e. a sphere with one handle, $g=1$. On the one hand, it is clear that
\be
\chi(T^2)=0
\ee
since an explicit geometry with everywhere vanishing curvature can easily be given. On the other hand, one can deform the geometry to a `pancake' with a handle attached in its upper flat region, cf.~Fig.~\ref{pancake}. If there were no handle, the curvature integral in (\ref{cint}) would give $\chi=2$, with the only contribution coming from the edge of the pancake. With the handle, we know we get zero. Thus, the hatched regions where the handle is attached give a negative contribution of $-2$ to the curvature integral defining $\chi$. Obviously, further handles will give an identical negative contribution, demonstrating the correctness of the term $-2g$ in (\ref{chg}). 

\subsubsection{Dilaton vs. String Coupling}
\index{dilaton}\index{string!coupling}

{\bf Task:} Give an argument for identifying the dilaton in the $\exp(-2\phi)$ prefactor of the 26d Einstein-Hilbert term with the dilaton defined as the coefficient of the Einstein-Hilbert term on the worldsheet.

\noindent
{\bf Hint:} Think of the loop expansion parameters in 26d-field-theory and on the worldsheet.

\noindent
{\bf Solution:} Think of graviton-graviton scattering in 26d quantum gravity as a low-energy effective field theory. Work in the string frame and treat the dilaton as fixed to some background VEV $\phi_0$, plus small fluctuations which we will not be interested in. Expanding the metric as $\eta_{\mu\nu}+h_{\mu\nu}$ and rescaling $h_{\mu\nu}\to h_{\mu\nu}\kappa e^{\phi_0}$, we see that 3-vertices are proportional to $\kappa e^{\phi_0}$ and 4-vertices to $\kappa^2 e^{2\phi_0}$. Hence, for example, a 1-loop contribution to a given process is suppressed relative to the tree level by $\kappa^2 e^{2\phi_0}$. (Draw a few example diagrams to be certain.) Moreover, we can set $\kappa^2\sim M_s^{-24}$, with $M_s\sim 1/l_s\sim 1/\sqrt{\alpha'}$ the string scale, as explained in the lecture. Finally, we assume that the string scale provides the cutoff $\Lambda$ for the UV divergent loop diagrams. Thus, the ratio of 1-loop to tree level is
\be
\kappa^2 e^{2\phi_0} \Lambda^{24} \sim e^{2\phi_0}\sim g_s^2\,,
\ee
consistently with the expectation from the amplitude formula of the worldsheet analysis, under the assumption that $\phi_0$ governs the worldsheet Einstein-Hilbert term. Up to an additive redefinition of $\phi$, this identifies the two a priori different definitions of the dilaton.

\subsubsection{Elementary exercises with 2d spinors}
\label{so2r}

{\bf Task:} Make the action of SO(1,1) and SO(2) on vectors and spinors completely explicit, paying particular attention to how the transformations of vectors and spinors differ in the Lorentz case.

\noindent
{\bf Hints:} Fix the normalisation of generators by analogy to the familiar 4d case. Recall what you know from undergraduate special relativity. 

\noindent
{\bf Solution:} In the non abelian case ($SO(1,d-1)$ with $d>2$), the normalisation of the generators $J_{ab}$ is unambiguously fixed by the non-trivial Lie algebra relations
\be
[J_{ab},J_{cd}]=i(\eta_{bc}J_{ad}-\eta_{ac}J_{bd}-\eta_{bd}J_{ac}+
\eta_{ad}J_{bc})\,,
\ee
which implies
\be
(J_{ab})_{cd}=i(\delta_{ac}\delta_{bd}-\delta_{ad}\delta_{bd})\,.
\ee
In the $SO(1,1)$ (and similarly in the $SO(2))$) case, the Lie algebra is trivial and does not fix the normalisation. We still use the general-$d$ definition, such that
\be
(J_{01})_{ab}=i\left(\begin{array}{rr} 0 & 1 \\ -1 & 0 \end{array}
\right)_{ab}\qquad\mbox{and}\qquad (J_{01})^a{}_b=i\left(\begin{array}{rr} 0 & -1 \\ -1 & 0 \end{array}
\right)^a\!\!\begin{array}{c}{} \\ {}_b \end{array}\,.
\ee
Hence a boost specified by $\epsilon^{01}=-\epsilon_{10}=\alpha/2$ explicitly reads
\be
\exp(i\epsilon^{ab}J_{ab})=\exp\left(\begin{array}{rr} 0 & \alpha \\ \alpha & 0
\end{array} \right)
=\left(\begin{array}{rr} \cosh\alpha & \sinh \alpha \\ \sinh \alpha & \cosh \alpha \end{array} \right) \,.
\ee
The last equality follows, e.g., from its obvious infinitesimal version together with the group property, which in turn follows from the well-known formulae for $\cosh(\alpha+\beta)$ and $\sinh(\alpha+\beta)$. This is where remembering undergraduate special relativity is useful.

Moreover, it is convenient to switch from the coordinates $x^{0,1}$ underlying the above formulae to light-cone coordinates, $x^{\pm}=x^0\pm x^1$.
One has $x'^+=x'^0+x'^1= x^0\cosh\alpha + x^1\sinh\alpha + x^0\sinh\alpha + x^1\cosh\alpha = x^+\exp\alpha$ and similarly for $x'^-$. Hence,
\be
\exp(i\epsilon^{ab}J_{ab})=\left(\begin{array}{cc} e^\alpha & 0 \\ 0 & e^{-\alpha} \end{array} \right)
\ee
in that basis.

Next, we have
\be
\frac{1}{4}[\gamma_0,\gamma_1]=\frac{1}{2}\left(\begin{array}{rr}
-1 & 0 \\ 0 & 1 \end{array} \right)
\ee
and hence
\be
S(\alpha)=\exp(i\epsilon^{ab}\{i[\gamma_0,\gamma_1]/4\})=
\left(\begin{array}{cc} e^{\alpha/2} & 0 \\ 0 & e^{-\alpha/2} \end{array} \right)\,.
\ee
We now see explicitly how $SO(1,1)$, here realised as $\mathbb{R}$ with addition as the group operation, is represented in two different ways on vectors and spinors.

Repeating the analysis for $SO(2)$, we now label the coordinates by $1,2$ rather than $0,1$ since no special role is played by $x^0=t$. The lower-index version of $J$, now called $(J_{12})_{ab}$, remains unchanged. The upper-lower version reads
\be
(J_{12})^a{}_b=i\left(\begin{array}{rr} 0 & 1 \\ -1 & 0 \end{array}
\right)^a\!\!\begin{array}{c}{} \\ {}_b \end{array}
\ee
and hence 
\be
\exp(i\epsilon^{ab}J_{ab})=
\left(\begin{array}{rr} \cos\alpha & \sin \alpha \\ -\sin \alpha & \cos \alpha \end{array} \right) \,.
\ee
The correct Clifford algebra is obtained if $\gamma_0$ is multiplied by `$i$', such that dilation by $e^{\alpha/2}$ becomes a phase rotation by half of the $SO(2)$ rotation angle:
\be
S(\alpha)=\exp(i\epsilon^{ab}\{i[\gamma_0,\gamma_1]/4\})=
\left(\begin{array}{cc} e^{i\alpha/2} & 0 \\ 0 & e^{-i\alpha/2} \end{array} \right)\,.
\ee
This was, of course, expected.

\subsubsection{SUSY algebra in 2d}

{\bf Task:} Check the 2d SUSY algebra given in the lecture using the explicit definitions of $Q$ and $\ol{Q}$.

\noindent
{\bf Hints:} Show that $\ol{\psi}\chi=\ol{\chi}\psi$ if we impose a Majorana condition on our 2d spinors. Then work out explicitly what $\ol{Q}^\alpha$ is in terms of $\ol{\theta}$ and $\partial/\partial\theta$. It is convenient to think of the action of bilinears like $(\partial/\partial\theta_\alpha)\epsilon_\alpha$. After these preliminaries, write down the commutator of $\ol{\epsilon}Q$ and $\ol{Q}\eta$, which is equivalent to the SUSY algebra (as you already learned in 4d). 

\noindent
{\bf Solution:} Let us start by checking that  $\ol{\psi}\chi=\ol{\chi}\psi$ for Majorana spinors. In our case, Majorana spinors are simply spinors with real components. Hence,
\be
\ol{\psi}\chi=\psi^\dagger\gamma^0\chi=
\left( \begin{array}{c} \psi_- \\ \psi_+ \end{array} \right)^T
\left( \begin{array}{rr} 0 & -i \\ i & 0 \end{array} \right)
\left( \begin{array}{c} \chi_- \\ \chi_+ \end{array} \right)
= i(\psi_+\chi_- - \psi_-\chi_+) = i (\chi_+\psi_- - \chi_-\psi_+)
= \ol{\chi}\psi\,.\label{majd}
\ee
One may think of this relation as of an alternative definition of what a Majorana spinor is.

Next, we want to understand how the formal $*$-operation must act on $\partial/\partial\ol{\theta}$ for $Q$ to be Majorana. For this purpose, consider
\be
\left(\ol{\epsilon}^\alpha\frac{\partial}{\partial\ol{\theta}^\alpha}\right)
(\ol{\theta}\psi) = \ol{\epsilon}\psi \label{maj1}
\ee
and
\be
\left(\frac{\partial}{\partial\theta_\alpha}\epsilon_\alpha\right)(\ol{\psi} \theta) = \epsilon_\alpha \ol{\psi}^\beta \delta_\beta{}^\alpha= - \ol{\psi}\epsilon = - \ol{\epsilon}\psi\,.\label{maj2}
\ee
We see that, with the definition 
\be
\ol{\left(\frac{\partial}{\partial\ol{\theta}}\right)}=-\left(
\frac{\partial}{\partial\theta} \right)\,,
\ee
the spinor $\partial/\partial\ol{\theta}$ is a Majorana spinor in the sense that it obeys relations like (\ref{majd}). Note that this peculiar minus-sign is similar to the one encountered in 4d SUSY.

Also, we have
\be
\ol{i(\slashed{\partial})\theta)}=-i\theta^\dagger\gamma^0\gamma^0 \slashed{\partial}^\dagger\gamma^0=-i\ol{\theta}\slashed{\partial}\,.
\ee
Using the definition of $Q_\alpha$ from the lecture, this gives
\be
\ol{Q}^\alpha=-\frac{\partial}{\partial\theta_\alpha}-i(\ol{\theta}
\slashed{\partial})^\alpha = -\frac{\partial}{\partial\theta_\alpha}-i(\ol{\theta}
\gamma^a)^\alpha\,\partial_a\,.
\ee
It is clear that our $Q$ is Majorana: The first term is Majorana according to what was said above, the second term is Majorana because $\theta$ is Majorana and both $\partial_a$ and $i\gamma^a$ are real.

Now the actual calculation is easy. Using two Majorana SUSY parameters $\epsilon$ and $\eta$, we have
\be
[\ol{\epsilon}Q,\ol{Q}\eta] = \left[ \ol{\epsilon}^\alpha
\frac{\partial}{\partial\ol{\theta}^\alpha}+i\ol{\epsilon}\slashed{\partial}\theta\,,\,
-\frac{\partial}{\partial\theta_\beta}\eta_\beta-i\ol{\theta}\slashed{\partial} \eta \right] = -2i\epsilon\slashed{\partial}\eta\,.
\ee
Crucially, since $Q$ and $\ol{Q}$ are not independent, there are no additional $QQ$ or $\ol{Q}\ol{Q}$ relations.

\section{10d actions and compactification}\label{comp}
\index{supergravity!10d}

\subsection{10d supergravities and Type IIB as an example}

The existence of a field-theoretic target space action and its fundamental relation to the worldsheet definition of the theory has already been discussed in the context of the bosonic string. All that was said there remains true. In particular, there are always the 10d graviton and $B_2$ field, coupling to the worldsheet (except in the case of the unoriented type I string, where it falls victim to the projection taking us from type IIB to type I). There is also always the dilaton, governing the convergence of perturbation theory. Together, dilaton, graviton and $B_2$ form the NS-NS sector (see above). As a novelty, one has the $C_{p+1}$ (or R-R) form fields and the corresponding D$p$-branes. These are dynamical objects, just like the string itself, but with different dimensionality and (at weak string coupling) larger tension. This analysis would be slightly different in the heterotic case, where there are no $C$-forms but rather gauge fields. Crucially, there are now also fermionic partners for all the fields above. 

What is very different from the bosonic case is the uniqueness status of the above five 10d theories. In the bosonic case, we found one of many possible, similar 26d field theories. For example, without the stringy definition, one could just add another scalar or modify some coupling. Here, by contrast, our five 10d theories are very special and (at the 2-derivative level) even unique, independently of their string-theoretic underpinning. This is due to supergravity. Indeed, supersymmetry (and supergravity) exists in various dimensions (cf.~Appendix of volume II of~\cite{Polchinski:1998rq}), but its realisation becomes harder and harder as the dimension grows. This can be roughly understood by noting that the spinor dimension grows exponentially with $D$, making it more difficult to find a matching bosonic structure. Thus, it turns out that there exist precisely four supergravity theories in 10d, and all of them can be viewed as coming from the five superstring constructions we discussed (recall that  heteretotic $SO(32)$ and type-I give the same field theory).

Let us start by noting that for even $D$ one has
\be
\Gamma_0,\cdots, \Gamma_{D-1} \qquad \mbox{and}\qquad \Gamma\equiv \Gamma_0\cdot \Gamma_1\cdot \cdots \cdot \Gamma_{D-1}\,,
\ee
allowing us to define chirality through the projector $(1+\Gamma)/2$. In the dimension $(D+1)$, which is now odd, $\Gamma$ becomes the highest `usual' gamma matrix and the product of all gamma matrices becomes $\Gamma\cdot\Gamma \sim \mathbbm{1}$. Hence, chirality\index{chirality} can not be defined. 

In some dimensions (see~\cite{Polchinski:1998rq}) there is a Majorana spinor and, if both Weyl\index{Weyl!spinor} and Majorana spinors\index{Majorana spinor} exist independently, it is sometimes possible to impose both constraints together. We have seen that this happens in $D=2$, where the naive spinor dimension is $2^{D/2}=2$, i.e. 4 real d.o.f., and we found spinors with one real component. 

This situation occurs again in $D=10$, where the Dirac spinor has 32 components and a 16-component real spinor exists. This spinor has 4 times the degrees of freedom of a minimal 4d spinor, hence the minimal 10d SUSY is referred to as ${\cal N}=4$ SUSY in 4d language. One may also characterise this as 10d ${\cal N}=1$ SUSY. We have encountered a gauge theory with this amount of supersymmetry when we quantised the open superstring in 10d. This gauge theory (more precisely its non-abelian version) can be coupled to supergravity and it is the SUSY of the heterotic and type I theories. It is also possible to have 10d ${\cal N}=2$ supergravity (corresponding to ${\cal N}=8$ in 4d language). Gauge fields can not be added to such theories. This is the SUSY of the type II string. As already noted, SUSY is so constraining that (under very reasonable assumptions) these four theories can be shown to be {\bf the only 10d supergravities}. This uniqueness includes the gauge group -- only $E_8\times E_8$ and $SO(32)$ are possible on anomaly cancellation grounds \cite{Green:1984sg}. It is very intriguing that precisely these four 10d supersymmetric field theories are realised in string theory. All these theories are unique also in the sense that no dimensionless parameters are present. An equally unique supergravity theory exists in 11d - it is the 11d theory linked to type IIA via compactification on $S^1$ as noted earlier. No other supergravity theories above 9 dimensions are known.

All of them have been tried as starting points for a stringy description of the real world. The landscape, i.e. a very large number of potentially suitable 4d models has been most convincingly established in type IIB (although there are still reasonable doubts, to which we will come). We hence focus on this theory.

In the widely used conventions of~\cite{Polchinski:1998rq, Giddings:2001yu}, the bosonic part of the string-frame {\bf type IIB} action reads
\be
S=\frac{1}{2\kappa_{10}^2}\int d^{10}x\sqrt{-g}\left\{
e^{-2\phi}\left[{\cal R}+4(\partial\phi)^2-\frac{1}{2\cdot 3!}H_3^2\right]-\frac{1}{2}F_1^2-\frac{1}{2\cdot 3!}\tilde{F}_3^2-\frac{1}{4\cdot  5!}\tilde{F}_5^2\right\}+S_{CS}+S_{loc}\,.
\ee
Here $2\kappa_{10}^2=(2\pi)^7\alpha'^4$ and
\be
\tilde{F}_3=F_3-C_0\wedge H_3\,\,,\qquad \tilde{F}_5=F_5-\frac{1}{2}C_2\wedge H_3+\frac{1}{2}B_2\wedge F_3\,.
\ee
The RR-form field strengths with a tilde are gauge invariant (as is $H_3$). This implies special gauge transformation properties of some of the potentials, e.g.
\be
C_2\to C_2+d\lambda_1\qquad\mbox{goes together with}\qquad C_4\to C_4+\frac{1}{2}\lambda_1\wedge H_3\,.
\ee
Furthermore, terms which do not involve the metric are often referred to as Chern-Simons terms\index{Chern-Simons term}. In our case it reads
\be
S_{CS}=-\frac{1}{4\kappa_{10}^2}\int C_4\wedge H_3\wedge F_3\,.
\ee
Finally, we collect the actions of the various branes (including extended fundamental strings) which may be present in the target space and are described by the `localised' part $S_{loc}$. We just display the example of a D3-brane
\be
S_{loc}\supset S_{D3}=\frac{1}{2\pi^3\alpha'^2}\int_{D3}C_4\,\,-\,\,\int_{D3} d^4\xi\, \sqrt{-g}\,T_3\qquad\mbox{with}\qquad T_3=\frac{e^{-\phi}}{(2\pi)^3\alpha'^2}\,.
\ee
The first part of $S_{D3}$ may also be called a Chern-Simons-type term since the metric does not enter. The coordinates $\xi$ parameterise the world-volume of the brane and the metric next to them is the pullback of the 10d metric. Analogous expressions for the other odd-dimensional D$p$-branes and the string have to be added. The general formula for the tension appearing in the $S_{Dp}$ is
\be
T_p=\frac{e^{-\phi}}{(2\pi)^p\alpha'{}^{(p+1)/2}}
\qquad \mbox{or, in 10d Einstein frame,}\qquad
T_{p, {\rm E}}=\frac{e^{(p-3)\phi/4}}{(2\pi)^p\alpha'{}^{(p+1)/2}}\,.
\ee
This $S_{loc}$ is not yet complete: It should be extended by including the brane-localised (open-string-derived) gauge fields and their fermionic partners. The resulting so-called {\bf DBI}\index{DBI action} or {\bf Dirac-Born-Infeld}\index{Dirac-Born-Infeld action} action\footnote{
The reader may be surprised to see these names from the pre-stringy era of physics. It is their work on the closely related non-linear extensions of electrodynamics which is honoured here.
} 
also includes the pullpack of $B_2$ to the brane. Moreover, there are further, brane-localised Chern-Simons terms. For us, it is sufficient to record that, at leading order, the gauge fields come in simply through a brane-localised ${\cal N}=4$ $U(1)$ or, in the case of a brane-stack, $SU(N)$ gauge theory lagrangian. For many more details see e.g.~\cite{Johnson:2000ch, ortin}.

We do not display the completely analogous expression for type IIA, where the relevant RR form fields are $C_1$ and $C_3$. We only note that the non-localised CS term takes the form
\be
\int B_2\wedge F_4\wedge F_4\,.
\ee

The action for type I follows from that of type IIB upon a so-called orientifold projection, i.e., a projection on states invariant under worldsheet-parity inversion.  In 10d, this implies the removal of $C_0$, $B_2$ and $C_4$. Furthermore, 32 D9-branes have to be added, also subject to a certain projection, which restricts the gauge group to SO(32). Thus, one basically includes the lagrangian of the corresponding 10d super-Yang-Mills (SYM) theory.

Finally, in the heterotic case one removes the $C$-forms (keeping $B_2$) and adds SYM lagrangians with groups $E_8\times E_8$ or $SO(32)$. It is then clear that the advertised duality between the type-I and the heterotic $SO(32)$ theory also involves the exchange of the $F_3$ and $H_3$. 

We recall again that the fermionic parts also differ strongly between the various theories, given in particular that SUSY is reduced to 10d ${\cal N}=1$ in all but the two type II theories.

\subsection{Kaluza-Klein compactification}\label{kkc}
\index{Kaluza-Klein!compactification}

One has thus arrived at a possibly fundamental and (involving the various dualities above) unique 10d theory. To describe a 4d world on this basis, the logical procedure is to employ the idea of Kaluza-Klein compactification. This method of obtaining lower from higher-dimensional theories is old and has, as we will see, some appeal in its own right~\cite{Nordstrom:1988fi, Klein:1926tv, Klein:1926fj, Appelquist:1987nr, Duff:1994tn, Overduin:1998pn}.

Let us start with what may be the simplest example: a 5d scalar field on $M=\mathbb{R}^4\times S^1$, where the $S^1$ has radius $R$ (such that $x^5\in (0,2\pi R)$):
\be
S=\int_M d^5x\,\frac{1}{2}(\partial_M\phi)(\partial^M\phi)\,,\,\qquad M\in\{0,1,2,3,5\}\,.
\ee
We take $\phi=0$ (in fact any other value, $\phi=\,$const., would be equally good) as our vacuum and parameterise fluctuations around this solution according to
\be
\phi(x,y)=\sum_{n=0}^\infty\phi^c_n(x)\cos(ny/R)+\sum_{n=1}^\infty\phi^s_n(x)
\sin(ny/R)\,.
\ee
Here we renamed $x^5$ according to $x^5\to y$ and we use the argument $x$ as $x=\{x^0,x^1,x^2,x^3\}$. One immediately finds
\be
S=2\pi R\int d^4x \left[ \frac{1}{2}(\partial\phi_0^c)^2+\frac{1}{4} \sum_{n=1}^\infty\left\{ (\partial\phi_n^c)^2+m_n^2(\phi_n^c)^2+
(\partial\phi_n^s)^2+m_n^2(\phi_n^s)^2 \right\} \right]\,,
\ee
with $m_n=n/R$. Hence, our model is exactly equivalent to a 4d theory with one massless field and a (doubly degenerate) tower of KK modes.\index{KK mode} The massless mode parameterises a `flat direction', i.e. it is not only massless but has no potential at all. It can hence take an arbitrary constant value, which would still be a solution. Such a field is called a {\bf modulus}.\index{modulus}

We will frequently encounter cases where the value of the modulus governs the masses and couplings of the rest of the 4d theory. To create such a situation in our toy model, enrich our theory by 5d fermions and introduce the 5d coupling
\be
\lambda \phi\ol{\psi}\psi\,.
\ee
It is an easy exercise to derive the 4d action as above and read off explicitly how the 4d fermion masses depend on the VEV of $\phi$. Now $\phi$ is more like one of the moduli we will encounter in more realistic cases below. 
We note, however, that this `modulus' has a serious problem: It will acquire a mass from loop corrections even in the 5d local lagrangian. In this sense it is really not a proper modulus. We will see better examples below. 

Indeed, let us now turn to the historical example which is most directly associated with the word Kaluza-Klein theory.\index{Kaluza-Klein!theory} Consider pure general relativity in 5d,
\be
S=\frac{M_{P,5}^3}{2}\int d^4x\,dy\,\sqrt{-g_5}\,{\cal R}_5\,,
\ee
parameterise the metric as
\be
(g_5)_{MN}=\left(\begin{array}{cc}g_{\mu\nu}+(2/M_P^2)\phi^2 A_\mu A_\nu & (\sqrt{2}/M_P)\phi^2 A_\mu\\ (\sqrt{2}/M_P)\phi^2 A_\nu& \phi^2 \end{array}\right)\,.\label{mean}
\ee
where
\be
M,N,\cdots\in\{0,1,2,3,5\}\qquad\mbox{and}\qquad \mu,\nu,\cdots\in \{0,1,2,3\}\,.
\ee
Here $M_{P,5}$ is the 5d reduced Planck mass while $M_P$ is at the moment just a parameter which, together with the fields $A_\mu$ and $\phi$ is used to characterise all components of the 5d metric.

As above, we assume that $y\in (0,2\pi R)$ parameterises an $S^1$ and we base our analysis on the solution $g_{\mu\nu}=\eta_{\mu\nu}$, $A_\mu=0$ and $\phi^2=g_{55}=1$. From on our scalar-field example, we expect that the Fourier decomposition of all fields as functions of $y$ will give a 4d theory with a tower of massive modes. Focussing on the zero-modes only corresponds to assuming that all fields are independent of $y$. Under this assumption, it is straightforward to work out the higher-dimensional action, i.e. the 5d Ricci scalar, in terms of the ansatz (\ref{mean}). The result reads~\cite{Appelquist:1987nr}
\be
S=\int d^4x\,\sqrt{-g}\,\phi\,\left(\,\frac{M_P^2}{2}\,{\cal R}-\frac{1}{4}\phi^2 F_{\mu\nu} F^{\mu\nu}+\frac{M_P^2}{3} \,\frac{(\partial \phi)^2}{\phi^2} \right)\,, \label{4da}
\ee
which can of course be brought to the Einstein frame by $g_{\mu\nu}\to g_{\mu\nu}/\phi$. 

The key lessons are that the (zero modes of the) 5d metric degrees of freedom have turned into the 4d metric, an abelian gauge field and a scalar. The appearance of a $U(1)$ gauge theory is not surprising since our starting point, the $\mathbb{R}^4\times S^1$ geometry, clearly has a global $U(1)$ symmetry. But, since we are in general relativity and our starting point is diffeomorphism invariant, we are also allowed to rotate the $S^1$ (i.e. shift $y$) differently at every point $x$. Hence, our symmetry must actually be a $U(1)$ gauge symmetry. 

Moreover, we have a solution of the 5d Einstein equations for every fixed radius $R$. Thus, we expect a scalar degree of freedom, corresponding to $R$, with an exactly flat potential. This degree of freedom is the scalar field $\phi$. Note that it is sometimes convenient to parameterise the $S^1$ by the dimensionless variable $y\in (0,1)$ and correspondingly to have $\phi=\sqrt{g_{55}}=2\pi R$ in the vacuum. We note that, while $M_P$ was originally introduced as a parameter in the metric ansatz, the result (\ref{4da}) used the identification
\be
M_P^2=2\pi R\, M_{P,5}^3\,.
\ee

Before closing this generic Kaluza-Klein section, it will be useful to consider yet another example: Let the geometry again be $\mathbb{R}^4\times S^1$ and let the 5d lagrangian contain a $U(1)$ gauge theory. For simplicity, we will focus on the Kaluza-Klein or dimensional reduction of this $U(1)$, ignoring the 5d gravity part which we just discussed. Thus, we start with
\be
S=\int d^4x\,dy\,\left(-\frac{1}{4g_5^2}\, F_{MN}F^{MN}\right)\,,
\ee
where $g_{MN}=\eta_{MN}$ and $y\in (0,2\pi R)$. With the ansatz 
\be
A_M=(A_\mu,\phi)
\ee
one finds, at the zero-mode level,
\be
S=\int d^4 x\,\left(-\frac{1}{4g^2}\,F_{\mu\nu}F^{\mu\nu}-\frac{1}{2g^2}(\partial \phi)^2 \right)\,.
\ee
Here $1/g^2\equiv 2\pi R/g_5^2$. The crucial lesson is that the 5d gauge field gives rise to a 4d gauge field and a scalar, the latter being associated to $A_5$ or, in a more geometrical language, to the {\bf Wilson line} integral\index{Wilson line}
\be
\oint A=\int dy\,A_5=2\pi R\,\phi(x)\,.
\ee
This Wilson line measures the phase which a charged particle acquires upon moving once around the $S^1$, just as in the Aharonov-Bohm experiment. Assuming that the minimally charged particle (we do not display the corresponding part of the lagrangian) has unit charge, the phase measured by $\phi$ becomes equivalent to zero for $\phi=1/R$. Thus, we have found an exactly massless (at the classical level) periodic scalar field, also known as an {\bf axion}\index{axion} or {\bf axion-like particle}\index{axion-like particle} or {\bf ALP}.\index{ALP}

Let us draw a lesson from the above which will also be important for string compactifications, to be discussed below: We have seen two types of moduli arise, one associated to the geometry of the compact space ($g_{55}$), the other to the gauge field configuration in the compact space ($A_5$). Both have no classical potential since in one case 5d diffeomorphism invariance, in the other case 5d gauge invariance forbid the corresponding potential term. Moreover, due to this symmetry argument 5d loop corrections do not induce such a potential. However, in both cases 4d loop corrections can provide a potential and hence give a mass to the above fields. This is not in contradiction to the symmetry argument just stated since 4d loop effects can in general not be written in terms of 5d local operators. However, in the presence of enough supersymmetry in the resulting 4d theory, these loop corrections may vanish such that the relevant moduli remain exactly massless or more precisely, their potential remains identically zero as an exact statement. This generally happens in 4d ${\cal N}=2$ SUSY.

\subsection{Towards Calabi-Yau manifolds}
\index{Calabi-Yau manifolds}

We now want to explain how the 10d SUGRA theories provided by the superstring can be compactified to 4d. There are two approaches: We could start by developing the toy model path started in the previous section, i.e., we could consider the geometry $\mathbb{R}^9\times S^1$. This would give us a 9d theory, without too many new features (except for supersymmetry, which would keep all moduli massless). Next, we could consider $\mathbb{R}^8\times T^2$. We would now encounter moduli corresponding to $g_{88}$, $g_{99}$ and $g_{89}$, characterising both size and the shape of the torus. Thus, we would get an 8d theory with (at least) 3 scalars corresponding to geometric moduli. Much more could be said about compactifications on tori and related simple geometries. 

However, we will take a different approach and first introduce a much more general and powerful set of examples - the Calabi-Yau geometries. These are the compactification spaces on which the landscape as we presently understand it is mostly built. Later on, we will return to tori to illustrate some of the more abstract concepts used. 

Our key starting point is the desire to find a {\bf solution} of the 10d equations of motion corresponding to a 4d world. Setting all fields except the metric to zero, this implies that we must have $({\cal R}_{10})_{MN}=0$ to solve Einstein's equations. This condition is called {\bf Ricci flatness}\index{Ricci flatness} and it is obviously satisfied for $S^1$ and the (flat) tori $T^n$ mentioned above.  The interesting and non-trivial fact is that there exists a large class of relatively complicated compact 6d manifolds which are also Ricci flat and hence represent suitable compactification spaces: {\bf the Calabi-Yau manifolds}.

Before giving the definition, we need a few geometric concepts. Our treatment will be extremely brief and hence, unfortunately, superficial. Much more material can be found e.g. in~\cite{Green:1987sp, Blumenhagen:2013fgp, Candelas:1987is, Greene:1996cy, hub, He:2018jtw, Anderson:2018pui}.

To begin, Calabi-Yau manifolds are complex manifolds\index{complex manifold}. This is a fairly straightforward generalisation of the familiar concept of a 2n-dimensional real differentiable manifold $X$. The key new point is that the charts
\be
(U_i,\phi_i)\,\,,\qquad \phi_i:\,U_i \to \phi_i(U_i)\subset \mathbb{C}^n\,,
\ee
are now maps from open sets $U_i$ of $X$ to $\mathbb{C}^n$, with the key compatibility condition being that the functions $\phi_j\circ \phi_i^{-1}$ are holomorphic. In other words, our manifold locally looks like $\mathbb{C}^n$ and coordinate changes are of the form 
\be
z'^i=z'^i(z^1,\cdots,z^n)\,,
\ee
with any appearance of $\ol{z}^{\ol{\imath}}$ in the argument of the new coordinate excluded. 

On a complex manifold, it makes sense to complexify tangent and cotangent space as well as all their higher tensor products. Thus, tensor fields are complex. For example, local bases of tangent and cotangent space are provided by
\be
\frac{\partial}{\partial z^i}\,,\,\,\frac{\partial}{\partial \ol{z}^{\ol{\imath}}}\qquad\mbox{and}\qquad dz^i\,,\,\,d\ol{z}^{\ol{\imath}}\,,
\ee
with $z^i=x^i+iy^i$ etc. It is natural to define the tensor
\be
J=i\,dz^i\otimes\frac{\partial}{\partial z^i}-i\,d\ol{z}^{\ol{\imath}}\otimes
\frac{\partial}{\partial \ol{z}^{\ol{\imath}}}\,,
\ee
which may be obviously be interpreted as a map $T_p^*\to T_p^*$ for every $p\in X$. It roughly speaking corresponds to `multiplication by $i$' in cotangent space. Its components are
\be
J=\left(\begin{array}{cr}i\mathbbm{1} & 0 \\ 0 & -i\mathbbm{1} \end{array}\right)
\ee
in a complex basis and
\be
J=\left(\begin{array}{cr} 0& \mathbbm{1} \\ -\mathbbm{1} & 0 \end{array}\right)
\ee
in a real basis. A crucial feature is $J^2=-\mathbbm{1}$. 

A real manifold with a tensor $J$ as above is called an almost complex manifold and $J$ is called an almost complex structure. If such a $J$ satisfies a certain integrability condition (vanishing of the Nijenhuis tensor), complex coordinates can be given and $J$ turns into the so-called complex structure of a complex manifold. We will only be interested in this latter case. 

Even more, we will demand that our manifold has a metric (is a Riemannian manifold) and that this metric is compatible with $J$. In other words, we demand that $J$ is covariantly constant. This turns the manifold into a Kahler manifold\index{Kahler!manifold}, a concept which we already used when discussing field spaces of supersymmetric field theories. We will not demonstrate this but give right away a stronger definition: A complex manifold with a metric is called Kahler if the metric can locally be written as
\be
g_{i\ol{\jmath}}=\frac{\partial^2 K}{\partial z^i\,\partial\ol{z}^{\ol{\jmath}}}
\,,
\ee
with $K$ a real function defined in every patch and with $g_{ij}=g_{\ol{\imath}\ol{\jmath}}=0$. We note that this last condition by itself would make the metric hermitian, but we are interested only in the stronger Kahler condition. 

We also note that the metric allows us to lower the second index of $J$, turning $J$ into a rank-2 lower-index tensor. This tensor turns out to be antisymmetric and hence defines a 2-form, the so-called {\bf Kahler form}\index{Kahler!form}
\be
J=ig_{i\ol{\jmath}}\,dz^i\wedge d\ol{z}^{\ol{\jmath}}\,.
\ee
We see that, given a complex structure, the 2-form $J$ determines the metric and vice versa. This will become important below when we will be discussing different metrics on the same differentiable manifold. 

Next, we need the concept of {\bf holonomy}\index{holonomy}. We know from basic differential geometry that, with a metric, one gets a unique Riemannian or Levi-Civita connection and hence the possibility to parallel-transport tangent vectors. Given any point $p\in X$ and any closed curve $C$ beginning and ending in $p$, 
we hence have a linear map 
\be
R(C):\,T_p\to T_p\qquad \mbox{or} \qquad R(C)\in SO(2n)\,.
\ee
The latter statement follows if we assume orientability (for complex manifolds this is guaranteed) and recall that the Riemannian parallel transport does not change the length of a vector. It can be shown that the set of all $R(C)$ forms a group and that this group does not depend on the choice of $p$ (assuming $X$ is connected). This is the holonomy group.

We are now in the position to give one (of the many equivalent) definitions of a Calabi-Yau manifold: {\bf A Calabi-Yau 3-fold\index{3-fold} (our case of interest) is a compact, complex Kahler manifold with $SU(3)$ holonomy.}\index{$SU(3)$ holonomy} More generally, for a complex $n$-fold one demands that the holonomy is $SU(n)\subset SO(2n)$. As we will argue in a moment, this implies that some of the 10d supersymmetry is preserved in the 4d effective field theory and that Einstein equations are solved without sources (Ricci flatness). 

Though the Einstein equations are maybe physically more important, we will start with SUSY. Very superficially, we expect that a 4d supersymmetric effective theory will have massless spinors. Hence spinors need to have zero-modes. In the simplest case, this corresponds to the existence of covariantly constant spinors on the compactification space. We will see in a moment that this covariantly constant spinor is intimately linked to $SU(3)$ holonomy.

But let us first give a more careful argument for why unbroken SUSY requires the compact space to have a covariantly constant spinor: While we have not given the supergravity transformations of the various fields in 10d, we may recall the 2d case of worldsheet-SUGRA: Here, we have seen that the transformation of the gravitino is proportional to the covariant derivative of the SUSY parameter, i.e.~of the spinor $\xi(\sigma)$:
\be
\delta_\xi\chi_a=\nabla_a\xi\,.
\ee
This is similar in 10d. Hence, to identify a 4d SUSY parameter under which the vacuum is invariant, one needs a covariantly constant spinor. On a curved manifold this is a non-trivial requirement. 

To see this in more detail, we need the group-theoretic fact that $SO(6)=Spin(6)/\mathbb{Z}_2$, $Spin(6)=SU(4)$ and that the Weyl spinor representation of $Spin(6)$ (i.e.~a 6d spinor in eucliedan signature) transforms in the $\bf{4}$ of $SU(4)$, using the previous isomorphism. Furthermore, we embed our 10d spinor in the tensor product of 4d spinor and 6d spinor. Since 4d space is flat, the critical issue for the constancy of our 10d spinor is the constancy of its 6d spinor part. In other words, we have to take the 6d spinor to be covariantly constant along $X$. Furthermore, without loss of generality we assume that in $SU(4)$ notation this spinor takes the form 
\be
\xi(p)=\left(\begin{array}{c}0\\0\\0\\ \xi_0(p)\end{array}\right)
\ee
at some point $p\in X$. Since it is part of a covariantly constant spinor\index{covariantly constant spinor} field, the parallel transport will follow this field and, in particular, bring $\xi(p)$ back to itself for any loop $C$. But this clearly means that the holonomy matrices may only act on the first 3 components, i.e. we need $SU(3)$ holonomy. 

The reverse is obvious: Given $SU(3)$ holomomy, a covariantly constant spinor can be constructed by parallel transporting $\xi(p)$ given above to any point of $X$. The only way in which this might fail is if the construction were ambiguous, i.e., if two different paths from $p$ to $p'$ gave rise to two different spinors $\xi(p')$. But this would imply that a closed path starting at $p$ exists along which the parallel transport of $\xi(p)$ is non-trivial. This would be in contradiction to $SU(3)$ holonomy. Thus, we have the equivalence between $SU(3)$ holonomy and the existence of a covariantly constant spinor, i.e. the survival of 4d SUSY.\footnote{
More
precisely, 4d ${\cal N}=2$ SUSY in the type II case and  4d ${\cal N}=1$ SUSY in the type I and heterotic case. The reason is the presence of two independent 10d SUSY generators in the former situation.
}

Next, we consider Ricci flatness. We first note that, on Kahler manifolds, the only non-zero components of the Levi-Civita connection are
\be
\Gamma_{ij}{}^k = g^{k\ol{l}}\partial_i g_{j\ol{l}}\qquad \mbox{and}\qquad 
\Gamma_{\ol{\imath}\ol{\jmath}}{}^{\ol{k}} = g^{\ol{k}l}\partial_{\ol{\imath}}
g_{l\ol{\jmath}}\,.
\ee
This leads to significant simplifications for the Riemann tensor and the Ricci tensor which we do not work out. For example, the only non-vanishing Riemann tensor components are of the form
\be
R_{i\ol{\jmath}k\ol{l}}
\ee
and those related by antisymmetry in the first and second index pair. In other words, the first and second as well the third and fourth index have to be of opposite type (holomorphic and antiholomorphic). Moreover, the Ricci tensor can be written as
\be
R_{i\ol{\jmath}}=\partial_i\partial_{\ol{\jmath}}\,\ln\,\mbox{det}\,g\,\,.
\ee
(This is standard mathematics material, see e.g.~\cite{Viaclovsky} and many other sources.)

As is well known, the significance of $R_{i\ol{\jmath}}{}^\alpha{}_\beta$ is that, if interpreted as a matrix with indices $\alpha$ and $\beta$, it describes the rotation of a covector upon parallel transport along a loop with orientation specified by $i$ and $\ol{\jmath}$. Here we use greek letters for the second index pair to emphasise that they can take either holomorphic or antiholomorphic values, e.g.~$\alpha=k$ or $\alpha=\ol{k}$. The previously noted restrictions on holomorphy/antiholomorphy of the second index pair means that either $(\alpha,\beta)=(k,l)$ or $(\alpha,\beta)=(\ol{k},\ol{l})$. This can straightforwardly be shown to imply that the corresponding rotation matrix is in the $U(n)$ subgroup of the general holonomy group $SO(2n)$. More generally, the conditions of a manifold being Kahler and having $U(n)$ holonomy are equivalent. 

Since $U(n)=SU(n)\times U(1)$, the spin connection of Kahler manifolds can be thought of as the sum of an $SU(n)$ and a $U(1)$ connection. The latter is just a standard $U(1)$ connection, like in the case of an abelian gauge theory. Its field strength $F_{i\ol{\jmath}}$ being non-zero characterises the holonomy not being restricted to $SU(n)$. 

Concretely, recall that the complex structure is defined as multiplication by `$i$' on the cotangent or tangent vector space. In components, the corresponding operator or matrix is $J^\alpha{}_\beta$, which is hence the generator of the $U(1)$. The $U(1)$ part of the $U(n)$ field strength encoded in $R_{i\ol{\jmath}}{}^\alpha{}_\beta$ can hence be determined from the projection on $J^\alpha{}_\beta$. An explicit definition is
\be
F_{i\ol{\jmath}}\equiv  \mbox{tr}[ \tilde{R}_{i\ol{\jmath}} J ] \equiv R_{i\ol{\jmath}}{}^\alpha{}_\beta J^\beta{}_\alpha 
= i R_{i\ol{\jmath}}{}^k{}_k - i R_{i\ol{\jmath}}{}^{\ol{k}}{}_{\ol{k}} 
= 2i R_{i\ol{\jmath}}{}^k{}_k  = -2i R_i{}^k{}_{\ol{\jmath}k} 
= -2iR_{i\ol{\jmath}}\,.
\ee
Here the symbol $\tilde{R}$ is used to denote the Riemann tensor with suppressed second index pair, as opposed to the Ricci tensor. The final manipulations leading to the Ricci tensor require the use of the symmetry properties of the Riemann tensor together with the Kahler property of our manifold. We leave that as a problem (see e.g.~\cite{Candelas:1987is}). Eventually, we see that the $U(1)$ field strength components equal those of the Ricci tensor up to a prefactor (note however the different symmetry properties of the two tensors). Thus, $SU(n)$ holonomy is equivalent to Ricci flatness.

A final important point concerns the definition of Calabi-Yau manifolds via the Chern class (see e.g.~\cite{nak, bert, nashsen, gs}). Note first that, due to the $U(n)$ holonomy (or equivalently because of the special index structure of the Riemann tensor), the tangent bundle of Kahler manifolds can be viewed as a complex vector bundle with the curvature specified by $R_{i\ol{\jmath}}{}^k{}_l$. In other words, one can consider the curvature 2-form 
\be
R(T_X)=dz^i\wedge d\ol{z}^{\ol{\jmath}}\,R_{i\ol{\jmath}}{}^k{}_l\,,
\ee
which takes its values in $Lie(U(n))$. It is possible to write down the multi-form 
\be
c(X)=\mbox{det}(\mathbbm{1}+R(T_X))\,,
\ee
where the determinant refers to the matrix indices and multiplication relies on the wedge product. It is then expanded according to
\be
c(X)=1+c_1(X)+c_2(X)+\cdots=1+\mbox{tr}\,R(T_X)+\mbox{tr}\left( R(T_X)\wedge R(T_X)-2(\mbox{tr}\,R(T_X)^2\right)+\cdots\,.
\ee
Here $c_k(X)$ is a $(2k)$-form, defining the $k\,$th {\bf Chern class}. Concretely, the 1st Chern class is said to be zero if $c_1$ is {\bf exact}, which means that $c_1=d\omega$ for some $\omega$. More formally, this means that $c_1$ is zero in {\bf cohomology}, a concept we will discuss next. Crucially, while $c_1$ was defined using the metric, it is invariant (up to exact pieces) under smooth variations of the latter. It hence represents a {\bf topological invariant}. Intuitively, since $c_1(X)$ corresponds to the $U(1)$ field strength $F_{i\ol{\jmath}}$ introduced above, one may think of a non-exact $c_1$ characterising a non-trivial $U(1)$ bundle associated with $F_{i\ol{\jmath}}$.

After these preliminaries, we can formulate the celebrated theorem by {\bf Yau}:\hspace*{.2cm} {\bf Let $X$ be a Kahler manifold and $J$ its Kahler form. If the 1st Chern class\index{Chern class} vanishes, then a Ricci flat metric with Kahler form $J'$ in the same cohomology class can be given. This so-called Calabi-Yau metric is unique.}

Being in the same cohomology class means that $J-J'$ is exact. The key point is that, in practice, finding the Calabi-Yau metric is very hard (it has not been achieved analytically in any example). However, checking the topological condition of vanishing 1st Chern class is easy and guarantees the existence of many (explicitly known) suitable complex manifolds on which we hence know that a Calabi-Yau metric exists. But one will in general not find the metric for which $c_1$ is zero as a 2-form, only one with $c_1=d\omega$.

\subsection{Homology and cohomology}
\index{homology}\index{cohomology}

We are overdue with developing a few more simple mathematical ideas concerning in particular differential forms and topology. We start with homology and define a {\bf $p$-chain} as the formal sum of $p$-dimensional submanifolds $S_{p,\,i}$ of some compact manifold $X$:
\be
c_p=\sum_i \alpha_i \,S_{p,\,i}\,.
\ee
Depending on whether the coefficients $\alpha_i$ are real, complex, integer etc.~one can be talking about homology over the real, complex or integer numbers. In the first two cases, the $p$-chains form real and complex vector spaces respectively.

\begin{figure}[ht]
\begin{center} 
\includegraphics[width=11cm]{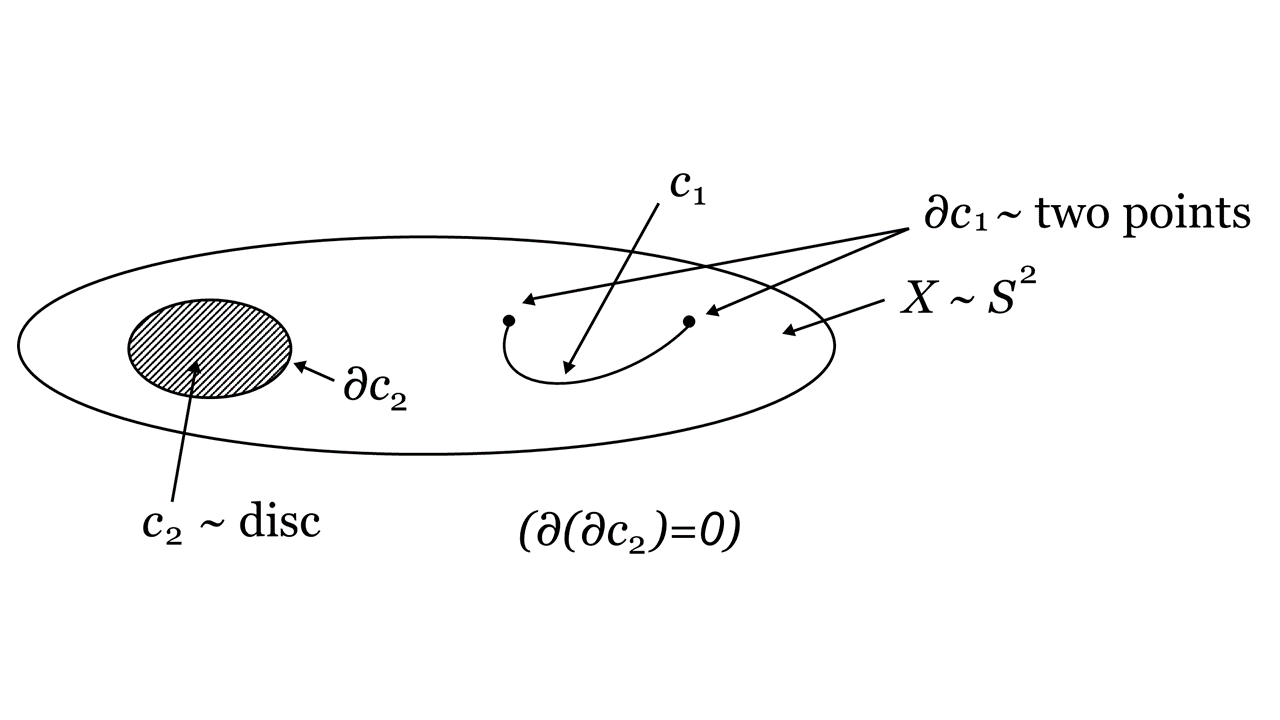}
\caption{Some simple submanifolds and their boundaries.}
\label{sexa} 
\end{center}
\end{figure}

One can consider the boundary of each $S_{p,\,i}$ and hence of $c_p$, which is a $(p-1)$-dimensional submanifold. Taking the boundary is denoted by the boundary operator $\partial$. A chain without boundary,
\be
\partial c_p=0\,,
\ee
is called a cycle. Crucially, $\partial^2$ is zero, in other words, a boundary has itself no boundary. A few simple examples are given in Fig.~\ref{sexa}. 

\begin{figure}[ht]
\begin{center} 
\includegraphics[width=7cm]{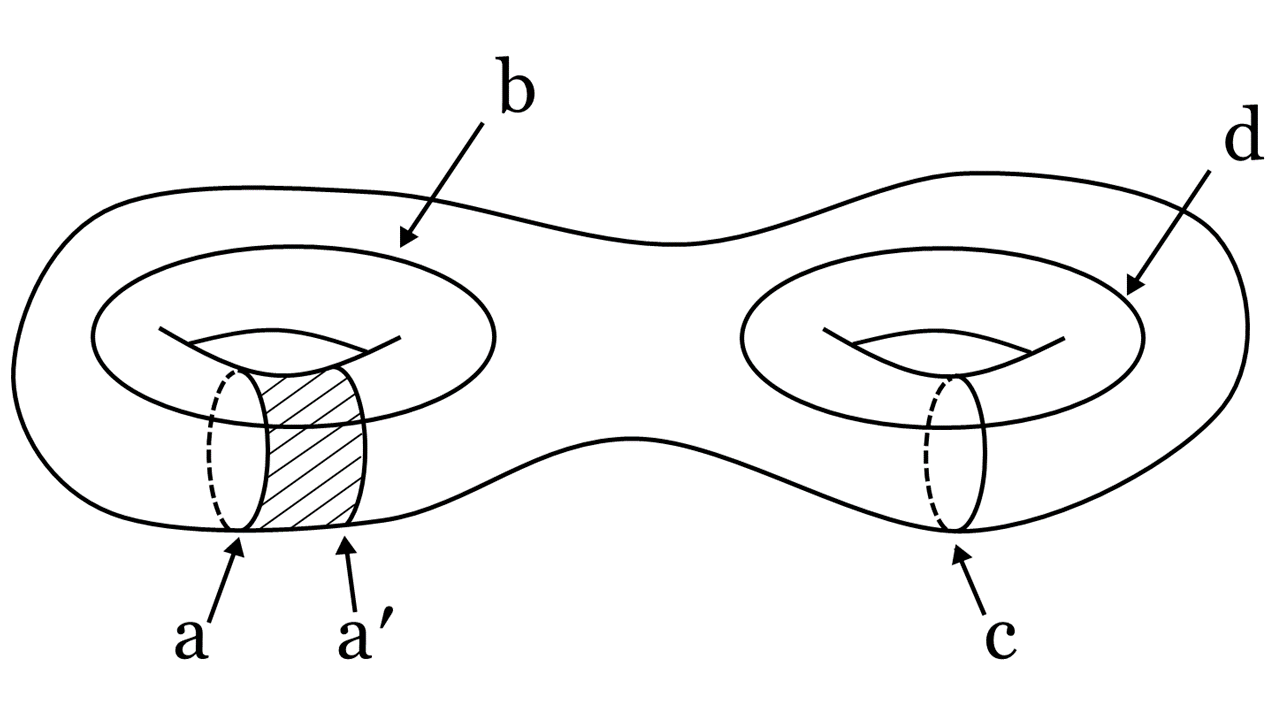}
\caption{Representatives of the four linearly independent homology classes in $H_1(R_2)$. The cycles $a$ and $a'$ are equivalent since their difference is a boundary. Concretely, the 1-cycle $a-a'$ represents the boundary of the hatched 2-dimensional submanifold.}
\label{r2} 
\end{center}
\end{figure}

Given the linear operator $\partial$ with $\partial^2=0$, it is natural to consider its homology groups:
\be
H_p=\frac{\mbox{Ker}(\partial_p)}{\mbox{Im}(\partial_{p+1})}=\frac{\mbox{$p$-cycles}}{\mbox{boundaries of $(p+1)$-chains}}\,.
\ee
The word group refers to addition, in the sense in which every vector space is an abelian group. The index $p$ of $\partial_p$ denotes the restriction of $\partial$ to the space of $p$-chains. We will suppress this index when it is clear from the context on which objects $\partial$ acts. As an example, we display certain 1-cycles\index{cycle} on the genus-2 Riemann surface $R_2$ in Fig.~\ref{r2}. It is easy to convince oneself that, working over the real numbers, $H_1(R_2)$ is 4-dimensional. Representatives $a$, $b$, $c$ and $d$ of the four linearly independent {\bf homology classes} (the elements of $H_p$) are shown. 

As another example, consider the 3-torus $T^3$ and convince yourself (at the intuitive level) that dim$(H_0)=1$ (which corresponds to $T^3$ being connected), 
dim$(H_1)=\,$dim$(H_2)=3$ and dim$(H_3)=1$. If the torus is thought of as $\mathbb{R}^3$ modulo discrete translations, representatives of $H_2$ can be thought of as three planes, each orthogonal to one of the three axes. 

One calls the above {\bf simplicial homology}.\index{simplicial homology}

Now we turn to $p$-forms as the dual objects with respect to the chains. So far, we use the word `dual' at an informal level, meaning simply that a chain\index{chain} $c_p$ and a form $\omega_p$ can be combined in a natural way to give a number:
\be
\omega_p(c_p)=\int_{c_p}\omega_p=\sum_i\alpha_i\int_{S_{p,\,i}}\omega_p\,.
\label{pair}
\ee
On the space of forms, we have an operator analogous to $\partial$, which also squares to zero: It is the exterior derivative $d$ or, restricted to $p$-forms, $d_p$:
\be
d_p: \omega_p\to \omega_{p+1}=d\omega_p\,.
\ee
Thus, it is natural to consider the cohomology groups of the {\bf de Rham cohomology}:\index{de Rham cohomology}
\be
H^p=\frac{\mbox{Ker}(d_p)}{\mbox{Im}(d_{p-1})}=\frac{\mbox{closed $p$-forms}}{\mbox{exact $p$-forms}}\,.
\ee
In the last expression, we use the definition that a $p$-form $\omega_p$ is called {\bf closed}\index{closed} if $d\omega_p=0$. Similarly, it is called {\bf exact}\index{exact} if it can be written as $\omega_p=d\omega_{p-1}$.

It is easy to convince oneself that the pairing (\ref{pair}) between chains $c_p$ and forms $\omega_p$, if restricted to cycles $c_p$ and closed forms $\omega_p$, induces a pairing between the corresponding homology and cohomology classes. The latter are sometimes denoted by $[c_p]$ and $[\omega_p]$. In other words, we claim that for a cycle $c_p$ and a closed form $\omega_p$ the integral
\be
\int_{c_p}\omega_p
\ee
is independent of the representative. For example, one has
\be
\int_{c_p}(\omega_p+d\omega_{p-1})=\int_{c_p}\omega_p+\int_{\partial c_p}\omega_{p-1}=\int_{c_p}\omega_p\,,
\ee
since $c_p$ has no boundary. Analogously, replacing $c_p$ by $c_p+\partial c_{p+1}$ does not affect the integral. 

One can furthermore show that this pairing between homology and cohomology classes is not degenerate and that hence 
\be
H_p(X)=H^p(X)^*\,,
\ee
i.e. they are dual vector spaces (cf. de Rham's theorems). In particular, their dimensions coincide, defining the so-called {\bf Betti numbers}\index{Betti numbers} 
\be
b_p(X)=\mbox{dim}\,H_p(X)=\mbox{dim}\,H^p(X)
\ee
of the manifold $X$. Intuitively, they count the number of inequivalent $p$-cycles. 

We note that, if dim$\,X=n$, there also exists a natural pairing between $p$-cycles and $(n-p)$-cycles: the intersection number. For example, given $T^3$, a 1-cycle (a line) and a 2-cyle (a plane), one can find out whether the two intersect (intersection number one) or don't (intersection number zero). For a Riemann surface, the pairing is between a 1-cycle and a 1-cycle. Here the meaning of `intersection number' is obvious from Fig.~\ref{r2}. It is intuitively clear that this lifts to a paring between homology classes.

The analogue of this on the cohomology side is
\be
[\omega_p]\cdot[\omega_{n-p}]=\int\omega_p\wedge\omega_{n-p}\,.
\ee
This pairing is also non-degenerate and hence turns $H^p$ into the dual of the vector space $H^{n-p}$. But since we already know that $H_{n-p}$ is the dual of $H^{n-p}$, we have found a canonical isomorphism
\be
H^p(X)\cong H_{n-p}(X)\,.
\ee
This is known as {\bf Poincar\'{e} duality}\index{Poincar\'{e} duality}. To say this more explicitly, a $p$-form $\omega_p$ is Poincare dual to an $(n-p)$-cycle $c_{n-p}$ if
\be
\int_{c_{n-p}}\omega_{n-p}=\int \omega_p\wedge \omega_{n-p}\,\,,\qquad \forall 
\omega_{n-p}\,.
\ee

More structure arises if a metric is present. In particular, with a metric comes the Hodge star\index{Hodge!star} operator, 
\be
*\,:\omega_p\,\,\mapsto\,\,(*\omega)_{n-p}\qquad \mbox{with}\qquad
(*\omega)_{\mu_{p+1}\cdots\mu_n}=\frac{\sqrt{g}}{p!}\,\omega^{\mu_1\cdots \mu_p}\,\epsilon_{\mu_1\cdots \mu_n}\,.
\ee
This gives rise to a scalar product on the space of $p$-forms,
\be
(\omega_p,\alpha_p)=\int_X \omega_p\wedge *\alpha_p\,.
\ee
As a result, one can define the adjoint of $d$, the so-called co-differential\index{co-differential} $d^\dagger$. On forms of degree $p$, it takes the explicit form
\be
d^\dagger=(-1)^p *^{-1}d*\,.
\ee
With this, one defines the Laplace operator
\be
\Delta = d^\dagger d+d\,d^\dagger\,.
\ee
A form is called {\bf harmonic}\index{harmonic} if $\Delta\omega=0$. This definition gives rise to the {\bf Hodge decomposition theorem}\index{Hodge!decomposition theorem}, which states that on a compact manifold $X$ any form has a {\it unique} decomposition in an exact, a coexact and a harmonic piece: 
\be
\omega=d\alpha+d^\dagger \beta+\gamma\qquad\mbox{with}\qquad \Delta\gamma=0\,.
\ee
It can furthermore be shown that $\beta$ vanishes if $\omega$ is closed. As a result, any representative of a given cohomology class has a unique decomposition in an exact and harmonic piece. In other words, there is a unique harmonic form in any cohomology class. Intuitively speaking, this is the constant form with the right integral on all cycles (these integrals being fixed by the class). To give a simple concrete example, consider $T^2$ being parameterised by $(x,y)\in [(0,1)\times (0,1)]$. The harmonic one form with integral zero on the $x$-cycle and integral $1$ on the $y$-cycle is obviously given by $\omega=dy$. A non-harmonic form in the same class would e.g. be 
$\omega = (1+\sin(2\pi y))\,dy$.

Finally, it is possible to take the above to the realm of complex manifolds. To do so, recall that on a complex manifold a 1-form may be written as 
\be
\omega(z,\ol{z})=\omega(z,\ol{z})_i dz^i+\omega(z,\ol{z})_{\ol{\imath}} d
\ol{z}^{\ol{\imath}}\equiv\omega_{(1,0)}+\omega_{(0,1)}\,.
\ee
In other words, we can decompose it in its $(1,0)$ and $(0,1)$ parts. The first corresponds to a tensor with one holomorphic and no antiholomorphic index, the second to a tensor with no holomorphic and one antiholomorphic index.

Such a decomposition carries over to higher forms (i.e. antisymmetric tensors) and to cohomology classes. For example,
\be
\omega_3 = \omega_{(3,0)}+\omega_{(2,1)}+\omega_{(1,2)}+\omega_{(0,3)}\,,
\ee
where, e.g.,
\bea
\omega_{(2,1)} &=& \omega_{ij\ol{k}}\,dz^i\wedge dz^j\wedge d\ol{z}^{\ol{k}}
+ \omega_{i\ol{\jmath}k}\,dz^i\wedge d\ol{z}^{\ol{\jmath}}\wedge dz^k
+ \omega_{\ol{\imath}jk}\,d\ol{z}^{\ol{\imath}}\wedge dz^j\wedge dz^k
\nonumber \\
&=& 3 \,\omega_{ij\ol{k}}\,dz^i\wedge dz^j\wedge d\ol{z}^{\ol{k}}\,.
\eea

To see the corresponding, refined cohomology construction more explicitly, recall that the exterior derivative has the particularly compact definition
\be
d=dx^a\,\frac{\partial}{\partial x^a}\,.
\ee
Here the partial derivative is supposed to act on the coefficients of any given form and, subsequently, $dx^a$ has to be multiplied with the form using the wedge product from the left. Let us consider specifically a manifold of complex dimension $n$ (real dimension $2n$), such that $a=1,2,\cdots,2n$. Then it is easy to check that
\be
d=dz^i\frac{\partial}{\partial z^i}+d\ol{z}^{\ol{\imath}} \frac{\partial}{\partial \ol{z}^{\ol{\imath}}}\,,
\ee
or
\be
d=\partial+\ol{\partial}\qquad\mbox{with}\qquad \partial=dz^i\frac{\partial}{\partial z^i}\qquad\mbox{and}\qquad
\ol{\partial}=d\ol{z}^{\ol{\imath}} \frac{\partial}{\partial \ol{z}^{\ol{\imath}}}\,.
\ee
Here $i=1,2\cdots,n$. Furthermore, the {\bf holomorphic and antiholomorphic exterior derivatives} square to zero:
\be
\partial^2=\ol{\partial}^2=0\,.
\ee
This permits the construction of a cohomology, which turns out to be independent of whether $\partial$ or $\ol{\partial}$ is used. The conventional choice is $\ol{\partial}$. Thus, one defines the {\bf Dolbeault cohomology}
\be
H^{p,q}=\frac{\mbox{Ker}(\ol{\partial}_{p,q})}{\mbox{Im} (\ol{\partial}_{p,q-1})} \,,
\ee
which contains finer information than the de Rham cohomology. One may say that it characterises the interrelation between the non-trivial cycles and the complex structure. We also note the so-called Hodge decomposition
\be
H^k=\oplus_{p+q=k}H^{p,q}\,.
\ee
The dimensions of Dolbeault cohomology groups are known as {\bf Hodge numbers}\index{Hodge!numbers},
\be
h^{p,q}(X)\equiv\,\mbox{dim}\,H^{p,q}(X)\,.
\ee
They are commonly arranged in a so-called {\bf Hodge diamond}\index{Hodge!diamond}. With a view to our application to Calabi-Yau manifolds, we display the general form for the case of a complex 3-fold:
\be
\begin{array}{ccccccc}
&&&h^{0,0}&&& 
\\ 
&&h^{1,0}&&h^{0,1}&&
\\
&h^{2,0}&&h^{1,1}&&h^{0,2}&
\\
h^{3,0}&&h^{2,1}&&h^{1,2}&&h^{0,3}
\\
&h^{3,1}&&h^{2,2}&&h^{1,3}&
\\
&&h^{3,2}&&h^{2,3}&&
\\
&&&h^{3,3}&&& 
\end{array}\,\,.\label{ghd}
\ee

\subsection{Calabi-Yau moduli spaces}
\index{moduli space}

Due to $SU(3)$ holonomy, the hodge diamond for a Calabi-Yau 3-fold is very special. Using the same arrangement as in (\ref{ghd}), it reads
\be
\begin{array}{ccccccc}
&&&1&&& 
\\ 
&&0&&0&&
\\
&0&&h^{1,1}&&0&
\\
1\,\,&&h^{2,1}&&h^{2,1}&&\,\,1
\\
&0&&h^{1,1}&&0&
\\
&&0&&0&&
\\
&&&1&&& 
\end{array}\,\,.
\ee
Here, the simplifications arising from the vertical and horizontal reflection symmetry of the Hodge diamond (e.g.~$h^{1,1}=h^{2,2}$) are generic - they hold for any complex $n$-fold. Furthermore, connectedness implies $h^{0,0}=h^{3,3}=1$. But some features are specific to Calabi-Yau spaces, such as $h^{1,0}=h^{2,0}=0$ and, crucially, $h^{3,0}=h^{0,3}=1$. The latter implies the existence of a unique holomorphic, harmonic 3-form which is conventionally denoted by $\Omega$:
\be
\Omega=\Omega_{ijk}(z)\,dz^i\wedge dz^j\wedge dz^k\,.
\ee
Its existence can be understood on the basis of the covariantly constant spinor $\psi$:
\be
\Omega_{ijk}\sim\ol{\psi}\Gamma_{ijk}\psi\,.
\ee
We will not argue for uniqueness. It is however useful to note that the existence of a harmonic, holomorphic $3$-form $\Omega$ can be used as a defining feature for Calabi-Yau spaces: More generally, a Calabi-Yau $n$-fold can be defined as a Kahler manifold with a {\bf trivial canonical bundle}\index{canonical bundle}. The latter is the $n$th exterior power of the cotangent bundle - this is the bundle in which $\Omega$ lives and which is trivial exactly if there is a nowhere vanishing section - in our case the $n$-form $\Omega$. 

Now, given a Calabi-Yau 3-fold, Yau's theorem guarantees the existence of a unique (given Kahler class and complex structure) Ricci flat metric $g_{i\ol{\jmath}}$ . A key question for physics is whether this metric can be deformed maintaining Ricci-flatness since this would imply the existence of moduli:
\be
g_{i\ol{\jmath}}\,dz^i\,d\ol{z}^{\ol{\jmath}}\quad\to\quad g_{i\ol{\jmath}}
\,dz^i\,d\ol{z}^{\ol{\jmath}}\,+\,\delta g_{i\ol{\jmath}}
\,dz^i\,d\ol{z}^{\ol{\jmath}}\,+\,\delta g_{ij}\,dz^i\,dz^j\,+\,
\mbox{h.c.}
\ee
The presence of such deformations has the potential to contradict the uniqueness part of Yau's theorem. To avoid such a contradiction, these deformations must be accompanied by a change of either the Kahler class or the complex structure. This is indeed the case: A change of the metric of type $\delta g_{i\ol{\jmath}}$ can be directly interpreted as a change of (the harmonic representative of) the Kahler form $J$. The number of such independent deformations, also called {\bf Kahler deformations}\index{Kahler!deformation} is counted by $h^{1,1}$. This number is at least unity since it is always possible to simply rescale the metric, making our manifold larger or smaller without changing its shape. 

By contrast, a deformation of type $\delta g_{ij}$ violates the hermiticity assumption and it must hence be accompanied by a change of the complex structure if one wants to restore explicitly the Calabi-Yau situation after adding this $\delta g$ to the original metric. To count these deformations it is useful to define a $(2,1)$ form
\be
\delta \chi=\Omega_{ij}{}^{\ol{k}}\,\delta g_{\ol{k}\ol{l}}\,dz^i\wedge dz^j\wedge d\ol{z}^{\ol{l}}\,\, \in\,\,H^{2,1}(X)
\ee
associated with $\delta g_{\ol{k}\ol{l}}$. Here the index $k$ of $\Omega_{ijk}$ has been raised using the Calabi-Yau metric. It can be shown that this represents a one-to-one map between distinct {\bf complex structure deformations}\index{complex structure deformations} (and hence corresponding metric deformations) and linearly independent Dolbeault cohomology classes of type $(2,1)$. Here by distinct we mean those {\it not} corresponding to reparameterisations $z^i\to z'^i$. 

There is another way of understanding the counting of complex structure deformations: Think of the complexified vector space of 3-cycles, with dimension $2h^{2,1}+2$. Two directions are distinguished by $\Omega$ and $\ol{\Omega}$, a feature only visible in Dolbeault but not in de Rham cohomology. Now, the change of complex structure is accompanied by a change of the direction of $\Omega$ (and hence of $\ol{\Omega}$) in this space. In other words, $\Omega$ is infinitesimally rotated and these possible rotations are parameterized by $h^{2,1}$ complex numbers. One may say that there are $h^{2,1}$ complex directions in which $\Omega$ can develop new, infinitesimal components. 

\begin{figure}[ht]
\begin{center} 
\includegraphics[width=13cm]{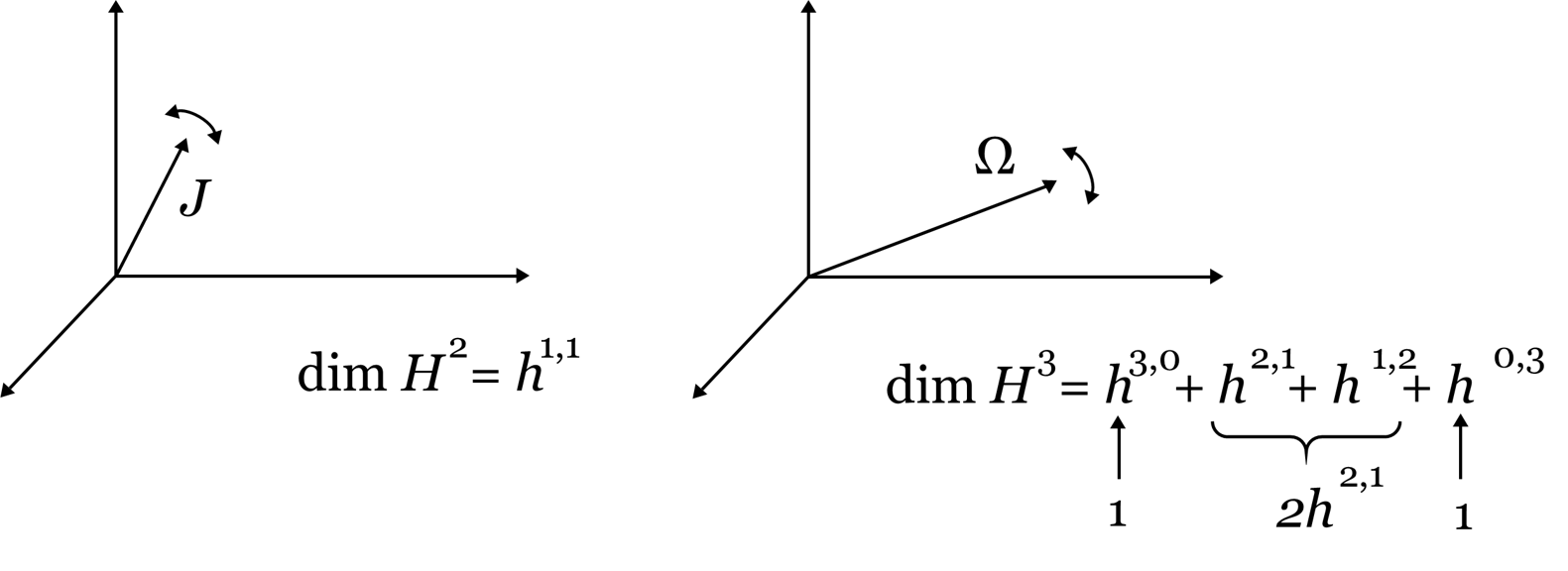}
\caption{A visualisation attempt of how $J$ and $\Omega$ move in the spaces $H^2(X)$ and (the complexification of) $H^3(X)$, thereby determining the metric on a Calabi-Yau manifold. Of course, the dimensions of these spaces are in general much higher than three.}
\label{cym} 
\end{center}
\end{figure}

One can also invert the equations above, i.e., given a harmonic $(2,1)$-form $\delta\chi$, one can explicitly write down how $\Omega$ and the metric change:
\be
\delta g_{\ol{\imath}\ol{\jmath}}=-\frac{1}{||\Omega||^2}\, \ol{\Omega}_{\ol{\imath}}\,{}^{kl}\,\delta\chi_{kl\ol{\jmath}}\,,\qquad\qquad
\delta\Omega=\delta \chi\,.
\ee
with the constant
\be
||\Omega||^2=\frac{1}{3!}\Omega_{ijk}\ol{\Omega}^{ijk}\,.
\ee
Together with the previously discussed relation 
\be
\delta g_{i\ol{\jmath}}=-i\delta J_{i\ol{\jmath}}\,,
\ee
we now see explicitly how the cohomology groups $H^{1,1}(X)$ and $H^{2,1}$ play a central role in describing allowed deformations of the metric and hence the moduli space of a Calabi-Yau manifold. Crucially, $H^{2,1}$ has to be viewed as a subspace of $H^3(X)$. This subspace moves as the complex structure changes. An illustration of this has been attempted in Fig.~\ref{cym}. In addition to the textbook literature given earlier, the reader may want to consult~\cite{Candelas:1990pi,Font:2005td} for more details.

Before characterising Calabi-Yau moduli spaces more quantitatively, we want to give at least the simplest example. To do so, let us start with an important set of examples for compact, complex Kahler manifolds: the so-called complex projective spaces. To begin, recall that a {\bf real projective space}\index{projective space} $\mathbb{R}P^n$ is $\mathbb{R}^{n+1} \!\setminus\! \ol{0}$ modulo the equivalence relation $\ol{x}\sim \lambda \ol{x}$ with $\lambda\in \mathbb{R}\setminus 0$. Intuitively speaking, this is the set of lines through the origin. Such a set can easily be given a differentiable structure. For the case of $\mathbb{R}P^2$, the real projective plane, we can equivalently think of $S^2/\mathbb{Z}_2$ -- a sphere with antipodal points identified. 

This has a natural complex generalisation: the complex projective spaces $\mathbb{C}P^n$. They are defined analogously as the set of all $(n+1)$-tuples of complex numbers (not all zero) with the equivalence relation
\be
(z^0,\cdots,z^n)\,\sim\,(\lambda z^0,\cdots,\lambda z^n)\qquad\mbox{with} \qquad \lambda\in \mathbb{C}\setminus 0\,.
\ee
For the subset $U_i$ of all equivalence classes in which $z^i\neq 0$, a chart is provided by
\be
\phi_i: \{\,\mbox{class of}\,\,(z^0,\cdots,z^n)\,\}\,\,\mapsto\,\,
\left(\frac{z^0}{z^i},\cdots,\frac{z^{i-1}}{z^i},\frac{z^{i+1}}{z^i},\cdots,
\frac{z^n}{z^i}\right)\in \mathbb{C}^n\,.
\ee
It is easy to show that these charts form an atlas and to give explicitly the (holomorphic) transition maps. A Kahler potential in $U_i$ is provided by 
\be
K^{(i)}(x)=\frac{1}{2}\,\ln\left(1+\sum_{j=1}^n\,|x^j|^2\right)\,\,, \qquad\mbox{with}\qquad \{x^1,\cdots,x^n\}=\left\{\frac{z^0}{z^i},\cdots,\frac{z^{i-1}}{z^i}, \frac{z^{i+1}}{z^i}, \cdots,\frac{z^n}{z^i}\right\}
\ee
the coordinates defined above. A straightforward calculation shows that this gives rise to a globally defined Kahler form and metric, the {\bf Fubini-Study metric}\index{Fubini-Study metric}. To be very concrete, it is easy to check that $\mathbb{C}P^1$ is the Riemann sphere. Crucially, all $\mathbb{C}P^n$ are compact. 

Quite generally, submanifolds of lower dimension can often be given as zero sets of polynomials. For example, the polynomial $x^2+y^2-1$ defines $S^1\subset \mathbb{R}^2$. The naive generalisation to (holomorphic) polynomials on $\mathbb{C}^n$ is not useful for us since the resulting submanifolds are always non-compact (for $n>1$). This is due to a generalisation of the maximum modulus theorem for analytic functions. However, starting from the compact space $\mathbb{C}P^n$\index{$\mathbb{C}P^n$}, compact submanifolds {\it can} be defined by polynomials. For the zero set to be well defined on the set of equivalence classes, the polynomials have to be homogeneous. Now, it can be shown that the crucial Calabi-Yau condition, the vanishing of the 1st Chern class, depends on the homogeneity degree of the polynomial. If we want to get a 3-fold, we must start from $\mathbb{C}P^4$. The Chern class vanishes if and only if the defining polynomial is of degree 5:
\be
P_5(z)=c_{i_1\cdots i_5}z^{i_1}\cdots z^{i_5}\,,
\ee
with indices running from 0 to 4 and labelling the projective coordinates of $\mathbb{C}P^4$. The so called {\bf quintic Calabi-Yau 3-fold}\index{quintic} is then defined as the zero set,
\be
P_5(z)=0\,\,,\qquad z\in \mathbb{C}P^4\,.
\ee
Here by $z$ we mean both the set of 5 numbers $\{z^i\}$ and the corresponding point in the projective space. As the coefficients of the polynomial vary, the complex structure changes. A concrete example is given, e.g., by
\be
P_5(z)=(z^0)^5+\cdots +(z^4)^5\,.
\ee

It is interesting to count the possible deformations of such a quintic hypersurface: One first notes that the number of different monomials in a homogeneous polynomial of degree $d$ in $n$ variables is given by the binomial coefficient\footnote{Try to prove this!}
\be
\left(\begin{array}{c} d+n-1 \\ n-1 \end{array}\right)\,.
\ee
In our case this gives
\be
\left(\begin{array}{c} 5+5-1 \\ 5-1 \end{array}\right)=
\left(\begin{array}{c} 9 \\ 4 \end{array}\right)=126\,.
\ee
From this, we have to subtract the 25 parameters of the symmetry group $GL(5,\mathbb{C})$ of $\mathbb{C}P^4$, giving us 101 parameters. Recalling what was said before about the interplay of Dolbeault cohomology and complex structure moduli spaces, we conclude that $h^{2,1}(quintic)=101$. Without derivation we also note that the Kahler form of $\mathbb{C}P^4$ is unique up to scaling, such that $h^{1,1}=1$. Thus, for the quintic the Hodge diamond reads
\be
\begin{array}{ccccccc}
&&&1&&& 
\\ 
&&0&&0&&
\\
&0&&1&&0&
\\
1\,\,&&101&&101&&\,\,1
\\
&0&&1&&0&
\\
&&0&&0&&
\\
&&&1&&& 
\end{array}
\ee
and the real dimension of the moduli space is $2\cdot 101+1=203$. (In fact, in string theory the volume modulus always comes with an axionic partner, such that the counting would be $2\cdot 101+2=204$.)

We note that the same construction goes through for the quartic polynomial in $\mathbb{C}P^3$, giving rise to the unique Calabi-Yau 2-fold, known as the K3-surface\index{K3-surface}. However, for 3-folds there are many more examples. First, one can generalise to the intersection of hypersurfaces (defined by polynomials) in products of projective spaces. This gives rise to so-called complete-intersection CYs\index{complete-intersection Calabi-Yau} or CICYs\index{CICY}. Then one can generalise from projective spaces to weighted projective spaces. In this case one still mods out by a rescaling with a complex parameter $\lambda$, but the different variables scale differently, i.e.~have different weight. Furthermore, one may mod out not just by the rescaling by one such complex parameter, but by several such scalings (with different parameters $\lambda_i$). This leads to the concept of toric geometry\index{toric geometry} and toric hypersurfaces, in which Calabi-Yaus can again be defined by polynomials of suitable degrees in the different variables (Batyrev's construction\index{Batyrev's construction} \cite{Batyrev:1994hm}).
See e.g.~\cite{Bouchard:2007ik} for a set of lecture notes starting at an elementary level and proceeding to toric geometry. Even more general Calabi-Yau constructions exist. The total number of known distinct examples is about half a billion: $\sim 5\times 10^8$.

\subsection{Explicit parameterization of Calabi-Yau moduli spaces}\label{epcy}
\index{moduli space}

\begin{figure}[ht]
\begin{center} 
\includegraphics[width=6cm]{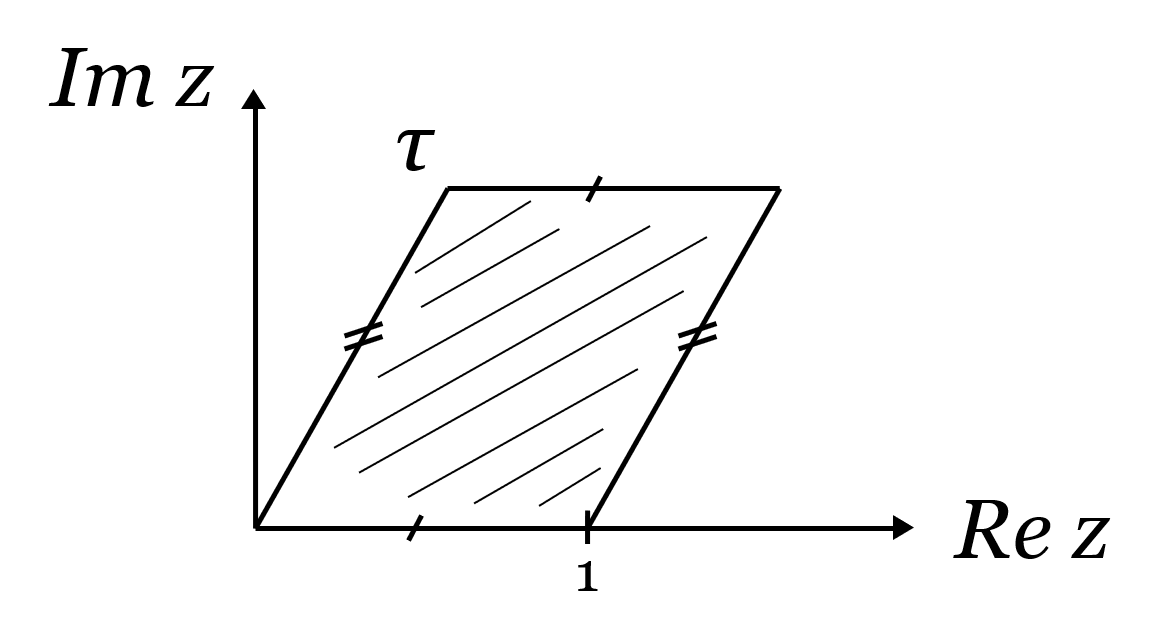}
\caption{Torus defined as $\mathbb{C}/\mathbb{Z}^2$.}
\label{t2z} 
\end{center}
\end{figure}

We start with an extremely simple toy model: $T^2$. We can give it a complex structure by defining it as $\mathbb{C}/\mathbb{Z}^2$. By this we mean starting from the complex plane and modding out a lattice of translations, generated by unity and $\tau\in\mathbb{C}$. The resulting set of independent points, the so-called fundamental domain, is shown in Fig.~\ref{t2z}. It is parameterised, on the one hand, by $z$ and, on the other hand, by $x,y\in [0,1)$, with the relation
\be
z=x+\tau y\,.
\ee
The complex number $\tau$ determines the complex structure. Note that tori with different $\tau$ are (in general) not isomorphic as complex manifolds.
The holomorphic $(1,0)$-form in this case is clearly
\be
\Omega=\alpha\,dz=\alpha\,dx+\alpha\,\tau\,dy\,,
\ee
with $\alpha\in \mathbb{C}$ an arbitrary constant. 

Now, in analogy to the proper Calabi-Yau case, the complex structure can be defined using the position of $\Omega$ in the complexification of $H^1(T^2)$. For this, it is sufficient to know the {\bf periods}\index{period}, i.e. the integrals of $\Omega$ over the integral 1-cycles:
\be
\Pi_1=\int_{y=\,\mbox{const.}}\Omega=\int_0^1\alpha\,dx=\alpha\,\,\,,\qquad
\Pi_2=\int_{x=\,\mbox{const.}}\Omega=\int_0^1\alpha\,\tau\,dy=\alpha\tau\,.
\ee
They can be combined in the period vector $\Pi=(\Pi_1,\Pi_2)$. Since the normalisation of $\Omega$ is arbitrary, only ratios of these periods are meaningful. Concretely, the (in this case single) complex structure parameter is given by $\tau=\Pi_2/\Pi_1$. 

Next, we come to the moduli (in this case the modulus) associated with the Kahler form. The Kahler form is harmonic and can be decomposed in a basis of harmonic 2-forms,
\be
J=t^i\omega_i\,.
\ee
Here the $\omega_i$ are in general chosen to represent an integral 2-form basis (where by integral we mean Poincare dual to the naturally defined integral basis of 4-cycles or, what is the same, the dual basis to the integral 2-cycle basis). In our case there is of course only one such 2-form:
\be
\omega_1=dx\wedge dy\,\,,\qquad \mbox{such that}\qquad 
J=t\,dx\wedge dy\,.
\ee

At the same time, we know that 
\be
J=ig_{i\ol{\jmath}}\,dz^i\wedge d\ol{z}^{\ol{\jmath}}=ig_{z\ol{z}}\,dz\wedge 
d\ol{z}=ig_{z\ol{z}}(dx\wedge\ol{\tau}\,dy+\tau\,dy\wedge dx)
=-i(\tau-\ol{\tau})\,g_{z\ol{z}}\,dx\wedge dy\,.
\ee
Hence, we identify $t$ as $t=-i(\tau-\ol{\tau})\,g_{z\ol{z}}$. We may also write the general metric as
\be
ds^2=2g_{z\ol{z}}\,dz\,d\ol{z}=2g_{z\ol{z}} \big[dx^2+|\tau|^2dy^2+(\tau+\ol{\tau})\,dx\,dy\big]\,.
\ee
Thus, we can finally explicitly give the matrix form of the metric in terms of the parameters $t$, $\Pi_1$, $\Pi_2$ which govern the position of $J$ and $\Omega$ in their respective cohomology groups:
\be
g_{ab}=2g_{z\ol{z}}\left(\begin{array}{cc} 1 & \mbox{Re}\,\tau \\ \mbox{Re}\,\tau & |\tau|^2 \end{array}\right) = \frac{t}{\mbox{Im}(\Pi_2/\Pi_1)} \left(\begin{array}{cc} 1 & \mbox{Re}(\Pi_2/\Pi_1) \\ \mbox{Re}(\Pi_2/\Pi_1) & |\Pi_2/\Pi_1|^2 \end{array}\right)\,.
\ee
With somewhat more writing, one can achieve the same level of explicitness for the toy-model 3-fold $T^2\times T^2\times T^2$, defined by modding out an appropriate lattice of translations from $\mathbb{C}^3$. Nevertheless, this is not a proper Calabi-Yau manifold since its holonomy group is trivial. By contrast, a Calabi-Yau should have holonomy group $SU(3)$ (not just a subgroup). However, this is clearly to some extent a matter of semantics. More importantly, $T^6$ is too simple for most physical applications and it does not give rise to the large landscape of solutions of string theory that we are after.

Thus, we now turn to the general case of proper Calabi-Yau 3-folds, such as the quintic and similar, even more complicated examples. The complete explicitness of metric parameterisation that we saw above can of course not be achieved in such cases. But our main goal for the moment will be a description in 4d supergravity language,
\be
{\cal L}=K_{i\ol{\jmath}}(\partial X^i)(\partial \ol{X}^{\ol{\jmath}})\,+\,\,\mbox{gauge, fermion, and other fields}\,,
\ee
where $K$ is the Calabi-Yau metric on moduli space, parameterised by the $X^i$, which include both Kahler and complex structure moduli. This {\it can} be given rather explicitly, even in the proper Calabi-Yau case. In principle, all that is needed is a careful Kaluza-Klein reduction using the cycle-structure of the Calabi-Yau space (see e.g.~\cite{Grimm:2004uq, Jockers:2004yj, Grimm:2004ua, Kerstan:2011dy} and refs.~therein). We will only report the results.

Let us start with the Kahler moduli. As we already explained,
\be
J=t^\alpha\omega_\alpha\qquad\mbox{with}\qquad \alpha=\,1,\cdots,h^{1,1}\,.
\ee
Moreover, the volume of the Calabi-Yau manifold can be given as 
\be
{\cal V}=\frac{1}{6}\int_X J\wedge J\wedge J = \frac{1}{6}\kappa_{\alpha\beta\gamma} t^\alpha t^\beta t^\gamma\,.
\ee
Here one may intuitively think of components of the vector $t^\alpha$ as measuring the volumes of the different 2-cycles present in the Calabi-Yau.  The integers $\kappa_{\alpha\beta\gamma}$ are the so-called triple intersection numbers of the 4-cycles Poincare dual to the $\omega_\alpha$.\footnote{
Note 
that two 4-cycles in a 6d manifold generically intersect in a 2d submanifold. The latter generically intersects the third 4-cycle in points. The total number of those, with orientation, is a function of the homology classes and is counted by the $\kappa_{\alpha\beta\gamma}$.
} 
The volumes of the dual 4-cycles, which are also labelled by the index $\alpha$, are given by 
\be
\tau_\alpha=\frac{1}{2}\int_{c_4^\alpha}J\wedge J=\frac{1}{2} \kappa_{\alpha\beta\gamma} t^\beta t^\gamma\,.
\label{taut}
\ee
Clearly, the variables $t^\alpha$ and $\tau_\alpha$ encode the same information. Using them as ${\cal N}=1$ SUGRA variables corresponds to choosing either of  two different ${\cal N}=1$ sub-algebras of the ${\cal N}=2$ SUSY of a Calabi-Yau compactification of type IIA or IIB string theory. We focus (for the purpose of our later discussion of a particularly well-understood model, called KKLT) on the IIB case and the $\tau$ variables. They are real but, in 4d SUSY, are complexified by adding the imaginary parts
\be
c_\alpha=\int_{c^\alpha}C_4\,.
\ee
Only as a side remark we note that, in the other SUSY, the $t^\alpha$ would be complexified by corresponding integrals of $B_2$ or $C_2$, depending on the particular model (for many more details see e.g.~\cite{Grimm:2004uq, Jockers:2004yj, Grimm:2004ua, Kerstan:2011dy}).

The relations (\ref{taut}) can in principle be solved for the $t^\alpha$:
\be
t^\alpha=t^\alpha(\tau_1,\cdots,\tau_{h^{1,1}})\,.
\ee
With $T_\alpha=\tau_\alpha+ic_\alpha$ and 
\be
\tau_\alpha=\frac{1}{2}(T_\alpha+\ol{T}_{\ol{\alpha}})\,,
\ee
the volume ${\cal V}$ can be expressed as real function of the variables $T_\alpha$ and $\ol{T}_{\ol{\alpha}}$. The type-IIB Kahler moduli Kahler potential\index{Kahler!potential} can then finally be written down as
\be
K_K=-2\ln{\cal V}\qquad\mbox{with}\qquad {\cal V}={\cal V}(T_\alpha,\ol{T}_{\ol{\alpha}})\,.
\ee

To describe the complex structure moduli space, we start by recalling the basis of $H_1(R_2)$ as given in Fig.~\ref{r2}. We rename the relevant cycles (representatives of the corresponding cohomology classes) as
\be
a \to A^1\,,\qquad b\to B_1\,,\qquad c\to A^2\,,\qquad d\to B_2\,.
\ee
It is clear that this carries over analogously to the 1-cycles of higher Riemann surfaces, giving rise to the basis $\{A^a,B_a\}$ and an intersection structure
\be
A^a\cdot A^b=0\,,\qquad B_a\cdot B_b=0\,,\qquad A^a\cdot B_b=\delta^a{}_b\,.
\ee
An analogous basis can be chosen for the (in this case naturally isomorphic) vector space $H_1$. Such bases are called {\bf symplectic bases}\index{symplectic bases}, on account of the antisymmetry of the only non-vanishing intersection numbers or, on the form side, wedge products:
\be
\int \omega^A_a\wedge \omega_B^b=\delta_a{}^b
=-\int \omega_B^b \wedge \omega^A_a\,.
\ee

The crucial point for us is that this represents a generic feature of the so-called {\bf middle homology} or {\bf cohomology} for manifolds where the dimensionalities of the relevant cycles/forms are odd. This is true for Riemann surfaces, with which we started, but it is equally true for complex 3-folds, our new case of interest. 

Thus, now in the context of Calabi-Yaus, we choose a symplectic 3-cycle basis as above and define the periods
\be
z^a=\int_{A^a}\Omega\,,\qquad {\cal G}_b=\int_{B_b}\Omega\,.
\ee
The complex parameters $z^a$ with $a=0,\cdots, h^{2,1}$ are sufficient to fully parameterise the position of $\Omega$ in $H^3(X)$. In fact, one of the parameters can be set to unity at the expense of a constant, complex rescaling of $\Omega$, which does not induce any physical (geometrical) change. Hence one may think of all the $z^a$s together as of `projective coordinates'. Alternatively, one can set $z^0=1$, with $h^{2,1}$ parameters left over. 

Crucially, the remaining periods ${\cal G}_b$ are not independent - they are in general complicated functions of the $z^a$:
\be
{\cal G}_b={\cal G}_b(z^0,\cdots,z^{h^{2,1}})\,.
\ee
One combines all of them in the period vector
\be
\Pi=(z^0,\cdots,z^{h^{2,1}},{\cal G}_0(z),\cdots,{\cal G}_{h^{2,1}}(z))\,.
\ee
The explicit form of the (dependent) periods can be obtained from appropriate differential equations (the {\bf Picard-Fuchs equations})\index{Picard-Fuchs equations} which can be formulated on the basis of certain topological features of the Calabi-Yau (see e.g.~\cite{Klemm:1992tx, fre} and refs.~therein). Crucially, they do not require the in general unavailable metric information. Thus, though with much work, the periods can in principle be explicitly obtained. 

With this, we are ready to give the complex structure Kahler potential:
\be
K_{cs}=-\ln(i\int_X\Omega\wedge \ol{\Omega})=-\ln(-i\Pi^\dagger\Sigma\Pi)=
-\ln(-i\ol{z}^a {\cal G}_a(z)+iz^a\ol{{\cal G}_a(z)})\,,
\ee
where
\be
\Sigma=\left(\begin{array}{rr}0 & \mathbbm{1} \\ -\mathbbm{1} & 0 
\end{array}\right)
\ee
is the symplectic metric. See e.g.~\cite{Giryavets:2005nf} for a nice summary and explanation of these and other, related formulae.

Finally, one non-geometric modulus related to the dilaton is generally present. It is known as the axio-dilaton (on account of the periodic scalar $C_0$):
\be
S=C_0+ie^{-\phi}=C_0+\frac{i}{g_s}\,.
\ee
With this, the full type-IIB moduli Kahler potential (corresponding to a so-called orientifold projection with $O3/O7$ planes - the projection to ${\cal N}=1$ mentioned earlier) reads 
\be
K=K_K(T^\alpha,\ol{T}^{\ol{\alpha}})+K_{cs}(z^a,\ol{z}^{\ol{a}}) -\ln(-i(S-\ol{S}))\,.
\ee
This defines a `ready-to-use' 4d supergravity model, so far without any scalar potential. The conventions are such that $M_{P,4}=1$, as usual in supergravity lagrangians, and that fields measuring distances or volumes in compact space (in our case the $T$'s) are doing so in string units, i.e. powers of $l_s\equiv 2\pi\sqrt{\alpha'}$. Note however that the 10d metric underlying our definition of the Kahler moduli is the 10d Einstein, not the string frame metric. Thus, 4-cycle volumes in units of $l_s^4$ are given explicitly by $g_s \tau_\alpha$. Apart from this subtlety, one may roughly say that the above is valid with $l_s=1$ concerning the Calabi-Yau geometry and $M_{P, 4}=1$ concerning 4d physics.

\subsection{An aside on string model building:
From heterotic compactifications to orientifold models with branes and F-theory}\label{asi}
\index{heterotic compactification}\index{F-theory}

This section serves two purposes: First, to give a very rough and entirely non-technical overview of particle-physics model building in string theory. In other words, we want to discuss briefly various approaches to engineering gauge group and matter content of the Standard Model in a string compactification. But second, as part of this discussion we will introduce the important concept of orientifold projections and orientifold planes. While our analysis will still be largely non-technical, this part of the section is more than just informative: A clear understanding of these ideas will be needed later on.

In the previous section, we discussed Calabi-Yaus almost entirely geometrically -- no fields other than the metric played a significant role. This view was in part biased towards the type-IIA/type-IIB framework, where the gauge fields crucial for 4d particle physics come in only in a second step, through the addition of branes. But, historically, once Calabi-Yaus were understood, a different perspective dominated:

The first successful attempts at semi-realistic string model building appeared in the context of {\bf heterotic compactifications}. They were based on compactifying the 10d ${\cal N}=1$ heterotic theory (with gauge group E$_8\times$E$_8$ or $SO(32)$) to 4d on Calabi-Yaus or torus orbifolds (to be explained momentarily). In this approach, it is possible to realise fairly straightforwardly 4d ${\cal N}=1$ SUSY EFTs very close to the MSSM. The gauge symmetry breaking comes from higher-dimensional gauge field strengths and/or Wilson lines. The 4d matter content comes from 10d gauginos, which are the only charged fermions in this context. We completely ignore this whole `universe' of string-theoretic model building opportunities in the present course (see \cite{Candelas:1985en, Dixon:1985jw, Ibanez:1986tp} for the foundational papers and \cite{Green:1987sp, Bailin:1999nk, Kobayashi:2004ya, Buchmuller:2005jr, Lebedev:2006kn, Lebedev:2008un, Braun:2005nv, Bouchard:2005ag, Blumenhagen:2006ux, Anderson:2011ns} for reviews and more recent work). This is due to time limitations and technical complexity as well as because the issue of moduli stabilisation and SUSY breaking is simpler to understand in the type-II context (see however \cite{Anderson:2011cza}).

With the understanding that D$p$-branes with even/odd $p$ are a natural part of type-IIA/IIB 10d supergravity (see in particular \cite{Polchinski:1995mt}), a whole new world of constructing standard-model-like EFTs opened up. Indeed, we have already learned that, in 10d, a D$p$ brane stack represents a dynamical submanifold with a certain tension on which a $(p\!+\!1)$-dimensional super-Yang-Mills theory lives. The highest dimensionality is obviously that of D9-branes, where we have a 10d gauge field and a 10d Majorana-Weyl gaugino. The number of supercharges (and hence of bosonic and fermionic degrees of freedom) remains the same for all $p$. For $p<9$, one has only a $(p\!+\!1)$-dimensional gauge field, and $9\!-\!p$ adjoint scalars carrying the remaining bosonic degrees of freedom. The fermionic degrees of freedom are carried by an appropriate set of lower-dimensional spinors. For example, for $p=3$ one is dealing with the famous, maximally supersymmetric 4d ${\cal N}=4$ super Yang-Mills theory with four 4d gauginos.

Thus, most naively and as was already briefly advertised in Fig.~\ref{brc}, one might think of simply compactifying either type-IIA or IIB string theory on a CY to 4d and wrap any desired number and type of D-brane stacks on the various cycles. In this way, one should be able to generate any desired gauge sector. This is almost true, but a crucial complication arises in the form of the necessary orientifold projections and orientifold planes, as we now explain.

Let us pick as our first toy model a type-IIB compactification from 10d to 8d on $T^2$. Then, add a D7-brane which fills out the eight non-compact dimensions and represents a point in the $T^2$. One might have hoped that this last step will only slightly modify the previously well-defined 8d model by adding a $U(1)$ gauge theory. However, it turns out that the whole construction becomes inconsistent for the following simple reason: Our brane couples to $C_8$, with $dC_8=F_9$, which possesses a standard dual description in terms of $F_1=dC_0$. Thus, from the point of view of the compact $T^2$, our brane represents a point carrying an `electric' gauge charge, detected by 
\be
\oint_{C_1} F_1 = 2\pi\,.\label{c1int}
\ee
Here $C_1$ is a 1-cycle encircling the D7-brane. This is completely  analogous to the charge of an electron being measured by an integral of the dual, magnetic field strength $\tilde{F}_2=*F_2$ over an $S^2$ enclosing the electron. Thus, just like in electrodynamics, (\ref{c1int}) makes it impossible to have a non-zero charge in a compact space: Indeed, $C_1$ is the boundary of what is left of the torus after the disc containing the D7 has been removed. But this 2d manifold contains no charge in our construction and hence it is inconsistent that the integral of $F_1=dC_0$ integral over its boundary is non-zero.

The way out must build on the fact that, in a consistent compactification, the total charge on the compact space must vanish. The most naive option of including a $\ol{\mbox{D7}}$ antibrane is problematic since the two branes attract each other and will eventually annihilate. Fortunately, a better route is offered by the possible presence of so-called {\bf orientifold planes}\index{orientifold plane}. These are objects which, in spite of their opposite RR-charge with respect to D-branes, are unable to annihilate them. Moreover, in contrast to antibranes, these {\bf O-planes}\index{O-plane} break supersymmetry consistently with the D-branes. In other words, a D-brane and a parallel O-plane break supersymmetry in the same way as the D-brane would by itself: to half of the flat-space SUSY of the 10d theory.

\begin{figure}[ht]
\begin{center} 
\includegraphics[width=8cm]{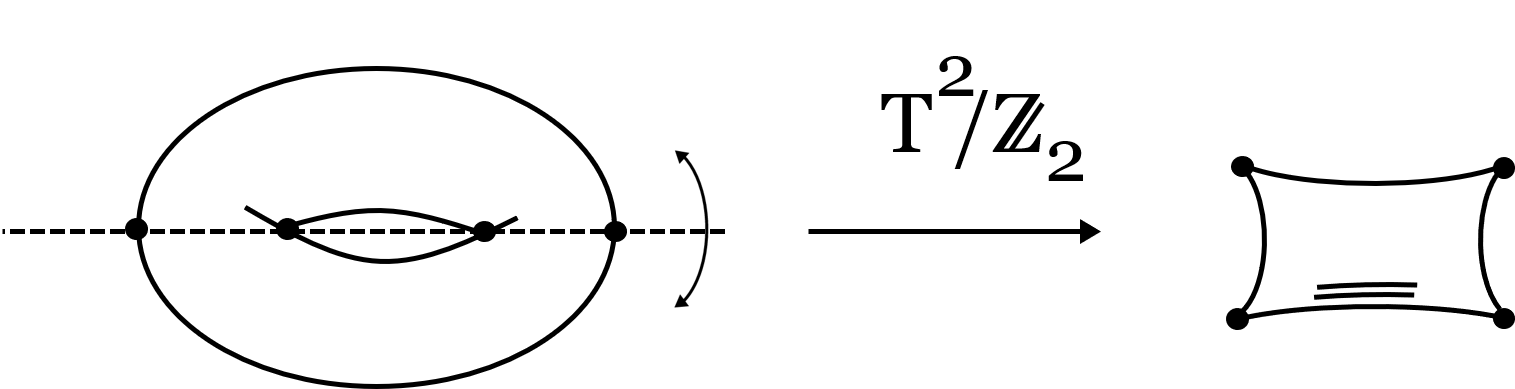}
\caption{Intuitive illustration of how in the procedure of modding out a $\mathbbm{Z}_2$ symmetry a torus is transformed into a space of half the volume and with four conical singularities at the points which are left invariant by the $\mathbb{Z}_2$ action. This resulting space has the topology of a sphere but the geometry of a `pillowcase', i.e. it is flat everywhere except at the four singularities.}
\label{orbi} 
\end{center}
\end{figure}

To understand orientifold planes or orientifold singularities, let us first briefly discuss the simpler, related concept of an {\bf orbifold}\index{orbifold}. For this purpose, let us view the $T^2$ in the above construction as $\mathbbm{C}/\mathbbm{Z}^2$, as explained in Sect.~\ref{epcy}. Let us further mod out the discrete group $\mathbb{Z}_2$ acting as $z\to -z$, i.e. a $\pi$-rotation in the compact space. It is easy to convince oneself that, under this transformation, four points of the $T^2$ remain invariant and that the resulting quotient space $T^2/\mathbb{Z}_2$ will have four conical singularities at these loci (the so-called fixed points\index{fixed point} of the orbifolding), cf.~Fig.~\ref{orbi}. We note in passing that many more options for orbifolding exist (e.g. $T^d/\mathbb{Z}_n$) with various $d$ and $n$). In many cases, supersymmetry is partially preserved and string theory continues to be well defined on such spaces in spite of the orbifold singularities.

The key novelty now arises if one does not just mod out a geometric action (in our case rotation) but combines it with worldsheet parity\index{worldsheet!parity} (for reviews and introductions see e.g.~\cite{Polchinski:1996fm, Johnson:2000ch}): In other words, one restricts the states of the original $T^2$ model to those invariant under $\pi$-rotation {\it together} with an orientation change of the string worldsheet.\footnote{
To 
be precise, one has to include $(-1)^{F_L}$ in the projection, with $F_L$ the left-moving fermion number.
}
 As a result, the fixed-point loci of the geometric $\mathbbm{Z}_2$ action become charged under RR and NS fields -- they are the orientifold planes. Explicitly, in the concrete case at hand each of the four singular points on the r.h.~side of Fig.~\ref{orbi} (corresponding to 8d hyperplanes in the full 10d geometry) develop a $C_8$ charge equivalent to 4 $\ol{\mbox{D7}}$ branes and a corresponding negative tension. As a result, type-IIB theory on $T^2/\mathbbm{Z}_2$, where the $\mathbbm{Z}_2$ acts as an orientifold projection, is only consistent if 16 spacetime-filling D7 branes are added. Moreover, if these are located in groups of four at the four orientifold planes, no deformation of the flat 10d geometry arises since neither RR nor NS fields are sourced. One says that `all tadpoles are cancelled locally'. 

The gauge group at each O7-plane is not, as one might have thought, $U(4)$ but rather $SO(8)$. This can be understood by first placing twice the required number of D7 branes on the $T^2$ in a $\mathbbm{Z}_2$-symmetric fashion and then modding out. In this procedure, the branes are pairwise identified. It can be shown that, for eight branes at one fixed point with original gauge group $U(8)$, the appropriate projection reduced the gauge symmetry to $SO(8)=U(8)/\mathbbm{Z}_2$. However, even after the projection the D7-branes may be displaced from the O7-planes and are, in fact, free to move around in the $T^2$ at zero energy cost. This breaks the gauge group generically from $SO(8)^4$ to $U(1)^{16}$. Many different intermediate gauge groups are also possible depending on which point in this `D-brane' or `open-string' moduli space one is located at.

With the basic concepts explained, we may now generalise. First, as we already noted, an orientifold projection without an accompanying space-time action is possible: Applied to the 10d type-IIB theory, it introduces a spacetime filling O9-plane, which requires the presence of 16 D9-branes to cancel the RR tadpole. This is nothing but a construction of the type-I open string through orientifolding. The gauge group is $SO(32)$, analogously to the $SO(8)$ above. Next, we may start with type IIA on $S^1$ and mod out a $\mathbbm{Z}_2$ reflection in the single, real compact dimension. The new compact space is an interval with two O8-planes at the boundaries. Each requires the presence of eight D8 branes, leading to an $SO(16)^2$ gauge theory. Analogously to the O7/D7 case, this is the situation with local tadpole cancellation.
If the D8 branes are moved, the gauge group changes.

Next, we may compactify type IIA to 7d on $T^3$ and mod out a total reflection, $y^i\to -y^i$ (for $i=1,2,3$), of the compact space together with a worldsheet orientation change. This leads to $2^3=8$ singularities with a `solid angle deficit', at each of which an O6-plane is located. Each O6 requires 2 D6-branes to cancel its tadpole. The gauge theory is $SO(4)^8$ in this simplest configuration and, as before, can change as the branes move. The reader may complete our discussion with O5, O4 and O3-planes in the obvious way.

Finally, coming closer to our our goal of phenomenology, one may  compactify type IIA on $T^6$ and mod out various $Z_2$ symmetries together with orientation changes, leading to the presence of intersecting O6 planes and intersecting branes. In particular, intersections of two D6 branes or brane stacks at points in the compact space are possible. At such points (which are in fact 4d hyperplanes filling out all of our non-compact dimensions), open string states ending on the different stacks are localised, cf.~the last picture in Fig.~\ref{brc}. They correspond to 4d chiral superfields, allowing for a particularly simple and intuitive road to particle-physics model building. These constructions are known as {\bf intersecting brane models}\index{intersecting brane models}. They have a geometrically more challenging counterpart in the Calabi-Yau context: Here, the Calabi-Yau which one starts with has to possess an appropriate $\mathbb{Z}_2$ symmetry to allow for an orientifold projection. The branes can only be wrapped on so-called lagrangian submanifolds to ensure that at least ${\cal N}=1$ supersymmetry is preserved. An excellent source for specifically this type of constructions (and for more details on all subjects of this section) is \cite{Ibanez:2012zz}. See also the reviews \cite{Blumenhagen:2005mu, Blumenhagen:2006ci} and the original papers \cite{Ibanez:2001nd, Blumenhagen:2001te, Cvetic:2001nr, Gmeiner:2005vz}.

A similar story can be developed in the type IIB context.
Here, the most promising road is that of orientifold projections introducing O3 and O7-planes. As before, they may act on $T^6$ or on an appropriate Calabi-Yau. As a crucial fact, we note that O3/O7-planes and the corresponding branes may be introduced {\it together} while still preserving some amount of supersymmetry. Similarly, the combination O5/O9 can be supersymmetric but is phenomenologically less interesting. By contrast, combinations of the type O$p$/O$(p\!+\!2)$, for example O5/O7, break SUSY completely leading in general to control issues.

The gauge group of O3/O7 orientifold models comes from D3 and D7-brane stacks. Chiral matter can live, for example, at intersections of two D7-brane stacks. Each of those represents locally a holomorphic submanifold (a divisor) and their intersection is in general a complex curve. Yukawa couplings can arise at loci where such curves meet. See e.g.~\cite{Blumenhagen:2008zz} for explicit GUT models constructed in this approach. The whole setting of such type-IIB O3/O7 Calabi-Yau or torus orientifolds is T-dual to an appropriate type-IIA model with O6 planes. To see this, it is crucial to note that T-duality also works with branes: If a brane on the IIA side wraps a compact $S^1$, the corresponding brane on the IIB side does {\it not} wrap the dual $S^1$ (and vice versa). This is consistent with the change of the dimensionality of D-branes between IIA and IIB. It clearly allows for chains of three T-dualities which relate certain $T^6$-based models in type IIA with D6-branes to type-IIB models with D3/D7 branes. 

It is even more interesting that such a duality also exists between type-IIA models on a Calabi-Yau-orientifold with O6-planes and type IIB models on a Calabi-Yau orientifold with O3/O7 planes. For this, note first that Calabi-Yaus can in general be viewed as $T^3$ fibrations (with the $T^3$ degenerating at various loci) over $S^3$. T-dualising the 3-torus takes one from a IIA model on one Calabi-Yau to a IIB model on another Calabi-Yau \cite{Strominger:1996it}. This second Calabi-Yau is referred to as being mirror-dual to the first one. The fact that Calabi-Yaus come in pairs related by {\bf mirror symmetry}\index{mirror symmetry} (which in particular exchanges $h^{1,1}$ and $h^{2,1}$) is an important mathematical fact known independently and long before the Strominger-Yau-Zaslow picture of `mirror symmetry as T-duality' \cite{Strominger:1996it}. For details on mirror symmetry see e.g.~\cite{Candelas:1993dm, Hosono:1994av, Hori:2002fa, Hori:2003ic} and refs.~therein.

Finally, type-IIB theory allows in its strong coupling regime ($g_s\sim {\cal O}(1)$) for co-dimension-two objects other that D7-branes and O7-planes. Compactifications of this type, based on type-IIB at strong coupling and including generic 7-branes (especially with gauge groups other than $SU(N)$ and $SO(N)$) are known as F-theory models \cite{Vafa:1996xn} (see also \cite{Sen:1996vd, Denef:2008wq, Weigand:2010wm, Heckman:2010bq}). The various 7-branes of F-theory are detected by the monodromy which the axio-dilaton $S=i/g_s+C_0$ undergoes if one encircles the brane. This is closely related to a discrete $SL(2,\mathbbm{Z})$ gauge symmetry of the type IIB theory, which we however have no time to describe. What is crucial for us is that this symmetry identifies certain values of $S$. As a result, $S$ does in fact not take values in the complex upper half plane as one might naively have thought but only in the so called fundamental domain of $SL(2,\mathbbm{Z})$. Interpreting $S$ as the complex structure parameter of a $T^2$, this is precisely the region in which it describes geometrically distinct tori (not related to each other by large diffeomorphisms, which are in turn characterised by $SL(2,\mathbbm{Z})$). As a result of all of this, F-theory models can be characterised by $T^2$ fibrations (more precisely `elliptic fibrations')\index{elliptic fibrations} over a complex-3-dimensional manifold (not necessarily a Calabi-Yau). The complex-structure parameter of this `artificially introduced' fibre torus specifies how the variable $S$ of the type-IIB theory varies over the base. The monodromies of $S$ characterise submanifolds of complex co-dimension one (7-branes) on which gauge theories are localised. It is quite remarkable that solutions of the type-IIB equations-of-motion in this setting arise precisely when the torus-fibration describes a Calabi-Yau 4-fold.\footnote{This, 
in turn, can be understood by starting with 11d supergravity and compactifying on this Calabi-Yau 4-fold to 3d. One then shrinks the $T^2$ fibre to zero volume and uses one of its $S^1$s to go to type-IIA supergravity. Subsequently, one appeals to T-duality on the other $S^1$ to go to type IIB with one new non-compact dimension emerging (recall that $R'=\alpha'/R\to \infty$ at $R\to 0$). This takes one to a type IIB compactification to 4d, as desired.}

The punchline is that F-theory models are arguably the most general and powerful setting for string phenomenology, including through various limits and dualities all that can be done in perturbative type IIB models with branes and much of what can be done in the type IIA and heterotic context. It has proven a particularly fruitful setting for constructing grand unified theories, especially because its 7-branes allow for exceptional gauge groups and, through their breaking, for realistic GUT models with the right Yukawa coupling structure. This has been explored relatively recently under the name of `F-theory GUTs' \cite{Beasley:2008dc, Donagi:2008ca} (see \cite{Weigand:2010wm, Heckman:2010bq} for reviews). In addition, as will become clearer at the end of Sect.~\ref{tfl}, F-theory is particularly powerful in that it presumably generates the largest number of the presently known landscape vacua.

\subsection{Problems}

\subsubsection{Dimensional reduction}
\index{dimensional reduction}\index{Kaluza-Klein!theory}
\index{Kaluza-Klein!reduction}

{\bf Task:} Perform the KK reduction\index{KK reduction} of the 5d lagrangian
\be
{\cal L}_5=\ol{\Psi}i\slashed{\partial}\Psi-M\ol{\Psi}\Psi
\ee
to 4d on $S^1$. Give your result in a compact, standard 4d notation as appropriate for a theory with towers of Dirac fermions.

Now gauge the fermion in the lagrangian above, adding also a standard gauge-kinetic term. Perform again the dimensional reduction, but disregard the higher modes of the gauge field (to avoid dealing with towers of massive vectors, which is interesting but not essential in our context). Show how the 4d scalar coming from $A_5$ couples to 4d fermions. It appears naively that 5d gauge invariance, which should be manifest as a discrete shift symmetry of $A_5$, is broken by such an interaction. Resolve this puzzle!

\noindent
{\bf Hints:} It is essential to use
\be
\gamma^5\equiv i\left(\begin{array}{rr}
-\mathbbm{1} & 0\
\\
0 & \mathbbm{1}
\end{array}\right)\,,
\ee
which differs by a prefactor from standard 4d conventions. This is clear since otherwise the 5d Clifford algebra relations\footnote{
Recall that we 
use the mostly-plus convention.
}
\be
\{\gamma^M,\gamma^N\}=-\eta^{MN}
\ee
would have an incorrect sign for the index choice $(MN)=(55)$. The rest is a straightforward analysis following the scalar case presented in the lecture. 
It is more convenient to use exponentials rather than sines and cosines when dimensionally reducing the 5d fields.

\noindent
{\bf Solution:} Let us make the ansatz
\be
\Psi(x,y) = \sum_{n=-\infty}^{+\infty} \psi^L_n(x)\,e^{iny/R}+\sum_{n=-\infty}^{+\infty}
\psi^R_n(x)\,e^{iny/R}\,,
\ee
where $x\equiv\{x^\mu\}$ and the indices $L/R$ denote left and right-handed 4d fermions. After a straightforward calculation, using in particular manipulations like
\be
\ol{\psi}^L_n(x)\,e^{-iny/R}\,i\gamma^5\partial_5\, \psi^R_n(x)\,e^{iny/R}
\,=\,\ol{\psi}^L_n(x)(-in/R)\,\psi^R_n(x)\,,
\ee
one arrives at
\bea
S&=&2\pi R\int d^4x\,\Bigg[\ol{\psi}^L_0 i\slashed{\partial}\psi^L_0
+\ol{\psi}^R_0 i\slashed{\partial}\psi^R_0 - M \ol{\psi}^L_0 \psi^R_0+\mbox{h.c.}
\nonumber \\
&&
\hspace*{1cm}+\sum_{n\neq 0} \left\{\ol{\psi}^L_n i\slashed{\partial}\psi^L_n + \ol{\psi}^R_n i\slashed{\partial} \psi^R_n - M\ol{\psi}^L_n \psi^R_n+\mbox{h.c.}
\right.
\nonumber \\
&&
\hspace*{2cm}\left. (-in/R)\ol{\psi}_n^L\psi_n^R
+(in/R)\ol{\psi}_n^R\psi_n^L\,\right\}
\Bigg]\,.\label{beq}
\eea
We can absorb the volume factor in a field redefinition and write this as a tower of pairs of l.h.~and r.h.~fermions,
\be
S=\int d^4 x \sum_{n=-\infty}^{+\infty} \left[ \ol{\psi}^L_n i\slashed{\partial}\psi^L_n + \ol{\psi}^R_n i\slashed{\partial} \psi^R_n - M_n\ol{\psi}^L_n \psi^R_n +\mbox{h.c.}\right]
\ee
with Dirac-type mass terms, but with complex mass parameters
\be
M_n=M+in/R\,.
\ee
Of course, the complex phases of the $M_n$ can be absorbed in a phase rotation of, for example, the right handed parts. The mass parameters now become real and the two terms with $M_n$ and $\ol{M}_n$ can be combined in Dirac mass terms. Thus, we obtain
\be
S=\int d^4 x \sum_{n=-\infty}^{+\infty} \ol{\psi}_n (i\slashed{\partial}
-M_n)\psi_n
\ee
with
\be
M_n=\sqrt{M^2+(n/R)^2}\,.
\ee

Introducing the gauging, one gets a 4d gauge theory and a real scalar coming from $A_5$, as explained in the lecture. Crucially, one also finds a coupling of the scalar to the fermions,
\be
\ol{\Psi}i\gamma^5iA_5\Psi\quad \to\quad -i\phi\,\ol{\psi}_n^L \psi_n^R
+\mbox{h.c.}
\ee

The fermions can again be rescaled to absorb the volume prefactor $(2\pi R)$ of the fermionic part of the action. If $M\ll n/R$, it is natural to focus on the zero-mode level of this Kaluza-Klein theory:
\be
S=\int d^4 x\,\left(-\frac{1}{4g^2}\,F_{\mu\nu}F^{\mu\nu}-\frac{1}{2g^2}(
\partial \phi)^2 + \ol{\psi}_0^L\,i\slashed{D}\,\psi_0^L
+ \ol{\psi}_0^R\,i\slashed{D}\,\psi_0^R - \psi_0^L(M+i\phi)\psi_0^R +\mbox{h.c.}\right)\,.
\ee
It is clear however that, to make the apparently broken shift symmetry $\phi\to \phi+1/R$ manifest, one needs to include higher fermion modes. Indeed, when the modulus $\phi$ continuously changes its value from zero to $1/R$, the mode with $n=-1$ takes the place of the former zero mode. Thus, the model as a whole returns to a physically equivalent situation, as it should be given that $\phi=0$ and $\phi=1/R$ are related by a gauge transformation.

\subsubsection{SO(2n) vs. U(n)}

{\bf Task:} In the lecture we used the fact that, if $R_{\alpha\beta}{}^\gamma{}_\delta$ is pure in the second index pair, then the holonomy is reduced to $U(n)$. (Here we use greek rather than latin indices to symbolise that, e.g., $\alpha$ may stand for either $i$ or $\ol{\imath}$.) This fact is in principle obvious and does not require any demonstration. Still, to make the simple underlying techniques more manifest, consider the following simple problem:

Let $v^\alpha$, $w^\beta$ specify a vector pair such that 
$v^\alpha w^\beta R_{\alpha\beta}{}^\gamma{}_\delta\equiv Q^\gamma{}_\delta$ describes an infinitesimal $SO(2n)$ rotation in the complex basis, corresponding to the appropriate parallel transport along an infinitesimal loop. According to the pure index structure, $Q$ takes the form 
\be
Q=\left(\begin{array}{cc} M & 0 \\ 0 & N
\end{array}\right)\,.
\ee
In other words, $Q$ is characterised by $v^\alpha w^\beta R_{\alpha\beta}{}^i{}_j\equiv M^i{}_j$ and 
$v^\alpha w^\beta R_{\alpha\beta}{}^{\ol{\imath}}{}_{\ol{\jmath}}\equiv N^i{}_j$.

Which properties of $M$ and $N$ follow from the fact that $Q$ corresponds to an infinitesimal $SO(2n)$ transformation?$\,$\footnote{Of course, the transformation described by $Q$ is in $Lie(SO(2n))$ by the very definition of $R$, such that these properties could also be derived from elementary differential geometry. But we want a purely algebraic derivation.}

\noindent
{\bf Hint:} Use the notation $z'^i=M^i{}_jz^j$ and $z=x+iy$ such that
\be
\left(\begin{array}{c} x \\ y\end{array}\right) 
\ee
is the column vector transforming under $SO(2n)$. 

\noindent
{\bf Solution:} The matrix $Q$ characterises a linear transformation in the $(z,\ol{z})$-basis. To translate this into the $(x,y)$ basis, write
\bea
\hspace*{-.7cm}x'^i &=& (z'^i+\ol{z}'^i)/2 = ( M^i{}_jz^j+N^j{}_j\ol{z}^j)/2
= (M^i{}_jx^j + iM^i{}_jy^j + N^i{}_jx^j - iN^i{}_jy^j)/2
\\
\hspace*{-.7cm}y'^i &=& (z'^i-\ol{z}'^i)/2i = ( M^i{}_jz^j-N^j{}_j\ol{z}^j)/2i
= (M^i{}_jx^j + iM^i{}_jy^j - N^i{}_jx^j + iN^i{}_jy^j)/2i\,.
\eea
From this, the real-basis form $Q_r$ of the transformation $Q$ is easily read off:
\be
Q_r=\frac{1}{2}\left(\begin{array}{cc} M+N & i(M-N) \\ -i(M-N) & M+N
\end{array}\right)\,.
\ee
Our requirement $Q_r\in Lie(SO(2n))$ implies that $Q_r$ is real antisymmetric, i.e.
\bea
\ol{M}+\ol{N}=M+N\,\,\quad&\quad M^T+N^T=-M-N\,\,,
\\
\ol{M}-\ol{N}=-M+N\,\,\quad&\quad M^T-N^T = M-N\,\,.
\eea
Adding the first and the third equation gives $N=\ol{M}$. The other two equations imply $M^T=-N$. Thus, 
\be
N=\ol{M}\quad,\qquad M=-M^\dagger\,,
\ee
and
\be
Q=\left(\begin{array}{cc} M & 0 \\ 0 & \ol{M}
\end{array}\right)\qquad\mbox{with}\qquad M\in Lie(U(n))\,.
\ee
We see that $Q$ does indeed describe an infinitesimal $U(n)$ rotation in the complex basis.

\subsubsection{Complex projective spaces}\index{projective space}
{\bf Task:} Consider $\mathbb{C}P^n$ with charts as defined in the lecture and obtain explicitly the transition functions $\phi_i\circ\phi_j^{-1}$. Give a general formula for the components $g_{i\ol{\jmath}}$ of the Fubini-Study metric\index{Fubini-Study metric} in some chart $\phi_k$. Show consistency between different charts. In the special case of $\mathbb{C}P^1\cong S^2$, show agreement with the round metric on the sphere (up to normalisation).

\noindent
{\bf Hints:} Deriving the transition functions is completely straightforward, but some care is needed concerning the indexing of the variables in the two charts. Getting the Fubini-Study metric in one chart requires just differentiation. To show that the Fubini-Study metric is well-defined, it is useful to first investigate how the Kahler potential transforms between coordinate patches. Try to make use of the (multi-variable generalisation) of the fact that $\partial_z\ol{\partial}_{\ol{z}}\ln(z\ol{z})=0$ for $z\neq 0$. If you get stuck use, e.g., the lecture notes by Candelas \cite{Candelas:1987is} or the Wikipedia page for `Fubini-Study metric'. In the last part, think of the stereographic projection.

\noindent
{\bf Solution:} The two sets of local coordinates in $\phi^i$ and $\phi^j$ may be chosen as
\be
(x^1,\cdots,x^n)\,=\,\left(\frac{z^0}{z^i},\cdots,\frac{z^{i-1}}{z^i},
\frac{z^{i+1}}{z^i},\cdots,\frac{z^n}{z^i} \right)
\ee
and
\be
(y^1,\cdots,y^n)\,=\,\left(\frac{z^0}{z^j},\cdots,\frac{z^{j-1}}{z^j},
\frac{z^{j+1}}{z^j},\cdots,\frac{z^n}{z^j} \right)\,.
\ee
The coordinate change is found by explicitly rewriting each of the $x^k$ in terms of the $y$-coordinates. For definiteness, let us assume $i<j$. Then we find for $k\leq i$:
\be
x^k=\frac{z^{k-1}}{z^i}=\frac{z^{k-1}}{z^j}\cdot\frac{z^j}{z^i}=y^k\cdot\frac{1}{y^{i+1}}\,.
\ee
For $i+1\leq k < j$:
\be
x^k=\frac{z^k}{z^i}=\frac{z^k}{z^j}\cdot\frac{z^j}{z^i}=y^{k+1}\cdot\frac{1}{y^{i+1}}\,.
\ee
Then comes a special case: For $k=j$,
\be
x^j=\frac{z^j}{z^i}=\frac{1}{y^{i+1}}\,.
\ee
Finally, for $j<k$:
\be
x^k=\frac{z^k}{z^i}=\frac{z^k}{z^j}\cdot\frac{z^j}{z^i}=y^k\cdot\frac{1}{y^{i+1}}\,.
\ee
We may summarize all of this in the compact expression
\be
(x^1(y),\cdots,x^n(y))\,=\,\frac{1}{y^{i+1}}\,(y^1,\cdots,y^i,y^{i+2},\cdots,
y^j,1,y^{j+1},\cdots,y^n)\,.
\ee

Obtaining the explicit form of the Fubini-Study metric is easy: Consecutive differentiation w.r.t.~$x^i$ and $\ol{x}^{\ol{\jmath}}$ gives
\be
2K^{(k)}_i=\frac{\ol{x}^{\ol{\imath}}}{1+x^l \ol{x}^{\ol{l}}\delta_{l\ol{l}}}
= \frac{\ol{x}^{\ol{\imath}}}{\sigma}\qquad \mbox{with}\qquad \sigma\equiv 1+x^l \ol{x}^{\ol{l}}\delta_{l\ol{l}} 
\ee
and
\be
2g_{i\ol{\jmath}}\equiv 2K^{(k)}_{i\ol{\jmath}} = \frac{\delta_{i\ol{\jmath}}}{\sigma} - \frac{\ol{x}^{\ol{\imath}}x^j}{\sigma^2}\,.
\ee
Here summation over $l$ and $\ol{l}$ is implicit in the first line.

To see invariance under coordinate change, recall that the two metrics in $U_i$ and $U_j$ are defined as
\be
\frac{\partial}{\partial x^k}\, \frac{\ol{\partial}}{\ol{\partial}\ol{x}^{\ol{l}}} \,K^{(i)}(x,\ol{x})
 \qquad\mbox{and}\qquad \frac{\partial}{\partial y^k}\, \frac{\ol{\partial}}{\ol{\partial}\ol{y}^{\ol{l}}} \,K^{(j)}(y,\ol{y})\,.
\ee
This obviously defines two tensors which will by definition agree if 
\be
\frac{\partial}{\partial y^k}\, \frac{\ol{\partial}}{\ol{\partial}\ol{y}^{\ol{l}}} \,K^{(j)}(y,\ol{y})\,=\,
\frac{\partial}{\partial y^k}\, \frac{\ol{\partial}}{\ol{\partial}\ol{y}^{\ol{l}}}\, K^{(i)}(x(y),\ol{x}(\ol{y}))\,.\label{kij}
\ee
Note that here we also have to use the fact that the coordinate change is holomorphic, such that holomorphic and antiholomorphic indices do not mix under reparameterization.

Now, Eq.~(\ref{kij}) will clearly hold if
\be
K^{(i)}=K^{(j)}(y,\ol{y})+f(y)+\ol{f}(\ol{y})\,.
\ee
This is, in fact, known as a Kahler transformation\index{Kahler!transformation} -- the natural way in which a Kahler potential\index{Kahler!potential} changes between patches on a Kahler manifold.\index{Kahler!manifold}

Showing that this holds is easy if one notes that
\be
2K^{(i)}(x,\ol{x})=\ln\sigma(x,\ol{x})\qquad \mbox{and}\qquad 
2K^{(j)}(y,\ol{y})=\ln\sigma(y,\ol{y})
\ee
with $\sigma$ as defined above. Moreover, let us think of a different way of labelling our coordinates as follows:
\be
x^k\equiv z^k/z^i\qquad \mbox{and}\qquad y^k\equiv z^k/z^j\,,
\ee
such that $k=0,\cdots,n$, but with the caveat that $x^i=1$ and $y^j=1$ and hence these two do not count as coordinates. In this notation, one has
\be
\sigma(x,\ol{x})=\sum_{k=0}^n|x^k|^2\qquad \mbox{and}\qquad \sigma(y,\ol{y})=\sum_{k=0}^n|y^k|^2
\ee
and 
\be
\sigma(x(y),\ol{x}(\ol{y}))=\sigma(y,\ol{y})\,|z^j/z^i|^2
=\sigma(y,\ol{y})\,|y^j/y^i|^2\,.
\ee
Hence, we obtain the above form of a general Kahler transformation with 
$f(y)=\ln(y^j/y^i)$. This completes the demonstration that the metric is well-defined.

Finally, let us consider the specific case of $\mathbb{C}\!P^1$. In the patch $U_0$, we have
\be
2g_{x\,\ol{x}}=\frac{1}{1+|x|^2}-\frac{|x|^2}{(1+|x|^2)^2}=\frac{1}{(1+|x|^2)^2}
\ee
and, with $x=r\,\exp(i\phi)$,
\be
ds_x^2=g_{x\ol{x}}dx\,d\ol{x}+g_{\ol{x}x}d\ol{x}\,dx=2g_{x\ol{x}}dx\,d\ol{x}=
\frac{dr^2+r^2d\phi^2}{(1+r^2)^2}\,.
\ee
This has to be compared with the round metric on the unit sphere,
\be
ds_1^2=d\theta^2+\sin^2\theta\,d\phi^2\,.
\ee
Now imagine that this sphere is centered at the origin in $\mathbb{R}^3$ and map it to the $x$-$y$-plane using rays originating in the north pole and intersecting the plane and the sphere (stereographic projection).\footnote{
Beware 
that an alternative form of the stereographic projection uses a unit sphere centered at $(0,0,1)\in \mathbb{R}^3$. This corresponds to scaling distances on the plane up by a factor of two.
} 
Elementary geometry proves that the ray which intersects the sphere at $(\theta,\phi)$  will enclose an angle $\theta/2$ with the negative vertical axis. Hence, parameterising the plane by the complex variable $x=r\,\exp(i\phi)$ as above, we have $r=\tan(\theta/2)$. Thus, $2dr/d\theta =1+\tan^2(\theta/2)=1+r^2$, which gives
\be
ds_x^2=\frac{d\theta^2}{4}+\frac{\tan^2(\theta/2)}{(1+\tan^2(\theta/2))^2} d\phi^2 = \frac{d\theta^2}{4}+\tan^2(\theta/2)\cos^4(\theta/2) d\phi^2=\frac{1}{4}(d\theta^2+\sin^2\theta\,d\phi^2)
\ee
and $ds_x^2=ds_1^2/4$, as proposed. Our complex $x$-coordinate covers the sphere without the north pole. The coordinate change $x\to 1/x$ takes us to the second coordinate patch, which covers the sphere without the south pole.

\section{The flux landscape}\label{tfl}
\index{flux landscape}

The general idea will be to consider compactifications with non-zero internal components of the RR and NS field strength tensors $F_p$ and $H_3$. This induces a non-zero superpotential depending on the moduli of the supergravity models discussed above and leads to moduli stabilisation\index{moduli stabilisation}. Moreover, the number of available distinct models jumps from $10^8$ to $10^{600}$ (or, in more general geometries -- roughly speaking including D-branes -- even much higher). A small fraction of them, which would still be an enormous number, are expected to have broken supersymmetry and a positive cosmological constant.
If true, this implies a paradigm shift in fundamental physics  comparable to the Copernican revolution: Our fundamental  physics parameters are not fundamental at all but drawn from a large set of solutions - the string theory landscape.
It may even become unavoidable to invoke anthropic considerations to come to terms with the implications \cite{Susskind:2003kw}. We will now discuss this in detail following the path of the Bousso-Polchinski model \cite{Bousso:2000xa}, the more realistic GKP setting \cite{Giddings:2001yu} and the KKLT \cite{Kachru:2003aw} and LVS \cite{Balasubramanian:2005zx} proposals.\footnote{See \cite{Schellekens:2006xz} for a discussion of earlier references emphasising the non-uniqueness of string compactifications and the resulting need for anthropic considerations.}

\subsection{Compact geometries with $p$-form fluxes}\label{cgwf}
\index{$p$-form fluxes}

Let us start with a few general comments on $p$-form gauge theories. Consider a $(p\!-\!1)$-form gauge theory in $d$ dimensions, with an action of type (we disregard purely numerical prefactors)
\be
\int \frac{1}{g^2}\,F_p\wedge * F_p + \int_{(p-2)-brane} A_{p-1}\,.
\ee
One can easily show that a {\bf dual}\index{duality} description is provided by a theory based on the $(d-p)$-form field strength $\tilde{F}_{d-p}$. The latter is defined as
\be
\tilde{F}_{d-p}=\frac{1}{g^2} * F_p\,,
\ee
which in turn leads to the definition of a dual gauge potential via
\be
\tilde{F}_{d-p}=d\tilde{A}_{d-p-1}\,.
\ee
In these new variables the action takes the form
\be
\int \frac{1}{\tilde{g}^2}\, \tilde{F}_{d-p}\wedge *  \tilde{F}_{d-p} + 
\int_{(d-p-2)-brane} \tilde{A}_{d-p-1}\qquad\mbox{with}\qquad \tilde{g}=\frac{1}{g}\,.
\ee
While the new kinetic term is just a rewriting of the old one, the charged objects coupling to the dual potential are different. In fact, both types of charged objects are present in the full theory. But the coupling of any one of them to the fields can only be explicitly given on one side of the duality.\footnote{
The purely classical dualisation above can also be performed at the quantum level, i.e. under the path integral. The basic idea is to implement the original Bianchi identity constraint $dF_p=0$ by a Lagrange multiplier, i.e. by adding a term $dF_p\wedge A_{d-p-1}$ to the action. Then one can integrate out $F_p$, arriving at the dual action. In the latter, the Lagrange multiplier $A_{d-p-1}$ has become the new dynamical field.}

The above is of course familiar from electrodynamics, where $d=4$ and $p=4-p=2$, such that the tilde is really necessary to distinguish the otherwise identical-looking dual descriptions. The charged objects on both sides are 0-branes, i.e. particles. 

Now let us consider the particularly simple case of $F_1$ in $d=4$, which is of course nothing but a scalar (axion) field model, with $A_0\equiv \phi$:
\be
\int f^2 (\partial \phi)^2\,+\,\,\phi(x_i)\,.\label{imo}
\ee
The last term is the coupling to an instanton, a tunnelling event localised at a point in spacetime. Clearly, such objects can not be included in an initial field configuration -- one has to sum over them and integrate over all the $x_i$ in the path integral. 

\begin{figure}[ht]
\begin{center} 
\includegraphics[width=5.5cm]{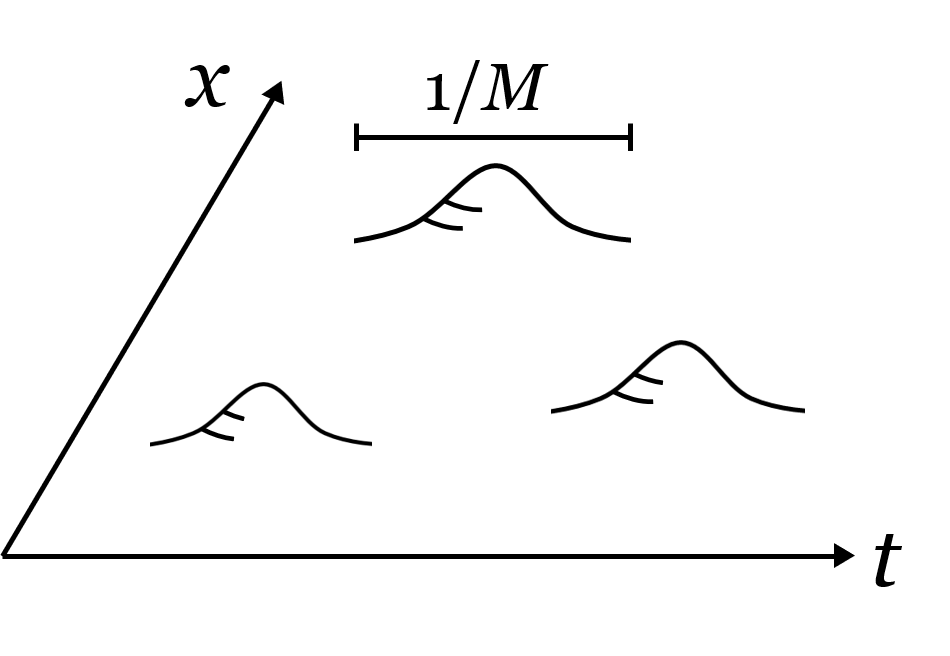}
\caption{Instantons as localised lumps of field strength (figure adapted from~\cite{Hebecker:2018ofv}).}
\label{instantons} 
\end{center}
\end{figure}

In case this concept is unfamiliar, here is a brief excursion concerning {\bf instantons}\index{instanton} (see \cite{Coleman:1985rnk, shifman, Blumenhagen:2009qh, Bianchi:2007ft} as well as many QFT textbooks). The `classical' setting is in fact that of a gauge theory coupled to a periodic pseudo-scalar or axion-like-field or axion for short. For definiteness, say the gauge group is $SU(2)$:
\be
{\cal L}=\frac{1}{2g^2}\,\mbox{tr}\,F_{\mu\nu}F^{\mu\nu}+\frac{\phi}{8\pi^2}\, \mbox{tr}F\wedge F\,. 
\ee
The term multiplying $\phi$ is a total derivative but there exist field configurations (which can not be smoothly deformed to the vacuum) on which the integral gives $8\pi^2n$ with $n\in\mathbb{Z}$. Very roughly speaking, the existence of such a field configuration is related to the fact that $SU(2)\cong S^3$ and the possibility of identifying this group-theoretic $S^3$ with the $S^3$ of radial coordinates in $\mathbb{R}^4$. In the euclidean path integral, one has to sum over all such `bumps of energy-density' (to be interpreted as local tunnelling events, leading from vacuum to vacuum). One also has to integrate over all their sizes $1/M$ and positions. The events are suppressed by their action -- $\exp(-S_i)$ -- and for large $S_i$ one uses the `dilute gas approximation' (cf.~Fig.~\ref{instantons}). It should now be clear in which sense our model of (\ref{imo}) corresponds to instantons of an $SU(2)$ (more generally $SU(N)$) gauge theory: The point at which the gauge-field-theoretic instanton is localised is identified with $x_i$ and the $F\wedge F$ term of the lump of field strength is replaced by an approximate $\delta$-function. 

\begin{figure}[ht]
\begin{center} 
\includegraphics[width=2.7cm]{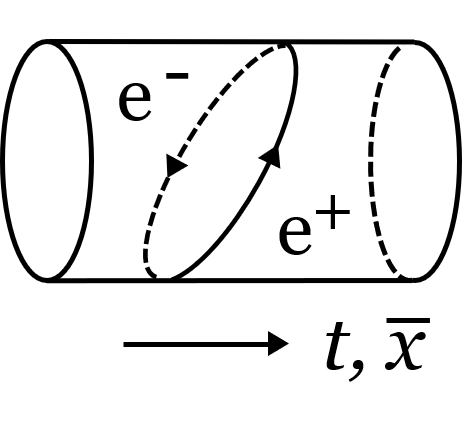}
\caption{Effective instanton arising from a particle-antiparticle fluctuation 
wrapping the compact space of an $S^1$ compactification (figure adapted from~\cite{Hebecker:2018ofv}).}
\label{strins} 
\end{center}
\end{figure}

Still within our excursion about instantons, we recall that a model with a periodic scalar like that of (\ref{imo}) can be derived by compactifying a 5d $U(1)$ gauge theory to 4d. Interestingly, this also has instantons, but of a very different type (cf.~Fig.~\ref{strins}). We leave it as an exercise for the reader to derive the correct coupling of this type of instanton to $\phi$. 
This ends our instanton excursion.

As a side remark, the dual theory has field strength $H_3=dB_2$ and couples to strings (which are here unrelated to any fundamental string theory). What interests us here is flux quantisation, which is particularly easy to understand in this case: If our spacetime has non-trivial one-cycles, the gauge potential $\phi$ does not need to be globally well-defined. Instead, assuming e.g. that $x^3$ parameterises an $S^1$, it may obey 
\be
\phi(x^3)=\phi(x^3+2\pi R)+2\pi n\,,\qquad n\in \mathbb{Z}\,.\label{flc}
\ee
The shift must be integer or else the exponential of the instanton coupling, $\exp(i\phi(x^3))$, would not be well-defined. Another way to formulate the same condition is
\be
\oint F_1=2\pi n\,.\label{f1str}
\ee
Here $n$ is a discrete choice one has to make when defining the theory on a spacetime with a non-trivial 1-cycle. An analogous non-trivial boundary condition would arise if the 5d topology were trivial but the loop in (\ref{f1str}) were wrapped around $n$ strings.

The above is clearly analogous to the familiar statement
\be
\oint_{S^2} F_2=2\pi n \label{f2n}
\ee
for electrodynamics and an $S^2$ enclosing $n$ magnetic monopoles. But this case is not our interest at present. What we care about is flux quantisation,
\be
\oint_{c_p}F_p\in 2\pi \,\mathbb{Z}\,,
\ee
which is simply a requirement of (quantum-mechanical) consistency of a $p$-form gauge theory (and its dual). We see that, in the absence of charges, the flux can only be non-zero if a non-trivial $p$ cycle exists in the geometry. If so, it is determined by a discrete choice one has to make for every such $p$-cycle.

\begin{figure}[ht]
\begin{center} 
\includegraphics[width=8cm]{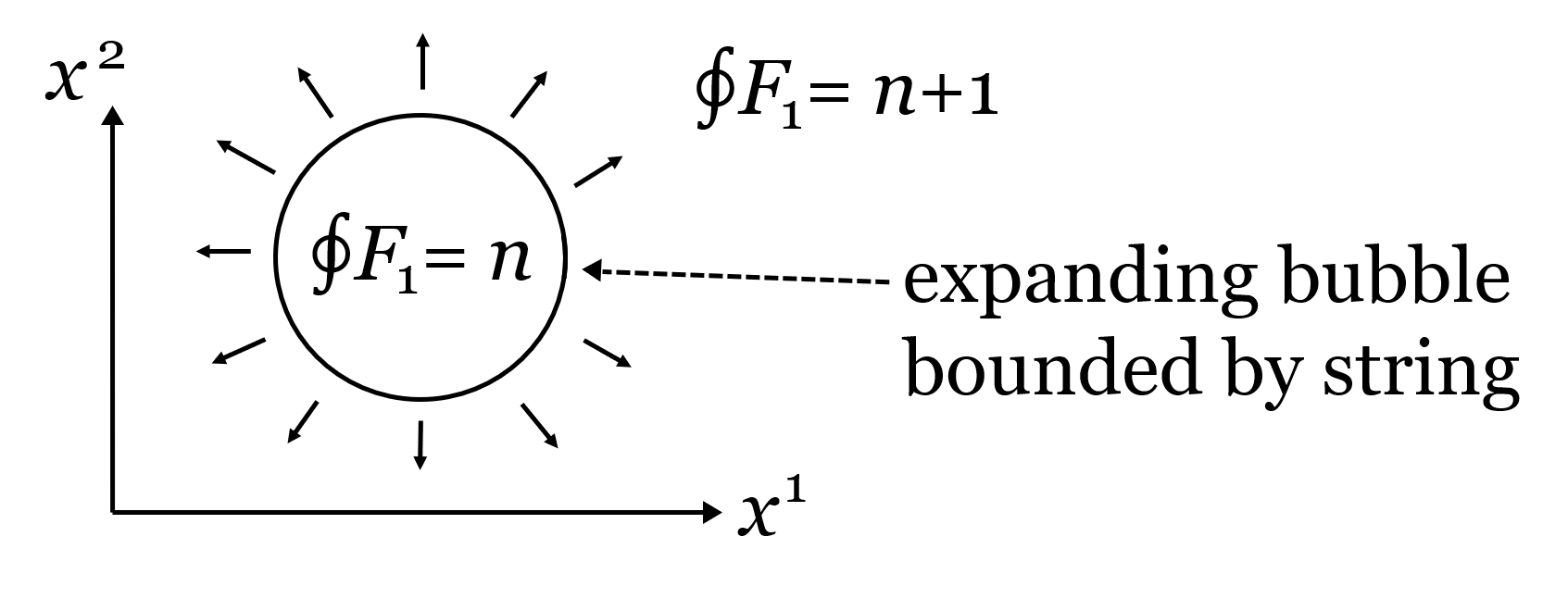}
\caption{Bubble nucleation in a 4d-to-3d toy model with 1-form flux.}
\label{bst} 
\end{center}
\end{figure}

Now let us compactify a 4d model with a $0$-form gauge theory (an axion) to 3d on $S^1$. The compact geometry has a single compact 1-cycle. This allows a choice of boundary condition or, equivalently, 1-form flux on the $S^1$. The freedom is precisely that of choosing $n\in \mathbb{Z}$ in (\ref{flc}). 
Thus, one obtains an infinity of 3d models with different vacuum energy: Indeed, from the perspective of the non-compact dimensions $\{x^0,x^1,x^2\}$ the gradient term $(\partial_3\phi)^2$ contributes to the cosmological constant. This already represents a small flux landscape. Moreover, the theory possesses strings. Let us include an infinite string in our compactification. This string is a point in the compact $x^3$-direction and hence still has two dimensions -- one time and one spacelike -- in the noncompact $(2+1)$-dimensional spacetime. It is hence a domain-wall in the noncompact 2d space. Once can convince oneself that, on the two sides of this wall, the flux on the $S^1$ differs by one unit. Hence our flux landscape is actually not just a collection of different theories, but it possesses a dynamics allowing one to change between those: This dynamics is {\bf bubble nucleation}\index{bubble!nucleation} (cf.~Fig.~\ref{bst}). The surfaces of the bubbles are the domain walls made of the higher-dimensional charged objects. This crucial feature will survive in the full-fledged string theory landscape. 

Clearly, an analogous situation may be considered if one compactifies, for example, a 6d gauge theory to 4d on $S^2$. The $S^2$ may be given 2-form flux in the sense of (\ref{f2n}), giving rise to a 4d landscape of vacua labelled by $n$. In this case, the flux quantisation is literally based on the same logic that forces the $F_2$ integral around a magnetic monopole to be quantised. One may also use the $U(1)$ principle bundle approach to gauge theories to think of this in terms of non-trivial fibrations of $U(1)$ over $S^2$, which are known to be labelled by an integer, our flux number. The case of unit flux corresponds to the famous Hopf fibration\index{Hopf fibration} (see e.g.~\cite{nak, bert, nashsen, gs}).

\subsection{Bousso-Polchinski model}\label{bpm}
\index{Bousso-Polchinski model}
We have just understood that the compactification of higher-dimensional $p$-form theories provides apparently rather general mechanism for creating 4d `landscapes'. Now let 
see how this is reflected in an effective field theory which works directly in 4d~\cite{Bousso:2000xa} (see also~\cite{Feng:2000if}). Such an effective description arises naturally if we consider the (somewhat special) case of a $(d-1)$-form gauge theory in $d$ dimensions. Our specific interest is of course in 3-form gauge theories in $d=4$:
\be
S= -\int \frac{1}{\Lambda^4}F_4^2+\int_{\mbox{\footnotesize domain wall}} A_3\,. \label{sfbp}
\ee
Without sources, the equation of motion $d*F_4=0$ implies that $F_4$ is constant, so there are no propagating degrees of freedom. The only dynamics is that of domain walls, which have some tension and hence move according to their own classical dynamics. Moreover, they couple to $A_3$ and hence source $F_4$. 

\begin{figure}[ht]
\begin{center} 
\includegraphics[width=5.5cm]{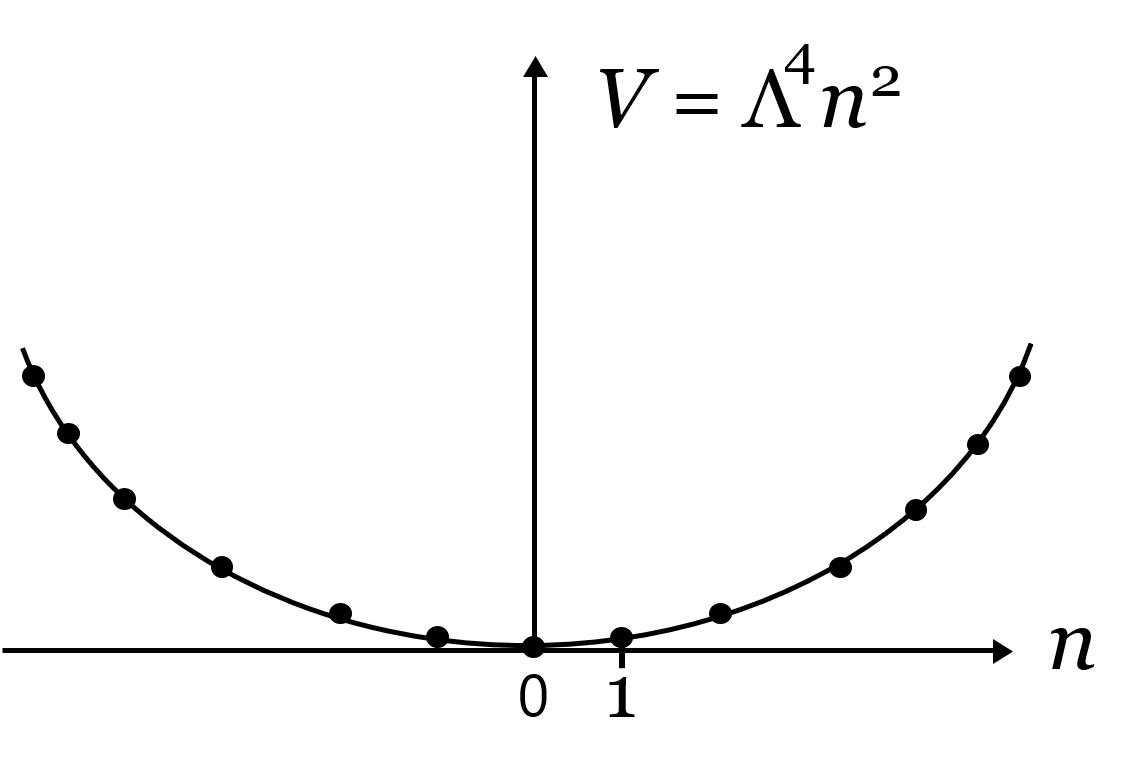}
\caption{Flux discretuum of a 3-form gauge theory.}
\label{bp} 
\end{center}
\end{figure}

With the domain wall comes a 1-form current, appearing in
\be
\frac{1}{\Lambda^4} d*F_4=j_1\,,
\ee
which is localised at the wall. As is generally the case in $p$-form gauge theories, the integral of the current counts the number of charged objects. In the most familiar case of 4d electrodynamics, the integral of the 3-form current over a spatial 3d volume counts the number of charged-particle worldlines crossing that volume. Here, our 1-form current should integrate to unity on any line 
that crosses the domain wall once. Concretely, consider a finite line, with beginning and end point on opposite sides of the wall, such that
\be
1 = \int_{Line} j_1 = \frac{1}{\Lambda^4}\int_{Line} d*F_4 = \frac{1}{\Lambda^4}(*F_4)\Big|_{x_1}^{x_2}\,.
\ee
From this, we see right away that the scalar $*F_4$ jumps by $\Lambda^4$ when crossing the wall. The dual description, though even more exotic, is simpler:
\be
S= -\int \Lambda^4\, F_0^2\,, \label{moft}
\ee
without any meaningful `$A_{-1}$' or sources. The $0$-form field strength is classically identified with $*F_4$, it is constant in spacetime by its Bianchi identity, $dF_0=0$, and it only takes discrete values. This follows from the solution for $F_4$ in the vicinity of a domain wall discussed above. It can also be viewed as a degenerate version of flux quantisation. The set of vacua following from the $F_0$ description is displayed in~Fig.~\ref{bp}. 

\begin{figure}[ht]
\begin{center} 
\includegraphics[width=8cm]{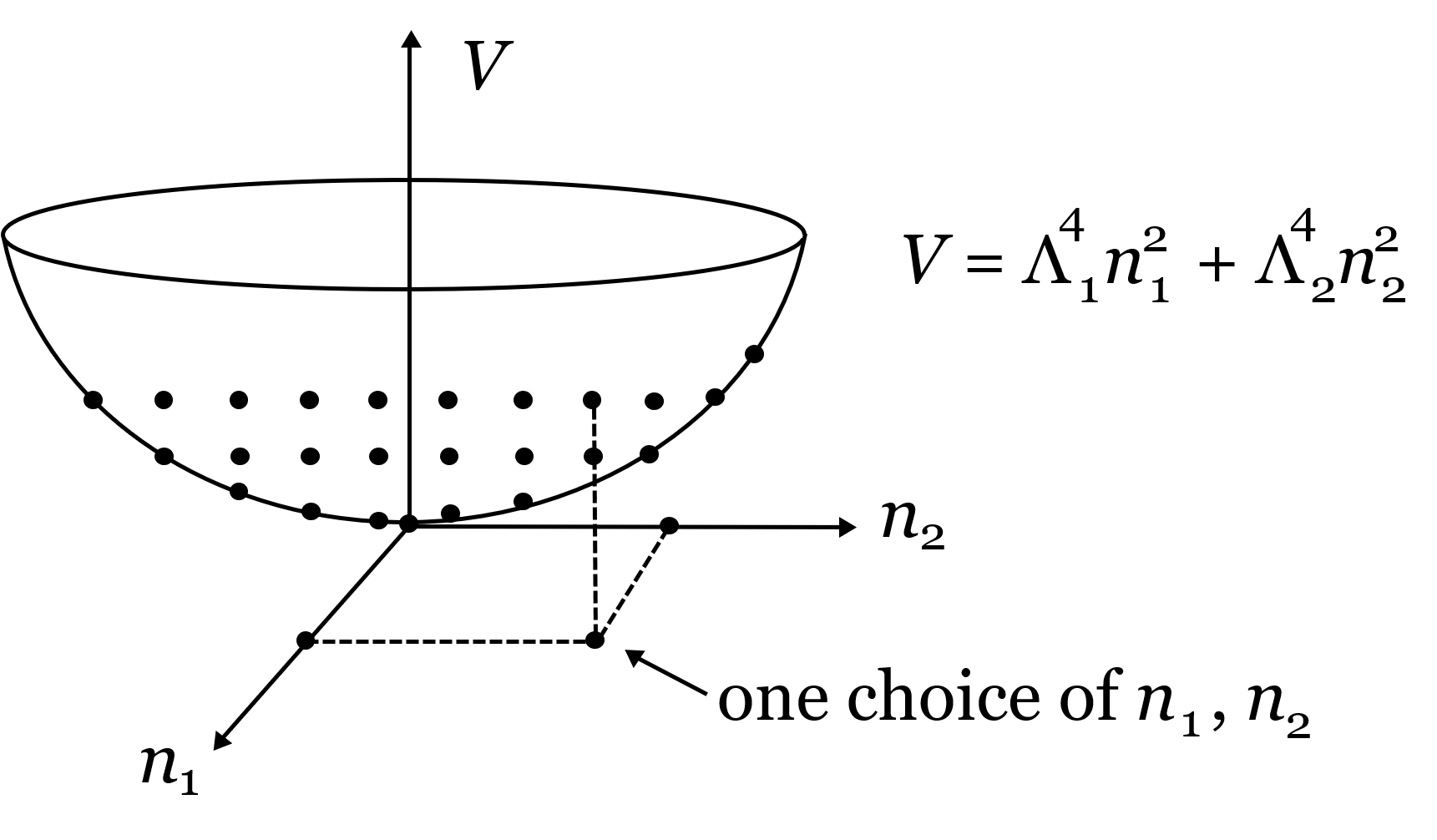}
\caption{Flux discretuum of a 3-form gauge theory with two fields.}
\label{bp2} 
\end{center}
\end{figure}

Now let us assume that our 4d theory possesses a large number of such 4-form fields,
\be
S= -\int \sum_{i=1}^N\Lambda_i^4\, F_{i,\,0}^2\,.\label{f0s}
\ee
This can arise, for example, if it originates from a compactification of a higher-dimensional $p$-form gauge theory on a compact space with $N$ $(p\!+\!1)$-cycles. The flux on each of those cycles then corresponds to the flux number $n$ in one of the $F_0$-models in (\ref{f0s}). The two-field case is illustrated in Fig.~\ref{bp2}.

Each flux choice $\ol{n}=\{n_1,\cdots,n_N\}$ gives rise to a particular cosmological constant 
\be
\lambda(\ol{n})\equiv V(\ol{n})=\sum_i \Lambda_i^4 n_i^2\,.
\ee
One may ask how many different flux choices lead to $\lambda(\ol{n})<\lambda_0$. To simplify the discussion, let us assume that all 4-form gauge couplings are equal: $\Lambda_i=\Lambda$. The number of flux choices is then simply the number of lattice points $\ol{n}\in\mathbb{Z}^N$ inside a ball of radius $\sqrt{\lambda_0}/\Lambda^2$. The lattice is $N$-dimensional, so the desired number is 
\be
K(\lambda_0)\sim (\sqrt{\lambda_0}/\Lambda^2)^N\,.
\ee
If $\sqrt{\lambda_0}>\Lambda^2$, this grows exponentially fast with $N$. In particular, the number $\delta K(\lambda_0,\delta\lambda)$ of points leading to
\be
\lambda\in [\lambda_0,\lambda_0+\delta\lambda]
\ee
will be extremely large for large $N$. This remains true if $N$ is only moderately large (say $N={\cal O}(100)$, as suggested by the number of 3-cycles of the quintic). It will also still be true if $\delta\lambda$ is chosen very small:
\be
\delta K(\lambda_0,\delta\lambda)\sim (\sqrt{\lambda_0}/\Lambda^2)^{N-1}(\delta\lambda/\sqrt{\lambda_0}\Lambda^2)\,.
\ee
Note that we do not have to be afraid that regularities in the distribution of $\lambda$-values could lead to intervals into which $\lambda$ never falls: Such possible regularities will be destroyed if we make all $\Lambda_i$ different, as expected in a more realistic situation. 

So far, we have a model with many solutions. These solutions give rise to a discretuum of cosmological constants, which becomes extremely dense in the region $\lambda\gtrsim \Lambda^4$ (where $\Lambda$ sets the typical scale for the couplings $\Lambda_i$). Now, by adding a negative cosmological constant
$\lambda_{AdS}<0$, such that
\be
S= -\left(\int \sum_{i=1}^N\Lambda_i^4\, F_{i,\,0}^2\,\,+\,\,\lambda_{AdS}\right) \,, \label{effl}
\ee
we can shift this dense discretuum downward. In this model, we are statistically guaranteed that vacua with an extremely small cosmological constant exists. Clearly, due to the possible bubble nucleation processes these
vacua will only be metastable, but they can be very long-lived. We will play with numbers later on to see {\it how} small $\lambda(\ol{n})$ in the model of (\ref{effl}) can really become.

\subsection{The type-IIB flux landscape (GKP)}

The key idea or observation is that, in type IIB Calabi-Yau compactifications, the 3-form fluxes of $H_3$ and $F_3$ can play roughly the role of the multiple fluxes of the Bousso-Polchinski model discussed above. The details are, however, more complicated and in part qualitatively different, mainly due to the central role of supersymmetry. 

We start with the intuitive observation that a non-zero flux on a compact cycle (say a 1-cycle) clearly has an energetic effect. Indeed, let us for simplicity assume that the compact space is $S^1_A\times S^1_B$ and one unit of 1-form flux sits on the $B$-cycle. Then
\be
\int dy_B\, F_1=1\qquad\mbox{and hence}\qquad F_1\sim 1/R_B\,. 
\ee
This gives rise to a contribution to the action
\be
S\supset -\int d^4x \int dy_A\,dy_B\, F_1\wedge *F_1\sim -\int d^4x\,(R_A R_B)\cdot\frac{1}{R_B^2}\sim  -\int d^4x\,\frac{R_A}{R_B}\,.
\ee
We learn that a flux on a cycle prevents this cycle from shrinking. More generally, if there are fluxes of various values on various cycles of a compact space, then these fluxes tend to stabilise the shape of the manifold in a certain way. Specifically, the ratio between the volumes of two cycles gets stabilised roughly according to the ratio of the flux numbers on these cycles. 

Concretely, we expect that 3-form fluxes will stabilise (give mass to) the complex structure moduli, which as we know govern the ratios of 3-cycle volumes. But this is not possible in a 4d supergravity model without superpotential since for $W=0$ no scalar potential is induced.

To make the right guess for the form of the expected flux-induced $W\!$, it is useful to observe that (already in 10d) one can use the complex scalar field
\be
S=C_0+ie^{-\phi}
\ee
to define a complex 3-form flux
\be
G_3=F_3-S\,H_3\,.
\ee
The kinetic terms of the two 3-form fields take the simpler form (suppressing constant prefactors)
\be
S\supset \int d^{10}x\,G_3\wedge *\ol{G}_3\,.
\ee

With this, one may guess the mathematically natural expression for the superpotential induced by 3-form fluxes:
\be
W=\int_X G_3\wedge\Omega_3\,.
\ee
This is known as (the type IIB version of) the Gukov-Vafa-Witten superpotential\index{Gukov-Vafa-Witten superpotential}~\cite{Gukov:1999ya}. The latter has first been postulated and mathematically justified (in an abstract way) for M-theory compactifications to 3d on Calabi-Yau 4-folds:
\be
W_{GVW}=\int_{X_4} G_4\wedge \Omega_4\,.
\ee
In the famous paper by Giddings, Kachru and Polchinski\index{Giddings-Kachru-Polchinski model} (GKP)\index{GKP}~\cite{Giddings:2001yu} (see also~\cite{Dasgupta:1999ss} and \cite{Grana:2005jc} for a review), this superpotential was used and justified explicitly by comparing the 4d scalar potentials derived from 4d ${\cal N}=1$ supergravity and directly from 10d. 

Now one can make this fully explicit by normalising the 3-form fields such that flux quantisation takes the form
\be
\frac{1}{2\pi\alpha'}\int F_3\in 2\pi\mathbb{Z}\,\,,\qquad 
\frac{1}{2\pi\alpha'}\int H_3\in 2\pi\mathbb{Z}
\ee
for integrals over integer cycles (see e.g.~\cite{Giryavets:2005nf} for more details and examples). Equivalently, one may decompose the fluxes in a symplectic integral form basis,
\be
F_3=-(2\pi)^2\alpha'(f^a\omega^A_a+f_{b_3/2+b}\omega_B^b)\,\,,\qquad
H_3=-(2\pi)^2\alpha'(h^a\omega^A_a+h_{b_3/2+b}\omega_B^b)\,,
\ee
where the entries of the coefficient vectors $f$ and $h$ now have to be integer. With this the superpotential, given in its simplest and mathematically natural form above, can be worked out explicitly:
\be
W=\int_X G_3\wedge \Omega_3=(2\pi)^2\alpha'\,(f-Sh)\cdot\Pi(z)\,.
\ee
The scalar potential reads, as usual,\footnote{
The 
ambiguity of the normalisation of $\Omega$ cancels out since $\Omega$ also appears in $K$. However, relative to the conventions of \cite{Giddings:2001yu} we have absorbed a factor of $\ln(2\pi)$ into the definition of $K$ to be consistent with our previous supergravity definition of the scalar potential.
}
\be
V=e^K \left( K^{i\ol{\jmath}}(D_i W)(D_{\ol{\jmath}}\ol{W})
+ K^{\alpha\ol{\beta}}(D_\alpha W)(D_{\ol{\beta}}\ol{W}) - 3|W|^2\right)\,.
\label{spot}
\ee
Here we have, for simplicity and since they all appear in $W=W(S,z)$, combined the axio-dilaton $S$ and the complex structure moduli $z^1,\cdots,z^{h^{2,1}}$ in one vector:
\be
z^i=\{S,z^1,\cdots,z^{h^{2,1}}\}\,.
\ee
We have furthermore redefined
\be
K_{c.s}-\ln(-i(S+\ol{S}))\,\,\to\,\,K_{c.s}\,,
\ee
absorbing the axio-dilaton Kahler potential into the complex-structure Kahler potential. 

It is essential that $W$ is independent of the Kahler moduli $T^\alpha$. Moreover, the Kahler modulus Kahler potential takes the special form $K_K=-\ln({\cal V}^2)$, with ${\cal V}^2$ a homogeneous function of the $T^\alpha$ of degree three. This constitutes a so-called {\bf no-scale model}\index{no-scale model}, implying the very special result that the last two terms in (\ref{spot}) exactly cancel \cite{Cremmer:1983bf}. This is discussed in more detail in Problem \ref{nsp}. In the simplest case of a single Kahler modulus,
\be
K_K=-2\ln({\cal V})=-2\ln((T+\ol{T})^{3/2})=-3\ln(T+\ol{T})\,,
\ee
the cancellation of the last two terms in (\ref{spot}) is easily observed. Hence, we have
\be
V=e^{K}\, K^{i\ol{\jmath}}(D_i W)(D_{\ol{\jmath}}\ol{W})\,.
\ee
Moreover, the equations for unbroken SUSY (the $F$-term conditions) 
\be
D_i W=0\qquad\mbox{for}\qquad i=1,\cdots,b_3/2\,,
\ee
represent $b_3/2$ equations for equally many complex variables. They will in general possess a solution (or a finite set of solutions). This fixes all $z^i$ to specific values. One may view these fields, which now have a large mass in this positive definite potential, as being integrated out. The result is a model depending just on $T$ (or, more generally, all Kahler moduli) in which 
\be
V=V(T,\ol{T})\equiv 0\,.
\ee
Since
\be
\ol{F}^{\ol{T}}=D^{\ol{T}} W=K^{\ol{T}T}K_T W=\left(\frac{3}{(T+\ol{T})^2}\right)^{-1}\frac{(-3)}{(T+\ol{T})}\,W
=-(T+\ol{T})\,W\neq 0\,,
\ee
supersymmetry is broken. The scale at which it is broken (e.g. the gravitino mass $e^{K/2}W$) is not fixed since $T$ is not fixed. The explains the name no-scale model. 

One of the key points of~\cite{Giddings:2001yu} (known as `GKP') is that they established this vanishing potential not only (as we just did) indirectly, via 4d SUGRA arguments, but by explicitly providing the 10d geometry. The term  `explicitly' is here interpreted as follows: One assumes that a Calabi-Yau metric is given (this is of course not explicit but rests on the famous existence theorem). Then, given in addition certain fluxes and other sources in the Calabi-Yau (e.g. O3-planes and D3 branes), one is able to write down differential equations determining the actual metric, including backreaction from fluxes. This metric corresponds to a flux compactification to 4d Minkowski space. In fact, there is a family of such solutions, corresponding to the flat direction characterised by the `no-scale modulus' $T$, as explained above.\footnote{
As emphasised in \cite{Sethi:2017phn}, the detailed situation is more complicated: Once $W\neq 0$, certain higher-derivative terms present in the 10d action necessarily induce a small 4d effective potential which in general leads to a runaway to small or large volume. However, once the non-perturbative effects discussed in the next section are included, the volume is stabilised and it turns out that, at least in the appropriate parametric regime, the higher-derivative corrections are not important. This viewpoint is widely accepted but vigorously disputed in \cite{Sethi:2017phn} on the ground that one may not proceed from a starting point which is not a solution for all times. We here take the pragmatic EFT attitude that a sufficiently slow runaway is as good as a static solution and correct the latter by non-perturbative effects governed by shorter time-scales.
}

\subsection{Kahler modulus stabilisation and SUSY breaking (KKLT)}
\label{kklt1}\index{KKLT}
The leading proposals for Kahler moduli stabilisation and the realisation of a positive cosmological constant are known as KKLT \cite{Kachru:2003aw} and LVS (Large Volume Scenario) \cite{Balasubramanian:2005zx}. We will discuss them in the next three sections. For further suggestions see e.g.~\cite{Westphal:2006tn, Balasubramanian:2004uy}.

We start with KKLT. As in the previous section, we focus on the simplest case $h^{1,1}=1$, such that
\be
K=-3\ln(T+\ol{T})\qquad\mbox{and}\qquad W=W_0=\mbox{const.}
\ee
The complex structure moduli have been integrated out and the corresponding flux choice (together with the VEVs of the $z^i$ which it prescribes) has fixed $W_0$.  At leading order, we have $V\equiv 0$ and SUSY breaking with $m_{3/2} = e^{K/2}W_0$.

Various (quantum) corrections will generically lift the flatness of $V$, breaking the no-scale structure. This can be $\alpha'$ corrections (corresponding to higher-dimension operators in 10d), loop-corrections, non-perturbative instanton effects or non-perturbative effects from (SUSY) gauge theory confinement (also known as gaugino condensation). The last two of these four qualitatively different effects lead to technically similar results. In particular, $W$ is corrected according to
\be
W_0\qquad\rightarrow \qquad W_0+A\,e^{-2\pi T/N}\,,
\ee
by either instantons (in this case $N=1$) or by gaugino condensation\index{gaugino condensation} (here $N$ is the dimension of the fundamental representation of the gauge group $SU(N)$). 
This type of corrections (see \cite{Blumenhagen:2009qh, Bianchi:2007ft} for reviews) is one of the basic ingredients of the KKLT-scenario for (complete) moduli stabilisation and SUSY breaking~\cite{Kachru:2003aw} which we will describe in the rest of this section.

We will not introduce the technology necessary to analyse the case of gaugino condensation (see \cite{Derendinger:1985kk, Dine:1985rz} for its original discussion in heterotic models). Suffice it to say that, if a stack of $N$ D7 branes is wrapped on a 4-cycle with volume $\sim\,$Re$\,T$, the 4d theory contains a corresponding ${\cal N}=1$ super-Yang-Mills theory. The latter exhibits confinement, as familiar from the non-SUSY QCD sector of the Standard Model. In the SUSY case, confinement is characterised by a non-zero $W_{\rm non-pert.}\sim \Lambda^3$. Here $\Lambda$ is the confinement scale and its relation to $W_{\rm non-pert.}$ follows on dimensional grounds. Using also the fact that the 4d gauge coupling squared is $\sim 1/$Re$\,T$ by the standard logic of Kaluza-Klein reduction from 8d to 4d, one may run from high to low energy scales and determine at which scale the gauge coupling reaches ${\cal O}(1)$ values. This fixes $\Lambda$ in terms of Re$\,T$, and by holomorphicity leads to $W_{\rm non-pert.}\sim \exp(-2\pi T/N)$, where $N$ comes in through the beta-function.

Since we have already introduced some of the ideas relevant for instanton effects, we will describe the instanton case in slightly more detail. For this, it is useful to recall how an instanton correction in a 5d-to-4d compactification of a gauge theory is related to the possibility of wrapping a closed electron worldline on a 1-cycle (in this case the unique 1-cycle $S^1$) of the compact space. This was illustrated in~Fig.~\ref{strins}, where the reader was invited to think of a particular type of $e^+e^-$ fluctuation of the vacuum. But this time-dependent picture is not necessary -- the simplest and dominant effect corresponds to just wrapping the worldline on the minimal volume cycle (at fixed 4d space-time point $x^\mu$), subsequently integrating over all $x^\mu$. 

\begin{figure}[ht]
\begin{center} 
\includegraphics[width=6.5cm]{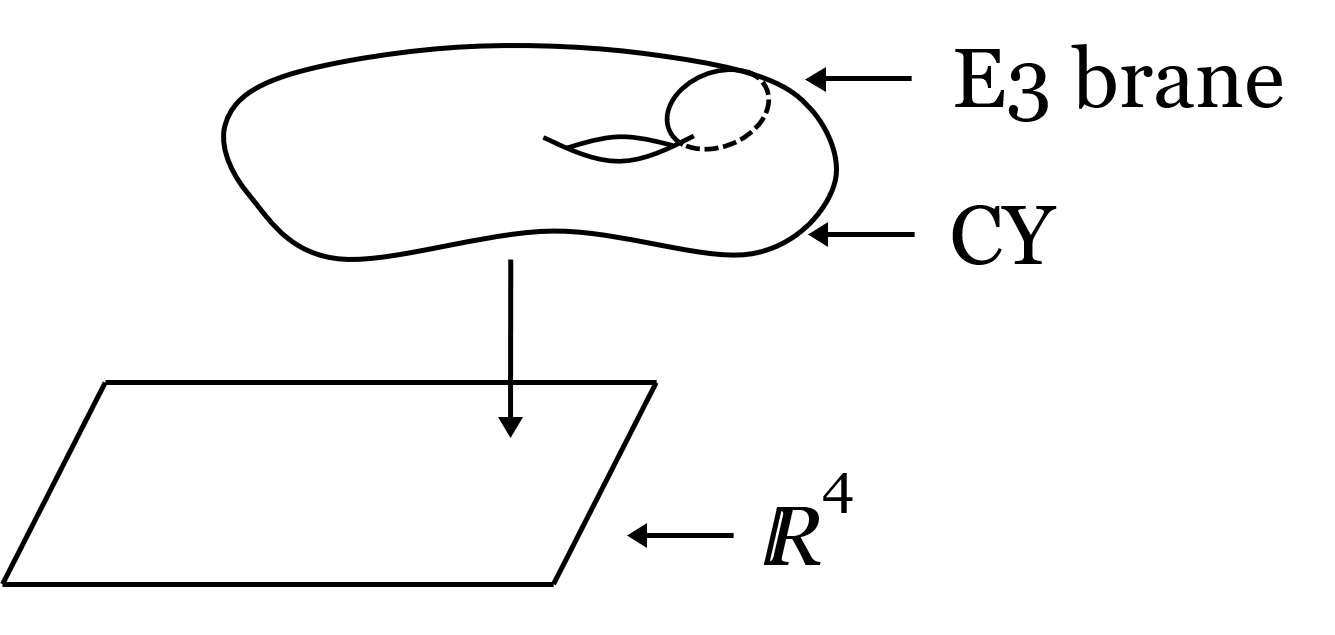}
\caption{An E3-brane\index{E3-brane} instanton, corresponding to a euclidean D3 brane wrapped on a 4-cycle of a CY over one of the points of the non-compact space-time $\mathbbm{R}^4$.}
\label{e3} 
\end{center}
\end{figure}

This type of instanton has an obvious analogue in compactifications of higher-form gauge theories. The effect occurs if the compact space possesses cycles the dimensions of which correspond to the dimensionality of the available charged objects. Our case of interest is type IIB with its 4-form $C_4$ and the corresponding D3 branes. We think of the electron worldline above as of a 0-brane and, once it is wrapped in euclidean signature to describe tunnelling, as of an E0 brane. Analogously, we can think of a D3 brane (now often called an E3 brane) as being wrapped at fixed $x^\mu$ on the minimal volume 4-cycle of our Calabi-Yau. This is the origin of the instanton correction we are after, cf.~Fig.~\ref{e3}. Such instantons are called stringy or exotic or D-brane instantons.

At the quantitative level, we recall that our complex Kahler modulus is $T=\tau+ic$, where
\be
\tau \,=\, \int_{4-cycle}\sqrt{g_{CY}} \,\sim\, R_{CY}^4\,.
\ee
The last expression is, up to the proper normalisation by the tension prefactor, the action of the wrapped brane. Furthermore, the wrapped brane couples to $C_4$ through
\be
2\pi \int_{4-cycle} C_4\,\equiv\, c\,,
\ee
which is just the 4d axionic scalar in $T$ (recall that we have set $2\pi\sqrt{\alpha'}=1$). Thus, a single instanton contributes to the 4d partition function as
\be
\sim e^{-2\pi\tau}\,e^{-2\pi i c}\,,
\ee
where the first factor is the tunnelling suppression by the euclidean brane action. The second factor comes from the part of the D3-brane action displayed in the previous line. It can equivalently be viewed purely in 4d as the coupling of the 0-form gauge field $c$ to its 0-dimensional charged object, the instanton.

Summing over all numbers of instantons and anti-instantons (which come with $e^{+2\pi i c}$) leads to an exponentiation:
\be
{\cal L}_{4d}\,\,\supset\,\,\exp\left[ \sim e^{-2\pi \tau}\,\cos(2\pi c)\right]\,.\label{v4di}
\ee
The term in the exponent is the instanton correction to the 4d effective action and it is precisely analogous to the possibly more familiar gauge theory case. Here, one gets corrections $\sim e^{-8\pi/g^2}\cos(2\pi\phi)$, where $g$ is for example the strong gauge coupling and $\phi$ the QCD axion, famously obtaining a cosine-potential from this effect. 

In SUSY, such instanton corrections can enter the 4d effective action only through either $K$ or $W$:
\be
W_0\quad\to\quad W_0+Ae^{-2\pi T}\qquad\qquad\mbox{or}\qquad\qquad 
K\quad\to\quad K+B e^{-2\pi T}+\mbox{c.c.}\,.\label{w4di}
\ee
Which of the two happens depends on the geometry of the wrapped brane and will not be discussed here \cite{Witten:1996bn} (see \cite{Blumenhagen:2009qh, Bianchi:2007ft} for reviews as well as \cite{Bianchi:2011qh, Palti:2020qlc} and refs. therein). For KKLT, we require that a correction to $W$ arises. We also note that the $\tau$ and $c$ dependences are such that they combine in a holomorphic function of $T$ (as required by SUSY), with the proper periodicity in Im($T$). Conversely, as shown in the problems, the evaluation of the scalar potential on the basis of $W$ from (\ref{w4di}) leads to a term of the type of (\ref{v4di}). 

\begin{figure}[ht]
\begin{center} 
\includegraphics[width=6.5cm]{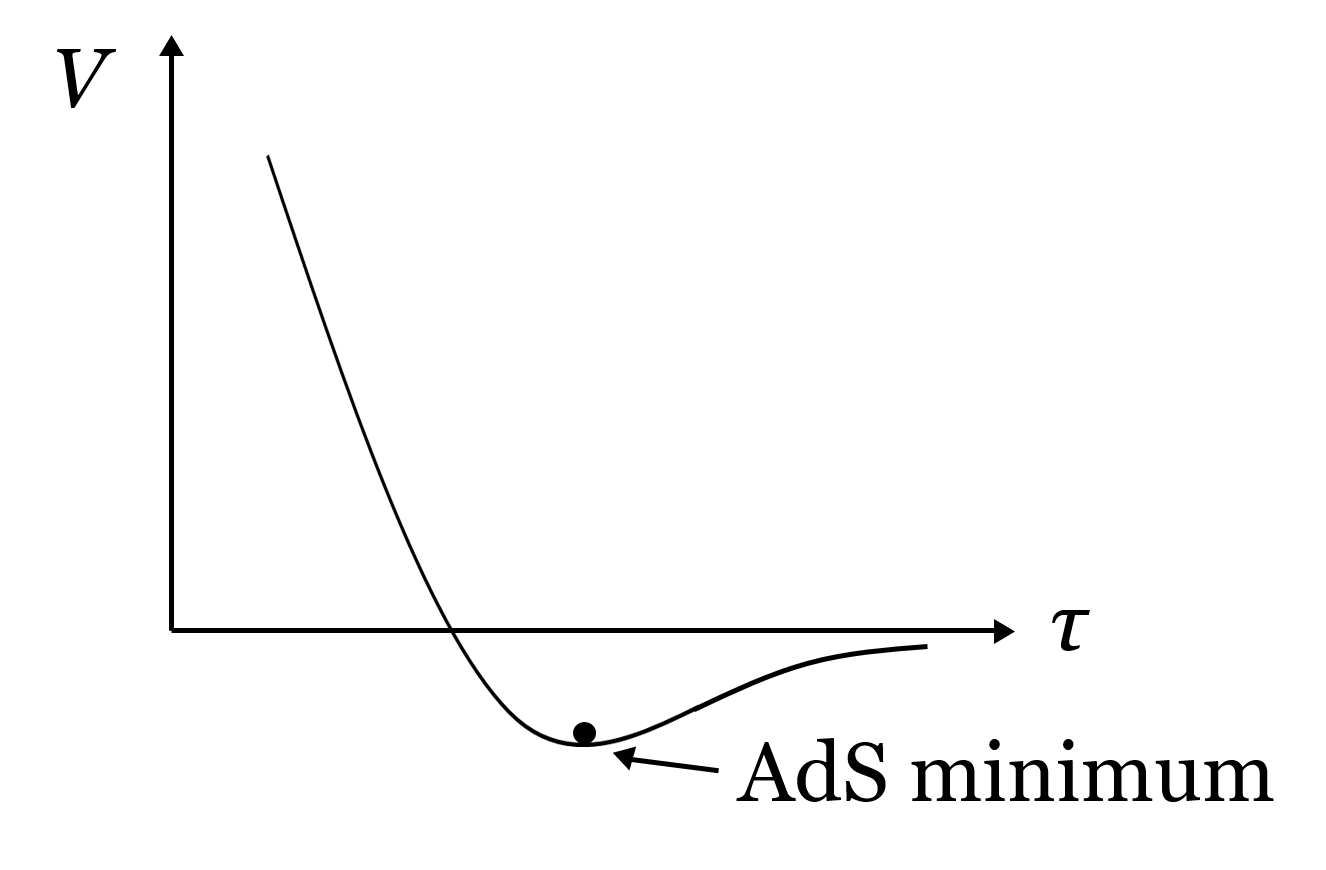}
\caption{Qualitative behaviour of the scalar potential arising after the inclusion of instanton or gaugino condensation effects in $W$.}
\label{kkltads} 
\end{center}
\end{figure}

We can now finally proceed with the analysis of the 4d effective theory, defined by 
\be
K=-3\ln(T+\overline{T})\qquad \mbox{and}\qquad W=W_0+Ae^{-aT}\,.
\ee
It is a straightforward exercise (Problem~\ref{nsp}) to derive the scalar potential $V(\tau,c)$, integrate out $c$ (by simply finding the minimum in $c$), and thus to obtain
\be
V=V(\tau)\,.
\ee
The qualitative behaviour of this potential at $W_0\ll 1$ is displayed in Fig.~\ref{kkltads}. It is easy to derive by analysing the standard supergravity formula for $V$ in the regimes $e^{-aT}\gg W_0$ and $e^{-aT}\ll W_0$. One checks that $V$ grows at small $\tau$ and approaches zero from below at large $\tau$. This is sufficient to conclude that the qualitative picture is that of Fig.~\ref{kkltads}.

Moreover, it is easy to prove in general that the supergravity scalar potential has an extremum at supersymmetric points, where the $F$-terms and hence the first, positive-definite term in the supergravity potential formula vanish,
\be
DW=-aAe^{-T}-\frac{3}{T+\overline{T}}\left(W_0+Ae^{-aT}\right)=0\,.
\ee
In our case this extremum is always a minimum. Assuming $c=0$, this vanishing-$F$-term condition is solved (implicitly in $\tau$) if
\be
W_0=-\left(1+\frac{2}{3}a\tau\right)\,Ae^{-a\tau}
\ee
holds. The conclusion that $W_0$ must be real and negative is a mere consequence of our simplifying assumption $c=0$. For a general phase of $W_0$ (and $A$), we would simply have found a non-zero value of $c$ at the minimum. This is not important for us. 

What {\it is} important is the conclusion that $W_0$ must be exponentially small for parametric control, i.e. to have $R_{CY}\gg 1$. Of course, making $W_0$ small should not be a problem since it depends on the flux choice -- it can hence be finely tuned in the landscape. In fact, to be sure that nothing goes wrong one needs to know that the statistical distribution of $W_0$ in the complex-$W_0$-plane for random flux choices has no special feature near the origin. This crucial fact, more precisely the flatness of the distribution of $|W_0|^2$ values near zero, has been established with some level of rigour in~\cite{Denef:2004ze}.

Thus, we have uncovered a landscape of supersymmetric vacua with a negative cosmological constant, so-called SUSY AdS vacua\index{AdS vacuum}. (Note that, in the `first step of KKLT' leading to these solutions the broken supersymmetry of GKP is restored in the minimum). But to describe the real world we need a positive (even though very tiny) cosmological constant and broken supersymmetry. Moreover, turning at least a small fraction of the SUSY-AdS vacua above into dS vacua is essential for eternal inflation, the presently leading mechanism for populating the landscape cosmologically (see~Sect.~\ref{eimp}). 

We first give a much simplified, `macroscopic' description of how dS vacua may arise on the basis of the above (see e.g.~\cite{Luty:2002hj, Choi:2005ge, Brummer:2006dg, Dudas:2006gr}). Let us assume that some further details of the model, such as branes with their gauge theories and charged matter fields, introduce extra light degrees of freedom $X$ and corresponding corrections to $K$ and $W$:
\be
K\quad\to\quad K(T\overline{T})+\delta K(X,\overline{X},T,\overline{T})\quad,\qquad \qquad
W\quad\to\quad W(T)+\delta W(X)\,.
\ee
Now, let us choose $\delta K$ and $\delta W$ in analogy to the one-field O'Raifeartaigh-type model\index{O'Raifeartaigh model} discussed in Sect.~\ref{susybr}:
\be
\delta K\sim X\overline{X}-(X\overline{X})^2\,\,\,,\qquad \delta W=\alpha X\,.
\ee
This will lead to $D_XW\neq 0$ in the vacuum. Moreover, let us choose the parameters such that the fluctuations of $X$ around this SUSY breaking vacuum have a very large mass. Then the upshot of the whole construction is that the scalar potential $V$ is supplemented by a so-called uplifting term 
\be
V\quad\to\quad V+\delta V\qquad \qquad\mbox{with}\qquad \qquad 
\delta V=e^K K^{X\overline{X}}|D_XW|^2\,.
\ee

\begin{figure}[ht]
\begin{center} 
\includegraphics[width=6cm]{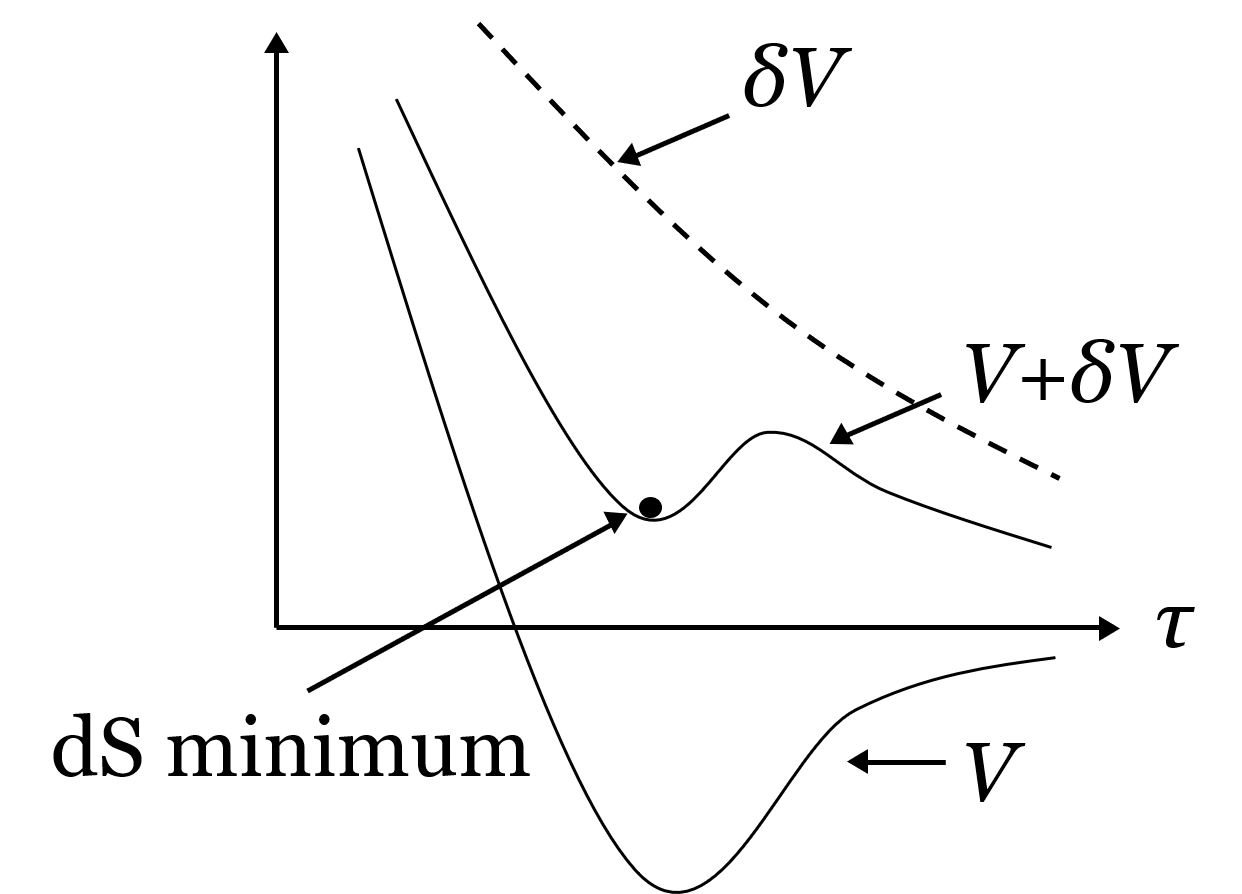}
\caption{Uplifting to a KKLT dS\index{dS vacuum} vacuum.\index{de Sitter vacuum}}
\label{kkltup} 
\end{center}
\end{figure}

At this generic level of analysis the uplifting term $\delta V$ could have any $T$ dependence, given our free choice of the $T$ dependence of $\delta K$. In concrete string constructions, for which the above is a toy model, $\delta V$ will always be decaying at large volume, cf.~Fig.~\ref{kkltup}. This can be understood if one imagines that (as is mostly the case) $\delta K$ and $\delta W$ are due to some local effect in the CY. Then, going to large volume, the SUSY-breaking and uplifting effects stay the same in string units, but the Planck mass diverges. Hence, in standard supergravity conventions with $M_P=1$, $\delta V$ will decay with growing $T$. 

One may expect that in the huge string theory landscape many options for such an uplift\index{uplift} exist. Yet, it turns out not to be easy to construct an uplift of the above O'Raifeartaigh type explicitly. Thus, the most explicit uplift has a somewhat different structure: It is the anti-D3-brane uplift originally suggested by KKLT, which arguably remains the most explicit (though nevertheless not uncontroversial$\,$\footnote{
There 
even exists the opinion that no uplift to a dS minimum can ever be constructed for fundamental reasons, challenging most ideas about how string theory might be relevant to the real world \cite{Danielsson:2018ztv,Ooguri:2018wrx}. We will return to this subject.
}
$\!$) possibility. We turn to this construction, which requires some more technology, next. As we will see, even though different in detail, the KKLT uplift behaves qualitatively as explained using the O`Raifeartaigh toy model above.

\subsection{The anti-D3-brane uplift of KKLT}\label{kklt2}
\index{anti-brane}

As we explained earlier, a so-called orientifold projection reduces the  supersymmetry of a type-II Calabi-Yau compactification from ${\cal N}=2$ to ${\cal N}=1$. Let us consider the example of an O3-plane projection, which can locally be thought of as the geometric action
\be
(z^1,z^2,z^3)\quad\to\quad (-z^1,-z^2,-z^3)\,,
\ee
to be combined with worldsheet orientation change.

Locally, this projection introduces a singularity at $\{z^i\}=0$, at which (due to the orientation change) a so-called O3-plane\index{O-plane} is localised.\index{orientifold plane} This is a negative-tension object which also has opposite $C_4$-charge\footnote{
This 
is often referred to as D3-charge.
} 
compared to a D3-brane. In a consistent compactification, an O3-plane always has to come with a certain number of D3 branes for total charge neutrality (tadpole cancellation). 
Concretely, the D3 charge of an O3-plane is -1/4. The fractionality is not a problem since the compact Calabi-Yau after orientifolding will usually have a large number, divisible by 4, of O3-planes. For example, it is easy to check that $T^6/\mathbb{Z}_2$, with the $\mathbb{Z}_2$ acting as above, has 64 O3-planes. 

Now, given a consistent Calabi-Yau with a number of O3-planes and a corresponding number of D3-branes, it is possible to replace some or all of the D3-branes by 3-form fluxes. This possibility arises since, through the Chern-Simons term, 3-form fluxes contribute to the total D3 tadpole. This takes us to the realm of flux compactifications \`{a} la GKP and, if we also allow for the non-perturbative effects $\sim e^{-aT}$ introduced above, we will find ourselves in an ${\cal N}=1$ SUSY setting with O3-planes, D3-branes and fluxes. The O3-planes, the D3-branes and the fluxes all break SUSY to the same ${\cal N}=1$ sub-algebra of the original ${\cal N}=2$ SUSY of the pure Calabi-Yau model. 

Next, we can think of breaking SUSY by adding a D3 and anti-D3 (for short: $\overline{D3}$) brane pair.\footnote{
See \cite{Alexander:2001ks, Dvali:2001fw, Burgess:2001fx} for the first suggestions of how to realise approximate de Sitter space using the positive energy of a brane-anti-brane pair.
}
The $\overline{D3}$ breaks ${\cal N}=2$ to the opposite ${\cal N}=1$ subalgebra, such that 4d SUSY is now completely broken. D3 tadpole cancellation is not violated since we added two oppositely charged objects. However, brane and anti-brane attract each other both gravitationally and through $C_4$, so they will quickly find each other and annihilate, releasing twice the energy density of the D3-brane tension. Our `uplift' is thus very short-lived and not practically useful.

However, we could avoid having any D3-branes by cancelling the tadpole of the O3-planes by flux alone. If we now add a $\overline{D3}$-brane and increase the flux appropriately to ensure tadpole cancellation, we appear to have the desired uplift. Now, the $\overline{D3}$ still breaks SUSY relative to flux and O3-planes but there is no D3 which it could attract and annihilate. 

\begin{figure}[ht]
\begin{center} 
\includegraphics[width=6.3cm]{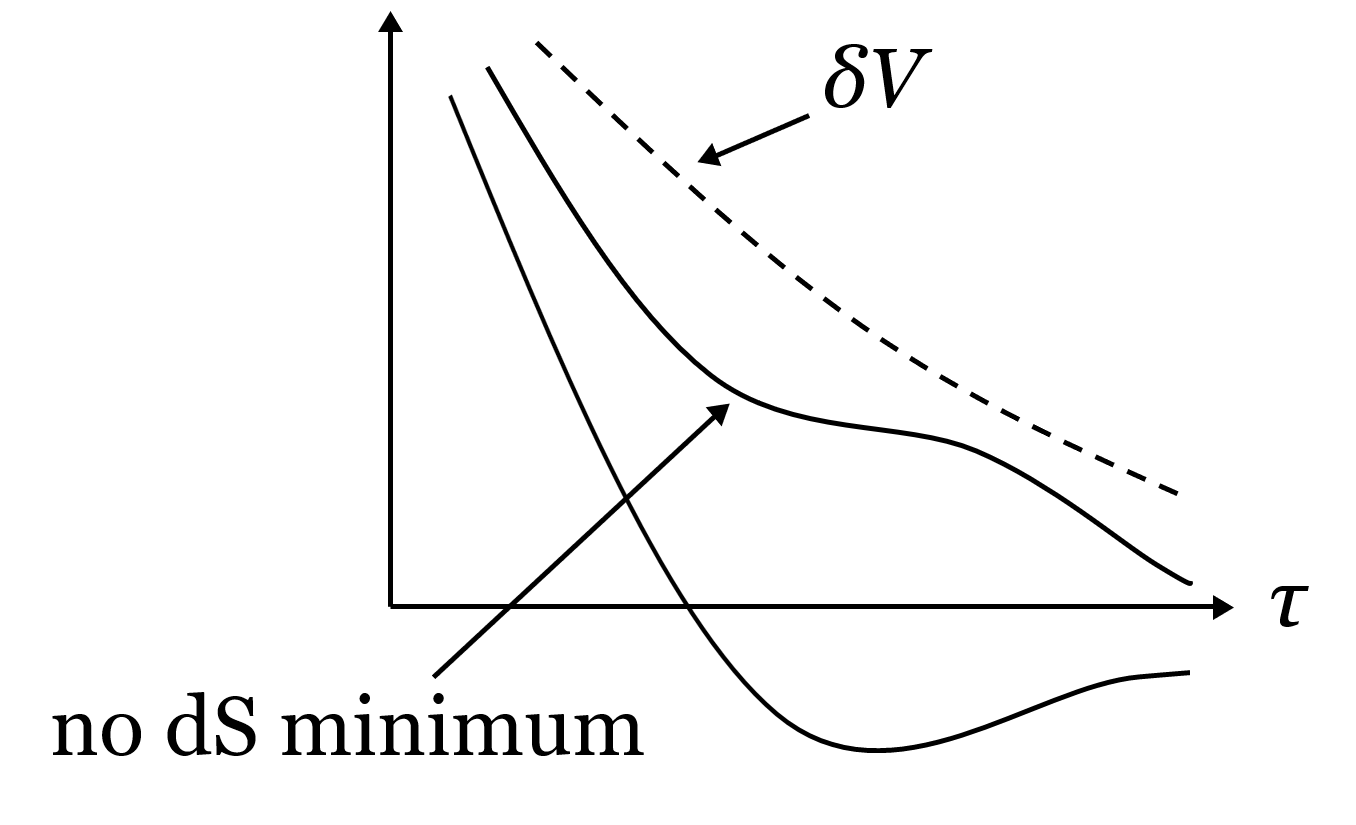}
\caption{If too high an uplift is added to a model with SUSY-AdS vacuum, no metastable\index{metastable} de Sitter minimum is generated.}
\label{highup} 
\end{center}
\end{figure}

Unfortunately, this is not yet good enough since this uplift (by twice the D3-brane tension, which is string-scale) is much too strong. Indeed, given that the non-perturbative effects and hence the depth of the original AdS minimum are exponentially small, the situation will be as in Fig.~\ref{highup}: The uplift is much too strong and no local dS minimum can be generated. 

Fortunately, the key to a resolution of this problem is already contained in the seminal work of GKP~\cite{Giddings:2001yu} discussed above. They show explicitly that the metric on a CY orientifold threaded by 3-form-flux is not of product type but {\bf warped}:
\be
ds^2=\Omega^2(y)\eta_{\mu\nu}dx^\mu dx^\nu + g_{mn}(y) dy^m dy^n\,. \label{wma}
\ee
Here $x^\mu$ (with $\mu=0,\cdots, 3$) and $y^m$ (with $m=1\cdots 6$) parameterise the non-compact $\mathbb{R}^4$ and compact $X_6$ part of our total space respectively. This space is, topologically and as a differentiable manifold, still of product type, $\mathbb{R}^4\times X_6$. However, the metric manifold built on this basis does not share this product structure. As we can see from the warped metric ansatz in (\ref{wma}), this breaking of the product structure is perfectly consistent with 4d Poincare invariance as long as $y$ enters in the prefactor of the non-compact metric but $x$ does {\it not} enter in the prefactor of the compact part of the metric. One refers to $\Omega(y)$ as the {\bf warp factor}.\index{warp factor}

\begin{figure}[ht]
\begin{center} 
\includegraphics[width=4.5cm]{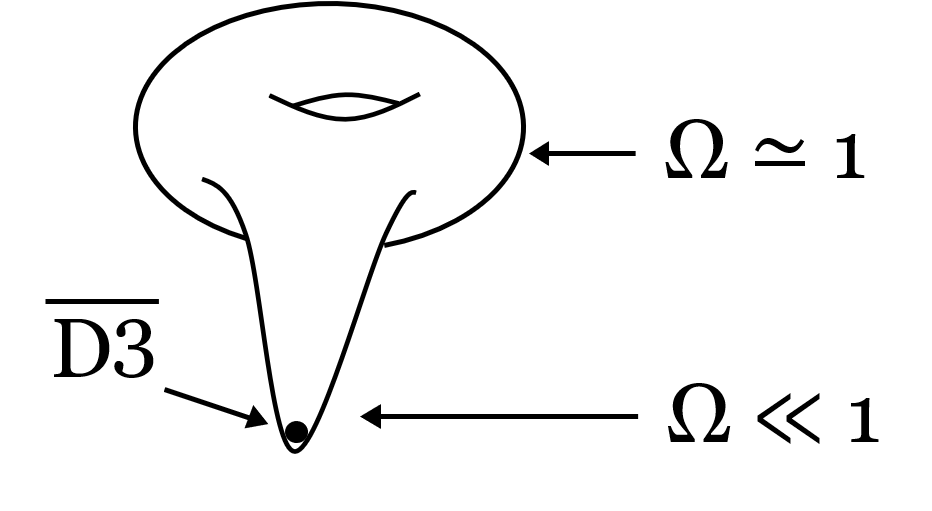}
\caption{Calabi Yau with warped throat. The Calabi-Yau is basically undeformed in the region of small warping ($\Omega\simeq 1$) and strongly deformed in the $\Omega\ll 1$ domain (where the $\overline{D3}$ brane is localised).
}
\label{cy+throat} 
\end{center}
\end{figure}

Moreover, GKP show that given certain (very common\footnote{
The 
feature we need is a so-called conifold singularity. The latter develops when a certain type of 3-cycle shrinks to zero volume (i.e.~$z\to 0$ if $z$ is the modulus parameterising the corresponding period). This is in fact a generic type of 3-cycle of a CY, so such a situation arises frequently. Conversely, the conifold singularity can be made smooth (`deformed') by `blowing up' a 3-cycle. For more details see e.g.~\cite{Candelas:1989js}.
}) 
features of the CY and a particular flux choice, the compact manifold develops a  strongly warped region. This region is also known as a {\bf Klebanov-Strassler throat}\index{Klebanov-Strassler throat}~\cite{Klebanov:2000hb} and\index{throat} is graphically often represented as in Fig.~\ref{cy+throat}. To understand that the compact geometry is strongly deformed at strong warping, one also needs to know that~\cite{Klebanov:2000hb, Giddings:2001yu}
\be
g_{mn}(y)=\Omega^{-2}\tilde{g}_{mn}(y)\,,
\ee
where $\tilde{g}$ is the Calabi-Yau metric. One says that the compact space is not `Calabi-Yau' but only `conformal-Calabi-Yau'\index{conformal-Calabi-Yau}. For our purposes, it is essential that strong warping can substantially change the energy effect of the SUSY-breaking $\overline{D3}$ brane placed in the Calabi-Yau orientifold. 

To understand this central aspect, recall the Schwarzschild black hole metric
\be
ds^2=-f(r)dt^2+f^{-1}(r)dr^2+r^2d\omega^2\,,
\ee
where $d\omega^2$ is the metric on the unit sphere. Clearly, $f(r)$ bears similarity to our $\Omega^2(y)$. As is well known, the vanishing of $f(r)$ as one approaches the horizon is responsible for the redshift effect and the force that pulls any massive object into the black hole. The same happens in our case: The $\overline{D3}$ brane represents a SUSY-breaking local energy density in the warped Calabi-Yau and this brane is pulled towards strong warping (where $\Omega\ll 1$). Once there, its energetic effect as seen from the unwarped `bulk'\index{bulk} of the Calabi-Yau is greatly reduced. In other words, the anti-brane naturally sits at the bottom of the warped throat and uplifts the total potential energy of the compactification only by 
\be
\Omega_{min}^4\times {\cal O}(1)
\ee
in string units. The fourth power of $\Omega$ arises since, as known from black hole physics, $f^{1/2}$ is the redshift factor and, in our context, we are redshifting an energy density, i.e. an object of mass dimension four. 

As shown in GKP, 
\be
\Omega_{min}\sim \exp(-2\pi K/3M g_s)\,,\label{oks}
\ee
where $K$ and $M$ are flux numbers associated with 3-cycles of the Klebanov-Strassler throat geometry and $g_s$ is the string coupling constant. The latter is governed by the modulus $S$ stabilised by fluxes. Thus, one apparently has enough freedom to choose fluxes in such a way that $\Omega_{min}$ is exponentially small.\footnote{
See
\cite{Bena:2018fqc, Blumenhagen:2019qcg} for recent, in part critical comments related to this point.
} 

Before moving on, it should be mentioned that a debate about the metastability of the anti-brane at the bottom of the throat has been going on for a number of years (see~\cite{Bena:2014jaa, Michel:2014lva, Cohen-Maldonado:2015ssa, Polchinski:2015bea, Bena:2016fqp, Danielsson:2016cit} and refs. therein). Indeed, as should be clear form the above the $\ol{\rm D3}$  breaks SUSY (in the absence of any D3 brane) against the fluxes in the throat. It can annihilate against these fluxes only at the price of overcoming an energy barrier, making the uplifted configuration at best metastable~\cite{Kachru:2002gs}. However, the backreaction of fluxes to the presence of the anti-brane is poorly understood and a barrier-free decay or outright instability has been claimed. In spite of many efforts to show the opposite, long-lived metastability as described in~\cite{Kachru:2002gs} has remained plausible~\cite{Michel:2014lva, Polchinski:2015bea}. On the other hand, a better, fully-backreacted understanding of the geometry with the anti-brane included would be highly desirable but remains challenging.

Let us now assume that the above $\ol{\rm D3}$ uplift\index{uplift} does indeed provide metastable SUSY breaking and estimate its magnitude. For simplicity, we disregard factors of $g_s$ such that the tension of the 3-brane in either the 10d string or Einstein frame is ${\cal O}(1)\times l_s^{-4}\sim {\cal O}(1)$. Here we also, as before, use conventions in which all dimensionful quantities in 10d are measured in units of the string scale or the inverse string scale. If there were no warping then, compactifying, the $\ol{\rm D3}$ brane tension (more precisely twice this number - see above) induces a 4d energy density $\sim {\cal O}(1)$. Note that we are still using string units and our 4d Planck mass is $M_P^2\sim {\cal V}$ (i.e. we are in a `Brans-Dicke frame'). Next, we Weyl rescale the 4d metric to go to the 4d Einstein frame. This amounts to using 4d Planck units (i.e. setting the Planck mass to unity) in the 4d effective action. Since, in this process, dimensionless ratios of physical observables do not change, we have
\be
\frac{\rho_{\ol{\rm D3}}^{Einstein}}{(M_P^4)_{Einstein}} \,\sim\,\frac{\rho_{\ol{\rm D3}}^{Brans-Dicke}}{(M_P^4)_{Brans-Dicke}}
\,\sim\,\frac{1}{{\cal V}^2}\qquad \mbox{or}\qquad 
\rho_{\ol{\rm D3}}^{Eistein}\,\sim\,\frac{1}{{\cal V}^2}\,\sim\,\frac{1}{\tau^3}\,.
\ee

Most naively, one would now like to include warping by multiplying with the fourth power of the redshift factor $\Omega_{min}^4$~\cite{Kachru:2003aw}. This is correct in principle, but at a quantitative level a further fine point has to be taken into account~\cite{Kachru:2003sx}: Indeed, the expression (\ref{oks}) is valid in the strongly warped region near the tip of a Klebanov-Strassler throat. It represents correctly the dependence of the warping on the relevant discrete flux choice. Yet, if the Calabi-Yau volume is taken to infinity, then eventually the fluxes become so diluted that their backreaction on the geometry is negligible and $\Omega\sim 1$, even at the lowest point of the throat. This can be quantified~\cite{Kachru:2003sx} (see also \cite{Giddings:2005ff}) and leads to the more precise warping suppression 
\be
\Omega_{min}^4\quad\to\quad\Omega_{min}^4\tau\,,
\ee
valid only as long as $\Omega_{min}^4\tau\ll 1$. 

Combining everything, one arrives at
\be
V_{KKLT}=e^{K}\left(K^{T\ol{T}}|D_TW|^2-3|W|^2\right)+V_{up}(\tau)\,,
\ee
with
\be
K=-3\ln(T+\ol{T})\,\,,\qquad W=W_0+Ae^{-aT}\,\,\qquad\mbox{and}\qquad
V_{up}(\tau)=c\frac{\Omega_{min}^4}{\tau^2}\,.
\ee
Here $A,a$ and $c$ are numerical ${\cal O}(1)$ factors and $W_0$ and $\Omega_{min}$ can be chosen extremely small by an appropriate flux choice. It is easy to convince oneself numerically or analytically that an uplifted situation with a metastable de Sitter or Minkowski vacuum as in Fig.~\ref{kkltup} can be achieved on the basis of the above potential. The reader is invited to verify this. The key non-trivial point is that the AdS minimum is very steep (based on the exponential behaviour of the non-perturbative superpotential $\sim e^{-aT}$) while the uplift has a relatively flat, power-like $\tau$ dependence. Hence, the local minimum survives the uplift to a value above zero.\footnote{
We 
note that a new round of criticism and defense of this construction has appeared relatively recently, related mainly to the question whether the non-perturbative effect (in this case gaugino condensation) and the subsequent uplift can also be understood directly in 10d~\cite{Moritz:2017xto, Hamada:2018qef, Kallosh:2019oxv, Hamada:2019ack, Gautason:2019jwq, Carta:2019rhx}. At this point it appears that, yet again, the success of the KKLT construction remains plausible~\cite{Hamada:2019ack}. An interesting novel criticism, concerned with geometric consistency, was raised in~\cite{Carta:2019rhx}. From this, a quantitative singularity problem was derived in \cite{Gao:2020xqh}. We will comment on these issues in slightly more detail in Sect.~\ref{swds}.}

\subsection{The Large Volume Scenario}
\index{Large Volume Scenario}

A very promising alternative to the KKLT proposal for Kahler moduli stabilisation in an AdS vacuum (before uplift) is provided by the Large Volume 
Scenario or LVS\index{LVS}~\cite{Balasubramanian:2005zx, Conlon:2005ki}. It has the disadvantage of being slightly more involved than KKLT but the advantage that the stabilised value of the volume ${\cal V}_{LVS}$ may be {\it exponentially} large -- a feature not available in KKLT due to the parametric behaviour ${\cal V}_{KKLT}\sim \ln(1/|W_0|)$. 

In the simplest realisation, two Kahler moduli $T_b$ and $T_s$ (with the indices standing for `big' and `small') are required. The volume is assumed to take the form
\be
{\cal V}(\tau_b,\tau_s)\,\,\sim\,\, \tau_b^{3/2}-k \tau_s^{3/2}\,,
\ee
with $2\tau_i=T_i+\ol{T}_i$ and $K_K=-2\ln{\cal V}$, as usual. Here the small 4-cycle, governed by $\tau_s$, is also known as a `blow-up cycle'. This is because one may say that it arises from `blowing up' (in the sense of making smooth) a singularity of a Calabi-Yau with a single large 4-cycle. Explicit geometries leading to a Kahler potential of the required structure and possessing other necessary features (see below) have been studied, see e.g.~\cite{Cicoli:2012vw, Cicoli:2013cha}.

As in the KKLT setup of the last section, the first step is to assume that complex structure moduli and axio-dilaton are supersymmetrically stabilised by fluxes. The only fields we consider are hence $T_s$ and $T_b$ with the Kahler potential given above and $W_0=\,\mbox{const.}$ At this level, the scalar potential is identically zero because of the no-scale structure of $K\equiv K_K$.

Two types of correction are considered: First, instanton corrections or a gaugino condensate lead to
\be
W=W_0+A e^{-a T_s}\,.\label{wlvs}
\ee
An analogous term with an exponential suppression in $\tau_b =\,$Re$ \,T_b$ will generically also be present. It has been neglected since, as we will see below, one eventually finds $\tau_b\gg \tau_s$.

Second, there are so-called $\alpha'$ corrections\index{$\alpha'$ correction}. The original meaning of the term are higher-dimension operators in the 10d action, which are suppressed by an appropriate power of $l_s\sim \sqrt{\alpha'}\sim 1/M_s$ (see \cite{Green:1999qt} for an introduction and review).
The relevant term in the present context is a particular contraction of four 10d Riemann tensors, the integral of which over the Calabi-Yau corrects the 4d theory through a modification of the Kahler moduli Kahler potential \cite{Becker:2002nn}:
\be
K_K=-2\ln{\cal V}\qquad\to\qquad K_K=-2\ln({\cal V}+\xi/2)\,,\qquad\mbox{with}\qquad 
\xi=-\frac{\chi(CY)\zeta(3)}{2g_s^{3/2}(2\pi)^3}\,.\label{klvs}
\ee
Here $\chi(CY)$ is the Euler number of the relevant 3-fold, $\zeta(3)\simeq 1.202$ is the appropriate value of the Riemann zeta function, and $g_s$ follows from the stabilised value of the dilaton, as usual.

The evaluation of the $F$-term scalar potential using the standard supergravity formula for the 2-field model based on $T_{b,s}$ and defined by (\ref{wlvs}) and (\ref{klvs}) gives
\be
V({\cal V},\tau_s)\simeq \frac{g_s e^{K_{cs}}}{2}\left(
\frac{\alpha^2 \sqrt{\tau_s} e^{-2a\tau_s}}{6 k {\cal V}}
- \frac{\alpha |W_0|\tau_s e^{-a\tau_s}}{{\cal V}^2} +
 \frac{3\xi|W_0|^2}{4{\cal V}^3} \label{fsp}
\right)\,.
\ee
Here $\alpha\equiv 4a|A|$ and $K_{cs}$ is the complex-structure Kahler potential with flux-stabilised arguments. The axionic modulus Im$\,T_s$ has already been integrated out, while the $T_b$-axion remains massless in this approximation. We leave the straightforward derivation to the reader (Problem~\ref{lvsd}). This scalar potential is easily seen to have a relatively steep minimum in $\tau_s$. After integrating out $\tau_s$, one finds
\be
V({\cal V})\simeq \frac{3g_s e^{K_{cs}}}{2}\cdot \frac{|W_0|^2}{{\cal V}^3}
\left(\frac{\xi}{4}-\frac{k}{2 a^{3/2}}\ln^{3/2}\left(\frac{\alpha{\cal V}}{3k|W_0|}\right)\right)\,.
\label{vlpot}
\ee
Crucially, as observed in \cite{Balasubramanian:2005zx}, this leads to the stabilisation of the last remaining modulus ${\cal V}$ at an exponentially large value and at negative cosmological constant. Supersymmetry is mainly broken by the large $F$ term of $T_b$.

To be slightly more precise, the parametric control of the final result is achieved as follows: Eventually, the small-cycle volume is stabilised at
\be
\tau_s\simeq \left(\frac{\xi}{2 k}\right)^{2/3}\,.
\ee
This needs to be at least somewhat large, either due to a large Euler number\index{Euler number} or a small flux-stabilised values of $g_s$. But then, due to the exponential dependence, the parameter $\exp(-a\tau_s)$ controlling the instanton expansion can easily be extremely small. Similarly, the stabilised value of ${\cal V}$ can easily turn out to be huge,
\be
{\cal V}\sim |W_0|\,e^{a\tau_s}\,,
\ee
where we have suppressed an ${\cal O}(1)$ coefficient. This is excellent news since the volume is the main control parameter of the supergravity expansion underlying the whole approach. Nevertheless, one has to be cautious since there is also the (relatively) small cycle volume $\tau_s$ and hence the curvature in its vicinity does not become exponentially small in string units. Related to this, the $\alpha'$ correction used in the analysis is calculated on a Calabi-Yau, while the realistic case will also have orientifold planes and branes, with additional $\alpha'$ corrections that are not yet fully understood.

We leave the detailed derivation of the formulae given above to the problems. Also, we will not discuss in any detail LVS-specific issues concerning the uplift.\index{uplift} An uplift is of course necessary since the mechanism discussed leads to a non-SUSY AdS vacuum. One option is to use the same $\overline{\mbox{D3}}$-uplift analysed in the context of KKLT. In addition, specifically in LVS constructions, the so-called $D$-term uplift\index{$D$-term uplift} appears to represent a promising possibility (see \cite{Cremades:2007ig} and \cite{Burgess:2003ic, Villadoro:2005yq} for earlier discussions of the underlying idea).

Let us end with a comment concerning loop effects: While $W$ is protected by supersymmetry, $K_K$ receives corrections which, in different regimes, may be best viewed as either 4d loop effects in a theory with KK modes, 10d loop effects in a compactified model, or full-fledged string loop corrections. Consider first a one-field model, where $K_K=-3\ln(T+\ol{T})$ at tree level: A simple scaling analysis shows that a correction $\delta K_K\sim 1/(T+\ol{T})$, if it arises, can be absorbed in a constant shift of $T$ under the log and hence does not affect the scalar potential \cite{vonGersdorff:2005bf}. This cancellation of the formally leading correction was named `extended no-scale structure' in the analysis of \cite{Cicoli:2007xp} and it continues to hold in the multi-field case \cite{Berg:2007wt} and for the explicitly derived string loop corrections of simple torus-orbifold models \cite{Berg:2005ja}. For the present discussion the upshot is that the $\alpha'$ correction with its parametric behaviour $\delta K\sim 1/(T+\ol{T})^{3/2}$ produces the dominant effect in the scalar potential ($\sim 1/{\cal V}^3$). This justifies a posteriori our neglect of loop effects in the bulk of this section. However, once several large moduli are present, their relative size remains unstabilised in an LVS analysis using only $\alpha'$ effects. Loop corrections are then the leading contribution which induces a potential, at order $1/{\cal V}^{10/3}$, stabilising all large Kahler moduli.

\subsection{Vacuum Statistics and the tuning of the cosmological constant}
\index{cosmological constant!problem}

Let us now assume that one of the moduli stabilisation and uplifting procedures discussed in the literature (the two main examples being KKLT and LVS) or some variant thereof works. Moreover, as we did not explain in detail but only sketched in Sect.~\ref{asi}, there should be no problem in finding a compactification which, at the same time, contains a standard-model-like sector. Together, this implies the existence of a landscape of realistic 4d EFTs with a certain random distribution of operator coefficients, including in  particular the cosmological constant $\lambda$ and the Higgs mass parameter $m_H^2$. Crucially, they would have broken supersymmetry and, at least in part, positive $\lambda$.

Two non-trivial questions can then be asked. First, is it clear that the landscape contains a vacuum with the apparently highly fine-tuned values of $\lambda$ and $m_H^2$ we observe? Second, can we understand why we find ourselves in a world described by such a very special vacuum?

In this section, we want to discuss, at least briefly, the first (and simpler) of these two questions. We focus on $\lambda$ and on KKLT. In this case, a partial answer can be given using a fundamental technical result of~\cite{Denef:2004ze} (see also~\cite{Denef:2004cf, Douglas:2003um, Ashok:2003gk}). In the analysis of~\cite{Denef:2004ze}, the focus is entirely on the flux stabilisation of complex structure moduli (and the axio-dilaton), i.e.~Kahler moduli are 
ignored. The setting is (for our purposes) that of type IIB Calabi-Yau orientifolds with O3/O7-planes. In this setting, the tadpole constraint on the flux vector $(f,h)$ can be calculated (for details see below) such that one knows precisely in which subset of the space of integer vectors this object takes its values. Each such value corresponds to a point in complex-structure moduli space at which the geometry (the variables $z_i$ and $S$) is then stabilised. If the dimension of the moduli space and hence of the vector $(f,h)$ is large, solving for the $z_i$ on the basis of a given flux value is practically impossible. But, assuming that the set of relevant flux choices is large, it is possible to talk about the resulting (approximately) statistical distribution of vacua in moduli space. In fact, in the strict mathematical limit of a large tadpole (taking the restriction on the length of $(f,h)$ to infinity), this becomes a precise mathematical question. 

\begin{figure}[ht]
\begin{center} 
\includegraphics[width=6cm]{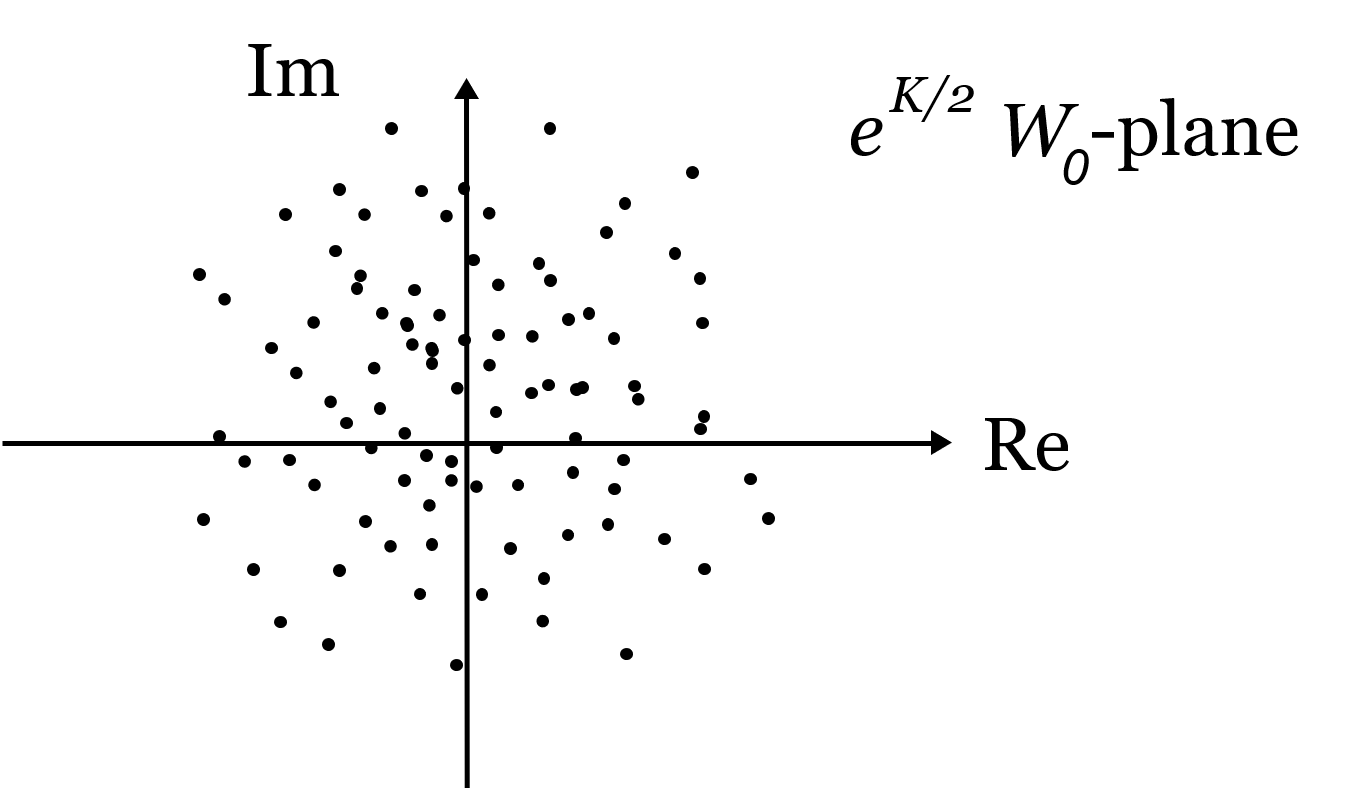}
\caption{The distribution of $e^{K/2}W_0$ in the complex plane has no special feature near the origin.}
\label{w0d} 
\end{center}
\end{figure}

\begin{figure}[ht]
\begin{center} 
\includegraphics[width=6cm]{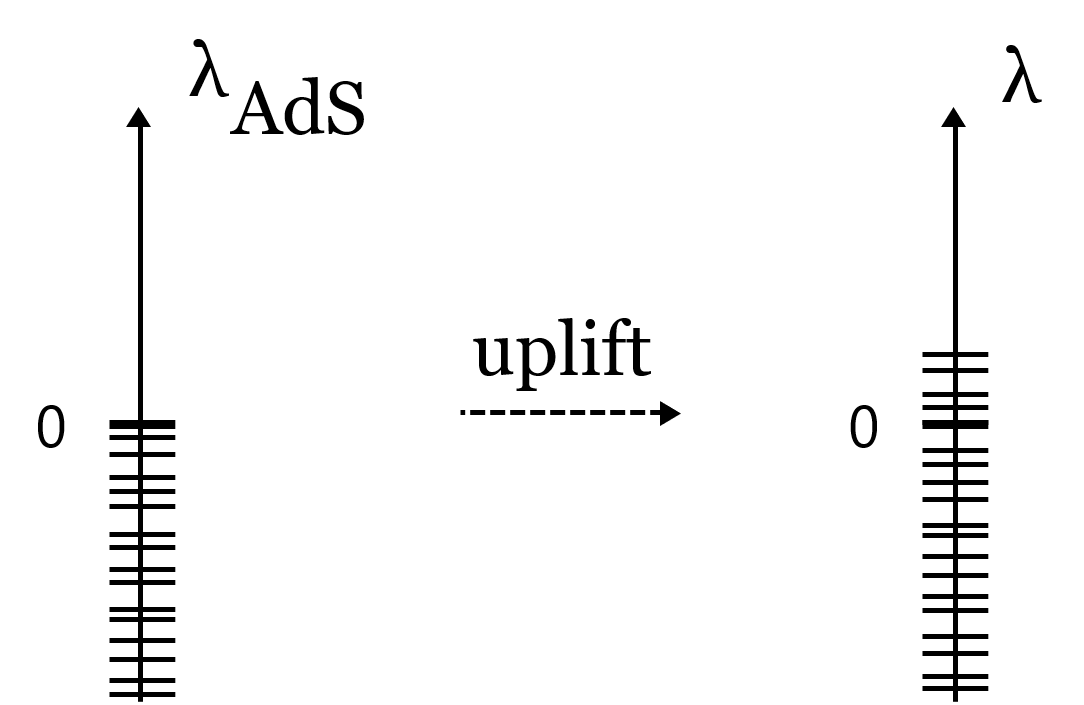}
\caption{Distribution of the cosmological constant before and after uplift.}
\label{ll} 
\end{center}
\end{figure}

The key answer given in~\cite{Denef:2004ze} concerns the distribution of a particular quantity, $e^{K/2}W_0$. It was shown that, under mild assumptions, the distribution of this number in the complex plane is flat near zero, cf.~Fig.~\ref{w0d}. This is not surprising since $W_0$ is by definition a sum of many terms of varying phase and there is nothing special about obtaining the value zero in total. We now want to include Kahler moduli (for simplicity a single Kahler modulus) assuming that, for a large subset of these vacua, instanton or gaugino condensation effects are present. This leads to
\be
W_0\quad \to\quad W_0+A e^{-aT}\,,
\ee
eventually giving rise to full moduli stabilisation in AdS with 
\be
\lambda_{AdS}\sim -e^K|W_0|^2\,.
\ee
The flat distribution of the complex number $e^{K/2}W_0$ now implies a flat distribution of $\lambda_{AdS}$, reaching up to zero from below (cf. the l.h. side of Fig.~\ref{ll}). After an uplift of the type described in Sect.~\ref{kklt1} or~\ref{kklt2}, a dense distribution of $\lambda$ values including the zero-point is obtained (cf. the r.h. side of Fig.~\ref{ll}).\index{uplift}

It is crucial in this logic that both the value of $W_0$ and the uplift energy can be extremely small. In the first case, the reason is the tuning in the flux discretuum, as described above. In the second case it is the exponential warping suppression. Thus, a value of $\lambda$ very close to zero can arise after a shallow AdS vacuum is uplifted by a small amount. The restriction to shallow AdS vacua and small uplifts is crucial for calculational control purposes. Specifically, small $W_0$ implies a relatively large volume and hence a suppression of various higher-order ($\alpha'$ and string loop) corrections.

Of course, it is important to quantify how dense the discretuum is and hence how finely spaced a distribution of $\lambda$ values in Fig.~\ref{ll} one can hope for. For this, we need to discuss tadpole cancellation for the $C_4$ potential. By this we mean that the coefficient of the action term linear in $C_4$ (the `tadpole') should be zero. The intuition behind this is best explained by an analogy to electrodynamics (see also Sect.~\ref{asi}): 

Imagine our space were not $\mathbb{R}^3$ but compact, say $S^3$. Then by Gauss' law a static solution of the Maxwell equations
\be
d * F_2=d * d A_1= j_3
\ee
clearly requires that the total number of sources add up to zero,
\be
\int_{S^3} j_3=0\,.
\ee
Even more intuitively, the number of electrons and positrons must be the same since there can not be more `beginnings' than `ends' of electric field lines on a compact manifold.

In our case, the Chern-Simons lagrangian 
\be
\int C_4\wedge F_3\wedge H_3\equiv \int C_4\wedge j^{flux}_6
\ee
implies that part of the sources for $C_4$ are provided by the 3-form flux. Moreover, the type IIB equations of motion also imply that $G_3$ is imaginary self-dual \cite{Giddings:2001yu},
\be
*_6\, G_3=iG_3\,,
\ee
which in turn implies that $\int_{CY}j^{flux}_6$ can not get different-sign contributions from different regions of the CY.
To see this, rewrite $F_3\wedge H_3$ in terms of $G_3\wedge\ol{G}_3$ and the latter in terms of the manifestly positive quantity $G_3\wedge\!*\ol{G}_3$.
The contribution of the fluxes to the $C_4$ source or the so-called `D3 tadpole' can be written as 
\be
\int_{CY}j^{flux}_6=\int F_3\wedge H_3 \sim (h,f)^2\,,
\ee
with an appropriately defined (symplectic) product on the space of flux vectors $(h,f)$.

This flux vector contribution to the D3 tadpole has to be cancelled by other charged objects. We are discussing this before the uplift, so $\ol{\rm D3}$ branes are not at our disposal. D3 branes contribute with the same sign as (supersymmetric) 3-form fluxes. The available options are then only O3-planes or O7-planes/D7-branes. The first contribute in an obvious way since they are charged oppositely w.r.t. D3 branes. By contrast, the O7-plane/D7-brane contribution is indirect and involves an integral over the curvature of the relevant co-dimension-2 submanifold in the 6d Calabi-Yau. We are not going to spell this out explicitly, but only report the results of a more general analysis:

Type IIB compactifications in the perturbative regime find their non-perturbative completion in so-called F-theory models~\cite{Vafa:1996xn} (cf.~the brief explanation at the end of Sect.~\ref{asi} and the reviews~\cite{Denef:2008wq, Heckman:2010bq, Weigand:2010wm}). We only recall here that F-theory models are based on the geometry of an elliptically-fibered (roughly torus-fibered) Calabi-Yau 4-fold. The fibre torus encodes the information that corresponds, in type IIB language, to the variation of the axio-dilaton $S$. In fact, $S$ is identified with the complex-structure parameter of the fibre torus.

In this F-theory setting, the tadpole\index{tadpole} contribution of the O7-planes and D7-branes above is encoded in the 4-fold geometry, more precisely in the Euler-number $\chi_4$ of the 4-fold. In fact, this result is much more general: It also includes contributions from 7-branes other than the standard D7-branes of perturbative type IIB string theory. In our context, the crucial constraint then becomes 
\be
N^T\Sigma N\leq L\equiv\frac{\chi_4}{24}\qquad \mbox{with}\qquad N=\left(
\begin{array}{c} h\\f \end{array}\right)\qquad \mbox{and}\qquad 
\Sigma\equiv\left(\begin{array}{cc}0 & \mathbbm{1} \\ \mathbbm{1} & 0\end{array}\right)\,.\label{nnl}
\ee
If this inequality is saturated, the fluxes precisely compensate the tadpole induced by 7-branes and O-planes. Otherwise, tadpole cancellation can be achieved by simply adding an appropriate number of D3 branes.

The key geometric input is the availability of 4-folds with Euler\index{Euler number} characteristics up to $\chi_4\sim 10^6$ (see e.g.~\cite{Klemm:1996ts}), leading to $L\sim 10^5$. The number of vacua can then be estimated as~\cite{Denef:2004ze}
\be
{\cal N}_{vac}\sim\frac{L^K}{K!}\,,
\ee
where $K$ is the number of 3-cycles of the Calabi Yau. This number is crucial in the present context since it determines the dimension of the flux vector $N$ to be $d=2K$. Thus, the estimate of ${\cal N}_{vac}$ above can roughly be understood as the volume of a $2K$-dimensional ball of radius $\sqrt{L}$. This is a natural expectation since we are dealing with a lattice with unit spacing on which the flux vector can end. Of this lattice only a certain subset, specified by the inequality in (\ref{nnl}), is available. The details are slightly more complicated since the metric $\Sigma$ is not positive definite, such that the `radius' $\sqrt{L}$ does not specify a ball but the interior of a hyperboloid (the non-compact directions of which are, however, cut off by physical arguments and do not lead to a divergence of ${\cal N}_{vac}$). 

In the end, using the (far from maximal) numbers $L=10^4$ for the 4-fold Euler number and $K=300$ for the number of 3-cycles of the corresponding Calabi-Yau orientifold, one arrives at
\be
{\cal N}_{vac}\sim (eL/K)^K\sim 100^{300}= 10^{600}\,.
\ee
Even after appropriate reductions for the geometric constraints implied by the gaugino condensation / instanton effect and by the warped throat required for the uplift, this is still more than sufficient to realise the desired fine tuning for the cosmological constant of $\sim 10^{-120}$. In fact, most naively (ignoring the reduction by geometric constraints) one expects that of the $10^{600}$ vacua about $10^{480}$ have a cosmological constant of the order of $10^{-120}$ or below.

At this point, a comment concerning a more recent development in the context of vacuum counting has to be made. It concerns the number of several hundred 3-cycles which we used and which is typical for a CY 3-fold. Clearly, an O7-orientifold of a CY 3-fold has more moduli due to the freedom of deforming the 4 D7 branes that originally lie on top of each O7 plane. Even more generally, similar situations can be analysed in the F-theory context, where more types of co-dimension-2 objects than just O7-branes and D7-planes are available. In this context, the 3-fold complex structure and D7-brane deformation moduli are unified as complex-structure moduli of the elliptically fibered CY 4-fold. In this F-theory setting, `the geometry with most flux vacua' (as far as presently known) has recently been identified~\cite{Taylor:2015xtz}. The size of the tadpole is consistent with what was discussed above, but the number $K$ has now roughly speaking to be replaced by the number of 4-fold complex structure moduli. In the maximal known case, this is $h^{3,1}=303,148$. The estimate of ${\cal N}_{vac}$ based on the volume of a sphere in flux space is not a good approximation any more. Instead, a more careful analysis leads to ${\cal O}(10^{272,000})$ flux vacua \cite{Taylor:2015xtz}. This exceeds all 3-fold-based estimates by far.

\subsection{Higgs mass and other landscape-related issues}
\index{landscape}

Let us now turn away from the need to fine tune the cosmological constant and focus on the Higgs mass and hence the electroweak hierarchy problem. The  Higgs mass parameter $m_H^2$ depends both on the $\mu$ term of the\index{Minimal Supersymmetric Standard Model} MSSM\index{MSSM} (if this model arises at an intermediate energy scale) and on SUSY breaking soft terms. Moreover, virtually all Standard Model parameters enter through the large loop effects. (Here we assume that SUSY is broken at a high scale.) All these parameters come from coefficients of operators in the landscape-derived 4d supergravity model. In the present course, we have discussed such models including only Kahler and complex structure moduli. However, matter fields coming from the D-brane sector can easily be added and the resulting structure is in principle well-understood (see e.g.~\cite{Ibanez:2012zz, Kaplunovsky:1993rd, Brignole:1998dxa, Jockers:2004yj, Conlon:2006tj, Kerstan:2011dy}). The values of complex-structure parameters, which are the main entities known to `scan very finely' in the landscape, enter in various ways. In many cases, they govern the coefficients of the effective matter-field lagrangian.

As a result of all of this, it is highly plausible that the fine distribution of landscape points in the complex structure moduli space will, through several calculational steps, translate into a fine distribution of values of $m_H^2$. Moreover, one expects this distribution to be in no way special (e.g. more dilute) near zero. Now, the highest SUSY breaking scale conceivable (and hence the highest natural scale for $|m_H^2|$) is $M_P$. Thus, to get down to the weak scale purely by tuning one has to pay the price of a suppression factor of $(100\,\mbox{GeV})^2/ (10^{18}\,GeV)^2\sim 10^{-32}$. Thus, even starting with the modest estimate of $10^{600}$, we would apparently still be left with 
\be
10^{600}\times 10^{-120} \times 10^{-32}\,\,\sim\,\,10^{448}
\ee
vacua with accidentally small cosmological constant {\it and} an accidentally light Higgs in a model with very high (near $M_P$) SUSY breaking. Even paying some high additional price for various model building requirements (gaugino condensate, throats, Standard Model field content and renormalisable couplings of the right magnitude), there appears to be no problem of finding `our world' in the landscape. There could of course be the problem that certain features (positive $\lambda$, a particular light fermion spectrum etc.) are simply {\it unavailable}. This would clearly invalidate the simplistic statistical reasoning we presented.

Needless to say, the problem of tuning the Higgs mass to a small value is alleviated if we also have low-scale (or at least relatively low-scale) SUSY. Such models with low-scale SUSY may also be available in the landscape. For example, in the case of the LVS the volume ${\cal V}$ can be stabilised at an exponentially large value, leading to a small gravitino mass. Moreover, the soft scale in the visible sector can be significantly smaller than $m_{3/2}$ in certain settings \cite{Aparicio:2014wxa}. An interesting question is now whether we are more likely to find ourselves in a world with purely fine-tuned light Higgs or with a light Higgs mostly due to SUSY (possibly with some extra tuning in addition). One part of the answer can be given by asking how many of the respective vacua are available. In other words, is it `cheaper' to directly tune for a small Higgs mass or to tune for a low SUSY breaking scale? The second option may be preferred if one could lower the SUSY-breaking scale through appropriate model building choices rather than simply through tuning. Yet another option would be to look for models with technicolor-like structure (see Sect.~\ref{lsstc}), lowering the Higgs mass in a non-SUSY-related dynamical way. The above questions have of course been studied from the very beginning of the string landscape to the present, but no widely accepted answer has so far emerged. The reader may consult e.g.~\cite{Denef:2004cf, Susskind:2004uv, Douglas:2004qg, Giudice:2006sn, Acharya:2008zi, Baer:2019zfl, Baer:2020kwz, Broeckel:2020fdz} and refs. therein.

Of course, the discussion just started has to remain highly incomplete within the limited scope of this course. Indeed, not just the SUSY\index{SUSY} breaking\index{supersymmetry (see SUSY)} scale but {\it all} sub-topics of string model building or string phenomenology are affected if not governed by vacuum statistics. This includes finding the right gauge group and matter content in the zillions of brane or gauge bundle configurations (see e.g.~\cite{Lebedev:2006kn, Anderson:2011ns, Gmeiner:2005vz}), understanding statistical aspects of the observed fermion mass patters, a possibly necessary tuning for a flat inflationary potential (see Sect.~\ref{sri}), and many others. Let us here only emphasise one aspect because it is particularly timely: The apparent genericity of axions and axion-like particles in the landscape.

As we have explained, in type-IIB models Kahler moduli govern brane volumes and hence the holomorphic gauge-kinetic functions\index{gauge-kinetic functions} of 4d SUSY gauge theories:
\be
f(T)\sim T =\tau+ic\,.
\ee
This leads to a 4d term $\sim c\,\mbox{tr}\,F\tilde{F}$, making $c$ a so-called QCD axion in the case of the $SU(3)$ gauge group of the Standard Model \cite{Weinberg:1995mt, Ringwald}. Axions are the most plausible solution of the strong CP problem:\index{strong CP problem}
$\theta_{QCD}$ becomes a dynamical field and its potential automatically drives it to zero. Thus, the natural presence of axions (they also arise if $f$ is governed by other moduli in the IIA and heterotic context) is a success of the stringy world view. There are also problems in that the axion mass tends to be too high in string models~\cite{Conlon:2006tq, Svrcek:2006yi}. 

The string landscape adds an important aspect to this topic. Namely, it can be shown that many, in fact possibly {\it extremely} many, periodic pseudo-scalars of the type of $c$ arise in some string compactifications. This idea of the so-called `String Axiverse'\index{Axiverse} has been introduced in \cite{Arvanitaki:2009fg}. The possible existence of such axion-like particles (ALPs) and other super-light fields has also been studied independently of the string landscape as well as in connection with it, see e.g.~\cite{Jaeckel:2010ni, Cicoli:2012sz}. A particularly interesting aspect of ALPs\index{ALP} is\index{axion-like particle} that they can play the role of dark matter, one of the central puzzles of beyond-the-standard-model physics. In this context, string theory has more to contribute than just axions since, quite generally, string theory compactifications tend to predict light sectors in addition to the field content of the Standard Model. In terms of the cosmological evolution, one then expects so called `Remnants'\index{Remnants} (see e.g. \cite{Halverson:2018vbo}). This is yet another aspect of string phenomenology which is strongly influenced by what we think is typical in the landscape.

However, vacuum counting alone is not sufficient to settle all the interesting questions above. Indeed, it is possible that many more vacua with low-scale SUSY rather than with purely fine-tuned non-SUSY light Higgs are available. But this would become irrelevant if cosmological dynamics prefers inflation to always end in vacua with high-scale SUSY-breaking. Thus, we need to turn to the dynamics which might be responsible for populating the landscape.

\subsection{Problems}

\subsubsection{No-scale Kahler potentials and KKLT}
\label{nsp}\index{no-scale model}\index{KKLT}

{\bf Task:}
Using the general supergravity formulae given earlier in the course, calculate the scalar potential of a one-field supergravity model with
\be
K(T,\ol{T})=-\ln\left[(T+\ol{T})^n\right]\qquad \mbox{and}\qquad W=W_0=\,\mbox{const.}
\ee
Which striking feature arises if $n\!=\!3\,$? Try to generalise this special result to the case of $m$ variables, with $e^{-K}$ being a general homogeneous function of the variables $(T^i+\ol{T}^{\ol{\imath}})$ of degree $n$.

Returning to the single-modulus case, analyse the so-called `KKLT potential' arising from the superpotential $W=W_0+Ae^{-aT}$ for $n=3$. Use the notation $T=\tau+ic$, set $A=a=1$ for simplicity and assume $|W_0|\ll 1$. To draw a qualitative plot of $V(\tau)$, after minimising in $c$, it is sufficient to understand the qualitative behaviour of $V$ in the two regimes $|e^{-T}|\gg |W_0|$ and $|e^{-T}|\ll |W_0|$. Throughout, assume $\tau \gg 1$. 

\noindent
{\bf Hints:} The first part is completely straightforward. For the general case, it is useful to prove the relation $(T^i+\ol{T}^{\ol{\imath}})K_i=-n$ and to consider its derivatives. 

The discussion of the KKLT potential is a straightforward exercise in parametrically analysing a given function. Note that, in the second regime, you also need to assume that the axionic variable Im$\,T=c$ takes the value minimising the scalar potential. The result is shown in Fig.~\ref{kkltads}.

\noindent
{\bf Solution:} First, we have
\be
K=-n\,\ln(T+\ol{T})\,,\qquad K_T=K_{\ol{T}}=\frac{-n}{T+\ol{T}}\,, \qquad
K_{T\ol{T}} = \frac{n}{(T+\ol{T})^2}=(K^{T\ol{T}})^{-1}
\ee
and hence
\be
V(T,\ol{T})=e^K(K^{T\ol{T}}|K_T\,W_0|^2-3|W_0|^2)=e^K|W_0|^2(n-3)\,.
\ee
We see that for $n=3$ the potential vanishes identically, implying that $T$ remains a modulus in spite of $W\neq 0$. This is the simplest form of the famous no-scale cancellation.

Now consider the multi-variable case, with 
\be
K=-\ln f(T^1+\ol{T}^1,\cdots,T^k+\ol{T}^k)
\ee
and
\be
f(\alpha(T^1+\ol{T}^1),\cdots,\alpha(T^k+\ol{T}^k))=\alpha^n f(T^1+\ol{T}^1,\cdots,T^k+\ol{T}^k)\,,
\ee
as proposed. By Euler's homogeneous function theorem, we have
\be
(T^i+\ol{T}^{\ol{\imath}})\partial_i(e^{-K})=n\,e^{-K}
\ee
and hence
\be
(T^i+\ol{T}^{\ol{\imath}})K_i=-n\,.\label{orel}
\ee
Differentiation w.r.t. $\ol{T}^{\ol{\jmath}}$ gives
\be
K_{\ol{\jmath}}+(T+\ol{T})^i K_{i\ol{\jmath}}=0\,,
\ee
where we used $K_j=K_{\ol{\jmath}}$. Multiplying by the inverse metric one obtains
\be
K^{i\ol{\jmath}}K_{\ol{\jmath}}+(T+\ol{T})^i=0\,, \label{kud}
\ee
and after further multiplication by $K_i$ and application of (\ref{orel}), 
\be
K_i K^{i\ol{\jmath}}K_{\ol{\jmath}} = n\,. \label{nsr}
\ee
Now one immediately finds the multi-variable result
\be
V=e^K(K^{i\ol{\jmath}} (K_i W_0)\,(K_{\ol{\jmath}}\ol{W}_0)-3|W_0|^2)=
e^K|W_0|^2(n-3)\,.
\ee

Finally, we turn to the discussion of the model with $n=3$ and superpotential $W_0+e^{-T}$. In the first regime, Re$\,T\ll \ln(1/|W_0|)$, we may set $W\simeq e^{-T}$. Then the second scalar potential term, $3|W|^2$, is suppressed by two powers of the large quantity $\tau$ with respect to the $F$-term squared. Similarly, $D_TW\simeq \partial_TW$. Hence,
\be
V\simeq e^K K^{T\ol{T}}|\partial_T\,e^{-T}|^2\sim \frac{1}{T+\ol{T}}|e^{-T}|^2
\sim \frac{e^{-2\tau}}{\tau}\,.
\ee
This is positive and monotonically falling.

In the second regime, Re$\,T\gg \ln(1/|W_0|)$, the naively leading term is obtained by setting $W=W_0$. But this vanishes by the no-scale property. Hence, we need to consider the formally subleading terms, which involve one power of $W_0$ and one power of $e^{-T}$. Such terms, $\sim W_0\,e^{-T}$, appear both in the $F$-term squared and in $-3|W|^2$. But the second contribution suffers a relative suppression by one power of the large quantity $\tau$. (This is due to the enhancement of the $F$-term squared by $K^{T\ol{T}}$, which is only partially compensated by $K_T$.) Thus, we find
\be
V\simeq e^K K^{T\ol{T}}\big[(\partial_T\,e^{-T})\,K_{\ol{T}}\ol{W}_0+\mbox{h.c.}\big]
\sim \frac{e^{-\tau}}{\tau^2}|W_0|\big[e^{i(c+{\rm Arg}\,W_0)}+\mbox{h.c.} \big]\sim 
-\frac{e^{-\tau}}{\tau^2}|W_0|\,.
\ee
In the last step, we assumed that $c$ takes the value minimising $\cos(c+{\rm Arg}\,W_0)$ at minus unity. We see that, at large $\tau$, $V$ is negative and approaches zero from below.

Our two results for large and `small' (still much larger than unity) values of $\tau$ guarantee the presence of a local minimum at negative value of $V$ and with $\tau\sim \ln(1/|W_0|)$.

\subsubsection{The LVS scalar potential and stabilisation mechanism}\label{lvsd}
\index{LVS}\index{Large Volume Scenario}

{\bf Task:}
Derive the formula for the LVS scalar potential and the two-step stabilisation procedure as discussed in the main text.

\noindent
{\bf Hints:} Correct the supergravity formula for the $F$-term potential in two ways: First, by the instanton effect in $W$ (you will need to keep the leading and subleading term), then by the $\alpha'$ effect in $K$. From the latter correction, only the leading term is needed, but to derive it some algebra along the lines of the multi-field derivation of the no-scale potential in Problem~\ref{nsp} is required. Basically, one has to correct this analysis by the no-scale breaking $\alpha'$ effect. Adding these two corrections, one obtains the desired scalar potential for ${\cal V}$ and $\tau_s$. The rest is simple algebra and elementary parametric analysis.

\noindent
{\bf Solution:}
The starting point is the basic supergravity formula 
\be
V=e^K\Big(K^{i\ol{\jmath}}\,(K_i W) \, (K_{\ol{\jmath}} \ol{W} )-3|W|^2\Big)\,,
\ee
with $i,j$ labelling the Kahler moduli $T_b$ and $T_s$. Let us for the moment ignore the $\alpha'$ correction and focus only on the effect of the instanton term $A e^{-aT_s}$ that is added to $W_0$. In its absence, $V$ would be identically zero. A non-zero result arises only because the $T_s$ derivative applied to $W$ gives a non-zero value. There are precisely three such terms, giving
\be
\delta V_1=e^K\Big(K^{s\ol{s}}a^2|A|^2 e^{-2a\tau_s}
-K^{s\ol{\jmath}}K_{\ol{\jmath}}\big[aAe^{-aT_s}\ol{W}_0+\mbox{c.c.}\big] \Big)\,,
\ee
where we disregarded the subleading instanton correction to $W$ in the square bracket and made use of the fact that the derivatives of $K$ are real in our approximation. Employing (\ref{kud}), this becomes
\be
\delta V_1=e^K\Big(K^{s\ol{s}}a^2|A|^2 e^{-2a\tau_s}
+2\tau_s\big[aAe^{-aT_s}\ol{W}_0+\mbox{c.c.}\big] \Big)\,,
\ee
which we can now easily minimise w.r.t.~the imaginary part $c_s$ of $T_s=\tau_s+ic_s$:
\be
\delta V_1=e^K\Big(K^{s\ol{s}}a^2|A|^2 e^{-2a\tau_s}
-4\tau_s a |A||W_0| e^{-a\tau_s}\Big)\,.
\ee

To determine $K^{s\ol{s}}$, one has to invert the matrix
\be
\left(
\begin{array}{cc}
K_{b\ol{b}} & K_{b\ol{s}}
\\
K_{s\ol{b}} & K_{s\ol{s}}
\end{array}
\right)\,.
\ee
As explained in the main text, stabilisation will eventually occur in the regime of exponentially large $\tau_b$ and modestly large $\tau_s$. Thus, it will be justified a posteriori that we use the relations $\tau_b\gg \tau_s\gg 1$ in the present analysis. One then finds that $K_{s\ol{s}}\gg K_{b\ol{b}} \gg
K_{b\ol{s}} = K_{s\ol{b}}$. This implies
\be
K^{s\ol{s}}\simeq (K_{s\ol{s}})^{-1}\simeq \left( \frac{3 k}{8 {\cal V}\sqrt{\tau_s}}
\right)^{-1}
\ee
and
\be
\delta V_1=e^K\left(\frac{8}{3k}{\cal V}\sqrt{\tau_s}a^2|A|^2 e^{-2a\tau_s} -4\tau_s a |A||W_0| e^{-a\tau_s}\right)\,.
\ee
We note in passing that this argument can be easily rerun with several rather than just one `big' Kahler moduli. In this case one has to invert a block-diagonal matrix with the above hierarchies characterising the different blocks.

When calculating the $\alpha'$ correction, we may replace $W$ by $W_0$ since we are not interested in quantities which are doubly small. The correction arises since the no-scale cancellation analysed in Problem~\ref{nsp} fails. It is hence quantified by
\be
\delta V_2=e^K\big( K^{i\ol{\jmath}}K_i K_{\ol{\jmath}}
-3 \big)|W_0|^2\,. \label{v2def}
\ee
Moreover, we have
\be
K=-2\ln({\cal V}+x)+\cdots \qquad \mbox{with}\qquad x\equiv \xi/2
\ee
and with the ellipsis standing for Kahler moduli independent terms. Thus,
\be
K_i=-\frac{2{\cal V}_i}{{\cal V}+x}=K_i^{(0)}\,\frac{\cal V}{{\cal V}+x}\simeq K_i^{(0)}\left(1-\frac{x}{\cal V}\right)\,,
\ee
where $K^{(0)}$ is the uncorrected Kahler potential. With this (\ref{orel}) takes the form
\be
(T^i+\ol{T}^{\ol{\imath}})K_i=-3\left(1-\frac{x}{\cal V}\right)\,.\ee
One may now follow the steps which lead to the no-scale result (\ref{nsr}) with $n=3$. The corrected formula turns out to be
\be
K_i K^{i\ol{\jmath}} K_{\ol{\jmath}} -3\left(1-\frac{x}{\cal V}\right) = K_i K^{i \ol{\jmath}}\left(\frac{3x}{\cal V}\right)_{\ol{\jmath}}\,. \label{kkkc}
\ee
The term on the r.h.~side can be rewritten according to
\be
K_i K^{i \ol{\jmath}}\left(\frac{3x}{\cal V}\right)_{\ol{\jmath}}
= - K_i K^{i \ol{\jmath}}{\cal V}_{\ol{\jmath}} \cdot \frac{3x}{{\cal V}^2} =K_i K^{i \ol{\jmath}} K_{\ol{\jmath}}
\,\cdot \frac{3x}{2 \cal V}\,.
\ee
Here we have disregarded the difference between $K$ and $K^{(0)}$ since the whole term is subleading in the $1/{\cal V}$ expansion (see \cite{Becker:2002nn} for more complete results). Combining this with (\ref{kkkc}) one finds, again at leading order in the large volume expansion,
\be
K_i K^{i\ol{\jmath}} K_{\ol{\jmath}}=3\left(1+\frac{x}{2 \cal V}\right) = 3\left(1+\frac{\xi}{4 \cal V}\right)\,.
\ee
This completes the determination of $\delta V_2$. It remains to check that $\delta V_1+\delta V_2$ corresponds precisely to the scalar potential of (\ref{fsp}).

Next, one may simultaneously minimise (\ref{fsp}) in ${\cal V}$ and $\tau_s$ (see the original papers~\cite{Balasubramanian:2005zx, Conlon:2005ki} and the appendix of \cite{Hebecker:2012aw}, which we mainly follow in this problem). But it may be simpler and more intuitive to adopt the EFT logic of integrating out the heavier field $\tau_s$ first. For this, one may focus on the first two terms of (\ref{fsp}) only. Disregarding terms suppressed by $1/\tau_s$, one finds that the minimum is approximately at
\be
e^{-a\tau_s}\simeq \frac{3k |W_0|\sqrt{\tau_s}}{\alpha {\cal V}}\,.
\label{expn}
\ee
A good approximation for $\tau_s$ is obtained by disregarding the non-exponential $\tau_s$ dependence and then solving the equation:
\be
\tau_s\simeq \frac{1}{a}\,\ln\left(\frac{\alpha{\cal V}}{3k|W_0|}\right)\,.\label{apst}
\ee
The effective potential for ${\cal V}$ is now obtained by first replacing the exponentials $\exp(-a\tau_s)$ in (\ref{fsp}) according to (\ref{expn}) and then using (\ref{apst}) for the non-exponential $\tau_s$ terms. The result is (\ref{vlpot}).

The approximate minimisation of (\ref{vlpot}) is an easy task: First, note that as ${\cal V}\to\infty$, the second term dominates and the potential approaches zero from below. Second, as ${\cal V}$ falls the logarithm becomes small enough for the first term to dominate -- the potential becomes positive. Thus, the minimum occurs when the two terms are approximately equal, i.e.~at
\be
a\,\left(\frac{\xi}{2k}\right)^{2/3}\simeq \ln\left(\frac{\alpha{\cal V}}{3k|W_0|}\right)\,.
\ee
Together with (\ref{apst}), this confirms the last two relations of the LVS section of the main text.

\section{Eternal Inflation and the Measure Problem}\label{eimp}
\index{inflation}

\subsection{From slow-roll inflation to the eternal regime}
\label{sri}\index{slow-roll!inflation}
The present course does, of course, assume General Relativity as a prerequisite. Since most relativity courses include some cosmology, it appears logical to assume that the reader will also be familiar with the most basic cosmology-related formulae. A selection of relevant textbooks is \cite{Wald:1984rg, wei, mtw, car, strau, Mukhanov:2005sc, weico, peeb, kolb}. We only summarise the results to set our notation:

The cosmological principle, with excellent support from data, postulates that space is homogeneous and isotropic on large scales. Together, these two features imply that spacetime can be represented as a one-parameter family of homogeneous spatial hypersurfaces $H_t$ (with $t\in\mathbb{R}$) which are threaded orthogonally by `observer curves'. Each of those is parameterised by the observer eigentime $t$. In terms of the 4d metric, this means
\be
ds^2=-dt^2+a^2(t)\,g_{ij}\,dx^i dx^j\,,
\ee
where $a$ is the scale factor and $g_{ij}$ is the metric on a maximally symmetric 3d space, i.e. on a sphere, on flat $\mathbb{R}^3$, or on a 3d hyperboloid. 

In the simplest case, matter comes in the form of a perfect fluid,
\be
T_{\mu\nu}=\rho\,u_\mu u_\nu+p\,(g_{\mu\nu}+u_\mu u_\nu)
\ee
with density $\rho$ and pressure $p$. Then the Einstein equations and the continuity equation reduce to\index{Hubble parameter}
\bea
3 M_P^2 (H^2 + k/ a^2) &=& \rho \label{freq}
\\
\dot{\rho}+3H\,(\rho+p)&=&0\,,\qquad\mbox{with the Hubble parameter}\qquad H=\dot{a}/a\,. \label{cont}
\eea
Here $k=+1,0,-1$ distinguishes the three cases of positive, zero and negative spatial curvature.

A case of particular interest is that of a scalar $\phi$ with potential $V(\phi)$. Using the standard result $\rho=T+V$ and $p=T-V$ (with $T=\dot{\phi}^2/2$), one then immediately sees that (\ref{cont}) takes the form
\be
\ddot{\phi}+3H\dot{\phi}+V'=0\,.
\ee

Standard slow-roll inflation \cite{Starobinsky:1980te, Guth:1980zm, Sato:1980yn, Linde:1981mu, Albrecht:1982wi, Mukhanov:1981xt} arises in the regime where the potential $V$ is sufficiently flat. This is conventionally quantified by requiring smallness of the two slow-roll parameters\index{slow-roll!parameters} ($M_P=1$ here and below):
\be
\epsilon\equiv\frac{1}{2}\left(\frac{V'}{V}\right)^2\ll 1\qquad\mbox{and}
\qquad \eta\equiv \frac{V''}{V}\ll 1\,.
\ee
Indeed, in this regime $\ddot{\phi}$ can be neglected in the equation of motion for $\phi$ and $\rho$ is dominated by the potential energy. Thus, cosmology is described by
\be
3H\dot{\phi}=-V'\qquad \mbox{with}\qquad H^2=V/3\qquad \mbox{and}\qquad
a=\exp(Ht)\,.\label{3eq}
\ee
Here we have disregarded the curvature term $k/a^2$ since it anyway quickly becomes subdominant as $a$ grows while $H$ remains approximately constant. This represents a so-called quasi-de-Sitter\index{quasi-de Sitter} situation, exact de-Sitter expansion corresponding to an exactly constant (rather than slowly changing) $H$ in the last equation of (\ref{3eq}).\index{de Sitter space}

\begin{figure}[ht]
\begin{center} 
\includegraphics[width=6.5cm]{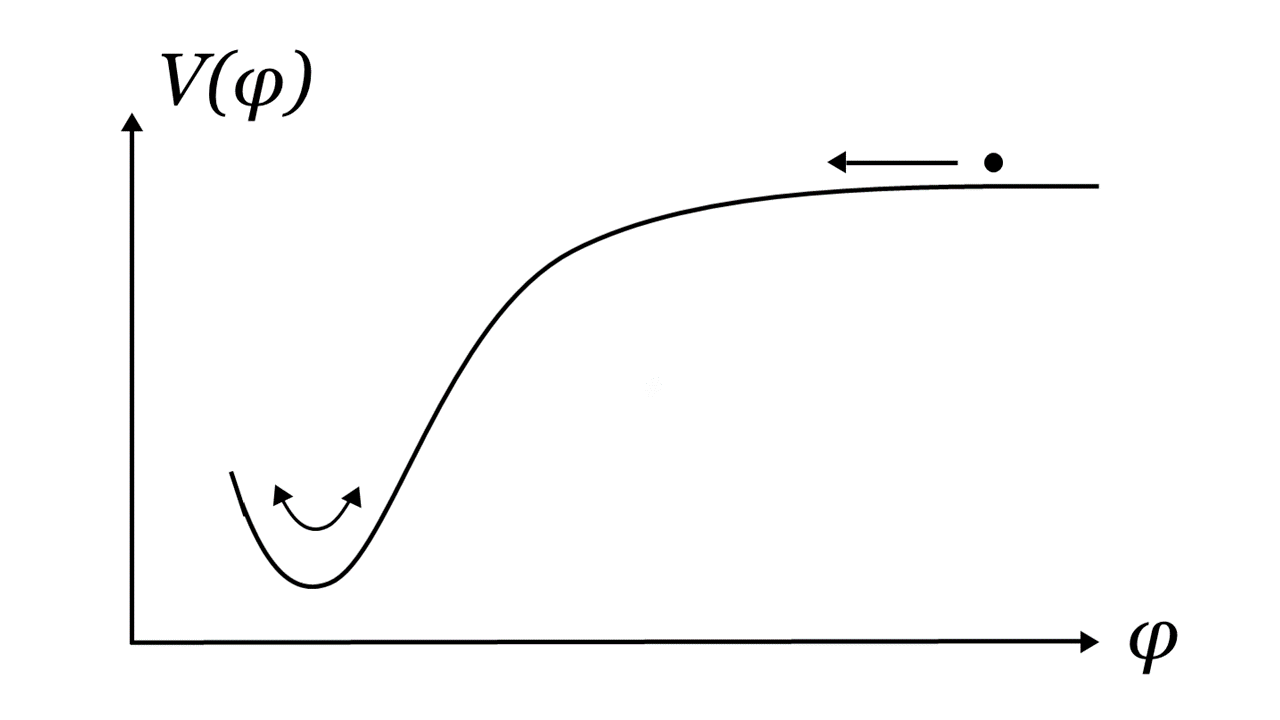}
\caption{Slow-roll inflation ending in field oscillations and reheating.}
\label{inf} 
\end{center}
\end{figure}

In standard inflationary cosmology one assumes that this situation lasts long enough to explain the flatness and homogeneity of our present-day universe. But eventually it ends since $\phi$ rolls into a region where the slow-roll conditions cease to hold. In the simplest case, $\phi$ oscillates about its minimum and eventually decays to Standard Model particles, reheating the universe (cf.~Fig.~\ref{inf}).

Crucially, while in the slow-roll regime, $\phi$ does not only roll classically but is, at the same time, subject to quantum fluctuations. To understand this qualitatively, it is useful to consider the simplified case of pure de Sitter ($V(\phi)=\mbox{const.}$ and hence $H=\mbox{const.}$). It is then easy to determine the inward-going geodesics in the relevant metric ($k=0$ for simplicity)
\be
ds^2=-dt^2+e^{2Ht}d\ol{x}^2\,.
\ee
One finds that, above some maximal radius $r_0$ (with $r^2\equiv d\ol{x}^2$), they never reach the origin. In other words, there exists a cosmological horizon. Its size is of the order of the {\bf de Sitter radius}\index{de Sitter space!radius} $1/H$. Each spatial slice falls into many so-called {\bf de Sitter patches}\index{de Sitter space!patches}, which are causally disconnected. As the universe evolves, the exponential expansion increases their number by $e^3$ in a Hubble time $t_H\equiv 1/H$. 

At the moment, all we need to conclude from the above is that a single geodesic observer, who by definition sits in the centre of his or her de Sitter patch, is surrounded by a horizon. This observer sees space expand and sees objects disappear forever as they cross the horizon. To be precise, the observer stops seeing them before they cross the horizon due to the diverging red-shift which affects their radiation (if emitted backward towards the central observer). We assume that the reader has at least some rudimentary familiarity with similar horizons and similar physical situations, either in the case of the Unruh effect (eternally accelerated observer, seeing an Unruh horizon) or in the black hole case (static observer near a black hole, seeing the black hole horizon). In both cases, the observer is subject to an approximately thermal radiation coming from the horizon. Its most naive explanation is virtual pair production, with one particle disappearing behind the horizon and the other hitting the observer's detector. This also happens in the de Sitter case and, for lack of another dimensionful parameter (the Planck scale can be taken to infinity at fixed $H$), we have $T\sim H$. This thermal radiation affects $\phi$ and, again on dimensional grounds, induces a random fluctuation $\delta\phi\sim H$ after each time interval $\delta t \sim 1/H$.\index{eternal inflation}

This allows us to delineate the boundary between the regimes of eternal and slow-roll inflation for a single scalar in a flat potential. Indeed, for the classical evolution to dominate the slow-roll displacement of $\phi$ should be larger than its random fluctuation during an interval $\delta t$. Thus, according to \eqref{3eq}
\be
|\dot{\phi}_{\rm roll}|\,\delta t\sim V'/H^2 \,\,\gtrsim\,\, |\delta\phi_{\rm fluct.}|
\sim H
\ee
is the condition for the rolling to win over the quantum diffusion which, as we explained, is an unavoidable feature of de Sitter space. We may rewrite this as
$V'/V\gtrsim \sqrt{V}$ or $\epsilon \gtrsim V$. As a side remark we note that for purely quadratic potentials, $V= m^2\phi^2/2$, this condition becomes $1/\phi^2\gtrsim m^2\phi^2$. Hence, for sufficiently small $m$ it can be violated at large $\phi$ while still in the controlled regime of small energy density, $m^2\phi^2\ll 1$. This is the famous model of `chaotic inflation'~\cite{Linde:1983gd}, which allows for eternal inflation, slow-roll inflation and reheating all to occur in different regions of the same simple, power-law potential.

In our more contrived potential of Fig~\ref{inf}, the above is clear even without any calculation: Continuing the potential to the right such it becomes more and more flat, it is apparent that all regimes exist: In the very flat region at large $\phi$, one basically has pure de Sitter. The field fluctuates as described above and, very occasionally, enters the intermediate regime where slow-roll dominates over fluctuations. In this region, while the universe still keeps exponentially expanding, the field now systematically rolls to the left, eventually reheating at $\phi$ near its minimum. As a result, a universe like our own forms in a small `pocket' inside the vast region of continued eternal expansion. In fact, infinitely many such pocket universes will form in the underlying infinite (approximate) de Sitter space with its fluctuating scalar $\phi$. 

The interested reader may want to consult \cite{Spradlin:2001pw} and \cite{Mukhanov:2005sc, Riotto:2002yw} for many more details of de Sitter space and slow-roll inflation respectively. For inflation specifically in the stringy context, see \cite{Baumann:2014nda, Westphal:2014ana}. Moreover, the earlier review of string cosmology in \cite{Quevedo:2002xw} includes the discussion of some more exotic alternative possibilities.

\subsection{Eternal inflation in the landscape}\index{eternal inflation}
For the purpose of these notes, we take slow-roll inflation in our past to be an essentially established observational fact. This may be justified on the basis of the excellent fit of its predictions for curvature fluctuations to cosmic-microwave-background or CMB data \cite{Akrami:2018odb}.\footnote{
Alternatives 
range from modifications of the simple slow-roll dynamics described earlier (see e.g.~\cite{Berera:1995ie, Lyth:2001nq, Alishahiha:2004eh}) to entirely different scenarios like string-gas or pre-big-bang cosmology~\cite{Brandenberger:1988aj, Gasperini:1992em}.
}
The previous section described how, in a simple single-field model, the epoch of slow-roll inflation in our universe can originate from an eternally inflating universe. While this is appealing from the perspective of solving (at least part of) the initial-condition problems, it is not the most obvious or most common way in which slow-roll inflation relates to the string landscape as it is presently understood. Indeed, in our present understanding de Sitter vacua are rare. Solutions with a positive energy density and a very flat potential, as required for slow roll, are more rare. Finally, solutions where the flatness is sufficient for the eternal regime (viz.~Fig.~\ref{inf} with $V$ becoming more and more flat at very large $\phi$) are the rarest of all. 

However, eternal inflation may arise in a much more natural, maybe even unavoidable way in a universe based on the string landscape. To see this, let us step back and forget for the moment about the phenomenological requirement of slow-roll inflation. Instead, focus on what the string landscape most naively predicts: It contains ${\cal N}=2$ SUSY Minkowski vacua (e.g.~from simple Calabi-Yau compactifications) as a very well established feature. Moreover, there are ${\cal N}=1$ AdS vacua (like in KKLT before the `uplift'), a feature that I would call established, although maybe with less mathematical rigour: One needs to combine Calabi-Yau string geometry with instantons (or the 4d non-perturbative phenomenon of gaugino condensation) and the fine-tuning of $W_0$.\footnote{
Most 
probably there also exist non-SUSY AdS solutions. It has been conjectured that those can only be metastable~\cite{Ooguri:2016pdq, Freivogel:2016qwc}. They would then not deserve the name vacua since, if AdS is metastable, any observer in it will encounter the decay after a time interval comparable to the AdS radius. Cosmologically, such `vacua' can nevertheless occur, but they appear not to add anything new to our discussion at a qualitative level and we will hence ignore them.

This may also be a suitable place to note that, if one does not insist on the AdS curvature scale being parametrically below the KK scale, the existence of SUSY AdS vacua would be as certain as that of SUSY Minkowski vacua. One example are compactifications of type IIB on an $S^5$ carrying $F_5$ flux to 5d AdS. This compactification may be the best established of all due to its possible fundamental definition via AdS/CFT~\cite{Maldacena:1997re, Aharony:1999ti}. However, we are here interested in EFTs in the non-compact dimensions and we hence insist on the scale separation between AdS-curvature and KK scale. 
}

Finally, and this is the most important category of vacua for the purposes of the present discussion, there are presumably metastable de Sitter vacua of the KKLT or LVS type discussed earlier (or some variants thereof). Their existence had been widely accepted since KKLT~\cite{Kachru:2003aw}, but has then more recently been challenged on fundamental grounds~\cite{Danielsson:2018ztv, Obied:2018sgi, Garg:2018reu, Ooguri:2018wrx}. This has led to a heated debate in the framework of the `Swampland paradigm', to which we will return in Sect.~\ref{swpr}. For now let us press ahead under the assumption that string theory has metastable\index{metastable} de Sitter vacua. In fact, while other options have been discussed (cf.~Sect.~\ref{swpr}), such dS vacua are presumably our best hope for relating string theory to the real world. Also, it has to be noted that no strong technical reasons have so far been provided for why KKLT or some variant thereof should not work.

\begin{figure}[ht]
\begin{center} 
\includegraphics[width=7.5cm]{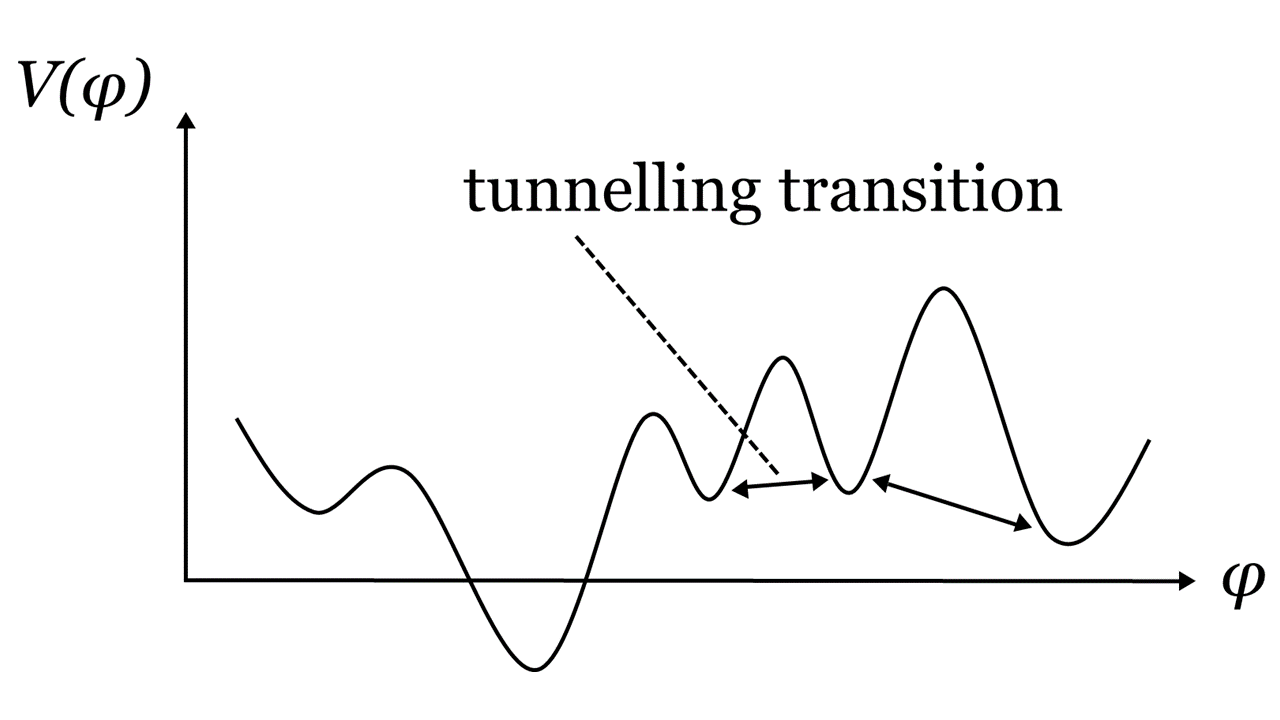}
\caption{A simple visualisation of the landscape over a 1-dimensional moduli space. Some possible down and up-tunnelling processes are illustrated.\index{tunnelling transition}}
\label{tunnelling} 
\end{center}
\end{figure}

We will ignore ${\cal N}=2$ vacua since they are presumably irrelevant for the real world. Thus, we have AdS and (with the caveats above) dS vacua, which can be visualised as points in a high-dimensional space. This space is parameterised by the moduli of the Calabi-Yau which, due to the non-zero scalar potential, have of course now ceased to be moduli and are simply scalar fields. Replacing the multi-dimensional moduli space by a single real line, the set of AdS and dS vacua may be visualised as in Fig.~\ref{tunnelling}. This figure or an analogous picture with a `mountain-range potential' over a 2-dimensional plane, is used widely to illustrated the string landscape. But one should always remember that this simple scalar field theory with a potential $V$ and many local minima (vacua) remains only a model.

Concretely, the reader should keep in mind that apart from the reduction of the dimensionality of field space for the purpose of drawing, there are at least two further (over)simplifications hidden in this picture: 

First, the transition between two vacua has nothing to do with climbing a smooth barrier in moduli space, at least this is not the generic case. Generically, two different vacua are associated with different 3-form flux and are hence separated by, e.g., a D5-brane or NS5-brane wrapping a 3-cycle of the compact space and representing a domain wall in the non-compact 4d spacetime (cf.~Sects.~\ref{cgwf} and \ref{bpm}). As a result of the flux change between the two sides of the domain wall, the values at which the moduli are stabilised also change. Hence the picture of different minima at different $\phi$ values is actually reasonable. Just the possibility of rolling over the smooth barrier must be replaced by more general tunnelling transitions. 

Second, the full string theory landscape does of course contain different Calabi-Yau geometries with different topologies. Dynamical transitions between some of them are possible and well-understood in many simple cases, (see e.g.~\cite{Greene:1996cy}) but it is in fact conjectured that {\it all} Calabi-Yau moduli spaces are, in a very well-defined mathematical sense, part of a single space. This is sometimes referred to as `Reid's fanatasy'\index{Reid's fanatasy} \cite{reid}. What happens to this statement in the case where one involves orientifolding, F-theory constructions or even compactifications of different perturbatively defined string theories together with M-theory, is less clear. But it is expected that all of them, even compactifications to space-time dimensions different from four, are part of the same dynamical structure and tunnelling transitions between all the vacua are possible
(see e.g.~\cite{Carifio:2017nyb} for a recent discussion). This, of course, makes Fig.~\ref{tunnelling} an enormous oversimplification. In fact, one should think of many such pictures, with different field-space dimensions of $\phi$, glued together.

Nevertheless, let us stick to the simple picture of scalar-field minima separated by potential barriers. Each of the minima at field values $\phi_i$ has a different cosmological constant $\lambda_i=V(\phi_i)$. If at least one of those minima has $\lambda_i>0$ and if the probability $T$ for tunnelling out of this minimum (per volume and time) is smaller than the fourth power of its expansion rate,
\be
T\lesssim H^4(\lambda)\,,\qquad \mbox{where}\qquad 3 M_P^2 H^2(\lambda) = \lambda\,,
\label{cond}
\ee
this already gives rise to eternal inflation. Condition (\ref{cond}) may roughly be understood as follows: It requires the density (in 4-volume) of nucleation points of bubbles of other vacua to be smaller than the typical scale $H$ of the underlying dS space. There is then no danger that the loss of volume to other vacua wins over the volume growth due to de Sitter expansion, which is governed by $H$.

\begin{figure}[ht]
\begin{center} 
\includegraphics[width=8.5cm]{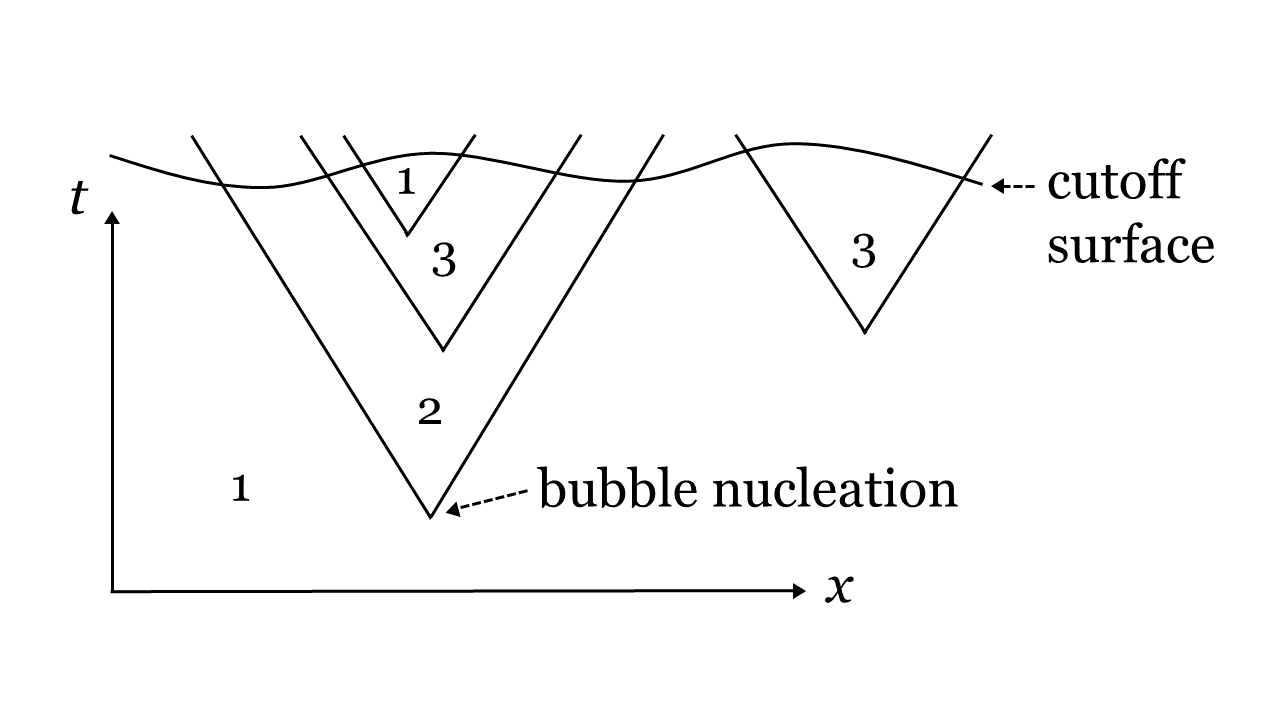}
\caption{Nucleation and speed-of-light expansion of bubbles in a `background' dS vacuum. The `cutoff surface' will be discussed later.\index{bubble!nucleation}}
\label{ds} 
\end{center}
\end{figure}

In the generic case there is more than one dS vacuum. There are then obviously tunnelling processes where a bubble of the energetically favoured, lower-lying de Sitter state is nucleated within a certain higher-energy de Sitter background. This could for example be the regions labelled `1' and `2' in the schematic (Penrose-type) diagram in Fig.~\ref{ds}. We have called the presence of such tunnelling events `obvious' since it corresponds to the familiar process of bubble nucleation in first-order phase transitions, where a bubble of the phase with smaller free energy nucleates in the phase with higher free energy in which the system is started. As a less obvious fact, in the cosmological context of tunnelling between different de Sitter vacua the inverse process is also possible. In other words, a bubble of the energetically disfavored, higher-lying de Sitter can be nucleated and grow in a low-lying background. In terms of Fig.~\ref{ds}, this may for example be the nucleation of phase `1' inside `3' together with nucleation of `3' inside `1' on the r.h.~side of the figure. The process of `up-tunnelling' is strongly suppressed, i.e.~much more rare than that of `down-tunnelling'. We will return to this at the quantitative level. For now, suffice it to say that the surprising fact that up-tunnelling is possible at all can be understood as a result of the exponential expansion of the background: While energetically the disfavoured bubble wants to shrink, for large enough bubbles the background expansion wins and `pulls' the bubble to larger size in spite of the apparent increase of (non-gravitational) energy associated with this. Recall that energy conservation is anyway not (at least not in the usual, straightforward way) a condition that can be used to constrain the allowed dynamical evolution in the general-relativistic context.

What is crucial for us at the moment is that, due to both up and down-tunnelling\index{tunnelling}, the whole landscape gets populated once eternal inflation is running, i.e. once a single Hubble-sized patch of any of the de-Sitter vacua exists. The continued process of the nucleation of bubbles within bubbles within bubbles sketched in Fig.~\ref{ds} is oversimplified since there are also AdS vacua. Nucleation of corresponding bubbles leads locally to `big crunches'\index{big crunch} since the energy density imprinted in such bubbles by the dynamics of bubble nucleation grows. This is clear from the fact that one may think of AdS as of a contracting space. This contraction does not interfere with empty AdS being a perfectly consistent solution of Einstein equations, yet it unavoidably leads to a crunch if a homogeneous energy density is present. Nevertheless, for appropriate tunnelling rates the continued appearance of such `terminal vacua'\index{terminal vacua} does not stop the overall process of eternal inflation. Finally, we note that the nucleation of Minkowski-space bubbles is in principle also possible.While these do no crunch, they also decouple from the eternal inflation process since no new de Sitter bubbles can be nucleated within them. The reason is conventional Minkowski-space energy conservation. We have not displayed AdS and Minkowski regions for simplicity.

\subsection{Tunnelling transitions in quantum mechanics}
\label{tqm}\index{tunnelling!quantum mechanics}
To put some meat on the largely qualitative discussion of the previous section, it is useful to understand the calculation of tunnelling rates between the different vacua. Let us start the discussion with tunnelling in quantum mechanics, for which there are references at the elementary textbook level. A particularly useful analysis, taking the reader all the way from quantum mechanics to the field theory case, is \cite{Coleman:1985rnk}. 

\begin{figure}[ht]
\begin{center} 
\includegraphics[width=10cm]{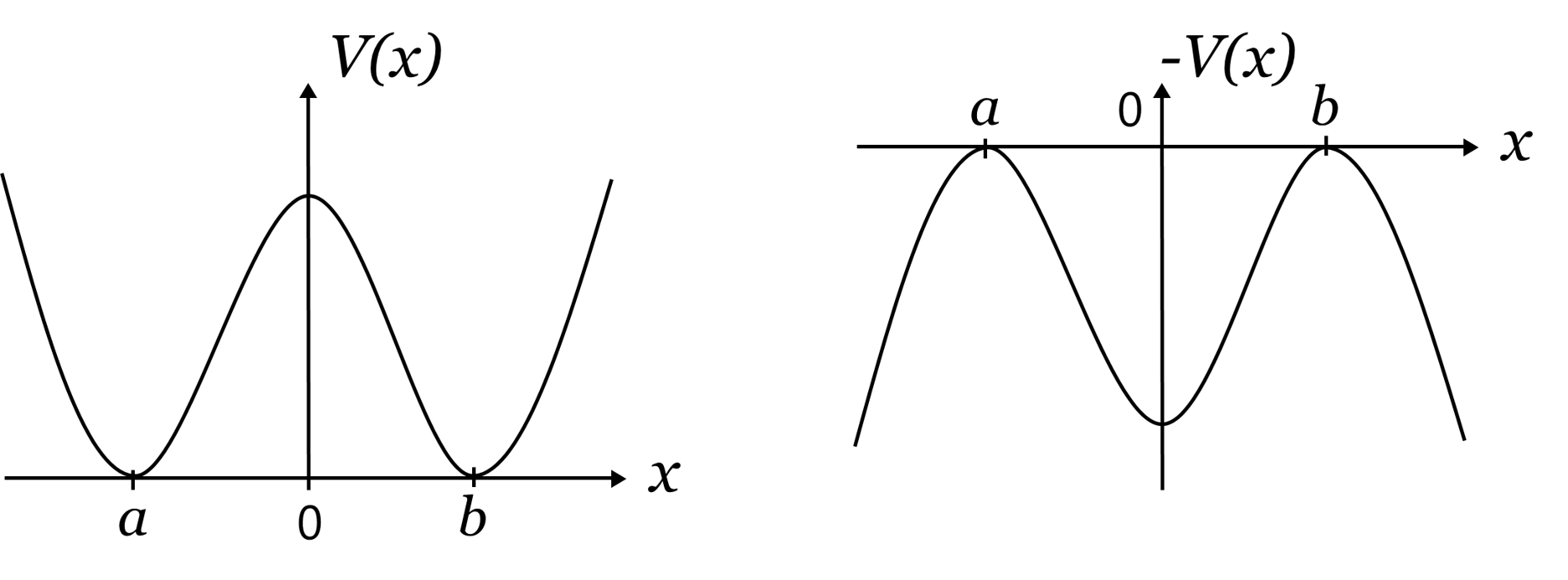}
\caption{Degenerate double well potential in quantum mechanics and the corresponding inverted potential relevant for the equation of motion of the euclidean theory.}
\label{dw} 
\end{center}
\end{figure}

In quantum mechanics, one of the simplest cases is that of tunnelling in the degenerate double well potential, cf.~the l.h.~side of Fig.~\ref{dw}. The reader is presumably familiar with the standard WKB calculation which shows that a particle with mass $m$ and energy $E$ hitting a generic barrier from the left has a non-zero transition probability governed by an amplitude
\be
A\,\sim\, \exp\left(-\int dx\,\sqrt{2m(V(x)-E)}\right)\,.\label{dw1}
\ee
Here the integration extends over the classically forbidden part of the barrier.

In the case of the double well, the relevant question to ask is that about the amplitude for a particle, originally localised on the l.h.~side in the ground state $|a\rangle$, to be observed in the ground state $|b\rangle$ after some time $T$. Assuming the ground-state energy is small compared to the potential height, the answer is simply
\be
A\,\sim\, T\,\exp\left(-\int dx\,\sqrt{2mV(x)}\right)\,.
\label{dw2}
\ee
The exponential follows from (\ref{dw1}) by neglecting $E$. One may argue for the prefactor $T$ using the following toy-model: Consider the ground states on the left and right as a two state system: $\{\,|a\rangle\,,\,|b\rangle\,\}$. The hamiltonian clearly has a small off-diagonal term, suppressed by the small exponent in (\ref{dw2}). It is then immediately clear that, at leading order in this exponent, the transition amplitude between $|a\rangle$ and $|b\rangle$ must be linear in $T$. (The reader is invited to think about this amplitude more carefully for $T\to 0$ and $T\to\infty$, where it turns out that the result is modified. See also the discussion below.)

Coleman \cite{Coleman:1985rnk} presents a very elegant path integral derivation for this result by considering the classical solution dominating the euclidean amplitude at very large $T$. This euclidean amplitude reads
\be
\langle b|\,e^{-HT}\,|a\rangle 
\sim \int\limits_{x(-T/2)=a}^{x(T/2)=b} Dx\,e^{-S}
\sim \int\limits_{x(-T/2)=a}^{x(T/2)=b} Dx \,\exp\left[-\int_{-T/2}^{T/2} dt\left\{\frac{m}{2}\dot{x}^2+V(x)\right\}\right]\,.
\ee
The path integral is dominated by the extremum of the euclidean action. Quite generally, the latter is given by the solution of the corresponding classical mechanics problem in the inverted potential, $V\to-V$. For infinitely large $T$, this solution consists of the following: A particle starts at $t=-\infty$ and with zero velocity on the maximum at $a$. It then very slowly accelerates and rolls through the minimum at $x=0$ to the maximum at $b$, where it comes to rest at $t=+\infty$.

\begin{figure}[ht]
\begin{center} 
\includegraphics[width=8cm]{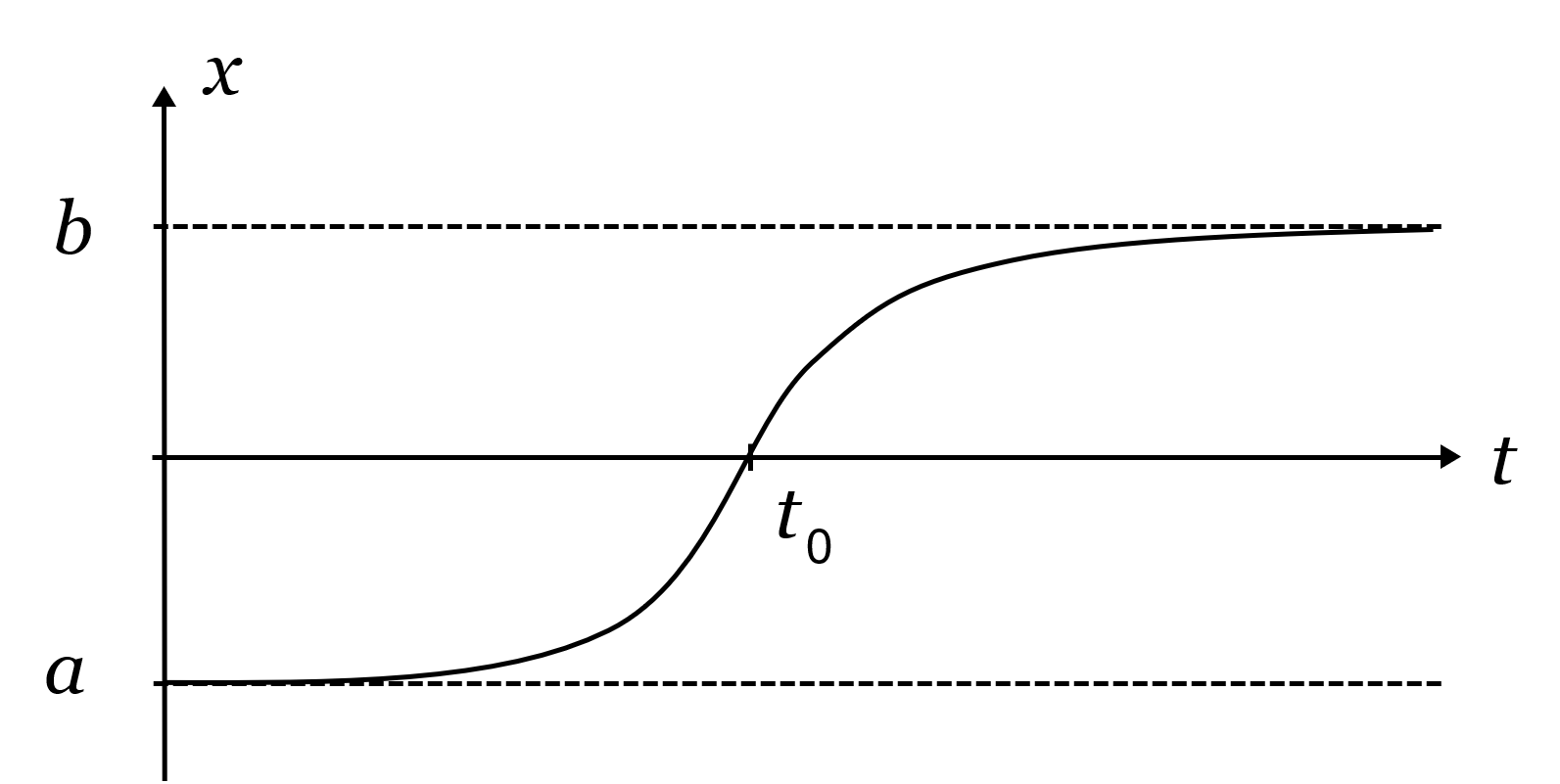}
\caption{Classical solution of the euclidean theory, which corresponds to a solution of the original theory with inverted potential. The large initial and final times $-T/2$ and $T/2$ are outside the plotted time range.}
\label{rolling} 
\end{center}
\end{figure}

At finite $T$, this process ceases to be an exact solution, but it is still an approximate one. More precisely, there is a continuous infinity of such approximate solutions, parameterised e.g.~by the time $t_0$ at which they cross $x=0$, cf.~Fig.~\ref{rolling}. Hence$\,$\footnote{
Here
we introduced an additive renormalisation of the original hamiltonian by the oscillator ground state energy $\omega/2$, where $\omega$ is the frequency for small oscillations around the minima (we assume $\omega_a=\omega_b$). Otherwise, an additional factor $\exp(-\omega T/2)$ would appear on the r.h.~side.
}
\be
\langle b|\,e^{-HT}\,|a\rangle \,\sim\,\int_{-T/2}^{T/2}dt_0\,\,e^{-S_{tunnel}(t_0)}\,\sim\, T\,e^{-S_{tunnel}}\,.
\ee
In going from the next-to-last to the last expression, we have disregarded the $t_0$ dependence of the classical action $S_{tunnel}$. Note that the final formula manifestly agrees with (\ref{dw2}) if one evaluates the action using energy conservation.

It is an easy exercise to repeat the analysis above, allowing for any number of `instanton' transitions between $a$ and $b$. Clearly, to contribute to the amplitude $\langle b|\,e^{-HT}\,|a\rangle$ this number must be odd, leading to the Taylor series for the hyperbolic sine:
\be
\langle b|\,e^{-HT}\,|a\rangle \,\sim\,\sinh\left[KT\,e^{-S_{tunnel}}\right]\,.
\ee
Here $K$ is a dimensionful prefactor of the amplitude which we suppressed in the preceding discussion. The analytic continuation to the physically relevant real-time case is also immediate:
\be
\langle b|\,e^{-iHT}\,|a\rangle \,\sim\,i\,\sin\left[KT\,e^{-S_{tunnel}}\right]\,.
\ee

\begin{figure}[ht]
\begin{center} 
\includegraphics[width=10.5cm]{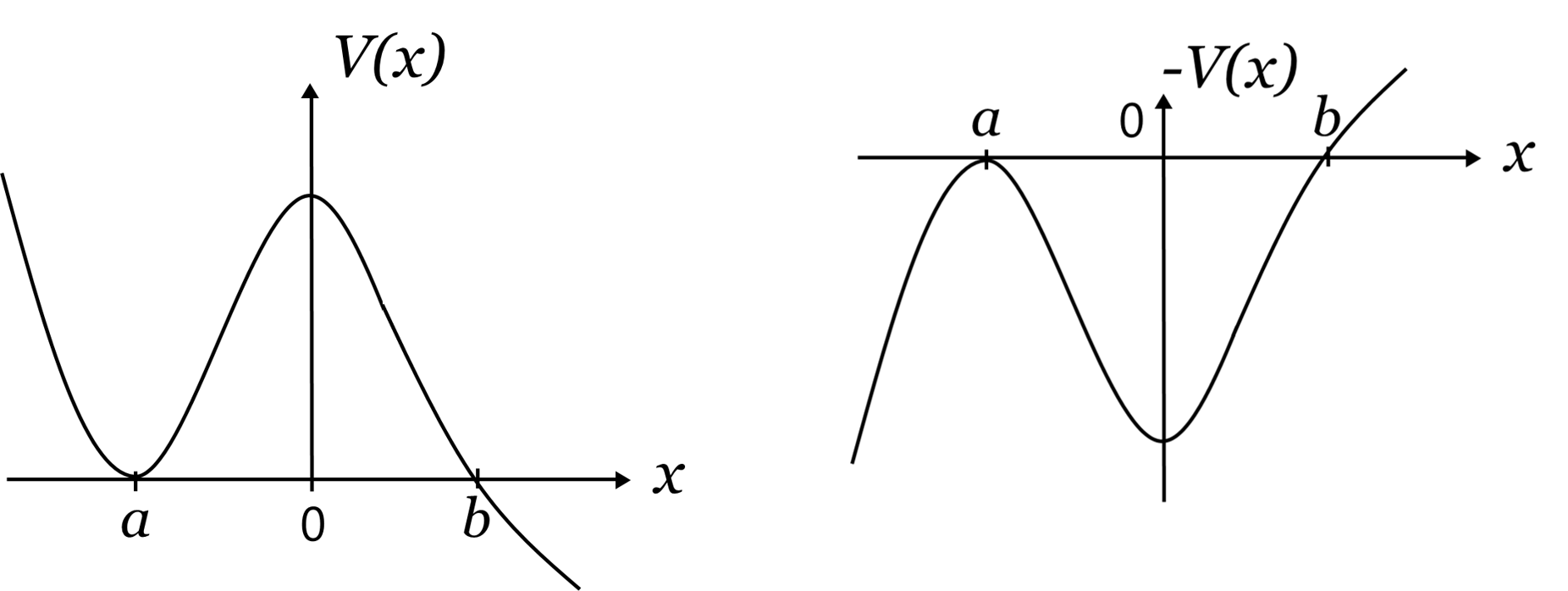}
\caption{On the left: Quantum mechanical potential allowing for a decay of a potentially long-lived `ground state' in the minimum at $x=a$. On the right: The corresponding inverted potential.}
\label{decay} 
\end{center}
\end{figure}

Now we turn to the more interesting case where a state, originally in the minimum at $a$, can decay to a negative-energy region which opens up to the right of the point $x=b$ (cf.~Fig.~\ref{decay}). Given the double-well analysis above, the most naive guess is that the amplitude for a state to decay after a time $T$ is again (at leading order in $T$)
\be
A\sim T\,e^{-S_{tunnel}}\,.\label{atun}
\ee
This time, $S_{tunnel}$ is the action for a process where the particle starts, in the inverted potential $-V$, on top of the maximum at $a$ and then runs through the minimum to arrive at $b$. Beyond $b$, there is no tunnelling suppression so one expects only the potential in between the points $a$ and $b$ in Fig.~\ref{decay} to contribute. This guess is correct, but will not be very useful for going to the field theory and gravity case later on. The problem is the abrupt and poorly defined way in which our calculation `ends' at the point $b$, where the particle returns from tunnelling and becomes a real particle rolling down a potential.

It is more effective to consider the euclidean classical solution in which the particle starts at $a$ at $t=-T/2$, rolls to $b$, and then returns to $a$ at $t=+T/2$. This process, called a {\bf bounce}\index{bounce} for obvious reasons, calculates a contribution to the amplitude 
\be
\langle a|\,e^{-HT}\,|a\rangle = e^{-E_0 T}+K T\,e^{-S_{bounce}}\,.\label{sbo}
\ee
Here $E_0$ is the leading-order or perturbative energy of $|a\rangle$. It may, as before, be set to zero by an appropriate redefinition of the Hamiltonian. Our interest is in the exponentially small (non-perturbative) correction on the r.h.~side of (\ref{sbo}). This correction comes from the fact that the potential has a second zero at $x>0$, leading to the existence of the classical bounce solution. The correction allows for a peculiar interpretation. To understand this, it is important to note that the prefactor $K$ is determined by fluctuations of $x(t)$ around the classical solution. More precisely, 
one writes $x(t)=x_{class.}(t)+\delta x(t)$. At leading order in $\delta x(t)$, the action then becomes
\be
S\simeq S_{bounce}-\int dt\,\delta x(t)\left[\frac{m}{2}\,\frac{d^2}{dt^2}+V''(x(t))\right]\,\delta x(t)\,.\label{sexp}
\ee
Expanding $\delta x$ in the complete set of eigenfunctions of the differential operator in (\ref{sexp}), one finds
\be
K \sim \prod_i\sqrt{\lambda_i}\,.
\ee
Thus, if one of the eigenvalues $\lambda_i$ is negative, $K$ becomes  imaginary and one sees that $\exp(-S_{bounce})$ does in fact govern the size of an {\it imaginary} correction to $E_0$. In other words, it actually governs the decay rate of the state $|a\rangle$, which is precisely what we want.

To complete this argument, we need to convince ourselves that one $\lambda_i$ is indeed negative, in other words, that our bounce solution has a negative mode.\index{negative mode} This can be shown rather generally, but a simple, non-rigorous argument is as follows: Consider first the tunnelling solution between two vacua $|a\rangle$, $|b\rangle$. This can not have a negative mode since, intuitively, it is simply the optimal (i.e. with smallest euclidean action) path between these vacua. Any deformation makes the action larger. By contrast, the bounce action can clearly become smaller if one deforms $x(t)$ appropriately. To be specific, thinking in terms of a particle rolling in the inverted potential in Fig.~\ref{decay}, we may slow this particle down slightly more than in the classical solution when it approaches the turning point at $b$. It then never reaches $b$ and returns prematurely, leading to a smaller action. For more careful arguments, both in the present quantum mechanical model and in the field theory case of the next section, see e.g.~\cite{Coleman:1985rnk}.

To summarise, we have learned that up to non-exponential effects, the decay rate through a barrier as in Fig.~\ref{decay} is
\be
\Gamma \sim \exp(-S_{bounce})\,,
\ee
where $S_{bounce}$ is the action for a classical bounce in the inverted potential. We finally note that $S_{bounce}=2 S_{tunnel}$ and $\Gamma\sim |A|^2$ such that the analysis based on the bounce is fully consistent with the naive guess of (\ref{atun}).

\subsection{Tunnelling transitions in field theory}
\label{tft}\index{tunnelling!field theory}\index{vacuum decay}
Before attempting to take this to the gravitational case, which is relevant for populating the landscape, it is useful to understand the generalisation to flat-space quantum field theory. As already in the previous section, we refer to \cite{Coleman:1985rnk} and the original papers listed therein for a more detailed treatment.

\begin{figure}[ht]
\begin{center} 
\includegraphics[width=6.5cm]{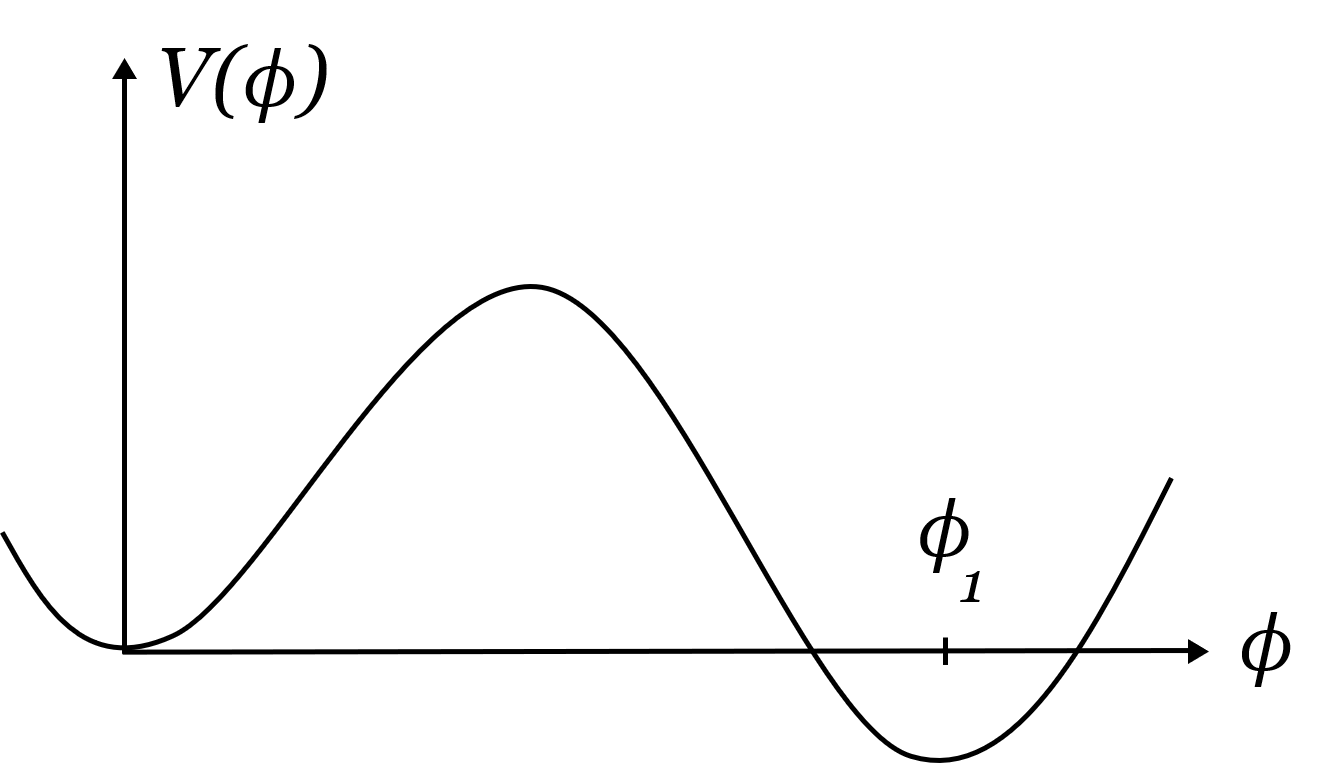}
\caption{Scalar-field double well potential with false vacuum\index{false vacuum} at $\phi=0$ and true vacuum\index{true vacuum} at $\phi=\phi_1$.}
\label{dwpot} 
\end{center}
\end{figure}

Maybe the simplest example is that of a real scalar $\phi$ with
\be
{\cal L}=\frac{1}{2}(\partial\phi)^2-V(\phi)\,,\label{lphi}
\ee
where $V$ is of an asymmetric double-well type, as sketched in Fig.~\ref{dwpot}. The decay of the metastable or false vacuum at $\phi=0$ to the stable or true vacuum at $\phi=\phi_1$ proceeds through bubble nucleation. To understand this, we need to first understand possible bubbles of the true vacuum inside of an infinitely extended false vacuum. 

For this, let us start with the bubble wall\index{bubble!wall}, focussing first on the simpler case where the two minima in Fig.~\ref{dwpot} are degenerate. Moreover, let us look for a solution where two half-spaces, say with $x^1\lesssim 0$ and $x^1\gtrsim 0$ (and for $\{x^2,x^3\}\in \mathbb{R}^2)$ are in the false and true vacuum respectively. More precisely, a solution with field profile $\phi(x^1)$ will exist such that $\phi(-\infty)=0$ and $\phi(+\infty)=\phi_1$, with the transition occurring mainly in the vicinity of the plane $x^1=0$. Let us define the energy per unit area of this configuration as the bubble wall tension $T$. The above stationary solution will, of course, cease to exist in the presence of an asymmetry 
\be
\Delta V\equiv V(0)-V(\phi_1)
\ee
between the two minima. Nevertheless, there will be an alternative solution in which the domain wall is accelerating  under the pressure of the energy density difference $\Delta V$ between l.h. and r.h. half-space. This solution will allow for an effective description in terms of a domain wall with a certain tension $T$. This latter concept will hence continue to be well-defined even in the case of an asymmetric double well.

Next, we note that the presence of a true-vacuum bubble of radius $R$ in a false-vacuum background leads to two energetic effects: On the one hand, there is an energy gain since the volume of the bubble is at a lower energy-density. On the other hand, the bubble wall with its positive tension comes with an energetic cost. The total energy of a bubble of radius $R$ is hence
\be
E_B(R)\simeq -\frac{4}{3}\pi R^3\,\Delta V + 4 \pi R^2\,T\,.
\label{ebr}
\ee
Here we used the so-called thin-wall approximation, i.e. the assumption that $R$ is much larger than the bubble wall thickness. The latter is defined as the typical length interval inside which most of the gradient energy of the bubble-wall is localised. The so-called {\bf critical bubble}\index{critical bubble} radius  $R_c$ is defined by $E_B(R_c)=0$. It is clear from \eqref{ebr} that
\be
R_c=3 T/\Delta V\,.\label{rc}
\ee
The critical bubble sits at the boundary between the regimes of small bubbles, which recollapse under the bubble-wall tension\index{bubble!wall tension}, and large bubbles, which grow indefinitely under the pressure induced by $\Delta V$. 

We note in passing that the concept of a critical bubble should be familiar from first-order thermal phase transitions. Here, for example in an overheated fluid, bubbles of all sizes continuously form due to thermal fluctuations. Supercritical bubbles then grow and lead to the emergence of extended regions of the energetically favoured (in this case gaseous) phase. In our zero-temperature context bubbles of different sizes form due to quantum fluctuations.

The basis for a quantitative understanding of the resulting false-vacuum decay rate has been laid in the previous section on tunnelling in quantum mechanics. We can now easily identify the analogue of the quantum mechanical potential of Fig.~\ref{decay}. The role of the variable $x$, plotted on the horizontal axis, is played by the bubble radius $R$. The potential plotted vertically is now replaced by $E_B(R)$. This function first rises since, for small $R$, a function $\sim R^2$ is always larger than a function $\sim R^3$. Then, of course, $R^3$ eventually wins and $E_B$ becomes negative as $R$ passes the critical radius $R_c$. 

This allows us to describe the field-theoretic analogue of the bounce as a process in $\mathbb{R}^4$, the euclidean version of $\mathbb{R}^{1,3}$: A small bubble emerges at some point $(t,\overline{x})$, grows to critical radius $R_c$, and then shrinks again to zero at some later euclidean time. Topologically, this clearly corresponds to a 4-dimensional ball of true vacuum inside a false-vacuum $\mathbb{R}^4$. It can be shown \cite{Coleman:1977th} that the solution of euclidean field-theory describing this process is in fact not only topologically such a ball. It has  perfect $O(4)$ symmetry, i.e. it is a 4d ball also geometrically. The 3d boundary of this ball is the bubble wall.

Let us supply a different perspective on the euclidean-field-theory version of the quantum mechanical bounce of the last section. This will at the same time provide an intuitive argument for its $O(4)$ symmetry. In quantum mechanics, the bounce is a solution which asymptotes (at $t\to\pm\infty$) to the metastable state. The field theory analogue is expected to be a solution asymptoting (at $|x|\to\infty$, with $x\in\mathbb{R}^4$) to the false vacuum. In quantum mechanics, the solution explores, in its centre, the region of the potential to which the particle can decay. Thus, we expect that the field-theory bounce also explores, in its centre, the decay region. In the field-theoretic case, this is the true vacuum and hence we expect the field theory bounce to contain a region of true vacuum in its centre. Thus we are indeed looking for a 4d ball of true vacuum in the false vacuum background. This ball should be a solution of the equations of motion following from the $O(4)$-symmetric euclidean action based on \eqref{lphi}. Thus, we expect $O(4)$ symmetry. Let us then estimate the action of such candidate field configurations, i.e. of balls of radius $R$ filled with true vacuum and centred on $x=0$. There are contributions from the (lower-energy) volume and the bubble-wall boundary, such that
\be
S_{ball}(R)\simeq -\frac{\pi^2}{2}R^4\,\Delta V+2\pi^2R^3\,T\,.
\ee
This has an extremum at $R=R_c$, with $R_c$ precisely the critical radius\index{critical radius} of \eqref{rc}.

As a result, we can finally write down the field-theoretic bounce action $S_{bounce}\equiv S_{ball}(R_c)$:
\be
S_{bounce}\simeq \frac{27\pi^2\,T^4}{2\,(\Delta V)^3}\,.
\label{ftb}
\ee
To translate this in a field-theoretic decay rate, we need to pay attention to one last important difference between the quantum-mechanical and field-theoretic analyses: The quantum mechanical rate characterises events per time, the field-theoretic rate characterises events per time and volume. This works out in the quantum mechanical case due to the explicit factor $T$ that appears in the last term of \eqref{sbo}. If we redo the analysis in field theory, taking our space to be $T^3\times \mathbb{R}$, we will have a continuum of bounces since the latter can occur at any point of the spatial torus $T^3$. Integrating over all of them will give a factor $V\!ol(T^3)$, accompanying the time factor $T$. Thus, we are justified in writing 
\be
\Gamma\sim e^{-S_{bounce}}\,,
\ee
with $S_{bounce}$ calculated above and with the interpretation of $\Gamma$ as a rate of events per time and volume.

\subsection{Tunnelling in gravitational theories}\label{cdls}
\index{tunnelling!gravity}
First, it is clear that in a limited parametric range the analysis of the last section continues to be valid even if our field-theoretic model is coupled to gravity. We may restrict attention to the case where $V_{false}\equiv V(0)\ge 0$ since in the opposite case one would be starting with AdS space, which cosmologically always crunches after a short time. Next, let us denote by $H_{false}=(V_{false}/3)^{1/2}/M_P$ and $H_{true}=(V_{true}/3)^{1/2}/M_P$ the Hubble parameters of the false and true vacuum. As long as $R_c\ll 1/H_{false/true}$, the bubble nucleation process is occurring essentially under flat-space conditions, such that our purely field-theoretic results for the rate continue to hold. This covers many interesting cases.

However, these conditions can also easily be violated. For example, if $\Delta V$ is very small the critical bubble can be very large, invalidating the flat-space approximation. More interestingly, in the gravitational case a qualitatively new possibility, namely that of up-tunnelling (from smaller to larger vacuum energy density) arises. Thus, a dedicated gravitational analysis is mandatory.

The classical paper on the subject is that of Coleman and De Luccia\index{Coleman-De Luccia}~\cite{Coleman:1980aw}, with a selection of subsequent analyses appearing in \cite{Parke:1982pm, Lindley:1984bj, Brown:1987dd, Lee:1987qc, SchwartzPerlov:2006hi, Johnson:2007jla} and refs.~therein (cf.~also the very recent discussion in \cite{Eckerle:2020opg}). To save time, we will start our presentation by directly generalising the crucial field-theoretic bounce of Sect.~\ref{tft} to the gravitational case. Comments concerning the intuitive interpretation will be provided subsequently.

\begin{figure}[ht]
\begin{center} 
\includegraphics[width=8.5cm]{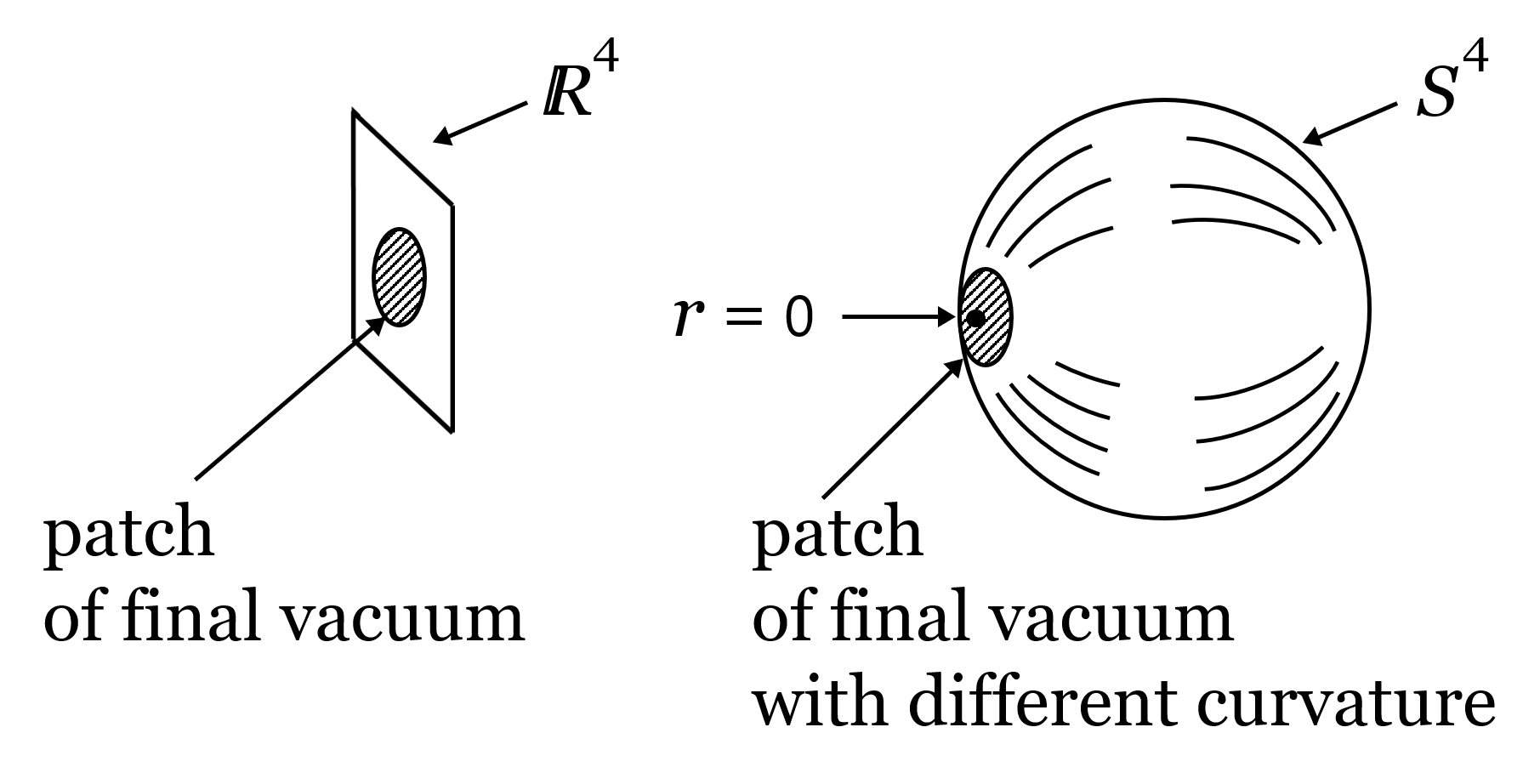}
\caption{On the left: Sketch of the field-theoretic bounce where a ball of the final-state vacuum is present inside flat $\mathbb{R}^4$. On the right: The gravitational analogue, where this ball is present inside of a 4-sphere (euclidean de Sitter\index{de Sitter space!euclidean} space). Crucially, the curvature in the final-vacuum patch is in general different from that of the surrounding $S^4$.}
\label{cdl} 
\end{center}
\end{figure}

The bounce of Sect.~\ref{tft} is a solution of the euclidean theory on $\mathbbm{R}^4$ which contains a spherical ($O(4)$-symmetric) domain-wall with the initial-state (false) vacuum outside and the final-state (true) vacuum inside, cf.~the l.h.~side of Fig.~\ref{cdl}. When generalising this to a gravitational theory, the absolute values of the respective energy densities become relevant. We first focus on the case where both $V_{false}$ and $V_{true}$ are non-negative, such that both the inside and the outside of the domain wall become positively curved. This is displayed on the r.h.~side of Fig.~\ref{cdl}. To understand this geometry, the key observation is that 4d euclidean de Sitter space is simply a 4-sphere, with the curvature characterising the value of the curvature scalar (which depends on the cosmological constant). Thus, all one needs is a geometry in which a portion of a small-radius sphere (the false vacuum) is cut out and replaced by a piece of a large-radius sphere (the true vacuum). Moreover this has to be a solution of Einstein's equations. The latter is obvious away from the domain wall, where simply the right choice of radius has to be made. What is non-obvious and will be discussed momentarily is the solution inside the wall, where both field-value and curvature change continuously.

Maybe the technically simplest and most straightforward approach remains that of~\cite{Coleman:1980aw}. It starts by parameterising the metric of the euclidean geometry on the r.h. side of Fig.~\ref{cdl} as
\be
ds^2=dr^2+f(r)^2 d\Omega_3{}\!^2\,,
\ee
where $d\Omega_3{}\!^2$ is the round metric on the unit-radius 3-sphere. The point $r=0$ is the centre of the true-vacuum patch. This metric is accompanied by a field profile $\phi_b(r)$, where the index `b' stands for `bounce'. The function $\phi_b$ is approximately constant and equal to the false-vacuum value, $\phi=\phi_f$, for $r\lesssim r_c$. For $r\gtrsim r_c$, it is again approximately constant and equal to the true-vacuum value, $\phi=\phi_t$. The reader may want to recall Fig.~\ref{dwpot}, where $\phi_f=0$ and $\phi_t=\phi_1$, but with the potential shifted upward such that both minima are de Sitter. We emphasise that $r_c$ is the value of the coordinate $r$ which corresponds to the location of the $O(4)$-symmetric domain wall. By contrast, $R(r_c)$ is the physical radius of the true-vacuum patch. The behaviour of $\phi_b(r)$ in the vicinity of $r=r_c$ depends on the precise form of the domain wall, which in turn depends on the details of the potential. This will not be important in the thin-wall approximation.

Let us first allow for a general $O(4)$-symmetric field profile $\phi$. The euclidean action then takes the form
\bea
S&=&\int d^4x\left[-\frac{1}{2}M_P^2{\cal R}+\frac{1}{2}(\partial \phi)^2+V(\phi)\right]
\nonumber \\
&=&2\pi^2\int f^3 dr\,\left[3M_P^2\left(\frac{f''}{f}+\frac{f'^2}{f^2}-\frac{1}{f^2}\right)+\frac{1}{2}\phi'^2+V\right]
\nonumber \\
&=&2\pi^2\int f^3 dr\,\left[-3M_P^2\left(\frac{f'^2}{f^2}+\frac{1}{f^2}\right)+\frac{1}{2}\phi'^2+V\right]\,,\label{lef}
\eea
where the prime denotes differentiation with respect to $r$. 
To obtain the expression in the last line, we used integration by parts together with our knowledge that $f(r)$ is vanishing sufficiently quickly at the minimal and maximal values of $r$. 

It is convenient to supplement the above with the $rr$-component of the Einstein equations,
\be
{\cal R}_{\mu}-\frac{1}{2}{\cal R}g_{\mu\nu}=M_P^2T_{\mu\nu}\,,
\ee
for the given matter action and geometry. This equation is, in fact, almost identical to the Friedmann equation which appeared in \eqref{freq}. In the present context, it reads
\be
3M_P^2\left[\left(\frac{f'}{f}\right)^2-\frac{k}{f^2}
\right]=\frac{1}{2}\phi'^2-V\,,\label{rree}
\ee
with $k=1$. The only changes are in the notation ($t\to r$ and $a(t)\to f(r)$) and in the relative sign between terms with and without $r$-derivatives. The latter arises due to the transition from lorentzian to euclidean signature. The combination of the action in the form of \eqref{lef}, the equation of motion for $\phi$ following from it, and the Einstein equation \eqref{rree} are sufficient to calculate the exponent in the tunnelling process we are after. 

To make this explicit, we first need to modify our field-theoretic result such that it allows for a non-zero vacuum energy in the false vacuum. This is of course essential if we want to be able to talk about tunnelling in de Sitter space. With a view on the l.h. side of Fig.~\ref{cdl}, one immediately sees that the correct modification is
\be
\Gamma\sim e^{-S_{bounce}}\qquad \to \qquad 
\Gamma\sim e^{-S_{bounce}+S_{false}}\,.
\ee
Here $S_{bounce}\equiv S[\phi_b]$ and $S_{false}\equiv S[\phi_f]$, with $\phi_b$ the field configuration of the bounce and $\phi_f$ the constant field configuration corresponding to the initial, false-vacuum state. We see that, with this generalised expression for the rate, we obtain the correct result even if $V(\phi_f)\neq 0$. Indeed, any possible constant contribution to the lagrangian simply cancels, such that only the effects of the true-vacuum region and the domain wall remain. 

Now the generalisation to the gravitational case is obvious:
\be
\Gamma=\exp\left(-S[\phi_b,g_b]+S[\phi_f,g_f]\right)\,.
\ee
Here the first term in the exponent is the action of the bounce geometry with the corresponding field configuration (cf.~the r.h.~side of Fig.~\ref{cdl}) and the second term is simply the action of the sphere with constant initial-state field value and appropriate curvature.

As noted, equation (\ref{lef}) and (\ref{rree}) contain enough information to evaluate the relevant actions and hence the decay rate. We leave this as an excercise (Problem~\ref{cdlp}), which consists essentially in following the analyses of \cite{Coleman:1980aw} and \cite{Parke:1982pm}. The result can be given in a particularly compact form as \cite{Johnson:2007jla}
\be
\Gamma=\exp(-B)\,\,,\qquad B= \left(\frac{27\pi^2\,T^4}{2\,(\Delta V)^3}\right)\,r(x,y)\label{bdef}
\,.
\ee
Here the first factor in the exponent $B$ is the field-theoretic bounce action already displayed in (\ref{ftb}), while the second factor, $r(x,y)$, characterises the gravitational correction. It reads explicitly
\be
r(x,y)=2\frac{1+xy-\sqrt{1+2xy+x^2}}{x^2(y^2-1)\sqrt{1+2xy+x^2}}\,,\label{rxy}
\ee
where
\be
x=\frac{3 T^2}{4 M_P^2\, \Delta V}\qquad,\qquad y=\frac{V_f+V_t}{\Delta V}\qquad \qquad\mbox{with}\qquad \qquad \Delta V=V_f-V_t\,,
\ee
and $M_P$ is the reduced Planck mass.

This formula was derived for the decay of a false de Sitter vacuum to a true de Sitter vacuum: $V_f\geq V_t>0$. Nothing changes if $V_t$ becomes zero or even negative: It is equally possible to glue a patch of true-vacuum AdS space into a false-vacuum de Sitter sphere. The geometric situation is analogous to that displayed on the r.h.~side of Fig.~\ref{cdl}. Also the derivation of the corresponding exponent $B$ is unchanged.

However, a new and possibly unexpected situation arises if one considers the tunnelling from true to false vacuum de Sitter. Let us first argue why such a process might at all be possible: As a quantum fluctuation, any state can form, even that of a higher-energy vacuum bubble in a lower-energy background. If the lower-energy background is Minkowski, then such a bubble can of course never materialise. It always has positive energy and hence would always violate energy conservation if it were to become a real state. 

By contrast, if the lower-energy vacuum is already de Sitter, then a sufficiently large virtual bubble with false vacuum inside can be pulled to larger and larger size by the background expansion. No energy-conservation argument forbidding such a process can be given in the (time-dependent!) global de Sitter geometry.

The corresponding rate is easily obtained without the need for a new calculation~\cite{Lee:1987qc, SchwartzPerlov:2006hi}. To see this, let us recall the rate for the decay of false to true de Sitter vacuum:
\be
\Gamma_{f\,\to\, t}=\exp(-B_{f\,\to\,t})\qquad
\mbox{with}\qquad B_{f\,\to\,t}=S[\phi_b,g_b]-S[\phi_f,g_f]\,.
\ee
As explained before, the bounce field configuration and geometry underlying $S[\phi_b,g_b]$ corresponds to a patch of true vacuum\index{true vacuum} $\phi_t$ glued into a sphere of false vacuum\index{false vacuum}, with $\phi=\phi_f$. But since both parts of this geometry are cutouts from spheres, it is clearly just a matter of convention which of them we call the `patch' or `bubble'\index{bubble} and which the background. If we invert this interpretation, the same action $S[\phi_b,g_b]$ may serve as a building block for the tunnelling rate from true to false vacuum:
\be
B_{t\,\to\,f}=S[\phi_b,g_b]-S[\phi_t,g_t]\,.
\ee
Crucially, we have now subtracted the true-vacuum rather than the false-vacuum full-sphere action. The subtraction is justified, as before, because this action represents the relevant background geometry. The above may be rewritten as 
\be
B_{t\,\to\,f}=\Big(S[\phi_b,g_b]-S[\phi_f,g_f]\Big)+S[\phi_f,g_f]-S[\phi_t,g_t]
\,.
\ee
Now, using (\ref{bdef}) as the definition of a function $B=B(V_f,V_t,T)$ and employing the expression
\be
S_{edS}(V)=-24\pi^2\frac{M_P^4}{V}
\ee
for the action of euclidean de Sitter space based on a potential $V$, one immediately derives\index{bounce}
\be
B_{t\,\to\,f}=B(V_f,V_t,T)+24\pi^2 M_P^4\left(\frac{1}{V_t}-\frac{1}{V_f}\right)\,.
\ee
Comparing this with our previous result
\be
B_{f\,\to\,t}=B(V_f,V_t,T)
\ee
one sees, not surprisingly, that tunnelling upwards comes at the cost of an extra exponential suppression.

\subsection{Our universe in the eternally inflating landscape}\label{sret}

Naively one might think that we live in one of the many bubbles, and ours just happens to have very small $\lambda$. This is roughly true, but important details are missing. First, given how small our $\lambda$ is, we naturally expect the previous vacuum's $\lambda$ to be much larger. But a corresponding tunnelling event would have endowed our vacuum with a large and negative spatial curvature. Our cosmological evolution would have been governed by the FRW equation
\be
3 H^2=\rho - 3k/a^2\qquad \mbox{with} \qquad k=-1
\ee
and with initial conditions where the curvature term (the 2nd term on the r.h. side) would be at least comparable to the matter term ($\rho$, which includes matter, radiation and $\lambda$) from the start. In such a situation, there can be either $\lambda$ domination or curvature domination succeeded by $\lambda$ domination, but no extended radiation or matter dominated epoch, as in our world. The reason is simply that, with expansion, matter and radiation densities decay faster than curvature.

\begin{figure}[ht]
\begin{center} 
\includegraphics[width=7.5cm]{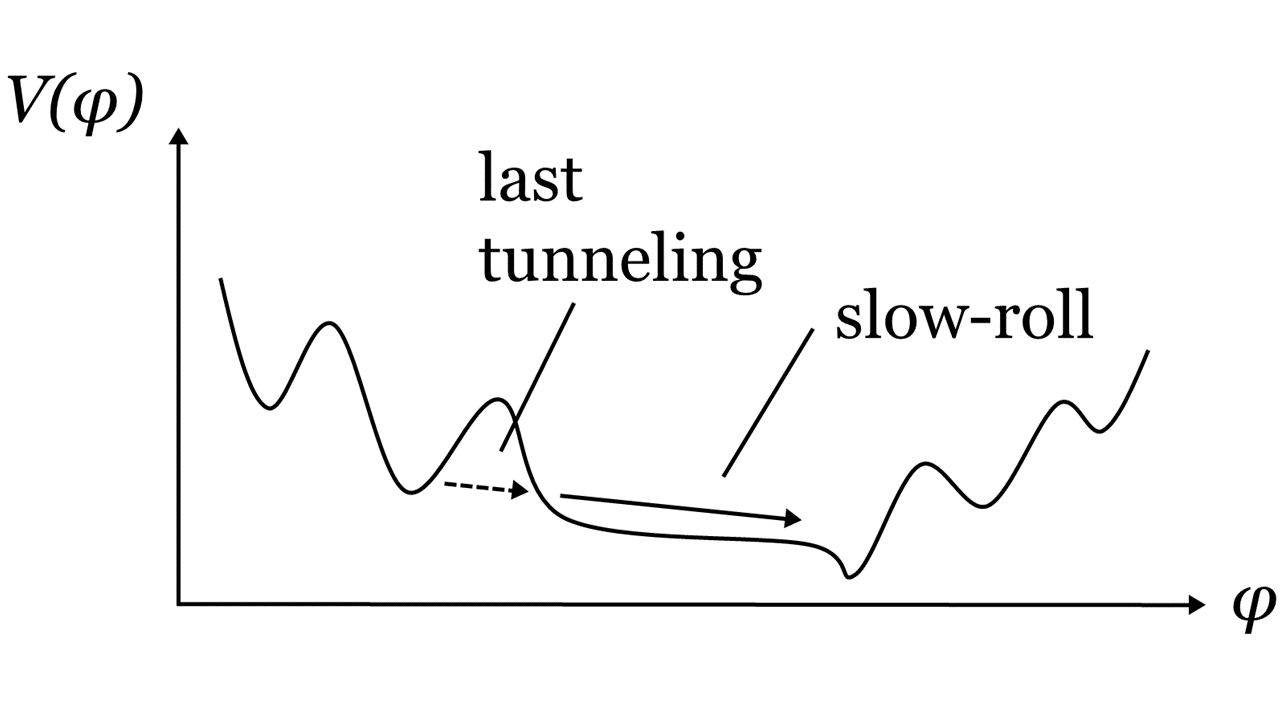}
\caption{Tunnelling to our vacuum, where a period of slow-roll inflation, reheating and structure formation preceed the dS phase.}
\label{slow-roll} 
\end{center}
\end{figure}

\begin{figure}[ht]
\begin{center} 
\includegraphics[width=9cm]{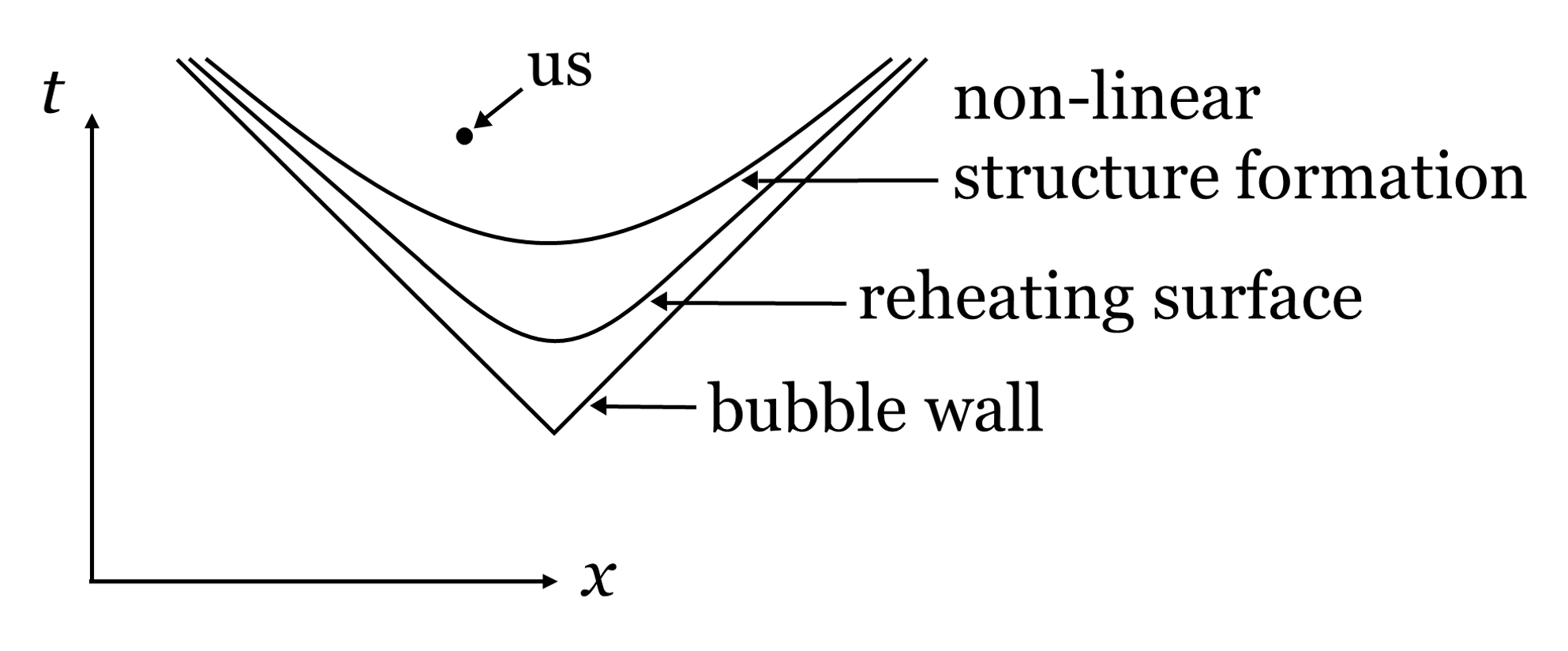}
\caption{Various `surfaces of constant energy density' following the initial tunnelling transition to our bubble\index{bubble}. This sketch is an adaptation of a figure from~\cite{Kleban:2011pg}, which deals with possible bubble collisions and their observational effects.\index{reheating surface}}
\label{reheating} 
\end{center}
\end{figure}

The way out is to postulate that the potential near our local minimum has the peculiar feature of an inflationary plateau, where cosmological inflation took place, cf.~Fig.~\ref{slow-roll}. (It does not actually have to be a plateau -- any sufficiently flat potential region would do.)
Together with making the universe highly homogeneous (i.e.~solving the horizon problem), this inflationary phase dilutes the curvature contribution (i.e.~it solves the flatness problem\index{flatness problem}). Inflation then ends in reheating and structure formation, with galaxies, stars, planets etc. Eventually, as $\lambda$ gets to dominate, the universe becomes empty and cold. (It could also crunch, if $\lambda$ were small and negative, but this is apparently not the case in our world.) Thus, observers of `our kind' actually always appear shortly after the\index{tunnelling transition} tunnelling transition.\footnote{
One 
may,  however, speculate that civilisations can survive {\it much longer} than it takes for galaxies and planetary systems to decay and stars to burn out~\cite{Dyson:1979zz}.
}
Nevertheless, in a given bubble of `our vacuum' their number is infinite, as are our reheating surface and our structure formation surface (see Fig.~\ref{reheating}). These surfaces (including presumably the surface of death of all stars and hence of all civilisations) follow the straight bubble wall surface all the way up to infinity. In the figure, these surfaces approach the bubble wall both to the left and to the right -- in reality they approach the light-cone in $\mathbb{R}^{1,3}$ which is defined by the bubble wall. (Of course, that could change if a collision with another bubble occurred.) One may say that the interior of any bubble is an open (infinite) Friedmann-Robertson-Walker universe\index{Friedmann-Robertson-Walker universe}. For many more details and references see e.g.~\cite{Freivogel:2005vv}.

\subsection{Making statistical predictions and the measure problem}
\index{measure problem}

Accepting the above landscape picture and eternal inflation as the process populating it, the measure problem is easy to state, at least at an intuitive level (see e.g.~the reviews~\cite{Vilenkin:2006xv ,Freivogel:2011eg, Schellekens:2013bpa}): We live in one of the vacua, but we do not know in which one. We would like to make a statistical prediction (given that we know certain features of our vacuum, but not all). Let us say the new observable which we are going to measure tomorrow can take the values $A$ or $B$. The most naive way to make a statistical prediction would be to say that the ratio of probabilities is
\be
p_A/p_B=N_A/N_B\,.
\ee
Here $N_{A/B}$ are the numbers of observers in the multiverse\index{multiverse} who have measured all that we have measured so far and who will, in the next measurement, find $A$ or $B$ respectively. But in eternal inflation, by definition, both numbers are infinite and their ratio is not well-defined. What is worse, if one cuts off the infinity in the future, the prediction becomes dependent on the precise type of cutoff. For example, one could restrict attention to measurements before some maximal time $t_{max}$, taking the limit $t_{max}\to \infty$ in the end. But such a maximal-time cutoff\index{time cutoff}, illustrated in Fig.~\ref{ds}, is not unique. This is due to the absence of a unique global time variable in de Sitter space or, more precisely, in the more complicated geometry of multiple de Sitter bubbles as they arise in eternal inflation. More generally speaking, the diffeomorphism invariance of general relativity prevents the existence of an unambiguously defined time cutoff.

Before discussing the various suggestions for how the measure problem might be overcome, one should note one pragmatic and historically very successful approach: One may assume that we are likely to observe what is common in the string theory landscape, independently of the dynamics populating the latter. More precisely, this amounts to the assumption that any bias in favor or against a certain observational feature induced by cosmology is small compared to the bias derived simply from counting vacua. The latter is, at least in principle, possible since to the best of our present understanding the landscape is finite (at least if one imposes a certain IR cutoff, excluding models with an arbitrarily low KK scale, but maybe even more generally) \cite{Acharya:2006zw}.

The historic success of this approach is Weinberg's prediction of the size of the observed cosmological constant \cite{Weinberg:1987dv} (see also~\cite{Linde:1984ir}). Crucially, the argument comes from a period when cosmology was well enough understood to provide a firm upper bound on our present expansion rate $H$. Yet, the time evolution of $H$ was not known precisely enough to determine whether this non-zero $H$ was predominantly due to matter, a cosmological constant or spatial curvature,
\be
3H^2=\rho_{matter}+\lambda-3k/a^2\,.
\ee
In other words, it was known that $|\lambda|\leq \lambda_0$ with some positive $\lambda_0$ which was very small compared to particle physics scales. Now, let us assume that we live in one universe in a multiverse with many different $\lambda$. Moreover, assume that these available values of $\lambda$ may be described by a statistical distribution which is smooth and dominated by large energy scales (like the Planck, the string, or the SUSY breaking scale). If the point $\lambda=0$ plays no special role, then one expects that projecting this distribution to the tiny interval $(-\lambda_0,\lambda_0)$
gives essentially a constant distribution on that interval. But this projection was exactly what observations at that time had achieved. Thus, the prediction for a future measurement of $\lambda$ was to be made using a constant distribution on the interval $(-\lambda_0,\lambda_0)$. This corresponds to simply drawing a $\lambda$ value from that interval. With overwhelming probability, the result should be a value (of either sign) comparable in magnitude to $\lambda_0$. A much smaller (absolute) value would be very unlikely. Famously, a non-zero $\lambda\sim \lambda_0$ was discovered only a few years later.

We should note that a closely related but stronger and more debatable argument predicting $\lambda$ can be made. Namely, fixing all other particle-physics and cosmological parameters (including in particular the initial curvature perturbations which have led to the formation of structure, including stars and planets), one may argue for a so-called {\bf anthropic} prediction\index{anthropic prediction} of $\lambda$. Indeed, if $\lambda$ were much larger than $\lambda_0$, exponential expansion would have set in earlier in the history of the universe, preventing the formation of any structure and hence of life. By contrast, too large negative $\lambda$ would have led to a big crunch before any observers could have emerged. Thus, the observed value of $\lambda$ can be said to represent an anthropic prediction (given the previously made assumption about the statistical distribution in the landscape), independently of the observational status at the moment of Weinberg's famous paper. In fact, Weinberg's paper emphasises this anthropic prediction rather than the one based on the observational situation, which we explained before.

We should emphasise that such anthropic arguments based on some form of multiverse are much older than the string theory landscape. Moreover, they can be applied to quantities other than the cosmological constant or electroweak scale. The reader may want to explore this line of thinking starting, e.g., with \cite{bt, Hogan:1999wh, Tegmark:2005dy, Hall:2007ja} and refs.~therein.

\subsection{Proposed measures}
This section draws very significantly on the relatively recent and very clear review~\cite{Freivogel:2011eg}. Following this analysis, one distinguishes global and local measures.\index{measure!local}\index{measure!global} The former count observers before some late cutoff time $t_{max}$ (or in some other way that keeps the total number of observers\index{observer} finite) and takes the cutoff to infinity in the end. By contrast, local measures count observers in  a way associated with a single timelike geodesic, in the simplest case following a single observer on their path through the tunnelling events of the multiverse.

Some of the oldest measures are global. For example, one may start with some spacelike surface and define time globally using geodesics originating in its every point. The initial surface should be finite, but this is not a problem: Global de Sitter space has the topology of `Time'$\times S^3$, so a spacelike cut can provide the required surface.

The most obvious choice is to use proper time~\cite{Linde:1993nz}. But this is claimed to be ruled out by observation on account of the `Youngness paradox'\index{Youngness paradox}~\cite{Linde:1993xx}: Think of us living inside a bubble\footnote{
To 
be precise, near the initial bubble wall of a small-$\lambda$ bubble, as explained in Sect.~\ref{sret}
}
which nucleated inside some high-$\lambda$ vacuum. Due to the very fast exponential expansion of this mother vacuum, and since there are bubbles of our type with all kinds of ages present, we are most likely to find ourselves to be as young as only physically possible. In other words, we should be the youngest observer on the youngest planet in the youngest galaxy. This appears not to be consistent with the place we occupy in our universe.

Another fairly obvious choice, one that apparently is not ruled out yet, is to use scaling time (i.e. a scale factor cutoff). In other words, one measures time (and hence introduces  a cutoff) on the basis of the number of e-foldings~\cite{Linde:1993nz}.

Finally, there is lightcone time and correspondingly a lightcone time cutoff~\cite{Garriga:2005av, Bousso:2009dm}. Here, given a point in the eternally inflating spacetime, one follows its lightcone to the future boundary. The part of the boundary inside this lightcone is then projected back to some initial surface (using the same congruence of geodesics discussed earlier). The size of this projection defines the time of the point.

\begin{figure}[ht]
\begin{center} 
\includegraphics[width=11cm]{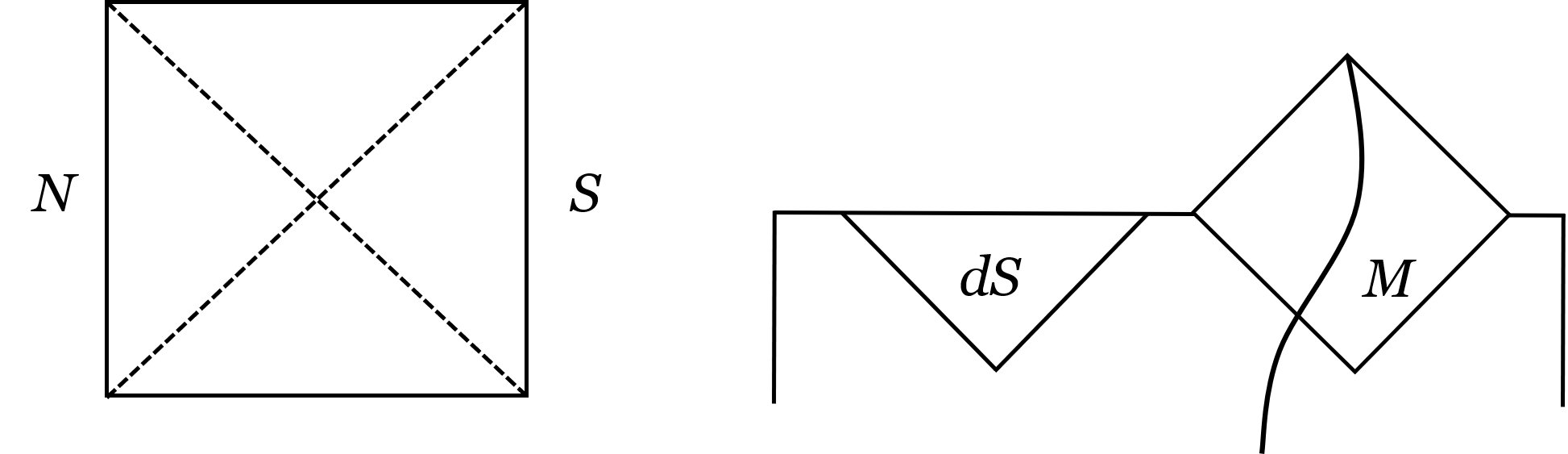}
\caption{On the left: Penrose diagram of de Sitter space\index{de Sitter space!Penrose diagram}. Here the spatial $S^3$ is represented as an $S^2$ fibered over an interval (the horizontal axis of the square). North and south pole are labelled by N and S and the horizons of corresponding observers are indicated as dashed lines. On the right: Upper portion of the same diagram with a de-Sitter bubble (`dS') and a Minkowski bubble (`M') added. A trajectory of an observer\index{observer!trajectory}\index{observer!worldline} ending up in the Minkowski bubble is also shown.}
\label{pdds} 
\end{center}
\end{figure}

Concerning local measures, one option is to count observers inside the causal patch (the so-called causal diamond) of a timelike geodesic. This is not automatically finite and, if it is, the late-time attractor behaviour of eternal inflation is lost. A variant without these problems is the `census taker's cutoff'\index{census taker}, which focuses on geodesics that end in a Minkowski vacuum~\cite{ctc, Susskind:2007pv}. Such geodesics are clearly infinite, but one may restrict attention to events the future lightcone of which crosses the central geodesic before some time $t$. In other words, one may count all events inside the causal diamond with apex at $t$, eventually taking the limit $t\to\infty$. The reader may consult Fig.~\ref{pdds} for an illustration.

Other local measures count observers in a cylinder of finite physical radius centered on a timelike geodesic (`fat geodesic measure' \cite{Bousso:2008hz}) or within its apparent horizon (`apparent horizon measure' \cite{Bousso:2010im}).

Many more possibilities and variants of the choices above are discussed in~\cite{Freivogel:2011eg}. At first sight, such a proliferation of measures looks rather discouraging. However, things are not quite as bad as they seem: It turns out that differently defined measures may be equivalent in the sense that they give the same (or in some cases very similar) results. In particular, there exist equivalences (referred to as `dualities') between certain local and global measures. As a result of this, one may basically focus on the following three options: \\[.2cm]
(1) Lightcone-time cutoff\index{cutoff!lightcone-time} -- equivalent to the causal diamond measure\index{measure!causal diamond};\\ 
(2) Scale factor cutoff\index{cutoff!scale factor} -- equivalent to the fat geodesic measure\index{measure!fat geodesic};\\
(3) Apparent horizon cutoff\index{cutoff!apparent horizon}.

A different approach is suggested in~\cite{Garriga:2005av}. Here, one determines the abundances of bubbles of different kinds by projecting them on some initial surface (using a congruence of timelike geodesics). One only counts bubbles that are larger than $\epsilon$, taking the limit $\epsilon\to 0$ in the end. Inside each bubble, one has an infinite open FRW universe, as explained above. Thus, if a certain type of bubble allows for observers, then the number of observers inside each specific bubble will always be infinite. These latter infinities are not used to further weight the relative probabilities between the different bubbles. As a result, this proposal represents a rather radical but logically consistent deviation from all the previous discussions (where the focus was on counting the number of observers in one way or another).

Let us end with an interesting argument against any measure based on some geometric cutoff. It is known as the {\bf Guth-Vanchurin paradox}\index{Guth-Vanchurin paradox}~\cite{gv} and goes as follows: Imagine an eternal de Sitter space in which observers are born at random points in spacetime. Each observer sets an alarm clock which will, with 50\% probability, let them sleep for a short or a long period of time. Then they go to bed. After waking up and before checking their watch, they ask themselves what the statistical prediction is that they slept for a short or a long time. The intuitive answer is clearly 50\%. However, if they look around themselves they will (given an appropriate choice of model parameters) clearly see many more short sleepers than long sleepers. The reason is that the latter tend to have left the horizon. The claim and the paradox is that this (apparently wrong) prediction will be shared by any geometric-cutoff measure.\index{measure!geometric-cutoff} Indeed, taking the prediction of such measures more seriously than the intuitive answer one can even arrive at the highly counterintuitive conclusion that `time will end'~\cite{Bousso:2010yn}. Such a sudden end of time resolves the issue by providing an objective reason for why it is more likely, after waking up, to find out that one has only slept a short time: For the long sleepers, the probability is higher that they run into the spacelike singularity during their nap.

\subsection{Predictions from first principles?}\label{pfp}

It is conceivable that further scrutiny of the proposed measures will show that all but one of them are either in some way inconsistent or are in conflict with observations. As a result, one would then find the one, correct measure for making statistical predictions in the string multiverse. However, there is no guarantee that this will happen. On the contrary, it is conceivable that over time even more different, consistent measures with varying predictions will be found. In either case, it appears highly unsatisfactory that an extra input in the form of a measure choice has to be added to the hopefully unique quantum gravity theory from which the multiverse follows. It would be desirable to derive a measure from first principles rather than postulate it.

Part of the problem for achieving this is, of course, that gravity in general (even apart from its notorious UV problems) and de Sitter space in particular lack a satisfactory quantum-mechanical understanding. One angle of attack concerning the quantum mechanics of de Sitter space is the idea of a so-called {\bf dS/CFT correspondence}\index{dS/CFT correspondence} \cite{Strominger:2001pn}. 

Before explaining this, it is unavoidable to devote at least a few words to the {\bf AdS/CFT correspondence}\index{AdS/CFT correspondence}, which is much better understood and established~\cite{Maldacena:1997re, Aharony:1999ti}. We have to restrict ourselves to stating the facts: First, $(d\!+\!1)$-dimensional Anti-de-Sitter space (homogeneous, negatively-curved space with Lorentzian signature) has the topology of a $B_d\times \mathbb{R}$. Here $B_d$ is a $d$-dimensional ball, with the boundary being at infinite distance. The real axis $\mathbb{R}$ represents time. The curvature of this space has the effect that, very roughly speaking, the centre of this `solid cylinder' is at lowest gravitational potential. In other words, objects tend to fall from the boundary towards this centre.

The boundary of AdS is clearly the `cylinder' $S^{d-1}\times \mathbb{R}$. AdS/CFT correspondence means that any consistent gravitational theory in the $(d\!+\!1)$-dimensional bulk has an equivalent description (is dual to) a $d$-dimensional CFT living on the boundary. Very roughly speaking, degrees of freedom of the CFT can be ordered according to their energy scale $\mu$. This scale parameter $\mu$ corresponds to the radial direction of the `solid cylinder' representing the bulk. Both in the CFT and in the bulk the path towards the UV (i.e. to towards the boundary) is infinite. But in the IR the CFT scale $\mu$ is restricted by the compactness of its spatial volume. The corresponding degrees of freedom of lowest energy scale are mapped to the bulk degrees of freedom near the centre, i.e.~at highest red-shift.

Now let us turn to the analogous logic in dS/CFT. First, we need to recall that $(d\!+\!1)$-dimensional dS space has the topology of $S^d\times \mathbb{R}$. This is very different from AdS since, in particular, spatial sections of this space are now simply spheres and as such have no boundary. The only boundaries are now those at past and future infinity. In contrast to AdS/CFT, these boundaries are spacelike. The idea of dS/CFT is to map the dynamics of the bulk to a $d$-dimensional (euclidean) CFT that lives on the future boundary $S^d$. This time, the energy scale parameter $\mu$ of the CFT is expected to correspond to the time evolution parameter of the dS bulk. The reader should be warned that, while AdS/CFT has become one of the cornerstones of modern field-theory and gravitational research, dS/CFT remains truly conjectural.

After these lengthy preliminaries it is easy to state what a dS/CFT-based first-principles approach to the measure problem might look like \cite{Garriga:2008ks}: One has to explicitly map the bulk of the eternally inflating dS spacetime to the boundary at future infinity and hence to the CFT. If one now introduces a UV cutoff in the CFT, this may correspond to a canonical or natural late-time cutoff in the bulk. One may hope that this gives rise to an unambiguous way of counting observers and hence to an a-priori definition of a measure. For an interesting toy model of the landscape that may be related to the dS/CFT measure proposal see \cite{Harlow:2011az}.
 
Another approach, suggested as an unambiguous first-principles definition of a measure in~\cite{Nomura:2011dt}, is the following: One considers the world from the perspective of a single, abstract observer. This observer sees various tunnelling events, `lives' through many big-bang cosmologies like our own, until he or she eventually ends up in a terminal vacuum. The sequence of events which the observer witnesses is subject to the usual uncertainty of quantum mechanics. In other words, the life of this observer is a superposition of all the possible sequences of tunnelling events. Thus, in a sense, the `many worlds' of Everett are identified with the many worlds of the multiverse (see also \cite{Bousso:2011up} for a related discussion). One now defines the statistical prediction for any observable using the quantum-mechanical expectation values for this observable on the basis of the {\bf single-observer worldline}\index{observer!worldline}\index{observer!trajectory} introduced above.

The approach just presented bears a certain similarity to the fat geodesic measure \cite{Bousso:2008hz} and its quantum version (the `quantum watcher measure') as discussed in~\cite{Vilenkin:2013loa}. It may moreover be problematic that terminal vacua, the quantum-mechanical significance of which is not understood, play a central role in the single-observer approach just introduced. Thus, it is probably fair to say that no consensus has so far been reached on whether dS/CFT correspondence or the single-observer perspective or some different approach are the leading candidate for a first-principles measure.\footnote{
We also note that the presentation of the single-observer approach in \cite{Nomura:2012zb} bears some formal similarity to the Wheeler-DeWitt equation (see below) and may hence be related to the approach to be discussed next.}

Yet another perspective, which has a strong aesthetic appeal but is probably not developed enough to be called a measure proposal, has been introduced in \cite{Hartle:2016tpo}: The idea is also to focus on a single observer, but in a way very different from counting events along the observer's worldline. Instead, one appeals to the concept of a {\bf wave function of the universe}\index{wave function of the universe} \cite{DeWitt:1967yk, Wheeler:1988zr, Hartle:1983ai}. To explain this, we have to introduce this possibly unfamiliar concept:

Let us consider the canonical quantisation of Einstein gravity, for simplicity in a spacetime $M\times \mathbb{R}$, where $M$ is a compact spacelike manifold. Now, recall that the hamiltonian $H$ generates time translations   $t \to t + \epsilon$. From the perspective of general relativity, this is just a particular diffeomorphism. However,  diffeomorphisms are gauged. Hence an operator like $H$ generating one of them must vanish on any physical state (since physical states are by definition gauge-invariant). Thus, there is no time evolution but rather a so called constraint equation,
\be
H\,\Psi=0\,.
\ee
In this so-called {\bf Wheeler-DeWitt equation}\index{Wheeler-DeWitt equation} $\Psi$ is a functional on the space of metrics on $M$:
\be
\Psi:\,M_g\,\,\to \Psi[M_g]\in\mathbbm{C}\,.
\ee
Here $M_g$ stands for the manifold $M$ with metric $g$. Hence $\Psi$ can be viewed as the gravitational analogue of the Schr\"odinger wave function of quantum mechanics or, better, of the Schr\"odinger wave functional\index{Schr\"odinger wave functional} of quantum field theory. Its physical role differs from the latter in that it does not evolve in time. Instead, it has to be interpreted as the wave function of the universe in the sense that it contains the probability for observing some relevant 3-manifold at {\it any} moment in time (which can not be in general defined). Note that this is not incompatible with conventional time dependent physics \cite{Banks:1984cw}: One may, for example, enrich the argument of $\Psi$ by non-gravitational fields, $\Psi[M_g]\,\to\,\Psi[M_g,\phi]$. Then one may consider physical situations with a clock (made of fields $\phi$) and ask for the probability to observe a given metric $g$ at a given time, encoded in the field configuration $\phi$. Standard physical questions about the occurrence of an event at a given time are hence encoded in questions about the correlation between values of $g$ and values of $\phi$.

Given these preliminaries, the suggestion of \cite{Hartle:2016tpo} may now be roughly formulated as follows: One should not ask about possible, approximately  classical histories of the universe as encoded in $\Psi$ and try to count observers which make a certain observation. Instead, one should adopt a coarse-grained perspective in which one ignores all information in $\Psi$ except that an observer makes, say, observation $A$ or that an identical observer makes observation $B$. The relative probability of these two observations should be encoded in a hopefully well-defined and finite form in $\Psi$. Crucially, in asking this question one ignores any irrelevant information about where in the multiverse the observation occurs and which of the many observers making identical observations one is considering. The hope is that this coarse-graining step would make the answer well-defined. As an alternative, it has more recently been suggested \cite{Halliwell:2018ejl} to implement the idea of coarse graining using specifically the so-called Hawking-Hartle no-boundary proposal\index{Hawking-Hartle no-boundary proposal} \cite{Hartle:1983ai} for the definition of the wave function of the universe. It has been argued that, as a result, the sensitivity to the exponentially large multiverse disappears if one asks the right questions.

In summary, it should have become clear that, while a first-principles definition of the measure is highly desirable, this subject is not settled. It is also conceivable that the measure has indeed to be viewed as fundamental new input, in addition to whatever the ultimate first-principles definition of string or M-theory will turn out to be. Furthermore, it is even possible that no unique and correct measure exists and that statistical predictions in the eternally-inflating  multiverse will remain impossible as a matter of principle. A justification of this pessimistic attitude might be the impossibility of repeating an experiment (such as the measurement of $\lambda$) many times. Hence, the usual physicist's definition of probabilities and of a probabilistic prediction does not work. On the contrary, one may also defend the possibility of a probabilistic prediction for a single observation as follows: View the 10-fold repetition of an experiment, each of which rules out a certain theory at 99\% confidence level, as one single experiment or observation. Such a viewpoint is clearly a matter of convention and is logically perfectly acceptable. Now, this `10-fold' measurement (assuming all 10 results agree) leaves only a probability of $10^{-20}$ that the theory is correct. We would clearly and rightfully dismiss such a theory and, as just argued, we do so on the basis of a single experiment. All that matters is a sufficiently high significance of the result.

While these are all very interesting and potentially important questions, we have now entered a field where very little is known with certainty. So we should maybe stop here, leaving it to the reader to explore recent, original papers and form their own opinion.

\subsection{Problems}
\subsubsection{Coleman-De-Luccia tunnelling}\label{cdlp}
\index{Coleman-De Luccia}

{\bf Task:} Fill in the calculational details leading to the expression for the exponent $B$ in the decay-rate $\Gamma\sim \exp(-B)$ of a false de Sitter vacuum that was given in Sect.~\ref{cdls}.

\noindent
{\bf Hints:} Use the Einstein equation to bring the action to a form without $r$-derivatives. Assume that the domain wall, i.e.~the interval in $f$ where most of the change of $\phi$ occurs, is defined by $f_{dw}<f<f_{dw}+\Delta$. Moreover, assume that the thin-wall approximation, $\Delta\ll f$, is valid. Then evaluate $B=S[\phi_b,g_b]-S[\phi_f,g_f]$ in three pieces: outside the  domain wall, in the domain wall region, and inside the domain wall. Since the resulting expression is, by assumption, extremal, the correct value of $f_{dw}$ can be determined by extremising $B$ w.r.t.~$f_{dw}$. If you get stuck, consult the original papers \cite{Coleman:1980aw, Parke:1982pm, Johnson:2007jla}.

\noindent
{\bf Solution:}
The desired form of the action is
\be
S = 4\pi^2 \int f(r)^3 dr\,\left[-\frac{3M_P^2}{f(r)^2}+V(\phi(r))\right]\,.
\ee
Based on this general expression, we now have to evaluate
\be
S[\phi_b,g_b]-S[\phi_f,g_f]\,.
\ee
We split the $r$-integration in three pieces. First, outside the domain wall the integrands in $S[\phi_b,g_b]$ and $S[\phi_f,g_f]$ are identical. So there is no contribution from that region. 

Second, in the domain-wall region the contribution from the $1/f(r)^2$-term can be neglected. The reason is that this region is small, $\Delta\ll f_{dw}$. As one enters the domain wall region from the outside, the function $f(r)$, which encodes the euclidean 4d geometry, only starts to distinguish between the bounce and the $S^4$ false-vacuum solution. The effect is second order in the small quantity $\Delta$ and can be neglected. Note that, by contrast, the contribution from the $V(\phi)$ term is not small. The reason is that $V$ has to change by a fixed amount as one passes the wall. Hence its variation grows as $\Delta$ is taken to zero. We then have
\bea
S[\phi_b,g_b]-S[\phi_f,g_f]\Bigg|_{domain\,\,wall} 
&\simeq&
4\pi^2 \int_{r_{dw}}^{r_{dw}+\delta} f(r)^3 dr\,\left[V(\phi_b(r))-V(\phi_f)\right]
\\
&\simeq&
4\pi^2 f(r_{dw})^3 \int_{r_{dw}}^{r_{dw} +\delta} dr\,\left[V(\phi_b(r))-V_f\right]\,.
\eea
Here the integral is over the interval in $r$ corresponding to the domain wall, defined by $f(r_{dw})=f_{dw}$ and $f(r_{dw}+\delta)=f_{dw}+\Delta$. Let us compare the above with the domain wall tension, which is by definition the sum of gradient and potential energy in the wall:
\be
T\simeq \int_{r_{dw}}^{r_{dw}+\delta} dr\,\left\{\frac{1}{2}\phi_b'^2+[V(\phi_b(r))-V_f]\right\}\,.\label{dwt}
\ee
Concerning the field profile in the wall, gravitational effects are subleading and we can appeal to our understanding of a field-theoretic domain wall or bubble wall\index{bubble!wall} from Sect.~\ref{tft}. The only difference is that, in (\ref{dwt}), we have subtracted the false-vacuum potential energy $V_f$, a quantity that was set to zero by definition in our previous field-theoretic analysis. Moreover, the dimensions parallel to the wall are irrelevant, such that we can go back even further and think of the euclidean tunnelling solution in quantum mechanics as discussed in Sect.~\ref{tqm} and visualised in Fig.~\ref{rolling}. There, we learned that gradient and potential energy are always equal during the transition between the two minima (since we can think of a dynamical rolling process in the inverted potential). As a result,
\be
S[\phi_b,g_b]-S[\phi_f,g_f]\Bigg|_{domain\,\,wall}\simeq 2\pi^2 f_{dw}^3\, T\,.
\ee

Third, inside the domain wall the field $\phi$ is constant and the Einstein equation (\ref{rree}) can be used to change the integration variable from $r$ to $f$:
\be
\frac{df}{dr}=\sqrt{1-\frac{Vf^2}{3M_P^2}}\,.
\ee
Thus
\bea
S[\phi_b,g_b]-S[\phi_f,g_f]\Bigg|_{inside} &\simeq &
-12\pi^2M_P^2\int_0^{f_{dw}}f\,df\,\left(1-\frac{V_t f^2}{3M_P^2}\right)^{1/2}\,\,-\,\,\Bigg\{ V_t\to V_f\Bigg\}
\nonumber\\
\nonumber\\
&=& \frac{12\pi^2M_P^4}{V_t}\left[\left(1-\frac{V_t f_{dw}^2}{3M_P^2}\right)^{3/2}-1\right] \,\,-\,\,\Bigg\{ V_t\to V_f\Bigg\}\,.
\eea

Recall that $f_{dw}$ is nothing but the physical radius of the ball of final-state vacuum. So let us somewhat simplify notation by writing $R$ instead of $f_{dw}$. Then, combining the contributions from the wall and the inside region, we have
\be
\frac{B(R)}{2\pi^2}=TR^3\,+\,\frac{6M_P^4}{V_t}\left[\left(1-\frac{V_t R^2}{3M_P^2}\right)^{3/2}-1\right] \,\,-\,\,\Bigg\{ V_t\to V_f\Bigg\}\,.\label{bbrb}
\ee
The algebra involved in solving $B(R)'=0$ for $R$ and inserting the result in (\ref{bbrb}) is less horrible than one might expect at first sight. Indeed, the equation $0=B'(R)$ may be brought to the form
\be
0=\zeta+\sqrt{1-a\zeta^2}-\sqrt{1-b\zeta^2}\qquad \mbox{with} \qquad \zeta=RT/2M_P^2\,,
\ee
and 
\be
a=\frac{4M_P^2}{3T^2}V_f\quad,\quad
b=\frac{4M_P^2}{3T^2}V_t\,.
\ee
Dividing this by $\zeta$ and rewriting the result in terms of $w=1/\zeta^2$ one finds 
\be
0=1+\sqrt{w-a}-\sqrt{w-b}\,,
\ee
which is easy to solve for $w$:
\be
w=\frac{1}{4}\big[1+2(a+b)+(a-b)^2\big]\,.
\ee

Next, we write the exponent $B$ as
\bea
\frac{B(R)}{2\pi^2}&=&\frac{8M_P^6}{T^2}\left[
\zeta^3-\frac{1}{a}\Big((1-a\zeta^2)^{3/2}-1\Big)+\Big\{
a\to b\Big\}
\right]
\\
&=& \frac{8M_P^6}{T^2}\left[\zeta^3-\frac{1}{a}
\Big(\zeta^3(w-a)^{3/2}-1\Big)+\Big\{
a\to b\Big\}\right]\,.
\eea
This simplifies upon noticing that (for $a-b>1$)
\be
\sqrt{w-a}\,=\,-\frac{1}{2}[1-(a-b)]\qquad \mbox{and}\qquad 
\sqrt{w-b}\,=\,\frac{1}{2}[1+(a-b)]\,.
\ee
One finds
\be 
\frac{B(R)}{2\pi^2} = \frac{8M_P^6}{T^2}\left[\frac{\zeta^3}{8ab}\Big\{8ab+\big[1+(a-b)\big]^3a+\big[1-(a-b)\big]^3b\Big\}+\frac{1}{a}-\frac{1}{b}\right]\,.\label{res1}
\ee
At this point, it is useful to spell out the relation to the variables $x,y$ introduced in the main text:
\be
a-b\,=\,\frac{1}{x}\,\,,\quad a+b\,=\,\frac{y}{x}\,\,,\quad
\zeta=\frac{2x}{\sqrt{1+2xy+x^2}}\,\,,\quad
ab=\frac{y^2-1}{4x^2}\,.
\ee
Moreover, we have $\Delta V=3T^2/4xM_P^2$, such the formula for $B$ from Sect.~\ref{cdls} (based on~\cite{Parke:1982pm}) takes the form
\be
\frac{B(R)}{2\pi^2} = \frac{8 M_P^6}{T^2}\,2x^3\,r(x,y)\,.
\label{res2}
\ee
Now the agreement between (\ref{res1}) and (\ref{res2}) follows more or less immediately.

\section{Concluding remarks and some alternative perspectives}\label{alt}
This last section is special in that it contains few equations and no exercises. It is to a large extent a brief tour through additional topics that could and maybe should have been covered but had to be left out for reasons of space or, more precisely, because they will most likely not fit in the time frame of a one-semester course. Moreover, part of the comments and ideas collected in this section lead away from the specific `string landscape perspective' that was advertised and taught throughout the core part of this course.

\subsection{Low-scale SUSY versus Technicolor}
\label{lsstc}\index{low-scale SUSY}\index{Technicolor}

Let us start with a brief discussion of alternatives to low-scale SUSY. Recall that {\bf low-scale SUSY} has played a central role in our course since it demonstrates that, in principle, gauged (or otherwise interacting) scalars can be `naturally' light. This could have been a perfect explanation for the large hierarchy between the quantum gravity scale and the electroweak scale. Maybe it still largely explains this hierarchy, but (in part because of the non-discovery of SUSY at the LHC) this does not work perfectly. It may still work partially and SUSY would then have to be `just around the corner' in the sense of energy scales.

However, roughly the same may be achieved by other means. The arguably main and historically first candidate is known as {\bf Technicolor} \cite{Weinberg:1975gm, Susskind:1978ms, Dimopoulos:1979es, Eichten:1979ah, Kaplan:1983fs, Bardeen:1989ds} (for reviews see e.g.~\cite{King:1994yr, Chivukula:1998if, Lane:2002wv, Hill:2002ap, Piai:2010ma, Cacciapaglia:2020kgq}). The term refers to a second version of `colour' (as in the QCD sector of the Standard Model), which is added to the Standard Model gauge theories for a purely `technical' reason. This reason is the creation of a technically natural light scalar.

To understand this, recall how pions arise in the low-energy EFT of the Standard Model (see e.g.~\cite{Cheng:1985bj, Peskin:1995ev, Weinberg:1995mt, Donoghue:1992dd, Kaplan:2005es}): If one neglects the small Yukawa couplings of $u$ and $d$ quarks, one can write the relevant lagrangian terms as 
\be
{\cal L}\,\,\supset\,\, i\ol{q}_L^T\ol{\sigma}^\mu D_\mu q_L \,+\, i\ol{q}_R^T\ol{\sigma}^\mu D_\mu q_R\,.
\label{udq}
\ee
Here we have combined the pairs of Weyl fermions $u_L/u_R$ and $d_L/d_R$ (which make up the up and down quark Dirac fermions $u$ and $d$) into doublets
\be
q_L=\left(\begin{array}{c} u_L \\ d_L \end{array} \right) \qquad \mbox{and} \qquad 
q_R=\left(\begin{array}{c} u_R \\ d_R \end{array} \right)\,.
\ee
At  energies above the weak scale, the $SU(2)$ acting on the first of these doublets is the familiar $SU(2)_L$ gauge symmetry. But this is irrelevant for now since we are at low energies. We also  disregard $U(1)_{em}$ because of its small coupling. Thus, the covariant derivatives in \eqref{udq} refer solely to the $SU(3)$ colour group.

Our lagrangian has, in addition to its $SU(3)$ gauge symmetry, a global $SU(2)_L\times SU(2)_R$ symmetry:
\be
(q_L)_i\,\to\, (U_L)_i{}^j(q_L)_j\quad,\qquad
(q_R)_i\,\to\, (U_R)_i{}^j(q_R)_j\,.
\ee
Due to the strong non-perturbative effects associated with $SU(3)$ gauge dynamics, it is conceivable that a non-zero vacuum expectation value of fermion bilinears is induced. Assuming that it respects the gauge symmetry and recalling that $q_L$ and $q_R$ transform as ${\bf 3}$ and ${\bf \ol{3}}$ under $SU(3)$, the only option is
\be
\langle (q_L)_i (q_R)_j\rangle \neq 0\,.
\ee
Here the contracted color indices have not been displayed.
Using a biunitary transformation, this expectation value can be made diagonal. Moreover, for symmetry reasons one expects the two eigenvalues to be equal, such that
\be
\langle (q_L)_i (q_R)_j\rangle =\lambda\,\delta_{ij}\,.
\label{qlqr}
\ee
This breaks $SU(2)_L\times SU(2)_R$ spontaneously to its diagonal subgroup, consisting of elements of the form $(U,U)\in SU(2)_L\times SU(2)_R$. The generators outside the Lie algebra of this subgroup act on the vacuum in a non-trivial way, but they clearly can not change the energy. Thus, we find a vacuum manifold, which can for example be parameterised by group elements of the form
\be
(\Sigma,\mathbbm{1})\in SU(2)_L\times SU(2)_R
\ee
acting on \eqref{qlqr}:
\be
\langle (q_L)_i (q_R)_j\rangle\quad\to\quad
\Sigma_i{}^k\langle (q_L)_k (q_R)_j\rangle =\lambda\,\Sigma_i{}^k\delta_{kj}\,.
\ee

The condensation effect above is known as chiral symmetry breaking\index{chiral symmetry breaking}. While it has not been rigorously derived from the QCD lagrangian, it is considered as well-established. This is based both on lattice studies and on the phenomenological success of calculations using the above assumptions (chiral perturbation theory).

Now one writes
\be
\Sigma=\exp(i\pi^a\sigma_a/f_\pi) \label{pide}
\ee
and identifies $\pi^a$ as the pion\index{pion} fields. The effective lagrangian for these massless fields, which are clearly Goldstone bosons of the spontaneously broken global symmetry, has to be invariant under
\be
\Sigma\quad\to\quad U_L\,\Sigma\, U_R^\dagger\,.
\ee
The lowest-dimensional such term is 
\be
{\cal L}\,\supset\,\frac{f_\pi^2}{4}\,\mbox{tr}\,[(\partial_\mu \Sigma) (\partial^\mu \Sigma)]\,=-\frac{1}{2}\delta_{ab}(\partial_\mu \pi^a)(\partial^\mu\pi^b)+\cdots\,, \label{pil}
\ee
where the normalisation of $\pi^a$ implicit in \eqref{pide} has been chosen such that the kinetic term is canonical.
The quantity $f_\pi$ is called pion decay constant. This name can be easily understood by promoting the partial derivatives in \eqref{pil} to $SU(2)_L$-covariant derivatives and calculating the mixing of the pions with the weak bosons and hence their decay rate (e.g.~$\pi^-\to W^-\to \mu^-\ol{\nu}_\mu$).

The fact that our real-world pions are massive is due to the non-zero fundamental up and down-quark masses. These explicitly break the chiral $SU(2)$ symmetry which would otherwise have only been broken spontaneously by the fermion condensate\index{fermion condensate}. As a result, the exactly massless  bosons of the Goldstone theorem are turned into pseudo-Goldstone bosons, which are allowed to have a small mass.

After these lengthy preliminaries, the idea of Technicolor is easy to state: Let us assume that, in addition to the $SU(3)_c$ gauge theory with its confinement scale $\Lambda_{QCD}\sim 0.2\,$GeV, there exists an $SU(N)_{tc}$ gauge theory confining at $\Lambda_{TC}\sim\,\mbox{few~TeV}$. Moreover, there are fermions (so-called techni-quarks) which are charged under both $SU(N)_{tc}$ and under the electroweak gauge group $SU(2)_L\times U(1)_Y$ of the Standard Model. Let us call them $(Q_L)_i$ and $(Q_R)_j$, transforming as $N$ and $\ol{N}$ of $SU(N)_{tc}$ respectively. The indices $i$ and $j$ signify some further (e.g.~electroweak) transformation properties.

In complete analogy to the well-understood case of QCD, one expects that a condensate $\langle (Q_L)_i (Q_R)_j\rangle\neq 0$ will form. This clearly has the potential to play the role of the standard-model Higgs and to give mass to $W$ and $Z$ bosons. For example, to be very concrete and following the QCD-example closely, one may introduce two techni-quark doublets
\be
Q_L=\left(\begin{array}{c} U_L \\ D_L \end{array} \right) \qquad \mbox{and} \qquad 
Q_R=\left(\begin{array}{c} U_R \\ D_R \end{array} \right)\,.
\ee
Here $Q_L$ is an $SU(2)_L$ doublet, uncharged under $U(1)_Y$. The fields $U_R$ and $D_R$ are $SU(2)_L$ singlets with $U(1)_Y$ charge $+1/2$ and $-1/2$ respectively. Clearly, a condensate of the expected type
\be
\langle (Q_L)_i (Q_R)_j\rangle\,\sim\,\delta_{ij}
\ee
will break the electroweak symmetry in the desired way. Indeed, it corresponds to two Higgs-doublet VEVs, both invariant under the same subgroup $U(1)_{em}\subset SU(2)_L\times U(1)_Y$.

Of course, this extremely simple-minded model is far from realistic, even if one only assumes the more limited data of the pre-LHC era. One of the main reasons is the need for Yukawa couplings, which in the present approach would most naturally come from operators like
\be
{\cal L}\supset \frac{1}{M^2}\,(\psi\cdot\psi)\,(Q\cdot Q)\,.
\ee
In this symbolic expression $\psi$ stands for Standard Model fermions and $Q$ for techni-quarks. All indices and their contractions have been suppressed. The only point the above expression intends to make is the following: Given that such four-fermion operators are present in the low-energy effective lagrangian and a non-zero condensate of techni-quarks develops, Standard Model fermion masses will in general be induced. It is also clear that now our theory will break down at the scale $M$. If one desires a renormalisable quantum field theory potentially valid up to the Planck or GUT scale, the model has to be extended. We will not study such constructions but note that they exist in principle. However, an obvious comment is that the largeness  of the top-Yukawa coupling forces the energy scale $M$ to be low, leading to phenomenological problems. Moreover, similarly to the situation with SUSY, both the well-established electroweak precision data and the more recent non-discovery of new physics at the LHC put technicolor under pressure.

For us, the main conceptual conclusion is the following: In addition to low-scale SUSY, technicolor offers in principle another perfectly viable, technically natural explanation of a low electroweak scale. Here by low we mean relative to the Planck scale. Both SUSY and technicolor share a growing `little hierarchy problem'\index{little hierarchy problem}. But assuming the small amount of tuning implied by this is accepted or better models avoiding it are found, the `large hierarchy problem' remains solved. The underlying main technical tools are very different: Non-renormalisation of  the Higgs mass vs. logarithmic running of a gauge coupling together with confinement dynamics.

While both types of model appear to fit reasonably well into what we know about the string landscape, there is at this point a significant difference: Low-scale SUSY comes, of course, from 4d SUSY at the compactification scale. The latter emerges (naturally but certainly not unavoidably) from the 10d SUSY. This, in turn, is apparently enforced on us when trying to consistently quantise the fundamental string. As a result, one may say that the low-scale-SUSY resolution of the hierarchy problem (as well as its modern and more modest version with SUSY at about 10 TeV) are directly related to the specifically stringy approach to quantum gravity.\footnote{
We 
should emphasise, however, that low-scale SUSY is certainly {\it not} a prediction of string theory. 10d stringy SUSY may be broken directly in the compactification process (e.g.~through a non-Calabi-Yau compactification) or at any energy scale between KK-scale and weak scale.
}

By contrast, technicolor requires only the right set of gauge groups and fermionic matter to be present at the high (e.g.~compactification) scale. Achieving such a field content in string theory looks perfectly reasonable, but there appears to be nothing specifically stringy about it. One might want to say that the relation between technicolor and string theory is a neutral one.

The last two paragraphs hint at a potential problem with the landscape resolution of the electroweak hierarchy problem: Indeed, to explain that low-scale SUSY has (so far) not been found, the string landscape has to prefer a higher SUSY breaking scale. Given the intimate relation between SUSY breaking, compactification and moduli-stabilisation, it is fairly easy to imagine that such a preference exists and can be quantified by a detailed study of the landscape, especially including the difficult subject of  stringy models of SUSY breaking and `uplifting'\index{uplift} from AdS to dS vacua \cite{Denef:2004cf, Susskind:2004uv, Douglas:2004qg, Giudice:2006sn, Acharya:2008zi, Baer:2019zfl}.\footnote{In fact, most work in this area is based on what is known about the statistics of the flux stabilisation of complex structure moduli. As emphasised e.g.~very recently in \cite{Broeckel:2020fdz}, the Kahler moduli stabilisation is also crucial. However, at that level of detail one must also consider the statistics of possible uplifts -- a hard subject that is not well understood.}

However, making the reasonable assumption that technicolor models can be found in the landscape as well, one would expect that a (technicolor-based) low electroweak scale should occur in a fraction of models which is not exponentially suppressed. Such models would then be preferred relative to models with a purely tuned small Higgs mass and even relative to low-scale SUSY models (if stringy moduli stabilisation implies a bias against those). Thus, the world above the weak scale should display a natural and generic variant of technicolor which, however, it does not. A possible way out may be a bias in the string landscape against large gauge groups with a chiral spectrum (which are needed for technicolor), but this is pure speculation.

\subsection{From the `Little Higgs' to large or warped extra dimensions}\label{lhm}
\index{Little Higgs}\index{extra dimensions!warped}

This is a good place to comment on a number of further model building ideas that have been proposed to resolve the naturalness problem of the electroweak scale (see \cite{Csaki:2018muy} for an introductory review).

To begin, let us remind the reader that the conceptual reason for the lightness of the pions in the Standard Model is the Goldstone-theorem: They are Goldstone bosons\index{Goldstone boson} of the spontaneously broken chiral $SU(2)$. The simplest versions of technicolor use this idea to generate a Standard Model Higgs as a Goldstone boson of a global symmetry acting on a set of techni-quarks. Moreover, technicolor separates the scale of this symmetry breaking from some high fundamental scale through the logarithmic running of a non-abelian gauge coupling.

{\bf Little Higgs models} implement and perfect the idea of a Goldstone-boson Higgs without the constraints and complications of the fermion-condensate \cite{ArkaniHamed:2001nc, ArkaniHamed:2002pa, ArkaniHamed:2002qx} (see \cite{Schmaltz:2005ky} for a review). The challenge attacked and overcome in these constructions is to keep the Higgs unexpectedly light  (hence the name `little') in spite of Standard-Model gauge and Yukawa interactions. In doing so, one does not ask at the same time for a UV completion that can be valid up to some very high scale. Things are then simpler since one does not need to realise the Goldstone scalars as fermion condensates. Instead, one can just start with a compact field space, like the quotient manifold $M=SU(5)/SO(5)$ analysed in \cite{ArkaniHamed:2002qy}. This can be viewed as the Little-Higgs-analogue of the group manifold of $SU(2)$ which arises in the low-energy EFT of the Standard Model and is parameterised by our familiar three pions. Here, by contrast, the number of scalars is larger: $SU(5)/SO(5)$ is 14-dimensional. Some of those scalars are removed by gauging, some become massive, but some stay light and can play the role of the observed light Higgs doublet. However, both the gauging by the electroweak group $SU(2)_L\times U(1)_Y$ and the Yukawa couplings to Standard Model fermions have to be introduced and violate the masslessness of all of the Goldstone bosons. The true model building challenge of Little Higgs models is then to realise a structure in which this effect is sufficiently small, e.g.~because a non-zero potential for the Higgs degrees of freedom arises only at the 2-loop level. We will not discuss the (very interesting) details of this.

The overall picture which eventually emerges has three basic energy scales: They can be characterised starting from the scale $f$ which sets the size of the compact field space of the Goldstone bosons. This scale is completely analogous to the pion decay constant of the Standard Model or to the axion decay constant which defines the volume $2\pi f$ of an $S^1$ parameterised by an axionic scalar. If some of the Goldstone bosons are removed by gauging, then they become massive (together with the corresponding vectors). The relevant mass scale is $f$, at least if the gauge couplings are ${\cal O}(1)$. The electroweak scale is much smaller, $m_{EW}\ll f$, with the hierarchy provided by loop suppression factors ($1/16\pi^2$) together with the smallness of Standard Model couplings governing the loops. Finally, the model is based on a compact field space, which implies the presence of higher-dimension operators suppressed by powers of the mass scale $f$. This is most easily seen in the case of standard-model pions by considering \eqref{pide} and \eqref{pil}. A loop expansion involving such higher-dimension operators breaks down at a scale $\Lambda\equiv 4\pi f$, where the $4\pi$ come from the 4d loop suppression factor $1/16\pi^2$. At this scale $\Lambda$, which is however significantly higher than the electroweak scale, e.g.~$\sim 10\,$TeV, a UV completion is required. It may be field-theoretic (in the spirit of technicolor), higher-dimensional or even stringy (see below).

A very different idea concerning the hierarchy problem has emerged in the late 90s under the name of {\bf Large Extra Dimensions}\index{extra dimensions!large}~\cite{ArkaniHamed:1998rs, Antoniadis:1998ig}.\footnote{
See e.g.~\cite{Rubakov:2001kp, Csaki:2004ay} for introductory reviews.
} 
Specifically, the scenario known as `ADD'\index{ADD} is extremely simple and builds on string-theoretic ideas, but without explicitly using any details of the stringy UV completion. It proposes that the world is $d$-dimensional, with $d=4+n$, and that the $n$ extra dimensions are large (in the sense of being much larger than the Planck length). The 4d Planck scale is then given by
\be
M_{P,\, 4}^2 \sim M_{P,\, d}^{2+n}\,R^n\,,
\ee
with $M_{P,\, d}$ the higher-dimensional Planck scale and $R$ a typical compactification radius. Based on this, one can envision a situation where $R$ is so large that $M_{P,\, d}$ is in the TeV domain. This resolves the large hierarchy problem since, in fact, the electroweak scale is now of the order of the fundamental UV scale $M_{P,\,d}$. Put differently, the quadratic Higgs mass divergence is cured by the fundamental quantum gravity cutoff (be that superstring theory or something else). Since this cutoff is low, no large hierarchy problem exists. However, one clearly needs a dynamical mechanism that explains why the compact space is stabilised at $R\gg 1/M_{P,\,d}$.

While the large hierarchy problem may well be solved in this way, the little hierarchy problem remains and may, in fact, be more severe than in other approaches. Indeed, given proton-decay, flavour and electroweak-precision-data constraints, it is certainly optimistic to assume that the fundamental quantum gravity scale can be as low as even 10$\,$TeV. This, of course, leaves a sizeable gap to the observed Higgs mass of $\sim 100\,$GeV and hence a significant little hierarchy problem. Concerning this issue, one may now imagine a combination of model-building ideas where the large hierarchy problem is overcome using the large-extra-dimensions approach and the little hierarchy problem is solved as in the Little Higgs models discussed above. We will return to a variant of this further down when talking about warped extra dimensions.

Before closing this large-extra-dimensions or ADD discussion, we need to comment on the number $n$ determining the dimensionality of the compact space. The choice $n=1$ is obviously excluded since $R$ would be way too large. Very intriguingly, setting $n=2$ and assuming $M_{P,\,6}\sim 1\,$TeV gives $R\sim1\, \mbox{meV}^{-1}\sim 1\,$mm. But this was just about the smallest distance at which gravity was at that time directly accessible to experiments. Hence, even such an extremely large compactification radius appeared to be on the one hand not excluded but, on the other, could be discovered in the foreseeable future in gravitational table-top experiments.

Of course, the attentive reader should immediately object that the 4-dimensional (rather than 6-dimensional) nature of the world was, even in 1998, already known to persist up to energy scales of at least 100$\,$GeV. However, this bound could be evaded by assuming that all Standard Model particles and gauge fields are confined on a brane, in this case a 3-brane filling out our 3$\,$+$\,$1 non-compact dimensions and being point-like in the 2d compact space. With that, the ADD scenario with $n=2$ is complete. Not surprisingly, it was perceived as extremely innovative and exciting at the time. Unfortunately, it became clear very fast that astrophysical and cosmological  constraints push the lower bound on $M_{P,\,6}$ way above 1$\,$TeV. Also, constraints on the short-distance behaviour of gravity developed fast, forcing $R$ way below a mm and, again, $M_{P,\,6}$ to values higher than TeV. This worsens the little hierarchy problem and disfavours the $n=2$ case. 

For $n\ge 3$, the compactification radius $R$ is much below 1$\,$mm even if $M_{P,\,d}$ is kept in the TeV domain. Then so-called fifth-force experiments, testing gravity in the sub-mm domain, provide no meaningful constraints. Also cosmological and astrophysical bounds become less prohibitive with growing $n$. Such scenarios with $n\ge 3$ are still constrained but not hopeless. They do, however, become less and less believable from the point of view of solving the hierarchy in the same way as SUSY, technicolor etc.: The LHC simply keeps pushing any exciting new physics to higher and higher energies, thereby making the little hierarchy problem more severe.

The next twist in this line of thinking is the very interesting idea of {\bf warped extra dimensions}, also known as the {\bf Randall-Sundrum model}\index{Randall-Sundrum model}~\cite{Randall:1999ee, Randall:1999vf} (see \cite{Rubakov:1983bz} for the first warped scenarios). The term `warping' has been discussed in quite some detail in Sect.~\ref{kklt2} and its meaning in the present context is the same as before: It denotes compactifications where the metric in the non-compact directions depends on the position in the compact space. More concretely, the present type of model is based on a 5d to 4d compactification on $S^1/\mathbb{Z}_2$, i.e. on an interval. The metric reads
\be
ds^2=e^{-2ky}dx^2+dy^2\,,
\ee
with the warp factor $\exp(-ky)$ and $y\in [0,y_{I\!R}]$ parameterising the extra dimension. One refers to the interval-boundaries or `end-of-the-world' branes at $y=0$ and $y=y_{I\!R}$ as the UV and IR brane respectively. The reason is, as in Sect.~\ref{kklt2}, that any mass-dimension quantity of fixed value in units of $M_{P,\,5}$ takes a higher or lower value from the perspective of the 4d observer depending  on whether it is located closer to $y=0$ or to $y=y_{I\!R}$. The model can be characterised as a slice of AdS$_5$ with two 4d-branes as boundaries. To make this geometry a solution of Einstein's equations, a 5d cosmological constant and appropriate 4d brane tensions have to be added -- we will not work out the details of this.

To be precise, the Randall-Sundrum model comes in two variants known as `RS1'\index{RS1}\cite{Randall:1999ee} and `RS2'\index{RS2}\cite{Randall:1999vf}. The brief description above refers to RS1, with RS2 corresponding essentially to the decompactification limit $y_{I\!R}\to\infty$. This variant, while conceptually very interesting, has nothing to say about the hierarchy problem and we will not discuss it.

The relevance of RS1 for the hierarchy problem arises as follows. Let us assume that the Standard Model is localised at the IR brane. By this we mean, very naively, adding a piece
\be
\int d^4 x\, dy\, \sqrt{-g}\,\,{\cal L}_{SM}[g_{\mu\nu},\psi]\,\,\delta(y-y_{I\!R})
\ee
to the 4d Einstein-Hilbert action. The metric $g_{\mu\nu}$ is the pullback of the 5d metric to the boundary and the Standard-Model fields $\psi$ are only defined at the boundary locus. Now, due to the warping, it turns out that a fundamental Higgs mass parameter $m_H^2\sim M_{P,\,5}^2$ (in the local action near the IR brane) would be perceived by a 4d observer as being much smaller than the 4d Planck scale. Parametrically, the 4d observer finds
\be
m_H\sim M_{P,\,5}\,e^{-k\,y_{I\!R}}\,\,\,,\quad
M_{P,\,4}^2\sim M_{P,\,5}^3L\,\,\,,\qquad\mbox{where}\qquad L\sim1/k\label{rsrel}
\ee
is the AdS curvature radius. If $k$, which is governed by the 5d cosmological constant, is chosen to be slightly smaller than $M_{P,\,5}$ for control purposes, and the interval can be stabilised such that $k\,y_{I\!R}$ is a largish ${\cal O}(1)$ number, then the smallness of the electroweak scale can be explained. The stabilisation problem has an elegant solution due to Goldberger and Wise \cite{Goldberger:1999uk}.\index{Goldberger-Wise mechanism}

The reader may at this point be confused about how $m_H$ can be comparable to $M_{P,\,5}$ in the microscopic 5d action and at the same time much smaller than $M_{P,\,5}$ according to the first relation in \eqref{rsrel}. The answer is simply that, due to warping, $M_{P,\,5}$ has no unambiguous meaning in 4d. For example, a 5d Planck-scale black hole would appear to the 4d observer as 4d-Planck-scale if it were located near the UV brane. By contrast, it would be perceived as slightly above the electroweak scale if it were found near the IR brane.

What one has achieved at this point is very similar to the `ADD solution' of the hierarchy problem: The large hierarchy between the Planck and electroweak scale is explained, but the little hierarchy problem is as severe as ever since, phenomenologically, we can not afford to take $M_{P,\,5}$ all the way down to 100$\,$GeV. An enormous amount of work has gone into attempts of improving the RS1 idea in such a way that it becomes realistic in view of precision, flavour and LHC data or, more generally, that its little hierarchy problem is ameliorated (see~\cite{Gherghetta:2000qt, Contino:2003ve, Agashe:2003zs, Agashe:2004rs} for some of the original papers and \cite{Csaki:2005vy, Gherghetta:2006ha, Kribs:2006mq, Rattazzi:2003ea, Contino:2010rs, Gherghetta:2010cj, vonGersdorff:2011rz} for a selection of reviews). One possible way forward is to combine RS1 with the little Higgs idea explained earlier: One may for example consider placing not just the Standard Model but its little-Higgs extension, with a UV scale in the 10$\,$TeV range, at the IR brane. Now, having $M_{P,\,5}$ near 10$\,$TeV looks much less impossible than near 100$\,$GeV. Moreover, significant advantages can be gained by turning the Standard Model into a mix of 4d and 5d fields, with the latter being visible to us only through their lowest-lying KK modes. Such KK modes can, depending on the 5d mass parameters of the underlying 5d fields, have a $y$-profile localised (exponentially) near the UV or IR brane. This opens up model building possibilities where, for example, the Higgs lives at or near the IR brane. Heavy fermions like the top quark may also be localised near the IR brane, naturally giving them a stronger coupling to the Higgs. The light quarks, by contrast, live mainly in the UV.  Consistently with phenomenological requirements, they are then less affected by large higher-dimension-operators induced in the IR-part of the model (an issue following from the low-lying local value of $M_{P,\,5}$).

The various model building ideas in the RS1 framework briefly described above have a dual CFT interpretation. To appreciate this, the reader has to recall our very short discussion of AdS/CFT\index{AdS/CFT correspondence} in Sect.~\ref{pfp}. There, we characterised AdS/CFT as a map between a gravitational theory in $(d\!+\!1)$-dimensional AdS and a $d$-dimensional CFT living on the boundary. In our case of interest, $d=4$ and the global boundary is $\mathbb{R}\times S^3$. The scale invariance of the CFT allows us to take the radius of the $S^3$ to infinity, considering instead an $\mathbb{R}^{1,3}$-boundary of AdS$_{1,4}$. This may, in a first step, be identified with the RS2 model, consisting just of an AdS space cut off by the UV brane. The crucial difference to the pure AdS/CFT correspondence is that the UV brane is at finite distance, which lets it play the role of a physical UV cutoff in the CFT language. Our variable $y$ is the analogue of the radial variable of formal AdS/CFT, which in turn corresponds to the energy-scale variable of the CFT.

Now, turning to RS1, the following interpretation can be given in CFT language: We start from the UV cutoff at $y=0$ and move into the $y$-direction. This corresponds to moving to smaller energy scales in the CFT. At some point, an IR cutoff is encountered in the form of the IR brane at $y=y_{I\!R}$. More precisely, in models where the position of the IR brane is stabilised a non-trivial bulk profile of a 5d scalar (the Goldberger-Wise scalar \cite{Goldberger:1999uk}) has to be present. This profile determines the value of $y=y_{I\!R}$ where the IR brane will be encountered. Thus, the 4d dual of the 5d bulk theory is not a CFT but a nearly conformal theory, in which the slow running of some coupling eventually leads to the dynamical generation of an IR cutoff. The natural mechanism to think about here is the running of a non-abelian gauge coupling leading to confinement at the energy scale corresponding to $y=y_{I\!R}$. We can now clearly appreciate that the mechanism by which RS1 explains the low-lying electroweak scale is actually the AdS-dual formulation of the technicolor idea. The close relation between these two ideas is explored in much of the literature on the subject cited earlier (see \cite{Gubser:1999vj, ArkaniHamed:2000ds} for the fundamental first steps).

While all of this is deeply connected with string theory, it remains unclear to which extent the RS1 approach to the little hierarchy problem can be really implemented in string model building. The basic setting is in fact well known to arise in the form of the Klebanov-Strassler throat glued to a compact Calabi-Yau, cf.~Sect.~\ref{kklt2}. Yet, the IR region of the Klebanov-Strassler throat is too simple to house a full-fledged Standard Model. One of the problems with more complete models (see e.g.~\cite{Cascales:2003wn, Cascales:2005rj}) is that the strongly warped geometry is not explicitly known.

\subsection{Cosmological selection and the Relaxion}
\index{cosmological selection}\index{relaxion}
Throughout this course, we have discussed two opposite ideas on how (apparent) fine-tunings in EFTs can arise: On the one hand, a hidden mechanism (SUSY, Technicolor etc.) may be present, such that the fine-tuning is only apparent. On the other hand, many parameter values may be realised in a landscape of vacua, to be found in different parts of a multiverse. We then observe  a certain parameter value for anthropic reasons or accidentally. One may want to call this a real fine tuning.

In this short section, we want to briefly mention a third option which may be viewed as a compromise between the previous opposite extremes. Namely, it is conceivable that a landscape of vacua with different parameter values exists, but not all of them are on the same footing cosmologically. More precisely, the special (apparently fine-tuned) value we observe may be due to the details of cosmological dynamics. One may call this option {\bf cosmological selection}.

Such an approach to the cosmological constant has been suggested long ago~\cite{Abbott:1984qf, Brown:1987dd, Feng:2000if} on the basis of subsequent brane nucleation events. The idea is to use a model with 4-form flux and membranes, as in \eqref{sfbp}, where the energy gap between the different vacua is chosen to be tiny. If the vacuum energy without flux is negative, one finds a dense discretuum near zero. For appropriate brane tension, the cosmological dynamics will consist of consecutive jumps to lower and lower energy until, just after crossing to negative values, the process stops.\footnote{
This 
is due to the fact that, in AdS, transitions to lower-energy vacua are impossible if the brane tension is too high. The reason is that, in contrast to flat space, both the volume and the surface area of an expanding bubble in AdS grow in the same parametric way, proportionally to $R^2$. Thus, for sufficiently high surface tension the expansion of a true-vacuum bubble never becomes energetically favourable.
}
Unfortunately, this model is not realistic: The exponential expansion in between the last jumps leads to an unacceptable dilution of matter and radiation in the late universe.

\begin{figure}[ht]
\begin{center} 
\includegraphics[width=7.5cm]{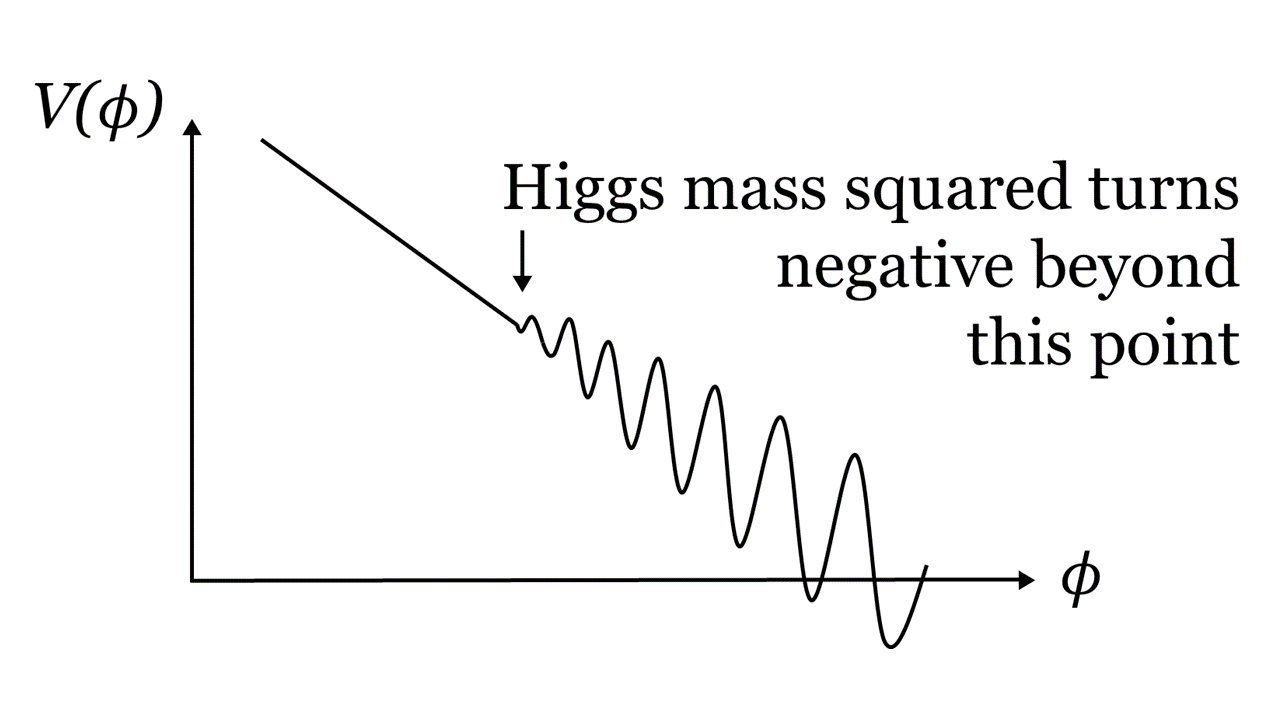}
\caption{Relaxion potential (adapted from \cite{Graham:2015cka}).}
\label{rel} 
\end{center}
\end{figure}

The idea of a cosmological selection of the electroweak scale has been around for a while \cite{Dvali:2003br, Dvali:2004tma} and has more recently received much attention in the context of the {\bf Relaxion model} \cite{Graham:2015cka}. The key ingredient is an axion-like scalar field $\phi$ which controls the Higgs mass. Specifically, one may assume that $m_H^2=m_H^2(\phi)$ is a monotonically falling function of $\phi$. This 
scalar rolls down a potential during cosmological history, for example during inflation. If multiple local minima are present, the field will eventually stop in one of them (cf.~Fig.~\ref{rel}). Crucially, if such minima are only present in the part of the field space of $\phi$ where $m^2_H(\phi)<0$, then the observed Higgs mass parameter will be negative. Moreover, if the dynamics is such that the field stops in one of the first minima it encounters, then $|m_H^2|$ will be much smaller than the `natural' scale determined by the UV cutoff. In short, the Higgs mass squared `relaxes' cosmologically to a value which is just below the threshold at which electroweak symmetry breaking first occurs.

Suppressing the (canonical) kinetic terms of Higgs field $H$ and relaxion $\phi$, the relevant part of the lagrangian reads
\be
{\cal L}\,\supset\, -(M_H^2-g\phi)|H|^2\,-\,V_0(\phi)\,+\,\Lambda^4(H)\cos(\phi/f)\qquad\mbox{with}\qquad
V_0(\phi)=-\kappa^3\phi+\cdots\,.\label{rell}
\ee
Here we recognise the $\phi$-dependent mass squared term for the Higgs doublet $H$, the perturbative $\phi$-potential $V_0$, and its non-perturbative correction producing a series of minima. The latter can be generated, for example, if $\phi$ couples to a non-abelian gauge group through the typical axionic coupling $\sim(\phi/f)\,\mbox{tr}F\tilde{F}$. The $H$-dependence of the prefactor of the cosine can arise if $H$ governs the masses of fermions charged under this group.\footnote{
This
is clearly modelled after QCD, with $\phi$ the QCD axion and the quark masses depending on the Higgs-VEV in the standard way. Yet, unfortunately, such a minimalist implementation does not work phenomenologically and an extra gauge group appears to be required.
} 
The key idea is that, if an $H$-VEV develops, the non-perturbative effect $\sim\cos(\phi/f)$ turns on, leading to the desired potential of Fig.~\ref{rel}. Here it is crucial to interpret $V(\phi)$ as resulting from \eqref{rell} after the Higgs has been integrated out. The reason why the cosine effect turns on only with a non-zero $H$-VEV rests on well-known instanton physics: As long as $H$ is zero, fermions are massless and, in the presence of massless fermions, no instanton potential is generated (see e.g.~\cite{Coleman:1985rnk, shifman}). Crucially, the parameters of this setting can be chosen such that the model is technically natural. A key role in this is played by the shift symmetry of $\phi$, which is broken only non-perturbatively and by the small parameters $g$ and $\kappa$.

This setting has been discussed intensely immediately after it appeared (see e.g.~\cite{Espinosa:2015eda, Hardy:2015laa, Patil:2015oxa, Antipin:2015jia, Jaeckel:2015txa}). It has also triggered a more general interest in cosmological selection, including for the cosmological constant. The reader may want to consult~\cite{Arkani-Hamed:2016rle, Arvanitaki:2016xds, Alberte:2016izw, Geller:2018xvz, Cheung:2018xnu, Graham:2019bfu, Strumia:2019kxg, Giudice:2019iwl, Bloch:2019bvc, Kaloper:2019xfj} and rethink the original ideas of~\cite{Dvali:2003br, Dvali:2004tma}. An objection one might have is that of a certain model-building complexity involved in creating precisely the desired type of landscape. By contrast, if superstring theory is the right theory of quantum gravity, the `standard' string landscape is simply there -- without any choice. Of course, one may also try to study explicitly whether cosmological selection arises on the basis of the string landscape \cite{McAllister:2016vzi}.

\subsection{The Swampland program}\label{swpr}\index{Swampland}
The Swampland is, by definition, the set of apparently consistent EFTs including gravity which are not found within the string landscape\index{landscape} \cite{Vafa:2005ui, Ooguri:2006in} (see \cite{Palti:2019pca, Brennan:2017rbf} for reviews). The qualification `apparently consistent' means that the EFT in question meets all consistency requirements which a low-energy observer not concerned with quantum gravity can impose. Thus, the Swampland program emphasises the following remarkable point: In spite of the enormous size of the landscape, not every field-theoretically reasonable model can be UV-completed in string theory.

A popular more general definition proposes that the Swampland consists of those low-energy EFTs which can not be UV completed in {\it any} model of quantum gravity, not just in string theory. The difficulty with this definition is that we have no overview of possible quantum gravity models and that the attempts that exist outside string theory are even less well understood than the string landscape.

At first sight, one might be very impressed with the strength of the claim that $10^{272,000}$  flux vacua \cite{Taylor:2015xtz} are not enough to realise any reasonable EFT. However, on second thought this is obvious since the landscape is discrete.\footnote{
At least if we count any ${\cal N}\ge 2$ SUSY moduli space as a single theory, which is presumably justified since the moduli are massless, dynamical fields rather than parameters.
}
By contrast, the space of EFTs is continuous due to the continuous choice of couplings or operator coefficients. Hence, almost any EFT is in the Swampland.

\begin{figure}[ht]
\begin{center} 
\includegraphics[width=8cm]{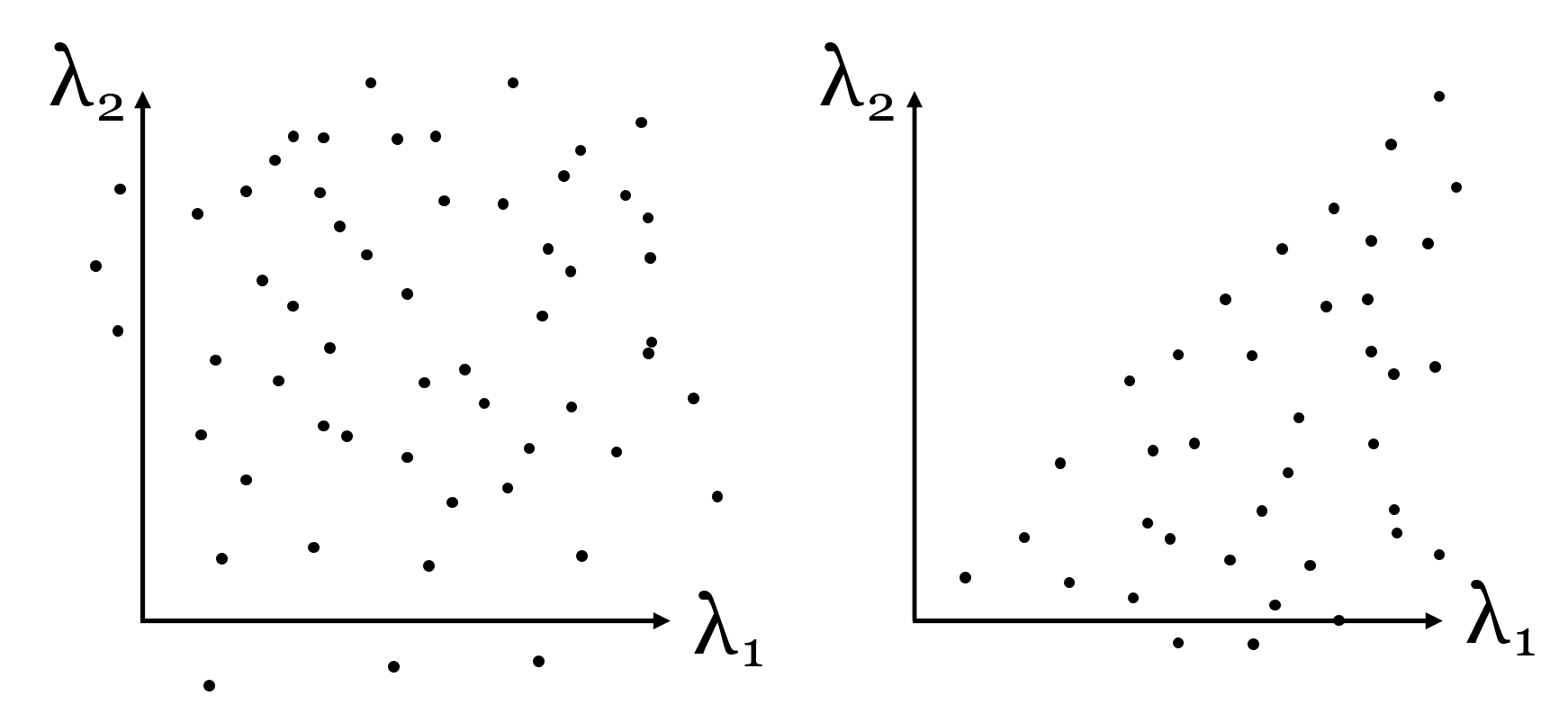}
\caption{On the left: String landscape discretuum filling out the whole 2-dimensional plane of EFT coupling constants $\lambda_1$ and $\lambda_2$. On the right: The region $\lambda_2>\lambda_1$ is forbidden.}
\label{lssw} 
\end{center}
\end{figure}

In fact, the Swampland paradigm attempts to make a slightly different and far less obvious point: It attempts to rule out whole classes of EFTs based on certain general features. An illustration is given in Fig.~\ref{lssw}: On the left, we see how the string landscape discretuum may essentially fill the whole plane of two EFT coupling constants $\lambda_1$ and $\lambda_2$. While not every combination of $\lambda_1$ and $\lambda_2$ is realised, the difference between the space of all EFTs and the landscape is clearly very hard to probe experimentally. By contrast, the r.h.~plot shows a situation where the region $\lambda_2>\lambda_1$ is forbidden, possibly due to a Swampland constraint. In this case, a single, not even very precise measurement of $\lambda_2>\lambda_1$ may in principle at once rule out string theory as the correct UV completion of quantum gravity in the real world.

To make our discussion somewhat more concrete, let us briefly go through the most popular swampland constraints. While some of them are extremely plausible, it is probably fair to say that, at the moment, all of them remain conjectures.

First, it is widely believed that {\bf no global symmetries}\index{no global symmetries conjecture} can occur in consistent theories of quantum gravity \cite{Banks:1988yz, Kamionkowski:1992mf, Holman:1992us, Kallosh:1995hi, Banks:2010zn, Harlow:2018tng}. This expectation is based on the violation of global charge in black hole evaporation and a CFT argument in perturbative string theory \cite{Banks:1988yz}. There is also a proof in AdS/CFT relying on the splittability assumption \cite{Harlow:2018tng}.

Then there is a set of conjectures about the geometry of moduli space, proposed in the seminal papers \cite{Vafa:2005ui, Ooguri:2006in} that defined the term `Swampland'. Among those, we emphasise the conjectures that the moduli spaces are infinite and that, as one moves to infinity, a tower of light states becomes exponentially light {\bf (Swampland Distance Conjecture)}\index{Swampland!Distance Conjecture}. These statements are to be read using the mathematically natural metric on moduli space, which is also the metric defining the kinetic term for the corresponding scalar fields in the 4d lagrangian. Given what we know about string compactifications, it is hard to imagine how the distance  conjecture could be false: For example, one can always take the volume modulus to infinity, in which case the tower of KK modes becomes exponentially light. As another option, one can take the dilaton to infinity, i.e.~$g_s\to 0$, implying that the tower of string excitations becomes light. In fact, these two options -- light KK modes or light string states -- appear to represent an exhaustive list \cite{Lee:2019wij}. For a more quantitative (refined) version of the distance conjecture see \cite{Klaewer:2016kiy}. 

The swampland conjectures described above are, however, not quite as appealing phenomenologically as envisioned in Fig.~\ref{lssw}, where EFT couplings are directly constrained. A conjecture of this latter type is the {\bf Weak Gravity Conjecture}\index{Weak Gravity Conjecture} \cite{ArkaniHamed:2006dz}, which states that gravity is always the weakest force. Concretely, the statement is that  in the presence of gravity any $U(1)$ gauge theory must come with a light charged particle of mass
\be
m_Q \le2\,g\,|Q|\,M_P\,. \label{wgc}
\ee
Here $g$ is the gauge coupling, $Q\in \mathbb{Z}$ the charge of the particle, and the precise numerical coefficient is chosen such that equality arises  (for $|Q|\gg 1$) precisely if the charged object is an extremal Reissner-Nordstrom black hole. An equal sign in (\ref{wgc}) would also mean that two particles of this type are attracted by gravity precisely as strongly as they are repelled by their common $U(1)$ charge. In this sense the inequality really deserves the name Weak Gravity Conjecture.

There are different motivations for this conjecture, one of the most popular being that, if it were false, then extremal black holes would be absolutely stable. This may be problematic \cite{Susskind:1995da}, though no sharp argument for the Weak Gravity Conjecture has so far emerged from this line of reasoning. Another argument for the conjecture is that the limit $g\to 0$ should be forbidden because it would generate a global symmetry. The Weak Gravity Conjecture then quantifies what exactly goes wrong with taking such a limit. Maybe most importantly, the conjecture is supported by all controlled stringy examples, and this is at least superficially easy to understand: Indeed, consider a D-brane model and try to take the gauge coupling to zero. The only ways in which this can be done is by either sending the brane-volume or the dilaton to infinity. But in both these cases the string scale goes to zero in Planck units,\footnote{
To be precise, one may also consider a double scaling limit where the space transverse to the brane shrinks such that the Calabi-Yau volume does {\it not} diverge together with the brane volume. One then needs to change duality frames to maintain control and see the emerging light states \cite{Lee:2018urn}.
} 
such that one may say that \eqref{wgc} is trivially fulfilled because the 4d EFT cutoff falls below the energy scale $g\, M_P$.

Note that our statement of the conjecture in \eqref{wgc} was an enormous oversimplification. We have disregarded that many different versions of the Weak Gravity Conjecture are discussed. For example, one may demand that {\it some} charged particle satisfying the conjecture exits (mild form) or that the {\it lightest} charged particle should do so (strong form).\footnote{
The 
maybe most naively expected formulation that the particle with $Q=1$ should satisfy the conjecture has counterexamples. However, in all such counterexamples the lowest charge $Q_{min}$ at which the conjecture holds exceeds unity only by an ${\cal O}(1)$ factor.
}
These options have already been considered in \cite{ArkaniHamed:2006dz}. More recently, a lot of additional effort has been devoted to the Weak Gravity Conjecture and its extensions. This revival of the Swampland discussion (see e.g.~\cite{Cheung:2014vva, delaFuente:2014aca, Rudelius:2015xta, Montero:2015ofa, Brown:2015iha, Bachlechner:2015qja, Hebecker:2015rya, Junghans:2015hba, Heidenreich:2015wga}) has in part been triggered by an increased interest in the observational signals of cosmological inflation, which have at that time started to constrain the magnitude of primordial gravitational waves and, through this, the field range of the inflaton. This, in turn, has a surprisingly direct connection to the Weak Gravity Conjecture and the Swampland program, as we now briefly explain.

The point here is that the Weak Gravity Conjecture has a natural extension to $p$-form gauge theories with $p\neq 1$. One then basically constrains the tension of the charged $(p\!-\!1)$-brane in terms of coupling strengths and $M_P$, in complete analogy to \eqref{wgc}. Specifically for an axion, viewed as a $0$-form gauge theory, the coupling strength is $\sim 1/f$ and the role of $m_Q$ is taken over by the instanton action $S_{inst}$. One then has
\be
S_{inst}\lesssim M_P/f\qquad \Rightarrow\qquad f\lesssim
M_P\,,\label{fcon}
\ee
where the implication rests on the (non-trivial) assumption that $S_{inst}\gtrsim 1$. This is motivated by the desire to use the dilute instanton gas approximation. One sees that, interpreted in this way, the Weak Gravity Conjecture limits the allowed field range of axions. This restricts the model of so-called `natural inflation', which in its simplest form relies on `superplanckian' axion field ranges to realise large-field inflation \cite{Freese:1990rb} (see \cite{Westphal:2014ana} for a review).

Such possible limitations of axionic field ranges in string theory (see also \cite{Banks:2003sx}) may in principle be overcome by so-called axion monodromy inflation \cite{Silverstein:2008sg, McAllister:2008hb, Kaloper:2008fb, Kaloper:2011jz} or its modern version, $F$-term axion monodromy \cite{Marchesano:2014mla, Blumenhagen:2014gta, Hebecker:2014eua}. The underlying idea here is to break the axion periodicity weakly. As a result, the circular field space turns into a spiral which rises slowly to higher and higher potential energy. Yet, these models may turn out to be in the Swampland, either because of concrete model building difficulties or due to generic constraints, such as a sufficiently strong version of the Swampland Distance Conjecture (see e.g.~\cite{Baume:2016psm, Klaewer:2016kiy}). Much further interesting work has recently been done in the context of developing and connecting various forms of the Weak Gravity and the Swampland Distance conjecture, see e.g~\cite{Ibanez:2015fcv, Grimm:2018ohb, Lee:2018urn}. Concerning the applicability to inflation, things remain unclear: On the one hand, one may indeed hope that the tower of light states coming down at superplanckian field excursions constrains models of inflation. On the other hand, realistic large-field inflation\index{inflation!large-field} needs only field ranges of the order $(few) \times M_P$. Such modestly transplanckian field ranges may turn out to be consistent with all reasonable conjectures.

Note that it is also conceivable to simply break the proposed inequality on the r.h. side of \eqref{fcon} in a concrete model. One idea is to start with the field space of two axions, say a $T^2$ with volume $(2\pi f)^2$ and $f<M_P$. All one needs is to realise a scalar potential on this $T^2$ which forces the lightest effective field on a spiralling trajectory \cite{Kim:2004rp}. This trajectory, while still periodic, may clearly be much longer than $2\pi f$. The required potential could be realised by the interplay of several instanton-induced cosine-terms. 

Even simpler, such a long spiralling or winding trajectory may be enforced by making a certain combination of the two axions massive by a flux choice \cite{Hebecker:2015rya}. This has an interpretation as the `Higgsing' of the axion or $0$-form gauge theory with the help of a $(-1)$-form gauge theory (as explained near \eqref{moft}), cf.~\cite{Dvali:2005an}. In fact, more generally, it has been pointed out that this method of Higgsing a $p$-form gauge theory with a $(p-1)$ form gauge theory apparently represents a field-theoretic method of breaking the strong form of the Weak Gravity Conjecture in the IR, including in the `classic' case of $p=1$ \cite{Saraswat:2016eaz}. Here by Higgsing we mean the substitution
\be
\frac{1}{g_p^2}|dA_p|^2+\frac{1}{g_{p-1}^2}|dA_{p-1}|^2\qquad \to\qquad 
\frac{1}{g_p^2}|dA_p|^2+\frac{1}{g_{p-1}^2}|dA_{p-1}+A_p|^2\,.
\ee
In the case $p=1$, this is clearly the standard meaning of the term Higgsing, where $A_0$ represents the radial direction of the conventional complex scalar which Higgses a $U(1)$ gauge theory. Now, starting with two gauge fields, say $A_p^{(1)}$ and $A_p^{(2)}$, one may Higgs the linear combination $A_p^{(1)}+N A_p^{(2)}$, where $N$ is a large integer. It is then easy to see that, for the surviving $p$-form gauge theory the Weak Gravity Conjecture in its strong form  will be broken in the IR \cite{Hebecker:2015rya, Saraswat:2016eaz}. The reason is basically that the effective gauge coupling is lowered from $g_p$ to $g_p/\sqrt{|N|}$. This effect can even be made exponentially strong using the so-called clockwork\index{clockwork} idea \cite{Kaplan:2015fuy, Choi:2015fiu}. If such constructions are possible in the landscape, this clearly weakens the phenomenological relevance of the Weak Gravity Conjecture as an IR constraint. It could also turn out that models of this type are in the Swampland and the Weak Gravity Conjecture remains strong.

\subsection{The Swampland and de Sitter}\label{swds}
\index{Swampland}\index{de Sitter space}\index{Swampland!de Sitter conjecture}

The Swampland program has many aspects. Proceeding largely in historical order, the previous section emphasised its more phenomenological side: constraints on global symmetries, weak gauge couplings and fields ranges, with a view on inflation. There is also a more mathematical or conceptual side, focussing on what models can arise from string theory as a matter of principle, without immediately asking for relevance to the real world (see e.g.~\cite{Brennan:2017rbf}). We will not discuss this here. 

Instead, we now turn to a more recent and possibly the most important aspect of the Swampland discussion. It is interesting conceptually but, in addition, has very far-reaching phenomenological implications. The conjecture we are referring to states that all de Sitter solutions, even metastable ones, are in the Swampland \cite{Danielsson:2018ztv, Obied:2018sgi, Garg:2018reu, Ooguri:2018wrx}.\footnote{
For
a critical review see e.g.~\cite{Akrami:2018ylq}.
} 
A particularly intense debate has initially surrounded the very strong conjecture that \cite{Obied:2018sgi}
\be
|V'|/V\ge c\,, \label{dec}
\ee
with $c$ an ${\cal O}(1)$ constant and $M_P=1$. This clearly rules out de Sitter minima, but is actually much stronger by also excluding de Sitter maxima, i.e. unstable de Sitter solutions. This is presumably too strong since it collides with the Standard Model Higgs potential \cite{Denef:2018etk, Cicoli:2018kdo}, the EFT of pions \cite{Choi:2018rze}, and with relatively well established string constructions \cite{Conlon:2018eyr}.

A refined form of the de Sitter conjecture \cite{
Garg:2018reu, Ooguri:2018wrx} and an attempt of a first-principles derivation \cite{Ooguri:2018wrx} (see however \cite{Hebecker:2018vxz, Junghans:2018gdb}) have subsequently appeared. The refined formulation states that either
\be
|V'|/V \ge c\qquad \mbox{or}\qquad V''/V \le c'\,, \label{rde}
\ee
which is roughly the opposite of the slow-roll requirement. The derivation of \cite{Ooguri:2018wrx} aims only at establishing the claim at asymptotically weak coupling (basically at large field distance, e.g. at asymptotically large volume). It uses the Swampland Distance Conjecture, but also relies on strong assumptions about the origin of de Sitter entropy. Unless this argument can be made water tight, counter examples are conceivable even asymptotically \cite{Hebecker:2018vxz, Junghans:2018gdb}. But more importantly, for the actual string landscape an asymptotic validity of \eqref{rde} is not threatening: The landscape as we know it certainly needs a very large set of metastable de Sitter vacua, but it is perfectly acceptable for this set to nevertheless be finite. A series of vacua extending to zero coupling is not required. The possibility of achieving arbitrarily weak couplings is essential for a rigorous mathematical proofs, but in physics this may simply be too much to ask for.

Let us start the more detailed discussion by explaining why a very strong conjecture such as (\ref{dec}) might appear appealing. Quite generally, compactifications lead to 4d effective lagrangians of the type
\be
{\cal L}\,\,\sim\,\,{\cal V}\left[{\cal R}_4-\frac{(\partial 
{\cal V})^2}{{\cal V}^2}-E\right]\,,
\ee
where we have suppressed ${\cal O}(1)$ coefficients for simplicity. The overall volume factor ${\cal V}$ multiplying the 4d Ricci scalar ${\cal R}_4$ and everything else comes from the integral over the compact space. But the volume also figures as a dynamical field, and its kinetic term features a logarithmic derivative. Moreover, we have included a positive energy source $E$ (one may think of a SUSY breaking effect). After Weyl-rescaling to the Einstein 
frame and introducing the canonical field $\phi=\ln({\cal V})$, one finds
\be
{\cal L}\,\,\sim\,\,\left[{\cal R}_4-(\partial \phi)^2-E\,e^{-\phi}
\right]\,.
\ee
The potential is $V(\phi)\sim e^{-\phi}$, which does indeed satisfy (\ref{dec}). This will remain the case if one goes more carefully through different types of simple explicit string compactifications.

\begin{figure}[ht]
\begin{center} 
\includegraphics[width=12cm]{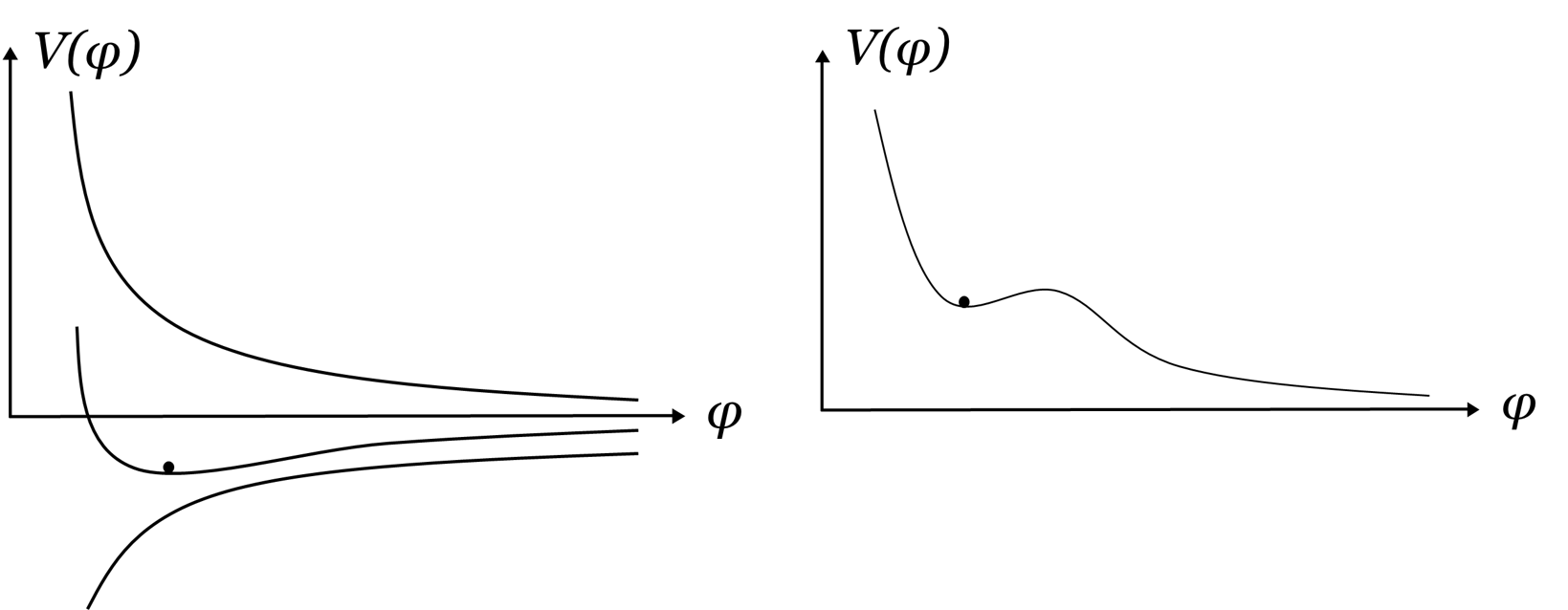}
\caption{Left: The sum of two simple falling potential terms allows only for AdS, not for dS minima. Right: Involving a third simple, monotonic term a metastabe dS minimum can be created with some tuning of coefficients.}
\label{adsds} 
\end{center}
\end{figure}

Let us make a slightly different but related point about creating dS space through a compactification. In the limit ${\cal V}\to \infty$, all potential terms (from fluxes, 10d curvature, SUSY-breaking effects) tend to zero. Indeed, in this limit one approaches uncompactified, flat 10d space, so one expects no energy density. Such an argument has first been given for asymptotically weak string coupling rather than for large volume \cite{Dine:1985he} and is known as the {\bf Dine-Seiberg problem}\index{Dine-Seiberg problem}. Now, even if one has two independent such terms, each of which approaches zero as $\phi=\ln({\cal V})\to \infty$, their sum will in general not create a dS minimum (Fig.~\ref{adsds}). Indeed, if both terms are monotonically falling and have the same sign, the sum will also be monotonic. If they have opposite sign, then depending on which term scales more strongly with ${\cal V}$ the potential may first fall and then approach zero from below, as shown in the figure. It may also first grow and then turn over to approach zero from above. This would give a de Sitter maximum. Thus, as long as every term has a simple scaling behaviour with ${\cal V}$, it requires the interplay of {\it at least three} such terms to realise a dS solution. Moreover, these three terms, each with a different fall-off behaviour in $1/{\cal V}$, need to be of the same order of magnitude to realise such a non-trivial potential. But since the coefficients are ${\cal O}(1)$ numbers (string theory not having free parameters), the de Sitter minimum can clearly not be at asymptotically large volume (see also \cite{Ooguri:2018wrx}).

However, as already emphasised above, phenomenology and the landscape in general may not need asymptotically large volume or weak coupling. The scenarios of KKLT \cite{Kachru:2003aw} and LVS \cite{Balasubramanian:2005zx} show precisely how, through the interplay of three different terms falling with $1/{\cal V}$, a non-trivial de Sitter minimum may in principle arise. Thus, we are in the end faced with the, admittedly hard, problem of evaluating the reliability of concrete proposed string constructions of 4d metastable de Sitter EFTs. To liberate us from that challenge and `kill' the dS landscape as a matter of principle one would need harder arguments against de Sitter in string theory than those given above.

This may be a good place to mention an older no-go argument about what can or can not be achieved in string compactifications. It has been shown in \cite{Maldacena:2000mw} that perturbative compactifications involving fluxes, positive-tension branes and warping can not lead to 4d de Sitter. Minkowski space can only be achieved if fluxes and warping are turned off, which is of course not interesting. The key loophole utilised by GKP \cite{Giddings:2001yu} is the existence of negative-tension objects, for example O3 planes, which make warped compactifications leading to no-scale Minkowski space possible. This is of course the basis on which then, adding non-perturbative effects and anti-branes in warped throats, KKLT suggested what still stands up as the simplest realistic proposal for string-derived de Sitter.

Let us note that KKLT has come under concrete criticism in the Swampland context on a basis of arguments related to the Maldacena-Nunez no-go theorem\index{Maldacena-Nunez no-go theorem} mentioned above \cite{Maldacena:2000mw}. It has been pointed out in \cite{Moritz:2017xto} that, re-running the logic of \cite{Maldacena:2000mw} while including the 10d effects of gaugino condensation apparently leads to problems. However, a more careful analysis of the relevant action of the brane stack on which gauginos condense \cite{Hamada:2018qef, Kallosh:2019oxv} shows that a 10d description of KKLT (usually only derived by 4d EFT methods) may be possible \cite{Hamada:2019ack, Carta:2019rhx} (see however \cite{Gautason:2019jwq}). This has been further developed, again with positive outcome for KKLT \cite{Kachru:2019dvo}
(or at least for the AdS part of the construction \cite{Bena:2019mte}). 

An interesting new KKLT criticism \cite{Carta:2019rhx} observes that in the relevant parametric regime the warped throat housing the $\ol{\rm D3}$ uplift  is in general too large to fit into an uwarped Calabi-Yau. It has been argued in \cite{Gao:2020xqh} that the resulting strong warping effects in the bulk of the compact space imply a `bulk-singularity problem', quantitatively ruling out at least the simplest, generic versions of KKLT. This is clearly too recent to treat it as a conclusive statement. Moreover, by their very nature the above issues do not extend to LVS-type models, which allow for a much larger compact volume and hence have no problems related to strong warping.\index{uplift}

Yet another recent set of Swampland conjectures, dealing with constraints on AdS compactifications, may affect de Sitter constructions \cite{Gautason:2015tig, Gautason:2018gln, Lust:2019zwm}. It states (in versions of varying strength) that scale-separated AdS compactifications are impossible. The  basis is the observation that, in many well-studied string compactifications to AdS, the AdS space and the compact space share the same radius.\index{AdS conjectures} For example, this is famously the case in AdS$_5\times S^5$, the geometry of the original AdS/CFT proposal. The absence of a separation between the AdS and KK scale implies that a proper, purely-4-dimensional EFT regime is missing. This may imply difficulties for the so-called uplift to de Sitter, which in the simplest case is conceived as the addition of small positive energy density to a well-defined AdS$_4$ EFT. The absence of a 4d EFT regime does not a priori exclude an uplift, but it forces one to analyse it in the full 10d theory. 

Concerning specifically KKLT, a conjecture which is strong enough to exclude scale-separated AdS in general does, of course, also exclude KKLT: Recall that, before the uplift, KKLT has a 4d AdS scale $\sim |W_0|$ and KK scale $\sim 1/\ln^{1/2}(1/|W_0|)$ with $|W_0|\ll 1$. Given that this first step of KKLT is rather well established and the possibility of fine-tuning $|W_0|$ to parametrically small values has been convincingly argued for \cite{Denef:2004ze} (see also \cite{Demirtas:2019sip}), this may speak against the scale-separated AdS conjecture rather than against KKLT. Moreover, a class of type-IIA compactifications to scale-separated AdS has been proposed some time ago \cite{DeWolfe:2005uu} (see also \cite{Marchesano:2020qvg, Junghans:2020acz}). These constructions are purely perturbative and do not rely on the tuning of fluxes.

In summary, while doubts about the existence and calculational control of stringy dS constructions are justified, they do not appear strong enough at present to overthrow the landscape idea as it has been developed since about the turn of the millenium. The doubts receive support from the relative complexity of the simplest concrete models. However, as we argued, the need for such a complexity follows simply from the parametric analysis of sums of decaying functions of the volume (cf.~Fig.~\ref{adsds}). If, which is certainly possible, metastable de Sitter vacua turn out to be inconsistent with string theory, a serious phenomenological problem arises due to the observed cosmological dark energy. While a rolling scalar or `Quintessence'\index{Quintessence} \cite{Wetterich:1987fm, Peebles:1987ek} is certainly an acceptable way out \cite{Obied:2018sgi}, the realisation of such a scenario in string theory is not without its own problems (see e.g.~\cite{Cicoli:2012tz} and \cite{Cicoli:2018kdo} for a review). In particular, the technical difficulty of realising sufficiently strong SUSY breaking in a moduli-stabilised compactification appears to be as severe in quintessence as in de Sitter constructions \cite{Hebecker:2019csg}. 
As a result, one may doubt that phenomenologically viable quintessence models {\it can} but metastable dS vacua {\it can not} be realised within the class of presently understood string compactifications. 

Needless to say, research towards establishing or disproving known de Sitter constructions or developing viable alternatives (quintessence or other \cite{Hardy:2019apu}) must go on.

\subsection{More direct approaches to quantum gravity}\index{quantum gravity}
Before coming to the end of this section and the whole set of lecture notes, we now want to change perspective drastically. Let us recall how far from established physics we have come in studying issues like the construction of 4d de Sitter space through the compactification of 10d superstring theories. We were, under certain assumptions, forced on this path by the desire to find predictivity in the UV and hence the need to control UV divergences in quantum gravity. But promoting point particles to strings may not be the only option for achieving this goal. Which implications for the hierarchy problems and for low-scale physics in general would follow if gravity could be quantised in a simpler, more direct approach?

As we already mentioned, one may treat quantum gravity as just another gauge theory, with a spin-2 particle called the graviton being the propagating degree of freedom \cite{vel}. In many respects this works perfectly below the scale $M_P$, but it proves impossible to raise the cutoff above the Planck scale. This is at least one way to characterise the problem. An alternative perspective is that it is perfectly acceptable to take the cutoff $\Lambda$ to infinity, but the price to pay is an infinite set of operators, suppressed by growing powers of $1/M_P^2$, with unknown coefficients. All these operators are necessary to absorb the divergences of perturbation theory and, as the net result, one is again limited to a quantum effective theory below $M_P$.

Now, the above impasse may clearly be a phenomenon of perturbation theory and a non-perturbative definition may lead to a well-defined theory, potentially even with the option of taking the limit $\Lambda\to \infty$. The preferred cutoff for a perturbative treatment of gauge theories, dimensional regularisation as used in \cite{vel}, is not suitable for a non-perturbative definition. One obvious alternative, supported by its success in fixed-space QFT, is the method of discretisation of spacetime, i.e.~`the lattice'. Of course, given that gravity makes space itself dynamical, it appears mandatory to make the lattice dynamical rather than using a fixed (e.g.~hypercubic) lattice as is common in QFT. Thus, one should be studying the dynamics of or a functional integral over {\bf triangulations}.\index{triangulations}

This class of approaches goes back to what is known as {\bf Regge calculus}\index{Regge calculus} \cite{Regge:1961px} (see \cite{Williams:1991cd} for a review and early references). The idea is clear if one visualises how a 2d manifold can be approximated by a collection of (flat) triangles glued at their edges. Curvature is now localised at the vertices, where 3 triangles meet, and can be quantified by the respective deficit angles. This clearly generalises to 3d manifolds, which can be analogously modelled by tetrahedra. The latter are now glued at faces (triangles) and additionally meet at edges and vertices. The general terminology would be that of {\bf simplices}\index{simplices}, in this case 3-simplices glued at 2-faces.

To describe 4d gravity, the above discretised 3-manifold has in some way to be supplemented with time. One option would be allow the discrete 3-manifold to depend dynamically on a continuous time variable. Another possibility is to make time steps discrete. A natural option in this latter case is to allow the number of 3-simplices to change from one time-slice to another. In this way, one is in effect triangulating a lorentzian or (3+1)-dimensional manifold. Similarly, one may consider discretised 4d riemannian manifolds if one is interested in euclidean gravity. Supplementing such a discrete description of spacetime with an action principle, one can use this as an approximation method for classical numerical relativity. But our interest here is based on the hope that this line of thought may define quantum gravity. Both canonical and functional-integral quantisation can be considered in this context. Finally, while this is not logically necessary, one may desire to take a continuum limit, in which the typical length of an edge becomes much smaller than the Planck length. Crucially, macroscopic dynamics should continue to be correctly described in this limiting procedure.

A central role in present-day research on lattice or discrete quantum gravity is played by the method of {\bf Causal Dynamical Triangulations}\index{Causal Dynamical Triangulations} or CDT\index{CDT} \cite{Ambjorn:1998xu} (see e.g.~\cite{Loll:2019rdj, Ambjorn:2012jv} for reviews). The key new term here is `causal', by which one means that the causal structure coming with flat Lorentz space is respected. In other words, the topology of all spatial slices of the 4d lattice is the same. Maybe more intuitively: The emission of baby universes is forbidden. By this we mean a process by which our universe, say with $S^3$ topology, emits a small 3-sphere, such that the late-time
spatial topology is $S^3\times S^3$. 

This is of course quite different from what happens on the worldsheet of the string: Here, the famous `trousers' or `pair of pants' geometry, corresponding to the decay of one particle into two, is a central part of the theory. Similarly, topological transitions are an important aspect of 10d dynamics, allowing for example the continuous deformation of one Calabi-Yau into another. Yet, it is also clear that with topology change comes potentially an explosion of the number of geometries to be considered in the path integral.\footnote{
We note in passing that there was at some point much excitement about the phenomenological relevance of baby universes \cite{Giddings:1987cg, Coleman:1988tj} (see \cite{Hebecker:2018ofv} for a review). But no completely convincing calculational approach emerged. By contrast, in the simpler 2-dimensional case, much progress concerning the euclidean path integral treatment of quantum gravity has been made, also outside the very special case of the critical string. This is in particular due to matrix model techniques \cite{Kazakov:1985ea, Ambjorn:1985az, David:1984tx} (for reviews see \cite{Klebanov:1991qa, Ginsparg:1993is}). It remains unclear how much of this carries over to 4d and we have no time to further comment on this rich field.
}
Thus, CDT is much more tractable than an unrestricted euclidean path integral over triangulations.
Still, even with this simplification there is so far no established result about 4d Einstein gravity emerging from a continuum limit of a discrete gravitational theory (see \cite{Ambjorn:2020rcn} for a very recent discussion).

A different though not unrelated approach to the quantisation of general relativity is {\bf Loop Quantum Gravity}\index{Loop Quantum Gravity} (LQG)\index{LQG}. The perspective adopted and the methods used differ very much from the rest of these notes, such that even a superficial introduction is impossible. We will only try to say a few words about the basic idea and refer the reader to the various introductory texts on the subject, see e.g.~\cite{Thiemann:2002nj, Nicolai:2005mc, Nicolai:2006id, Ashtekar:2007tv, Dona:2010hm,  Rovelli:2011eq, Rovelli:2014ssa}.

The approach is based on what is widely known as {\bf Ashtekar variables}\index{Ashtekar variables} for the canonical formulation of gravity and its subsequent quantisation \cite{Sen:1982qb, Ashtekar:1986yd}. To explain this, one has to look at 4d spacetime as a slicing, with each slice being a 3d spatial manifold with the pullback metric. The general holonomy group on the latter is $SU(2)$, such that the Hilbert space may be built using the parallel transports along loops in these spatial hypersurfaces. These are not just elements of $SU(2)$ but, since we are dealing with quantum mechanics, functions on $SU(2)$. The Hilbert space of the latter can be described using the series of all $SU(2)$ representations, classified by spin. The total Hilbert space of course also involves all possible loop configurations. The canonically conjugate classical variables are related to the embedding of the spatial surfaces in the 4d spacetime, involving in particular the extrinsic curvature. We will not attempt to explain this and the related construction of a Hamiltonian. But we should recall that we have already superficially met a situation of this type when we mentioned the Wheeler-DeWitt equation and wave function of the universe in Sect.~\ref{pfp}.

In the present context, it is crucial that Ashtekar variables make the problems of canonical quantisation of gravity more manageable, at least at some initial level. The resulting theory of LGC is nevertheless complicated. As was already the case with the triangulation-based approaches mentioned earlier, the crucial limiting procedure by which an approximately flat 4d spacetime with the familiar dynamics of the Einstein-Hilbert action should emerge remains problematic.

We recall that, not surprisingly, it is much simpler to approach quantum gravity from the perspective of standard low-energy EFT. Here, the Hilbert space is the Fock space of spin-2 particles, with interactions introduced in perturbation theory. It is this approach which relates most directly to string theory where (let us say for simplicity in 10d) the string interpretation resolves the UV problem of loop corrections. However, the criticism that may be raised at this point is that of so-called {\bf background dependence}\index{background dependence}. In other words, in the perturbative approach and its stringy UV completion, one starts on a given background, in the simplest case 10d Minkowski space. The string worldsheet relies on this background for its very existence, for example because its fields, including the 2d metric, come from the embedding in target space.\footnote{
Of 
course, the quantisation of the string also provides the massless 10d graviton. Then, condensates or coherent states of the relevant string excitations are capable of describing, at least in principle, the full dynamics of 10d target space. We leave it to the reader to explore the relevant literature, keeping the keywords {\bf string field theory} and {\bf tachyon condensation} in mind.
}
In this sense, approaches to quantum gravity like CDT and LQG may to some extent claim {\bf background independence}\index{background independence} as an important merit. Yet, the price that has so far to be paid is the difficulty of connecting to Einstein gravity at large length scales.

But one goal of this section is to arrive at another important distinction between the canonical approaches just mentioned and string theory. It is related to phenomenology and the hierarchy problem. Namely, if one of the former approaches were fully successful, i.e. if it could derive a low-energy EFT from a simple quantised model at the Planck scale, another problem is expected to arise: 

One the one hand, one could arrive at a unique low energy theory which is not the Standard Model. This would simply be the end of the route taken. On the other hand, one could discover ambiguities (such as the choice of matter fields and their couplings) in the UV, enabling one to fit the low-energy EFT to the Standard Model. This would in some sense be satisfactory, but it would also leave key questions about the fundamental laws unanswered. In particular, it may then be viewed as highly unsatisfactory that certain UV parameters would have to be tuned with the enormous precision required to describe the small cosmological constant and Higgs mass. Of course, we can not rule out the third possibility that the low-energy EFT will be unique and it will be {\it precisely} the Standard Model, with just the right apparent fine-tuning. From what we presently know about how a UV model produces low-energy observables, this would appear miraculous.

Indeed, visualise a set of simple formulae without free parameters predicting all operator coefficients in an EFT at some high scale $\mu$:
\be
m_H^2(\mu)/M_P^2(\mu)=f(\mu)\,,\qquad \lambda(\mu)/M_P^4(\mu)=g(\mu)\,, \qquad m_{\nu,R}(\mu)/M_P(\mu)=h(\mu)\,, \quad \cdots\,.
\ee
Here we have explicitly displayed Higgs mass parameter, cosmological constant and right-handed neutrino mass, and we have chosen to make all these parameters dimensionless using the Planck scale. By thinking in terms of an EFT at scale $\mu$ we assume that a UV cutoff at or slightly above the scale $\mu$ is imposed. We may moreover think of our theory as being compactified, for example, on a 3-torus of radius just slightly below $1/\mu$.

Now, we also know that there are highly non-trivial, all-loop formulae relating the above parameters to the eventually interesting quantities $m_H^2(0)/M_P^2(0)$ etc. This last step of high-scale to low-scale evolution is entirely independent of the quantum gravity theory producing the functions $f$, $g$, etc. It is very hard to see how, through the disturbances of this last EFT evolution from energy scale $\mu$ to energy zero, the required fine tuning or apparent fine tuning should be produced by the fundamental theory. It is at this point where one might be tempted to prefer a fundamental theory producing a landscape to one producing a unique field content and set of functions $f$, $g$, etc. Clearly, in the latter case these fundamentally predicted functions would have to be very special. For example, they must conspire to ensure that $\lambda(0)/M_P^4(0)$ vanishes with very high precision.
In particular, this high-precision almost-zero result must appear {\it after} integrating out the electroweak sector and QCD.

\subsection{Asymptotic safety and the hierarchy problem}
\index{asymptotic safety}

Finally, let us come to yet another perspective on quantum gravity which, as we will discuss, may be viewed as being closely related to the triangulation approach discussed above. It goes back to Weinberg \cite{Weinberg:1976xy} and can be formulated in a generic way, viewing gravity as a quantum EFT with UV cutoff, without yet committing to a specific technical implementation of the latter.

To explain this, recall that our best examples of well-defined quantum field theories are, like QCD, asymptotically free: They run to a trivial RG fixed point in the UV, where the theory becomes non-interacting. One may say that it is this fixed point which allows one to remove the cutoff completely, $\Lambda\to\infty$, thus making the theory well-defined and predictive on all energy scales. Of course, this is only a special and particularly simple example for a well-defined QFT: One may equally well have a non-trivial or `interacting' UV fixed point, as in the case of 4d ${\cal N}=4$ Super-Yang Mills theory. This theory is conformal, which means in particular that the beta functions of all operators in the lagrangian vanish.\footnote{
This 
is by now a standard fact.
The reader may explore the original references using e.g.~\cite{Brink:2015ust}.
} 
To make the situation more interesting, one may add some relevant operator, such as a mass term, to this theory. One will then have a model with a non-trivial RG evolution and a UV-definition in terms of an interacting fixed point. Such situations can also arise in field theories without supersymmetry, which in some cases simply happen to flow to a non-trivial fixed point in the UV \cite{Litim:2014uca}.

With this is mind, one may now ask (as Weinberg did much before all of the above examples), whether a similar situation might arise in gravity. In other words, could gravity, which is clearly not asymptotically free, instead be {\bf asymptotically safe} by running into a non-trivial UV fixed point. More generally, this could be a so-called {\bf fixed surface}\index{fixed surface}: A set of scale-invariant theories, parameterised by a (hopefully finite) set of parameters. The low-energy theory that we observe would then be defined by one RG trajectory taken from a continuum of such trajectories. For example, if the fixed surface were just a fixed line, it could be that we are free to choose the ratio of $\lambda/M_P^4$ in the IR. All other (higher-dimension) operators would then be predicted. This prediction would follow from the requirement that, in the UV, the RG trajectory of our theory of gravity hits the fixed line just described.

So far, this is general enough to include the case of a gravitational theory defined by some form of triangulation. The set of possible continuum limits would correspond to the above fixed surface. However, today the term asymptotic safety is frequently used for a specific and rather different implementation of these general physics ideas. This implementation is based on the concept of the {\bf Exact Renormalisation Group}\index{Exact Renormalisation Group} or ERG\index{ERG}. To explain the idea, recall the standard textbook knowledge that a QFT is defined by a (Wilsonian) effective action $S_\Lambda[\phi]$. By this we mean that a path integral with cutoff $\Lambda$ and the above action in the exponent defines all correlation functions. Keeping the theory unchanged, one may vary the cutoff, in which case the functional $S_\Lambda$ will vary or run with $\Lambda$. The general idea of an RG evolution of a whole action rather than separate coefficients can be implemented in various explicit forms, e.g. as the Polchinski equation\index{Polchinski equation} \cite{Polchinski:1983gv} (cf. the discussion in \cite{Ellwanger:1993mw, Morris:1993qb}). A related form which has proven to be particularly useful in the present context is the Wetterich equation\index{Wetterich equation} \cite{Wetterich:1992yh} for the so-called effective average action, which we now explain. We follow the particularly compact and clear discussion in \cite{Litim:2008tt} (see e.g.~\cite{Reuter:2019byg, Reuter:2012id, Niedermaier:2006ns} for pedagogical introductions):

Let the theory be defined by some microscopic action, the interacting part of which is denoted by $S_{int}$:
\be
Z[j]=\int D\phi\,\exp\left(-\frac{1}{2}\int_p \phi(-p)(p^2+m^2)\phi(p)-S_{int}[\phi] - \phi\cdot j\right)\,.
\ee
Suppressing IR fluctuations, one may write a partition function with IR cutoff $k$,
\be
Z_k[j]=\int D\phi\,\exp\left(-\frac{1}{2}\int_p \phi(-p)(p^2+m^2+R_k(p^2))\phi(p)-S_{int}[\phi] - \phi\cdot j\right)\,.
\ee
Here the cutoff function $R_k(p^2)$ vanishes for $p^2\gg k^2$ and diverges for $p^2\ll k^2$. Then, one may in standard fashion define an effective action $\Gamma_k[\phi]$ by taking the logarithm of $Z_k$ and performing a Legendre transformation. Intuitively, one can think of $\Gamma_k$ as of a `coarse-grained effective action': it encodes the information about the theory after dynamics on length scales below $1/k$ has been integrated out. The definitions imply that $\Gamma_k$ approaches the microscopic UV action as $k\to \infty$ and the standard quantum effective action, including fluctuations on all scales, as $k\to 0$. Thus, the dependence of $\Gamma_k$ on $k$ represents one way of characterising the RG evolution of the theory. 

It turns out that one may derive a simple evolution equation for $\Gamma_k$ \cite{Wetterich:1992yh}:\footnote{By
slight abuse of notation, $R_k$ now denotes the diagonal matrix $R_k(p^2)\delta^4(p-q)$.}
\be
\frac{d}{d\ln k}\,\Gamma_k=\frac{1}{2}\,\mbox{tr}\left[\left(\Gamma^{(2)}_k+R_k\right)^{-1}\!\frac{d}{d\ln k}R_k\right]\qquad\mbox{with}\qquad
\Gamma_k^{(2)}[\phi](p,q)\equiv \frac{\delta \Gamma[\phi]}{\delta\phi(p)\,\delta\phi(q)}\,.
\ee
Of course, this simplicity is somewhat misleading since in practice one has to use some form of approximation for $\Gamma_k$, for example a derivative expansion truncated after a finite number of terms. One is then dealing with a growing system of differential equations, depending on how many terms one wants to keep. The observation of \cite{Reuter:1996cp}, underlying much of the ongoing work in the literature (for a very recent review see \cite{Bonanno:2020bil}), is that a non-trivial UV fixed point arises on the basis of the first few terms in $\Gamma_k$: the cosmological constant, the Einstein-Hilbert term, and the first higher-curvature terms. More precisely, the dimensionless couplings $\lambda/k^4$, $M_P^2/k^2$, etc. appear to run to a finite ${\cal O}(1)$ values as $k\to \infty$. The interesting trajectories are those where this happens together with realistic limits for the corresponding coefficients at $k\to 0$, such that one also finds Einstein gravity in the IR. As described before, one would ideally hope that a finite-dimensional set of such trajectories exists. The dimension would be that of the UV fixed surface. All further parameters (i.e. operator coefficients) of the IR theory would then be predicted.

It is probably fair to say that the above picture has a number of open issues. One is the unavoidable truncation of $\Gamma_k$. The problem is that, while some form of truncation is technically necessary, this does not represent a controlled approximation: one is missing a small expansion parameter in the strong coupling regime $(k\gtrsim M_P)$ of gravity. Next, the diffeomorphism invariance of the cutoff is clearly an issue (though it may be traded for background dependence). Finally, one may be concerned at a more conceptual level that, excluding topological and black hole fluctuations in the UV, one is missing fundamental ingredients for a UV completion of gravity.\footnote{
The possibility has been entertained that, at some higher energy scale, asymptotic safety comes together with string theory \cite{deAlwis:2019aud}. In this case, such more extreme spacetime fluctuations, which are essential in the stringy UV completion, would enter the stage after all.
} 
To be very concrete, one may formulate the following complaint: In the asymptotic safety scenario, gravity in the UV is treated quite similarly to a standard CFT. This suggests that arbitrarily small-size and hence high-energy, localised fluctuations are part of the Hilbert space. But taking the (experimentally established) perspective of the IR observer, those small fluctuations should collapse to black holes  \cite{Dvali:2010bf, Dvali:2010jz, Dvali:2014ila}. In other words, the black hole collapse should prevent us from considering the QFT-like UV limit on which everything was built. For further critical discussions see \cite{Donoghue:2019clr}.

However, such possible criticism is not our concern here. We want for the moment to adopt the point of view that gravity is UV completed through a non-trivial UV fixed point and, moreover, that this also holds for gravity together with certain matter fields, like for example the Standard Model (see e.g.~\cite{Percacci:2003jz, Dona:2013qba, Meibohm:2015twa, Eichhorn:2017egq, Hamada:2017rvn, Christiansen:2017cxa}). This allows us to ask, as we did already in the previous section, what the consequences for the hierarchy problems might be.

The answer depends on which matter content is allowed (in the sense that the fixed point is not lost) and what the predictions for the operator coefficients at $k=0$ are. The present understanding does not suggest that the matter content is extremely constrained. On the contrary, it appears that various theories `of the right type' (with gauge groups, fermions, scalars) may emerge in the IR. Since the fixed-surface is expected to be finite-dimensional, a certain amount of predictivity should arise. However, it does not appear to be the case that the Standard Model parameters come out uniquely (i.e. that there is a unique model which runs into the fixed point).\footnote{A 
noteworthy exception is the argument presented in \cite{Shaposhnikov:2009pv} that the Higgs quartic coupling must vanish at the transition point between weakly-coupled IR and UV regime . This has lead to a successful prediction of the Higgs mass value after perturbative running to the weak scale. However, from today's perspective the prediction does not work perfectly any more since the central top mass value has shifted \cite{Degrassi:2012ry}.}

In summary, it may be a reasonable expectation that the Standard Model with gravity (supplemented by further fields, e.g. at the neutrino seesaw scale) runs into a quantum gravity UV fixed point. Some of its parameters may be constrained by this requirement. However, it remains completely unclear why the two crucial, dimensionful parameters $\lambda(0)$ and $m_H^2(0)$ are so incredibly tiny in Planck units.\footnote{
In 
this close connection with $m_H^2$ it is tempting to misread $\lambda$ as the Higgs quartic coupling. We hence remind the reader that, to distinguish it from the cutoff $\Lambda$, we use the symbol $\lambda$ for the cosmological constant.
} 
The problem is that, at the high scale $M_P$, when the assumed near-fixed-point evolution transits to the well-understood perturbative evolution of a low-energy EFT with gravity, one would need $\lambda(k)/k^4$ and $m_H^2(k)/k^2$ to be extremely small. More than that, they need to have just the right size to compensate for loop and non-perturbative effects from scales between $k=M_P$ and $k=0$. Achieving this remains a challenge (see however \cite{Wetterich:2019qzx}). Without such a mechanism, and assuming that tiny values of these parameters are consistent with the UV fixed point, one arrives at the following situation: 

The hierarchy problems can be solved, but there is a price to be paid: We would live in one of continuously many UV-consistent theories. Ours just happens to have these peculiar parameters. One may call this a fine-tuning or refuse to use this term: after all, there is only one theory realised in nature. Compared to the string (or any other) landscape, one has lost the option of explaining the very special observed parameter values using cosmology plus anthropics. But this may be just fine since, given the absence a natural measure on the critical surface, it is hard to claim that the observed parameters are special in the first place.

\section{Summary}\label{summ}
Let us recapitulate what one may have learned by going through these notes. First, we have learned to view the Standard Model as an effective field theory with a number of issues: too many (and too random) parameters, especially in the Yukawa sector; no dark matter candidate; no mechanism for producing the baryon asymmetry of the universe; no obvious inflaton candidate. Moreover, and this was our central theme, one faces a hierarchy problem concerning the smallness of the Higgs mass parameter $m_H^2$ relative to the cutoff scale $\Lambda^2$. Finally, including gravity as a low-energy effective field theory (EFT) in our framework, a very similar second hierarchy problem between the cosmological constant $\lambda$ and the relevant cutoff scale $\Lambda^4$ was seen to arise. We were careful to spell out these problems more quantitatively: They are related to an enormous sensitivity of the low-energy theory to the exact value of any new-physics parameters and mass scales that can (and probably have to) be present between the electroweak and Planck scale.

While no widely accepted solution to the second hierarchy problem (the cosmological constant problem) exists, the electroweak hierarchy problem can at least be drastically reduced. The essence of this resolution is to introduce an intermediate cutoff scale, let us call it $\Lambda_{BSM}$, at which the Standard Model is replaced by a theory in which a light scalar (like the Higgs) may naturally coexist with a much higher fundamental cutoff scale $\Lambda$. Our focus was on low-energy supersymmetry as a concrete realisation of this idea, i.e. $\Lambda_{BSM}\equiv \Lambda_{SUSY}\gtrsim m_{ew}$, where $m_{ew}$ stands for the electroweak scale. The problem is that this solution works less and less well as experiments (most recently the LHC) push $\Lambda_{SUSY}$ significantly above $m_{ew}$.

We have also discussed how, in the presence of gravity, supersymmetry must be replaced by supergravity. Here we had our first encounter with complex geometry since the scalar fields now live on a Kahler manifold. The scalar potential is a function on that manifold and derives (at least partially) from the superpotential. The latter is a section in a line bundle over the aforementioned Kahler manifold. Unfortunately, this elegant description of the world through quantities like the Kahler potential $K$ and superpotential $W$ does not help with the cosmological constant problem: Loop corrections to $K$ are roughly speaking as bad as those to $\lambda$. Concerning the electroweak hierarchy, things do not improve compared to rigid SUSY.
Maybe most importantly, the infinite-dimensional continuous freedom of choosing EFTs is not reduced -- it is only repackaged in the freedom of choosing $K$ and $W$.

Given this unsatisfactory state of affairs, it is natural to keep pushing for a more fundamental understanding at the highest scale accessible in this line of thought: the quantum gravity scale. Here, an at first sight rather far-fetched idea offers surprising insights: It is the suggestion to identify elementary particles with loops of fundamental string. This approach naturally cuts off the perturbative infinities of quantum gravity, but only at the price of 26 spacetime dimensions and, moreover, an unstable (tachyonic) vacuum. The latter problem is solved in the supersymmetric version of the theory, which has a stable supersymmetric 10d Minkowski vacuum. The problem of too many spatial dimensions can be overcome by compactifying six of them (in the simplest case on Calabi-Yau manifolds). This clearly leads to ambiguities concerning the resulting 4d EFT. Moreover, even in 10d five different highly supersymmetric EFTs result from slightly different ways of defining the superstring. Yet, miraculously, the concept of dualities ties all constructions (together with a highly supersymmetric 11d model with fundamental 2-branes) into a single theory.
We recall here that by `duality' we mean a situation where two different mathematical formulations define exactly the same physical theory.

At this point, one has arrived at a possibly fundamental and essentially unique theory (or model) of quantum gravity which has no free parameters. The only exception is the string scale, which however simply sets the fundamental energy scale of the world. But there is a large number of 4d solutions, related to the concrete choice of 10d theory (one in five) and of Calabi-Yau space (one in about half a billion). Very elegantly, the 3d complex geometry of the Calabi-Yaus is governed by the (${\cal O}(100)$-dimensional) complex geometry of the Calabi-Yau moduli spaces. These are precisely the Kahler manifolds of the 4d supergravity models describing the low-energy EFTs of these compactifications. The massless scalar fields or moduli encode 10d metric degrees of freedom and determine the precise size and shape of the Calabi-Yau geometry.

But this is only the beginning of the actual String Landscape: The 10d theory possesses non-perturbative objects (such as D-branes and orientifold planes) and allows for non-zero expectation values of $p$-form gauge fields (fluxes). Equipping the Calabi-Yaus with those leads to an explosion of the number of 4d solutions, a recent estimate being as high as $10^{272,000}$. This so-called flux landscape plausibly contains many 4d models (flux vacua) with a realistic, Standard-Model-like matter sector. It is more complicated to show explicitly but plausible that vacua with broken supersymmetry and a positive cosmological constant are present in this large set of solutions. The best concrete examples are known under the acronyms KKLT and LVS.

Once this is accepted as a matter of principle (or better: rigorously established in the future), one expects that {\it extremely many} such vacua, with the right structure to describe the real world, exist. Thus, all EFT parameters and in particular the cosmological constant and the Higgs mass squared are very finely scanned in the landscape. One still has a unique fundamental theory. But, through its sheer abundance of solutions, it may accommodate what the low-energy observer perceives as extremely fine-tuned parameters. In the case of the electroweak hierarchy problem, this tuning may bridge a small or even very large gap between SUSY breaking scale and weak scale, i.e. both $\Lambda_{SUSY}\gg m_{ew}$ and $\Lambda_{SUSY}\gtrsim m_{ew}$ are compatible with this picture.

But to accommodate is not the same as to explain or even predict. Indeed, one may feel surprised and unsatisfied by finding oneself in a `very special' landscape vacuum, with (at least) two parameters chosen in a highly non-generic way.  To quantify such a surprise or unease, one needs a measure on the landscape. Since the landscape is discrete (under mild conditions probably even finite), a natural choice might be that of giving each vacuum an equal weight. Then, by all that we know only a very small fraction of vacua is closer to the special point $(\,\lambda=0, m_H^2=0\,)$ than our observed Standard-Model-EFT. One could say that we have hence not avoided the fine-tuning but shown that it can be explicitly realised. 

However, one may also say that finding ourselves in one of these very special vacua is not surprising after all: Indeed, it is easy to argue that any kind of observer (limited by what we can presently imagine) can only form if some structure in the universe forms first. One presumably also requires some scale separation between the energy scales of chemistry and the Planck scale. Involving this so-called anthropic argument, one may say that we find ourselves near this particular point in the landscape simply because other vacua have no observers.

Clearly, this is very rough and not at all quantitative. Also, the proposed measure of counting vacua lacks justification. It would be much better to have a theory of how the various vacua get populated cosmologically and how to ask an observer-dependent question as we just did in a more objective manner. One option that has been explored in some detail is that of eternal inflation during which, starting with one of the metastable de Sitter vacua, all of them get realised cosmologically through consecutive tunnelling processes. This eternal evolution of `bubbles within bubbles within bubbles', while in principle capable of populating the whole landscape, comes with its own issues. Maybe the most problematic is the so-called measure problem: Namely, due to the divergence of the number of bubbles (and hence of potential observers) at late times, one needs a cutoff. With such a late-time cutoff imposed, one may count observers and hence derive a measure. However, due to diffeomorphism invariance of general relativity, there appears to be no unambiguously preferred choice of cutoff and hence no established first-principles measure.

To illustrate what is at stake, let us imagine the landscape were understood well-enough to know precisely which vacua with which features it contains and what the transition rates between them are. Moreover, let us also assume an a-priori, quantum-gravity-derived measure could then be established on this basis. As a result, one may ask questions of the following type: Given all that we know about our vacuum, how many observers share all these observations and live in a vacuum with a low SUSY breaking scale? By contrast, how many observers share all our observations and live in a vacuum with high-scale SUSY breaking? The resulting numbers predict probabilistically what we expect to find in our EFT at the next energy frontier. While we have only a single experiment to perform, such a probabilistic prediction may still be meaningful since the ratio of these two probabilities can be exponentially large. Under such circumstances we may get to rule out a theory at many standard deviations based on a single observation.

The last two paragraphs attempted to take the string landscape idea (and specifically its implications for the electroweak hierarchy problem) to an idealised endpoint of purely statistical predictions about future measurements. Thinking that far may be interesting and important, but it is also relatively speculative and far-fetched. Many more modest and more approachable questions can be asked. First, key aspects of the string landscape are not understood. We are very far from having an overview of the landscape as a whole. Next, we may use the landscape not to make statistical predictions but to draw inspiration for what could be observed in cosmology and particle phenomenology in the future. This may be a more immediate way in which string theory can relate to experiment. Finally, it is interesting to investigate whether there exist consistent EFTs which we could find realised in our universe and which do {\it not} represent any of the string landscape vacua.

This last point deserves a more careful discussion. It is remarkable that, due to an observation of this type, a single experiment could in principle rule out string theory with certainty, without the need to appeal to statistics. Moreover, the approach of studying which EFTs are not present in the landscape is very popular at the moment of this writing. It is known as the Swampland program and its claims about the inconsistency of certain EFTs are characterised as Swampland conjectures. Some of them even try to exclude metastable de Sitter vacua in complete generality. While this does not immediately rule out string theory (since the observed exponential expansion of our  universe could be due to dynamical dark energy), it certainly clashed with most of what we thought we have learned about string theory phenomenology in recent years. Thus, it is of immense importance to either establish those claims or, on the contrary, to further develop our quantitative understanding of the proposed constructions of metastable de Sitter vacua. In this context, much more precision and explicitness is certainly desirable.

The landscape approach to thinking about and maybe resolving hierarchy problems, especially once anthropic arguments are invoked, has received a fair amount of criticism. This is understandable given how enormous a paradigm change is involved. It also has to be admitted that restricting the predictivity of fundamental scientific thought to (at least at low energies) only probabilistic statements may be perceived as frightening. However, as we have tried to argue in the previous subsection, the alternatives have significant shortcomings as well. Indeed, let us assume a unique theory of quantum gravity is established which, as opposed to string theory, does not possess a landscape of solutions. Then one clearly expects some simple formula for e.g. the operator coefficient that we call the cosmological constant to be provided by this theory at some energy scale $\mu$ just below $M_P$. But now, at least according to all that we know, it is very hard to see how such a simple fundamental formula would combine with the known loop effects to give the observed value $\lambda(\mu=0)\sim 10^{-120} M_P^4$. Nevertheless, this possibility can not be ruled out.

Alternatively, it is conceivable that the correct theory of quantum gravity is not unique but comes with continuous parameters. This would allow one to literally tune those fundamental parameters to realise the hierarchies observed in our low-energy EFTs. One may call this unsatisfactory, but it is hard to make an objective point against this option.

At the moment, we have to let these different attitudes to the hierarchy problems coexist and compete. In addition to studying string theory, it is certainly interesting to look for fundamental theories capable of predicting Higgs mass parameter and cosmological constant or to at least explicitly realise the tuning. We should also not forget that dynamical resolutions of the hierarchy problems have not been proved to be impossible. Just because no convincing version for the cosmological constant has been found and because the simplest SUSY models are under pressure, one can not be certain that the resolution will not, after all, come from a very particular EFT and its dynamics. Finally, recent ideas of cosmological selection, which combine elements of EFT dynamics and a landscape, may turn out to be correct.

While all the options above are very interesting, they were of course not our main subject. Our goal was to develop, in some technical detail, specifically the string landscape view on the hierarchy problems. In two sentences, the result is as follows: Through its immense number of solutions, string theory as  a very concrete model of quantum gravity may be capable of accommodating low-energy EFTs with an extremely fine-tuned appearance. Depending on how much `anthropics' and eternal-inflation cosmology one is willing to involve, this may even be promoted to an explanation or prediction. 

The readers will make up their own mind and decide which of the above directions to pursue or which new ideas to propose. Hopefully, these notes can be useful for making such choices on the basis of a somewhat more technical understanding of the string landscape.

\subsection*{Acknowledgements}
Many friends and colleagues deserve my deepest gratitude for helping me to learn und understand the material covered in this course. Among those, I am particularly indebted to Wilfried Buchm\"uller, Miriam Cvetic, Jan Louis,  Dieter L\"ust, John March-Russell, Timo Weigand, and Alexander Westphal. Especially the time during which Timo Weigand was my colleague in Heidelberg was extremely fruitful and enjoyable. Moreover, I am grateful to my younger colleagues Daniel Junghans and Sascha Leonhardt who were involved in the actual teaching of this course. Special thanks go to Bj\"orn Friedrich, Daniel Junghans and Alexander Westphal for their careful reading of the manuscript and their corrections and comments. In addition, I am grateful to Janning Meinert for producing professional figures.
There is a long list of colleagues and friends whose help and advice were very important for me and eventually for these notes. They include Andreas Braun, Felix Br\"ummer, Joe Conlon, Laura Covi, Gia Dvali, Ben Freivogel, Gero von Gersdorff, Benedict von Harling, Joerg Jaeckel, Olaf Lechtenfeld, Hans-Peter Nilles, Eran Palti, Jan Pawlowski, Tilman Plehn, Fernando Quevedo, Riccardo Rattazzi, Michael Ratz, Fabrizio Rompineve, Micheal G. Schmidt, Gary Shiu, Pablo Soler, Stefan Theisen, Gianmassimo Tasinato, Michele Trapletti, Enrico Trincherini, Roberto Valandro, Johannes Walcher, James Wells, Julius Wess, Christof Wetterich, Lukas Witkowski, Timm Wrase and many others. Each of them clearly deserves to be mentioned more personally and I apologise for not being able to do so for reasons of space. Also, I ask for the forgiveness of those friends and colleagues whom I have missed. Finally, I would like to thank my parents for encouragement throughout my whole life and, most importantly, my wife, Andrea, for her love and for tolerating the excessive working hours that were sometimes unavoidable in preparing these notes. This work was supported by Deutsche Forschungsgemeinschaft (DFG, German Research Foundation) under Germany's Excellence Strategy EXC-2181/1 - 390900948 (the Heidelberg STRUCTURES Cluster of Excellence).



\begin{thebibliography}{999}

\bibitem{Peskin:1995ev}
M.~E.~Peskin and D.~V.~Schroeder, ``An Introduction to Quantum Field Theory,'' Westview Press, 1995.

\bibitem{Wald:1984rg}
R.~M.~Wald, ``General Relativity,'' The University of Chicago Press, 1984.

\bibitem{Cheng:1985bj}
T.~Cheng and L.~Li, ``Gauge Theory of Elementary Particle Physics'' and ``Gauge Theory of Elementary Particle Physics -- Problems and Solutions'', Oxford University Press, 1984 and 2000, respectively.

\bibitem{Donoghue:1992dd}
J.~Donoghue, E.~Golowich and B.~R.~Holstein,
``Dynamics of the Standard Model,''
Camb. Monogr. Part. Phys. Nucl. Phys. Cosmol. \textbf{2} (1992), 1-540.

\bibitem{Wess:1992cp}
J.~Wess and J.~Bagger, ``Supersymmetry and Supergravity,'' Princeton University Press, 1991.

\bibitem{Freedman:2012zz}
D.~Z.~Freedman and A.~Van Proeyen, ``Supergravity,'' Cambridge University Press, 2012.

\bibitem{Polchinski:1998rq}
J.~Polchinski, ``String Theory'', Vol. 1 and 2, Cambridge University Press, 2001.

\bibitem{Blumenhagen:2013fgp}
R.~Blumenhagen, D.~L\"ust and S.~Theisen,
``Basic Concepts of String Theory,'' Springer, 2013.

\bibitem{Ibanez:2012zz}
L.~E.~Ibanez and A.~M.~Uranga,
``String Theory and Particle Physics: An Introduction to String Phenomenology,'' Cambridge University Press, 2012.

\bibitem{Denef:2008wq}
  F.~Denef, ``Les Houches lectures on constructing string vacua,'' Les Houches {\bf 87} (2008) 483 [arXiv:0803.1194]. 
  
\bibitem{Schellekens:2013bpa}
  A.~N.~Schellekens,
  ``Life at the interface of particle physics and string theory,''
  Rev.\ Mod.\ Phys.\  {\bf 85} (2013) no.4,  1491
  [arXiv:1306.5083 [hep-ph]].

\bibitem{Weinberg:1995mt}
S.~Weinberg, ``The Quantum Theory of Fields'', Vol. 1 and 2, Cambridge University Press, 2005.

\bibitem{Itzykson:1980rh}
C.~Itzykson and J.~Zuber, ``Quantum Field Theory,''  McGraw-Hill, 1980.

\bibitem{Srednicki:2007qs}
M.~Srednicki, ``Quantum Field Theory,'' Cambridge University Press, 2007.

\bibitem{Schwartz:2013pla}
M.~D.~Schwartz, ``Quantum Field Theory and the Standard Model,'' Cambridge University Press, 2014.

\bibitem{Nachtmann:1990ta}
O.~Nachtmann, ``Elementary Particle Physics: Concepts and Phenomena, Springer, 1990.

\bibitem{sche} Schellekens: Beyond the Standard Model, 
\url{https://www.nikhef.nl/~t58/lectures.html}

\bibitem{Blanke:2017ohr}
  M.~Blanke,
  ``Introduction to flavour physics and CP violation,''
  CERN Yellow Rep.\ School Proc.\  {\bf 1705} (2017) 71
  [arXiv:1704.03753 [hep-ph]].

\bibitem{Grinstein:2017pvg}
B.~Grinstein, ``Lectures on flavor physics and CP violation,''
[arXiv:1701.06916 [hep-ph]].

\bibitem{kt} Kooijman/Tuning: Lectures on CP violation,\\
\url{https://www.nikhef.nl/~h71/Lectures/2015/ppII-cpviolation-19032018.pdf}

\bibitem{Bigi:2000yz}
I.~I.~Bigi and A.~Sanda, ``CP violation,''
Camb. Monogr. Part. Phys. Nucl. Phys. Cosmol. \textbf{9} (2009), 1-485.

\bibitem{Fleischer:2006fx}
R.~Fleischer, ``Flavour physics and CP violation,'' Lectures at the 2005 European School of High-Energy Physics [arXiv:hep-ph/0608010 [hep-ph]].

\bibitem{Branco:1999fs}
G.~C.~Branco, L.~Lavoura and J.~P.~Silva,
``CP violation,'' Int. Ser. Monogr. Phys. \textbf{103} (1999), 1-536

\bibitem{Anselm:1993uj}
A.~A.~Anselm and A.~A.~Johansen,
  ``Can electroweak theta term be observable?,'' Nucl.\ Phys.\ B {\bf 412}
  (1994) 553 [hep-ph/9305271].

\bibitem{Cao:2017ocv}
  C.~Cao and A.~Zhitnitsky,
  ``Axion detection via topological Casimir effect,''
  Phys.\ Rev.\ D {\bf 96} (2017) no.1,  015013 [arXiv:1702.00012 [hep-ph]].

\bibitem{Georgi:1994qn}
H.~Georgi,
``Effective field theory,''
Ann. Rev. Nucl. Part. Sci. \textbf{43} (1993), 209-252.

\bibitem{Manohar:1996cq}
A.~V.~Manohar,
``Effective field theories,''
Lect. Notes Phys. \textbf{479} (1997), 311-362
[arXiv:hep-ph/9606222 [hep-ph]].

\bibitem{Pich:1998xt}
A.~Pich,
``Effective field theory: course,'' Lectures at Les Houches Summer School 1998
[arXiv:hep-ph/9806303 [hep-ph]].

\bibitem{Luty:2005sn}
  M.~Luty, ``2004 TASI lectures on supersymmetry breaking,''
  hep-th/0509029.

\bibitem{Kaplan:2005es}
  D.~B.~Kaplan,
  ``Five lectures on effective field theory,'' nucl-th/0510023.

\bibitem{Cohen:2019wxr}
T.~Cohen,
``As scales become separated: Lectures on effective field theory,''
PoS \textbf{TASI2018} (2019), 011
[arXiv:1903.03622 [hep-ph]].

\bibitem{Wells:2012rla}
J.~D.~Wells,
``Effective Theories in Physics -- From Planetary Orbits to Elementary Particle Masses'', Springer, 2012.

\bibitem{Minkowski:1977sc}
P.~Minkowski,
``$\mu \to e\gamma$ at a rate of one out of $10^{9}$ muon decays?,''
Phys. Lett. B \textbf{67} (1977), 421-428.

\bibitem{Yanagida:1979as}
T.~Yanagida,
``Horizontal gauge symmetry and masses of neutrinos,''
Conf. Proc. C \textbf{7902131} (1979), 95-99, 
KEK-79-18-95.

\bibitem{GellMann:1980vs}
M.~Gell-Mann, P.~Ramond and R.~Slansky,
``Complex spinors and unified theories,''
Conf. Proc. C \textbf{790927} (1979), 315-321
[arXiv:1306.4669 [hep-th]].

\bibitem{Gildener:1976ai}
E.~Gildener, ``Gauge symmetry hierarchies,''
Phys. Rev. D \textbf{14} (1976), 1667.

\bibitem{Veltman:1980mj}
M.~Veltman, ``The infrared - ultraviolet connection,''
Acta Phys. Polon. B \textbf{12} (1981), 437.

\bibitem{tHooft:1980xss}
  G.~'t Hooft, C.~Itzykson, A.~Jaffe, H.~Lehmann, P.~K.~Mitter, I.~M.~Singer
  and R.~Stora,
  ``Recent Developments in Gauge Theories. Proceedings, Nato Advanced Study 
  Institute, Cargese, France, August 26 - September 8, 1979,''
  NATO Sci.\ Ser.\ B {\bf 59} (1980) pp.1.

\bibitem{Barbieri:1987fn}
  R.~Barbieri and G.~F.~Giudice,
  ``Upper Bounds on Supersymmetric Particle Masses,''
  Nucl.\ Phys.\ B {\bf 306} (1988) 63.

\bibitem{Ellis:1986yg}
  J.~R.~Ellis, K.~Enqvist, D.~V.~Nanopoulos and F.~Zwirner,
  ``Observables in Low-Energy Superstring Models,''
  Mod.\ Phys.\ Lett.\ A {\bf 1} (1986) 57.

\bibitem{Wells:2018sus}
  J.~D.~Wells,
  ``Naturalness, Extra-Empirical Theory Assessments, and the Implications of
  Skepticism,'' arXiv:1806.07289 and 
  ``Finetuned Cancellations and Improbable Theories,'' arXiv:1809.03374.

\bibitem{Azhar:2018lzd}
F.~Azhar and A.~Loeb,
``Gauging Fine-Tuning,''
Phys. Rev. D \textbf{98} (2018) no.10, 103018
[arXiv:1809.06220 [astro-ph.CO]].

\bibitem{Tanabashi:2018oca}
  M.~Tanabashi {\it et al.} [Particle Data Group],
  ``Review of Particle Physics,''
  Phys.\ Rev.\ D {\bf 98} (2018) no.3,  030001.
  
\bibitem{Kass:1995loi}
R.~E.~Kass and A.~E.~Raftery,
``Bayes Factors,''
J. Am. Statist. Assoc. \textbf{90} (1995) no.430, 773-795.  

\bibitem{Trotta:2008qt}
R.~Trotta,
``Bayes in the sky: Bayesian inference and model selection in cosmology,''
Contemp. Phys. \textbf{49} (2008), 71-104
[arXiv:0803.4089 [astro-ph]].  

\bibitem{Allanach:2007qk}
B.~C.~Allanach, K.~Cranmer, C.~G.~Lester and A.~M.~Weber,
``Natural priors, CMSSM fits and LHC weather forecasts,''
JHEP \textbf{08} (2007), 023
[arXiv:0705.0487 [hep-ph]].

\bibitem{Cabrera:2008tj}
M.~E.~Cabrera, J.~A.~Casas and R.~Ruiz de Austri,
``Bayesian approach and Naturalness in MSSM analyses for the LHC,''
JHEP \textbf{03} (2009), 075
[arXiv:0812.0536 [hep-ph]].

\bibitem{Fichet:2012sn}
S.~Fichet,
``Quantified naturalness from Bayesian statistics,''
Phys. Rev. D \textbf{86} (2012), 125029
[arXiv:1204.4940 [hep-ph]].

\bibitem{Fowlie:2014xha}
A.~Fowlie,
``CMSSM, naturalness and the "fine-tuning price" of the Very Large Hadron Collider,''
Phys. Rev. D \textbf{90} (2014), 015010
[arXiv:1403.3407 [hep-ph]].

\bibitem{Wells:2009kq}
J.~D.~Wells, ``Lectures on Higgs Boson Physics in the Standard Model and Beyond,''
[arXiv:0909.4541 [hep-ph]]

\bibitem{Giudice:2013yca}
G.~F.~Giudice,
``Naturalness after LHC8,''
PoS \textbf{EPS-HEP2013} (2013), 163\\
$\mbox{[arXiv:1307.7879]}$ and
``The Dawn of the Post-Naturalness Era,''
[arXiv:1710.07663].

\bibitem{Craig:2022uua}
N.~Craig,
``Naturalness: A Snowmass White Paper,''
[arXiv:2205.05708 [hep-ph]].

\bibitem{wei} S.~Weinberg: ``Gravitation and Cosmology,''
John Wiley and Sons, 1972.

\bibitem{mtw} C.$\,$S.~Misner, K.$\,$W.~Thorne, J.$\,$A.~Wheeler: ``Gravitation,'' Princeton University Press, 2017.

\bibitem{car} S.~M.~Carroll: ``Spacetime and Geometry,'' Cambridge University Press, 2019.

\bibitem{strau} N.~Straumann: ``General Relativity,'' Springer, 2013.

\bibitem{Gibbons:1978ac}
G.~Gibbons, S.~Hawking and M.~Perry,
``Path Integrals and the Indefiniteness of the Gravitational Action,'' Nucl. Phys. B \textbf{138} (1978), 141-150.

\bibitem{vel} M. Veltman, ``Quantum theory of gravitation'', in: {\it Methods in Field Theory, Les Houches 1975}, edited by R. Balian and J. Zinn-Justin (North-Holland Publ., Amsterdam, the Netherlands, 1976), Course 5.

\bibitem{Weinberg:1988cp}
S.~Weinberg,
``The Cosmological Constant Problem,''
Rev. Mod. Phys. \textbf{61} (1989), 1-23.

\bibitem{Weinberg:2000yb}
S.~Weinberg,
``The Cosmological constant problems,''
Talk at 4th International Symposium ``Dark Matter 2000''
[arXiv:astro-ph/0005265 [astro-ph]].

\bibitem{Padmanabhan:2002ji}
T.~Padmanabhan,
``Cosmological constant: The Weight of the vacuum,''
Phys. Rept. \textbf{380} (2003), 235-320
[arXiv:hep-th/0212290 [hep-th]].

\bibitem{Padmanabhan:2006ag}
T.~Padmanabhan,
``Dark energy: mystery of the millennium,''
AIP Conf. Proc. \textbf{861} (2006) no.1, 179-196
[arXiv:astro-ph/0603114 [astro-ph]].

\bibitem{Georgi:1999wka}
H.~Georgi, ``Lie algebras in Particle Physics,''
Front. Phys. \textbf{54} (1999), 1-320.

\bibitem{Slansky:1981yr}
R.~Slansky, ``Group Theory for Unified Model Building,''
Phys. Rept. \textbf{79} (1981), 1-128.

\bibitem{Ross:1985ai}
G.~G.~Ross, ``Grand Unified Theories'', Westview Press, 2003.

\bibitem{Nath:2006ut}
P.~Nath and P.~Fileviez Perez,
``Proton stability in grand unified theories, in strings and in branes,'' Phys. Rept. \textbf{441} (2007), 191-317
[arXiv:hep-ph/0601023 [hep-ph]].

\bibitem{Raby:2017ucc}
S.~Raby, ``Supersymmetric Grand Unified Theories,'' Lect. Notes Phys. \textbf{939} (2017), 1-308.

\bibitem{hh} 
A.~Hebecker and J.~Hisano, ``Grand Unified Theories'', Review article in Ref.~\cite{Tanabashi:2018oca}.

\bibitem{Croon:2019kpe}
D.~Croon, T.~E.~Gonzalo, L.~Graf, N.~Kosnik and G.~White, ``GUT Physics in the era of the LHC,''
Front. in Phys. \textbf{7} (2019), 76
[arXiv:1903.04977 [hep-ph]].

\bibitem{Coleman:1967ad}
  S.~R.~Coleman and J.~Mandula,
  ``All Possible Symmetries of the S Matrix,''
  Phys.\ Rev.\  {\bf 159} (1967) 1251.

\bibitem{Haag:1974qh}
  R.~Haag, J.~T.~Lopuszanski and M.~Sohnius,
  ``All Possible Generators of Supersymmetries of the s Matrix,''
  Nucl.\ Phys.\ B {\bf 88} (1975) 257.

\bibitem{west} P.~West, ``Introduction to Supersymmetry and Supergravity'', World Scientific, 1990.

\bibitem{weinberg3} S.~Weinberg, ``Quantum Field Theory'', Vol. 3, Cambridge University Press, 2005.

\bibitem{turning} J.~Terning, ``Modern Supersymmetry'', Oxford University Press, 2006.

\bibitem{shifman} M.~Shifman, ``Advanced Topics in Quantum Field Theory'', Cambridge University Press, 2012.

\bibitem{Wess:1974tw}
J.~Wess and B.~Zumino,
``Supergauge Transformations in Four-Dimensions,''
Nucl. Phys. B \textbf{70} (1974), 39-50.

\bibitem{Volkov:1973ix}
D.~Volkov and V.~Akulov,
``Is the Neutrino a Goldstone Particle?,''
Phys. Lett. B \textbf{46} (1973), 109-110.

\bibitem{ORaifeartaigh:1975nky}
L.~O'Raifeartaigh,
``Spontaneous Symmetry Breaking for Chiral Scalar Superfields,''
Nucl. Phys. B \textbf{96} (1975), 331-352.

\bibitem{Fayet:1974jb}
P.~Fayet and J.~Iliopoulos,
``Spontaneously Broken Supergauge Symmetries and Goldstone Spinors,''
Phys. Lett. B \textbf{51} (1974), 461-464.

\bibitem{Martin:1997ns}
  S.~P.~Martin,
  ``A Supersymmetry primer,''
  Adv.\ Ser.\ Direct.\ High Energy Phys.\  {\bf 21} (2010) 1
  [hep-ph/9709356].

\bibitem{Giudice:1998bp}
  G.~F.~Giudice and R.~Rattazzi,
  ``Theories with gauge mediated supersymmetry breaking,''
  Phys.\ Rept.\  {\bf 322} (1999) 419
  [hep-ph/9801271].

\bibitem{Plehn:2017fdg}
M.~Bauer and T.~Plehn,
``Yet Another Introduction to Dark Matter,''
Lect. Notes Phys. \textbf{959} (2019), pp.
[arXiv:1705.01987 [hep-ph]].

\bibitem{bertone}
Gianfranco Bertone (ed.), `` Particle Dark Matter'', Cambridge University Press, 2010.

\bibitem{Hooper:2009zm}
D.~Hooper,
``Particle Dark Matter,'' TASI lectures
[arXiv:0901.4090 [hep-ph]].

\bibitem{Olive:2003iq}
K.~A.~Olive,
``TASI lectures on dark matter,''
[arXiv:astro-ph/0301505 [astro-ph]].

\bibitem{Seiberg:1993vc}
  N.~Seiberg, ``Naturalness versus supersymmetric nonrenormalization theorems,''
  Phys.\ Lett.\ B {\bf 318} (1993) 469 [hep-ph/9309335].

\bibitem{Giudice:1988yz}
G.~Giudice and A.~Masiero,
``A Natural Solution to the mu Problem in Supergravity Theories,'' Phys. Lett. B \textbf{206} (1988), 480-484.

\bibitem{Dimopoulos:1981yj}
S.~Dimopoulos, S.~Raby and F.~Wilczek,
``Supersymmetry and the Scale of Unification,''
Phys. Rev. D \textbf{24} (1981), 1681-1683.

\bibitem{Dimopoulos:1981zb}
S.~Dimopoulos and H.~Georgi,
``Softly Broken Supersymmetry and SU(5),''
Nucl. Phys. B \textbf{193} (1981), 150-162.

\bibitem{Ibanez:1981yh}
L.~E.~Ibanez and G.~G.~Ross,
``Low-Energy Predictions in Supersymmetric Grand Unified Theories,''
Phys. Lett. B \textbf{105} (1981), 439-442.

\bibitem{Sakai:1981gr}
N.~Sakai, ``Naturalness in Supersymmetric Guts,''
Z. Phys. C \textbf{11} (1981), 153.

\bibitem{Amaldi:1991cn}
U.~Amaldi, W.~de Boer and H.~Furstenau,
``Comparison of grand unified theories with electroweak and strong coupling constants measured at LEP,''
Phys. Lett. B \textbf{260} (1991), 447-455.

\bibitem{ArkaniHamed:2004fb}
  N.~Arkani-Hamed and S.~Dimopoulos,
  ``Supersymmetric unification without low energy supersymmetry and 
  signatures for fine-tuning at the LHC,'' JHEP {\bf 0506} (2005) 073
  [hep-th/0405159].

\bibitem{Giudice:2004tc}
  G.~F.~Giudice and A.~Romanino,
  ``Split supersymmetry,''
  Nucl.\ Phys.\ B {\bf 699} (2004) 65
  Erratum: [Nucl.\ Phys.\ B {\bf 706} (2005) 487] [hep-ph/0406088].

\bibitem{Denef:2004cf}
  F.~Denef and M.~R.~Douglas,
  ``Distributions of nonsupersymmetric flux vacua,''
  JHEP {\bf 0503} (2005) 061
  [hep-th/0411183].

\bibitem{Hebecker:2012qp}
  A.~Hebecker, A.~K.~Knochel and T.~Weigand,
  ``A Shift Symmetry in the Higgs Sector: Experimental Hints and Stringy
  Realizations,''
  JHEP {\bf 1206} (2012) 093 [arXiv:1204.2551 [hep-th]].

\bibitem{bk} Buchbinder/Kuzenko, ``Ideas and Methods of Supersymmetry and Supergravity'', Institute of Physics Publishing, 1995.

\bibitem{Quevedo:2010ui}
  F.~Quevedo, S.~Krippendorf and O.~Schlotterer,
  ``Cambridge Lectures on Supersymmetry and Extra Dimensions,''
  arXiv:1011.1491 [hep-th].
  
\bibitem{Villadoro:2005yq}
G.~Villadoro and F.~Zwirner,
``De-Sitter vacua via consistent D-terms,''
Phys. Rev. Lett. \textbf{95} (2005), 231602
[arXiv:hep-th/0508167 [hep-th]].

\bibitem{Komargodski:2009pc}
Z.~Komargodski and N.~Seiberg,
``Comments on the Fayet-Iliopoulos Term in Field Theory and Supergravity,''
JHEP \textbf{06} (2009), 007
[arXiv:0904.1159 [hep-th]].

\bibitem{Dienes:2009td}
K.~R.~Dienes and B.~Thomas,
``On the Inconsistency of Fayet-Iliopoulos Terms in Supergravity Theories,''
Phys. Rev. D \textbf{81} (2010), 065023
[arXiv:0911.0677 [hep-th]].

\bibitem{Green:1987sp}
 M.~B.~Green, J.~Schwarz and E.~Witten, ``Superstring theory'', vol.~I and II, Cambridge University Press, 1987.

\bibitem{bbs} K.~Becker, M.~Becker and J.~H.~Schwarz, ``String Theory and M-theory'', Cambridge University Press, 2007.

\bibitem{kir} E.~Kiritsis, ``String Theory in a Nutshell'', Princeton University Press, 2007.

\bibitem{zw} B.~Zwiebach, ``A First Course in String Theory'', Cambridge University Press, 2009.

\bibitem{Deser:1976rb}
S.~Deser and B.~Zumino,
``A Complete Action for the Spinning String,''
Phys. Lett. B \textbf{65} (1976), 369-373.

\bibitem{Brink:1976sc}
L.~Brink, P.~Di Vecchia and P.~S.~Howe,
``A Locally Supersymmetric and Reparametrization Invariant Action for the Spinning String,''
Phys. Lett. B \textbf{65} (1976), 471-474.

\bibitem{Polyakov:1981rd}
A.~M.~Polyakov,
``Quantum Geometry of Bosonic Strings,''
Phys. Lett. B \textbf{103} (1981), 207-210.

\bibitem{Johnson:2000ch}
C.~V.~Johnson,
``D-brane primer,''
[arXiv:hep-th/0007170 [hep-th]] and ``D-branes'', Cambridge University Press, 2003.

\bibitem{Dai:1989ua}
J.~Dai, R.~G.~Leigh and J.~Polchinski,
``New Connections Between String Theories,''
Mod. Phys. Lett. A \textbf{4} (1989), 2073-2083.

\bibitem{Leigh:1989jq}
R.~G.~Leigh,
``Dirac-Born-Infeld Action from Dirichlet Sigma Model,''
Mod. Phys. Lett. A \textbf{4} (1989), 2767.

\bibitem{Blumenhagen:2005mu}
R.~Blumenhagen, M.~Cvetic, P.~Langacker and G.~Shiu,
``Toward realistic intersecting D-brane models,''
Ann. Rev. Nucl. Part. Sci. \textbf{55} (2005), 71-139
[arXiv:hep-th/0502005 [hep-th]].

\bibitem{Blumenhagen:2006ci}
R.~Blumenhagen, B.~Kors, D.~Lust and S.~Stieberger,
``Four-dimensional String Compactifications with D-Branes, Orientifolds and Fluxes,''
Phys. Rept. \textbf{445} (2007), 1-193
[arXiv:hep-th/0610327 [hep-th]].

\bibitem{Ibanez:2001nd}
L.~E.~Ibanez, F.~Marchesano and R.~Rabadan,
``Getting just the standard model at intersecting branes,''
JHEP \textbf{11} (2001), 002
[arXiv:hep-th/0105155 [hep-th]].

\bibitem{Blumenhagen:2001te}
R.~Blumenhagen, B.~Kors, D.~Lust and T.~Ott,
``The standard model from stable intersecting brane world orbifolds,''
Nucl. Phys. B \textbf{616} (2001), 3-33
[arXiv:hep-th/0107138 [hep-th]].

\bibitem{Cvetic:2001nr}
M.~Cvetic, G.~Shiu and A.~M.~Uranga,
``Chiral four-dimensional N=1 supersymmetric type 2A orientifolds from intersecting D6 branes,''
Nucl. Phys. B \textbf{615} (2001), 3-32
[arXiv:hep-th/0107166 [hep-th]].

\bibitem{Belavin:1984vu}
  A.~A.~Belavin, A.~M.~Polyakov and A.~B.~Zamolodchikov,
  ``Infinite Conformal Symmetry in Two-Dimensional Quantum Field Theory,''
  Nucl.\ Phys.\ B {\bf 241} (1984) 333.

\bibitem{fms} P.~Di~Francesco, P.~Mathieu and D.~Senechal, ``Conformal Field Theory'', Springer, 1997.

\bibitem{Schottenloher:2008zz}
M.~Schottenloher,
``A Mathematical Introduction to Conformal Field Theory,''
Lect. Notes Phys. \textbf{759} (2008), 1-237.

\bibitem{Blumenhagen:2009zz}
R.~Blumenhagen and E.~Plauschinn,
``Introduction to Conformal Field Theory,''
Lect. Notes Phys. \textbf{779} (2009), 1-256.

\bibitem{Ginsparg:1988ui}
  P.~H.~Ginsparg, ``Applied Conformal Field Theory,'' Lectures at Les Houches Summer School 1988, hep-th/9108028.

\bibitem{sche1} A.N.~Schellekens, Conformal Field Theory, lecture notes,\\
\url{https://www.nikhef.nl/~t58/Site/Lectures.html}

\bibitem{Antoniadis:1988vi}
I.~Antoniadis, C.~Bachas, J.~R.~Ellis and D.~V.~Nanopoulos,
``An Expanding Universe in String Theory,''
Nucl. Phys. B \textbf{328} (1989), 117-139.

\bibitem{Tseytlin:1991xk}
A.~A.~Tseytlin and C.~Vafa,
``Elements of string cosmology,''
Nucl. Phys. B \textbf{372} (1992), 443-466
[arXiv:hep-th/9109048 [hep-th]].

\bibitem{Silverstein:2001xn}
E.~Silverstein,
``(A)dS backgrounds from asymmetric orientifolds,''
Clay Mat. Proc. \textbf{1} (2002), 179
[arXiv:hep-th/0106209 [hep-th]].

\bibitem{Hellerman:2006nx}
S.~Hellerman and I.~Swanson,
``Cosmological solutions of supercritical string theory,''
Phys. Rev. D \textbf{77} (2008), 126011
[arXiv:hep-th/0611317 [hep-th]].

\bibitem{Ramond:1971gb}
P.~Ramond,
``Dual Theory for Free Fermions,''
Phys. Rev. D \textbf{3} (1971), 2415-2418.

\bibitem{Neveu:1971rx}
A.~Neveu and J.~H.~Schwarz,
``Factorizable dual model of pions,''
Nucl. Phys. B \textbf{31} (1971), 86-112.

\bibitem{Green:1983wt}
M.~B.~Green and J.~H.~Schwarz,
``Covariant Description of Superstrings,''
Phys. Lett. B \textbf{136} (1984), 367-370
and ``Properties of the Covariant Formulation of Superstring Theories,''
Nucl. Phys. B \textbf{243} (1984), 285-306.

\bibitem{Schwarz:1982jn}
J.~H.~Schwarz,
``Superstring Theory,''
Phys. Rept. \textbf{89} (1982), 223-322

\bibitem{Berkovits:2002zk}
N.~Berkovits,
``ICTP lectures on covariant quantization of the superstring,''
ICTP Lect. Notes Ser. \textbf{13} (2003), 57-107
[arXiv:hep-th/0209059 [hep-th]].

\bibitem{Buscher:1987qj}
T.~Buscher,
``Path Integral Derivation of Quantum Duality in Nonlinear Sigma Models,''
Phys. Lett. B \textbf{201} (1988), 466-472 and
``A Symmetry of the String Background Field Equations,''
Phys. Lett. B \textbf{194} (1987), 59-62.

\bibitem{Green:1984sg}
M.~B.~Green and J.~H.~Schwarz,
``Anomaly Cancellation in Supersymmetric D=10 Gauge Theory and Superstring Theory,''
Phys. Lett. B \textbf{149} (1984), 117-122.

\bibitem{ortin}
T.~Ortin, ``Gravity and Strings'', Cambridge University Press, 2015.

\bibitem{Giddings:2001yu}
  S.~B.~Giddings, S.~Kachru and J.~Polchinski,
  ``Hierarchies from fluxes in string compactifications,''
  Phys.\ Rev.\ D {\bf 66} (2002) 106006
  [hep-th/0105097].

\bibitem{Nordstrom:1988fi}
  G.~Nordstrom, ``On the possibility of unifying the electromagnetic and 
  the gravitational fields,'' 
  Phys.\ Z.\  {\bf 15} (1914) 504 [physics/0702221 [physics.gen-ph]].

\bibitem{Klein:1926tv}
  O.~Klein, ``Quantum Theory and Five-Dimensional Theory of Relativity.''
  Z.\ Phys.\  {\bf 37} (1926) 895 [Surveys High Energ.\ Phys.\  {\bf 5} 
  (1986) 241].

\bibitem{Klein:1926fj}
  O.~Klein, ``The Atomicity of Electricity as a Quantum Theory Law,''
  Nature {\bf 118} (1926) 516.

\bibitem{Appelquist:1987nr}
  T.~Appelquist, A.~Chodos and P.~G.~O.~Freund,
  ``Modern Kaluza-klein Theories,'', FRONTIERS IN PHYSICS, 65 (1987).

\bibitem{Duff:1994tn}
  M.~J.~Duff, ``Kaluza-Klein theory in perspective,'' talk at The Oskar Klein
  Centenary, Stockholm, 1994, hep-th/9410046.

\bibitem{Overduin:1998pn}
  J.~M.~Overduin and P.~S.~Wesson, ``Kaluza-Klein gravity,'' 
  Phys.\ Rept.\  {\bf 283} (1997) 303 [gr-qc/9805018].

\bibitem{Candelas:1987is}
  P.~Candelas, ``Lectures On Complex Manifolds,''
  in ``Trieste 1987, Proceedings, Superstrings 1987'', pp. 1-88.

\bibitem{hub} T.~H\"ubsch: Calabi-Yau manifolds: A Bestiary for physicists, World Scientific, 1991.
  
\bibitem{Greene:1996cy}
  B.~R.~Greene, ``String theory on Calabi-Yau manifolds,'' lectures at TASI 1996, hep-th/9702155.
  
\bibitem{He:2018jtw}
Y.~H.~He,
``The Calabi-Yau Landscape: from Geometry, to Physics, to Machine-Learning,'' [arXiv:1812.02893 [hep-th]].

\bibitem{Anderson:2018pui}
L.~B.~Anderson and M.~Karkheiran,
``TASI Lectures on Geometric Tools for String Compactifications,''
PoS \textbf{TASI2017} (2018), 013
[arXiv:1804.08792 [hep-th]].

\bibitem{Viaclovsky}
J.~Viaclovsky, ``Lectures on Kahler geometry, Ricci curvature, and hyperkahler metrics'', Lecture notes, 2019,\\ \url{https://www.math.uci.edu/~jviaclov/lecturenotes/lecturenotes.html}

\bibitem{Candelas:1990pi}
  P.~Candelas and X.~de la Ossa,
  ``Moduli Space of {Calabi-Yau} Manifolds,''
  Nucl.\ Phys.\ B {\bf 355} (1991) 455;\\
  P.~Candelas,
  ``Yukawa Couplings Between (2,1) Forms,''
  Nucl.\ Phys.\ B {\bf 298} (1988) 458.

\bibitem{Font:2005td}
  A.~Font and S.~Theisen,
  ``Introduction to string compactification,''
  Lect.\ Notes Phys.\  {\bf 668} (2005) 101.

\bibitem{Batyrev:1994hm}
V.~V.~Batyrev,
``Dual polyhedra and mirror symmetry for Calabi-Yau hypersurfaces in toric varieties,''
J. Alg. Geom. \textbf{3} (1994), 493-545
[arXiv:alg-geom/9310003 [math.AG]].

\bibitem{Bouchard:2007ik}
V.~Bouchard,
``Lectures on complex geometry, Calabi-Yau manifolds and toric geometry,''
[arXiv:hep-th/0702063 [hep-th]].

\bibitem{Grimm:2004uq}
  T.~W.~Grimm and J.~Louis,
  ``The Effective action of N = 1 Calabi-Yau orientifolds,''
  Nucl.\ Phys.\ B {\bf 699} (2004) 387 [hep-th/0403067].

\bibitem{Jockers:2004yj}
  H.~Jockers and J.~Louis,
  ``The Effective action of D7-branes in N = 1 Calabi-Yau orientifolds,''
  Nucl.\ Phys.\ B {\bf 705} (2005) 167 [hep-th/0409098].

\bibitem{Grimm:2004ua}
  T.~W.~Grimm and J.~Louis,
  ``The Effective action of type IIA Calabi-Yau orientifolds,''
  Nucl.\ Phys.\ B {\bf 718} (2005) 153 [hep-th/0412277].

\bibitem{Kerstan:2011dy}
  M.~Kerstan and T.~Weigand,
  ``The Effective action of D6-branes in N=1 type IIA orientifolds,''
  JHEP {\bf 1106} (2011) 105 [arXiv:1104.2329 [hep-th]].

\bibitem{Klemm:1992tx}
A.~Klemm and S.~Theisen,
``Considerations of one modulus Calabi-Yau compactifications: Picard-Fuchs equations, Kahler potentials and mirror maps,''
Nucl. Phys. B \textbf{389} (1993), 153-180
[arXiv:hep-th/9205041 [hep-th]].

\bibitem{fre}
P.~Fre and P.~Soriani, ``The N=2 wonderland: From Calabi-Yau manifolds to topological field theories'', World Scinetific, 1995.

\bibitem{Giryavets:2005nf}
  A.~Giryavets, ``New attractors and area codes,''
  JHEP {\bf 0603} (2006) 020 [hep-th/0511215].
  
\bibitem{Candelas:1985en}
P.~Candelas, G.~T.~Horowitz, A.~Strominger and E.~Witten, ``Vacuum Configurations for Superstrings,''
Nucl. Phys. B \textbf{258} (1985), 46-74.

\bibitem{Dixon:1985jw}
L.~J.~Dixon, J.~A.~Harvey, C.~Vafa and E.~Witten,
``Strings on Orbifolds,''
Nucl. Phys. B \textbf{261} (1985), 678-686
and ``Strings on Orbifolds. 2.,''
Nucl. Phys. B \textbf{274} (1986), 285-314.

\bibitem{Ibanez:1986tp}
L.~E.~Ibanez, H.~P.~Nilles and F.~Quevedo,
``Orbifolds and Wilson Lines,''
Phys. Lett. B \textbf{187} (1987), 25-32.

\bibitem{Bailin:1999nk}
D.~Bailin and A.~Love,
``Orbifold compactifications of string theory,''
Phys. Rept. \textbf{315} (1999), 285-408.

\bibitem{Kobayashi:2004ya}
T.~Kobayashi, S.~Raby and R.~J.~Zhang,
``Searching for realistic 4d string models with a Pati-Salam symmetry: Orbifold grand unified theories from heterotic string compactification on a Z(6) orbifold,''
Nucl. Phys. B \textbf{704} (2005), 3-55
[arXiv:hep-ph/0409098 [hep-ph]].

\bibitem{Buchmuller:2005jr}
W.~Buchmuller, K.~Hamaguchi, O.~Lebedev and M.~Ratz,
``Supersymmetric standard model from the heterotic string,''
Phys. Rev. Lett. \textbf{96} (2006), 121602
[arXiv:hep-ph/0511035 [hep-ph]].

\bibitem{Kaplunovsky:1993rd}
V.~S.~Kaplunovsky and J.~Louis,
``Model independent analysis of soft terms in effective supergravity and in string theory,''
Phys. Lett. B \textbf{306} (1993), 269-275
[arXiv:hep-th/9303040 [hep-th]].

\bibitem{Brignole:1998dxa}
A.~Brignole, L.~E.~Ibanez and C.~Munoz,
``Soft supersymmetry breaking terms from supergravity and superstring models,''
Adv. Ser. Direct. High Energy Phys. \textbf{18} (1998), 125-148
[arXiv:hep-ph/9707209 [hep-ph]].

\bibitem{Conlon:2006tj}
J.~P.~Conlon, D.~Cremades and F.~Quevedo,
``Kahler potentials of chiral matter fields for Calabi-Yau string compactifications,''
JHEP \textbf{01} (2007), 022
[arXiv:hep-th/0609180 [hep-th]].

\bibitem{Lebedev:2006kn}
O.~Lebedev, H.~P.~Nilles, S.~Raby, S.~Ramos-Sanchez, M.~Ratz, P.~K.~Vaudrevange and A.~Wingerter,
``A Mini-landscape of exact MSSM spectra in heterotic orbifolds,''
Phys. Lett. B \textbf{645} (2007), 88-94
[arXiv:hep-th/0611095 [hep-th]].

\bibitem{Lebedev:2008un}
O.~Lebedev, H.~P.~Nilles, S.~Ramos-Sanchez, M.~Ratz and P.~K.~Vaudrevange,
``Heterotic mini-landscape. (II). Completing the search for MSSM vacua in a Z(6) orbifold,''
Phys. Lett. B \textbf{668} (2008), 331-335
[arXiv:0807.4384 [hep-th]].

\bibitem{Braun:2005nv}
V.~Braun, Y.~H.~He, B.~A.~Ovrut and T.~Pantev,
``The Exact MSSM spectrum from string theory,''
JHEP \textbf{05} (2006), 043
[arXiv:hep-th/0512177 [hep-th]].

\bibitem{Bouchard:2005ag}
V.~Bouchard and R.~Donagi,
``An SU(5) heterotic standard model,''
Phys. Lett. B \textbf{633} (2006), 783-791
[arXiv:hep-th/0512149 [hep-th]].

\bibitem{Blumenhagen:2006ux}
R.~Blumenhagen, S.~Moster and T.~Weigand,
``Heterotic GUT and standard model vacua from simply connected Calabi-Yau manifolds,''
Nucl. Phys. B \textbf{751} (2006), 186-221
[arXiv:hep-th/0603015 [hep-th]].

\bibitem{Anderson:2011ns}
L.~B.~Anderson, J.~Gray, A.~Lukas and E.~Palti,
``Two Hundred Heterotic Standard Models on Smooth Calabi-Yau Threefolds,''
Phys. Rev. D \textbf{84} (2011), 106005
[arXiv:1106.4804 [hep-th]].

\bibitem{Anderson:2011cza}
L.~B.~Anderson, J.~Gray, A.~Lukas and B.~Ovrut,
``Stabilizing All Geometric Moduli in Heterotic Calabi-Yau Vacua,''
Phys. Rev. D \textbf{83} (2011), 106011
[arXiv:1102.0011 [hep-th]].

\bibitem{Polchinski:1995mt}
J.~Polchinski,
``Dirichlet Branes and Ramond-Ramond charges,''
Phys. Rev. Lett. \textbf{75} (1995), 4724-4727
[arXiv:hep-th/9510017 [hep-th]].

\bibitem{Polchinski:1996fm}
J.~Polchinski, S.~Chaudhuri and C.~V.~Johnson,
``Notes on D-branes,'' Lectures presented by J.~Polchinski in 1995 [arXiv:hep-th/9602052 [hep-th]].

\bibitem{Gmeiner:2005vz}
F.~Gmeiner, R.~Blumenhagen, G.~Honecker, D.~Lust and T.~Weigand,
``One in a billion: MSSM-like D-brane statistics,''
JHEP \textbf{01} (2006), 004
[arXiv:hep-th/0510170 [hep-th]].

\bibitem{Blumenhagen:2008zz}
R.~Blumenhagen, V.~Braun, T.~W.~Grimm and T.~Weigand, ``GUTs in Type IIB Orientifold Compactifications,''
Nucl. Phys. B \textbf{815} (2009), 1-94
[arXiv:0811.2936 [hep-th]].

\bibitem{Strominger:1996it}
A.~Strominger, S.~T.~Yau and E.~Zaslow,
``Mirror symmetry is T duality,''
Nucl. Phys. B \textbf{479} (1996), 243-259
[arXiv:hep-th/9606040 [hep-th]].

\bibitem{Candelas:1993dm}
P.~Candelas, X.~De La Ossa, A.~Font, S.~H.~Katz and D.~R.~Morrison,
``Mirror symmetry for two parameter models. 1.,''
AMS/IP Stud. Adv. Math. \textbf{1} (1996), 483-543
[arXiv:hep-th/9308083 [hep-th]]
and
``Mirror symmetry for two parameter models. 2.,''
Nucl. Phys. B \textbf{429} (1994), 626-674
[arXiv:hep-th/9403187 [hep-th]].

\bibitem{Hosono:1994av}
S.~Hosono, A.~Klemm and S.~Theisen,
``Lectures on mirror symmetry,''
Lect. Notes Phys. \textbf{436} (1994), 235-280
[arXiv:hep-th/9403096 [hep-th]].

\bibitem{Hori:2002fa}
K.~Hori, ``Trieste lectures on mirror symmetry,''
ICTP Lect. Notes Ser. \textbf{13} (2003), 109-202.

\bibitem{Hori:2003ic} K.~Hori, S.~Katz, A.~Klemm, R.~Pandharipande, R.~Thomas, C.~Vafa, R.~Vakil and E.~Zaslow, ``Mirror symmetry,'' Published by the American Mathematical Society for the Clay Mathematics Institute, 2003.

\bibitem{Vafa:1996xn}
  C.~Vafa, ``Evidence for F theory,''
  Nucl.\ Phys.\ B {\bf 469} (1996) 403 [hep-th/9602022].

\bibitem{Sen:1996vd}
A.~Sen, ``F theory and orientifolds,''
Nucl. Phys. B \textbf{475} (1996), 562-578
[arXiv:hep-th/9605150 [hep-th]].

\bibitem{Beasley:2008dc}
C.~Beasley, J.~J.~Heckman and C.~Vafa,
``GUTs and Exceptional Branes in F-theory - I,''
JHEP \textbf{01} (2009), 058
[arXiv:0802.3391 [hep-th]].

\bibitem{Donagi:2008ca}
R.~Donagi and M.~Wijnholt,
``Model Building with F-Theory,''
Adv. Theor. Math. Phys. \textbf{15} (2011) no.5, 1237-1317
[arXiv:0802.2969 [hep-th]].

\bibitem{Weigand:2010wm}
  T.~Weigand, ``Lectures on F-theory compactifications and model building,''
  Class.\ Quant.\ Grav.\  {\bf 27} (2010) 214004 [arXiv:1009.3497 [hep-th]].
  
\bibitem{Heckman:2010bq}
  J.~J.~Heckman, ``Particle Physics Implications of F-theory,''
  Ann.\ Rev.\ Nucl.\ Part.\ Sci.\  {\bf 60} (2010) 237
  [arXiv:1001.0577 [hep-th]].

\bibitem{Susskind:2003kw}
L.~Susskind, ``The Anthropic landscape of string theory,''
[arXiv:hep-th/0302219 [hep-th]].

\bibitem{Bousso:2000xa}
  R.~Bousso and J.~Polchinski,
  ``Quantization of four form fluxes and dynamical neutralization of the cosmological constant,''
  JHEP {\bf 0006} (2000) 006
  [hep-th/0004134].

\bibitem{Kachru:2003aw}
  S.~Kachru, R.~Kallosh, A.~D.~Linde and S.~P.~Trivedi,
  ``De Sitter vacua in string theory,''
  Phys.\ Rev.\ D {\bf 68} (2003) 046005
  [hep-th/0301240].

\bibitem{Balasubramanian:2005zx}
  V.~Balasubramanian, P.~Berglund, J.~P.~Conlon and F.~Quevedo,
  ``Systematics of moduli stabilisation in Calabi-Yau flux compactifications,''
  JHEP {\bf 0503} (2005) 007 [hep-th/0502058].
  
\bibitem{Schellekens:2006xz}
A.~Schellekens,
``The Landscape 'avant la lettre',''
[arXiv:physics/0604134 [physics]].  

\bibitem{Feng:2000if}
J.~L.~Feng, J.~March-Russell, S.~Sethi and F.~Wilczek,
``Saltatory relaxation of the cosmological constant,''
Nucl. Phys. B \textbf{602} (2001), 307-328
[arXiv:hep-th/0005276 [hep-th]].

\bibitem{Coleman:1985rnk}
  S.~Coleman,
  ``Aspects of Symmetry : Selected Erice Lectures,'' Cambridge University Press, 1985.

\bibitem{Blumenhagen:2009qh}
R.~Blumenhagen, M.~Cvetic, S.~Kachru and T.~Weigand,
``D-Brane Instantons in Type II Orientifolds,''
Ann. Rev. Nucl. Part. Sci. \textbf{59} (2009), 269-296
[arXiv:0902.3251 [hep-th]].

\bibitem{Bianchi:2007ft}
M.~Bianchi, S.~Kovacs and G.~Rossi,
``Instantons and Supersymmetry,''
Lect. Notes Phys. \textbf{737} (2008), 303-470
[arXiv:hep-th/0703142 [hep-th]].

\bibitem{Hebecker:2018ofv}
A.~Hebecker, T.~Mikhail and P.~Soler,
``Euclidean wormholes, baby universes, and their impact on particle physics and cosmology,''
Front. Astron. Space Sci. \textbf{5} (2018), 35
[arXiv:1807.00824 [hep-th]].

\bibitem{nak} M.~Nakahara, ``Geometry, topology and physics'', Institute of Physics Publishing, 2003.

\bibitem{bert} R.~A.~Bertlmann, ``Anomalies in Quantum Field Theory'', Oxford University Press, 1996.

\bibitem{nashsen} C.~Nash and S.~Sen, ``Topology and Geometry for Physicists'', Academic Press, 1983.

\bibitem{gs} M.~G\"ockeler and T.~Sch\"ucker, ``Differential geometry, gauge theories, and gravity'', Cambridge University Press, 1987.

\bibitem{Gukov:1999ya}
  S.~Gukov, C.~Vafa and E.~Witten, ``CFT's from Calabi-Yau four folds,''
  Nucl.\ Phys.\ B {\bf 584} (2000) 69
  Erratum: [Nucl.\ Phys.\ B {\bf 608} (2001) 477] [hep-th/9906070].

\bibitem{Witten:1996bn}
E.~Witten,
``Nonperturbative superpotentials in string theory,''
Nucl. Phys. B \textbf{474} (1996), 343-360
[arXiv:hep-th/9604030 [hep-th]].

\bibitem{Bianchi:2011qh}
M.~Bianchi, A.~Collinucci and L.~Martucci,
``Magnetized E3-brane instantons in F-theory,''
JHEP \textbf{12} (2011), 045
[arXiv:1107.3732 [hep-th]].

\bibitem{Palti:2020qlc}
E.~Palti, C.~Vafa and T.~Weigand,
``Supersymmetric Protection and the Swampland,''
JHEP \textbf{06} (2020), 168
[arXiv:2003.10452 [hep-th]].

\bibitem{Dasgupta:1999ss}
  K.~Dasgupta, G.~Rajesh and S.~Sethi,
  ``M theory, orientifolds and G - flux,''
  JHEP {\bf 9908} (1999) 023 [hep-th/9908088].

\bibitem{Grana:2005jc}
M.~Grana, ``Flux compactifications in string theory: A Comprehensive review,'' Phys. Rept. \textbf{423} (2006), 91-158 [arXiv:hep-th/0509003 [hep-th]].

\bibitem{Cremmer:1983bf}
E.~Cremmer, S.~Ferrara, C.~Kounnas and D.~V.~Nanopoulos,
``Naturally Vanishing Cosmological Constant in N=1 Supergravity,'' Phys. Lett. B \textbf{133} (1983), 61.

\bibitem{Sethi:2017phn}
S.~Sethi,
``Supersymmetry Breaking by Fluxes,''
JHEP \textbf{10} (2018), 022
[arXiv:1709.03554 [hep-th]].

\bibitem{Westphal:2006tn}
A.~Westphal,
``de Sitter string vacua from Kahler uplifting,''
JHEP \textbf{03} (2007), 102
[arXiv:hep-th/0611332 [hep-th]].

\bibitem{Balasubramanian:2004uy}
V.~Balasubramanian and P.~Berglund,
``Stringy corrections to Kahler potentials, SUSY breaking, and the cosmological constant problem,''
JHEP \textbf{11} (2004), 085
[arXiv:hep-th/0408054 [hep-th]].

\bibitem{Derendinger:1985kk}
J.~Derendinger, L.~E.~Ibanez and H.~P.~Nilles,
``On the Low-Energy d = 4, N=1 Supergravity Theory Extracted from the d = 10, N=1 Superstring,''
Phys. Lett. B \textbf{155} (1985), 65-70.

\bibitem{Dine:1985rz}
M.~Dine, R.~Rohm, N.~Seiberg and E.~Witten,
``Gluino Condensation in Superstring Models,''
Phys. Lett. B \textbf{156} (1985), 55-60.

\bibitem{Denef:2004ze}
  F.~Denef and M.~R.~Douglas,
  ``Distributions of flux vacua,''
  JHEP {\bf 0405} (2004) 072
  [hep-th/0404116].

\bibitem{Luty:2002hj}
M.~A.~Luty and N.~Okada,
``Almost no scale supergravity,''
JHEP \textbf{04} (2003), 050
[arXiv:hep-th/0209178 [hep-th]].

\bibitem{Choi:2005ge}
K.~Choi, A.~Falkowski, H.~P.~Nilles and M.~Olechowski,
``Soft supersymmetry breaking in KKLT flux compactification,''
Nucl. Phys. B \textbf{718} (2005), 113-133
[arXiv:hep-th/0503216 [hep-th]].

\bibitem{Brummer:2006dg}
F.~Brummer, A.~Hebecker and M.~Trapletti,
``SUSY breaking mediation by throat fields,''
Nucl. Phys. B \textbf{755} (2006), 186-198
[arXiv:hep-th/0605232 [hep-th]].

\bibitem{Dudas:2006gr}
E.~Dudas, C.~Papineau and S.~Pokorski,
``Moduli stabilization and uplifting with dynamically generated F-terms,''
JHEP \textbf{02} (2007), 028
[arXiv:hep-th/0610297 [hep-th]].

\bibitem{Danielsson:2018ztv}
  U.~H.~Danielsson and T.~Van Riet,
  ``What if string theory has no de Sitter vacua?,''
  Int.\ J.\ Mod.\ Phys.\ D {\bf 27} (2018) no.12,  1830007
  [arXiv:1804.01120 [hep-th]].

\bibitem{Obied:2018sgi}
  G.~Obied, H.~Ooguri, L.~Spodyneiko and C.~Vafa,
  ``De Sitter Space and the Swampland,''
  arXiv:1806.08362 [hep-th].

\bibitem{Garg:2018reu}
S.~K.~Garg and C.~Krishnan,
``Bounds on Slow Roll and the de Sitter Swampland,''
JHEP \textbf{11} (2019), 075
[arXiv:1807.05193 [hep-th]].

\bibitem{Ooguri:2018wrx}
  H.~Ooguri, E.~Palti, G.~Shiu and C.~Vafa,
  ``Distance and de Sitter Conjectures on the Swampland,''
  Phys.\ Lett.\ B {\bf 788} (2019) 180
  [arXiv:1810.05506 [hep-th]].

\bibitem{Alexander:2001ks}
S.~H.~S.~Alexander,
``Inflation from D - anti-D-brane annihilation,''
Phys. Rev. D \textbf{65} (2002), 023507
[arXiv:hep-th/0105032 [hep-th]].

\bibitem{Dvali:2001fw}
G.~R.~Dvali, Q.~Shafi and S.~Solganik,
``D-brane inflation,''
[arXiv:hep-th/0105203 [hep-th]].

\bibitem{Burgess:2001fx}
C.~P.~Burgess, M.~Majumdar, D.~Nolte, F.~Quevedo, G.~Rajesh and R.~J.~Zhang,
``The Inflationary brane anti-brane universe,''
JHEP \textbf{07} (2001), 047
[arXiv:hep-th/0105204 [hep-th]].

\bibitem{Klebanov:2000hb}
  I.~R.~Klebanov and M.~J.~Strassler,
  ``Supergravity and a confining gauge theory: Duality cascades and chi 
  SB resolution of naked singularities,''
  JHEP {\bf 0008} (2000) 052
  [hep-th/0007191].

\bibitem{Candelas:1989js}
  P.~Candelas and X.~C.~de la Ossa,
  ``Comments on Conifolds,''
  Nucl.\ Phys.\ B {\bf 342} (1990) 246.

\bibitem{Bena:2018fqc}
  I.~Bena, E.~Dudas, M.~Graña and S.~Lüst,
  ``Uplifting Runaways,''
  Fortsch.\ Phys.\  {\bf 67} (2019) no.1-2,  1800100
  [arXiv:1809.06861 [hep-th]].

\bibitem{Blumenhagen:2019qcg}
  R.~Blumenhagen, D.~Klaewer and L.~Schlechter,
  ``Swampland Variations on a Theme by KKLT,''
  arXiv:1902.07724 [hep-th].

\bibitem{Kachru:2002gs}
  S.~Kachru, J.~Pearson and H.~L.~Verlinde,
  ``Brane / flux annihilation and the string dual of a nonsupersymmetric 
  field theory,''
  JHEP {\bf 0206} (2002) 021
  [hep-th/0112197].

\bibitem{Bena:2014jaa}
  I.~Bena, M.~Graña, S.~Kuperstein and S.~Massai,
  ``Giant Tachyons in the Landscape,''
  JHEP {\bf 1502} (2015) 146
  [arXiv:1410.7776 [hep-th]].

\bibitem{Michel:2014lva}
  B.~Michel, E.~Mintun, J.~Polchinski, A.~Puhm and P.~Saad,
  ``Remarks on brane and antibrane dynamics,''
  JHEP {\bf 1509} (2015) 021
  [arXiv:1412.5702 [hep-th]].

\bibitem{Cohen-Maldonado:2015ssa}
  D.~Cohen-Maldonado, J.~Diaz, T.~van Riet and B.~Vercnocke,
  ``Observations on fluxes near anti-branes,''
  JHEP {\bf 1601} (2016) 126
  [arXiv:1507.01022 [hep-th]].

\bibitem{Polchinski:2015bea}
  J.~Polchinski,
  ``Brane/antibrane dynamics and KKLT stability,''
  arXiv:1509.05710 [hep-th].

\bibitem{Bena:2016fqp}
  I.~Bena, J.~Blåbäck and D.~Turton,
  ``Loop corrections to the antibrane potential,''
  JHEP {\bf 1607} (2016) 132
  [arXiv:1602.05959 [hep-th]].

\bibitem{Danielsson:2016cit}
  U.~H.~Danielsson, F.~F.~Gautason and T.~Van Riet,
  ``Unstoppable brane-flux decay of $ \overline{\mathrm{D}6} $ branes,''
  JHEP {\bf 1703} (2017) 141
  [arXiv:1609.06529 [hep-th]].

\bibitem{Kachru:2003sx}
  S.~Kachru, R.~Kallosh, A.~D.~Linde, J.~M.~Maldacena, L.~P.~McAllister 
  and S.~P.~Trivedi, ``Towards inflation in string theory,''
  JCAP {\bf 0310} (2003) 013 [hep-th/0308055].

\bibitem{Giddings:2005ff}
S.~B.~Giddings and A.~Maharana,
``Dynamics of warped compactifications and the shape of the warped landscape,''
Phys. Rev. D \textbf{73} (2006), 126003
[arXiv:hep-th/0507158 [hep-th]].

\bibitem{Moritz:2017xto}
  J.~Moritz, A.~Retolaza and A.~Westphal,
  ``Toward de Sitter space from ten dimensions,''
  Phys.\ Rev.\ D {\bf 97} (2018) no.4,  046010
  [arXiv:1707.08678 [hep-th]].

\bibitem{Hamada:2018qef}
  Y.~Hamada, A.~Hebecker, G.~Shiu and P.~Soler,
  ``On brane gaugino condensates in 10d,''
  JHEP {\bf 1904} (2019) 008
  [arXiv:1812.06097 [hep-th]].

\bibitem{Kallosh:2019oxv}
  R.~Kallosh,
  ``Gaugino Condensation and Geometry of the Perfect Square,''
  Phys.\ Rev.\ D {\bf 99} (2019) no.6,  066003
  [arXiv:1901.02023 [hep-th]].

\bibitem{Hamada:2019ack}
  Y.~Hamada, A.~Hebecker, G.~Shiu and P.~Soler,
  ``Understanding KKLT from a 10d perspective,''
  arXiv:1902.01410 [hep-th].

\bibitem{Gautason:2019jwq}
  F.~F.~Gautason, V.~Van Hemelryck, T.~Van Riet and G.~Venken,
  ``A 10d view on the KKLT AdS vacuum and uplifting,''
  arXiv:1902.01415 [hep-th].

\bibitem{Carta:2019rhx}
  F.~Carta, J.~Moritz and A.~Westphal,
  ``Gaugino condensation and small uplifts in KKLT,''
  arXiv:1902.01412 [hep-th].

\bibitem{Gao:2020xqh}
X.~Gao, A.~Hebecker and D.~Junghans,
``Control issues of KKLT,''
arXiv:2009.03914.

\bibitem{Conlon:2005ki}
J.~P.~Conlon, F.~Quevedo and K.~Suruliz,
``Large-volume flux compactifications: Moduli spectrum and D3/D7 soft supersymmetry breaking,''
JHEP \textbf{08} (2005), 007
[arXiv:hep-th/0505076 [hep-th]].

\bibitem{Cicoli:2012vw}
M.~Cicoli, S.~Krippendorf, C.~Mayrhofer, F.~Quevedo and R.~Valandro,
``D-Branes at del Pezzo Singularities: Global Embedding and Moduli Stabilisation,''
JHEP \textbf{09} (2012), 019
[arXiv:1206.5237 [hep-th]].

\bibitem{Cicoli:2013cha}
M.~Cicoli, D.~Klevers, S.~Krippendorf, C.~Mayrhofer, F.~Quevedo and R.~Valandro,
``Explicit de Sitter Flux Vacua for Global String Models with Chiral Matter,''
JHEP \textbf{05} (2014), 001
[arXiv:1312.0014 [hep-th]].

\bibitem{Green:1999qt}
M.~B.~Green,
``Interconnections between type II superstrings, M theory and N=4 supersymmetric Yang-Mills,''
Lect. Notes Phys. \textbf{525} (1999), 22
[arXiv:hep-th/9903124 [hep-th]].

\bibitem{Becker:2002nn}
K.~Becker, M.~Becker, M.~Haack and J.~Louis,
``Supersymmetry breaking and alpha-prime corrections to flux induced potentials,''
JHEP \textbf{06} (2002), 060
[arXiv:hep-th/0204254 [hep-th]].

\bibitem{Cremades:2007ig}
D.~Cremades, M.~P.~Garcia del Moral, F.~Quevedo and K.~Suruliz,
``Moduli stabilisation and de Sitter string vacua from magnetised D7 branes,''
JHEP \textbf{05} (2007), 100
[arXiv:hep-th/0701154 [hep-th]].

\bibitem{Burgess:2003ic}
C.~Burgess, R.~Kallosh and F.~Quevedo,
``De Sitter string vacua from supersymmetric D terms,''
JHEP \textbf{10} (2003), 056
[arXiv:hep-th/0309187 [hep-th]].

\bibitem{Hebecker:2012aw}
A.~Hebecker, S.~C.~Kraus, M.~Kuntzler, D.~Lust and T.~Weigand,
``Fluxbranes: Moduli Stabilisation and Inflation,''
JHEP \textbf{01} (2013), 095
[arXiv:1207.2766 [hep-th]].

\bibitem{vonGersdorff:2005bf}
G.~von Gersdorff and A.~Hebecker,
``Kahler corrections for the volume modulus of flux compactifications,''
Phys. Lett. B \textbf{624} (2005), 270-274
[arXiv:hep-th/0507131 [hep-th]].

\bibitem{Cicoli:2007xp}
M.~Cicoli, J.~P.~Conlon and F.~Quevedo,
``Systematics of String Loop Corrections in Type IIB Calabi-Yau Flux Compactifications,''
JHEP \textbf{01} (2008), 052
[arXiv:0708.1873 [hep-th]].

\bibitem{Berg:2007wt}
M.~Berg, M.~Haack and E.~Pajer,
``Jumping Through Loops: On Soft Terms from Large Volume Compactifications,''
JHEP \textbf{09} (2007), 031
[arXiv:0704.0737 [hep-th]].

\bibitem{Berg:2005ja}
M.~Berg, M.~Haack and B.~Kors,
``String loop corrections to Kahler potentials in orientifolds,''
JHEP \textbf{11} (2005), 030
[arXiv:hep-th/0508043 [hep-th]]
and 
``On volume stabilization by quantum corrections,''
Phys. Rev. Lett. \textbf{96} (2006), 021601
[arXiv:hep-th/0508171 [hep-th]].

\bibitem{Douglas:2003um}
  M.~R.~Douglas,
  ``The Statistics of string / M theory vacua,''
  JHEP {\bf 0305} (2003) 046 [hep-th/0303194].

\bibitem{Ashok:2003gk}
  S.~Ashok and M.~R.~Douglas,
  ``Counting flux vacua,'' JHEP {\bf 0401} (2004) 060 [hep-th/0307049].

\bibitem{Klemm:1996ts}
  A.~Klemm, B.~Lian, S.~S.~Roan and S.~T.~Yau,
  ``Calabi-Yau fourfolds for M theory and F theory compactifications,''
  Nucl.\ Phys.\ B {\bf 518} (1998) 515 [hep-th/9701023].

\bibitem{Taylor:2015xtz}
  W.~Taylor and Y.~N.~Wang,
  ``The F-theory geometry with most flux vacua,''
  JHEP {\bf 1512} (2015) 164 [arXiv:1511.03209 [hep-th]].

\bibitem{Ringwald}
A.~Ringwald, L.J.~Rosenberg and G.~Rybka, ``Axions and Other Similar Particles'', Review article in Ref.~\cite{Tanabashi:2018oca}.

\bibitem{Conlon:2006tq}
J.~P.~Conlon,
``The QCD axion and moduli stabilisation,''
JHEP \textbf{05} (2006), 078
[arXiv:hep-th/0602233 [hep-th]].

\bibitem{Svrcek:2006yi}
P.~Svrcek and E.~Witten,
``Axions In String Theory,''
JHEP \textbf{06} (2006), 051
[arXiv:hep-th/0605206 [hep-th]].

\bibitem{Arvanitaki:2009fg}
A.~Arvanitaki, S.~Dimopoulos, S.~Dubovsky, N.~Kaloper and J.~March-Russell,
``String Axiverse,''
Phys. Rev. D \textbf{81} (2010), 123530
[arXiv:0905.4720 [hep-th]].

\bibitem{Jaeckel:2010ni}
J.~Jaeckel and A.~Ringwald,
``The Low-Energy Frontier of Particle Physics,''
Ann. Rev. Nucl. Part. Sci. \textbf{60} (2010), 405-437
[arXiv:1002.0329 [hep-ph]].

\bibitem{Cicoli:2012sz}
M.~Cicoli, M.~Goodsell and A.~Ringwald,
``The type IIB string axiverse and its low-energy phenomenology,''
JHEP \textbf{10} (2012), 146
[arXiv:1206.0819 [hep-th]].

\bibitem{Halverson:2018vbo}
J.~Halverson and P.~Langacker,
``TASI Lectures on Remnants from the String Landscape,''
PoS \textbf{TASI2017} (2018), 019
[arXiv:1801.03503 [hep-th]].

\bibitem{Mukhanov:2005sc}
  V.~Mukhanov,
  ``Physical Foundations of Cosmology,'' Cambridge University Press, 2005.

\bibitem{weico}
S.~Weinberg, ``Cosmology'', Oxford University Press, 2008.

\bibitem{peeb}
P.~J.~E.~Peebles, ``Principles of Physical Cosmology'', Princeton University Press, 1993.

\bibitem{kolb}
E.~W.~Kolb and M.~S.~Turner, ``The Early Universe'', Westview Press, 1990.

\bibitem{Starobinsky:1980te}
A.~A.~Starobinsky,
``A New Type of Isotropic Cosmological Models Without Singularity,'' Phys.\ Lett.\ {\bf 91B} (1980) 99.

\bibitem{Guth:1980zm}
A.~H.~Guth,
``The Inflationary Universe: A Possible Solution to the Horizon and Flatness Problems,'' Phys.\ Rev.\ {\bf D23} (1981) 347.

\bibitem{Sato:1980yn}
K.~Sato, ``First Order Phase Transition of a Vacuum and Expansion of the Universe,''
Mon. Not. Roy. Astron. Soc. \textbf{195} (1981), 467-479.

\bibitem{Linde:1981mu}
A.~D.~Linde,
``A New Inflationary Universe Scenario: A Possible Solution of the Horizon, Flatness, Homogeneity, Isotropy and Primordial Monopole Problems,''
Phys.\ Lett.\ {\bf 108B} (1982) 389.

\bibitem{Albrecht:1982wi}
A.~Albrecht and P.~J.~Steinhardt,
``Cosmology for Grand Unified Theories with Radiatively Induced Symmetry Breaking,''
Phys.\ Rev.\ Lett. {\bf 48} (1982) 1220.

\bibitem{Mukhanov:1981xt}
V.~F.~Mukhanov and G.~V.~Chibisov,
``Quantum Fluctuations and a Nonsingular Universe,''
JETP Lett. \textbf{33} (1981), 532-535.

\bibitem{Linde:1983gd}
  A.~D.~Linde,
  ``Chaotic Inflation,''
  Phys.\ Lett.\  {\bf 129B} (1983) 177
and
  ``Scalar Field Fluctuations in Expanding Universe and the New Inflationary Universe Scenario,''
  Phys.\ Lett.\  {\bf 116B} (1982) 335.

\bibitem{Spradlin:2001pw}
  M.~Spradlin, A.~Strominger and A.~Volovich,
  ``Les Houches lectures on de Sitter space,''
  hep-th/0110007.

\bibitem{Riotto:2002yw}
  A.~Riotto,
  ``Inflation and the theory of cosmological perturbations,''
  ICTP Lect.\ Notes Ser.\  {\bf 14} (2003) 317
  [hep-ph/0210162].

\bibitem{Baumann:2014nda}
  D.~Baumann and L.~McAllister,
  ``Inflation and String Theory,''
  arXiv:1404.2601 [hep-th] and
  ``Inflation and String Theory,'' Cambridge Monographs on Mathematical 
  Physics, 2015.
  
\bibitem{Westphal:2014ana}
A.~Westphal,
``String cosmology -- Large-field inflation in string theory,''
Int. J. Mod. Phys. A \textbf{30} (2015) no.09, 1530024
[arXiv:1409.5350 [hep-th]].

\bibitem{Quevedo:2002xw}
F.~Quevedo,
``Lectures on string/brane cosmology,''
Class. Quant. Grav. \textbf{19} (2002), 5721-5779
[arXiv:hep-th/0210292 [hep-th]].

\bibitem{Akrami:2018odb}
  Y.~Akrami {\it et al.} [Planck Collaboration],
  ``Planck 2018 results. X. Constraints on inflation,''
  arXiv:1807.06211 [astro-ph.CO].

\bibitem{Berera:1995ie}
  A.~Berera, ``Warm inflation,''
  Phys.\ Rev.\ Lett.\  {\bf 75} (1995) 3218
  [astro-ph/9509049].

\bibitem{Lyth:2001nq}
  D.~H.~Lyth and D.~Wands,
  ``Generating the curvature perturbation without an inflaton,''
  Phys.\ Lett.\ B {\bf 524} (2002) 5 [hep-ph/0110002].

\bibitem{Alishahiha:2004eh}
  M.~Alishahiha, E.~Silverstein and D.~Tong,
  ``DBI in the sky,''
  Phys.\ Rev.\ D {\bf 70} (2004) 123505 [hep-th/0404084].

\bibitem{Brandenberger:1988aj}
  R.~H.~Brandenberger and C.~Vafa,
  ``Superstrings in the Early Universe,''
  Nucl.\ Phys.\ B {\bf 316} (1989) 391.

\bibitem{Gasperini:1992em}
  M.~Gasperini and G.~Veneziano,
  ``Pre - big bang in string cosmology,''
  Astropart.\ Phys.\  {\bf 1} (1993) 317 [hep-th/9211021].

\bibitem{Maldacena:1997re}
  J.~M.~Maldacena,
  ``The Large N limit of superconformal field theories and supergravity,''
  Int.\ J.\ Theor.\ Phys.\  {\bf 38} (1999) 1113
   [Adv.\ Theor.\ Math.\ Phys.\  {\bf 2} (1998) 231]  [hep-th/9711200].
  
\bibitem{Aharony:1999ti}
  O.~Aharony, S.~S.~Gubser, J.~M.~Maldacena, H.~Ooguri and Y.~Oz,
  ``Large N field theories, string theory and gravity,''
  Phys.\ Rept.\  {\bf 323} (2000) 183 [hep-th/9905111].
  
\bibitem{Ooguri:2016pdq}
  H.~Ooguri and C.~Vafa, ``Non-supersymmetric AdS and the Swampland,''
  Adv.\ Theor.\ Math.\ Phys.\  {\bf 21} (2017) 1787
  [arXiv:1610.01533 [hep-th]].

\bibitem{Freivogel:2016qwc}
B.~Freivogel and M.~Kleban, ``Vacua Morghulis,''
[arXiv:1610.04564 [hep-th]].

\bibitem{reid}
M.~Reid, ``The moduli space of 3-folds with $K = 0$ may nevertheless be irreducible'', Math.\ Ann.\ {\bf278} (1987) 329.

\bibitem{Carifio:2017nyb}
J.~Carifio, W.~J.~Cunningham, J.~Halverson, D.~Krioukov, C.~Long and B.~D.~Nelson,
``Vacuum Selection from Cosmology on Networks of String Geometries,''
Phys. Rev. Lett. \textbf{121} (2018) no.10, 101602
[arXiv:1711.06685 [hep-th]].

\bibitem{Freivogel:2011eg}
  B.~Freivogel,
  ``Making predictions in the multiverse,''
  Class.\ Quant.\ Grav.\  {\bf 28} (2011) 204007
  [arXiv:1105.0244 [hep-th]].

\bibitem{Linde:1993nz}
  A.~D.~Linde and A.~Mezhlumian,
  ``Stationary universe,''
  Phys.\ Lett.\ B {\bf 307} (1993) 25
  [gr-qc/9304015].
  
\bibitem{Linde:1993xx}
  A.~D.~Linde, D.~A.~Linde and A.~Mezhlumian,
  ``From the Big Bang theory to the theory of a stationary universe,''
  Phys.\ Rev.\ D {\bf 49} (1994) 1783
  [gr-qc/9306035].
    
\bibitem{Coleman:1977th}
  S.~R.~Coleman, V.~Glaser and A.~Martin,
  ``Action Minima Among Solutions to a Class of Euclidean Scalar Field Equations,''
  Commun.\ Math.\ Phys.\  {\bf 58} (1978) 211.

\bibitem{Coleman:1980aw}
  S.~R.~Coleman and F.~De Luccia,
  ``Gravitational Effects on and of Vacuum Decay,''
  Phys.\ Rev.\ D {\bf 21} (1980) 3305.
 
\bibitem{Parke:1982pm}
  S.~J.~Parke,
  ``Gravity, the Decay of the False Vacuum and the New Inflationary Universe Scenario,''
  Phys.\ Lett.\  {\bf 121B} (1983) 313.
 
 \bibitem{Lindley:1984bj}
  D.~Lindley,
  ``The appearance of bubbles in de Sitter space,''
  Nucl.\ Phys.\ B {\bf 236} (1984) 522.

\bibitem{Brown:1987dd}
  J.~D.~Brown and C.~Teitelboim,
  ``Dynamical Neutralization of the Cosmological Constant,''
  Phys.\ Lett.\ B {\bf 195} (1987) 177
and 
  ``Neutralization of the Cosmological Constant by Membrane Creation,''
  Nucl.\ Phys.\ B {\bf 297} (1988) 787.
  
\bibitem{Lee:1987qc}
  K.~M.~Lee and E.~J.~Weinberg,
  ``Decay of the True Vacuum in Curved Space-time,''
  Phys.\ Rev.\ D {\bf 36} (1987) 1088.

\bibitem{SchwartzPerlov:2006hi}
  D.~Schwartz-Perlov and A.~Vilenkin,
  ``Probabilities in the Bousso-Polchinski multiverse,''
  JCAP {\bf 0606} (2006) 010 [hep-th/0601162].
  
 \bibitem{Johnson:2007jla}
  M.~C.~Johnson, ``Vacuum Transitions and Eternal Inflation,'' PhD Thesis, Univ. of California in Santa Cruz,
\url{http://inspirehep.net/record/1263739/files/thesis.pdf}

\bibitem{Eckerle:2020opg}
K.~Eckerle,
``A Simple System For Coleman-De Luccia Transitions,''
[arXiv:2003.04365 [hep-th]].

 \bibitem{Dyson:1979zz}
  F.~J.~Dyson,
  ``Time without end: Physics and biology in an open
  universe,'' Rev.\ Mod.\ Phys.\  {\bf 51} (1979) 447.

\bibitem{Freivogel:2005vv}
B.~Freivogel, M.~Kleban, M.~Rodriguez Martinez and L.~Susskind,
``Observational consequences of a landscape,''
JHEP \textbf{03} (2006), 039
[arXiv:hep-th/0505232 [hep-th]].

\bibitem{Kleban:2011pg}
  M.~Kleban, ``Cosmic Bubble Collisions,''
  Class.\ Quant.\ Grav.\  {\bf 28} (2011) 204008\\{}
  [arXiv:1107.2593].   

\bibitem{Vilenkin:2006xv}
  A.~Vilenkin,
  ``A Measure of the multiverse,''
  J.\ Phys.\ A {\bf 40} (2007) 6777
  [hep-th/0609193].

\bibitem{Acharya:2006zw}
  B.~S.~Acharya and M.~R.~Douglas,
  ``A Finite landscape?,'' hep-th/0606212.
  
\bibitem{Weinberg:1987dv}
  S.~Weinberg,
  ``Anthropic Bound on the Cosmological Constant,''
  Phys.\ Rev.\ Lett.\  {\bf 59} (1987) 2607.
  
\bibitem{Linde:1984ir}
  A.~D.~Linde,
  ``The Inflationary Universe,''
  Rept.\ Prog.\ Phys.\  {\bf 47} (1984) 925.
  
\bibitem{bt}
J.~D.~Barrow and F.~J.~Tipler, ``The Anthropic Cosmological Principle'', Oxford University Press, 1986.

\bibitem{Hogan:1999wh}
C.~J.~Hogan, ``Why the universe is just so,''
Rev. Mod. Phys. \textbf{72} (2000), 1149-1161
[arXiv:astro-ph/9909295 [astro-ph]].
  
\bibitem{Tegmark:2005dy}
M.~Tegmark, A.~Aguirre, M.~Rees and F.~Wilczek,
``Dimensionless constants, cosmology and other dark matters,''
Phys. Rev. D \textbf{73} (2006), 023505
[arXiv:astro-ph/0511774 [astro-ph]]. 
  
\bibitem{Hall:2007ja}
L.~J.~Hall and Y.~Nomura,
``Evidence for the Multiverse in the Standard Model and Beyond,'' Phys. Rev. D \textbf{78} (2008), 035001
[arXiv:0712.2454 [hep-ph]].
  
\bibitem{Garriga:2005av}
  J.~Garriga, D.~Schwartz-Perlov, A.~Vilenkin and S.~Winitzki,
  ``Probabilities in the inflationary multiverse,''
  JCAP {\bf 0601} (2006) 017
  [hep-th/0509184].

\bibitem{Bousso:2009dm}
  R.~Bousso, ``Complementarity in the Multiverse,''
  Phys.\ Rev.\ D {\bf 79} (2009) 123524
  [arXiv:0901.4806 [hep-th]].

\bibitem{ctc}
A. Maloney, S. Shenker, and L. Susskind, unpublished,\\
R. Bousso, B. Freivogel, S. Shenker, L. Susskind, I. S. Yang, unpublished.
  
\bibitem{Susskind:2007pv}
L.~Susskind,
``The Census taker's hat,''
[arXiv:0710.1129 [hep-th]].

\bibitem{Bousso:2008hz}
  R.~Bousso, B.~Freivogel and I.~S.~Yang,
  ``Properties of the scale factor measure,''
  Phys.\ Rev.\ D {\bf 79} (2009) 063513
  [arXiv:0808.3770 [hep-th]].
  
\bibitem{Bousso:2010im}
  R.~Bousso, B.~Freivogel, S.~Leichenauer and V.~Rosenhaus,
  ``Geometric origin of coincidences and hierarchies in the landscape,''
  Phys.\ Rev.\ D {\bf 84} (2011) 083517
  [arXiv:1012.2869 [hep-th]].
  
 \bibitem{gv} A.~Guth and V.~Vanchurin, unpublished.

\bibitem{Bousso:2010yn}
  R.~Bousso, B.~Freivogel, S.~Leichenauer and V.~Rosenhaus,
  ``Eternal inflation predicts that time will end,''
  Phys.\ Rev.\ D {\bf 83} (2011) 023525
  [arXiv:1009.4698 [hep-th]].
  
\bibitem{Strominger:2001pn}
  A.~Strominger,
  ``The dS / CFT correspondence,''
  JHEP {\bf 0110} (2001) 034 [hep-th/0106113].
  
\bibitem{Garriga:2008ks}
J.~Garriga and A.~Vilenkin,
``Holographic Multiverse,''
JCAP \textbf{01} (2009), 021\\
$\mbox{[arXiv:0809.4257 [hep-th]]. }$

\bibitem{Harlow:2011az}
  D.~Harlow, S.~H.~Shenker, D.~Stanford and L.~Susskind,
  ``Tree-like structure of eternal inflation: A solvable model,''
  Phys.\ Rev.\ D {\bf 85} (2012) 063516
  [arXiv:1110.0496 [hep-th]].  
  
\bibitem{Nomura:2011dt}
  Y.~Nomura,
  ``Physical Theories, Eternal Inflation, and Quantum Universe,''
  JHEP {\bf 1111} (2011) 063
  [arXiv:1104.2324 [hep-th]].
    
\bibitem{Bousso:2011up}
  R.~Bousso and L.~Susskind,
  ``The Multiverse Interpretation of Quantum Mechanics,''
  Phys.\ Rev.\ D {\bf 85} (2012) 045007
  [arXiv:1105.3796 [hep-th]].

\bibitem{Vilenkin:2013loa}
  A.~Vilenkin, ``A quantum measure of the multiverse,''
  JCAP {\bf 1405} (2014) 005 \\{} 
  [arXiv:1312.0682].
  
\bibitem{Nomura:2012zb}
Y.~Nomura,
``The Static Quantum Multiverse,''
Phys. Rev. D \textbf{86} (2012), 083505\\
$\mbox{[arXiv:1205.5550 [hep-th]].}$ 
 
 \bibitem{Hartle:2016tpo}
  J.~Hartle and T.~Hertog,
  ``One Bubble to Rule Them All,''
  Phys.\ Rev.\ D {\bf 95} (2017) no.12,  123502
  [arXiv:1604.03580 [hep-th]].
  
 \bibitem{DeWitt:1967yk}
  B.~S.~DeWitt,
  ``Quantum Theory of Gravity. 1. The Canonical Theory,''
  Phys.\ Rev.\  {\bf 160} (1967) 1113.

\bibitem{Wheeler:1988zr}
  J.A. $\!$Wheeler,
  ``Superspace and the Nature of Quantum Geometrodynamics,'' $\!$in {\it Battelle ren\-contres - 1967 lectures in mathematics and physics} (Seattle),
pp. 242 - 307, ed. C. DeWitt and J.A. Wheeler (Benjamin, New York, 1968) and Adv.\ Ser.\ Astroph.\ Cosm.\  {\bf 3} (1987) 27.

\bibitem{Hartle:1983ai}
  J.~B.~Hartle and S.~W.~Hawking,
  ``Wave Function of the Universe,''
  Phys.\ Rev.\ D {\bf 28} (1983) 2960
   [Adv.\ Ser.\ Astrophys.\ Cosmol.\  {\bf 3} (1987) 174].
 
\bibitem{Banks:1984cw}
  T.~Banks,
  ``T C P, Quantum Gravity, the Cosmological Constant 
  and All That...,''  Nucl.\ Phys.\ B {\bf 249} (1985) 332.

\bibitem{Halliwell:2018ejl}
  J.~J.~Halliwell, J.~B.~Hartle and T.~Hertog,
  ``What is the No-Boundary Wave Function of the Universe?,''
  Phys.\ Rev.\ D {\bf 99} (2019) no.4,  043526
  [arXiv:1812.01760 [hep-th]].

\bibitem{Weinberg:1975gm}
S.~Weinberg,
``Implications of Dynamical Symmetry Breaking,''
Phys. Rev. D \textbf{13} (1976), 974-996.

\bibitem{Susskind:1978ms}
L.~Susskind, ``Dynamics of Spontaneous Symmetry Breaking in the Weinberg-Salam Theory,''
Phys. Rev. D \textbf{20} (1979), 2619-2625.

\bibitem{Dimopoulos:1979es}
S.~Dimopoulos and L.~Susskind,
``Mass Without Scalars,'' Nucl.~Phys.~B~155 (1979) 237-252.

\bibitem{Eichten:1979ah}
E.~Eichten and K.~D.~Lane,
``Dynamical Breaking of Weak Interaction Symmetries,''
Phys. Lett. B \textbf{90} (1980), 125-130.

\bibitem{Kaplan:1983fs}
D.~B.~Kaplan and H.~Georgi,
``SU(2) x U(1) Breaking by Vacuum Misalignment,''
Phys. Lett. B \textbf{136} (1984), 183-186.

\bibitem{Bardeen:1989ds}
W.~A.~Bardeen, C.~T.~Hill and M.~Lindner,
``Minimal Dynamical Symmetry Breaking of the Standard Model,'' Phys. Rev. D \textbf{41} (1990), 1647.

\bibitem{King:1994yr}  
S.~F.~King,
``Dynamical electroweak symmetry breaking,''
Rept. Prog. Phys. \textbf{58} (1995), 263-310
[arXiv:hep-ph/9406401 [hep-ph]].

\bibitem{Chivukula:1998if}
R.~Chivukula, ``Models of electroweak symmetry breaking: Course,'' Lectures at NATO Advanced Study Institute on Quantum Field Theory (1998) and Les Houches Summer School in Theoretical Physics, Session 68: Probing the Standard Model of Particle Interactions (1997)
[arXiv:hep-ph/9803219 [hep-ph]].
  
\bibitem{Lane:2002wv}
K.~Lane, ``Two Lectures on Technicolor,'' preprints 
FERMILAB-PUB-02-040-T and BUHEP-02-15 [arXiv:hep-ph/0202255 [hep-ph]].  

\bibitem{Hill:2002ap}
C.~T.~Hill and E.~H.~Simmons,
``Strong Dynamics and Electroweak Symmetry Breaking,''
Phys. Rept. \textbf{381} (2003), 235-402
[arXiv:hep-ph/0203079 [hep-ph]].

\bibitem{Piai:2010ma}
M.~Piai, ``Lectures on walking technicolor, holography and gauge/gravity dualities,''
Adv. High Energy Phys. \textbf{2010} (2010), 464302
[arXiv:1004.0176 [hep-ph]].  

\bibitem{Cacciapaglia:2020kgq}
G.~Cacciapaglia, C.~Pica and F.~Sannino,
``Fundamental Composite Dynamics: A Review,''
[arXiv:2002.04914 [hep-ph]].

\bibitem{Aparicio:2014wxa}
L.~Aparicio, M.~Cicoli, S.~Krippendorf, A.~Maharana, F.~Muia and F.~Quevedo,
``Sequestered de Sitter String Scenarios: Soft-terms,''
JHEP \textbf{11} (2014), 071
[arXiv:1409.1931 [hep-th]].

 \bibitem{Susskind:2004uv}
L.~Susskind,
``Supersymmetry breaking in the anthropic landscape,''
Published in: M.~Shifman (ed.) et al.: From fields to strings, vol. 3, p.~1745 [arXiv:hep-th/0405189 [hep-th]]. 
  
 \bibitem{Douglas:2004qg}
M.~R.~Douglas,
``Statistical analysis of the supersymmetry breaking scale,''
[arXiv:hep-th/0405279 [hep-th]]. 

\bibitem{Giudice:2006sn}
G.~Giudice and R.~Rattazzi,
``Living Dangerously with Low-Energy Supersymmetry,''
Nucl. Phys. B \textbf{757} (2006), 19-46
[arXiv:hep-ph/0606105 [hep-ph]].

\bibitem{Acharya:2008zi}
B.~S.~Acharya, K.~Bobkov, G.~L.~Kane, J.~Shao and P.~Kumar,
``The G(2)-MSSM: An M Theory motivated model of Particle Physics,''
Phys. Rev. D \textbf{78} (2008), 065038
[arXiv:0801.0478 [hep-ph]].

\bibitem{Baer:2019zfl}
H.~Baer, V.~Barger and D.~Sengupta,
``Landscape solution to the SUSY flavor and CP problems,''
Phys.\ Rev.\ Res.\  \textbf{1} (2019) no.3, 033179
[arXiv:1910.00090 [hep-ph]].

\bibitem{Baer:2020kwz}
H.~Baer, V.~Barger, S.~Salam, D.~Sengupta and K.~Sinha,
``Midi-review: Status of weak scale supersymmetry after LHC Run 2 and ton-scale noble liquid WIMP searches,''
[arXiv:2002.03013 [hep-ph]].

\bibitem{Broeckel:2020fdz}
I.~Broeckel, M.~Cicoli, A.~Maharana, K.~Singh and K.~Sinha, ``Moduli Stabilisation and the Statistics of SUSY Breaking in the Landscape,''
[arXiv:2007.04327 [hep-th]].

\bibitem{Csaki:2018muy}
Csaki, Csaba, S.~Lombardo and O.~Telem,
``TASI Lectures on Non-supersymmetric BSM Models,''
[arXiv:1811.04279 [hep-ph]].

\bibitem{ArkaniHamed:2001nc}
N.~Arkani-Hamed, A.~G.~Cohen and H.~Georgi,
``Electroweak symmetry breaking from dimensional deconstruction,''
Phys. Lett. B \textbf{513} (2001), 232-240
[arXiv:hep-ph/0105239 [hep-ph]].

\bibitem{ArkaniHamed:2002pa}
N.~Arkani-Hamed, A.~G.~Cohen, T.~Gregoire and J.~G.~Wacker,
``Phenomenology of electroweak symmetry breaking from theory space,''
JHEP \textbf{08} (2002), 020
[arXiv:hep-ph/0202089 [hep-ph]].

\bibitem{ArkaniHamed:2002qx}
N.~Arkani-Hamed, A.~Cohen, E.~Katz, A.~Nelson, T.~Gregoire and J.~G.~Wacker,
``The Minimal moose for a little Higgs,''
JHEP \textbf{08} (2002), 021
[arXiv:hep-ph/0206020 [hep-ph]].

\bibitem{Schmaltz:2005ky}
M.~Schmaltz and D.~Tucker-Smith,
``Little Higgs review,''
Ann. Rev. Nucl. Part. Sci. \textbf{55} (2005), 229-270
[arXiv:hep-ph/0502182 [hep-ph]].

\bibitem{ArkaniHamed:2002qy}
N.~Arkani-Hamed, A.~Cohen, E.~Katz and A.~Nelson,
``The Littlest Higgs,'' JHEP \textbf{07} (2002), 034 [arXiv:hep-ph/0206021 [hep-ph]].

\bibitem{ArkaniHamed:1998rs}
N.~Arkani-Hamed, S.~Dimopoulos and G.~Dvali,
``The Hierarchy problem and new dimensions at a millimeter,'' Phys. Lett. B \textbf{429} (1998), 263-272
[arXiv:hep-ph/9803315 [hep-ph]].

\bibitem{Antoniadis:1998ig}
I.~Antoniadis, N.~Arkani-Hamed, S.~Dimopoulos and G.~Dvali,
``New dimensions at a millimeter to a Fermi and superstrings at a TeV,''
Phys. Lett. B \textbf{436} (1998), 257-263 [arXiv:hep-ph/9804398 [hep-ph]].

\bibitem{Rubakov:2001kp}
V.~A.~Rubakov,
``Large and infinite extra dimensions: An Introduction,''
Phys. Usp. \textbf{44} (2001), 871-893
[arXiv:hep-ph/0104152 [hep-ph]].

\bibitem{Csaki:2004ay}
C.~Csaki,
``TASI lectures on extra dimensions and branes,'' (published in: In ``Shifman, M. (ed.) et al.: From fields to strings, vol. 2'') [arXiv:hep-ph/0404096 [hep-ph]].

\bibitem{Randall:1999ee}
L.~Randall and R.~Sundrum,
``A Large mass hierarchy from a small extra dimension,''
Phys. Rev. Lett. \textbf{83} (1999), 3370-3373
[arXiv:hep-ph/9905221 [hep-ph]].

\bibitem{Randall:1999vf}
L.~Randall and R.~Sundrum,
``An Alternative to compactification,''
Phys. Rev. Lett. \textbf{83} (1999), 4690-4693
[arXiv:hep-th/9906064 [hep-th]].

\bibitem{Rubakov:1983bz}
V.~A.~Rubakov and M.~E.~Shaposhnikov,
``Extra Space-Time Dimensions: Towards a Solution to the Cosmological Constant Problem,''
Phys. Lett. B \textbf{125} (1983), 139.

\bibitem{Goldberger:1999uk}
W.~D.~Goldberger and M.~B.~Wise,
``Modulus stabilization with bulk fields,''
Phys. Rev. Lett. \textbf{83} (1999), 4922-4925 [arXiv:hep-ph/9907447 [hep-ph]].

\bibitem{Gherghetta:2000qt}
T.~Gherghetta and A.~Pomarol,
``Bulk fields and supersymmetry in a slice of AdS,''
Nucl. Phys. B \textbf{586} (2000), 141-162 [arXiv:hep-ph/0003129 [hep-ph]].

\bibitem{Contino:2003ve}
R.~Contino, Y.~Nomura and A.~Pomarol,
``Higgs as a holographic pseudoGoldstone boson,''
Nucl. Phys. B \textbf{671} (2003), 148-174 [arXiv:hep-ph/0306259 [hep-ph]].

\bibitem{Agashe:2003zs}
K.~Agashe, A.~Delgado, M.~J.~May and R.~Sundrum,
``RS1, custodial isospin and precision tests,''
JHEP \textbf{08} (2003), 050 [arXiv:hep-ph/0308036 [hep-ph]].

\bibitem{Agashe:2004rs}
K.~Agashe, R.~Contino and A.~Pomarol,
``The Minimal composite Higgs model,''
Nucl. Phys. B \textbf{719} (2005), 165-187 [arXiv:hep-ph/0412089 [hep-ph]].

\bibitem{Csaki:2005vy}
C.~Csaki, J.~Hubisz and P.~Meade,
``TASI lectures on electroweak symmetry breaking from extra dimensions,'' [arXiv:hep-ph/0510275 [hep-ph]].

\bibitem{Gherghetta:2006ha}
T.~Gherghetta,
``Les Houches lectures on warped models and holography,''
[arXiv:hep-ph/0601213 [hep-ph]].

\bibitem{Kribs:2006mq}
G.~D.~Kribs,
``TASI 2004 lectures on the phenomenology of extra dimensions,'' [arXiv:hep-ph/0605325 [hep-ph]].

\bibitem{Rattazzi:2003ea}
R.~Rattazzi,
``Cargese lectures on extra-dimensions,'' (published in ``Cargese 2003, Particle physics and cosmology'') [arXiv:hep-ph/0607055 [hep-ph]].

\bibitem{Contino:2010rs}
R.~Contino,
``The Higgs as a Composite Nambu-Goldstone Boson,''\\
$\mbox{[arXiv:1005.4269 [hep-ph]]}$.

\bibitem{Gherghetta:2010cj}
T.~Gherghetta,
``A Holographic View of Beyond the Standard Model Physics,''\\
$\mbox{[arXiv:1008.2570 [hep-ph]]}$.

\bibitem{vonGersdorff:2011rz}
G.~von Gersdorff,
``Electroweak Symmetry Breaking in Warped Extra Dimensions,'' (contribution to 46th Rencontres de Moriond on Electroweak Interactions and Unified Theories) [arXiv:1107.1989 [hep-ph]].

\bibitem{Gubser:1999vj}
S.~S.~Gubser, ``AdS / CFT and gravity,''
Phys. Rev. D \textbf{63} (2001), 084017 [arXiv:hep-th/9912001 [hep-th]].

\bibitem{ArkaniHamed:2000ds}
N.~Arkani-Hamed, M.~Porrati and L.~Randall,
``Holography and phenomenology,''
JHEP \textbf{08} (2001), 017 [arXiv:hep-th/0012148 [hep-th]].

\bibitem{Cascales:2003wn}
J.~Cascales, G.~del Moral, M.P., F.~Quevedo and A.~Uranga,
``Realistic D-brane models on warped throats: Fluxes, hierarchies and moduli stabilization,''
JHEP \textbf{02} (2004), 031 [arXiv:hep-th/0312051 [hep-th]].

\bibitem{Cascales:2005rj}
J.~F.~Cascales, F.~Saad and A.~M.~Uranga,
``Holographic dual of the standard model on the throat,''
JHEP \textbf{11} (2005), 047 [arXiv:hep-th/0503079 [hep-th]].

\bibitem{Abbott:1984qf}
L.~F.~Abbott,
``A Mechanism for Reducing the Value of the Cosmological Constant,''
Phys. Lett. B \textbf{150} (1985), 427-430.

\bibitem{Dvali:2003br}
G.~Dvali and A.~Vilenkin,
``Cosmic attractors and gauge hierarchy,''
Phys. Rev. D \textbf{70} (2004), 063501 [arXiv:hep-th/0304043 [hep-th]].

\bibitem{Dvali:2004tma}
G.~Dvali, ``Large hierarchies from attractor vacua,''
Phys. Rev. D \textbf{74} (2006), 025018
[arXiv:hep-th/0410286 [hep-th]].

\bibitem{Graham:2015cka}
P.~W.~Graham, D.~E.~Kaplan and S.~Rajendran,
``Cosmological Relaxation of the Electroweak Scale,''
Phys. Rev. Lett. \textbf{115} (2015) no.22, 221801 [arXiv:1504.07551 [hep-ph]].

\bibitem{Espinosa:2015eda}
J.~Espinosa, C.~Grojean, G.~Panico, A.~Pomarol, O.~Pujolas and G.~Servant,
``Cosmological Higgs-Axion Interplay for a Naturally Small Electroweak Scale,''
Phys. Rev. Lett. \textbf{115} (2015) no.25, 251803 [arXiv:1506.09217 [hep-ph]].

\bibitem{Hardy:2015laa}
E.~Hardy,
``Electroweak relaxation from finite temperature,''
JHEP \textbf{11} (2015), 077
[arXiv:1507.07525 [hep-ph]].

\bibitem{Patil:2015oxa}
S.~P.~Patil and P.~Schwaller,
``Relaxing the Electroweak Scale: the Role of Broken dS Symmetry,''
JHEP \textbf{02} (2016), 077 [arXiv:1507.08649 [hep-ph]].

\bibitem{Antipin:2015jia}
O.~Antipin and M.~Redi,
``The Half-composite Two Higgs Doublet Model and the Relaxion,'' JHEP \textbf{12} (2015), 031 [arXiv:1508.01112 [hep-ph]].

\bibitem{Jaeckel:2015txa}
J.~Jaeckel, V.~M.~Mehta and L.~T.~Witkowski,
``Musings on cosmological relaxation and the hierarchy problem,'' Phys. Rev. D \textbf{93} (2016) no.6, 063522
[arXiv:1508.03321 [hep-ph]].

\bibitem{Arkani-Hamed:2016rle}
N.~Arkani-Hamed, T.~Cohen, R.~T.~D'Agnolo, A.~Hook, H.~Kim, Do and D.~Pinner,
``Solving the Hierarchy Problem at Reheating with a Large Number of Degrees of Freedom,''
Phys. Rev. Lett. \textbf{117} (2016) no.25, 251801 [arXiv:1607.06821 [hep-ph]].

\bibitem{Arvanitaki:2016xds}
A.~Arvanitaki, S.~Dimopoulos, V.~Gorbenko, J.~Huang and K.~Van Tilburg, ``A small weak scale from a small cosmological constant,'' JHEP \textbf{05} (2017), 071 [arXiv:1609.06320 [hep-ph]].

\bibitem{Alberte:2016izw}
L.~Alberte, P.~Creminelli, A.~Khmelnitsky, D.~Pirtskhalava and E.~Trincherini,
``Relaxing the Cosmological Constant: a Proof of Concept,'' JHEP \textbf{12} (2016), 022
[arXiv:1608.05715 [hep-th]].

\bibitem{Geller:2018xvz}
M.~Geller, Y.~Hochberg and E.~Kuflik,
``Inflating to the Weak Scale,''
Phys. Rev. Lett. \textbf{122} (2019) no.19, 191802
[arXiv:1809.07338 [hep-ph]].

\bibitem{Cheung:2018xnu}
C.~Cheung and P.~Saraswat,
``Mass Hierarchy and Vacuum Energy,''
[arXiv:1811.12390 [hep-ph]].

\bibitem{Graham:2019bfu}
P.~W.~Graham, D.~E.~Kaplan and S.~Rajendran,
``Relaxation of the Cosmological Constant,''
Phys. Rev. D \textbf{100} (2019) no.1, 015048
[arXiv:1902.06793 [hep-ph]].

\bibitem{Strumia:2019kxg}
A.~Strumia and D.~Teresi,
``Cosmological constant: relaxation vs multiverse,''
Phys. Lett. B \textbf{797} (2019), 134901
[arXiv:1904.07876 [gr-qc]].

\bibitem{Giudice:2019iwl}
G.~Giudice, A.~Kehagias and A.~Riotto,
``The Selfish Higgs,''
JHEP \textbf{10} (2019), 199
[arXiv:1907.05370 [hep-ph]].

\bibitem{Bloch:2019bvc}
I.~M.~Bloch, C.~Csaki, M.~Geller and T.~Volansky,
``Crunching Away the Cosmological Constant Problem: Dynamical Selection of a Small $\Lambda$,''
[arXiv:1912.08840 [hep-ph]].

\bibitem{Kaloper:2019xfj}
N.~Kaloper and A.~Westphal,
``A Goldilocks Higgs,''
[arXiv:1907.05837 [hep-th]].

\bibitem{McAllister:2016vzi}
L.~McAllister, P.~Schwaller, G.~Servant, J.~Stout and A.~Westphal,
``Runaway Relaxion Monodromy,''
JHEP \textbf{02} (2018), 124
[arXiv:1610.05320 [hep-th]].

\bibitem{Vafa:2005ui}
C.~Vafa, ``The String landscape and the swampland,''
[arXiv:hep-th/0509212 [hep-th]].

\bibitem{Ooguri:2006in}
H.~Ooguri and C.~Vafa,
``On the Geometry of the String Landscape and the Swampland,''
Nucl. Phys. B \textbf{766} (2007), 21-33
[arXiv:hep-th/0605264 [hep-th]].

\bibitem{Brennan:2017rbf}
T.~D.~Brennan, F.~Carta and C.~Vafa,
``The String Landscape, the Swampland, and the Missing Corner,''
PoS \textbf{TASI2017} (2017), 015
[arXiv:1711.00864 [hep-th]].

\bibitem{Palti:2019pca}
E.~Palti, ``The Swampland: Introduction and Review,''
Fortsch. Phys. \textbf{67} (2019) no.6, 1900037
[arXiv:1903.06239 [hep-th]].

\bibitem{Banks:1988yz}
T.~Banks and L.~J.~Dixon,
``Constraints on String Vacua with Space-Time Supersymmetry,'' Nucl. Phys. B \textbf{307} (1988), 93-108.

\bibitem{Kamionkowski:1992mf}
M.~Kamionkowski and J.~March-Russell,
``Planck scale physics and the Peccei-Quinn mechanism,''
Phys. Lett. B \textbf{282} (1992), 137-141
[arXiv:hep-th/9202003 [hep-th]].

\bibitem{Holman:1992us}
R.~Holman, S.~D.~Hsu, T.~W.~Kephart, E.~W.~Kolb, R.~Watkins and L.~M.~Widrow,
``Solutions to the strong CP problem in a world with gravity,''
Phys. Lett. B \textbf{282} (1992), 132-136
[arXiv:hep-ph/9203206 [hep-ph]].

\bibitem{Kallosh:1995hi}
R.~Kallosh, A.~D.~Linde, D.~A.~Linde and L.~Susskind,
``Gravity and global symmetries,''
Phys. Rev. D \textbf{52} (1995), 912-935
[arXiv:hep-th/9502069 [hep-th]].

\bibitem{Banks:2010zn}
T.~Banks and N.~Seiberg,
``Symmetries and Strings in Field Theory and Gravity,''
Phys. Rev. D \textbf{83} (2011), 084019
[arXiv:1011.5120 [hep-th]].

\bibitem{Harlow:2018tng}
D.~Harlow and H.~Ooguri,
``Symmetries in quantum field theory and quantum gravity,'' [arXiv:1810.05338 [hep-th]].

\bibitem{Lee:2019wij}
S.~J.~Lee, W.~Lerche and T.~Weigand,
``Emergent Strings from Infinite Distance Limits,''
[arXiv:1910.01135 [hep-th]].

\bibitem{Klaewer:2016kiy}
D.~Klaewer and E.~Palti,
``Super-Planckian Spatial Field Variations and Quantum Gravity,''
JHEP \textbf{01} (2017), 088
[arXiv:1610.00010 [hep-th]].

\bibitem{ArkaniHamed:2006dz}
N.~Arkani-Hamed, L.~Motl, A.~Nicolis and C.~Vafa,
``The String landscape, black holes and gravity as the weakest force,''
JHEP \textbf{06} (2007), 060
[arXiv:hep-th/0601001 [hep-th]].

\bibitem{Susskind:1995da}
L.~Susskind, ``Trouble for remnants,'' [arXiv:hep-th/9501106 [hep-th]].

\bibitem{Cheung:2014vva}
C.~Cheung and G.~N.~Remmen,
``Naturalness and the Weak Gravity Conjecture,''
Phys. Rev. Lett. \textbf{113} (2014), 051601
[arXiv:1402.2287 [hep-ph]].

\bibitem{delaFuente:2014aca}
A.~de la Fuente, P.~Saraswat and R.~Sundrum,
``Natural Inflation and Quantum Gravity,''
Phys. Rev. Lett. \textbf{114} (2015) no.15, 151303
[arXiv:1412.3457 [hep-th]].

\bibitem{Rudelius:2015xta}
T.~Rudelius,
``Constraints on Axion Inflation from the Weak Gravity Conjecture,''
JCAP \textbf{09} (2015), 020
[arXiv:1503.00795 [hep-th]].

\bibitem{Montero:2015ofa}
M.~Montero, A.~M.~Uranga and I.~Valenzuela,
``Transplanckian axions!?,''
JHEP \textbf{08} (2015), 032
[arXiv:1503.03886 [hep-th]].

\bibitem{Brown:2015iha}
J.~Brown, W.~Cottrell, G.~Shiu and P.~Soler,
``Fencing in the Swampland: Quantum Gravity Constraints on Large Field Inflation,''
JHEP \textbf{10} (2015), 023
[arXiv:1503.04783 [hep-th]].

\bibitem{Bachlechner:2015qja}
T.~C.~Bachlechner, C.~Long and L.~McAllister,
``Planckian Axions and the Weak Gravity Conjecture,''
JHEP \textbf{01} (2016), 091
[arXiv:1503.07853 [hep-th]].

\bibitem{Hebecker:2015rya}
A.~Hebecker, P.~Mangat, F.~Rompineve and L.~T.~Witkowski,
``Winding out of the Swamp: Evading the Weak Gravity Conjecture with F-term Winding Inflation?,''
Phys. Lett. B \textbf{748} (2015), 455-462
[arXiv:1503.07912 [hep-th]].

\bibitem{Junghans:2015hba}
D.~Junghans,
``Large-Field Inflation with Multiple Axions and the Weak Gravity Conjecture,''
JHEP \textbf{02} (2016), 128
[arXiv:1504.03566 [hep-th]].

\bibitem{Heidenreich:2015wga}
B.~Heidenreich, M.~Reece and T.~Rudelius,
``Weak Gravity Strongly Constrains Large-Field Axion Inflation,''
JHEP \textbf{12} (2015), 108
[arXiv:1506.03447 [hep-th]].

\bibitem{Freese:1990rb}
K.~Freese, J.~A.~Frieman and A.~V.~Olinto,
``Natural inflation with pseudo - Nambu-Goldstone bosons,''
Phys. Rev. Lett. \textbf{65} (1990), 3233-3236.

\bibitem{Banks:2003sx}
T.~Banks, M.~Dine, P.~J.~Fox and E.~Gorbatov,
``On the possibility of large axion decay constants,''
JCAP \textbf{06} (2003), 001
[arXiv:hep-th/0303252 [hep-th]].

\bibitem{Silverstein:2008sg}
E.~Silverstein and A.~Westphal,
``Monodromy in the CMB: Gravity Waves and String Inflation,''
Phys. Rev. D \textbf{78} (2008), 106003
[arXiv:0803.3085 [hep-th]].

\bibitem{McAllister:2008hb}
L.~McAllister, E.~Silverstein and A.~Westphal,
``Gravity Waves and Linear Inflation from Axion Monodromy,''
Phys. Rev. D \textbf{82} (2010), 046003
[arXiv:0808.0706 [hep-th]].

\bibitem{Kaloper:2008fb}
N.~Kaloper and L.~Sorbo,
``A Natural Framework for Chaotic Inflation,''
Phys. Rev. Lett. \textbf{102} (2009), 121301
[arXiv:0811.1989 [hep-th]].

\bibitem{Kaloper:2011jz}
N.~Kaloper, A.~Lawrence and L.~Sorbo,
``An Ignoble Approach to Large Field Inflation,''
JCAP \textbf{03} (2011), 023
[arXiv:1101.0026 [hep-th]].

\bibitem{Marchesano:2014mla}
F.~Marchesano, G.~Shiu and A.~M.~Uranga,
``F-term Axion Monodromy Inflation,''
JHEP \textbf{09} (2014), 184
[arXiv:1404.3040 [hep-th]].

\bibitem{Blumenhagen:2014gta}
R.~Blumenhagen and E.~Plauschinn,
``Towards Universal Axion Inflation and Reheating in String Theory,''
Phys. Lett. B \textbf{736} (2014), 482-487
[arXiv:1404.3542 [hep-th]].

\bibitem{Hebecker:2014eua}
A.~Hebecker, S.~C.~Kraus and L.~T.~Witkowski,
``D7-Brane Chaotic Inflation,''
Phys. Lett. B \textbf{737} (2014), 16-22
[arXiv:1404.3711 [hep-th]].

\bibitem{Baume:2016psm}
F.~Baume and E.~Palti,
``Backreacted Axion Field Ranges in String Theory,''
JHEP \textbf{08} (2016), 043
[arXiv:1602.06517 [hep-th]].

\bibitem{Ibanez:2015fcv}
L.~E.~Ibanez, M.~Montero, A.~Uranga and I.~Valenzuela,
``Relaxion Monodromy and the Weak Gravity Conjecture,''
JHEP \textbf{04} (2016), 020
[arXiv:1512.00025 [hep-th]].

\bibitem{Grimm:2018ohb}
T.~W.~Grimm, E.~Palti and I.~Valenzuela,
``Infinite Distances in Field Space and Massless Towers of States,''
JHEP \textbf{08} (2018), 143
[arXiv:1802.08264 [hep-th]].

\bibitem{Lee:2018urn}
S.~Lee, W.~Lerche and T.~Weigand,
``Tensionless Strings and the Weak Gravity Conjecture,''
JHEP \textbf{10} (2018), 164
[arXiv:1808.05958 [hep-th]].

\bibitem{Kim:2004rp}
J.~E.~Kim, H.~P.~Nilles and M.~Peloso,
``Completing natural inflation,''
JCAP \textbf{01} (2005), 005
[arXiv:hep-ph/0409138 [hep-ph]].

\bibitem{Dvali:2005an}
G.~Dvali,
``Three-form gauging of axion symmetries and gravity,''
[arXiv:hep-th/0507215 [hep-th]].

\bibitem{Saraswat:2016eaz}
P.~Saraswat,
``Weak gravity conjecture and effective field theory,''
Phys. Rev. D \textbf{95} (2017) no.2, 025013
[arXiv:1608.06951 [hep-th]].

\bibitem{Choi:2015fiu}
K.~Choi and S.~H.~Im,
``Realizing the relaxion from multiple axions and its UV completion with high scale supersymmetry,''
JHEP \textbf{01} (2016), 149
[arXiv:1511.00132 [hep-ph]].

\bibitem{Kaplan:2015fuy}
D.~E.~Kaplan and R.~Rattazzi,
``Large field excursions and approximate discrete symmetries from a clockwork axion,''
Phys. Rev. D \textbf{93} (2016) no.8, 085007
[arXiv:1511.01827 [hep-ph]].

\bibitem{Akrami:2018ylq}
Y.~Akrami, R.~Kallosh, A.~Linde and V.~Vardanyan,
``The Landscape, the Swampland and the Era of Precision Cosmology,''
Fortsch. Phys. \textbf{67} (2019) no.1-2, 1800075
[arXiv:1808.09440 [hep-th]].

\bibitem{Denef:2018etk}
F.~Denef, A.~Hebecker and T.~Wrase,
``de Sitter swampland conjecture and the Higgs potential,''
Phys. Rev. D \textbf{98} (2018) no.8, 086004
[arXiv:1807.06581 [hep-th]].

\bibitem{Cicoli:2018kdo}
M.~Cicoli, S.~De Alwis, A.~Maharana, F.~Muia and F.~Quevedo,
``De Sitter vs Quintessence in String Theory,''
Fortsch. Phys. \textbf{67} (2019) no.1-2, 1800079
[arXiv:1808.08967 [hep-th]].

\bibitem{Choi:2018rze}
K.~Choi, D.~Chway and C.~S.~Shin,
``The dS swampland conjecture with the electroweak symmetry and QCD chiral symmetry breaking,''
JHEP \textbf{11} (2018), 142
[arXiv:1809.01475 [hep-th]].

\bibitem{Conlon:2018eyr}
J.~P.~Conlon,
``The de Sitter swampland conjecture and supersymmetric AdS vacua,''
Int. J. Mod. Phys. A \textbf{33} (2018) no.29, 1850178
[arXiv:1808.05040 [hep-th]].

\bibitem{Hebecker:2018vxz}
A.~Hebecker and T.~Wrase,
``The Asymptotic dS Swampland Conjecture ? a Simplified Derivation and a Potential Loophole,''
Fortsch. Phys. \textbf{67} (2019) no.1-2, 1800097
[arXiv:1810.08182 [hep-th]].

\bibitem{Junghans:2018gdb}
D.~Junghans,
``Weakly Coupled de Sitter Vacua with Fluxes and the Swampland,''
JHEP \textbf{03} (2019), 150
[arXiv:1811.06990 [hep-th]].

\bibitem{Dine:1985he}
M.~Dine and N.~Seiberg,
``Is the Superstring Weakly Coupled?,''
Phys. Lett. B \textbf{162} (1985), 299-302.

\bibitem{Maldacena:2000mw}
J.~M.~Maldacena and C.~Nunez,
``Supergravity description of field theories on curved manifolds and a no go theorem,''
Int. J. Mod. Phys. A \textbf{16} (2001), 822-855
[arXiv:hep-th/0007018 [hep-th]].

\bibitem{Kachru:2019dvo}
S.~Kachru, M.~Kim, L.~McAllister and M.~Zimet,
``de Sitter Vacua from Ten Dimensions,''
[arXiv:1908.04788 [hep-th]].

\bibitem{Bena:2019mte}
I.~Bena, M.~Grana, N.~Kovensky and A.~Retolaza,
``Kahler moduli stabilization from ten dimensions,''
JHEP \textbf{10} (2019), 200
[arXiv:1908.01785 [hep-th]].

\bibitem{Gautason:2015tig}
F.~F.~Gautason, M.~Schillo, T.~Van Riet and M.~Williams,
``Remarks on scale separation in flux vacua,''
JHEP \textbf{03} (2016), 061
[arXiv:1512.00457 [hep-th]].

\bibitem{Gautason:2018gln}
F.~F.~Gautason, V.~Van Hemelryck and T.~Van Riet,
``The Tension between 10D Supergravity and dS Uplifts,''
Fortsch. Phys. \textbf{67} (2019) no.1-2, 1800091
[arXiv:1810.08518 [hep-th]].

\bibitem{Lust:2019zwm}
D.~L\"ust, E.~Palti and C.~Vafa,
``AdS and the Swampland,''
Phys. Lett. B \textbf{797} (2019), 134867
[arXiv:1906.05225 [hep-th]].

\bibitem{Demirtas:2019sip}
M.~Demirtas, M.~Kim, L.~Mcallister and J.~Moritz,
``Vacua with Small Flux Superpotential,''
[arXiv:1912.10047 [hep-th]].

\bibitem{DeWolfe:2005uu}
O.~DeWolfe, A.~Giryavets, S.~Kachru and W.~Taylor,
``Type IIA moduli stabilization,''
JHEP \textbf{07} (2005), 066
[arXiv:hep-th/0505160 [hep-th]].

\bibitem{Marchesano:2020qvg}
F.~Marchesano, E.~Palti, J.~Quirant and A.~Tomasiello,
``On supersymmetric AdS$_4$ orientifold vacua,''
[arXiv:2003.13578 [hep-th]].

\bibitem{Junghans:2020acz}
D.~Junghans,
``O-plane Backreaction and Scale Separation in Type IIA Flux Vacua,''
[arXiv:2003.06274 [hep-th]].

\bibitem{Wetterich:1987fm}
C.~Wetterich,
``Cosmology and the Fate of Dilatation Symmetry,''
Nucl. Phys. B \textbf{302} (1988), 668-696
[arXiv:1711.03844 [hep-th]].

\bibitem{Peebles:1987ek}
P.~Peebles and B.~Ratra,
``Cosmology with a Time Variable Cosmological Constant,''
Astrophys. J. \textbf{325} (1988), L17.

\bibitem{Cicoli:2012tz}
M.~Cicoli, F.~G.~Pedro and G.~Tasinato,
``Natural Quintessence in String Theory,''
JCAP \textbf{07} (2012), 044
[arXiv:1203.6655 [hep-th]].

\bibitem{Hebecker:2019csg}
A.~Hebecker, T.~Skrzypek and M.~Wittner,
``The $F$-term Problem and other Challenges of Stringy Quintessence,''
JHEP \textbf{11} (2019), 134
[arXiv:1909.08625 [hep-th]].

\bibitem{Hardy:2019apu}
E.~Hardy and S.~Parameswaran,
``Thermal Dark Energy,''
Phys. Rev. D \textbf{101} (2020) no.2, 023503
[arXiv:1907.10141 [hep-th]].

\bibitem{Regge:1961px}
T.~Regge,
``General Relativity without Coordinates,''
Nuovo Cim. \textbf{19} (1961), 558-571.

\bibitem{Williams:1991cd}
R.~M.~Williams and P.~A.~Tuckey,
``Regge calculus: A Bibliography and brief review,''
Class. Quant. Grav. \textbf{9} (1992), 1409-1422.

\bibitem{Ambjorn:1998xu}
J.~Ambjorn and R.~Loll,
``Nonperturbative Lorentzian quantum gravity, causality and topology change,''
Nucl. Phys. B \textbf{536} (1998), 407-434
[arXiv:hep-th/9805108 [hep-th]].

\bibitem{Loll:2019rdj}
R.~Loll,
``Quantum Gravity from Causal Dynamical Triangulations: A Review,''
Class. Quant. Grav. \textbf{37} (2020) no.1, 013002
[arXiv:1905.08669 [hep-th]].

\bibitem{Ambjorn:2012jv}
J.~Ambjorn, A.~G\"orlich, J.~Jurkiewicz and R.~Loll,
``Nonperturbative Quantum Gravity,''
Phys. Rept. \textbf{519} (2012), 127-210
[arXiv:1203.3591 [hep-th]].

\bibitem{Ambjorn:2020rcn}
J.~Ambjorn, J.~Gizbert-Studnicki, A.~G\"orlich, J.~Jurkiewicz and R.~Loll,
``Renormalization in quantum theories of geometry,''
[arXiv:2002.01693 [hep-th]].

\bibitem{Giddings:1987cg}
S.~B.~Giddings and A.~Strominger,
``Axion Induced Topology Change in Quantum Gravity and String Theory,''
Nucl. Phys. B \textbf{306} (1988), 890-907. 

\bibitem{Coleman:1988tj}
S.~R.~Coleman,
``Why There Is Nothing Rather Than Something: A Theory of the Cosmological Constant,''
Nucl. Phys. B \textbf{310} (1988), 643-668.

\bibitem{Kazakov:1985ea}
V.~Kazakov, A.~A.~Migdal and I.~Kostov,
``Critical Properties of Randomly Triangulated Planar Random Surfaces,''
Phys. Lett. B \textbf{157} (1985), 295-300. 

\bibitem{Ambjorn:1985az}
J.~Ambjorn, B.~Durhuus and J.~Frohlich,
``Diseases of Triangulated Random Surface Models, and Possible Cures,''
Nucl. Phys. B \textbf{257} (1985), 433-449.

\bibitem{David:1984tx}
F.~David,
``Planar Diagrams, Two-Dimensional Lattice Gravity and Surface Models,''
Nucl. Phys. B \textbf{257} (1985), 45.

\bibitem{Klebanov:1991qa}
I.~R.~Klebanov,
``String theory in two-dimensions,'' published in ``Trieste Spring School 1991'' [arXiv:hep-th/9108019 [hep-th]].

\bibitem{Ginsparg:1993is}
P.~H.~Ginsparg and G.~W.~Moore,
``Lectures on 2-D gravity and 2-D string theory,'' published in ``Boulder 1992, Proceedings, Recent directions in particle theory'' [arXiv:hep-th/9304011 [hep-th]].

\bibitem{Thiemann:2002nj}
T.~Thiemann,
``Lectures on loop quantum gravity,''
Lect. Notes Phys. \textbf{631} (2003), 41-135
[arXiv:gr-qc/0210094 [gr-qc]].

\bibitem{Nicolai:2005mc}
H.~Nicolai, K.~Peeters and M.~Zamaklar,
``Loop quantum gravity: An Outside view,''
Class. Quant. Grav. \textbf{22} (2005), R193
[arXiv:hep-th/0501114 [hep-th]].

\bibitem{Nicolai:2006id}
H.~Nicolai and K.~Peeters,
``Loop and spin foam quantum gravity: A Brief guide for beginners,''
Lect. Notes Phys. \textbf{721} (2007), 151-184
[arXiv:hep-th/0601129 [hep-th]].

\bibitem{Ashtekar:2007tv}
A.~Ashtekar,
``An Introduction to Loop Quantum Gravity Through Cosmology,''
Nuovo Cim. B \textbf{122} (2007), 135-155
[arXiv:gr-qc/0702030 [gr-qc]].

\bibitem{Dona:2010hm}
P.~Dona and S.~Speziale,
``Introductory lectures to loop quantum gravity,'' 3rd School on Theoretical Physics ``Gravitation: Theory and Experiment'', Jijel, 2009 [arXiv:1007.0402 [gr-qc]].

\bibitem{Rovelli:2011eq}
C.~Rovelli,
``Zakopane lectures on loop gravity,''
PoS \textbf{QGQGS2011} (2011), 003 \\
$\mbox{[arXiv:1102.3660 [gr-qc]]}$.

\bibitem{Rovelli:2014ssa}
C.~Rovelli and F.~Vidotto,
``Covariant Loop Quantum Gravity,'' Cambridge University Press, 2015.

\bibitem{Sen:1982qb}
A.~Sen,
``Gravity as a Spin System,''
Phys. Lett. B \textbf{119} (1982), 89-91.

\bibitem{Ashtekar:1986yd}
A.~Ashtekar,
``New Variables for Classical and Quantum Gravity,''
Phys. Rev. Lett. \textbf{57} (1986), 2244-2247.

\bibitem{Weinberg:1976xy}
S.~Weinberg,
``Critical Phenomena for Field Theorists,'' in: Zichichi A. (ed.) ``Understanding the Fundamental Constituents of Matter'', The Subnuclear Series, vol 14, Springer, Boston, 1976.

\bibitem{Brink:2015ust}
L.~Brink,
``Maximally supersymmetric Yang-Mills theory: The story of ${\cal N}= 4$ Yang-Mills theory,''
Int. J. Mod. Phys. A \textbf{31} (2016) no.01, 1630002
[arXiv:1511.02971 [hep-th]].

\bibitem{Litim:2014uca}
D.~F.~Litim and F.~Sannino,
``Asymptotic safety guaranteed,''
JHEP \textbf{12} (2014), 178
[arXiv:1406.2337 [hep-th]].

\bibitem{Polchinski:1983gv}
J.~Polchinski,
``Renormalization and Effective Lagrangians,''
Nucl. Phys. B \textbf{231} (1984), 269-295.

\bibitem{Ellwanger:1993mw}
U.~Ellwanger,
``FLow equations for N point functions and bound states,''
Z.~Phys.~C {\bf 62} (1994) 503-510 
[arXiv:hep-ph/9308260 [hep-ph]].

\bibitem{Morris:1993qb}
T.~R.~Morris,
``The Exact renormalization group and approximate solutions,''
Int. J. Mod. Phys. A \textbf{9} (1994), 2411-2450
[arXiv:hep-ph/9308265 [hep-ph]].

\bibitem{Wetterich:1992yh}
C.~Wetterich,
``Exact evolution equation for the effective potential,''
Phys. Lett. B \textbf{301} (1993), 90-94
[arXiv:1710.05815 [hep-th]].

\bibitem{Litim:2008tt}
D.~F.~Litim,
``Fixed Points of Quantum Gravity and the Renormalisation Group,''
PoS \textbf{QG-Ph} (2007), 024
[arXiv:0810.3675 [hep-th]].

\bibitem{Reuter:1996cp}
M.~Reuter,
``Nonperturbative evolution equation for quantum gravity,''
Phys. Rev. D \textbf{57} (1998), 971-985
[arXiv:hep-th/9605030 [hep-th]].

\bibitem{Bonanno:2020bil}
A.~Bonanno, A.~Eichhorn, H.~Gies, J.~M.~Pawlowski, R.~Percacci, M.~Reuter, F.~Saueressig and G.~P.~Vacca,
``Critical reflections on asymptotically safe gravity,'' [arXiv:2004.06810 [gr-qc]].

\bibitem{Reuter:2019byg}
M.~Reuter and F.~Saueressig,
``Quantum Gravity and the Functional Renormalization Group,'' Cambridge University Press, 2019.



\bibitem{Reuter:2012id}
M.~Reuter and F.~Saueressig,
``Quantum Einstein Gravity,''
New J. Phys. \textbf{14} (2012), 055022
[arXiv:1202.2274 [hep-th]].

\bibitem{Niedermaier:2006ns}
M.~Niedermaier,
``The Asymptotic safety scenario in quantum gravity: An Introduction,''
Class. Quant. Grav. \textbf{24} (2007), R171-230
[arXiv:gr-qc/0610018 [gr-qc]].

\bibitem{deAlwis:2019aud}
S.~de Alwis, A.~Eichhorn, A.~Held, J.~M.~Pawlowski, M.~Schiffer and F.~Versteegen,
``Asymptotic safety, string theory and the weak gravity conjecture,''
Phys. Lett. B \textbf{798} (2019), 134991
[arXiv:1907.07894 [hep-th]].

\bibitem{Dvali:2010bf}
G.~Dvali and C.~Gomez,
``Self-Completeness of Einstein Gravity,''
[arXiv:1005.3497 [hep-th]].

\bibitem{Dvali:2010jz}
G.~Dvali, G.~F.~Giudice, C.~Gomez and A.~Kehagias,
``UV-Completion by Classicalization,''
JHEP \textbf{08} (2011), 108
[arXiv:1010.1415 [hep-ph]].

\bibitem{Dvali:2014ila}
G.~Dvali, C.~Gomez, R.~S.~Isermann, D.~L\"ust and S.~Stieberger,
``Black hole formation and classicalization in ultra-Planckian $2\to N$ scattering,''
Nucl. Phys. B \textbf{893} (2015), 187-235
[arXiv:1409.7405 [hep-th]].

\bibitem{Donoghue:2019clr}
J.~F.~Donoghue,
``A Critique of the Asymptotic Safety Program,''
Front. in Phys. \textbf{8} (2020), 56
[arXiv:1911.02967 [hep-th]].

\bibitem{Percacci:2003jz}
R.~Percacci and D.~Perini,
``Asymptotic safety of gravity coupled to matter,''
Phys. Rev. D \textbf{68} (2003), 044018
[arXiv:hep-th/0304222 [hep-th]].

\bibitem{Dona:2013qba}
P.~Dona, A.~Eichhorn and R.~Percacci,
``Matter matters in asymptotically safe quantum gravity,''
Phys. Rev. D \textbf{89} (2014) no.8, 084035
[arXiv:1311.2898 [hep-th]].

\bibitem{Meibohm:2015twa}
J.~Meibohm, J.~M.~Pawlowski and M.~Reichert,
``Asymptotic safety of gravity-matter systems,''
Phys. Rev. D \textbf{93} (2016) no.8, 084035
[arXiv:1510.07018 [hep-th]].

\bibitem{Eichhorn:2017egq}
A.~Eichhorn,
``Status of the asymptotic safety paradigm for quantum gravity and matter,''
Found. Phys. \textbf{48} (2018) no.10, 1407-1429
[arXiv:1709.03696 [gr-qc]].

\bibitem{Hamada:2017rvn}
Y.~Hamada and M.~Yamada,
``Asymptotic safety of higher derivative quantum gravity non-minimally coupled with a matter system,''
JHEP \textbf{08} (2017), 070
[arXiv:1703.09033 [hep-th]].

\bibitem{Christiansen:2017cxa}
N.~Christiansen, D.~F.~Litim, J.~M.~Pawlowski and M.~Reichert,
``Asymptotic safety of gravity with matter,''
Phys. Rev. D \textbf{97} (2018) no.10, 106012
[arXiv:1710.04669 [hep-th]].

\bibitem{Shaposhnikov:2009pv}
M.~Shaposhnikov and C.~Wetterich,
``Asymptotic safety of gravity and the Higgs boson mass,''
Phys. Lett. B \textbf{683} (2010), 196-200
[arXiv:0912.0208 [hep-th]].

\bibitem{Degrassi:2012ry}
G.~Degrassi, S.~Di Vita, J.~Elias-Miro, J.~R.~Espinosa, G.~F.~Giudice, G.~Isidori and A.~Strumia,
``Higgs mass and vacuum stability in the Standard Model at NNLO,''
JHEP \textbf{08} (2012), 098
[arXiv:1205.6497 [hep-ph]].

\bibitem{Wetterich:2019qzx}
C.~Wetterich, ``Quantum scale symmetry,''
[arXiv:1901.04741 [hep-th]].

\end{thebibliography}
\end{document}